# The Lustre Storage Architecture

Peter J. Braam

(with others)

Cluster File Systems, Inc.

http://www.clusterfs.com

09/29/2005, HEAD

# Contents

# Glossary

### 0.1. Lustre Specific Acronyms

**Catalogue:** A directory managed on the MDS containing a list of inodes stored on a logical object volume (LOV).

**CFS:** Client FileSystem - a system running the Lustre or Lustre Lite filesystem client code .

**CFS:** Cluster FileSystems, Inc.

**CFS Meta-data Writeback Cache:** Many local and network filesystems have a cache of file data which applications have *written* but which has not yet been flushed to storage devices. A *meta-data writeback cache* is a cache of meta-data updates (*mkdir*, *create*, *setattr*, other operations) which an application has performed and which have not yet been flushed to a storage device or server. InterMezzo is one of the first *network* filesystems to have a meta-data write back cache.

**COBD:** Collaborative Cache

**Collaborative Cache:** A read cache shared between a subset of the client systems to enable client to client data transfer, enabling enormous scalability benefits for mostly read-only situations.

**Load Balancing MDS:** A system of MDS's that perform load balancing of the requests among the systems .

**Lock Client:** A module making lock RPC's to a lock server and handling revocations from the server .

**Lock Server:** A system managing locks on certain objects. It also issues lock revocation calls while servicing lock requests for already locked objects.

**Lustre:** Name of the project and the ultimate cluster filesystem incorporating clustered meta-data and full security functionality.

**Lustre Lite:** An intermediate milestone on the Lustre road map which by itself will provide a scalable cluster filesystem with high performance I/O.

**MDC:** The meta-data client code module which interacts with the MDS over the network.

**MDD:** Meta-Data device, this will be available in future to have a division of function similar to the OST/OBD pair

**MDS:** Meta-Data Server - a system offering Lustre meta-data services.

**MDS Client:** Same as MDC.

**MDS Server:** Same as MDS.

**MDT:** Meta-Data target

**NIO API Networking Layer:** A library for sending small network requests and for moving large buffers.

**OBDFS:** Object Based File System - a single node filesystem storing data and meta-data on object devices.

**OSC:** Object Storage Client - the client unit talking to an OST.

**OST:** Object Storage Target - a system offering object storage service through an instance of an OBD server on the system.

**RDB:**

**Request Processing Network Layer:** A library issuing and handling network requests, involving suspending processes until timeouts or responses happen.

### 0.2. Filesystem, Clustering, and Networking Terminology

**ACL:** Access Control Lists - an authorization mechanism for access to files and directories that provides fine granularities of control.

**Barriers:** A mechanism to synchronize a group of processes across multiple systems. Typically used to let all members of a group of systems proceed up to a certain point (the barrier) before proceeding with the next phase of recovery.

**CS:** Client-server mode

**CPSD:** Current protection sub domain

**DCache:** The Linux VFS cache of dentry's.

**Dentry:** Linux VFS data structure representing the name of a file and its association to an inode.

**Fileset:** An exported file tree which is treated as an administrative unit within the global namespace. Operations within file sets have stronger transactional guarantees than those that span fileset boundaries.

**Global Namespace:** A mechanism to unify the exported filesystems offered by multiple file servers into a single directory tree. Such mechanisms are offered by AFS, DCE/DFS, Microsoft dfs, Coda, InterMezzo, and NFS v4.

**Lock Revoke:** A message sent by a server to a client to ask a client to release a lock it is holding.

**Membership:** A code module that maintains a list of systems that is actively participating in a group of collaborating systems

**OS bypass file I/O:** A mechanism for a user level application to do I/O without invoking the read/write system calls. The read/write system calls involve a copy of a buffer from kernel to user memory and OS bypass I/O can eliminate this copy. OS bypass I/O can also avoid putting the calling process asleep and instead let it spin on a bit change to detect arrival of the response.

**PAG:** Process authentication groups

**PAM:** Pluggable authentication module

**Portals Networking Layer:** A Sandia designed message passing layer exposing the API, defined on Sandia's website at http://www.cs.sandia.gov/~bright/papers/portals3/portals3.html.

**Portal NAL:** A NAL is a Networking Abstraction Layer. A NAL can be linked with Portals (through a method table) to provide Portals networking support for a particular layer.

**Quorum:** A measurement on membership to determine if the membership is large enough to assume an authoritative group. Lack of quorum can happen for small groups of systems that have become isolated from a cluster.

**QSW:**

**RPC Networking Layer:** The layer in the network stack that implements a request/response model similar to executing procedures in code.

**UDP:** User Datagram Protocol

**VPN:** Virtual private network

**WB:** Writeback

## 0.3. Industry Products and Protocol Acronyms

**ADIO:** A low-level API allowing for an efficient implementation of MPI-I/O over a filesystem. The object storage API is well adapted for this purpose. ADIO MMLustreNameoftheprojectandtheulti //www-unix.mcs.anl.gov/~thakur/adio/paper/paper.html

**ADIO-OBD:** An adaptor.

**AES:**

**ANSI:** American National Standards Institute (ANSI T10 committee group). ANSI http://www.ansi.org

**CIFS:**

**Coda:** A high availability file system developed at Carnegie Mellon University, featuring most of the AFS and adding features related to server replication and disconnected operation. It is very useful to check out the Coda development website at http://www.coda.cs.cmu.edu.

**DAFS:** Direct Access FileSystem. DAFS http://www.dafscollaborative.org

**DAFS-style Request Format:** DAFSAPI http://www.dafscollaborative.org/tools/dafsapi.pdf

**DHCP:** Dynamic Host Configuration Protocol DHCP http://www.dhcp.org

**Ensemble:** A project at Cornell that offers a group communication infrastructure that includes membership API's and could form the basis of a clustering infrastructure. Ensemble http://www.cs.cornell.edu/Info/Projects/Ensemble

**Event Delivery:** Recovery manager, see Kimberlite Kimberlite http://oss.missioncriticallinux.com/projects/kimberlite

**FSPDB:** File system protection database

**IBM DLM:** DLM stands for Distributed Lock Manager. IBM-DLM http://oss.software.ibm.com/dlm

**IETF:** Internet Engineering Task Force - an IP storage working group. IETF http://www.ietf.org

**Infiniband:** Infiniband http://www.intel.com/technology/infiniband/index.htm

**InterMezzo:** A high availability file system with 100% write back caching for metadata updates and disconnected operation. See the Intermezzo website at http://www.intermezzo.org.

**ISCSI:** Internet SCSI iSCSI http://www.snia.org/tech_activities/ip_storage/iscsi

**KDC:**

**Kerberos X509:** Kerberos http://www.ietf.org/rfc/rfc1510.txt

**Kimberlite:** Kimberlite http://oss.missioncriticallinux.com/projects/kimberlite

**LDAP:** Lightweight Directory Access Protocol. See http://www.openldap.org.

**LDAP/DCE Integration:**

**Linux Extended Attributes:** Application- or kernel-managed persistent storage of inode meta-data.

**LIPKEY:**

**Lock Conversion:**

**LVS:** Linux Virtual Services - an infrastructure to manage load balancing and redirection of services over IP networks, widely used for building clusters of http servers. LVS http://www.linuxvirtualserver.org

**Meta-data Update Records:**

**Meta-data Updates:**

**NAS:** Network Attached Storage.NAS http://www.sun.com/storage/white-papers/nas.html

**NASD:** Network Attached Secure Disk NASD http://www.pdl.cmu.edu/NASD/

**NASD-style Capability:**

**NFSv4:** Network FileSystem NFSv4 http:/www.ietf.org/rfc/rfc3010.txt

**NSS:**

**NIS:**

**Operation-based Locking:**

**OSD Technical work group:** An SNIA work group that works closely with ANSI T10/OSD committee to define the standards for Object Based Storage Devices OSD http://www.snia.org/tech_activities/workgroups/osd

**Page-based Directory Data:**

**RDMA:** Remote DMA.

**Service Guard:** A Hewlett Packard product with similar functionality as Kimberlite. XXX URL

**SNIA:** Storage Network Industry Association SNIA http://www.snia.org

**Tree/Hashed Directories:** Directories in which the layout of data follows that of a tree or hash, to enable a fast lookup of entries by name even when the directory is fast. Hashed directories are offered by advanced file systems, such as ReiserFS, XFS, JFS, and Ext3.

**TUX:** A high performance extremely scalable Linux web server that runs in kernel space. It is tightly integrated with the IP stack to enable performance. The optimizations used by TUX might be useful to Lustre. TUX http://www.redhat.com/docs/manuals/tux/

**VI/VIA:**

**WARP:**

**XDSM:** Was DMAPI. XDSM http://openxdm.sourceforge.net/xdsm/index.html

**Lock (Pinning, Releasing, Acquiring, Revoking):**
**Enhancement to Linux VFS for Locking:** VFSlock http://www.lustre.org/docs/fslocking.pdf
**Filtering MDS System:**
**Mission Critical Linux Kimberlite:** Kimberlite http://oss.missioncriticallinux.com/projects/kimberlite/
**Kimberlite Failover:**
**Cross-realm :**
**TCP ULP Framing:**
**Membership and Quorum:**

CHAPTER 1

# Foreword

This document has seen the contributions of many people. Many people have contributed substantially: Andreas Dilger co-authored the object API, Terry Heidelberg has provided a lot of timely feedback and guidance. Phil Schwan, Alex Tomas, M. Satynarayanan, Lee Ward, Stephen Tweedie, Al Viro, Ted T'so, Chris Malakappalli, Michael Callahan, Ron Minnich, Zach Brown, Rumi Zahir, and Daniel Phillips have been excellent discussion partners for years on getting Lustre to reality. I have probably forgotten others that also seriously contributed to this project.

Software doesn't reside in papers - the implementation of Lustre has greatly benefitted with funding from Seagate, the National Laboratories, and some unnamed industrial visionairies. Much unpaid time and fun was contributed to the project by Cluster File Systems and Stelias Computing. The implementation so far has been done by first rate software people; primarily Andreas Dilger and Phil Schwan, who did the bulk of the work with me. We hope to see many other programmers, voluntary or paid, contribute to this project. And without Amanda Coe this document would have no figures and the Cluster File Systems office would have collapsed into chaos.

# Part 1

# Architecture

CHAPTER 2

# Introduction

Lustre provides a novel modular storage framework including a variety of storage management capabilities, networking, locking, and mass storage targets all aiming to support scalable cluster filesystems for small to very large clusters.

The name Lustre embodies "Linux" and "Cluster". Lustre focuses on scalability for use in large computer clusters, but can equally well serve smaller commercial environments through minor variations in the implementation and deployment of the modules that make up the system. Lustre runs over different networks, including at present Ethernet and Quadrics.

Lustre originated from research done in the Coda project at Carnegie Mellon. It has seen interest from several companies in the storage industry that have contributed to the design and funded some implementation. Soon after the original ideas came out, the USA National Laboratories and the DOD started to explore Lustre as a potential next generation filesystem. During this stage of the project we received a lot of help and insight from the Los Alamos and Sandia National Laboratories, most significantly from Lee Ward.

Lustre provides many new features and embodies significant complexity. In order to reach a usable intermediate target soon, Mark Seager from Lawrence Livermore pushed forward with Lustre Lite. Lustre Lite is an intermediate point on the Lustre roadmap which should be fully usable by itself.

## 2.1. Lustre Components

In Lustre clusters there are three major types of systems: the Clients, the Object Storage Targets (OST), and Meta-Data Server (MDS) systems. Each of the systems internally has a very modular layout. For many modules, such as locking, the request processing and message passing layers are shared between all systems and form a part of the framework. Others are unique, such as the Lustre Lite client module on the client systems. Figure 2.1.1 gives an introductory impression of the interactions that are to be expected.

Lustre `http://www.lustre.org` provides a clustered filesystem which combines features from scalable distributed filesystems such as AFS [12], Coda [16], InterMezzo `http://www.inter-mezzo.org`, and Locus CFS [19], with ideas derived from traditional shared storage cluster filesystems like Zebra [11], Berkeley XFS which evolved into Frangipani Petal [23], GPFS [20], Calypso [8], InfiniFile [21], and GFS [22]. Lustre clients run the Lustre filesystem and interact with Object Storage Targets (OST's) for file data I/O and with Meta-Data Servers (MDS) for namespace operations.

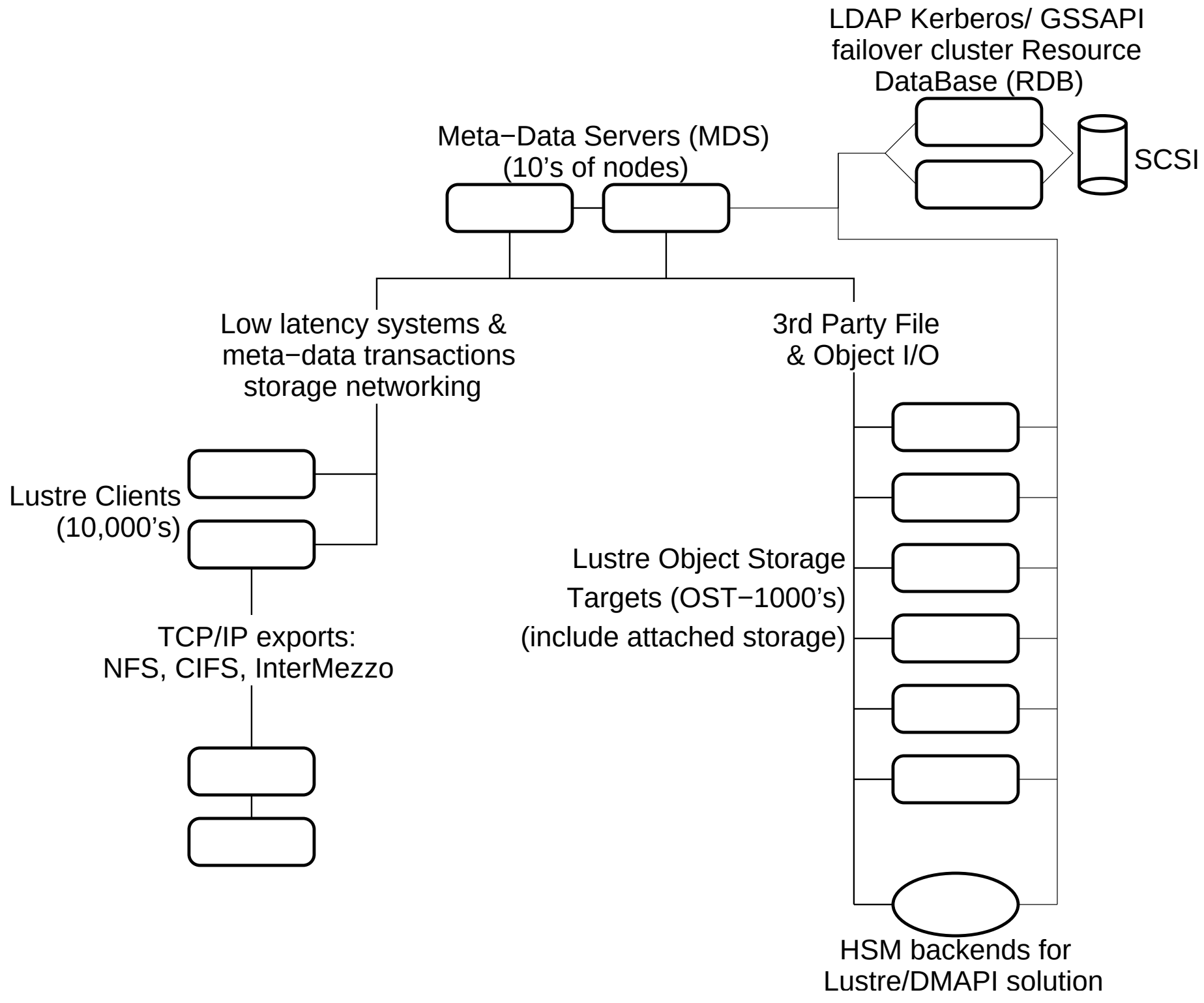

FIGURE 2.1.1. A Lustre Cluster

When Client, OST, and MDS systems are separate, Lustre appears similar to a cluster filesystem with a file manager, but these subsystems can also all run on the same system, leading to a symmetrical layout. The main protocols are described in figure 2.1.2.

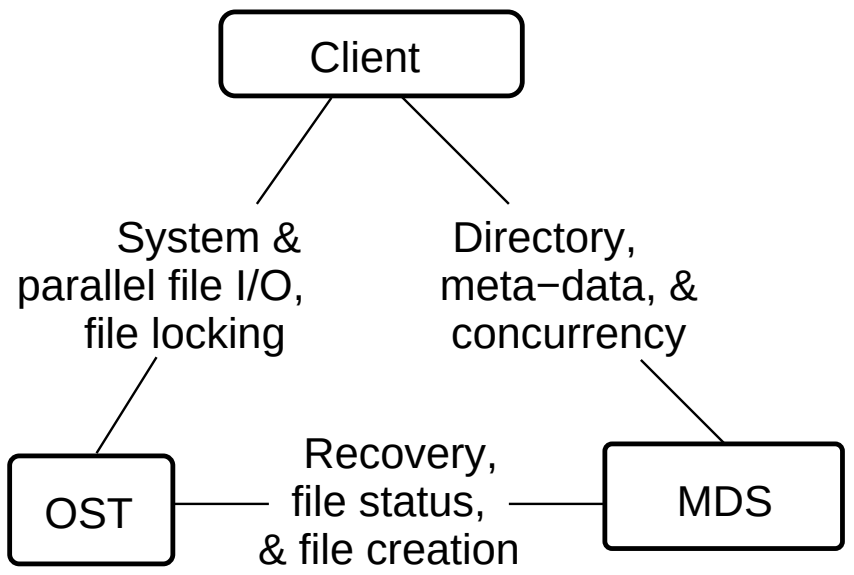

FIGURE 2.1.2. Outline of Interactions Between Systems

## 2.2. Object Storage Targets

At the root of Lustre is the concept of **object storage**. Objects can be thought of as inodes and are used to store file data. Access to these objects is furnished by OST's which provide the file I/O service in a Lustre cluster. The namespace is managed by meta-data services which manage the Lustre inodes. Such inodes can be directories, symbolic links, or special devices in which case the associated data and meta-data is stored on the meta-data servers. When a Lustre inode represents a file, the meta-data merely holds references to the file data objects stored on the OST's.

Fundamental in Lustre's design is that the OST's perform the block allocation for data objects, leading to distributed and scalable allocation meta-data. The OST's also enforce security regarding client access to objects. The client-OST protocol bears some similarity to systems like DAFS in that it combines request processing with remote DMA. The software modules in the OST's are indicated in figure 2.2.1. Object storage targets provide a networked interface to other object storage. This second layer of object storage, so-called direct object storage drivers, consists of drivers that manage objects, which can be thought of as files, on persistent storage devices. There are many choices for direct drivers which are often interchangeable. Objects can be stored as raw ext2 inodes by the *\emph{obdext2}* driver, or as files in many journal filesystems by the filtering driver, which is now the standard driver for Lustre Lite. More exotic compositions of subsystems are possible. For example, in some situations an OBD filter direct driver can run on an NFS filesystem (a single NFS client is all that is supported).

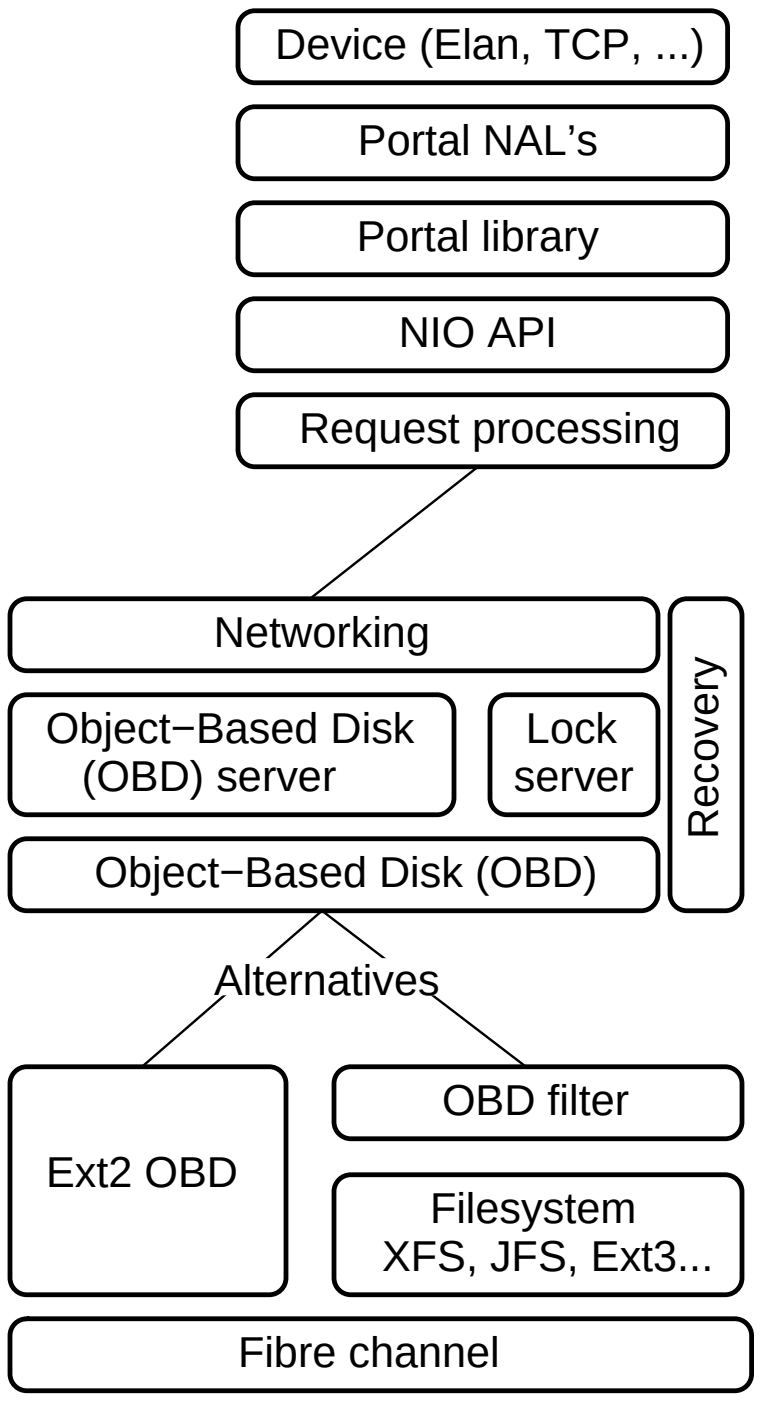

FIGURE 2.2.1. Object Storage Target (OST) Software Modules

In the OST figure we have expanded the networking into its subcomponents. Lustre request processing is built on a thin API, called the Portals API, which was developed at Sandia. Portals inter-operates with a variety of network transports through Network Abstraction Layers (NAL). This API provides for the delivery and event generation in connection with network messages and provides advanced capabilities such as using Remote DMA (RDMA) if the underlying network transport layer supports this.

### 2.3. Meta-data Service

The **meta-data servers** are perhaps the most complex subsystem. They provide backend storage for the meta-data service and update this transactionally over a network interface. This storage presently uses a journal filesystem, but other options such as shared object storage will be considered as well. Figure 2.3.1 illustrates this function.

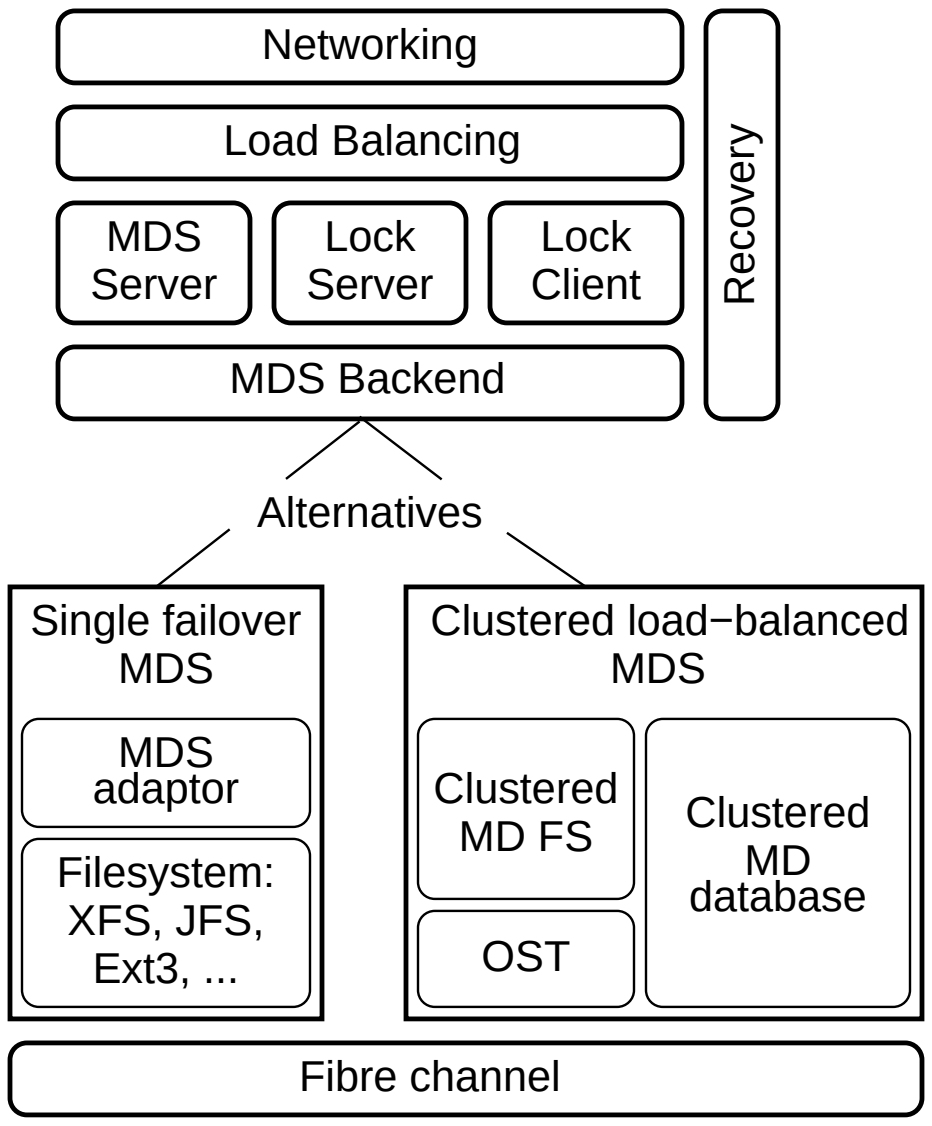

FIGURE 2.3.1. Meta-data Server Software Modules

The MDS's contain locking modules and heavily exercise the existing features of journal filesystems, such as Ext3 or XFS. In Lustre Lite the complexity is limited as just one single meta-data server is present. The system still avoids single points of failure by offering failover meta-data services, based on existing solutions such as Kimberlite.

In the full Lustre system meta-data processing will be load balanced, which leads to significant complexity related to the concurrent access to persistent meta-data.

The meta-data handling in Lustre will evolve to a well balanced API. While remotely offered meta-data services will remain most common, we will incorporate a stackable framework where logical meta-data services can be offered, which, for example, manage the dispatch of meta-data requests to replicated meta-data servers. Similarly, a logical driver can manage the dispatch to a local persistent cache for meta-data in conjunction with the dispatch to a remote service, much as is done in the Andrew FileSystem. How soon implementations of these will appear will depend on demand for them, but the framework will support these extensions in the near future.

## 2.4. Client Filesystem

The **client** meta-data protocols are transaction-based and derive from the AFS, Coda, and InterMezzo filesystems. The protocol features authenticated access and writebehind caching for all meta-data updates while maintaining strict meta-data and data coherency. Data coherency can optionally be relaxed when applications manage coherency itself. The client again has multiple software modules as shown in figure 2.4.1.

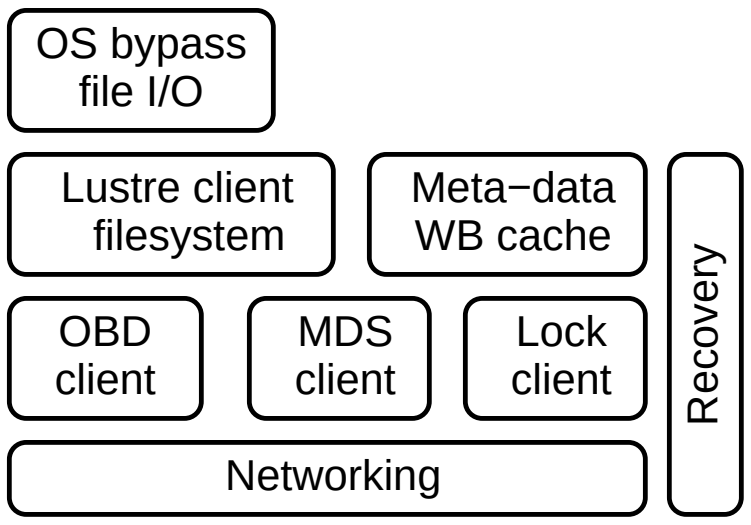

FIGURE 2.4.1. Client Software Modules

Lustre can provide UNIX semantics for file updates. Lock management in Lustre supports coarse granularity locks for entire files and subtrees when contention is low, as well as finer granularity locks. Finer granularity locks appear for extents in files and as pathname locks to enable scalable access to the root directory of the filesystem. All subsystems running on Lustre clients can transparently fail over to other services.

The Lustre and Lustre Lite filesystems provide explicit mechanisms for advanced capabilities such as scalable allocation algorithms, security, and meta-data control. In traditional cluster filesystems such as IBM's GPFS, many similar mechanisms are found but are not independent abstractions, instead being part of a large monolithic filesystem.

## 2.5. Storage Management

Lustre provides numerous ways of handling storage management functions, such as data migration, snapshots, enhanced security, and quite advanced functions such as active disk components for data mining. Such storage management is achieved through stacks of object modules interacting with each other. A general framework is provided for managing and dynamically changing the driver stacks.

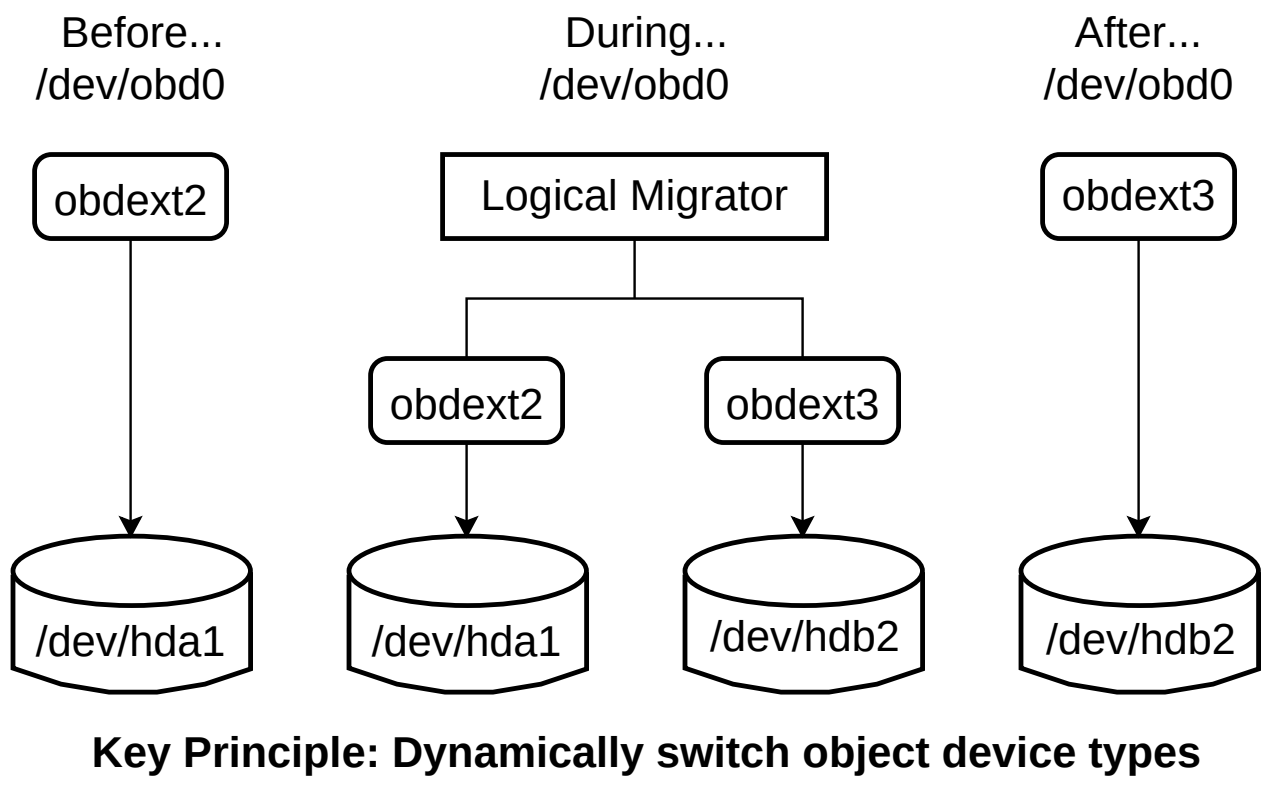

FIGURE 2.5.1. Hot Data Migration

An example of stacking object modules is shown in figure 2.5.1 for the case of hot data migration from one storage target to another.

### 2.6. Lustre and SAN's

Lustre employs two kinds of backend storage: for data and for meta-data. The preferred protocols to address the storage is through the meta-data services and object storage targets as particularly the latter provide extra security features not found on traditional SAN running the SCSI protocol over fibre channel.

However, Lustre incorporates extensive SAN compatibility. First there is a client target pair called the **SANOSC** and **SANOST**, which allow the object storage targets to use a control network for parts of the commands and use a traditional SAN to allow for direct data transfer from clients to block storage. The object storage target would continue to manage security and allocation for requests but delegates only the data transfers for objects to the SAN.

Secondly, there is an integrated meta-data server and OST, the **OSTMDS** module, which exports the meta-data and object storage API using a single backend store. This store can be an ordinary Linux ext3 filesystem and this solution allows the clustering of commodity Linux boxes without any reformatting of filesystems.

Combining the SANOST and OSTMDS leads to a first class cluster filesystem that can utilize SAN's and share Linux filesystems. This solution will lack the scalability of Lustre Lite and Lustre, but might have significant relevance for commercial clusters.

## 2.7. The Lustre Framework

As we have seen above Lustre embodies a modular architecture, and the flexibility afforded by stacking and using storage modules is important. This beckons the question of what is part of the framework, i.e. the more static environment in which the modules reside, and what classifies as a modular, pluggable component.

## 2.8. Changelog

**Version 5.0 (Dec. 2002)**

(1) P.D. Innes - updated & resized figures

**Version 4.0 (Nov. 2002)**

(1) P.D. Innes - updated text, URL's added

**Version 3.0 (Jun. 2002)**

(1) P.D. Innes - figure float insertion, labelling, and cross-referencing, proofed, and edited

**Version 2.0**

(1) P. Braam - rewrite
(2) A. Coe - figure insertion

**Version 1.0**

(1) P. Braam - original draft

CHAPTER 3

# Global Namespaces for Filesystems - with Lee Ward

A global namespace for a distributed filesystem provides a globally valid single directory tree to all clients of the filesystem. Typically the tree contains folder collections, also known as filesets or volumes, from many different servers which are grafted into the namespace. An important characteristic of such trees is that they are self-describing, i.e. they require little or no configuration information on the clients. We describe how global namespaces require mount-point objects stored in the file system. We address the path name traversal of mount-point objects and compare approaches from AFS/DCE DFS and Coda to automounting filesystems such as **autofs**, which are based on directory information. Two further cases which require special attention are persistent, filesystem-based caches for filesets on clients and their relation to the global namespaces. Finally, when global filesystems are accessed on the servers where some of the filesets are stored in local filesystems, their interaction with the namespace is of particular interest. We describe how InterMezzo and Lustre use a Linux namespace module implemented by the author to provide the required features.

## 3.1. Introduction

A global namespace for a filesystem unifies the directory trees available from multiple file servers. This paper describes a global namespace system which overcomes some difficulties with existing systems. The intent is to eliminate configuration data on clients and to provide a directory tree which is valid from every workstation in the installation and remains valid when configurations are updated.

Global namespaces were perhaps introduced first in the Andrew FileSystem (AFS) and have since been used in many different forms in other environments. In AFS, the global namespace provides access to filesets (called volumes in AFS lingo) and typically each user has a unique fileset as a home directory as well as possible filesets associated with projects. An AFS namespace can unify 10,000's of filesets, distributed over different administrative AFS domains called cells. The traversal of a fileset mount-point takes place within a Unix mounted AFS filesystem. AFS does not rely on configuration data on the client to traverse mount-points, but contacts a server to obtain fileset location data. Coda, DCE/DFS, and Microsoft DFS use similar mechanisms.

Another mechanism for providing global namespaces is provided by the automounter family (**amd** and **autofs**). These provide magical file systems that detect the traversal of mount-points. When the traversal takes place, a mount map is consulted and a Unix mount is executed to bring the new filesystem into the directory tree. Mount maps can be held in configuration files or stored in NIS or LDAP directory services. The automounter offers global namespaces for heterogeneous filesystems, e.g. it can combine NFS and local filesystem mounts. The initial **amd** automount daemons were special NFS servers, but with Solaris 2.6, SUN introduced an **autofs** filesystem and an advanced automount daemon.

The purpose of this paper is to describe a global namespace design that is slightly more general. It combines the advantages of the AFS system with the newer features offered by the automounter.

The implementation details for global namespaces are in fact somewhat involved. We indicate a solution somewhat different from existing implementations and provide a roadmap incorporation of the global namespaces in Linux environments building on local filesystems, NFS, InterMezzo, and Lustre.

**Acknowledgment:** This paper was written with support from Tacitus Systems, Inc.

### 3.2. Fundamental Definitions and Requirements

Unix filesystems can be mounted on directories. There are no special requirements on directories acting as mount-points; they can be empty, non-empty, and can have any name. The *mount* command parses the command line options or `/etc/fstab` file and issues a system call which provides the kernel with the device, filesystem driver, and other parameters required to mount a filesystem on a directory. The root directory of the filesystem covers the mount-point directory when the mount has taken place. In the kernel the virtual filesystem is responsible for handling the traversal of mount-points. From this we see that (1) there is no mount-object in a Unix filesystem, they are described in `/etc/fstab`, and (2) there is no automatic mounting on directories when mount-points are entered.

Mounting filesystems is done by covering the so-called mount-point with the root directory of the new filesystem. Inside the VFS a special mechanism, typically a *vfs_mount* structure, is used to link the covered mount-point directory to the root directory of the mounted filesystem. It is important to note that typically the name of the root of a filesystem is inherited from the parent filesystem, while the inode belongs to the mounted filesystem. Special handling of mount-points happens, for example, when `..` is encountered in a pathname, at a filesystem root.

The Andrew FileSystem contains mount-objects which it represents as symbolic links whose data contains the mounting information. While one can argue that a separate object to represent mount-points might be more natural, there are significant advantages in making mount-objects ordinary filesystem objects. Such objects can be created, modified, and removed easily without changes to system software, even when the filesystem is accessed through NFS or SMB servers. While AFS, Coda, and DCE/DFS store mount-points in symbolic links, they nevertheless manipulate such objects with a special *ioctl* on the filesystem.

The simplest implementation of a global namespace would introduce mount-objects that carry the name of the mount-point and mount filesets or filesystems on these objects when a lookup operation on the object is performed. This turns out to be inefficient. In many cases a lookup is only performed for getting the attributes of the object. If a directory such as *home* were to contain thousands of mount-objects for users home directories, a mount storm would ensue. This is a deficiency of the AFS system. As we will see, recent versions of autofs have overcome this problem.

In order to build flexible global namespaces the following requirements appear reasonable:

**Fileset and filesystem support:** The global namespace should be able to provide a global namespace which includes Unix filesystem mount-points and fileset mount-points. The ability to mount filesets and filesystems in multiple locations is desirable.

**Local and networked configuration data:** Configuration data should be fetched from network servers, but the system should also be able to use locally stored data.

**Activation, caching, and releasing:** Mount-points should be cached objects which can be released after having been idle. Retry storms after failures and cluster state changes must be avoided by limiting the retry frequency. Invalidations of mount-points should be possible on a timeout or callback basis. Mounting a mount-object should not be triggered merely by lookups of the mount-point itself.

**Mount-objects:** Mount-objects should preserve full Unix semantics, i.e. they are directories and contribute to the link count of the parent directories. Invisible mount-objects, such as indirect mount maps, are to be avoided. Traversal of symbolic links to reach the roots of filesets and filesystems is to be avoided. Handling of mount-objects in a filesystem should be a mount option for that filesystem. Preferably, mount-objects should be ordinary filesystem objects which can be created without special administrative tools.

**Mount-object manipulation:** Mount-objects should be filesystem objects that can be created through standard API's.

**Authorization:** Mounted mount-objects should be protected by their access control list or another filesystem protection mechanism. The process of mounting a mount-object should require authentication and authorization, using the identity and authentication of the process which triggers the mount.

**Update handling:** Modifications to mount-objects should trigger events within a reasonable amount of time on all systems that have the mount-object.

### 3.3. The Andrew FileSystem and Autofs

In this section we will review the functionality of the **AFS** fileset handling and the transient features of **autofs**. Both the autofs/automounter and the AFS global namespaces provide important insights and mechanisms which we exploit in our system.

**Autofs** is a filesystem which transparently mounts filesystems. There are implementations for Linux and Solaris (and likely for other systems), and the Solaris implementation is at present more fully featured. A good overview of this system is found in [17].

Autofs is a filesystem that contains mount-objects. It intercepts lookup calls on directories and interacts with a user level automount daemon to automatically mount filesystems. Earlier versions of autofs intercepted the VFS issued *lookup* calls for the mount-point itself. Prior to the return of the correct inode to the VFS layer, autofs requested the automount daemon to mount the directory.

The automount daemon uses configuration files or network databases then locates the mount information associated with the directory name which autofs presents. Next it creates the directory under the autofs mount-point, if it doesn't exist yet, and executes the mount command for that directory. Now it returns control to autofs, which gives the mount-point inode to the VFS. Autofs can be used to build global namespaces by incorporating symbolic links into the filesystem which point to directories under an autofs mount-point. Upon traversal such symbolic links transparently lead into a mounted filesystem. Such situations are referred to as indirect maps and are suitable for mounting, for example, a family of home directories. Autofs can also be attached to a single existing directory, which it mounts upon traversal. Autofs is capable of caching mounted filesystems, and releasing filesystems that have not been used for a given amount of time.

In recent versions, several refinements were made that are important. First, in the case of indirect maps, autofs can execute a *readdir* operation, based on the entries in the mount map. In this way, directory listings appear normally, while in early versions only the already mounted entries would appear. Early versions of autofs mounted the filesystem when the mount-point was looked up. This gave rise to mount storms in common situations (such as doing *ls -l* under an autofs mounted directory). This was solved by only invoking the mount when a lookup is done for an entry in the mounted directory, or when the mounted directory is opened with an *opendir* system call. Finally, autofs gained support for dealing with trees of mount-points and also performing multiple mounts in parallel.

Autofs still has a few shortcomings:

- The autofs solution introduces a separate filesystem type, which has to be mounted for mount-objects to become available.
- The objects that need mounting under autofs are not self-describing but, instead, need mount maps.

**AFS** filesets, also known as volumes or folder collections, are distinguished subsets of the Unix mounted AFS filesystem. A fileset has a root directory and contains a standard directory tree, augmented with mount-point objects. Some limitations are present when AFS filesystem operations traverse filesets: hard links cannot span multiple filesets and rename operations can only work within filesets. The mount-point objects are represented as symbolic links which point to names that are recognized by AFS as an identifier of another fileset. When such objects are traversed, AFS contacts a Volume Location DataBase (VLDB) and finds the server of the fileset in question. The root directory of the fileset undergoes a *lookup* operation and the new fileset is grafted into the namespace. To avoid administrative difficulties, operations involving mount-objects, such as creation and removal, have to be done using *ioctl* commands issued by a filesystem utility. Coda and DCE/DFS filesets are handled quite similarly.

A few shortcomings of the AFS mount-objects are:

- AFS cannot graft other filesystems into its namespace.
- AFS mount-objects are subject to (fileset) mount storms.
- AFS mount-objects undergo complicated transitions from symbolic links to directories at mount time and then back again at unmount.

- The link count of the parent directory of mount-points is frequently not equal to the number of subdirectories plus two.

We will discuss the requirements for global namespaces in more detail, heading for a system that combines some of the benefits of the AFS and autofs solutions.

### 3.4. Implementing and Handling Mount-Objects

The mount-objects stored in the filesystem provide the glue to traverse from one fileset into another and from one filesystem into another inside a global namespace. When the objects are **entered**, which we define to be an *opendir* or *lookup* operation **inside** the directory, the namespace implementation must turn the mount-object into a local mount, proper. The local mount continues to be monitored by an automount daemon.

We will break with the AFS tradition (now also in use in Microsoft's DFS) to use symbolic links to represent mount-objects by exposing them as directories instead. By using directories, we avoid the problems of having to change object types when mount-objects become mounted and we correctly maintain the link count of the directory containing a mount-object.

As the global namespace handling involves overhead, the direct and simple representation of mount-objects as a directory is not particularly efficient: there are many directories in typical filesystems and usually directories will not represent a mount-object. The mount-object, cum directory, must carry extra information within it's own attribute set which is not commonly used, and provides a hint that the object **might** be a mount-object.

We will use directories which have the *setuid* mode bit set (which is not interpreted by the kernel) and store the mount information in a file named *mntinfo* inside such a directory. The file contains sufficient information for a user level mount daemon to perform a full mount of the fileset or filesystem. As mount directives we propose that the format of this file be:

**For filesets:** *fileset://fileset_name[@cell.domain.name][/[...sub-directory]]*
**For filesystems:** *unix://mount command arguments*

In either case, it is important that the effects of mount options can be exploited in two ways:

(1) Making them part of the `mntinfo` file.
(2) Making them part of site-specific default options for particular workstations.

For instance, in the case of user generated mount-objects, honoring the *setuid* bit when present on an executable file from an untrusted host is probably not desired.

An implementation should not casually attempt to reuse successful mounts for two distinct mount-objects. Credentials of the process driving the operations may be different or the two mount-objects may request dissimilar options.

In order to activate mount-object handling, the filesystem containing the mount-objects needs a mount option which indicates that mount-objects must be honored. One implementation of such an option is to mount the filesystem under a global namespace filter driver. The filter driver can install new filesystem methods that intercept those operations that are related to global namespaces. Alternatively, the VFS can be modified to contain mount-object handling, but we prefer to impose no or minimal changes on the VFS for portability.

If we rely on the VFS to intercept the mount-objects we would mount a filesystem with:

`mount -o mountobjs,otheropts,otherflags` device directory

If one uses a filesystem filter, the mount command which would activate mount-object handling in a filesystem would look like:

`mount -t gnsfs -o bottom_fs=somefs,otheropts` device directory

The following summarizes the transient features of the mount-object handling:

1. **Detect mount-objects:** The filtering layer's *lookup* operation should call *lookup* in the filtered filesystem and install special handling if the inode found is a directory with *setuid* mode bit set.
2. ***Lookup* and *opendir* for mount-object:** The mount-object itself should have a new *lookup* and *opendir* method, both of which should trigger a mount before handing over to the root directory of the newly mounted filesystem.
3. ***lookup* and *opendir*:** This should make sure the filtered *lookup* and *opendir* methods for the newly mounted filesystem get executed when the automounter completes its work.

   Before the mount can be executed, the filesystem will have to read the *mntinfo* file in the mount-object. This is passed to the automount daemon to process the mount command. Also passed is the identity of the user triggering the mount-event. The mount commands executed by the mount daemon should allow for considerable flexibility to allow for the case of filesets, namespace bindings, and simpler mounts. Also important is that, for example in NFS, the mount commands should be the fully enabled mount commands which can establish connections to servers.

   We will use the infrastructure found in autofs to keep track of how long mount-objects have been unused and mounted, and unmount them appropriately.
4. **VFS fileset mount support:** Fileset mounts should be arranged without a new superblock for each mount. This may require support from the VFS as these are not namespace bindings, but submounts that introduce a fileset root inode into the directory tree.

An important distinction between this solution and the AFS fileset mounting is that this system can handle mount-objects in any filesystem. For example, it is possible to graft a local `/tmp` filesystem into an NFS tree and have a global `/tmp/global` directory NFS mounted underneath. A distinction between the autofs solution and this system is that the mount-objects are part of the underlying filesystem, not part of an autofs filesystem. As well, no map files are needed to identify the mount-objects and no external configuration data is needed to drive the process.

A final issue is that of invalidations of cache mount-point objects. Invalidations arising from modifications of the server directory shown under a mount-point are part of the standard update propagation from servers to clients, available in a clustered infrastructure. It is necessary to intercept such updates, even when a filesystem or fileset is mounted on top of the mount-object; this might well require some special filesystem methods.

The delicate issue here is what to do if the mount-object is changed. Once Unix processes start executing filesystem operations in a mounted fileset or filesystem, it is not possible to unmount that fileset. There are two proposals on how to handle mount-object updates:

(1) The filesystem or fileset is only unmounted when it is not busy.
(2) An aggressive solution could turn all inodes below a mount-point into bad inodes. This will likely lead to processes releasing the inodes sooner. The new information in the `mntinfo` file under the mount-object should be used to mount the new fileset or filesystem as soon as possible.

### 3.5. Namespace Bindings

The mechanisms outlined in the previous section show how to graft filesets and filesystem mount-points into a global namespace for a filesystem. We must still discuss what is needed to mount filesets and filesystems at multiple points. On file servers and on clients with persistent caches, different problems surface. The key concept to solving these issues is a binding in a namespace.

Linux 2.4 saw the introduction of the –bind option on the mount command. It allows a directory mounted in a filesystem to be re-mounted in another location. The implementation is straightforward but subtle:

- The dentry defining the filesystem name must be that of the mount-point.
- The inode defining the root directory must be the root of the namespace that is bound to the new location.

Normally dentries point to their inodes, and at the junctions in the namespace provided by bindings a vfs mount data structure is used to provide the namespace traversal. This data structure links the mount-point and the root of the new tree.

Using bindings, mounting filesets and filesystems at multiple points is not a problem.

The server considerations are best treated with an example. Consider an NFS file server augmented with *lookup* operations to support mount-objects. NFS exports can be made part of a global namespace using the mount-objects. On the file server for an export it is possible to mount the global namespace containing the NFS fileset, but on this server it is less optimal than using the local filesystem containing the NFS export. There are two issues that require attention.

First, we need OS support to mount the root of the NFS export into the global namespace. Linux 2.4 provides support for this with the –bind option on the mount command.

Second, it is important that the mount-objects stored in the NFS export receive different treatment when traversed within the global namespace than in the NFS export, where they are merely directories. This is solved by using the global namespace option at mount time. Without the option, mount-objects are passed through and remain simple directories, to be interpreted on the client as appropriate. With the option set, the file server has the opportunity to perform the mount and thus expose the fileset tree to NFS clients that cannot or should not do the job themselves.

On filesystem clients with persistent caches, similar considerations apply. The caches belonging to different filesets are most easily maintained in their own trees, but require visibility inside the global namespace too. These problems are solved in an analogous manner, outlined below.

### 3.6. The Fileset Location Database and the Automounter

The Linux autofs infrastructure provides the necessary enhancements to handle mount events triggered by *lookup* and *opendir* operations on mount-objects. When the corresponding requests reach the automount daemon the following should happen:

The daemon locates the necessary fileset information, based on the information found in the `mntinfo` file. The fileset mount information will in some cases be obtained from the server and may contain more detailed information than what is found in the `mntinfo` file. For example, a system may wish to have the flexibility to store the identity of the server for a fileset in a networked database and not in the `mntinfo` file. In other cases the information found in `mntinfo` may be sufficient for the automount daemon to directly execute its command.

The daemon locates and applies pertinent local information, allowing the system administrator to preclude access to other hosts or networks and to limit or override the user specified mount options.

In the case where the automounter needs to find information on a server, this mechanism could be implemented as an executable autofs/automount map. We expect to support LDAP-based mount maps in this fashion.

The daemon next executes the mount command using appropriate credentials. The credentials used may come from the process driving the operation, optional local information, or, in an advanced implementation, from directory meta-data when and where supported.

Expiries and invalidations are initiated by the daemon but performed by the kernel, through *ioctls*, as in autofs.

### 3.7. The Global Namespace of the InterMezzo Filesystem

InterMezzo is a distributed filesystem with a global namespace. InterMezzo wraps around disk filesystems which it uses both as server-based persistent storage and as a persistent cache on the client. InterMezzo refers to such storage space as a *cache.* Caches can contain full replicas of filesets, as happens on a server of a fileset, or can contain partial caches of filesets, typical in the case of clients in installations with a large amount of servers.

The global namespace solution for InterMezzo needs to address three issues:

(1) Combine filesets from multiple servers into a single directory tree for clients, by exploiting mount-points.
(2) On an InterMezzo server for a fileset, combine the global namespace of filesets from other servers with the locally stored filesets. Avoid installing an unnecessary cache for the filesets located on this server.
(3) If a system is a client for a fileset, provide a directory tree for that fileset in the global namespace, as well as a directory tree outside of the global namespace which is not subject to automatic mounting of mount-objects.

InterMezzo in fact provides further flexibility by exporting multiple namespaces.

As alluded to above, there are two filesystem layouts of an InterMezzo cache. We refer to the layout provided by the disk filesystem as the physical layout and to that provided by the global namespace mechanism as the logical layout.

The physical layout is normally only used by InterMezzo daemons, such as cache purging daemons, but is in principle accessible to normal users of the filesystem as well. In the physical layout, the mount-points are not covered and receive updates normally. Updates to these objects should trigger messages to the mount daemon to adjust mount situations accordingly, using the filter mechanisms.

InterMezzo uses the namespace bindings discussed earlier to provide the physical and logical views of the cache. We'll now show how the mechanisms discussed are used to provide a global namespace on clients. Figures 3.7.1 & 3.7.2 show the layout of the physical and logical namespaces, as well as the mount commands issued to achieve these views of the filesystem.

The following output from the mount command shows how the /bind/ option on the Linux 2.4 mount command has been used to build this namespace:

```
$> mount
....
/dev/sda1 on /cache type intermezzo (rw)
/cache/cell1/fset1/ROOT on /izo type none (rw,bind)
/cache/cell1/fset2/ROOT on /izo/dir/fset2-mtpt type none(rw,bind)
/cache/cell2/fset1/ROOT on /izo/dir/fset1@cell2-mtpt type none(rw,bind)
```

On servers we face a somewhat different situation. Servers provide filesets from other servers, which are usually partial caches, but have InterMezzo caches which are full replicas for locally stored filesets. For example, consider the server of `fset1@cell2` in figure 3.7.1.

The output from the mount command shows the mount arrangement:

```
$> mount
/dev/sda1 on /cache type intermezzo (rw)
/dev/sda2 on /cache2 type intermezzo (rw)
/cache/cell1/fset1/ROOT on /izo type none (rw,bind)
/cache2/cell2/fset1/ROOT on /izo/dir/fset1@cell2-mtpt type none (rw,bind)
/cache/cell1/fset2/ROOT on /izo/dir/fset1@cell2-mtpt/dirdir/fset2@cell1-mtpt type
none (rw,bind)
```

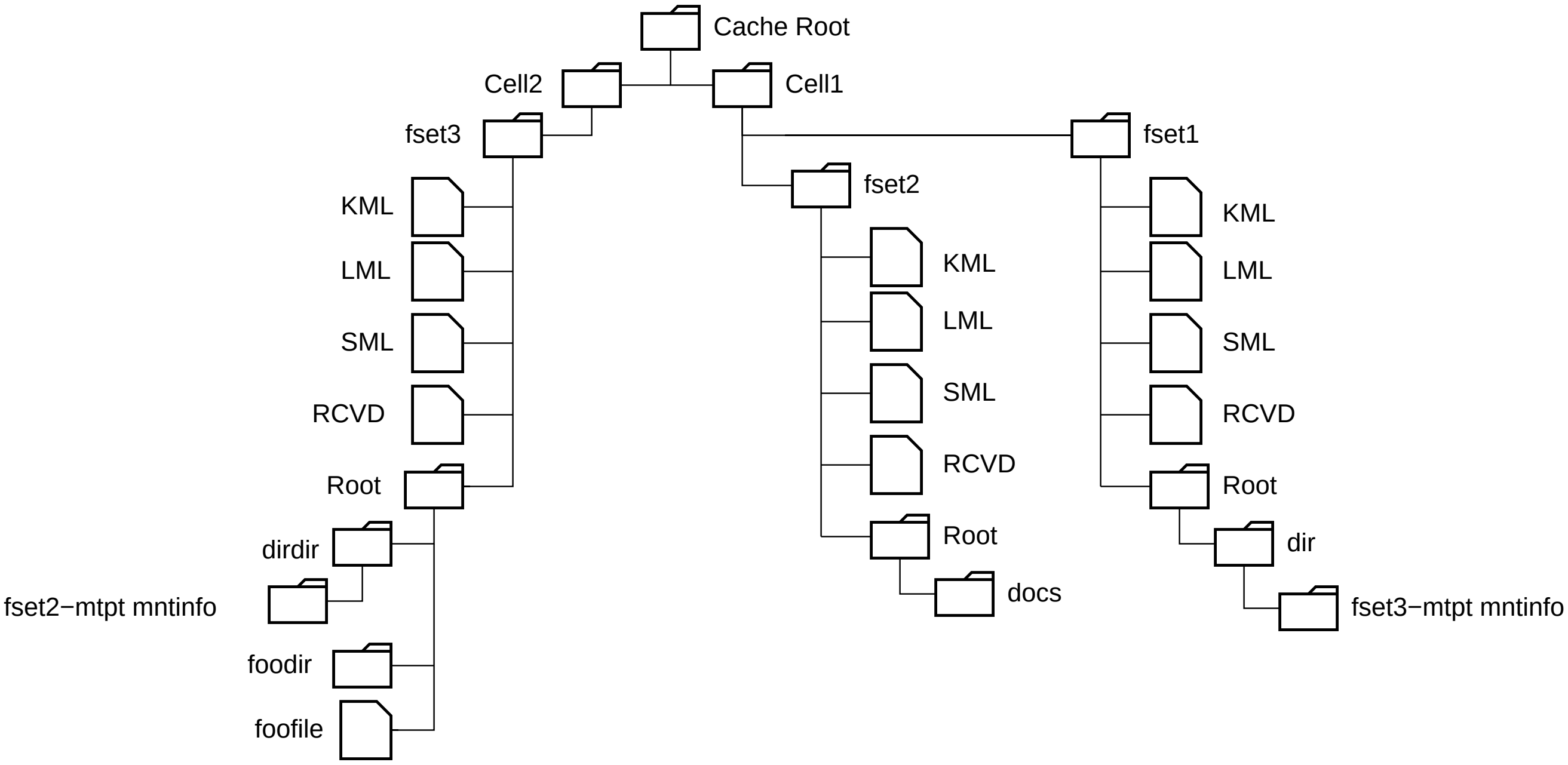

FIGURE 3.7.1. The Physical Layout of the InterMezzo Global Namespace

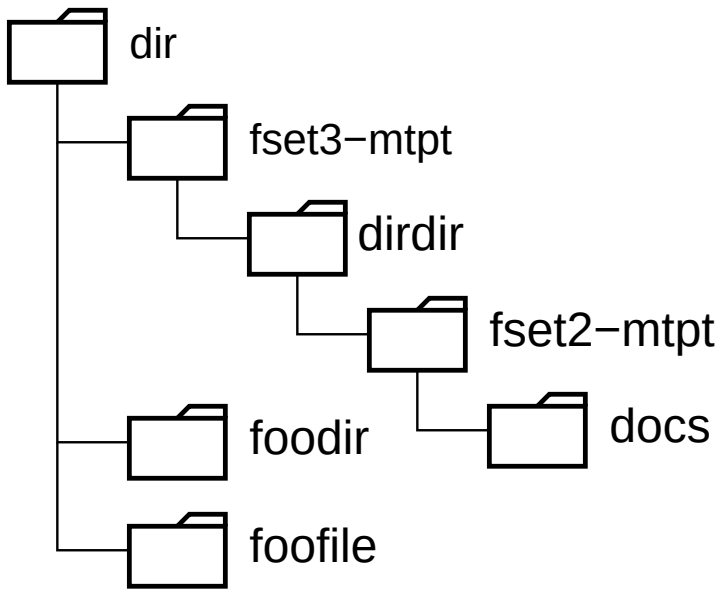

FIGURE 3.7.2. The Logical Layout of the InterMezzo Global Namespace

Note that the bindings now combine physical fileset namespaces from more than one cache into the global namespace found under `/izo`.

The caches now have a different layout. There are two caches, one for the locally stored `/fset1@cell2/` fileset, mounted on `/cache2` (figure 3.7.4), and a partial cache for the imported filesets, mounted on `/cache1` (figure 3.7.3).

We believe that this provides all the flexibility InterMezzo filesets require.

Our mechanisms also allow the mix of different filesystem types. In InterMezzo installations where the root filesystem is of type InterMezzo, it can be attractive to have a `/tmp` directory backed by storage on a local disk. A subdirectory `/tmp/global` might be very useful to provide globally visible temporary files. The following mount commands show how this can be accomplished.

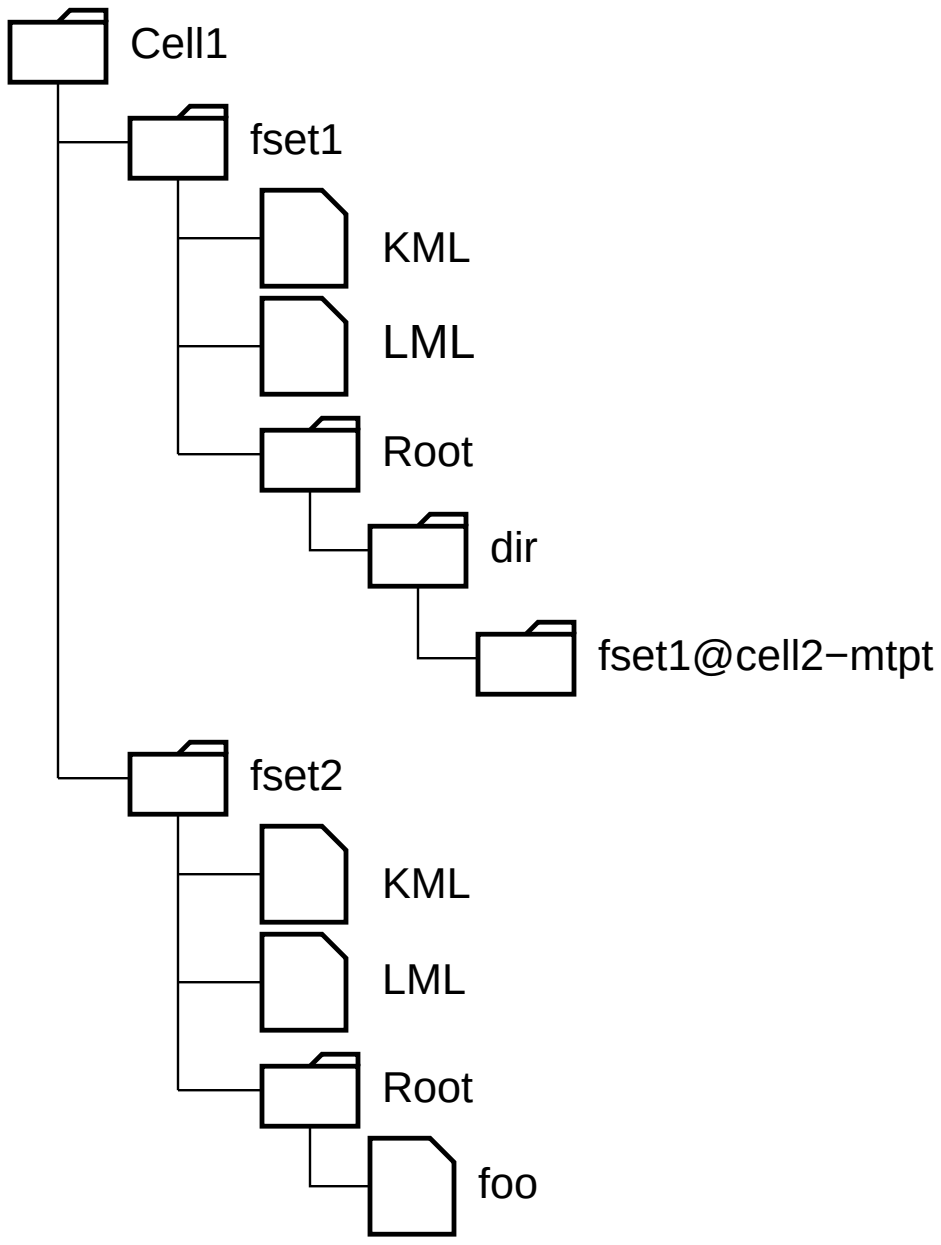

FIGURE 3.7.3. Layout of `/cache1`

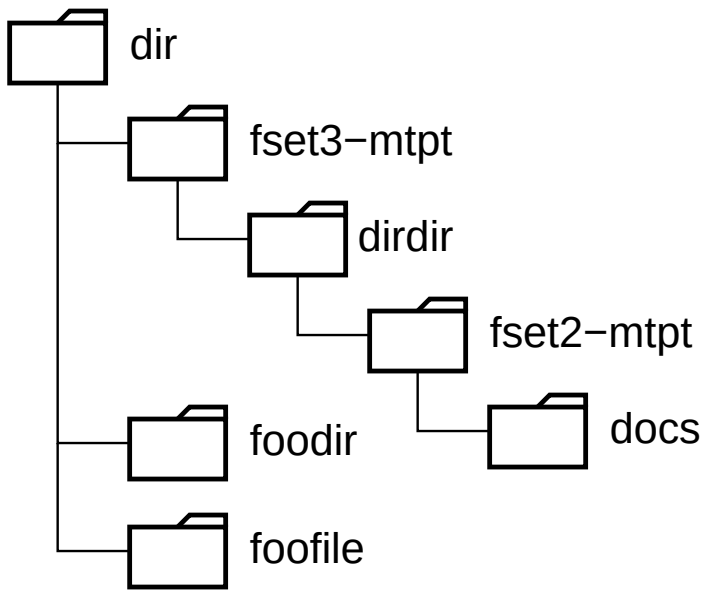

FIGURE 3.7.4. Layout of `/cache2`

## 3.8. Mount-objects in Lustre

Describe here that Lustre has filesets XXX. Explain that traversing a mount-object may involve a query to LDAP to locate the meta-data server for a fileset.

### 3.9. Implementation

We have begun implementation of a filtering global namespace mechanism. XXX describe the implementation here.

### 3.10. Changelog

**Version 3.0 (Dec. 2002)**

(1) P.D. Innes - updated and resized figures, added Changelog

**Version 2.0 (Nov. 2002)**

(1) P.D. Innes - updated text

**Version 1.0**

(1) L. Ward - original draft

CHAPTER 4

# Networking for Lustre

This chapter gives an overview of the entire request processing stack used in Lustre. The stack accommodates industry trends toward Remote DMA (RDMA) transport for storage networks such as found in FC, VIA, and InfiniBand. It incorporates a scalable message passing layer based on Sandia Portals version 3. Finally, it offers a flexible request processing layer, supporting multi-RPCs for example.

## 4.1. Introduction

The Lustre filesystem needs a networking infrastructure that can benefit from different networks, yet offers a uniform high level request processing API to the core modules that implement the filesystem. In this document we describe a design and implementation plan for Lustre's networking.

Basic requirements for the networking are:

(1) Maximum portability across operating systems.
(2) Scalability to clusters of 10,000's of nodes.
(3) Easy porting to different network infrastructures.
(4) Utilize performance of TCP/IP and fast networks such as Quadrics, Myrinet [3], I/B.
(5) OS bypass for parallel I/O from user space.
(6) Best of breed request processing infrastructure for the filesystem.
(7) Infrastructure for recovery and error handling.
(8) A well understood security model.

**Acknowledgments:** This paper has been influenced heavily by Rumi Zahir, Lee Ward, Zach Brown, and the Portals group at Sandia. Rumi contributed the basic API's to move compound data. Lee explained how Portals provides scalable and flexible mechanisms to implement these. Zach explained how in the Linux kernel these ideas could become a reality over commodity networking. The Portals group helped to clarify numerous details.

## 4.2. Layering the Networking

The purpose of this paper is to present a design for the full networking stack for Lustre. The top of this stack is made up by the interfaces for request processing. The bottom of the stack are the device drivers. Based on experience we introduce several layers in between:

(1) Device drivers
(2) Portals message passing layer
(3) Niobuf data movement layer
(4) Request processing layer

The purpose of this section is to explain how all of these relate together. The figure 4.2.1 shows how these layers stack.

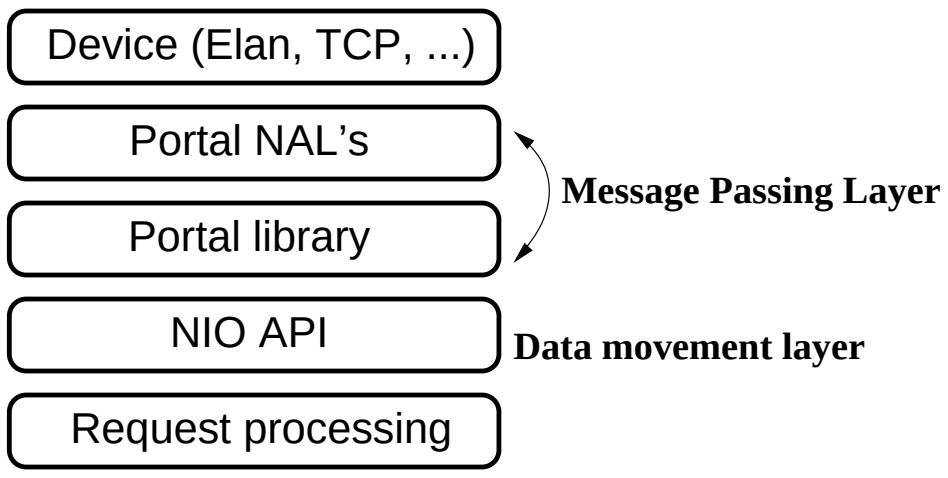

FIGURE 4.2.1. Network Layers

At the bottom of the stack we need **device support** for particular interfaces. In most cases device drivers for hardware are available in Linux, but extensions to interface with a low level message passing layer are still required to benefit from remote DMA and other advanced facilities.

Among the devices of interest to us are:

(1) Ethernet drivers
(2) Myrinet
(3) Non-interrupting memory bus-based networking
(4) Quadrics
(5) InfiniBand
(6) VIA enabled devices
(7) Traditional SAN networks

Most, but not all, of the interfaces used will interrupt the CPU when traffic handling reaches certain stages. Those that don't interrupt will require some kind of polling-based handling. The nature of the networking interfaces Lustre needs to support and the scale of the clusters we target introduces some fundamental constraints into our design:

(1) The networks are so fast that interrupt-driven resource management, such as buffer allocation, is not an option.
(2) The network can carry some 10,000 request packets per scheduler cycle (10ms). Such request packets could come from one client or from 10,000 different ones. The infrastructure needs to handle both extremes well.

Our second layer is the **message passing layer**. Driven by requirements from the scientific computing community we want a lightweight message passing API to underlie our networking. At present it seems that the Portals [5] package developed at Sandia for CPlant is a good candidate. The message passing API provides us with **asynchronous *put*** and ***get*** methods to move packets, **matching** mechanisms to deliver packets to buffers, and **event queues** to provide for notifications and handle completions. Portals also provides mechanisms for OS-bypass including memory management interfaces for 0-copy I/O and NIC resident drivers.

Portals has a modular structure and relies on a support module, called a **Network Abstraction Layer (NAL)** which defines the interaction with the devices mentioned above. Portals also allows for crossing of protection domains, such as required for OS bypass networking.

The *put* and *get* operations are too low level to conveniently define data movement. Our next layer is the **network data movement layer**, responsible for moving vectors of I/O buffers from one system to another.

Above this layer we need a **request processing layer**. This provides an API to dispatch requests and replies. Our model is that requests and reply packets consist of a small **command** or **response** packet and that associated with a request can be a collection of bulk data buffers that will be transferred as part of the request handling. In some cases it is not known how much data will be received and in some cases it is. Further, we make requests to a single or multiple hosts and need facilities to optionally wait for a response, run a callback function upon completion, and to also receive notifications of timeouts to handle errors and cluster transitions.

## 4.3. Portals

The Sandia Portals package is a carefully crafted lightweight reliable low level messaging API; see the Portals API documentation [5] [4]. We start the discussion of the networking stack with Portals because Portals includes an abstraction to handle multiple network interfaces. Lustre makes every effort to use Portals in an unmodified fashion but has required minor changes to the API.

**4.3.1. Portals Data Structures and Mechanisms.** The fundamental object is a **portal table** associated with a **process**, an **interface,** and a **node**. The Portals library can transmit and receive packets and these are dispatched to a particular portal table. As we see, the portal table is named by a **network interface**, a **network id**, and a **process identifier**, the so-called **ni/nid/pid**. The word process here is used more broadly than a Unix process - for example, kernel level usage of Portals is done with a single process id of 0.

Each portal array has multiple portal entries (the number of which can be specified at setup time). Each message is explicitly sent to a particular entry. Attached to a portal entry is a list of **match entries**, each of which points to a **memory descriptor**(See figure 4.3.1). When a message is dispatched to a particular portal entry, **match bits in the packet** are compared with match entries in the list. When a suitable match entry is found, this determines the memory descriptor which will receive the message.

Memory descriptors have a variety of features. For example, they can be automatically freed after having received a certain number of messages, and they can feature receiver-managed offsets.

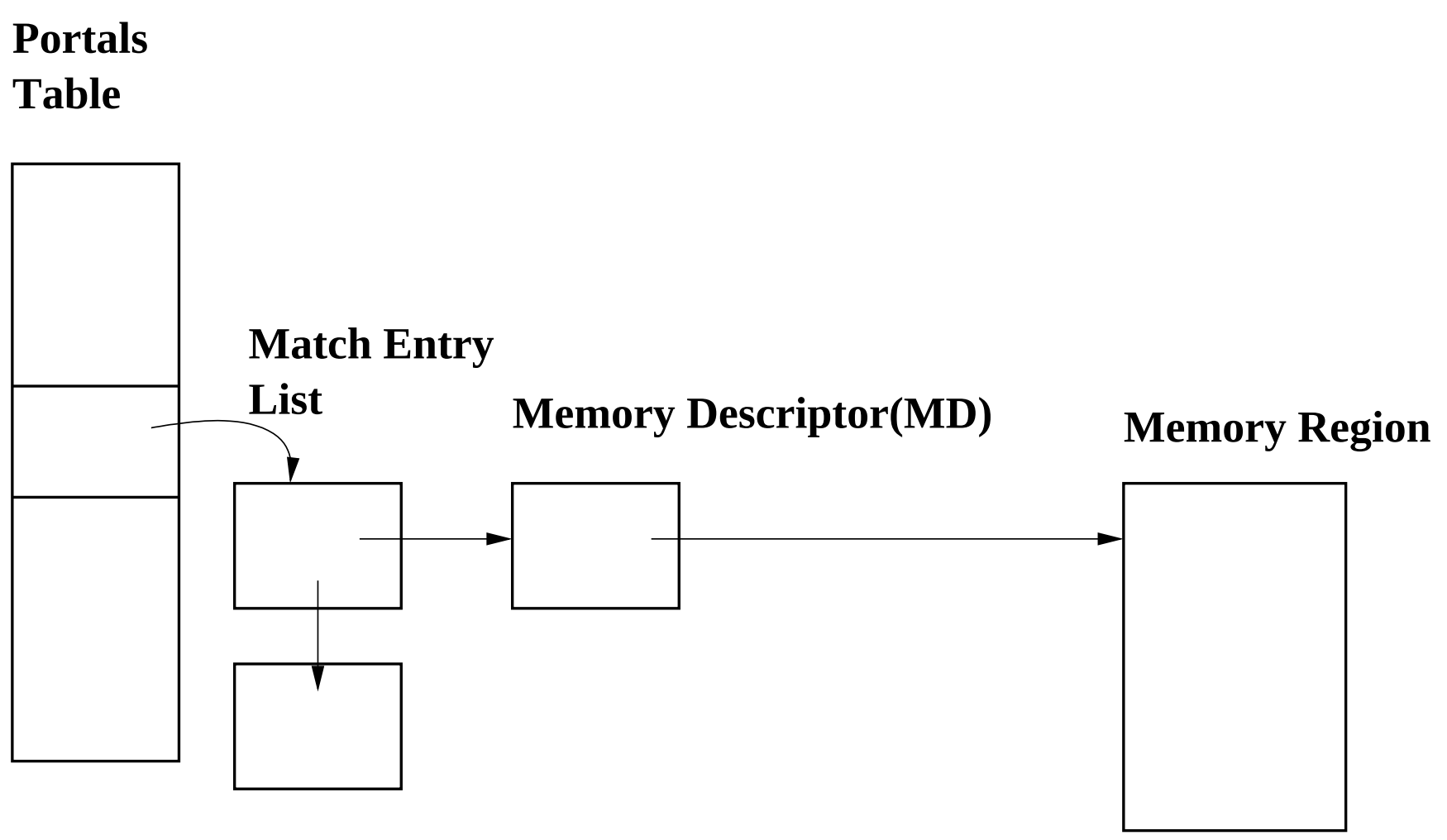

FIGURE 4.3.1. Portals Addressing

When events take place, such as dispatch of packets, receipt of messages, and filling up memory descriptors, Portals can write events into **event queues**. The event queues can be monitored by processes.

The Portals API resides in the space of the application, which for a filesystem would be the kernel. However, a particular application performing MPI message passing or parallel OS-bypass I/O into Lustre files, may want to use the Portals API from user space. Portals has a mechanism (the *forward* method of the NAL) that enables a particular application or the kernel to coordinate with the library layer that controls the networking. We plan to have a kernel and process level interface to the Portals library, leading to a Portals API for the filesystem code in the kernel as well as for user level applications that perform MPI messaging and parallel I/O with OS bypass.

A portal comes with the following capabilities:

(1) *put* and *get* operations to send and receive buffers.
(2) Packet matching and dispatch to buffers.
(3) Completion event handling.
(4) A combination of kernel and user level infrastructure that allows multiple Portals to be used simultaneously from kernel and user levels.

Portals supports unsolicited packets, but assumes all buffers are pre-posted. The dispatching mechanisms are extremely suitable for request/reply processing, particularly in the presence of compound requests involving bulk transfers or vectors of buffers.

Portals is licensed under the GNU GPL, has a proven success record, and appears very closely aligned with our objectives; barring strong alternatives this will be our choice.

**4.3.2. Portals Internals.** The Portals implementation was very carefully designed to allow great flexibility in the runtime environments associated with the API. The key concepts are:

**Protection domains:** The Portals API can be useful to user level systems like MPI or filesystem libraries, but also to kernel level subsystems; this environment defines the protection domain of the API. The API will universally need library (lib) support for particular networking subsystems and this support will normally run in a different protection domain. Two important library protection domains are a kernel level library interacting with a device driver or networking library and a library resident on a network interface card.

**NAL's:** Network Abstraction Layers exist for both the API and lib domains. These abstraction layers are two virtual classes, one for use by the API, one for use by the lib domain. The derived classes are implemented by drivers usually associated with particular network systems, and instances of these classes are associated with a network interface offered by a network.

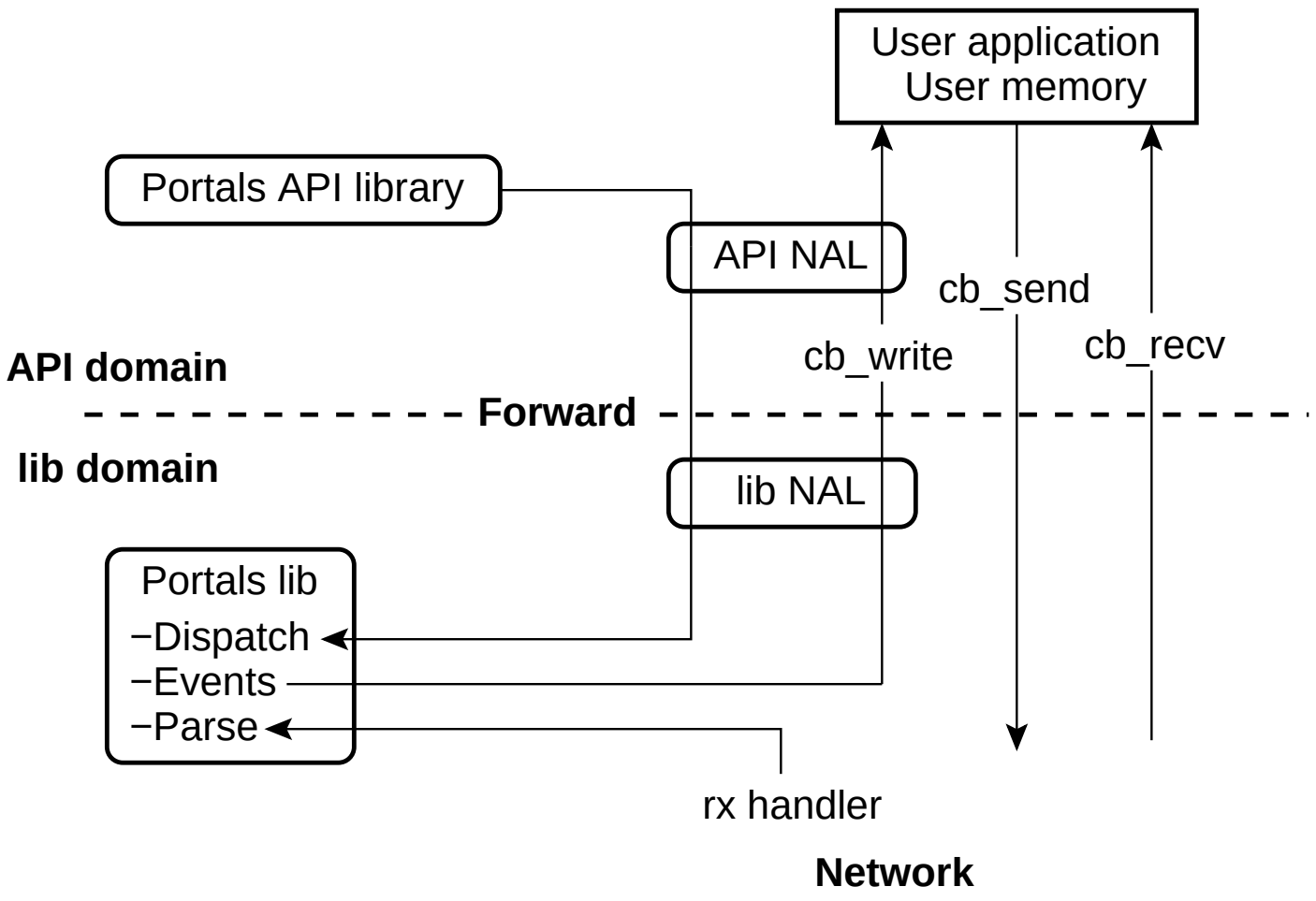

FIGURE 4.3.2. Portals

An instance of the Portals service therefore breaks into four components, as shown in figure 4.3.2:

1. **Portals API library:** It offers the Portals API to applications and interacts with the API NAL.
2. **The Portals library:** The library offers a method table offering the following fundamental functions: *lib_dispatch* executes functions forwarded from the API library, *lib_finalize* is responsible for generating events visible to the API library. Finally, *lib_dispatch* is offered to find the purpose and destination of incoming packets.
3. **The API NAL:** Supplies a method table to the API library. One of the methods is the forward method. The API library uses the forward method as a function call interface to the Portals library.

**4. The lib NAL:** The final component. It dispatches requests forwarded by the API NAL to the Portals library and has several other functions. It has an autonomous handler associated with incoming traffic. It offers several methods to the Portals library: *cb_send* is used to send packets and *cb_write* crosses the protection domain and delivers both data and events to memory areas in use in the API domain.

The interactions between the portals library and the NALs are further detailed in figure 4.3.3. The *lib_parse* parses a packet based on the type of message and invokes the *cb_send* or *cb_recv* funtion of the relevant NAL. As an example, for TCP, the *cb_send* function corresponds to *ksocknal_send* function.

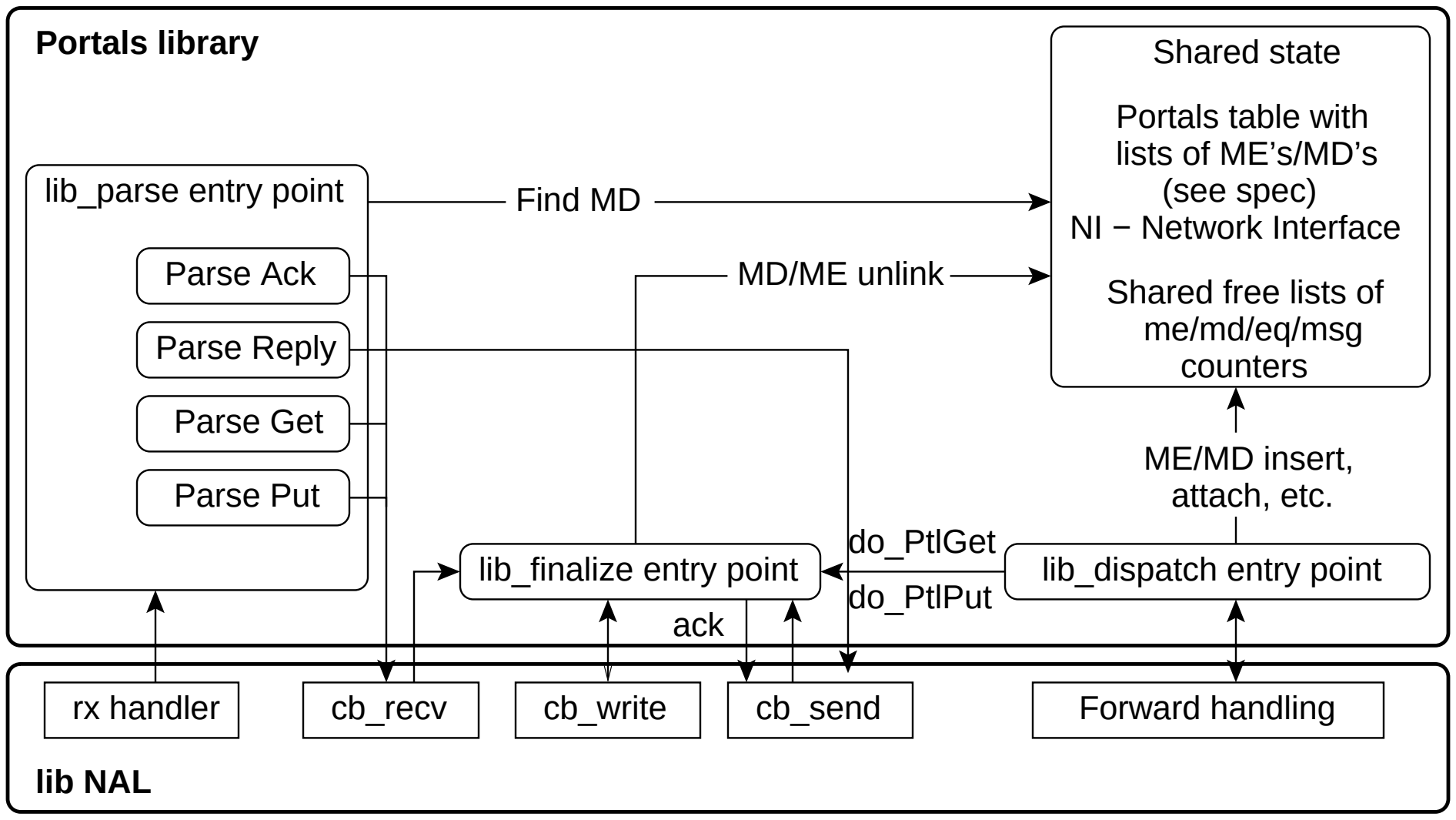

FIGURE 4.3.3. Portals lib Interaction

**4.3.3. RDMA Support in Portals NAL's.** RDMA support commonly relies on registering a memory region with the device and obtaining a handle for this memory that a remote system can use to access the extent. It is necessary for such handles to be communicated to the remote station.

Portals can support RDMA through a **long send**. A long send exchanges a RTS/CTS message with the peer during which memory can be registered for remote DMA. Typically the hardware returns a handle for the memory buffer which can be used by a remote system. The RTS packet transmits this to the peer that wishes to access remote memory and once the CTS packet is received the packet is sent.

It is attractive to use **early binding**. In this model when a memory buffer is prepared for use by a remote host (for reading or writing), it is registered and the NAL is invoked to generate a handle for the descriptor which we store in the memory descriptor structure as an opaque NAL dependent field.

During the request processing the higher levels will inform the remote station about these memory descriptors and transmit the opaque fields to the peer. The *put* and *get* methods can now use the handles to invoke RDMA.

At present Portals offers the NAL an opportunity to register such data in the *addrkey* field of the memory descriptor during the *cb_validate* callback made to the NAL in the function *lib_md_build*. However, this merely assigns a pointer in the MD structure to possibly hold the handle, and the size of the structure is not explicit. We propose that Portals instead allows for the remote handle information to be stored in the memory descriptor, so that this can be communicated to the remote side.

There is also no possibility to feed these cookies back into ***put*** and ***get*** commands. We propose to pass in opaque fields to the ***put*** and ***get*** commands that can be used by the NAL for RDMA, perhaps as an extension of the match bits parameter.

This model matches the common form of RDMA processing by network interfaces. It does not affect the higher level Portals API as such.

**4.3.4. TCP - the Socknal.** The socknal translates a streaming protocol from a message protocol. The socknal has made accomodation to distribute its attention fairly over many different sockets, i.e. it will read one request from one socket and then go on to the next socket. Throughput numbers are lower than Elan, but not bad.

Portals is initialized with a **Network Abstraction Layer (NAL) object**. This object should amalgamate all the TCP connections that have been established between a particular node and others (see [**6**] for motivation). The NAL object will need the ability to add and remove TCP connections that were accepted or connected. This is not a Portals method.

There are two tables of methods that need to be initialized and which offer the core of the Portals library to *send* and *receive* packets as well as manage memory and some auxiliary operations.

The TCP NAL relies on a state machine to assemble packets from the stream. The *receive* handler builds the packets and moves the state forward from assembling headers to assembling body to reading so-called slop (unused data on the wire).

On the *send* side we would like the NAL to use 0-copy *sends*. If general hardware assisted receive API's become available these could lead to 0-copy *receives*.

**4.3.5. Quadrics Elan.** Quadrics ships its network interface cards with a Kernel Communications Library. This library has features to register memory with the NIC and register callback functions for receipt of data. To exploit DMA capabilities of the card, memory needs to be registered through the Kernel Communications Library. For small messages it is more advantageous to copy the memory into a permanently registered buffer while for larger buffers the DMA registration is cheaper.

Communication of the QSW DMA handles between clients and servers can lead to a desirable early bind feature.

The Elan NIC is programmable and a more scalable implementation which avoids the connection-oriented nature of Elan can possibly be built by programming the NIC.

The NAL has seen a fair amount of tuning to handle requests from up to 6,000 concurrent threads. Tuning has involved adjusting the envelope handling, and setting buffer sizes to appropriate numbers.

4.3.5.1. *Sending Data.* The QSWNAL has a sophisticated transmit handling system. There is a pool of transmittor threads. When a *cb_send* method is invoked, a free transmittor is located; if none is available the work is queued. It is asked to map the pages into the card, and then send the data out. When it completes, the thread checks for more work.

4.3.5.2. *Receiving Data.* Similarly, there is a thread per CPU handling incoming packets.

**4.3.6. VIA.** VIA supports a remote DMA model where memory is registered with the driver to be used in RDMA. The driver returns a handle for the memory buffer and the remote can use this handle to address memory at any offset in this buffer.

We believe that a VIA NAL can be implemented efficiently using the Portals/NAL DMA support routines and through connection registration as for TCP/IP connections.

VIA has some limitations documented in [6] related to its connection-oriented nature and the affinity of buffers and connections. In practice, most VIA hardware suppliers allow a buffer pool to be shared among multiple circuits. However it remains unclear how well VIA will scale to clusters of 1000's of nodes. A further issue with VIA is that the execution of an event handler upon receipt or transmission of data is possible but the provided infrastructure is very bare bones for request processing purposes.

## 4.4. Portals Routing

In order to allow systems attached to different types of networks to share Lustre file systems, two mechanisms are provided. On the one hand, a node can use multiple interfaces and accept and initiate sessions over each.

Also available are routing or gateway nodes. These nodes route Portals packets from one net to another.

**4.4.1. Gateway API.**

**4.4.2. Gateway Failover.** When a single lustre filesystem runs on more than one cluster, the lustre RPC messages must be forwarded between the clusters. This forwarding takes place on so-called gateway nodes. For example, two elan clusters might be connected to each other by gigE. The nodes in both clusters that connect to the gigE network(s) have to "shovel" lustre RPC messages between their elan and gigE network connections (i.e. between the qswnal and socknal).

This mail is about how to exploit redundant gateways so that a lustre filesystem spread on multiple clusters remains functional in the presence of individual gateway failures.

At this time, I have implemented redundant gateways in the portals router. This implements

(1) load balancing over equivalent routes and
(2) a means to enable and disable particular gateways.

The idea is that when gateway failure is detected, all relevent nodes disable routes which use that gateway until the gateway is rebooted.

The remaining issues are

(1) how do we detect failure of a gateway and trigger the disabling of routes that use it,
(2) how do we detect the return to service of a gateway and trigger re-enabling the routes that use it and
(3) how does this interact with lustre.

Note that this discussion restricts itself to multi-cluster configurations in which only immediate network neighbours of gateways need to be notified of failure; other configurations probably aren't relevent since portals message forwarding is a big latency hit and such networks should be avoided anyway.

4.4.2.1. *Gateway Failure.* I think the best place to detect gateway failure is by the relevent NAL in the immediate neighbours of the gateway.

Currently the elan and socket NALs can detect peer failure reliably and could

(1) notify the local router that this gateway should be disabled and
(2) notify the world via an upcall.

Note that detecting peer failure necessarily involves one or more messages (incoming or outgoing) getting dropped, so local router notification ensures that all further messages can avoid the failed gateway and the upcall can be used to pro-actively inform other nodes of the problem, rather than letting them find out for themselves. This should also avoid any possibility of a "chained timeout" problem.

4.4.2.2. *Gateway Reboot.* Spookily, the gateway itself knows when it has rebooted, so it seems sensible that the script that configures lustre on the gateway should also trigger notification on the network neighbours (of all the gateway's NALs), rather than coming up with some clever scheme for detecting when the gateway has returned to service on these neighbours.

4.4.2.3. *Impact on Lustre.* When a gateway fails, at least 1 message gets dropped; think of it as portals networking providing a very good (but obviously not perfect) "best efforts" message delivery mechanism.

Whenever a message gets dropped a lustre client or server will time out. This triggers lustre's "normal" recovery actions which will be able to proceed immediately if the failed gateway has been disabled (and an alternative route exists).

The key issue here is "Can we ensure that the failed gateway has been disabled when lustre notices something amiss?"

Unfortunately, the answer has to be "no", since

- lustre timeouts can be set arbitrarily with no reference to the NALs and
- NALs can only guarantee that messages complete (possibly with failure) in finite, but not bounded time.

However this is of no consequence provided

(1) lustre can insulate itself from anything the network throws at it after it has abandoned a particular RPC attempt and
(2) lustre can recover from failed recovery attempts.

I think (1) is almost implemented already, since we unlink all ME/MDs associated with an RPC after a timeout occurs. However, we should ensure we never re-use the same matchbits for the RPC reply, even on a retry of the same transaction. Personally I'd also like to _not_ reuse the same matchbits for bulk, even if bulk is idempotent; IMHO it's just good practice to use unique matchbits for each RPC attempt. (I think this might mean we have to change the protocol to stop overloading transaction number and reply matchbits).

I'd also have thought that (2) should work right now since we can never guarantee we don't experience failures during recovery in any case.

4.4.2.4. *Summary of the scheme.* Here's the scheme for portals gateway failover.

(1) Everywhere can get to where they need to go by at least 2 gateways.
(2) When the peer of a gateway detects its death...
    (a) The peer stops using the gateway and upcalls.
    (b) Site admin uses this upcall to trigger all nodes running 'lctl set_gw <failed_nid> down'.

    All nodes stop using the failed gateway (and use the remaining alternative)
(3) When the failed gateway reboots and reloads its config, site admin causes all nodes to run 'lctl set_gw <resored_nid> up'. All nodes may start using the gateway again.

Note that lustre sees this as if it's running on a network that is usually very reliable, but once in a while can lose messages (i.e. the messages in transit via the failed gateway). It detects these lost messages by timeouts and I assume it recovers from them by doing something like 'lctl –device ? recover'.

**contentious point** There are no guarantees that lustre's network heals itself within the lustre timeout. However it would be most sensible for site admin to ensure that all nodes are notified of a gateway's death before lustre recovery is attempted on any of them.

4.4.2.5. *Error handling.* The qswnal seems wonderfully robust at detecting and recovering from errors (I blame the beautiful EKC scalable cluster membership code).

The same isn't true for the socknal. Currently, we consider a socknal peer dead (or unreachable) when I/O on the last connection to the peer returns an error or times out. Mostly, it's the elan-side peers of a gateway that detect I've powered the gateway down; it's the TCP/IP side only occasionally. At a guess socket buffering is obscuring things; unless a packet 'wedges' half in or half out of the socket, socknal doesn't detect an error, and won't until it tries to send another message.

So if connections remain open after peer death (seemingly the normal case), the socknal basically relies on the notification in (2b) above to clear connections to the failed node and stop routing via it before anyone else tries to use them, including (and importantly) before lustre recovery is attempted (which expects the network to have healed).

This gives rise to a couple of thoughts.

(1) Maybe socknal should heartbeat its peers (if I understood SO_KEEPALIVE, or with a portals HELLO message) for more prompt detection. Then the NAL notification stuff wouldn't be so critical since peer death will be detected in bounded time in all cases.
(2) If I stick with what I have now, maybe notification should become general (implemented in the portals module rather than the router), and not just used for gateway failover.

Large messages sent using zero copy time out (in the socknal itself) quite reliably. However small messages get copied into the socket buffer and appear to have been sent OK although they are just sitting there doing nothing. The TCP keepalives seem not to help at all, at the small timeouts (10s of seconds) I was testing with.

I considered (and discussed with Phil) the idea of sending a portals HELLO message if nothing had been sent for some time, expecting the same in return and timing out the connection if it remained idle for some period. However the case of a server with hundreds of normally dormant clients seems a bad one. Idle clients could possibly disconnect, but that might just compound this problem.

Another alternative would be to somehow detect that the socket buffer isn't draining. This should be reliable since zero-copy messages in-flight to a failed peer _do_ time out properly as they don't complete until they have actually made it across the wire.

Given that the lustre pinger is going to be checking the health of critical nodes anyway, it seems that including any gateways in the set of nodes it checks provides the back-stop; i.e. NAL peer failure detection is just the first line of defence and failure avoidance.

So I'm going to proceed assuming the lustre pinger always detect gateway failure, while peer NALs just make best efforts. When this works, all the components will be in place, and I can then return to making the socknal better at detecting peer failure if it seems worthwhile.

**4.4.3. Practical issues.** Radhika, Some brief notes; I can fill out as required.

(1) 1/ The portals router spreads messages over all gateways that can reach a particular destination. There is a new command...

    lctl –net <nal> set_gw <nid> {up|down} ... to enables and disables individual gateways.

    When a gateway is disabled, the portals router avoids it and notifies the relevent NAL so it can clear any connections it may have with the gateway. When it is re-enabled, the portals router resumes using it.

Note that the portals routing tables are not affected (i.e. no topology information is changed), just which entries in the routing table will be considered for any particular routing decision.

Also note that redundant notifications are harmless.

(2) When a NAL detects a peer has died, it notifies the router. If the dead peer is a gateway, the router notifies the world via an upcall.
This upcall should be used to notify all nodes about the failed gateway using the lctl command described in (1). Any nodes, which haven't already detected the death of the gateway for themselves will now avoid it in future.
Note it uses the standard portals upcall (set via /proc/sys/portals/upcall and '/usr/lib/lustre/portals_upcall' by default) as follows....
/usr/lib/lustre/portals_upcall ROUTER_NOTIFY <nal> <nid> {up|down}
...where <nal> is a decimal number and <nid> is the NID of the gateway in hex with a leading 0x.

(3) When the dead gateway reboots and reconfigures, it should notify all nodes that it is back online using the lctl command described in (1). All notified nodes will now be able to use the gateway again.

(4) NAL peer death detection is invariably accompanied by the loss of one or more messages, so lustre will experience a transmit failure or a timeout when a gateway dies.
The notification described in (2) above will heal the network after a gateway dies, but lustre will still have to take its own recovery action when it sees an error caused by gateway failure.
Lustre's first attempt at recovery should be a simple retry, on the assumption that the network failed momentarily, and healed itself.

- lctl set_route <nid> <up/down> enables or disables particular portals routers (i.e. gateways) without forgetting network topology (i.e. adding/deleting routes).
- Socknal and qswnal automatically notify their local router and make an upcall when they detect peer death. * portals router load balances over equivalent routes
- ENETUNREACH returned when a NAL thinks the router isn't loaded
- Improved socknal network failure detection.
- /proc/sys/socknal/* interface
  - timeout is the socknal I/O timeout (50 by default) in seconds .
  - eager_ack is a boolean (set by default) that enables setting TCP_QUICKACK after every incoming message to ensure peer zero-copy sends complete quickly.
  - zero_copy is the size (2k by default) in bytes, below which message fragments are copied into a socket, rather than using zero-copy sends. Setting this above PAGE_SIZE will disable zero copy.
- Socknal autoconnect option to create all peer connections eagerly or not. If more than one autoconnect reaches the same peer NID, with the eager option (this was the previous default), all connections will be made when one is required, otherwise only one at a time will be attempted (the new default). NB socknal still load balances over all established connections.

## 4.5. Lustre Data Movement and Portals

In this section we will describe how Lustre is using the Portals API for its request processing. There are several key aspects at this level:

(1) Naming.
(2) API's for moving data.
(3) Event generation and dispatch upon data departure and arrival.

The Lustre naming scheme will be discussed in the next section. It can be summarized that Lustre includes an address translation mechanism that provides Portals usable ni/nid/pid addresses.

**4.5.1. Portals & Packet Types.** Lustre offers several network protocols, each formulated as a client-server protocol. For each protocol three packet types can be exchanged:

(1) Request packets.
(2) Reply packets.
(3) Bulk packets.

For each protocol and packet type, Lustre statically assigns a Portal index for the receiving end. For example, at present the object storage protocol has:

```
#define OSC_REPLY_PORTAL 4
#define OSC_BULK_PORTAL 5
#define OST_REQUEST_PORTAL 6
#define OST_BULK_PORTAL 8
```

So any request packet sent to an OST will be addressed to Portal entry 6.

**4.5.2. Request/reply Identification.** Every request will be sent out with a request id, *xid*, generated at a high level in the system. The xid's form an increasing sequence of integers for each connection. Reply packets are dispatched to the correct memory descriptor and invoke callbacks on the right request based on their destination portal and xid (which is stored in a match list entry).

So in the case of an OSC-OST request when a request with xid 400 is sent out, the reply is sent back to the OSC system, portal 4 with match bits set to 400. When it arrives, Portals delivers it based on a match entry that matches on 400.

For bulk transport, described below, extra xid's are transmitted to the sending node, one per bulk page. This mechanism makes the correct identification of the data that has arrived independent of the order in which it's delivered.

#### 4.5.3. Event Handling.

There are many events that can be handled:

***request_in*:** This is the most complex event handling system in Lustre. The request is added to a pool of buffers, which is re-arranged when a buffer is full. The request is then passed on to a handler thread.

***request_out*:** These can be generated to free a packet that has been put on the wire. To recover protocols, requests are retained for re-transmission and this event is not used anymore.

***reply_in*:** When replies come in, the process that is waiting for this needs to either wake up or a callback function needs to be run for the process. This is done with the *reply_in_eq*.

***reply_out*:** After replies are sent, they are freed. The *reply_out_eq* sees an event that indicates the outgoing reply packet can be freed.

***bulk_source*:** These events increment a page counter; when all bulk pages have left the system, a callback is run that allows pages to be unlocked or freed as appropriate.

***bulk_sink*:** A similar page counter is incremented; when the counter reaches the full count of the buffers to be transmitted, a callback is executed.

Some of the event handling is uniform, e.g. all protocols share the *reply_out* event queue. Others are specific to a particular service, such as the incoming *request* event queue. For events deposited in each of these event queues, Lustre registers an event handler with Portals. The event handler decides whether to update state in line, execute a callback, or wake up a thread.

The yields to other threads are very explicit and it should not be hard to move over to an interrupt free request processing scheme.

#### 4.5.4. Context Management.

Portals can run an event handler or a thread can pick up events from event queues. In each case the event provides a pointer to the memory descriptor that received the data. This memory descriptor in turn has a user supplied pointer that can relate the memory descriptor to, for example:

- A service in the case of an incoming request.
- A waiting *reply* thread in the case of an incoming reply.
- A bulk descriptor when bulk completes.

In the memory descriptor for a *reply* packet, the user pointer will point to the *request* structure to which the *reply* belongs. When a *reply* packet arrives an event is added to the *reply* event queue.

#### 4.5.5. Buffer Management.

As indicated in an earlier section, we are relying on **pre-allocated buffers**. In many cases it is easy to predict what buffer sizes are required for responses and allocating the buffer early is not problematic. However, incoming requests are of unknown size and require careful management.

Portals supports buffers with **receiver-managed offsets**. In such buffers, packets are appended to the end of a buffer and the event generated includes the offset of the packet that was delivered. A critical aspect of handling requests is what to do when the buffer is full. Portals allows multiple match entries per buffer and this allows a new buffer to automatically take over from a full buffer when the latter overflows. Moreover, Portals optionally generates an event when a buffer is full.

Lustre employs a *request*/*reply* model, like many other networked systems. *request* and *reply* packets are handled as a single buffer. Lustre packs *requests* in place, i.e. it uses structures that are aligned on all architectures of interest and transforms byte order in place when required. As such, *request* dispatch and handling avoids copies of the *request* buffers.

To avoid allocating and freeing such memory, large buffers are used and new *requests* and *replies* are appended to the end of the buffers. The buffers can get full after which they are replaced with others; the buffers track how many *requests* are still using extents in the buffer. When no *request* is using the buffer, it can be returned to the ring of available buffers.

We believe that the VI architecture requires that buffers are associated with a particular connection. This raises a scalability problem as it is not known on which connection the packets will be coming in, and very many connections may be needed. If this is the case, a possible solution would be for the *receive* handler to re-assign buffers to particular VI connections when the preallocated buffers for any connection are exhausted. Most VIA capable interfaces have introduced a method extending the VIA protocol to share a buffer across multiple connections. Please note that the mismatch with our requirements is not at the Portals level, but below.

**4.5.6. Bulk Transport.** Some Lustre operations, such as *reads* and *writes* of object data on the OST, involve the transfer of large buffers during *request* processing; see RPC2 [9] and Zahir [25]. This bulk data movement is made in conjunction with a RPC, but uses Portals independently. The bulk movement is symmetric and supports data movement in both directions: from the RPC client to the server, as for file *writes*, and from server to client, as in file *reads*. To avoid confusion, the systems roles in the bulk movement are described as *sink* and *source*.

The bulk transport is described by three basic data structures: the *sink_niobuf*, the *source_niobuf*, and a *logical_niobuf*. The node that will receive the data, the bulk sink, reserves the necessary memory area and builds a bulk descriptor, the *sink_niobuf*. This setup of this descriptor prepares all of the information that Portals needs in order to receive the data directly into that memory, as well as a mechanism for notifying the sink when the transfer completes. This event delivery mechanism relies on the Portals ACK facility to receive notification upon successful message delivery, and is the only Lustre component to do so. The sending node, the bulk source, builds a nearly identical descriptor, the *source_niobuf*.

In order to communicate enough information between *sink* and *source* a third niobuf is needed, the *logical_niobuf*. Each niobuf represents a single extent. When Lustre needs to transfer multiple extents of one or more objects, e.g. when flushing a dirty client page cache, it sends a vector of logical niobufs, corresponding to non-overlapping extents. *logical_niobufs* have two components:

(1) A logical object extent which enables the remote side to prepare its *sink_niobuf* or *source_niobuf*. An example descriptor could contain nine 4k pages in inode 4: (inode=4, extent=[4096,40960]).
(2) Space for additional delivery information, e.g. tokens to enable remote DMA. When a client wishes to engage in remote DMA transfers, it registers a memory buffer with the network interface card and receives an opaque token which is added to the niobuf. The remote peer can use that token to directly access the memory on the client and perform faster I/O with lower overhead. Depending on support in the underlying transport layer, such delivery information may contain a DMA handle for each page in the niobuf.

In a typical Lustre object *write*, the source will create *source_niobufs* pointing to kmapped pages, send remote niobufs with offset and length information, and commence bulk delivery. As the buffers are delivered the source receives ACK's which trigger page unmapping and niobuf cleanup. The sink will receive the remote niobufs, map the relevant pages into memory, and then wait for the data to arrive. As each buffer is received, a handler is notified and can perform any necessary finishing and cleanup. In the case of the filesystem, the *sink_niobufs* for file *writes* trigger events that release the pages to the page cache flush daemons. The *sink_niobufs* for file *reads* mark cached pages as up to date and unlocked. The preparation of the *source_niobuf* for *reads* involves reading the pages from the storage subsystems.

**4.5.7. Data Structures and API Summary.** Data structures mentioned in this chapter are detailed in *lustre_net.h* and *lustre_idl.h*. The API implementation is defined in these files and can be implemented in the *ptlrpc* directory.

***ptlrpc_request*:** The organizing center for request processing is this structure. It contains pointers to request and reply buffers, match entry and memory descriptor for the incoming reply and pointers to the service, client, and connection structures. Much of this will be discussed later in this chapter in more detail.
***ptlrpc_bulk_desc*:** The local descriptor for a bulk movement. It holds a list of
***ptlrpc_bulk_page*:** Provide the buffers that are moved in the bulk transport.
***niobuf_remote*:** Describes the logical niobuf.
***niobuf_local*:** A descriptor formed by the remote for such logical descriptors.

The following API calls define the data movement layer:

***int ptlrpc_bulk_put(struct ptlrpc_bulk_desc *);*** This call initiates the movement of data from source to sink. The bulk descriptor is the source descriptor.
***int ptlrpc_bulk_get(struct ptlrpc_bulk_desc *);*** This call initiates the movement of data from source to sink in case of writes.
***int ptlrpc_register_bulk_put(struct ptlrpc_bulk_desc *);*** This registers a bulk descriptor on the *sink* side.
***int ptlrpc_register_bulk_get(struct ptlrpc_bulk_desc *);*** This registers a bulk descriptor on the *sink* side.
***int ptlrpc_abort_bulk(struct ptlrpc_bulk_desc *bulk);*** This cleans up a sink bulk descriptor.
***int ptlrpc_reply(struct ptlrpc_service *svc, struct ptlrpc_request *req);*** This sends out the reply contained in the request structure.
***int ptlrpc_error(struct ptlrpc_service *svc, struct ptlrpc_request *req);*** This sends an error message back to the client.
***void ptlrpc_resend_req(struct ptlrpc_request *request);*** This retransmits a request, used in protocol recovery.

***int ptl_send_rpc(struct ptlrpc_request *request);*** This sends out the request message.

### 4.6. Naming and structures used in Lustre Network

Lustre provides a request processing structure suitable for a complex clustered filesystem. Lustre networking is managed as a data structure of which multiple instances can be created. When we speak of initialization below, we mean the initialization of an instance.

**4.6.1. Nodes and Addresses.** Lustre names every **node** with a **UUID** and a **common name**. Lustre can handle multiple **networks** attached to nodes. Each network is described by a **type**, **common name**, and a **UUID**. Attached to a node can be multiple **network interfaces**, of different types. Each such interface is described by a UUID, indicating what network the interface is attached to, a name, and one or more network addresses.

Upon startup the Lustre networking subsystem is told its UUID. If a Lustre system is attached to persistent storage relevant to the system, a boot count is given to the system at initialization time.

To reach a remote node, Lustre needs to know its address, which is stored in a **peer** structure. The peer contains a **network interface handle** and a **network id**. The interface handle is derived directly from the network to which the interface is attached; the network id is the address of the interface in a form usable for Portals message passing. Peers can be found by UUID.

**4.6.2. Lustre Handles.** In many places in Lustre, **handles** are used to locate objects. A handle is associated to an object and can be used on the node where the object resides to locate the object without searching. Handles can be given to authorized remote nodes for later use in retrieving objects. When a remote node presents the handle to the node mastering the handle's object, the correctness and validity of the handle can be verified. Remote nodes never interpret the contents of handles, hence the implementation of handles can be a per platform choice.

In the Linux implementation, handles incorporate two fields: a 64 bit integer containing the address of the object and a random cookie which is also stored in the object. A handle is validated by checking (i) that it is a valid pointer to an object of the type the handle describes and (ii) that the cookie in the handle equals a copy held in the object. Part (i) is done by using slab cache mechanisms already present in Linux, part (ii) by using a random number generator. Handles cannot be forged because of the cookie. As long as the system using handles to locate objects is careful not to introduce race conditions between mapping handles to objects while objects are being freed, the use of handles is straightforward and very fast.

**4.6.3. Connections.** Lustre manages **connections** which indicate endpoints of circuits between network interfaces on nodes. Connections have a remote UUID field indicating the node that is at the other end of the connection and a peer structure indicating the address to be used to reach the remote. Connections manage a message counter, called the **xid**, which indicates the sequence number of the last message sent and received over the circuit. Connections also include the handle of the remote connection for the circuit.

Finally, connections have a generation, epoch, and remote bootcount number used for recovery purposes.

Connections are used with a *get*/*put* API which increases/decreases a refcount. Unused connections may be garbage collected.

**4.6.4. Target.** A **target** is an object through which network encapsulated access is offered to device API's. The target combines a **target mode driver** and a **device**.

At the core of a target mode driver are the handling functions. The target mode driver has handling functions for the interfaces it offers, i.e. the MDS has handling functions for management, meta-data requests, and locking. The handling functions are capable of decoding and servicing incoming requests, and executing them by using the API's exported by the device.

Lustre offers services for low level connection management, meta-data (MDS), and object storage (OST). These targets combine management, object storage, meta-data, and locking API's.

Examples: We have object storage targets offering remote access to direct object devices and meta-data targets offering network access to meta-data devices. Targets are similar in nature to the entries a NFS /etc/exports file or Windows shares.

Targets are posted in the resource database and labeled by UUID.

**4.6.5. Export.** An **export** is a server-based object managing the use of a target by an import. Each target has a list of exports associated with it. Each export points to a connection which it used and each connection has a list of exports attached to it.

Exports contain server resident client specific data associated with the use of a target. For example, the meta-data stores a list of files open on an import in the export structure.

**4.6.6. Services.** A **service** is a data structure that connects a **target** to **network resources** to offer network access to a device. The low level networking dispatches packets to the service based on the portals on which packets come in. By decoding the initial part of the packet and dispatching it appropriately, the service is responsible for (i) identifying the target that needs to run the request and (ii) identifying the API for which a handler in the target needs to be invoked. The service manages the threads handling incoming requests for the targets and the buffers and event queues used by the service. Services typically handle many targets. Services are also responsible for providing the reply and bulk portals and provide handling functions for low level service functionality such as freeing sent reply requests.

In spirit a service is similar to an NFS daemon offering remote file services over NFS for its export.

**4.6.7. Client.** A **client** is the counterpart of the service. It provides the client side networking data for accessing exports available under a service. Associated with clients are a list of imports.

Clients are organized as a tree to organize recovery. The root of the tree is the low level connection service. A system recovers by descending the tree.

**4.6.8. Import.** An **import** is a structure associated with a device which offers a local API dispatched through a particular connection to an export of a service offering a target. A node can have multiple imports of a single target. The import uses a client driver to encapsulate requests in network format. Imports are in one-on-one correspondence with exports on the nodes offering service.

Imports instantiate devices that offer an API. These devices provide remote access to targets.

Imports use a particular connection and a client. A connection has a list of imports.

Each import is operating at a certain level. Requests can only go across to the export when they are placed at a level not higher than that of the import.

**4.6.9. Requests.** A **request** is a structure managing the remote execution and data transfers between an import and its export.

## 4.7. Request Processing

This section illustrates how Lustre prepares clients and server systems for request processing and how it relies on the API to actually send requests and wait for replies.

**4.7.1. Setting Up a Client.**

**4.7.2. Setting Up a Service.**

**4.7.3. Writing the Server Handling Code.**

**4.7.4. Request Processing on the Client.**

**4.7.5. The Connection Handshakes.** The connection handshake is first made for every circuit between two systems in the *ptlrpc_connect* method. This allows the systems to authenticate each other at a system level. This first connect call is not a special case but fits the general framework:

- The client sends a secure pointer to its connection structure.
- The server responds with the same.
- The systems can exchange a generation number of the connection, a boot count, etc.

When subsequently a meta-data or object storage client initializes, it calls the connect method for that service. That connect call will (in the near future):

- Map a UUID for the target, sent by the client to the service, to a device number for the export: see *mds_connect* below.
- Generate a new connection id for the client (id++ will work fine).
- Create an export structure for the client. A secure pointer to the export structure should be returned to the client. This pointer will be filled into the first 16 bytes of the connection handle on future Lustre messages.

- Allow a future mapping from connection handle to export data (probably the connection handle in the Lustre message will point to the export data at that point). The export data will maintain per client state, such as pre-allocated object lent to the client, last received data for client, etc. The MDS has an example of this.
- Return the generation of the connection, the boot cycle, and possibly a cluster epoch of the target.

## 4.8. Error Handling

Portals provides a number of error handling issues that we can make use of. These error handling routines are primarily interacting with the recovery system which can take appropriate actions.

**4.8.1. NACK.** A packet with an ACK request on it, which is dropped by the receiver, should cause a NACK packet to the sender. This is a proposed change in the Portals API which will help cleaning up.

**4.8.2. Failed Sends.** If a packet fails to get transmitted it is important that an event indicates this failure, and that an event is generated when the packet can be freed. Some further design considerations here are in progress.

## 4.9. Changelog

**Version 6.0 (Mar. 2003)**

(1) Radhika Vullikanti - Added a new figure and made some minore changes/corrections

**Version 5.0 (Dec. 2002)**

(1) P.D. Innes - updated and resized figures

**Version 4.0 (Nov. 2002)**

(1) P.D. Innes - updated text

**Version 3.5 (Jun. 2002)**

(1) P.D. Innes - tables changed to floats with captions and cross-referencing added.
(2) Updates to DLM protocol.

**Version 3.0**

(1) Added DLM network protocol, first draft.
(2) Corrections to OST protocol.

**Version 2.6**

(1) P.D. Innes - new references added

**Version 2.5**

(1) P.D. Innes - edited and proofed, bibliography added

**Version 2.0 (June 2002)**

(1) A complete overhaul of the document to reflect the state of the system as of June 2002.
(2) Many speculative statements removed, more concrete descriptions of the internals.
(3) Include the high level picture of services, exports etc.

**Version 1.1**

(1) First sparse draft of wire-level protocol documentation

**Version 1.0**

(1) Clarify the RDMA situation further.
(2) Explain the double binding needed for user/kernel Portals instances to co-exist.
(3) Portals to accept scatter/gather arguments in put/get.

**Version 0.9**

(1) Include a discussion of RDMA
(2) Security section
(3) Vectors in put/get operations
(4) Policy handling in higher layers
(5) Data structures describing kernel memory

CHAPTER 5

# Object Storage

This document gives an overview of interfaces needed to exploit object-based storage. We cover the entire stack between applications exploiting OBD's, filesystems mounting OBD devices, and aggregation and management techniques involving OBD's.

## 5.1. Overview of the Architecture

At the bottom of the stack there is an object storage device, a device driver interfacing with a storage device offering a particular kind of storage. We call such drivers **direct OBD drivers** to indicate that they interact with a storage device. In contrast, **logical OBD drivers** may interact with other OBD drivers.

Direct OBD drivers can be **simulated** direct drivers exploiting a block-based storage device and filesystem-like software or **physical**, firmware-based drivers, possibly with optimized code. A storage device controlled by a direct driver will be called an **Object-Based Storage Device** or **OBSD**.

Object-based storage management exploits a stack of software modules layered over storage devices. At the top of the stack there is a filesystem or specialized application, such as an object-based database or optimized file server, cache manager, etc. We will refer to both filesystems and other applications as **Storage Object Applications (SOA's)** or **applications** for short.

A SOA will typically interface with logical object devices or with an object storage device directly. A logical object device is defined as an operating system-based device offering an interface, exported by a device driver, identical to that offered by an object-based storage device. When device drivers for logical object devices are migrated to the firmware of an object storage device, the distinction between the logical devices and storage devices can become blurred.

Between the top layer SOA and direct driver(s), one can insert a variety of indirect or **logical object drivers**. The name indicates the analogous functions as compared with logical volumes. For example, object drivers can be built that arrange for the aggregation of objects over multiple object storage devices. Such an intermediary object driver may address several direct drivers to *fetch*/*put* data to persistent storage. Obviously the various RAID variants are examples of this.

Other intermediary drivers allow different images of a single logical or physical device to be exported to applications. Examples of this type of logical device are object stores holding snapshots (also known as *read*-only clones) taken at various points in time, as well as devices holding multiple versions of objects. Figure 5.1.1 gives some examples of the stacking of these software layers.

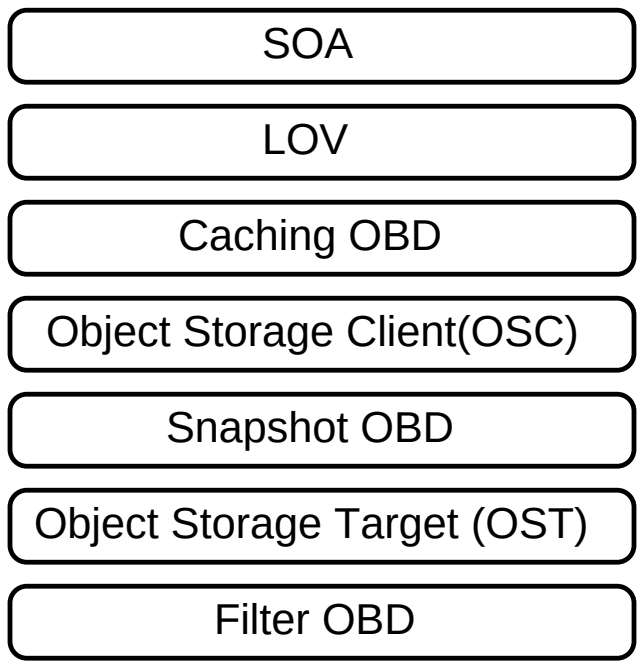

FIGURE 5.1.1. Layering of Object Drivers

It is important that object-based devices can be reached over a variety of **interfaces**. The concepts described above allow for software interfaces to address certain object devices. Certain object device drivers will be network **OBD client drivers** speaking to **OBD targets**. OBD targets are device drivers exporting a network interface, and are attached to a physical or logical object device.

The purpose of an OBD client is to package OBD commands in a format suitable for a network protocol and to dispatch such command packets to a network interface. The purpose of a target is to receive these packets from a network client OBD driver, and to translate them into methods dispatched to an object device. Obvious candidates for network protocols supported by clients and targets are IP, VIA, and SCSI. These can then be exploited over a variety of transport layers, such as Ethernet, System I/O, SCSI, fibre channel, and Myrinet, to mention a few.

With the introduction of targets and network OBD drivers we see the possibility of clustered object-based storage. Clustered layouts of object storage devices will use clients and targets to implement the OBD protocol across an interface.

We have introduced several software modules here. We have mentioned object-based protocols which are modeled on a client/server interface, i.e. request processing interface. Several protocols will be exploited, such as software protocols and a variety of network components. The stacking of drivers and protocols is important to understand.

The stacking of drivers describes from where a module may receive commands and where it may send other object commands, to process requests. Table 1 below summarizes some of the protocol behavior of our components while table 2 depicts the relationship between the software modules.

## 5.2. Fundamental Definitions

**5.2.1. Driver Organization and Initialization.** All the devices are managed by the OBD class driver interface. The OBD **class driver** has the function of the VFS layer (but it is considerably simpler) and has been implemented as a character device.

| Protocol name | Served by... | Clients are... | Propagated through... |
|---|---|---|---|
| Software | Direct OBD Drivers<br>Logical OBD Drivers<br>OBD Client Drivers | Applications<br>Logical OBD's<br>OBD Targets | Software Interfaces |
| Network | ONC Targets<br>VIA Targets | ONC Client<br>VIA Client | SUN RPC TCP/IP<br>InfiniBand |

TABLE 1. Protocols in Object-based Storage

| Module Name | Accepts Requests from.... | Makes Requests to... |
|---|---|---|
| SOA | Users of the SOA | Direct OBD Drivers<br>Logical OBD Drivers<br>OBD Client Drivers |
| OBD Targets | OBD Clients | Direct OBD Drivers<br>Logical OBD Drivers<br>OBD Client Drivers |
| OBD Clients | Logical OBD Drivers<br>SOA's<br>Targets | OBD Targets |
| Direct OBD Drivers | Logical OBD Drivers<br>SOA's<br>OBD targets | Storage Devices |
| Logical OBD Drivers | SOA's | Direct OBD Drivers<br>Logical OBD Drivers<br>OBD Clients |

TABLE 2. Interaction of Object Storage Software Modules

An OBD **type** is a data structure specifying specific methods for a particular type of OBD device. It is similar to a filesystem type. Just as a filesystem module registers itself with the VFS layer, an OBD type registers with the OBD class driver.

OBD **drivers** are the software modules defining an OBD type. They attach to the OBD class driver to form a named instance of an **object device** that exports a table of methods suitable of storage object interaction. Such methods are suitable to initialize the device and to fetch objects from the device. The driver's data structures are defined in a *struct obddev*.

This structure is analogous to that of the super-block of filesystems: it refers to an OBD device of a specific type. It has generic data for OBD devices, methods specific to the particular OBD driver, and data specific to the particular OBD device of this type.

**5.2.2. Naming and Identification of Devices.** The naming and identification of OBD devices needs careful consideration given a history of problems in this arena. Each direct driver should have a UUID to uniquely identify it to the system. The UUID may be accompanied by a user readable string. When drivers discover devices or make themselves known to other parts of the system, they will use the UUID for identification.

Logical drivers express their dependencies on other drivers using the UUID and are themselves defined through a UUID.

A system will need to be designed which allows for the storage configuration of inter dependent drivers on the storage devices used by direct drivers. For example, a RAID1 logical driver may want to leave its configuration data on every drive it is mirroring.

In a running system, object devices need to be identified in an efficient way. We see two simple solutions. The class driver can maintain the device attachments in a running system and name devices by integers. It could export the association of UUID's with its device numbering scheme. Note that this scheme is transient and will be reset when the system restarts.

Alternatively, but less attractive, is for a character device with major and minor device numbers to represent OBD devices, which we will designate through pathnames `/dev/obdX`. The OBD class driver is a device driver module controlling this device. Each minor device has a type, which characterized it as a specific direct OBD driver, logical OBD driver, or OBD client. OBD driver types are registered through loading or initializing in the kernel so-called OBD type modules with the appropriate definitions. Such modules register themselves with the class driver, thereby making their type known through a character string and their methods through pointers.

**5.2.3. Object Namespaces.** The NSIC OBD specification suggests that the namespace for objects not be flat but divided into object groups. This is highly desirable, since it will often be very advantageous to be able to create object with given object *id*'s. For example, filesystem snapshots would benefit from object groups: each snapshot could have its own group *id*. A group *id* could be a 32-bit or 64-bit integer.

Objects are identified by naming an object device controlled by an OBD driver, naming an object group, and a *struct OBDid* structure. The opinion of experts is that this should be a 128-bit structure to ensure UUID's fit into it.

The NSIC spec is not sufficiently clear about this topic. For example, we propose that a single object *id* may appear under multiple object groups, but the spec does not address this issue.

The introduction of object groups as we suggest has major implications for drive implementation, since the object store is not a flat indexed name space. The benefits of having object groups far outweighs the disadvantage of having greater complexity.

Object-based drives should provide a mechanism through which information about object namespaces can be retrieved from the device and control exerted over them. Many systems will find it difficult to handle object *id*'s that are not presentable as 32 bit integers, and others will want to know what range of object *id*'s may be allocated on the drive. The following list is a proposal for some of the desirable knobs on the namespace:

(1) Limit the size of object *id*'s that the drive will generate to 32-bit, or 64-bit, etc.

(2) Provide information about the lowest, highest values of the object and group *id*'s that the drive will generate.
(3) Also, provide information how many objects and object groups can be stored on the drive.

Another point of disagreement we have with the NSIC specification is that it states that the object *id* will be generated by the drive. Many storage management applications will benefit tremendously from the possibility to create objects with a given object *id*. Of course the drive can return an error if the requested object *id* already exists.

The reason why this is so desirable can be seen by looking at filesystems. It is reasonable to ask that an OBD-based storage facility can move filesystems by moving objects. Unfortunately, object *id*'s are written into directory contents. So if directory *foo*> lists file *bar* as having OBD *id* 4711, then after copying this to a new OBSD we would still want *bar* to have *id* 4711. The only way we think this can be achieved is by allowing software to attempt the creation of 4711 specifically on the new drive, as opposed to leaving OBD *id* allocation to the drive. Without this, we would have to reparse the directory and lose an important feature of OBSD storage management.

5.2.3.1. *The Role of ReiserFS.* ReiserFS is a filesystem which has a storage layout totally based on balanced trees. It is a sophisticated environment, which would lend itself to a very good, scalable implementation of a fully featured object-based storage device.

The code to ReiserFS can be obtained under commercial licenses which allow proprietary elements to be added to the publicly available package.

The emphasis on trees in ReiserFS should make it straightforward to implement object groups as described here. Journaling is present in the filesystem and can be adapted to provide journaling support for the Object-Based Device (OBD).

**5.2.4. Meta-data Handling.** The biggest challenge in the object based storage architecture is the proper handling of **meta-data** or **attributes**.

The key questions are:

- Subdivisions of attributes...
  What attributes are private to the OBD, which belong with logical drivers, and which belong with SOA's?
- What attributes to store?
  We found it very hard to even get close to an answer. Attributes for filesystems are a nightmare to agree on. For logical object drivers, the meta-data is almost entirely dependent on the driver in question.
- Leasing attributes...
  When attempting to build logical object drivers, we found that leasing attributes to other drivers, for example leasing logical object driver attributes to the object-based filesystem, led to many difficulties regarding stale data. This is a topic that we will need to continue to explore until a good scheme has been found. Not leasing data will be a performance killer.

Inside a system a struct OBDO will be the type representing a storage object. The counterpart on the kernel side should be an object that holds the *OBDid* of the object, but also has the attributes, and other meta-data of the object incorporated.

Such storage objects can contain several distinct types of meta-data:

(1) Name of the object and device it resides on.
(2) Meta-data managed on the object storage device.
(3) Meta-data for use by logical OBD drivers.
(4) Meta-data and data for use by filesystems and other applications.

We will again draw an analogy with filesystems. Inodes in the kernel are instantiated through the *read_inode* super-block method, which is responsible (together with the VFS) for instantiating an intricate *struct* inode, which is totally self-describing. The identity of a file, the operations possible on it, and the associated data such as caches, super-blocks, etc. are all part of the *struct* inode.

**Struct OBDO's** should have the same properties. They should at the minimum have:

- Methods to *read*/*write* attributes.
- Methods to *read*/*write* data.
- A method to unlink the object.
- A connection to the *obd_device* which holds a pointer to the device structure.

OBDO's should be fully self-describing.

The rules followed by modules when interacting with the data are illustrated in table 3.

| **Module** | **Process** | **Pass Through** |
|---|---|---|
| Direct Driver | 1,2 | 3,4 |
| Logical Driver | 3 | 1,4 |
| Application | 4 | 1,3 |
| Client | - | 1,3,4 |
| Target | - | 1,3,4 |

TABLE 3. OBDO Meta-data Interaction

Struct OBDO's would have types, just as inodes do. There is a **U area** in the inode, which is used by the filesystems to store information unique to that filesystem. Similarly a struct OBDO has a **U area** the content of which is interpreted differently by different OBD drivers.

Objects may have in-line data. The existence of in-line data in an object should be checkable in a type-independent fashion.

## 5.3. Journal Recovery of the Object-based Disk Filesystem

In this section we will describe how, in outline, journal recovery of OBDFS can be achieved. It should be emphasized that this is a delicate project, which would take a very long time to totally stabilize in view of memory pressure issues, particularly on low memory machines.

Journal recovery also subtly interacts with aggregation, and only working through an actual design and implementation will reveal the full extent of the problems.

Despite these caveats, journal recovery for OBDFS is possible and attractive, and we proceed to describe how.

**5.3.1. Overview of the Journaling Mechanism.** The principle of journal transactions is fairly well known and covered in many papers (see Stephen Tweedie's paper in Linux Expo 1998).

A typical filesystem operation will modify several data structures, and unless the modifications to these as a group are atomic, the filesystem can end up in an inconsistent state. The simplest example is the modification of a directory and the removal or addition of an allocated inode. More subtle ones involve the modification of a directory, as above, but also taking into account the changes made to allocation bitmaps for inodes and blocks in the directory file.

To preserve consistency, the related operations are grouped into a **journal transaction**. Such transactions do not possess all the ACID properties: the durability is lacking, in favor of doing I/O asynchronously. The *journal_start* operation will designate a journal file and the size of the data the caller expects to need in the transaction log. It returns a handle, and every operation modifying the data structures will be passed the handle.

During journal transaction processing, I/O will send memory buffers to two distinct locations. One is the final location of the data, the other is the journal. The crux of the mechanism is that all data belonging to a transaction is first written to the journal and only when that is on stable storage the I/O to the final location may be started.

To prevent a lot of synchronous I/O, the transaction log itself is not flushed to disk immediately, but this is done lazily, and the atomicity is achieved through ordering (durability is not present).

The **log flush**, i.e. the I/O from memory regions to the log, can only be done on buffers that are not involved in transactions. This means that once flushing of the buffers associated with the transaction has been initiated buffers may not be overwritten in memory (and thus be made part of another transaction). Buffers are made *copy* on *write*. In certain cases up to three copies of a single buffer may be present.

When the log flush is done, I/O to the final destinations of buffers can be done normally, i.e. asynchronously.

**5.3.2. Journal Modifications to the Ext2OBD Driver.** First, the OBD API needs to change. The OBD interface should allow a transaction *id* to be allocated, and passed in with operations. As for every other allocation, pre-allocation of a transaction structure with a handle is necessary for good performance. Of course we also need an operation to commit the transaction and intermediate operations should pass in the handle to indicate to what transaction the operation belongs.

The direct storage driver exploiting Ext2 format is called the Ext2OBD driver. This driver needs to be adapted to use journaling. Based on the transaction handle that is passed in, the driver should handle a group of operations as a journal transaction. That is easily done, since the Ext2OBD driver can be adapted to act like an Ext3 driver and now operations can be made part of transactions, at the level of the direct OBD driver. This guarantees that at the level of the direct OBD drivers, updates are grouped in transactions.

More serious is that at the level of the page cache, multiple versions of a single page belonging to an object may need to exist. These pages must be marked *copy* on *write* and be grouped according to transactions. This involves rather non-trivial modifications to the page cache. It is possible, but not certain, that the journaling implementations for both ReiserFS and Ext3FS will face the same issue when they are ported to Linux 2.5. At present only one decision has been made, namely that the journaling mechanisms will be independent of the buffer cache.

A sample implementation of journaling support in the page cache could instantiate multiple *struct address_space* structures to handle the grouping of versions of pages due to transactions. These address space structures form the per inode page cache and are used for I/O. The arrangement of the address space structures is that they could hold snapshots of pages associated with certain transactions. The COW on pages would move pages to those *address_space* structures that hold the frozen pages associated with a transaction that is to be written to disk.

Such address spaces can be reused to coalesce multiple transactions, but when a log flush is initiated, the pages belonging to those transactions that are being flushed out must be protected from further *writes*. Any new transaction that is started in the meantime can create further *copy* on *write* pages, leading to the expected maximum of three copies max per page.

Further serious issues will arise from memory pressure and the need to flush transactions out. This has not yet been addressed in the Linux kernel, and likely mechanisms we use for this would adapt such mechanisms. Logical object volumes also raise important questions. Each logical OBD driver would need to group its operations by transaction, and to make matters worse, this can mean that distributed transactions enter the picture. Here, much thought will need to be given to the design, and likely optimistic concurrency methods such as explored by Cheops need to be brought into play.

## 5.4. The Snapshot Logical Driver

In this section we will describe the functioning of the snapshot logical object driver. Such volumes will be clients of the class driver and attach as usual. This is a case study in logical volume management with object-based disks and has revealed a lot of requirements for the interface. All of these were taken up and implemented in the code.

Object stores can easily support snapshots. Here we describe how to build snapshots within a single object volume. A different construction is possible, which uses a separate volume for each snapshot. The latter allows the primary object volume to remain a consistent filesystem, while storing snapshots in the same volume creates a volume that can only be accessed through snapshot-aware drivers. If the good old filesystem has to get back to the volume at some point, it can be desirable to keep an object volume pure.

The snapshot object volume is characterized by a sequence of snapshot times:

T1 < T2 < ..... < Tk. These times may lie in the future, *written* to current time.

The volume will contain two types of objects:

- Direct objects; ones that hold filesystem data.
- Redirecting objects (in data array of object).

For example, the redirecting object might be laid out as follows:

0 -> currently active direct object with current data.

1 -> object in snapshot at T1 (oldest snapshot).

2 -> object in snapshot at T2.

3 ->.

4 -> — perhaps no snapshots yet —.

5 ->.

6 -> — no snapshot here until future snapshot time passes —.

7 ->.

### 5.4.1. The Behavior of the Snapshot Volume.

The**:** behavior of the snapshot volume is as follows:

**Attaching the volume:**

- The snapshot number S which shall be used is given to the device.
- The list of snapshot times (T1, ... Tk) is passed to the drive.

**The volume will be:**

- *read*/*write* if S is 0. We speak of the primary volume.
- *read*-only if S>0For write/setattr operations.
- Check {mc}time T of object n.

  **For *read*/*getattr* operations:**

  - If accessed object is direct return the appropriate data.
  - If object is redirector:
    - If snapshot number = 0, use attributes of redirector & use data of direct object 0.
    - If snapshot number > 0, use attributes & data of correct direct object.

  **For write/setattr operations:**

- Check {mc}time T of object n. Suppose:

  T1 < ... Ti < T < = T i = 1 ... < current time < Tk. This means the object definitely does NOT belong in the snapshots T1 ... Ti, since it was already modified after Ti. If the object is a direct object, we know that the object was created after Ti, since otherwise the modification that occurred after Ti would have created an indirect object.

  But from now on the correct value for the indirect object for Ti+1 ... Tk-1 is object "n".

  Proceed as follows:

  * Make a new direct frozen object "M" say. Copy the current data of "n" to "M". This data is data in "n" if n is direct else, when n is a redirector, data is in N, which is pointer 0 of n.

  If n is a direct object do the following:

  * Create a new direct object "N" - give N the data of "n"
  * Turn "n" into a redirector
  * Set pointer 0 to N.

  Finally update the pointers:

  * Let n point to "M" for (i+1) ... (k-1)
  * Leave alone pointers for 1....i.

The COW copying is now complete, so write new object data to N, meta-data to n.

**Create operations:**

Always create a direct object.

**Delete operations:**

Consider *ctime* of object:

- If object is new:
  - Delete current 0 pointer or direct object when:
- If the object is indirect:
  - Make pointer 0 equal to 0.
- If object is older:
- For direct objects:
  - Make a copy "N" with data of "n".
- Now for any object:
  - point 0 to 0 (the non existent object)
  - point (i+1) ... (k-1) at "N".

Further issues are *the removal of a snapshot* and *rolling back to a certain snapshot (snap restore)*. Both have been implemented with an iterator.

Our implementation requires a client with the following behavior upon attach:

- Give underlying object volume.
- Give the snapshot number under which this should be accessed.
- Give a list of timestamps at which snapshots occurred.

E.g. the setup sequence for using an Ext2 object store as a snapshot volume:

```
obdcontrol
% attach ext2_obd /dev/obd0 /dev/hda2
% attach snap_obd /dev/obd1 /dev/obd0 0 3 time1 time2 time3
% attach snap_obd /dev/obd2 /dev/obd0 3 3 time1 time2 time3
% setup /dev/obd1
% snap add /dev/obd1 `date`
% snap list /dev/obd1
  * 0 current <------ used by /dev/obd1
  1 last month
  2 last week
  3 yesterday
% snap list /dev/obd2
  0 current
  1 last month
  2 last week
  *3 yesterday <------ used by /dev/obd2
% snap del /dev/obd1 2
  Busy (?)
% quit
mount -t obdfs /dev/obd1 /mnt/obd
mount -t obdfs /dev/obd2 /mnt/obd-yesterday
```

### 5.5. Caching OBD Driver

An important aspect of a filesystem is read scalability. Many client filesystems may try to contact the file servers to access data at the same time, e.g. when a cluster boots that has shared root and filesystems. If every client has to read data from the same single server, the network load at the server would be extremely high and the available network bandwidth would pose a bottleneck for the system. We are introducing a collaborative cache (COBD) driver that will distribute this read load over multiple nodes which can be dedicated cache servers or clients. In a filesystem using COBD, there could be copies of the same data at multiple nodes, so *read* for the same data could be serviced by these various nodes. This can result in an unprecedented improvement in scalability for *reads*.

Caching nodes use existing client (OSC) and server (OST) infrastructure but introduce a new object driver, the Caching Object-Based Driver (COBD). The COBD can be introduced on client nodes or run on dedicated caching servers and such nodes can now service the client *read* requests. This plan would considerably reduce the number of requests that need to be serviced by the target OST's.

In this section we will explore the various issues pertaining to the design of such collaborative caches; some of the important issues are failure recovery, collaboration management, and cache coherency.

**5.5.1. Read Cache.** This could be running on a dedicated cache server or a peer node (Client1 in figure 5.5.1) that would export an OST so that it can receive and process *read* requests. If COBD was on a peer, i.e. client node, it will be used by the filesystem to service requests. Lock requests are still made to the target OST, so we will disable any lock granting ability at the caching OST (on top of COBD). The COBD should simply service the *read* request. When accessed by the client filesystem, other requests are passed through to the OST by the COBD, but when accessed by the OST, the COBD only honors *read* requests.

The COBD needs to maintain a cache of mapped pages of objects. This cache needs to shrink when kernel memory comes under pressure. We will implement this by associating cached objects with (in memory) inodes. Inodes have an associated address space to manage a page cache on which the kernel can exert pressure and which include suitable callbacks for cleanup of related data (such as locks) in the COBD. If this page cache is backed by swap it will be possible to use persistent storage for large caches.

Even more attractively, but not planned immediately, we might try to make the persistent cache use a filesystem so that it might be usable after a reboot. In this case a direct OBD would be used for storage.

The COBD needs to maintain *read* locks on the data it caches to ensure its consistency. It fetches such *read* locks from the target OST, as the filesystem is presently doing, but there is one modification. The target OST, when faced with a client lock request, will grant as big a lock as possible. For a request initiated by the COBD, the target OST will grant a lock that reflects the actual data that the COBD caches. This puts the target OST in a position to have reliable information about which referrals of *read* requests can be served from cached data.

**5.5.2. Referral Module.** The target OST will receive a read request and must select a caching OST to service the request. It will use the information from the lock manager on the target OST to search for caching OST's that have the file extent cached in the COBD.

As shown in figure 5.5.1, the referral decision is made in the new Referral/Policy module between the OST and OBD. This new module will take as input a resource name (file & extents) and determine the nodes that have a lock on it and the type of locks held. The request should be referred to COBD only if *'PR'* compatible locks are in the granted list for the resource. The referral module has to choose one caching OST from the list of caching OST's available; it can have a policy module that would determine which caching OST to select (a simple scheme would be selection in round-robin manner).

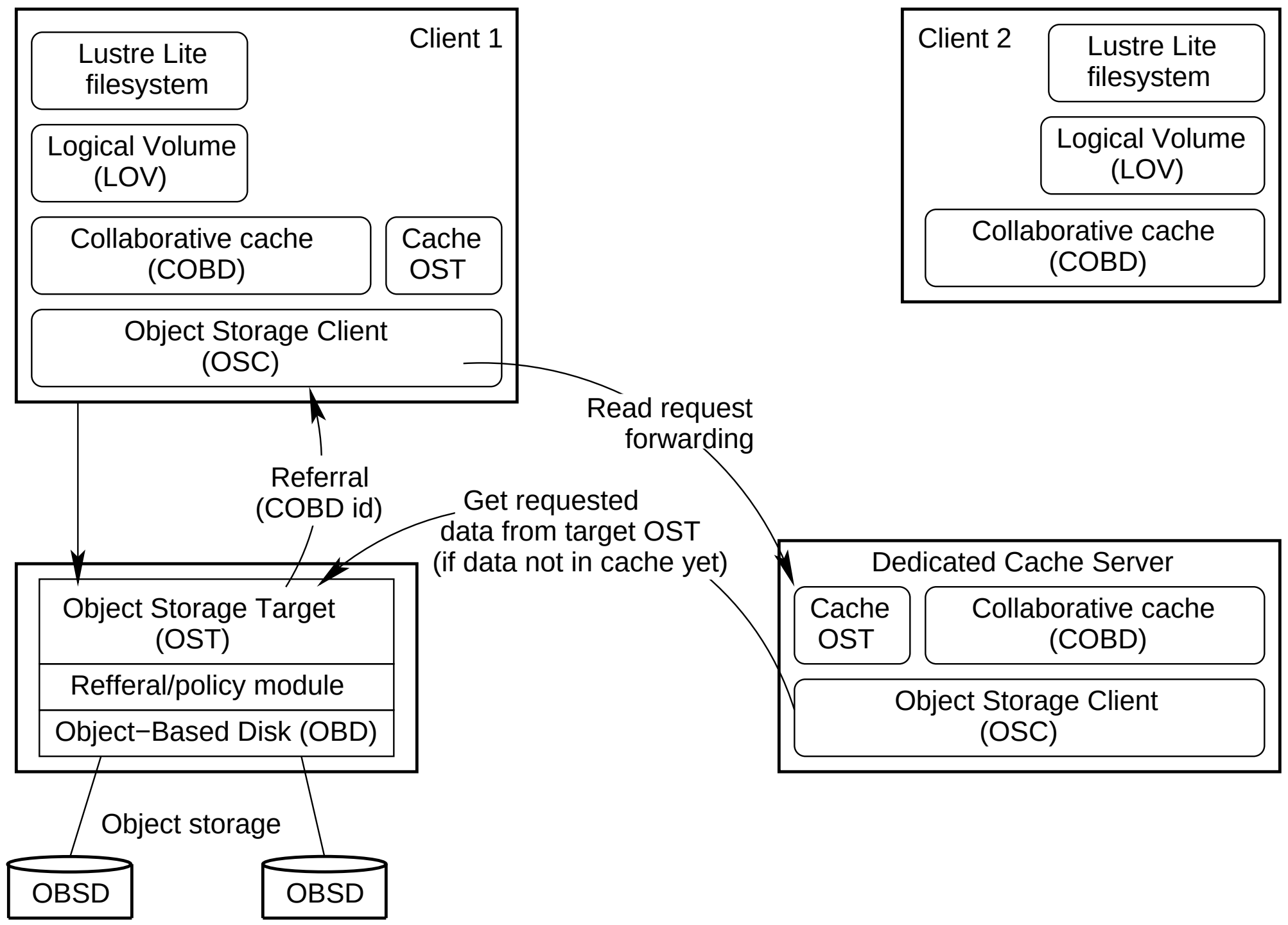

FIGURE 5.5.1. COBD Architecture

If a caching OST with the requested data is found, the target OST will return a referral which consists of the caching OST UUID. This can be used by the OSC to forward the read request to the COST. If no caching nodes are found, the target OST makes a cache population policy decision and may still direct a client to a caching OST. When the caching OST receives the *read* request, the COBD will not find it in the cache, take out a lock, and use the standard OSC to populate its cache.

## 5.6. Changelog

**Version 3.0 (Dec. 2002)**

(1) P.D. Innes - updated and resized figures

**Version 2.5 (Nov. 2002)**

(1) P.D.Innes - updated, tables turned into floats

**Version 2.0 (Nov. 2002)**

(1) Radhika Vullikanti - made some revisions in the document, added missing figure, added a new section on COBD

**Version 1.0 (Dec. 1999)**

(1) P. Braam - original draft

CHAPTER 6

# The Lustre Metadata Service

## 6.1. Introduction - Subsystems and Protocols

The Lustre file system uses two API's to reach services and manage persistent storage of data. The Object Storage API is primarily used for file data. The metadata API is capable of managing the namespace of the file system. It is through the metadata API that new files, directories, symlinks and other namespace elements are added to, removed from or renamed within the file system. The metadata API's can be exported by a Lustre device and can run on the same or a separate device as the device exporting the object API to the file system.

The API for metadata bears similarity with the API's found in other networking file systems, but at a more detailed level many differences can be found. Lustre offers two execution modes for metadata operations: one uses write behind caching of metadata another a client server model where file system transactions are typically reduced to a single RPC. A second difference is that Lustre offers clustered metadata service, where multiple metadata servers cooperate to load balance a typical stream of requests received from a cluster of client nodes. The Lustre metadata service works with the distributed lock service to manage synchronization of data cached by the clients and other metadata servers in the case of clustered services. The lock services require a fairly tight integration with the namespace of the file system to support metadata write back caching.

This chapter will describe the metadata service, first from an API perspective. The API is what a client system requires to update and access the file system. After that we focus on the recovery properties required from the service. We then focus on possible architectures for the implementation, describing both the client-server approach, the write back caching and the clustering of metadata services. The management of file sizes and update times ties together the metadata service and the object storage service and is covered in this chapter.

## 6.2. The Meta Data Namespace

**6.2.1. The Name Space Elements.** The metadata API provides the mechanisms to ensure a *single view* of the namespace of the file system. A *single view* requires that all clients in a cluster see exactly the same namespace for the file system directory tree.

The **inodes** described by the namespace are *objects* (not to be confused with data objects although those could provide an implementation of metadata inodes) which are uniquely identified by a *file identifier* or *fid* provided by the metadata service. The **elements** that build up the namespace are triples which describe a node in the namespace. A *parent directory*, which is identified by a parent fid, a *name* and the fid of the inode pointed to by the name are the key elements of the namespace. The lookup mechanisms found in file systems or libraries provide the clients of the metadata services with these namespace elements containing:

(1) a parent fid
(2) a name
(3) the fid of the namespace inode associated with the name.

It is possible to build different descriptions of the namespace. For example, the namespace could be described through a database of full path names and fids. This hides more of the internal structure which standard file system infrastructure utilizes, such as browsing through the namespace, and using the parent child relationship of namespace elements. However, other file system operations, particularly the creation of new inodes and performing I/O to them are relatively independent of the detailed internal directory tree structure of the namespace.

The *lookup* process establishes the existence or non-existence of elements in the namespace. This process happens very frequently on client systems. Maintaining an efficient coherent view of namespace elements is done through a cache of lookup results on the client systems in conjunction with a lock management system. In addition to caching existing elements as a triple described above, it can be very beneficial to cache the non-existence as a special triple of a parent fid, a name a *null* fid for the non-existent inode.

Lustre maintains one important invariant for namespace elements: a *fid* has the property that -

(1) It will be used at most once. The *fids* are never re-used
(2) An inode is uniquely identified by its *fid*.

These properties simplify the coherency of the namespace.

It is not common for other file systems to expose unique identifiers to clients, but in Lustre it is quite important that the unique *fids* are exposed to the clients. There are implications for recovery associated with this requirement that will be discussed later.

**6.2.2. API support for lookup.** There are many different approaches that can be taken to perform lookup operations. A client can request the *fid* and other attributes for a name in a directory through a *getattr_lock* operation. A successful return of this call indicates that the server found a namespace element and returned it to the client. This approach establishes a single namespace element, through a single API call when coupled atomically with the acquisition of a lock guaranteeing the validity of the element to the client until such time that the lock sees a revocation callback. A *getattr* API call can instead use a *fid* to return the attributes for that *fid* if the server has an inode with that *fid* in its namespace.

A different mechanism to establish namespace elements is for the client to obtain parts of the directory file through a *readpage* operation. If the directory file data is known to be valid through a lock on a part of the directory data, the client can establish the existence and *fids* of elements without further invocations of the API. This can lead to significant reduction of client latency for lookup and permit the use of negative elements, but when directory data sees frequent updates and is cached by many systems, the invalidations of the locks can become prohibitively expensive and limit scalability.

**6.2.3. Synchronization of the namespace elements.** Lustre will support multiple mechanisms to ensure namespace synchronization across multiple client systems. In the first model, each element is managed separately. A client can use an element in the namespace only if it holds a lock for the element and the lock guarantees that the entry as cached is still valid. In order to avoid race conditions between revocation callbacks of locks and the use of elements or attributes, Lustre strictly enforces an ordering which mandates that locks are acquired *before* the associated data the lock protects is used. This still causes locks and data to be provided to clients through a single RPC using the *mdc_enqueue* API. This *mdc_enqueue* API call is made to the lock service on the metadata server, which invokes a metadata intent policy function. The policy function analyzes the full request, obtains locks and returns data with locks to the client. When locks cannot be granted synchronously the lock server returns the attributes as part of the lock completion callback made to the client when locks have been acquired.

The write back model is the second model. Here a single lock protects a subtree of the namespace. In order to establish the existence of such a lock, the child / parent relationships in the namespace need to be used to find an ancestor that holds a lock. Even more complicated is the acquisition of such a lock, as it requires locks held on any element in the subtree by other nodes to be revoked. This process forces a tie between the locks on individual elements and the namespace structure maintained by the MDS so that descending the namespace will make all locks available for cancellation callbacks.

## 6.3. Attributes

**6.3.1. Attributes.** In the previous section we already discussed the *fid* as the identifying attribute of an inode in the namespace. The inodes have many other attributes which are discussed here:

- **protection attributes:** owner, group owner, mode bits and access control lists are attributes of inodes that are primarily used for authorization.
- **directory content:** directory content serves many important purposes. The server modifies directories persistent storage when making updates. Clients use directory data for lookups when write back caching is enabled and for listing directories under all circumstances. The directory data protocol is subject to review for integration with other file systems or database services offering persistent storage for metadata inodes and their attributes.
- **size and modification times:** when file inodes or directories are modified their sizes change and their update times need adjustment. As we will see later, Lustre distributes data over multiple nodes in a cluster and the management of these attributes is quite distinct from others.
- **location attributes:** Regular files can store their data in one or more objects that reside on object storage targets. The metadata inode for the file contains a descriptor of the location and usage of the data objects. Large directories have a similar pattern and segments of the directory, called *buckets,* these can get scattered over multiple metadata servers. A similar extended attribute describes the location and purpose of the directory buckets.

**other extended attributes:** Lustre provides mechanisms for applications to store extended attributes in Lustre inodes, either on persistent storage or merely in memory. This provides important features for clustered file services such as CIFS and NFSv4 file servers to synchronize service metadata through the Lustre file system between the nodes on which the service is running. The detailed API's here remain to be determined.

The target of a symbolic link inode is a very special attribute. No API's are commonly used to modify the target name of a symbolic link, the only operations on symbolic links are to read the target, create and unlink the inode. This makes the management of symbolic link targets quite simple, they are rigidly associated with elements in the namespace.

**6.3.2. API support for managing attributes.** Lustre metadata service provides a mechanism to obtain attributes of inodes through the same API call that is used to establish elements in the namespace. The *getattr* call can obtain the *fid* and all other attributes of inodes on the server. The key data structures here are:

**attribute descriptors:** The attribute descriptors fall into two classes. One contains the standard attributes associated with inodes in file systems, including file identifiers, mode bits, owners, modification time etc. All other attributes such as location attributes, ACL's and other extended attributes are returned to the client as opaque buffers. Directory data is sent using the bulk protocol.

**file name, fid:** A *getattr* call always takes a *fid* argument. If a filename is not given, the attributes for that *fid* are returned. If a filename is included, the MDS service interprets the *fid* as the *fid* of a directory, then searches for the name in the directory and returns the attributes associated with the element defined by the name. In this case the server also returns the *fid* of the element to the caller.

**valid flags:** The valid flags indicate what attributes should be returned to the client.

Normally clients should not use attributes without having a lock on the attributes. Accordingly the attributes are typically returned to the client with a lock through the *enqueue* API discussed above.

**6.3.3. Synchronization of attributes.** Lustre Lite (aka Lustre 1.0) handled all attributes and element synchronization with a single lock for an inode. In some cases this is less than optimal. For example, a client repeatedly modifying the attributes of an inode should have the element cached. But the metadata service would benefit from a lock giving it exclusive access to the attributes associated with the *fid*. If these two locks coincide a *ping-pong* effect will be seen. The RPC model in Lustre still largely eliminates *ping-pong* effects but multiple locks are desirable.

In client server mode Lustre will have different locks for the 5 types of attributes listed above and for the elements. In write back caching mode, a single lock will be used to protect all aspects of the subtree.

### 6.4. Updates to the namespace

The validity of elements changes when they undergo a *rename, unlink* or *rmdir* operation. New elements are introduced when *create* operations insert symbolic links, and special files or files.. *Link* operations introduce hard links and *mkdir* operations introduce new directories. The *setattr* operation changes the attributes of an inode in the namespace. Of special interest is *open* which not only provide facilities to create new file inodes but also includes handles to these open files, and provides atomicity guarantees when an *EXCL* flag is passed in conjunction with create.

**6.4.1. Server based update operations.** Lustre separates most metadata updates in separate components that run on the metadata server and routines that run on the client. In client server mode, the clients for the most part issue requests to the server and do not update their cached components of the namespace during the update. The changes to the namespace are discovered later through lookup calls. The exception to this call is *open* which we will discuss later.

In write back cache mode, the primary operation is the opposite, namely the client updating it's cached namespace. A record is prepared for the corresponding MDS request that is transmitted when the metadata cache is flushed. Except for *open* calls the records that are prepared during write back caching are identical to those found during client server mode. Opens are purely local operations when write-back mode is active.

**6.4.2. mds_reint.** The following operations are reintegration operations:

**mds_reint_create:** create files, symbolic links, directories, special files. The arguments are a *parent fid*, a *name*, a *child fid* and extra attributes like the mode, target (for symbolic links) and device number (for special files).

**mds_reint_unlink:** remove files or directories. Arguments are the *parent fid* and a *name*. In this case the server returns the extended attributes to the client when the last link of a file is removed so that the client can remove the corresponding data objects. In the case of unlinking a regular file an exclusive lock on that file object is granted to the client to avoid a race condition.

**mds_reint_rename:** rename operations on files. This takes two *parent fid's* and a *source* and *target* name.

**mds_reint_link:** hard links. This takes a *fid* for the file to link, a *parent fid* and *name* for the destination.

**mds_reint_setattr:** update attributes. The arguments are the *fid* of the object to be updated and the attributes to be set. Setattr calls, except for the *lchown* call involve following symbolic links on the client. The server cannot do this, therefore the client performs the lookup and ships the *fid* for the setattr operation to the server. Setattr does not update the attributes on the client, a new lookup operation will do this. This could be optimized if we want maximal performance for pairs of calls setting attributes followed by *stat* calls.

All of these operations perform the corresponding changes to the namespace on the server. The return consists primarily of error codes. While these routines could return attributes for objects that were created this is not normally done, because the client would need to atomically acquire a lock as well. So the client detects namespace updates subsequently though normal *mds_getattr_lock* calls used for lookup.

The MDS itself will take a write or exclusive lock on the parent directories in fid order in the rename case. It will then perform lookups for the children and lock the fids that were established if necessary (not necessary for creates). The namespace update proper is then done, the recovery information is journaled with the update and the locks are released.

**6.4.3. mds_open.** The open call cannot easily be shipped to the server since the client has to support the server by following symbolic links, just like in setattr above. In this situation the intent solutions appear to be the only really good ones. The intents allow us to keep the decision path and the atomicity guarantees identical on the client and the server.

The open code path in the MDS is somewhat involved. Open involves lookup of the path, including following symbolic links, possible creation of the new file and then a file open. The mds open path resembles the *open_namei* code found in typical VFS systems very closely, including checks for creates, O_EXCL handling, detecting the opening of a symbolic link etc.

The client system will typically do a lookup on the file to be opened. This lookup requires a lock for the validity of the dentry that is returned. As a result the open request is piggybacked on a lock enqueue through the intent mechanism.

The status returned by *mds_open* consists of several pieces of information:

**disposition:** contains a bitmap that documents exactly what code paths on the MDS ran and what decisions where taken. E.g. the three phases, lookup, create, open are set, and further a negative or positive lookup result is documented.
**status:** this contains the error returned by the last phase that ran on the server
**attributes:** the attributes for the inode found, and if present the extended attributes with stripe information.
**lock:** if objects need to be created for a new file, the mds file object is returned to the client with a protected write lock so that only that client will create the objects for this file.

## 6.5. Write Back and Client Server Mode

The Lustre MDS offers clients two distinct locking protocols. In write-back mode a single lock is protecting an entire subtree which is given exclusively to a single client. In client-server mode objects can be shared among many nodes.

This section describes some of the mechanisms required to switch modes between write-back caching and client-server operation on the MDS.

**6.5.1. Switching Policies.** Several facts combine the policies Lustre implements in granting write back locks, or offering client server locks. Write back cache locks impose a very low resource usage on the client and MDS, at the expense of a somewhat higher synchronization latency.

When an object on which a write back lock has been granted is starting to see higher concurrency, this is easy to detect, and it is easy to give up the write back lock. Giving up a write-back lock means flushing all updates stored cached for the subtree. It is important that if another client requests a lock on an object below the write back lock of another, the updates associated with the entire write back lock need to be flushed. It is much harder is to assess if a collection of client-server locks on objects in a subtree make the subtree a good candidate for conversion to a write back lock.

To navigate between the modes Lustre will follow the following policies:

- The default lock to grant is a read or write subtree lock. If the resources associated with an object have seen frequent lock callback traffic recently a client server lock is granted.
- When this lock is granted other locks on the same object are revoked.
- When the client traverses the subtree below the write-back lock and fills its cache, other locks held by other clients need to be revoked. These other locks could be write-back locks.

In order to measure the traffic on a resource, it may be wise to cache resources for a short time before expunging them from the DLM to query the usage information.

**6.5.2. Mechanisms.** The MDS links write subdirectory locks on metadata objects to *dentries* in the *dcache* of the MDS namespace. When any lock on an object is enqueued a *dentry* is passed in and the parents are searched for subtree locks and these need to be revoked unless compatible with the lock mode requested.

In this way the MDS can keep an overview of the activity in a particular subtree by a any client. The MDS can keep counters around for locks that are used by particular clients. It can also track how many locks were granted to clients without conflicts. From these two pieces of data a simple policy function will decide that a particular client is better off working with a write-back lock.

**6.5.3. Reintegration.** The reintegration of update records associated with a subtree lock is straightforward. A batch of records can be sent to the MDS and they can be integrated one by one. There is no need for individual replies or acks and the standard infrastructure generating the transaction sequence makes this easy to recover in case of failures.

The write back cache metadata updates are extremely similar to InterMezzo (see [**1**]) reintegration and the details there might be helpful.

Another issue (already recognized by Coda) is that it is dangerous to have multiple users in a single update sequence. If a capability is refused by the server, the user owning that capability may block updates generated by another user. There are two reasons that capabilities might be refused: **token timeouts** and **administrative revocation authorization**. In the former case, update failures are needlessly annoying; in the latter, they are a beneficial feature and refusing all further updates from a client holding administratively revoked capabilities could be considered beneficial.

This rule doesn't have a big impact on the design, except in the case that an update is wedged behind another which might lead to a reboot of the client. A busy NFS/CIFS server working on a shared `file/directory` will see lots of updates on one object originating from different users. This is a case which merits further discussion.

## 6.6. MDS Recovery

The metadata service provides high availability. This means that sufficiently many features are available in the protocol to mask server failures to clients and client failures to other clients. Lustre has focused on implementing such transparent recovery in the case of single point MDS failures. Based on experience with large clusters, it is rare for multiple nodes to fail simultaneously. As we will see, this assumption allows for dramatic performance increases.

The MDS recovery infrastructure is required to ensure that a client is always exposed to a consistent namespace and does not ever see the results of partial or incomplete transactions. The system should be able to recover transparently in the presence of a single MDS failure or temporary network failures.

During recovery, clients should be able to replay all requests that the applications saw as completed but were never flushed to persistent store. Clients should be able to determine exactly what to do with each request during recovery and also ensure that the replay happens in the same order as the original execution. For example, before a failure client A completed transaction T1 and then client B completed transaction T2, the same order should be recreated during the recovery after failure.

The recovery infrastructure along with the various recovery protocols and APIs used are described in further details in the recovery chapters of the book.

## 6.7. Clustering Metadata

In order to provide enhanced scalability and performance, Lustre offers clustered metadata servers. This section will give an outline of the architecture.

The main challenge we face is to provide a substantial gain in scalability of the metadata performance of Lustre through great parallelism of common operations. This involves finding mechanisms which distribute operations evenly over the metadata cluster, while avoiding a more complex protocol involving further RPC's. The current trend in distributed file system design is to do such clustering by allowing clients to pre-compute the location of the correct services.

A second challenge is to provide good load balancing and resource allocation properties both for large installations where the metadata cluster acts in effect as a metadata server and in the case of small clusters in which the metadata cluster itself will access metadata on other nodes in the cluster.

Our architecture accomplishes this by heavily leveraging existing building bricks, primarily existing file systems and their metadata interfaces.

Finally the key challenge is to provide good scalability and simple recovery within the metadata cluster itself.

**6.7.1. Summary of metadata clustering configurations.** Overall the clustered metadata handling is structured as follows.

- A cluster of metadata servers manage a collection of inode groups. Each inode group is a Lustre device exporting the usual metadata api, augmented with a few operations specifically crafted for metadata clustering. We call these collections of inodes inode groups.
- Directory formats for file systems used on the MDS devices are changed to introduce a allow directory entries to contain an inode group and identifier of the inode.
- A logical metadata volume (LMV) driver is introduced below the client Lustre file system write back cache driver that maintains connections with the MDS servers.
- There is a single metadata protocol that is used by the client file system to make updates on the MDS's and by the MDS's to make updates involving other MDS's.
- There is a single recovery protocol that is used by the clients - MDS and MDS-MDS service.
- Directories can be split across multiple MDS nodes. In that case a primary MDS directory inode contains an extended attribute that points at other MDS inodes which we call directory objects.

6.7.1.1. *Modular design.* Client systems will have the write back client (WBD) or client file system directly communicate with the LMV driver: it offers themetadata api to the file system and uses the metadata api offered by a collection of MDC drivers. Each MDC driver managed the metadata traffic to one. The function of the LMV is very simple: it figures out from the command issued what MDC to use. This is based on:

(1) the inode groups in the request
(2) a hash value of names used in the request, combined with the EA of a primary inode involved in the request.
(3) for readdir the directory offset combined with the EA of the primary inode
(4) the clustering descriptor

In any case every command is dispatched to a single metadata server, the clients will not engage more than one metadata server for a single request. The api changes here are minimal and the client part of the implementation is very trivial.

6.7.1.2. *Basics of the operations.* For the most part, operations are extremely similar or identical to what they were before. In some cases multiple mds servers are involved in updates. Getattr, open, readdir, setattr and lookup methods are unaffected. Methods adding entries to directories are modified in some cases:

(1) "'mkdir"' always create the new directory on another MDS
(2) "'unlink, rmdir, rename"': may involve more than one MDS
(3) "'large directories"' all operations making updates to directories can cause a directory split. The directory split is discussed below.
(4) "'other operations"' If no splits large directories are encountered all other operations proceed as they are executed on one MDS.

6.7.1.3. *Directory Split.* A directory that is growing larger will be split. There is a fairly heavy penalty associated with splitting the directory and also with renames in within split directories. Moreover, at the point of splitting, inodes become remote and will incur a penalty upon unlink.

Probably it is best to delay the split until the directory is fairly large, and then to split over several nodes, to avoid further splits being necessary soon afterwards.

6.7.1.4. *Locking.* Locking can be done in fid order as it is currently done on the MDS. In order to obtain cluster wide ordering of resources, clients must chose the correct coordinating MDS, so that locks taken there initiate the lock ordering sequence to be followed. This is particuarly important for rename, which has to be started at the target or source directory, depending on which the highest order resource occurs.

**6.7.2. Resources.** The MDS handles the persistent storage of metadata objects and directory data. Internal to the metadata service is a large amount of allocation management.

The use of resources is easily summarized as follows:

**Names:**
(1) Look up the name in a directory
(2) insert / remove names in a directory

**FID:**
(1) get attributes for a fid
(2) create, remove the corresponding object

The ownership of resources varies among file systems. In local file systems a single node owns all resources. No parallelism can be achieved with this. In traditional clustering file systems, nodes own individual inodes or disk blocks. This leads to fine grained ownership of resources, but involves frequent collisions and poor locality of reference.

For Lustre we propose that each node owns a moderately large group of objects. There would be a large shared storage pool, which would be subdivided into relatively small file systems, this is shown in figure 6.7.1. We call the small file systems an **inode group.** Each inode group has its own journal for recovery, is formatted as a file system and can fail-over to another node for availability or adjustment of resources. We will make the load on the inode groups evenly distributed through randomness.

Clients will get a logical clustered metadata driver which exploits multiple MDC clients (see figure 6.7.2). Just like the logical object volume, the file system itself does not need to know the details of the object distribution, that can be left to a small logical metadata volume driver, invoked by the file system through the same API. The MDS system will get clustering and policy adaptations. The key to this is to add an **inode group** identifier to the fid, this marks the inode group to which an inode belongs. The resource database for the cluster will provide every client with a load balancing map which indicates on which MDS server a particular inode group is currently mounted.

The resource location will be managed as follows:

**File inodes:**

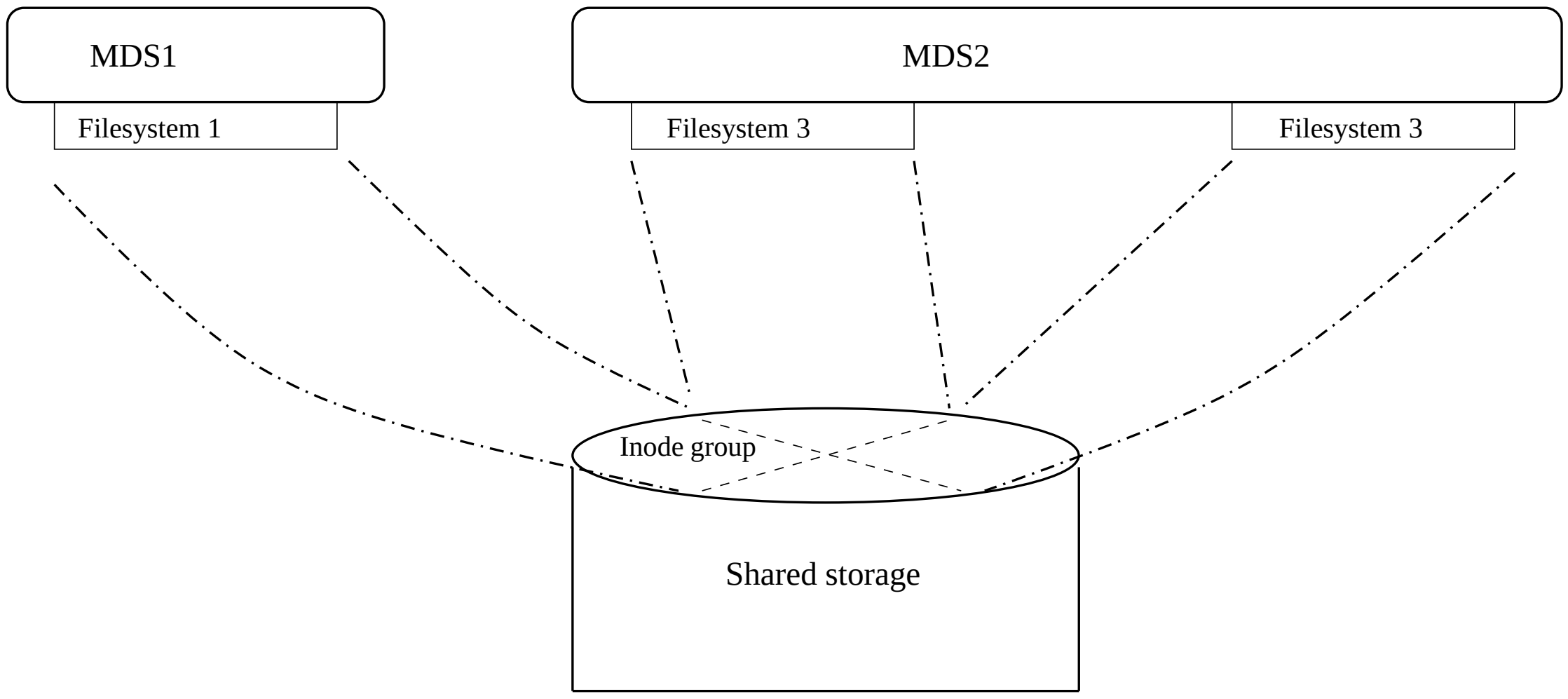

FIGURE 6.7.1. Metadata cluster using a single shared storage

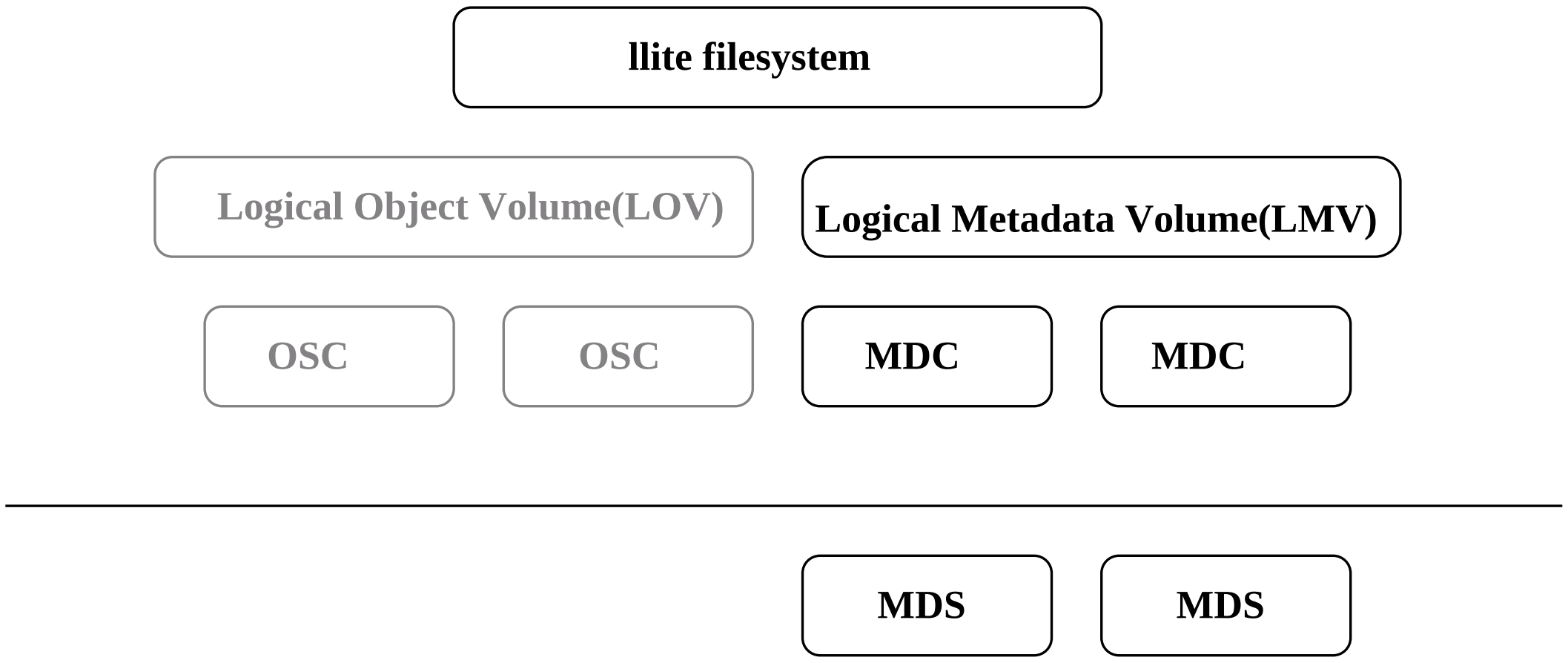

FIGURE 6.7.2. Logical metadata volume driver

- Create the file inode in the inode group of the directory inode holding the name

**Directory inodes:**

- Create in a new inode group
- The policy on which group to pick could be round robin, random, most space available etc. Probably every MDS reply packet should contain some status information to give clients policy information.

**Directory data:**

- While the directory is small, keep it with the inode
- When it grows fan it out.

**6.7.3. Clustered directories.** When directories grow we will split them up into **directory data objects** which are placed on multiple MDS servers, the figure 6.7.3shows this transition from a single directory to multiple directory objects. This is quite analogous to striped files, which are placed in data objects on multiple servers.

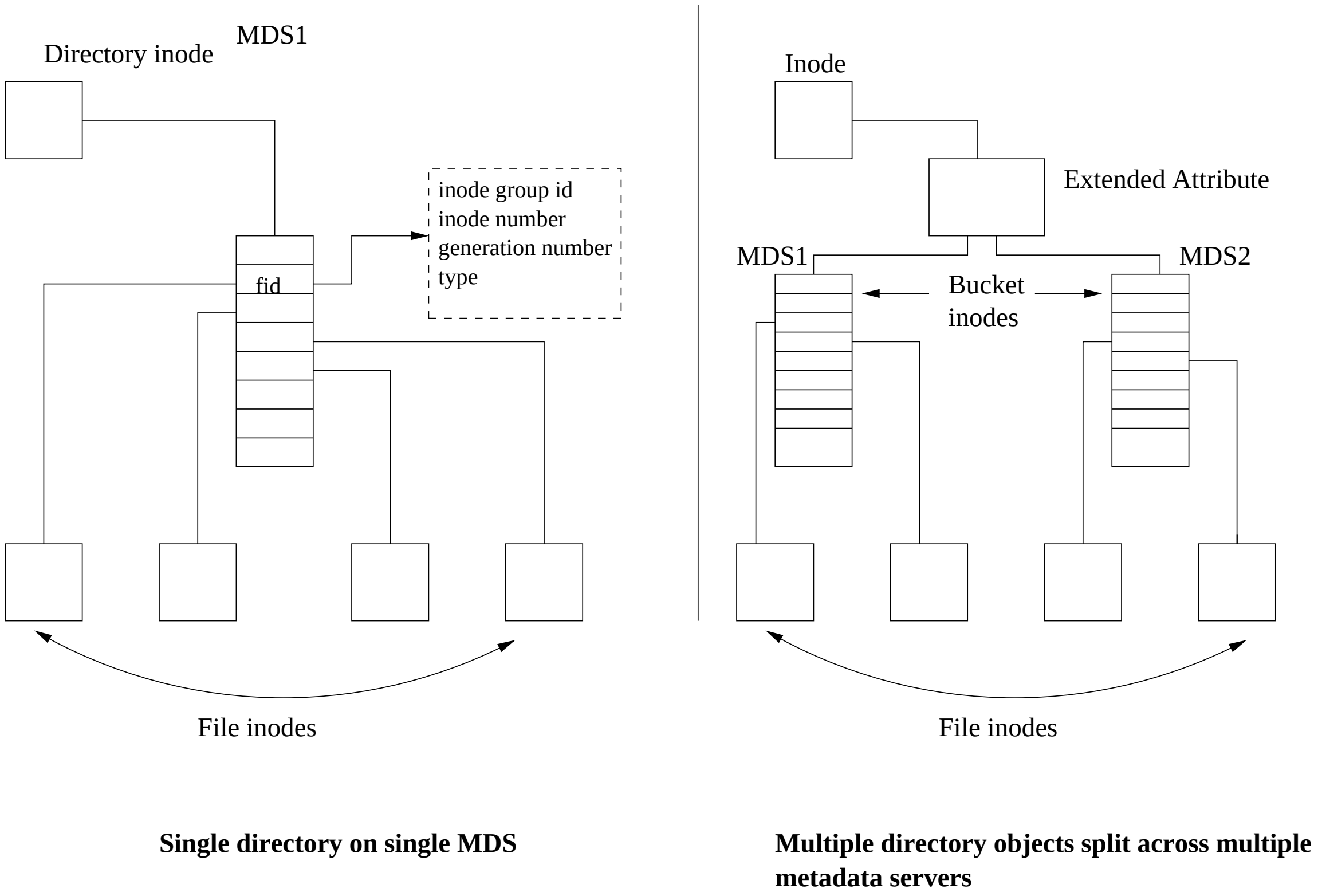

FIGURE 6.7.3. Transition from a single directory object on a single MDS to multiple directory data objects across metadata servers

Directory entries will hold a inode group identifier and inode number, compared to traditional entries holding merely a name and inode number. So once a name is found in directory data the inode group and inode number in this group is known.

**getattr_lock(parent_fid,:** name) To find the directory entry itself, the algorithm is similar to that of finding a file stripe. When a directory inode is located, the inode will either contain directory data in which case it is treated as a traditional directory. It can also contain an extended attribute describing what inode buckets exist, by specifying a fid for each bucket, each fid specifying its inode group, inode number and generation. A hash will then map the name to a particular bucket based on this metadata. A normal name lookup in the bucket will proceed to find the entry.

The worst case here is that this requires 3 RPC's. The first one to do a getattr on the directory inode which would give the extended attribute, the second to find the directory entry on the server holding the bucket, and the 3rd to find the inode attributes in the inode group associated with the entry. However, the common case is that a single RPC is sufficient, since normally the directory inode will be cached already, so the first RPC will go to the server containing the bucket. Furthermore, usually the inode is located on that server and will be fetched in the same RPC. The number of disk reads is identical or one higher than that for large non-clustered directories.

The process of creating a clustered directory is triggered by the directory growing beyond a certain size. The splitting of a directory occurs quite as early as possible, there might be a small effect to performance in the beginning when a directory is split. But the aggregate performance would be good since parallel operations can be done.

**6.7.4. Directory inodes and clustered metadata.** Directory inodes come in two variants:

**small directories:** An ordinary directory inode in a single inode group.
**large directories:**
- **master directory inode:** with an EA pointing to the buckets in other inode groups
- **bucket inodes:** in other inode groups. The buckets are associated with an inode that manages the space allocation for the bucket directory data. The bucket directory data describes the directory data covering a range of hash values. It provides a map from name to (group, inode number) to identify the fid up to the generation number.

The fanout operation, triggered by a directory growing beyond a certain size creates the buckets. This involves a new RPC in the MDS service that allows the creation of a remote bucket, and to populate it with directory entries.

This is a simple RPC that brings no complications to recovery since the buckets are exclusively visible to the the inode group of the master. It is possible that buckets are orphaned, and this requires cleanup.

Removal of a fanned out directory is similar in complexity. Here it is important to use an MDS to MDS reconnect handshake, identical to the client - mds handshake, between the master inode server and mds's holding the inode groups holding the buckets to handle the failure of MDS servers that have buckets that need to be removed.

The security of such MDS-MDS interaction is probably most easily managed with a capability model similar to that found between the clients and OST.

The attributes of clustered directories are most easily managed in a distributed fashion as we do for the file data objects.

**size:** sum of all the bucket sizes
**link count:** sum of all the bucket link counts **mtime:** latest of all mtimes

**6.7.5. Clustered MDS protocol.** The clustered MDS protocol involves a few changes to the API implementation found above. Most of the changes involve some new API calls between MDS servers. The goal is to use a single recovery infrastructure among the MDS servers and the clients, as described earlier in this chapter. Some detailed works remains to be done for the design to avoid cyclic lock dependencies or acknowledgment graphs (refer to section 11.3.6). As described previously in section 11.3.6, we now enforce *ACKs* for replies. The MDS takes locks on the resources it modifies, these locks are canceled once *ACKs* are received. In the clustered MDS scenario, it is important to ensure that a deadlock is not caused as a result of the various systems waiting for *ACKs* from each other.

**mds_create:** This call needs modifications when creating a new directory, because the new directory inode and new directory data will be created on another MDS server than the parent.

The node holding the parent directory data will do a lookup, find it's negative and hold a lock. Now it will make an MDC RPC to create a remote inode. When that call returns, the directory data can be filled in. The key issue here is recovery of the remote inode creation, which either requires writing the fid of the created inode in the commit log or using preallocated inodes.

It is easy to see that in the normal case of file creations the code path is equally efficient for a clustered metadata service and a single node one.

**mds_rename/mds_link:** These calls are probably the most interesting of all. It will involve three nodes. The source and target nodes holding the directory data and the node holding the inode which has a link that is to be renamed. An important invariant is that bucket inodes and directory inodes are always on the same node as the node holding the associated data.

This call pattern involves the mds making a *remote link* RPC to another MDS and a *remote setattr* RPC to the MDS holding the inode to be renamed. The calls appear to be easily recovered in case of failures.

**mds_unlink:** This is also a two stage call. Both for creation and unlinking the management of orphans is important. This orphan management is entirely analogous between the MDS and OST data objects. The *orphaned objects* can be created during the object creation/removal, objects might be created on the OSTs, but the MDS could fail before recording these in the extended attributes on a persistent store. Similarly, during deletion, its possible that the record of the objects is deleted on the MDS but the corresponding objects are not deleted on the OSTs before some failure occurs. These first situation can only be prevent by requiring the OSTs to log every object creation, the MDS would send an asynchronous message to the OSTs once the objects information has been stored on persistent store. The OSTs can then delete the corresponding logs. Similarly, in the second case, the MDS can keep logs of object deletion, if an OST fails before removing the corresponding objects, it could check with the MDS upon recovery and delete the required objects.

### 6.7.6. Clustered MDS recovery.

6.7.6.1. *Client - MDS replay protocol.* The clustered MDS - client recovery protocol is very similar to the single MDS - client protocol. In this case also, the MDS servers need to track whether a client request was executed, replied or committed. The MDS also regards other MDS systems that make requests as part of clustered metadata updates as a client for recovery purposes. If a request is committed, a replay is not required, the metadata server can simply forget the state associated with that request, except that it needs to be capable to reproduce the reply until the client has ack'd that. For a request was not executed, the client can simply retransmit it upon recovery; Lustre uses the word *resending* for this part of recovery. For requests that were executed and saw replies but lost on persistent storage the retransmission mechanism is called *replay*.

6.7.6.2. *Replay.* To order transaction sequences Lustre uses reply ack's: the acks server only one purpose to release a lock that enforces ordering of the transaction sequence. In the case where MDS operations involve more than server, the reply "ack" from the primary to secondary servers should only be sent after the client has sent the ack to the first server. This MDS-MDS reply ack is now not really an ack anymore but a simple lock cancelation review. Clients will replay lost transactions to the mds which they originally engaged for the request. Orphaned children will be cleaned up only after replay completes to allow orphaned objects to be re-used during replay.

6.7.6.3. *Failures of multiple MDS nodes.* The handling of recovery of orphan objects between clustered metadata servers is identical to that of the single MDS case.

A new problem arises from multiple metadata server failures, such as present in the case of power-off. In this case the MDS should be rolled back to a consistent state.

**Example:** In transaction one, a node X creates directories a. Then in transaction 2 a cross MDS node rename moves a file with a directory entry on node Y into this directory. It is now possible for this file to lose its directory entry on Y and for the transaction on X not to commit. More complex examples exist.

We do this with a standard algorithm known as a consistent cut in causal time or snapshot (see Birman [] or other books on distributed algorithms). A consistent snapshot is a state of the MDS that could have been reached through full execution of requests coming from cilents, in other words, a consistent snapshot is a state of the MDS file systems that represents a valid file system. After multiple simultaneous MDS failures the state of the MDS's must be rolled back to a consistent snapshot. We say that a transaction on an MDS1 depends on a transactions on MDS2 when the completion of a request to MDS1 has the transaction on MDS2 as a component.

Each MDS retains logs of transactions, sufficiently detailed that they can be undone. Each log record contains a transaction number corresponding to the transaction on this node and the transaction numbers of transactions that were started on other MDS to complete this transaction. The log records can be used for two operations. Log records can be canceled when the MDS cluster as a whole has committed the transactions that relate a particular log record. Also records can be used to undo operations that were already performed.

Every few seconds, the cluster computes a snapshot by first electing a leader. First leader asks all MDS's to give their last committed transaction numbers. The MDS's respond and also provide the transaction numbers for other MDS's they depend on for this transaction. If an MDS provided a dependency higher than what was committed, that MDS should be asked to resend its transactions and dependencies to account for this. This algorithm then repeats and it converges because it produces a strictly decreasing set of transaction numbers. When the transaction numbers have reached a consistent snapshot, all MDS's are told what their current last committed transaction for the snapshot is. Clients can be told to discard all requests held for replay that are older than those found in the snapshot.

The coordinating MDS of a client initiated transaction will first establish that the transaction can commit on all nodes, by acquiring locks on directories and checking for available space existing entries with the same name etc. It may also first perform a directory split if the size is becoming too large, and more MDS nodes are still available.

All nodes involved in the transaction need to have a transaction sequence number to place the transaction into their sequence and allow correctly replay. At this point the coordinator will:

- start a transaction locally.
- It will then report the transaction sequence number to all other nodes involved in the transaction.
- These nodes will commit (in memory as usual), write a journal record for replay and reply to the coordinator.
- The coordinator will then commit its own transaction.
- The MDS created metadata undo log records, which are subject to normal log commit cancelation messages, but on the coordinator commit messages must be received from the *leader* before the record will be canceled.

6.7.6.4. *Failover rings.* The configuration data can designate a standby MDS that will take over from a failed MDS. By organizing the servers in one or more rings, the nearest working left neighbor MDS can be the failover node. This leads to a simple scheme with multiple failover nodes, avoiding quorum and other complications beyond what is needed for two node clusters.

## 6.8. MDS Implementation

The MDS back-end is the system which stores meta-data on persistent storage. As we have shown above, our model is that the MDS maintains a collection of persistent objects which has a similar layout as a file system. The major difference is that the file inodes contain references to other storage objects, not data for the files.

We will make only a few requirements on the MDS systems - first, after recovery they form a consistent meta-data file system. Secondly, they should allow us to use pre-allocate fids to create objects. The fid pre-allocation would be specially required in the write back caching mode. In the write back cache mode, as soon as the transaction is completed in local cache, the client will be notified that the operation is indeed complete even though the updates/changes will move to the persistent store on the server at a later time. In such a case, it is important to make sure that the operation does not fail on the server when it is executed there. The recovery requirement will mean that all file system transactions were executed atomically with the transaction stream maintenance.

Even though there are these various options available for the MDS back-end, finally its the APIs that the server exports that matter as that is what the clients are exposed to.

There are a number of possible choices for such meta-data storage:

(1) A meta-data file system.
(2) A clustered database.

**6.8.1. Meta-data File system.** Perhaps our most natural approach is to use file systems for meta-data. The benefits are that the recovery, lock, and object storage protocols used between clients and OST systems can be re-used to implement shared object file systems. By using objects we also inherit the flexible API's for storing meta-data, e.g. those associated with security.

A possible drawback is that unless sophisticated logical modules are written that can execute transactions across multiple OST devices, the meta-data for a single file-set will be stored on a single OST. While such OST's can be made redundant and use heavy-duty block storage back-ends, there is the potential for a bottleneck in cases of extremely heavy meta-data traffic on a single file-set. File system transactions do not cross file-set boundaries: i.e. *link* and *rename* file system operations fail. We feel this limitation is acceptable.

**6.8.2. Database-managed Meta-data.** A second proposal would store meta-data in a high end clustered database, much as SQL Server is rumored to underly the file system in future versions of Windows 2000. It is certainly true that databases can handle different formats of meta-data easily and have high transaction throughput capability, however, the glue between the meta-data server and the back-end might involve some pretty unusual design. Given that our MDS is really intended as a transaction engine, this is worthy of further exploration.

## 6.9. ODDS and ENDS

This section contains a few odds and ends that can help the metadata server.

The first issue that we will discuss atime and mtime updates. We also pursue the implications for the server infrastructure arising from the use of file *id*'s. We explain why it is desirable to have multiple name-spaces for file systems, by file *id* and by name.

**6.9.1. The atime and mtime updates.** The **atime** updates are very frequent. These are mostly not supported in distributed file systems and Lustre proposes the same. It is probably possible to build a write back cache of *atime update records*, these move the **atime** forward when the cache is flushed.

On the other hand, **mtime** is important to applications and **mtime** updates satisfy *POSIX* semantics. The only operations that change **mtime** on files are *writes* and *truncates.* The requirement is that the *MDS* should normally have the correct **mtime,** when a file is opened for *write* or accessed to be *truncated*, the **mtime** will be managed by the OSTs on the individual data objects that make up a file. Then at a later time when a *setattr* or *close* is done by the client, the MDS will get the mtime updates. Some of the questions that need to be answered here are :

(1) How does a client know (by looking at an inode) if it should get the correct **mtime** from the MDS or from the OST?
(2) How can the **mtime** from the OSTs be sent to the clients with the least possible overhead?

These questions are explored and answered in the following section.

6.9.1.1. *The basic mtime access and update mechanism.* A client needs to know whether the **mtime** it is looking at in the inode is correct or old. This can be taken care of by having a **flag** in the inode to indicate if the **mtime** is valid or not. The flag is set upon opening the file on the MDS or at the beginning of a truncate operation. The client will now perform a truncate/write operation, when this completes the client and the OSTs will have the correct **mtime**. The network overhead is minimal since the **mtime** updates are piggybacked on the write/truncate replies from the OSTs to the clients. Next, when the client does a close/setattr and sends the correct **mtime** to the MDS, the flag will be cleared to indicate that the MDS has the correct **mtime**.

There is another issue that needs to be handled here, there should be some way for the MDS to indicate to the OST that it has committed the **mtime** updates to persistent store(disk). This will allow the OSTs to then clear there records regarding this. This can be done using a commit callback on the MDS, for the *setattr* calls, this would result in an asynchronous and batched RPC to the OSTs to inform them that the **mtime** updates were successful. It would mean that a single *truncate* call may now involve 3 calls : one to the MDS to begin the truncate, the second to the OSTs to execute the truncate, finally a third call from client to MDS to update the **mtime**. The third call can be done asynchronously since the client does not need any reply back, the subsequent commit message from MDS to OST can also be done asynchronously.

6.9.1.2. *Recovery aspects of mtime update.* It is essential that **mtime** updates remain consistent across a failure. For instance, if the MDS fails before it can update the **mtimes** on the disk, it should contact the OST after recovery to get the **mtime** updates again. In order to achieve this, the OST needs an accurate record, even after crashes, of

(1) Which files it was managing the mtime for?
(2) The mtime value itself.

The OST must create a persistent record indicating that it is updating mtime, this record must exist before the **mtime** is going to be changed (it need not be absolutely atomic with the operation). We can do this on an OST with a nested transaction writing a log in a file, accompanying the first write after open or the first part of a truncate transaction on the OST. This record can be removed only when the MDS has committed the new mtime to disk. So MDS's should make gratuitous connect calls to the OST to communicate the objects for which they have committed sizes and mtimes. This can trigger an asynchronous cleanup function for old records.

During an MDS recovery, the MDS needs to contact all OST's and find out which inodes have **mtimes** managed by the OST. This introduces a global flag on the MDS, which is only cleared when the MDS has contacted all the OST's to supply their list of inodes for which they are managing the **mtime**. The MDS can indicate in gratuitous connect calls (or otherwise) to the OST that it has processed and committed the pending sizes upon which the OST can clear its records. After this, the MDS can clear the global flag to indicate that the recovery associated with **mtime** update is completed. When the flag has cleared the MDS can proceed to use the per inode flags once more.

There is no requirement for synchronous recovery of this collection of inodes, but performance is degraded until the process completes.

**6.9.2. Server-based Lock Management.** Standard lock sequences involve revocations of locks, and separate lock acquisition and conversion from file system updates. There are several optimizations in lock acquisition and revocation which Lustre utilizes:

**Lock versioning:** Lustre's lock will hold version information about objects. If locks are re-acquired, the system granting the lock will indicate what version of the data is still valid, so that it may not need refreshing. This is a fairly standard concept described in [7].

**Operation-based lock fixup:** When revoking a lock, the holder may be offered sufficient information to retain the lock provided a change is implemented.

**Operation-based lock acquisition:** When a lock is acquired the operation for which it is acquired is included in the lock request. The nodes can grant the lock and/or perform the operation to deal with situations of higher and lower concurrency.

(1) Node sends lock request which includes modification request to MDS.
(2) MDS validates that it has locks to perform modification and sends revocation messages including the update information to other nodes, and confirmation of operation to the requesting node.

This protocol is fundamentally quicker but has more complicated recovery properties in case the cluster sees network or system failures.

**6.9.3. Fileid Caching.** The servers are going to see request sequences for directory and object meta-data, lock request traffic, and update records. Such a pattern is best served when frequently used objects are aggressively cached, but also receive messages from the VM system when shrinking is advisable. For this we intend to leverage existing caches in the OS, notably the inode and dentry caches.

It is of paramount importance to regard **file identifiers and names as two families of primary objects** in the server and client infrastructure. On the clients, usage of the file system will generate requests to the file system in terms of names. These are translated to requests for **file *id*'s** by the infrastructure sending requests to the server. During *read*-only file system traversal the MDS does not see names at all, except in directory pages which it doesn't parse.

Updates processed by the MDS are given in terms of pairs of *file id* and *name*. The *file id* is for the parent objects and the name is for the affected children. Similarly, revocations of locks are sent to clients in terms of *file id*'s.

The caches will rely on an object file system, or an object interface to existing file systems. The names of the objects are the file *id*'s. The objects themselves, including their attributes, are stored in inodes. Directory inodes have pages for directory entries and file inodes have pages with I/O object information.

### 6.10. Changelog

(1) Radhika Vullikanti(03/06/03) - Integrated this architecture 2 document into the book.
(2) Radhika Vullikanti(06/19/03) - Removed the recovery discussion from this chapter since that is available in the recovery chapters.
(3) Peter Braam (08/24/04) - Minor update of recovery overview for CMD. Rework and include the contents of the Clustered Metadata Design into this chapter.

CHAPTER 7

# A Distributed Lock Manager for Lustre

Lustre requires a distributed lock manager for synchronization of file and meta-data access. This lock manager is organized in much the same manner as the classical VAX Cluster Distributed Lock Manager, with some important modifications.

In distributed filesystems, the management of file extents is of critical importance to the system. File extent locking has different semantics from the locking of resources, as there are relationships between the extents that describe the locks. For example, if a client has a lock on a certain extent that implies it has locks on smaller extents contained therein. We found that by declaring locks to be of a certain type and to have policy functions that extend lock compatibility and acquisition we could integrate **extent locks** into the traditional model.

Lustre also has **intent locks**, locks that include an intent and allow the lock manager to execute the intent instead of granting a lock. Again, intents can be included into the framework by using policies.

This chapter is a guide to this distributed lock manager. We first review the key points of the VAX Cluster DLM and as we go along highlight some design decisions for the Lustre DLM. We include an overview of the API's offered by the lock manager, and a few sections that describe features we will be supporting as time goes on.

The references for this chapter are VAX Clusters by Roy Davis [7] and Programming Locking Applications for HCMP [24].

## 7.1. Resource Directories and Masters and Lock Owners

When a software subsystem wishes to use locks, it groups all the locks in a **lock namespace**. As part of acquiring a lock on an object, the object needs to be named inside the namespace. **Resources** name objects one wants to lock, and resources belong to a namespace. Access by applications to a resource's namespaces can be protected.

A *resource* is a string with an agreed maximum length, currently:

```
struct ldlm_res_id {
__u64 name[RES_NAME_SIZE];
}
```

is used to hold resource names, a variable length string may come in the future. VaxClusters pack this data differently. A 31-byte string is available for the name; the first byte probably holds the length. Perhaps this is a useful future optimization.

***Resources*** are managed as a forest of resource trees. Each name in a tree is unique to a parent. To acquire a lock on a resource, a system must first acquire a lock on all ancestors in the resource tree. All locks on resources in a particular tree in the forest are managed by a single system. Usually the resource manager is the first system to acquire a lock on the root of that tree.

When all locks in a namespace are managed by one system, this system acts as a ***lock server*** for the system. When a lock may be managed on any of a group of systems we speak of a ***distributed lock manager***.

In the case where a namespace is managed in a distributed fashion to locate the system managing a particular resource, a distributed resource directory is consulted. The resource directory does not manage locks, its sole purpose is to find the systems that do. The system with knowledge of the master for a particular resource is found through hashing the resource name.

The hash function will give us an index in the Lock Directory Weight Vector (LDWV), which lists the managing system for each hash value. It should be note that systems can be listed multiple times in the lock directory weight vector allowing more powerful systems to manage a larger resource directory.

If the lock directory does not find a system managing a particular resource then it should allocate the calling system as the new manager.

When the last lock on a resource tree is removed, the resource directory should be informed that the resource is "free" again.

Remarks:

- We will initially use a hash function similar to that in the dcache.
- The VMS [**18**] lock manager had a maximum size of 31-byte for resource names; it appears to have used the first byte of a 32-byte structure as the length of the resource.
- How do we identify systems in the cluster? This is done through a membership database which holds the addresses of each system. There will be multiple channels between systems. See Stephen Tweedie's core cluster design documents for more details.
- The resource directory is not a separate service but can easily be run inside the lock daemons.
- Note that none of the structures in this document refer to persistent storage; all are held in virtual memory.

## 7.2. Resource Directory API (Internal to the Lock Manager)

The resource directory API is used from inside the lock manager only. The following API is called by a system which needs to find the resource master for a specific resource:

```
res_getmgr(struct lck_resnm *, struct sysid *);
```

If the resource directory already has the resource, it returns the *sysid* of the mastering node. Otherwise it returns 0, indicating that the calling system will master the resource tree.

The API shown below is called by the resource master when the last lock or lock request for a resource is released. The resource directory releases the resource.

```
res_free(struct lck_resnm *);
```

The **master** of a resource will be the system that manages all lock requests for a resource. In many instances a single master is used for all locks in a namespace. Lock requests are separate data structures from the lock data. For every resource mastered, queues of granted, converting, and waiting lock requests are managed, as well as the master copy of the lock data. This mastering system manages locks both for itself and for other systems requesting a lock on a resource in the same tree.

Locks can be accessed on a system through a lock *id* (*struct ldlm_lock*), the lock *id* is local to the system. The locking subsystem converts the lock *id* to handles. The handles contain the address of the lock *id* structure and a random cookie that is used to verify the validity of a handle. The lock handle is sent to the lock manager along with any lock requests.

If a system has a lock on a resource, but is not the master of that resource, then the system will hold a copy of the resource and lock data and request information. This is a duplicate copy, the other copy of the lock information is on the master. If a cluster transition takes place, locks need to be remastered and these duplicate copies allow this to happen when a resource master leaves the cluster.

Locks on resources are handled as a tree. Before a process can lock a resource it must lock all resources which are ancestors of this lock. There would be a **master lock** on the parent and **child lock** on the descendant; as an example, there could be a **master lock** on an inode and **child locks** on the file extents. No lock can be dequeued when descendants remain.

The benefits of a tree organization of locks are two-fold:

(1) Low level locks can be held at coarse granularity. In the example of an inode presented earlier, the **master lock** on the inode could be a **NULL** lock allowing multiple clients to get a lock on the inode, and the **child locks** on the file extents can be held in different locking modes.
(2) Resource names need to be unique only to the parent.

## 7.3. Foundations of the Locking Infrastructure

Every system in the cluster will provide lock service and each cluster member can request a lock on a resource, remove a lock, or convert a lock to a different type of lock, provided it can get a connection to the resource namespace. The system mastering the resource will maintain lock management data.

#### 7.3.1. Lock Types, Compatibility.

Lock management data involves:

- Being able to find lock data given the resource name.
- Maintaining a list of granted lock requests on a resource .
- Maintaining a list of requests for lock conversions (ie. a change of mode of the lock) on a resource.
- Maintaining a list of waiting new lock requests on a resource.

| MODE | NAME | ACCESS GRANTED | MEANING |
|---|---|---|---|
| EX | Exclusive | RW | No other process can get R or W access. |
| PW | Protected Write | Write | No other process can get a W access. |
| PR | Prot. Read | R | No other process can get a W access. |
| CW | Concurrent Write | W | No restrictions on other processes. |
| CR | Concurrent Read | R | No restrictions on other processes. |
| NL | Null | none | To indicate an interest in a resource. |

TABLE 1. Lock Modes

-

Like VAX Clusters we have six modes of locks, listed and named in table 1. Informally, the locks request *read* and *write* permissions on a resource. Formally, there is a conversion table determining what locks can exist concurrently. All locks on the granted list of a resource should be compatible with each other. Lock mode compatibility is described in table 2.

Compatibility governs new lock requests and conversions, i.e. conversion or new locks are granted if the new mode is compatible with all granted modes.

| | EX | PW | PR | CW | CR | NL |
|---|---|---|---|---|---|---|
| **EX** | 0 | 0 | 0 | 0 | 0 | 1 |
| **PW** | 0 | 0 | 0 | 0 | 1 | 1 |
| **PR** | 0 | 0 | 1 | 0 | 1 | 1 |
| **CW** | 0 | 0 | 0 | 1 | 1 | 1 |
| **CR** | 0 | 1 | 1 | 1 | 1 | 1 |
| **NL** | 1 | 1 | 1 | 1 | 1 | 1 |

TABLE 2. Lock Mode Compatability

It is interesting to note that this table is symmetric. By associating a level with a lock mode, a necessary condition for conversion is that the level is not increasing. A simple utility function determines the lock compatibility:

```
int ldlm_lock_compat(lck_mode a, lck_mode b)
```

It is often useful to think in terms of lock severity. The lock levels, illustrated in figure7.3.1, are:

(1) EX
(2) PW
(3) PR and CW
(4) CR
(5) NL

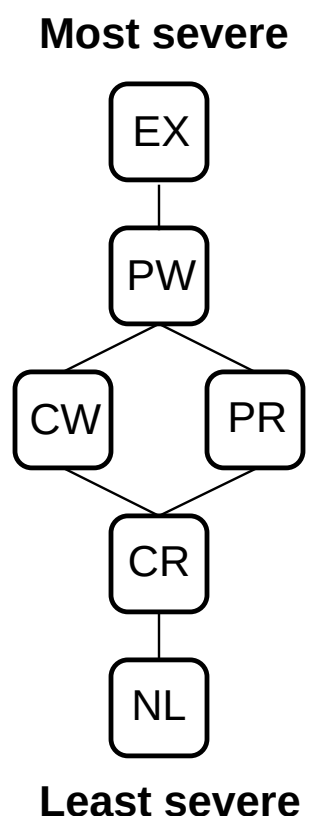

FIGURE 7.3.1. Lock Severity

**7.3.2. Lock Requests and Processing the Lock Queues.** There are several types of requests for locks:

- Get a new lock on a resource.
- Convert a lock on a resource.
- Cancel a lock.

When new locks are requested, the lock manager will grant this if only **the following 3 conditions hold:**

(1) The queue of ungranted conversion requests is empty.
(2) The queue of ungranted new lock requests is empty.
(3) The granted locks on the resource all have a mode compatible with the mode which is requested.

A conversion request can immediately be granted on one condition: **the mode requested is compatible with modes on all the granted locks**.

If the conversion request cannot be granted immediately, the lock manager places it at the end of the conversion queue. When that happens, the request will be reconsidered when all other conversion requests have been granted or cancelled, as illustrated in figure7.3.2. A lock which is in the conversion queue is still granted in the mode from which it needs to convert.

The lock manager places lock requests in one of three queues **- conversion, granted, or ungranted** - and remembers both the granted and requested mode for locks in the conversion queue. Hence, two conversion requests on a granted lock cannot be placed simultaneously in the queues associated with the lock resource. We still need to define what we do with a second conversion request coming in for a lock. It should probably overwrite the first, but with a warning.

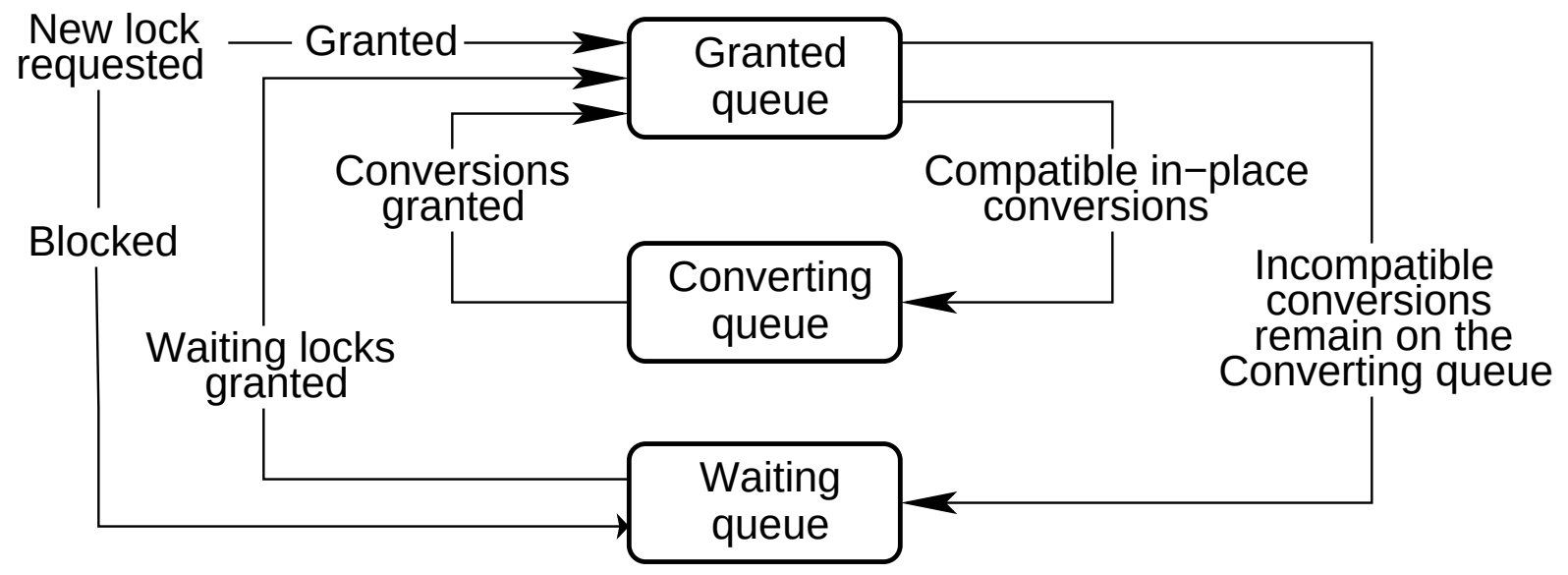

FIGURE 7.3.2. Lock Queues

Three events lead the lock manager on a system to visit the conversion and waiting queue:

(1) A lock in the granted queue is cancelled.
(2) A conversion request succeeds.
(3) Another ungranted request in the waiting or conversion queue is granted.

When one of these events takes place, all requests in the conversion queue are visited first, followed by those in the waiting queue.

The VAX Clusters Book highlights two problems associated with the above scheme:

(1) There is an "indefinite postponement problem" with this algorithm. Successful immediate conversion requests can hold up queued conversion requests.
(2) New NL requests are queued and postponed if conversion requests exist.

There are two optional flags to change the behaviour when this problem is expected and the application wants to explicitly avoid it:

1. ***QUEUECONV*:** Pass a flag to queue the conversion (even if it can be granted immediately). This leads to first-come-first-serve on ALL conversion requests, not just for those that cannot be granted immediately.
2. ***EXPEDITE*:** Pass a flag to expedite a new NL lock request.

Another flag on the request, ***NOQUEUE***, allows conversion requests to either be granted immediately, or to be rejected when they cannot be granted.

**7.3.3. More Details about Lock Processing.** There is a great variety in the exact behaviour of the lock manager when a request is processed and granted. This is discussed below in details.

Lock blocks contain the information held by the master of the lock and the owner. If the owner of the lock equals the master, there is only one copy of the block. The lock block contains "links" for a variety of lists on which it sits: the *sysid* of the system owning the lock, a process *id* for the process which requested the lock, the requested mode of the lock, and the granted mode of the lock. The lock owner holds a completion routine address and a blocking trigger address, as well as a semaphore. The node mastering the lock has to notify owners of events that need action.

Lustre uses a standard slab allocation for locks and resources. The handle API discussed in the networking chapter provides a remote access API to locks. The handles encode the pointer to a lock efficiently and securily and can be passed to clients.

**7.3.4. An Example Lock Sequence.** In the example shown in figure 7.3.3, two CR locks are being held on resource ***x***, lock 1 was granted first and then lock 2 was also granted as it was compatible to the first lock being held on the resource.

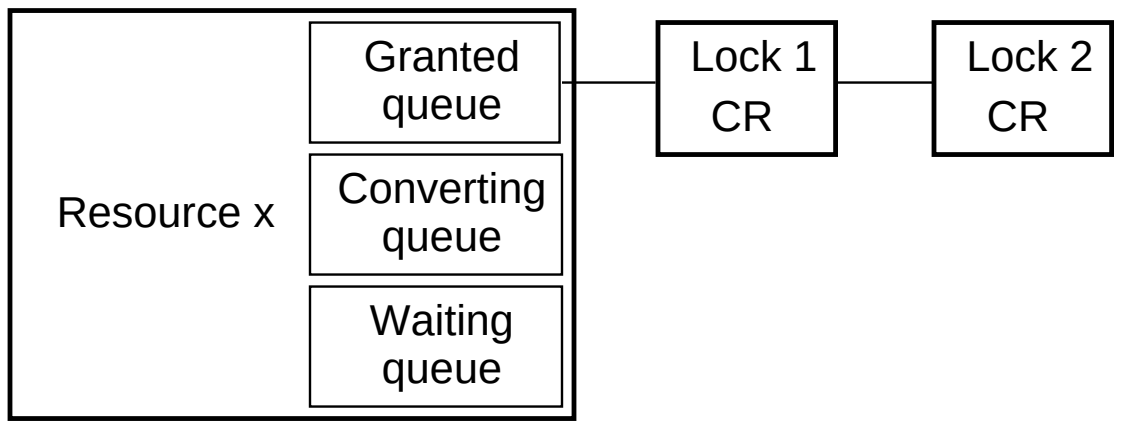

FIGURE 7.3.3. Lock Example 1

In figure 7.3.4, a third CR lock is granted on the same resource. The CR and CW are also compatible lock modes, that is they can co-exist, figure 7.3.5 illustrates this.

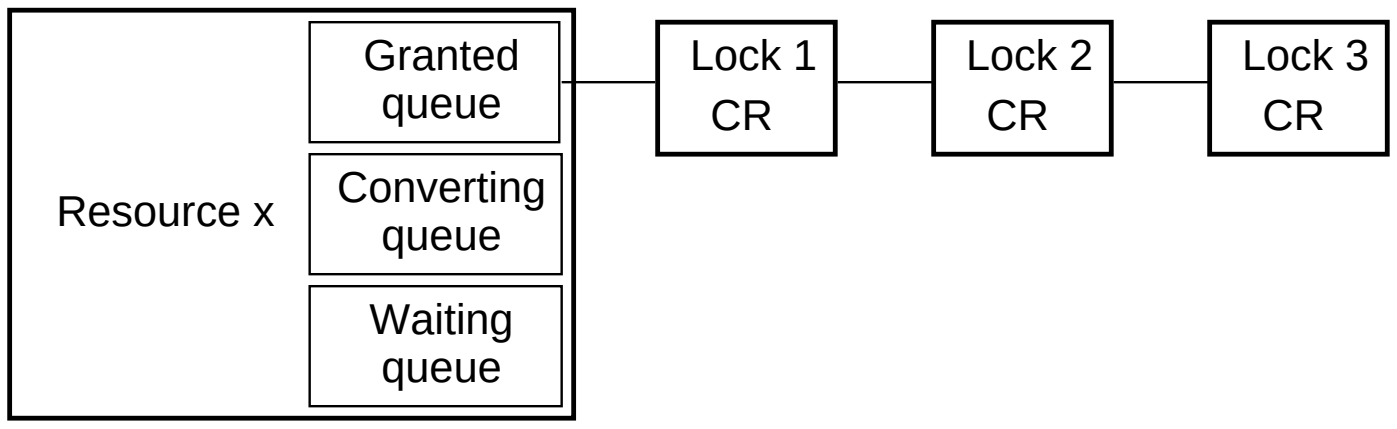

FIGURE 7.3.4. Lock Example 2

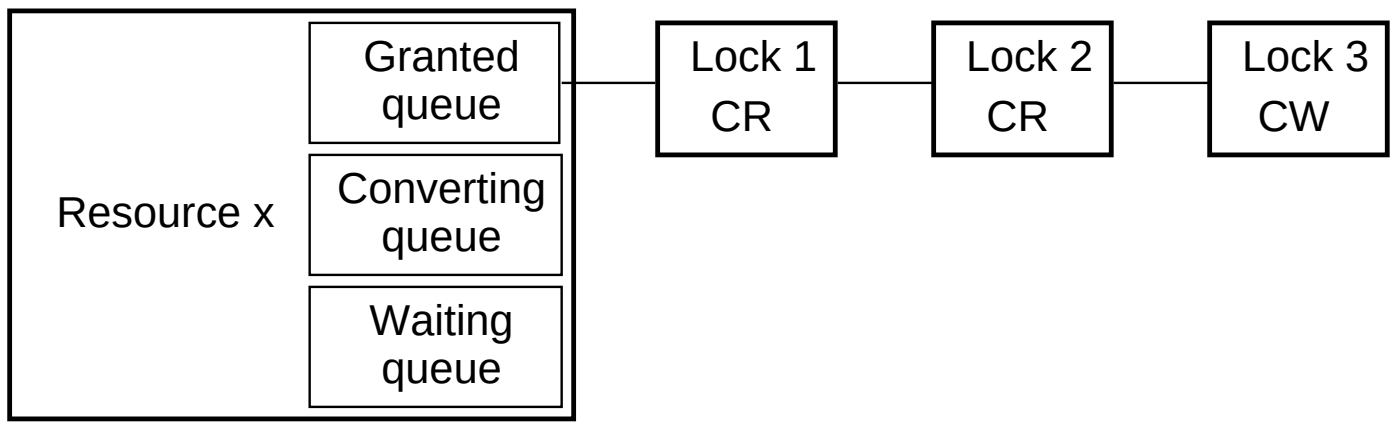

FIGURE 7.3.5. Lock Example 3

Now, suppose the client holding lock 1 wants to convert it to an EX lock, she needs to wait till all the other locks held on the same resource are cancelled. In the meantime, this conversion request is kept on the ***converting queue*** as shown in figure 7.3.6.

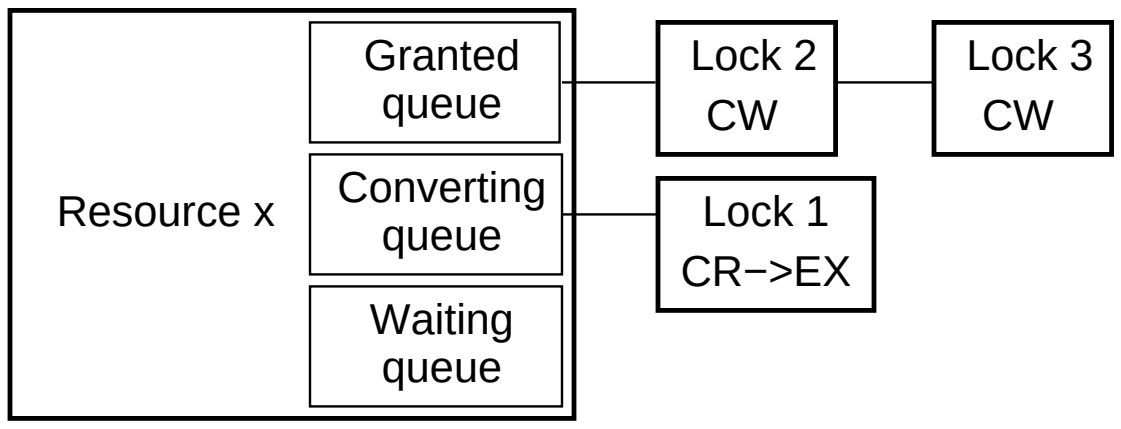

FIGURE 7.3.6. Lock Example 4

## 7.4. The Completion and Blocking Callbacks

**7.4.1. Completion Routine.** There could be some delay between the time a lock was requested and the time a reply is received from the lock server. So, with lock acquisition or conversion, a completion routine can be passed. This routine is run when the request is granted or converted. When a lock request is blocked, the completion routine is called twice, once when the lock request was received by the lock server and again when the lock is granted.

This function is also delivered when the lock or conversion request is canceled (dequeue with the cancel flag set).

The routine can be passed a parameter (such as the address of the lock status, although that should perhaps be included by default.)

**7.4.2. Blocking Routine.** This routine is called for locks that have been granted already and are holding up another lock acquisition or conversion request. The parameter for the completion routine is also given to the blocking routine.

The Lustre lock manager uses the completion on the client to interact with subsystems that have cached data for locks. A lock structure has two reference counts associated with it, one to indicate the number of readers on the resource and the second to count the number of writers. The subsystems check if these reference counts are zero (to verify that the lock is no longer in use) and then typically flush cached data and cancels the lock upon a callback. On the lock server side, the callback routines are used to deliver remote notifications of similar events.

## 7.5. Policies: Intents and Extents

The Lustre lock service deviates from the standard examples to offer richer features. When a lock is enqueued, a **lock type** can be passed. We support 3 lock types:

**Plain:** These locks behave completely as the VAX locks on resources.

**Extent:** These locks are present to lock file extents. When enqueuing such locks or considering conversions, the lock server calls extra functions to make optimal decisions. Some of these decisions are similar to those taken by GPFS [20].

**Intent:** Intent locks are used for meta-data service. They allow a client to indicate the intent of acquiring a lock. When the intent is received, the meta-data server is invoked before the lock manager treats the request. The meta-data server may execute the intent and return a different lock resource. For example, a node creating a file will ask for a *write* lock to modify the directory. The meta-data server may instead create the file for the client and return a lock on the new object instead.

### 7.6. Lock Services

Lock services should be available in three forms:

- As a kernel level library.
- As a kernel level daemon.
- Through a new system call.

Lock services need some way of identifying locks, passing a reference to clients when locks are granted or returning a status if a lock request was unsuccessful. The VAX Clusters made use of a lock status block, this is used in lock requests, conversion requests or cancellations:

```
struct lck_status
{ uint16 lckst_cond;
uint16 lckst_reserved;
uint32 lckst_id;
char lckst_val[16]; }
```

This appears to be quite sufficient, all the locking APIs would make use of this structure to pass references to a lock.

When a lock is or a conversion is requested, the caller passes in a pointer to an allocated *lck_status* block. The lock *id* is filled in by the lock service on the system requesting the lock, before anything else is done, and returned to the caller in all cases. So the lock *id* may be used for further conversion or cancellation requests. The *lcst_val* block is for the use of applications to exchange information. The *lckst_cond* gives the condition in which the request resulted, i.e. if the lock was granted, converted etc.

Table 3 lists status values returned by the VAX DLM for lock conversion and acquisition lists status values returned by the VAX DLM for lock conversion and acquisition.

Note that we fill in the lock *id* as part of the synchronous component of lock request processing, i.e. before any blocking operation occurs, we allocate the lock *id*. The advantage of doing this is that lock requests which have not yet reached the queues can be cancelled, table 4 lists the status values used when lock cancellation requests are sent.

| **VALUE** | **INDICATION** |
|---|---|
| *NORMAL* | Success |
| *ABORT* | The lock was dequeued before ENQ could grant the lock |
| *CANCEL* | The conversion request that was cancelled before it could be granted |
| *DEADLOCK* | A deadlock was detected |
| *ILLRSDM* | No adequate permissions on this resource domain |
| *NODOMAIN* | Not a valid resource domain for this system |
| *VALNOTVALID* | The lock value is not valid (see xref linkend=“lock-dequeue“>) |

TABLE 3. Lock status values used by VAX DLM for lock conversion or acquisition requests

| **VALUE** | **INDICATION** |
|---|---|
| *NORMAL* | Success |
| *ACCVIO* | The value block specified by the caller cannot be accessed |
| *CANCELGRANT* | The cancellation request was unsuccessful because the conversion request was already granted |
| *ILLRSDM* | No adequate permissions on this resource domain |
| *IVLOCKID* | Invalid lock *id* was specified or not enough privileges |
| *SUBLOCKS* | Lock has sublocks and cannot be dequeued |

TABLE 4. Status values used for lock cancellation requests

New and conversion lock requests may have the following as parameters:

**Requested mode:** Mode to acquire or convert to.
**Flags:** Indicate special action to be undertaken by lock manager.
**Resource name:** What needs to be locked.
**Resource domain *id* and access mode:** We probably ignore this in the first instance.
***lck_status* block:** With the *id* of a lock to convert.
***id* of a parent lock:** Optional, for acquiring a sublock.
**Semaphore:**
**Completion routine:**
**Blocking trap:**
**Pointer to parameters for either trap:**

Lock requests return integers indicating the status information described in table 5. The return value is computed immediately without blocking by the lock manager, and contains condition information that can be acquired by the lock manager on the system, without having to contact the resource manager on another system.

Table 6 describes the status codes in the lock status block for fully processed requests.

| VALUE | INDICATION |
|---|---|
| *L_NORMAL* | Lock was granted or queued |
| *L_SYNC* | Lock or conversion was granted synchronously |
| *EACCESS* | Error accessing the lock status block or resource name (segv) |
| *L_BADMODE* | Bad mode specified |
| *L_CVTNOTGR* | Conversion on ungranted lock |
| *L_CVTNOPAR* | Parent not granted on sublock |
| *L_NOTQUEUED* | The lock could not be granted immediately and the NOQUEUE flag was given |
| *L_BADRES* | The resource length exceeds the system bound |
| *L_BADST* | The lock status block is not present |
| *L_BADLCK* | No lock *id* was given when required (for conversion) |

TABLE 5. Lock Request Return Values

| VALUE | INDICATION |
|---|---|
| *L_SUCCESS* | Lock was granted or converted |
| *L_ABORT* | Lock was dequeued before it could be acquired |
| *L_CANCEL* | Lock was cancelled before the request could be granted |

TABLE 6. Status Block Values for Completed Requests

The flags described in table 7 can give additional information to the lock manager (the last three flags in the table will be useful once deadlock detection and lock quota are implemented, but not initially.)

Further error conditions can occur when the resource management is more complete, when we have quota or access control on resources. For each locking call we will have a blocking and non-blocking version.

The completion routine should be fired when locks are granted or canceled and the blocking trigger when a new lock cannot be granted or an existing one converted due to other granted locks.

Another flag, not present in VMS, might be useful to add:

***LCK_BLCKRETRY*:** If a new or conversion request comes in, and this flag is set for every lock for which a blocking trigger is fired, the lock mananger will retry the acquisition or conversion when the blocking triggers have run. With this flag on a lock request, a system indicates that it will dequeue the lock during the blocking trigger. This is very useful for filesystems and avoids a second lock request.

If a user level process makes lock requests, these routines should be queued for execution in user mode in the process, otherwise the kernel can execute them without delay.

Some other questions that might need to be answered are:

| VALUE | INDICATION |
|---|---|
| *LCK_NOQUEUE* | Don't queue the request, return a failure status. |
| *LCK_SYNCSTS* | Return L_SYNC if the lock was synchronously acquired or converted; if so, do not fire a completion routine. Otherwise the return code is L_SUCCESS, and the completion routine and sempaphore are triggered when the conversion/acquisition completes. |
| *LCK_SYSLCK* | The resource name is a system resource name (otherwise a process/user resource is assumed.) |
| *LCK_CONVERT* | The lock is a conversion request. |
| *LCK_CVTSYS* | Convert the process owned lock to a system lock. |
| *LCK_EXPEDITE* | Grant NL lock requests without queueing. |
| *LCK_QUEUECONV* | Place the conversion request on the queue. |
| *LCK_NOQUOTA* | The lock is not charged quota for the lock request. |
| *LCK_NODLCK* | The process is okay if the lock is merely queued (switches off deadlock detection.) |
| *LCK_NOBLCKDLCK* | The process will release its lock when asked to give it up, and the firing of the blocking completion may block but should not imply deadlock. |

TABLE 7. Lock Flags

(1) Do we do this as for signal handlers? Which comes first, the signal or the lock traps?
(2) Does the lock manager wait for the triggers and completions to have run before returning to the client?
(3) If a process sleeps on syncrhonous lock acquisition, what signals do we let in?
(4) We are going to hang a list of acquired locks & queued requests in the process structure. Process cleanup should dequeue granted locks or convert them to system owned locks. Which depends on the application - opinions here? Queued requests should be canceled - this should likely be done before dequeuing the granted locks.

Lock cancellation takes a few forms. Lock requests can be canceled and granted locks can be dequeued.

Variants of dequeueing are:

- Cancel/dequeue a particular request.
- Cancel all sublocks of a parent.
- Cancel all locks of a process.

The VMS dequeuing mechanism passes the lock *id* of the lock to be released as a parameter to the lock dequeue request. Since this is not known until the lock has been queued, it seems impossible to cancel a queued lock request before it has been queued (this is a race condition that is problematic in process termination).

### 7.7. Lock Version/ Value Blocks

Lock blocks can contain version information. This version information can be used by applications to communicate what version of the resource they want to lock. Lock value blocks are 16 bytes long in VMS.

Versions can be:

**Written:** The lock value info are copied from the lock status block in the request to the resource lock value field held by the lock manager.
**Returned:** The lock value block is copied from the resource to the lock request lock value field.
**Ignored:** The lock value fields are not changed.

Note that this table is essentially saying that whenever a protect *write* or *read* lock has been released the value is written into the resource. When a lock is promoted (say from NL to CR) the lock value information is returned to the requestor.

| | **NEW MODE** | **NL** | **CR** | **CW** | **PR** | **PW** | **EX** |
|---|---|---|---|---|---|---|---|
| **HELD MODE** | | | | | | | |
| **NL** | | ret | ret | ret | ret | ret | ret |
| **CR** | | none | ret | ret | ret | ret | ret |
| **CW** | | none | none | ret | ret | ret | ret |
| **PR** | | none | none | none | ret | ret | ret |
| **PW** | | write | write | write | write | write | ret |
| **EX** | | write | write | write | write | write | write |

TABLE 8. **Lock Version Assignment**

### 7.8. Cluster Transitions and the DLM

The basis for the algorithm is the following sequence of crude operations:

(1) When the new cluster membership is known, each system can compute a new LDWV to locate resource directories. Discard existing resource directories.
(2) Each node mastering a resource should discard the locks it doesn't own itself. For each resource this node still has locks, it should register itself as the resource master with the resource directory.
(3) Each node should re-acquire each lock it owns but does not master.

The problem with this algorithm is that it is doing much more than is needed in many cases and it can lead to unfortunate mastery. Trees will be remastered by the first system to touch them. This is bad if that system is slow, or has many fewer locks in the resource tree than another system, which now has to use this as a master.

This situation can be improved by first letting systems with a nonzero weight in the LDWV to remaster their locks and then other systems.

Another problem with the basic algorithm is that too much work is often done.

Namely:

- If a system with weight 0 leaves or enters, nothing needs to be done.
- When a system joins the cluster, only the directory needs to be affected not the lock database.
- When a system leaves the cluster, the minimum that needs to be done is:
  (1) Delete locks held by the system
  (2) Adjust directory information if the weight was not 0
  (3) Remaster trees mastered by the leaving system
  (4) Grant locks blocked by lock held on the leaving system

Changes in this direction are possible and greatly reduce the rebuild times of the lock database. Systems with lock directory weight 0 do not get to master trees if there are at least 2 systems with non-zero weight in the cluster.

**???:** I'd like to discuss how methods like these can use the cluster infrastructure to organise things like:
first let nonzero weight systems do such and so - then let zero weight systems do their bit.

A further improvement to rebuilding comes for organizing clean exits from the system, by asking the node leaving the system to remaster trees it masters.

**???:** It seems reasonable to me to let the cluster controller control the recovery of the lock database.

### 7.9. Dynamic Remastering

The next improvement to the lock manager is to move lock mastering on the basis of power of the system and activity.

The rules for mastering become:

- A resource is initially mastered by the first system to acquire a lock.
- If multiple systems acquire a lock on a resource, systems with higher lock directory weight should master the resource.
- If multiple systems of the same weight acquire locks, the system with the highest activity count should master the resource.

The rules are subject to thresholds. Activity is measured every second by forming a geometric average of lock acquisition, conversion and dequeueing operations. When a system mastering a resource is servicing a request from another system, it can mark the lock for remastering if:

- The weight of the other system is bigger than its own weight. This causes a "check for better master" flag to be set.
- The activity on the lock is at least xx (ops/sec) and the activity of the remote system is at least xx (ops/sec) higher than its own activity. This sets the "remaster pending" flag.

Every 8 seconds the lock manager looks at resource trees and can move up to 5 of them. It moves up to 5 trees during each such run.

A tree it masters is moved if one of the following two situations is true:

- The check for better master flag is set.
- The "remaster pending" flag is set and:
  (1) The benefit outweighs the cost.
  (2) The benefit is not outweighing the cost, but the same system has set the "remaster pending flag" three times during the previous 24 seconds.

The cost is the number of locks that need to move; the benefit is 16 times the difference in activity between the new master and the current master.

A different form of remastery concerns trees that are mastered by a node, but used by just a single other node. In this case it maskes sense to move the mastery of that tree to the node using the tree. VMS does this once a minute.

**???:** The remastering of locks may require some form of two phase commit between the resource, the current master and the new master. How can this be organized?

### 7.10. Changelog

**Version 3.6 (Mar. 2003)**

(1) Radhika Vullikanti - Added some details, made some minor corrections and changes

**Version 3.5 (Dec. 2002)**

(1) P.D. Innes - updated and resized figures, added Changelog

**Version 3.0 (Nov. 2002)**

(1) P.D. Innes - updated text, made tables into floats

**Version 2.5 (Jun. 2002)**

(1) Radhika Vullikanti - Updated/added some details to the DLM design

**Version 2.0 (Jun. 2002)**

(1) P.D. Innes - edited, spell-checked, proofed

**Version 1.0**

(1) P. Braam - original draft

## 7.11. The Lustre VFS library

Lustre will access file systems from a variety of contexts. The Linux kernel drivers in the MDS and OST are two examples. User level servers for Lustre would also require access, and ports on other OS's will require this. More adventurous constructions such as a SQL based metadata server would again benefit from this.

The LVFS - Lustre VFS - library gives a universal api to enable such access.

The library is modeled on the api's found in the Linux VFS and the extensions Lustre added to that.

## 7.12. The structure of the library

The library work in a **context**. The context indicates:

- What the library should use as the root of the file system. For this purpose the context contains a vfsmount and working directory.
- The address space in which the operations should be executed.
- The type of the file system that is being accessed. This includes operations offering transactional semantics and extended attribute storage associated with this file system.
- What are the user credentials, e.g. user id and group membership, for the operations that are about to be performed
- a variety of operations provided by the context, such as how to translate a fid to a dentry.

The relevant data structures are discussed in the next section.

There are several types of calls in the library:

(1) Low level file system operations. These are largely inherited from the previous construction named fsfilt.
(2) Medium level functions, to look up dentries, open files etc. These are modeled on the Linux VFS interfaces in the kernel
(3) High level generic functions that are expressed in terms of 1 and 2.

**7.12.1. Header structure.** The header structure is dictated by the supported architectures. Currently we are aiming at:

(1) Linux kernel
(2) User level file systems

For kernel compilation lvfs.h conditionally includes lvfs_linux.h. In turn lvs_linux.h conditionally includes the fsfsfilt_<fstype>.h headers depending on what file systems are being compiled into the kernel.

For user level compilation, activated by defining LIBLUSTRE, lvfs.h also includes lvfs_user.h.

## 7.13. Data Structures

### 7.13.1. VFS structures.

7.13.1.1. *struct l_dentry.*

7.13.1.2. *struct l_file.*

7.13.1.3. *struct l_inode.*

7.13.1.4. *struct l_vfsmnt.*

7.13.1.5. *struct l_sb.*

### 7.13.2. Transactional information.

7.13.2.1. *struct obd_trans_info.* This structure contains a pointer to a handle, storage room for one and an array of cookies, an array of ack locks (handle and mode), a transaction number and object id.

```
struct obd_trans_info {
        __u64                    oti_transno;
        __u64                   *oti_objid;
        /* Only used on the server side for tracking acks. */
        struct oti_req_ack_lock {
                struct lustre_handle lock;
                __u32                mode;
        }                        oti_ack_locks[4];
        void                    *oti_handle;
        struct llog_cookie       oti_onecookie;
        struct llog_cookie      *oti_logcookies;
        int                      oti_numcookies;
};
```

What is slightly puzzling about this structure is how to make it indepedent of obd's.

### 7.13.3. Context and credentials.

7.13.3.1. *obd_run_ctxt (NOTE: should be lvfs_ctxt).*

```
struct obd_run_ctxt {
        struct vfsmount *pwdmnt;
        struct dentry   *pwd;
        mm_segment_t     fs;
        struct obd_ucred ouc;
        int              ngroups;
        struct lvfs_callback_ops cb_ops;
};
```

7.13.3.2. *obd_ucred (should be lvfs_cred).*

```
struct obd_ucred {
        __u32 ouc_fsuid;
        __u32 ouc_fsgid;
        __u32 ouc_cap;
        __u32 ouc_suppgid1;
        __u32 ouc_suppgid2;
};
```

7.13.3.3. *lvfs_callback_ops.*

```
struct lvfs_callback_ops {
        struct dentry *(*l_fid2dentry)(__u64 id_ino, __u32 gen, __u64 gr, void *data);
};
```

## 7.14. Operations

### 7.14.1. Lustre operations.

7.14.1.1. *fid2dentry.* This is a direct callback operation from the context, offered by the caller.

7.14.1.2. *fid2locked_dentry.* Used in the MDS not yet imported in LVFS.

7.14.1.3. *locking operations.* The pdirops should be imported into Lustre's lvfs for use by the MDS.

7.14.1.4. *lvfs_update_server data.* This is used by the MDS and obdfilter. Needs more structure.

**7.14.2. Linux VFS operations.** Several of the key operations offered by the LVFS api are derived from the Linux VFS. For a kernel implemenation they are simply defined as their equivalent kernel fucntion, for the user library they need to be implemented in terms of the libc api.

7.14.2.1. *push_ctxt.*

7.14.2.2. *pop_ctxt.*

7.14.2.3. *simple_mknod.*

7.14.2.4. *simple_mkdir.*

7.14.2.5. *lustre_fwrite.*

7.14.2.6. *lustre_fread.*

7.14.2.7. *lustre_fsync.*

7.14.2.8. *l_dentry_open.*

7.14.2.9. *l_filldir.*

7.14.2.10. *l_readdir.*

7.14.2.11. *l_dput.*

7.14.2.12. *ll_lookup_one_len.*

**7.14.3. FsFilt api.** Lustre has employed an fsfilt api. We review this here briefly, it should be modified slightly to integrate better with the fsfilt approach.

```
struct fsfilt_operations {
        struct list_head fs_list;
        struct module *fs_owner;
        char   *fs_type;
        void   *(* fs_start)(struct inode *inode, int op, void *desc_private);
        void   *(* fs_brw_start)(int objcount, struct fsfilt_objinfo *fso,
                         int niocount, void *desc_private);
        int     (* fs_commit)(struct inode *inode, void *handle,int force_sync);
        int     (* fs_commit_async)(struct inode *inode, void *handle,
                                void **wait_handle);
        int     (* fs_commit_wait)(struct inode *inode, void *handle);
        int     (* fs_setattr)(struct dentry *dentry, void *handle,
                       struct iattr *iattr, int do_trunc);
        int     (* fs_iocontrol)(struct inode *inode, struct file *file,
                         unsigned int cmd, unsigned long arg);
        int     (* fs_set_md)(struct inode *inode, void *handle, void *md,
                      int size);
        int     (* fs_get_md)(struct inode *inode, void *md, int size);
        ssize_t (* fs_readpage)(struct file *file, char *buf, size_t count,
                        loff_t *offset);
        int     (* fs_add_journal_cb)(struct obd_device *obd, __u64 last_rcvd,
                              void *handle, fsfilt_cb_t cb_func,
                              void *cb_data);
        int     (* fs_statfs)(struct super_block *sb, struct obd_statfs *osfs);
        int     (* fs_sync)(struct super_block *sb);
        int     (* fs_prep_san_write)(struct inode *inode, long *blocks,
                              int nblocks, loff_t newsize);
        int     (* fs_write_record)(struct file *, void *, int size, loff_t *,
                            int force_sync);
        int     (* fs_read_record)(struct file *, void *, int size, loff_t *);
        int     (* fs_setup)(struct super_block *sb);
};
```

The api allows to:

```
typedef void (*fsfilt_cb_t)(struct obd_device *obd, __u64 last_rcvd,
                            void *data, int error);
```

```
struct fsfilt_objinfo {
        struct dentry *fso_dentry;
        int fso_bufcnt;
};
extern int fsfilt_register_ops(struct fsfilt_operations *fs_ops);
extern void fsfilt_unregister_ops(struct fsfilt_operations *fs_ops);
extern struct fsfilt_operations *fsfilt_get_ops(const char *type);
extern void fsfilt_put_ops(struct fsfilt_operations *fs_ops);
```

(1) Find an fsfilt given the type
(2) Register an fsfilt in the list
(3) fsfilt_get_ops(char *type) will find operations based on file system type and if necessary demand load a module for that purpose.

For each operation there is a wrapper function taking an obd as the first argument. The obd has an struct fsfilt_operations * field, obd_fsfilt.

**NOTES:**

(1) Instead these operations should take the fsfilt operations as an argument, to avoid undue import of abstractions.
(2) The names should start with lvfs_ not fsfilt_
(3) Where are the journaled writes write_rec and read_rec?

CHAPTER 8

# Lustre Logging API

## 8.1. Introduction

Lustre needs logging API in numerous places - orphan recovery, RAID1 synchronization, configuration, all associated with update of persistent information on multiple systems. Generally, logs are written transactionally and cancelled when a commit on another system completes.

Log records are stored in log objects. Log objects are currently implemented as files, and possibly, in some minor ways, the APIs below reflect this. In this discussion, we speak of log objects and sometimes of llogs (lustre-logs).

## 8.2. API Requirements

Some of the key requirements of these APIs that defines their design are:

- The API should be usable through methods
- The methods should not reveal if the API is being used locally or invoked remotely
- Logs only grow
- Logs can be removed, remote callers may not assume that open logs will remain available
- Access to logs should be through stateless APIs that can be invoked remotely
- Access to logs should go through some kind of authorization/authentication system

## 8.3. Fundamental data structures

### 8.3.1. Logs.

(1) Log objects can be identified in two ways
  (a) Through a name - The interpretation of the name is upto the driver servicing the call. Typical examples of named logs are files identified by a path name, text versions of the UUIDs, profile names.
  (b) Through an *object identifier* or *llog-log identifier* - A directory of *llogs* which can lookup a name to get an id can provide translation from naming system to an id based system. In our implementation, we use a file system directory to provide this catalog function.
(2) Logs only contain records
(3) Records in the logs have the following structure:

- llog_rec_hdr - a header, indicating the index , length and type. The header is 16 bytes long
- Body which is opaque, 32-bit aligned blob
- llog_rec_tail - length and index of recors for walking backwards, it is 16 byte long

(4) The first record in every log is a 4K long *llog_log_rec.* The body of this record contains:
- a bitmap of records that have been allocated; bit 0 is set immediately because the header itself occupies it
- A collection of log records behind the header

(5) Records can be accessed by :
- iterating through a specific log
- providing a *llog_cookie,* which contains the *struct llog_logid* of the log and the offset in the log file where the record resides.

(6) Some logs are potentially very large, for example replication logs, and require a hierarchical structure. A catalog of logs is held at the top level. In some cases the catalog structure is two levels deep:
- A catalog API is provided which exploits the lower lustre log API
- Catalog entries are log entries in the catalog log which contain the log id of the log file concerned.

**8.3.2. Logging contexts.** Each obd device has an array of logging contexts (*struct llog_ctxt).* The contexts contain:

(1) The generation of the logs. This is a 128 bit integer consisting of the mount count of the origianating device and the connection count to the replicators.
(2) A handle to an open log (*struct llog_handle *loc_handle)*
(3) A pointer to the logging commit daemon (*struct llog_canceld_ctxt *loc_llcd)*
(4) A pointer to the containing obd *(struct obd_device *loc_obd)*
(5) An export to the storage obd for the logs (*struct obd_export *loc_exp)*
(6) A method table (*struct llog_operations *loc_logops)*

**lop_destroy:** destroy a log
**lop_create:** create/open a log
**lop_next_bloc:** read next block in a log
**lop_close:** close a log
**lop_read_header:** read the header in a log
**lop_setup:** set up a logging subsystem
**lop_add:** add a record to a log
**lop_cancel:** cancel a log record
**lop_connect:** start a logging connection. This is called by the originator to initiate cancellation handling and log recovery processing on the replicators side. The originator calls this from a few places in the recovery state machine.
**lop_write_rec:** write a log record

## 8.4. Llog connections and the Cancellation API

This section describes the typical use of the logging API to manage distributed commit of related persistent updates. The next section describes the recovery in case of netowrk or system failures. We consider systems that make related updates and use the following definitions:

**Originator:** - the first system performing a transaction
**Replicators:** - one or more other systems performing a related persistent update

The key requirement is that the replicators must complete their updates if the originators do, even if the originating systems crash or the replicators roll back. Note that we do not require that the the system remains invariant under rollback of the originator.

This goal is achieved by transactionally recording the originators action in a log. When the replicators related action commits, it cancels the log entry on the originator. In the subsequent sections, we describe the handshake and protocols involved.

### 8.4.1. Llog connections.

In order to process cancellation and recovery actions, the originators and replicators use a *ptlrpc* connection to execute remote procedure calls. The connection can be set up on the originator or the replicator and we call the system setting up the connection the **initiator** and the target of that connection event the **receptor**.

The connection is used symmetrically, that is, the originator and replicator can either be the initiator or the receptor. The obd device structure has an optional *llog_obd_ctxt* which holds a pointer to the import to be used for queuing rpc's.

- The originator and the replicator establish a connection. These are the usual connections used by other subsystems.
- The logging subsystem on the originator uses the **lop_connect** method to the replicator. The lop connect call sends the logid's of the open catalog from the originator to the replicator.
- Just prior to sending this the originator context increases its generation, and includes the generation and the logid in the **lop_connect** method, usually calling **llog_orig_connect**.
- The replicator now receives a *llog_connect RPC*. The handler is the replicators **lop_connect** (usually **llog_repl_connect**). This method first increases the llcd's generation then initiates processing of the logs.

### 8.4.2. The cancellation daemon.

A replicator runs a subsystem responsible for collecting pages of cookies and sending them to the originator for cancellation of the origin log records. This is done as a side effect of committing the replicating transaction on the replicator.

A key element in the cancellation is to distinguish between old and new cookies. Old cookies are those that have a generation smaller than the current generation, new cookies have the current generation. The generation is present in the llog_context, hence it is both on the server and on the client.

The cancellation context is responsible for the queueing of cancel cookies. For each originator it is in one of two states:

(1) Accepting cookies for cancellation
(2) Dropping cookies for cancellation

The context switches from 1 to 2 if a timeout occurs on the cancellation rpc. It switches from 2 to 1 in two cases:

(1) A cookie is presented with an llog_generation bigger than the one held in the context
(2) The replicator receives a llog_connect method (which will also carry a new llog_generation)

The llog_generation is an increasing sequence of 128 bit integers with highest order bits the boot count of the originator and the lower bits the obd_conncnt between the originator and the replicator. The originator increases its generation just before sending the llog_connect call, the replicator increases it just prior to beginning the handling of recovery when receiving an llog_connect call.

**8.4.3. Normal operation.** Under normal operation, the originator performs a transaction and as a part of the transaction, writes a log record for each replicator. The following steps are then followed to ensure that the replicator is updated with a copy:

- The log record creation, done with **lop_add** produces a *log_cookie*
- The *log_cookie* is sent to the replicator, through a means that we do not discuss here.
- The replicator performs the related transaction and executes a commit callback for that. The callback indicates that the *log_cookie* can be put up for cancellation. The function **lop_cancel** is responsible for this queuing of the cancellation.
- When the replicator has a page full of cancellation cookies, it sends the cookies to the originator
- The originator cancels the the log records associated with the cookies and cleans up the empty log files. The handling function is **llog_handle_cancel** and it invokes the originators **lop_cancel** functions to remove the log record**.**

The replication scenarios are closely related to commit callbacks and RPCs, the key differences are:

- The commit callbacks with transaction numbers involve a volatile client and a persistent server
- The transaction sequence is determined by the server in the voilatile-persistent case by the originator in the replicating case

**8.4.4. Examples.**

8.4.4.1. *Deletion of files.* Change needs to be replicated from MDS (originator) to OST's (replicators):

- The OSC's used by the LOV on the MDS act as originator for the change log, using the storage and disk transactions offered by the MDS:
  - OSC's write log records for file unlink events. This is done through an obd api which stacks the MDS on the LOV on the OSC's.
    Such events are caused by unlink calls, by closing open but unlinked files, by removing orphans (which is recovery from failed closes) and by renaming inodes when they clobber.
  - The OSC's create cookies to be returned to OSTs. These cookies are piggy backed on the replies of unlink, close and rename calls. In the case of removing orphans the cookies are passed to *obd_destroy* calls executed on the MDS.
- OST's act as replicators, they must delete the objects associated with the inode.
  - Remove objects
  - Pass OSC generated cookies as parameters to obd_destroy transactions
  - Collect cookies in pages for bulk cancellation RPCs to the OSC on MDS
  - Cancel records on the OSCs on MDS

8.4.4.2. *File size changes.*

- Changes originate on OSTs, these need to be implemented on the MDS
  - Upon the first file size change in an I/O epoch on the OST:
    - Writes a new size changes record for new epoch
    - Records the size of the previous epoch in the record
    - Records the object id of the previous epoch in the record
    - It generates a cancellation cookie
  - When MDS knows the epoch has ended:
    - It obtains the size at completion of the epoch from client (or exceptionally from the OST)
    - It obtains cancellation cookies for each OST from the client or from the OSTs
    - It postpones starting a new epoch untill the size is known
    - It starts a setattr transaction to store the size
    - When it commits, it cancels the records on the OSTs

8.4.4.3. *RAID1 OST.*

- The primary is the originator, the secondary is the replicator
  - Writes on the primary are accompanied by a change record for an extent

**8.4.5. Cancellation timeouts.** If the replicator times out during cancellation, it will continue to process the transactions with cookies. The cancellation context will drop the cookies.

The timeout will indicate to the system that the connection must be recovered.

## 8.5. Llog recovery handling

When the replicator recieves an llog_connect rpc, it increases the llcd's generation, and then spawns a thread to handle the processing of catalogs for the context. For each of the catalogs it is handling, it fetches the catalog's *log_id* through an *obd_get_cat_info* call. When it has received the catalog logid, the replicator calls sync and proceeds with *llog_cat_process*

- It only processes records in logs from previous log connection generations.
- The catalog processing repeats operations that should have been performed by the initiator earlier
  - The replicator must be able to distinguish:
    **Done:** If the operation already took place. If so it queues a commit cancellation cookie which will cancel the log record which it found in the catalog's log that is being processed. Because sync was called there is no question that this cancellation is for a committed replicating action.
    **Not done:** The operation was not performed, the replicator performs the action, as it usually does, and queues a commit cookie to initiate cancellation of the log record.
- When log processing completes, an obd-method is called to indicate to the system that logs have been fully processed. In the case of size recovery, this means that the MDS can resume caching file sizes and guarantee their correctness.

**8.5.1. Log removal failure.** If an initiator crashes during log removal, the log entries may re-appear after recovery. It is important that the removal of a log from a catalog and the removal of the log file are atomic and idempotent. Upon re-connection, the replicator will again process the log.

**8.5.2. File size recovery.** The recovery of orphan deletion is adequately described by 1.5.1. In the case of file size recovery, things are more complicated.

## 8.6. Llog utility and OBD API

**8.6.1. Llog OBD methods.** There is only one obd method related to llog which *llog_init.*

**8.6.2. llog_init.** This obd method initializes the logging subsystem for an obd. It sets the methods and propages calls to dependent obd's.

**8.6.3. llog_cat_initialize.** There is a simple master function **llog_cat_initialize** for catalog setup that uses and array of object id's stored on the storage obd of the logging. The logids are stored in an array form and given to the llogging contexts during the **lop_setup** calls made by **llog_init.** It uses support from lvfs to read and write the catalog entries and create or remove them.

## 8.7. Log method table API

Logs can be opened and/or created, this fills in a log handle. the log handle can be used through the log handle API.

### 8.7.1. llog_create.

8.7.1.1. *Prototype.*

```
int llog_create(struct obd_device *obd, struct llog_handle **, struct llog_logid *, ch
```

8.7.1.2. *Parameters.*

8.7.1.3. *Return Values.*

8.7.1.4. *Description.* If the log_id is not null, open an existing log with this ID. If the name is not NULL, open or create a log with that name. Otherwise open a nameless log. The object id of the log is stored in the handle upon success of opening or creation.

### 8.7.2. llog_close.

8.7.2.1. *Prototype.*

```
int llog_close(struct llog_handle *loghandle);
```

8.7.2.2. *Parameters.*

8.7.2.3. *Return Values.*

8.7.2.4. *Description.* Close the log and free the handle. remove the handle from the catalog's list of open handles. If the log has a flag set of destroy if empty, the log may be zapped.

### 8.7.3. llog_destroy.

8.7.3.1. *Prototype.*

```
int llog_destroy(struct llog_handle *loghandle);
```

8.7.3.2. *Parameters.*

8.7.3.3. *Return Values.*

8.7.3.4. *Description.* Destroy the log object and close the handle.

### 8.7.4. llog_write_rec.

8.7.4.1. *Prototype.*

```
int llog_write_rec(struct llog_handle *handle, struct llog_reec_hdr *rec, struct llog_
```

8.7.4.2. *Parameters.*

8.7.4.3. *Return Values.*

8.7.4.4. *Description.* Write a record in the log. If *buf* is NULL, the record is complete. If *buf* is not NULL, it is inserted in the middle. Records are multiple of 128bits in size and have a header and tail. Write the cookie for the entry into the cookie pointer.

### 8.7.5. llog_next_block.

8.7.5.1. *Prototype.*

```
int llog_next_block(struct llog_handle *h, int curr_idx, int next_idx, __u64 *offset,
```

8.7.5.2. *Parameters.*

8.7.5.3. *Return Values.*

8.7.5.4. *Description.* Index curr_idx is in the block at **offset*. Set **offset* to the block offset of recort *next_idx*. Copy *len* bytes from the start of that block into the buffer *buf*.

### 8.7.6. lop_read_header.

8.7.6.1. *Prototype.*

```
int *lop_read_header(struct llog_handle *loghandle);
```

8.7.6.2. *Parameters.*

8.7.6.3. *Return Values.*

8.7.6.4. *Description.* Read the header of the log into the handle and also read the last *rec_tail* in the log to find the last index that was used in the log.

### 8.7.7. llog_init_handle.

8.7.7.1. *Prototype.*

```
int llog_init_handle(struct llog_handle *handle, int flags, struct *obd_uuid);
```

8.7.7.2. *Parameters.*

8.7.7.3. *Return Values.*

8.7.7.4. *Description.* Initialize the handle, try to read it from the log file. But if the log does not have a header built, build it from the arguments. If the header is read, verify the flags and UUID in the log equal those of the arguments.

### 8.7.8. llog_add_record.

8.7.8.1. *Prototype.*

```
int llog_add_record(struct llog_handle *cathandle, struct llog_trans_hdr *rec, struct
```

8.7.8.2. *Parameters.*

8.7.8.3. *Return Values.*

8.7.8.4. *Description.*

### 8.7.9. llog_delete_record.

8.7.9.1. *Prototype.*

```
int llog_delete_record(struct llog_handle *loghandle, struct llog_handle *cathandle);
```

8.7.9.2. *Parameters.*

8.7.9.3. *Return Values.*

8.7.9.4. *Description.*

### 8.7.10. lop_cancel.

8.7.10.1. *Prototype.*

```
int llog_cancel_record(struct llog_handle *cathandle, int count, struct llog_cookie *c
```

8.7.10.2. *Parameters.*

8.7.10.3. *Return Values.*

8.7.10.4. *Description.* For each cookie in the cookie array, we clear the log in-use bit and either:

- Mark it free in the catalog header and delete it if its empty
- Just write out the log header if the log is not empty

The cookies maybe in different log files, so we need to get new logs each time.

### 8.7.11. lop_next_block.

8.7.11.1. *Prototype.*

```
int llog_next_block(struct llog_handle *handle, int curr_idx, int next_idx, __u64 *cur
```

8.7.11.2. *Parameters.*

8.7.11.3. *Return Values.*

8.7.11.4. *Description.* Return the block in the log that contains record with index *next_idx*. The *curr_idx* at the offset *curr_offset* is used to optimize the search.

## 8.8. Sample Method Table Descriptions

The obd_llog api

The obd_llog api has several methods, setup, cleanup, add, cancel, as part of the OBD operations. These operations have 3 implementations:

**mds_obd_llog_*:** simply redirects and uses the method mds_osc_obd, which is normally the LOV running on the MDS to reach the OST's.

**lov_obd_llog_*:** calls the method on all relevant OSC devices attached to the LOV. A parameter including striping information of the inode is included to determine which OSC's should generate a log record for their replicating OST.

A more interesting implemenation is the collection of methods that is used by the OSC on the MDS and by the OBDFILTER:

**llog_obd_setup:** sets up a catalog entry based on a log id.
**llog_obd_cleanup:** cleans up all catalog entries in the array
**llog_obd_origin_add:** adds a record using the catalog in the llog_obd_ctxt array of handles
**llog_obd_repl_cancel:** queues a cookie for cancellation on the replicator.

**8.8.1. obd_llog_setup(struct obd_device *obd, struct obd_device *disk_obd, int index, int count, struct llog_logid *idarray).** To activate the catalogs for logging and make their headers and file handles available is fairly involved. Each system that requires catalogs manages an array of catalogs. This function is given an array of logid's and an index. The index pertains to the array of logs used by an originator, the array of logid's is an array with an entry for each osc in the lov stripe descriptor.

**8.8.2. obd_llog_cleanup(struct obd_device *).** Cleans up all initialized catalog handles for a device.

8.8.2.1. *int llog_obd_origin_add(struct obd_export *exp, int index, struct llog_rec_hdr *rec, struct lov_stripe_md *lsm, struct llog_cookie *logcookies, int numcookies).* Adds a record to the catalog at index *index*. The lsm is used to identify how to descend an LOV device. The cookies are generated for each record that is added.

**8.8.3. int llog_obd_repl_cancel(struct obd_device *obd, struct lov_stripe_md *lsm, int count, struct llog_cookie *cookies, int flags).** Queue the cookies for cancellation. Flags can be 0 or LLC_CANCEL_NOW for immediate cancellation.

### 8.9. Configuration Logs

Configuration of Lustre is arranged by using llogs with records that describe the configuration.

The first time a configuration is written it is given a version of 0. Each record is numbered. Configurations can then be updated, which results in:

(1) a new configuration log
(2) a change descriptor with the previous configuration

Configurations are then recorded on the configuration obd. At any time there are stored:

(1) One full configuration log (for the current version)
(2) A collection of change descriptors for every change made since the initial configuration.

A client uses the configuration logs in two ways:

- On startup it fetches the full current configuration log from the configuration obd and processes the records to complete the mount command
- A client can also receive a signal that it needs to refresh its configuration. This signal can be an ioctl, /proc/sys file or lock revocation callback. When the client gets this signal it:
  - Determines its current version of the configuration
  - Asks the config obd for the latest version
  - Fetches the change logs to change the current configuration to the latest one

The last operation is done with llog_process, using a suitable callback function, as well as the logs that the client has in memory.

### 8.10. Size Recovery

This section contains a discussion of the recovery of MDS cached sizes from OST's.

The MDS sees open calls which precede any I/O on a file. When an open request reaches the MDS the file inode is in one of two states:

**quiescent:** No I/O is currently happening on the inode
**I/O epoch:** The inode is in I/O epoch *k*.

If no I/O epoch is active the MDS starts a new one. The epoch number will be a random number from boot time which is increased each time a new epoch is started.

A fairly complicated sequence of events involving the inode may now ensue, such a many other openers. Eventually the clients will all close the file and flush their data. The simplest epoch management scheme is:

(1) **open** file is opened for write
(2) **closed and flushed** all clients have closed and flushed data
(3) **mds** changes file size and ends epoch

When a client closes the file, has no dirty data outstanding and knows the file size and OST size update cookies authoritatively it will include them with the close call to the MDS. The MDS will initiate the setattr to update its cached file size and use the MDS cookies.

When a client closes but doesn't satisfy some of these conditions it will still make a close call to the MDS. The MDS will know if this is the last client closing the file. If so, it will indicate in its response to the client that it requires the client to obtain the file size and cookies and make an additional setattr call to the MDS with the cookies.

The client can flush its data and force a flush of other clients data through the DLM. An obd_getattr call will obtain the file size and cookies for a particular epoch. A slightly more lax scheme is to allow the client to update the MDS even when it has not yet flushed all dirty data to the inode.

The epoch ends when the MDS receives the setattr call.

The OST should pin the inode in memory and remember the MDS epoch in volatile data. Perhaps it takes a refcount for each client writing to the inode. Each client can indicate to the OST when it

CHAPTER 9

# The Lustre Filesystem

## 9.1. Introduction - Subsystems and Protocols

**9.1.1. The Principal Subsystems.** The Lustre cluster filesystem has three software subsystems at its core and several peripheral systems. First there is the **Client FileSystem (CFS)**. This works in conjunction with the **Meta-Data Servers (MDS)** and the **Object Storage Targets (OST)**. Each of these subsystems can run on separate systems but multiple subsystems can run on a single system too.

Accordingly, there is a protocol for each pair of subsystems, and protocols involving the auxiliary subsystems.

The role of the client filesystem is to provide a directory tree, subdivided into filesets, which provides cluster-wide Unix file sharing semantics. The client filesystem interacts with the meta-data servers for meta-data handling, i.e. for the acquisition and updates of inodes and directory information. File I/O, including the allocation of blocks, striping, and security enforcement, is contained in the protocol between the client filesystem and the object storage targets. A third protocol exists between the OST and the MDS, largely for pre-allocation and recovery purposes.

MDS provides a load balanced clustered service to client systems which is used in meta-data updates. It is high performance and scales to large numbers of client nodes. The MDS systems transform client requests into journaled, batched meta-data updates on persistent storage. An MDS can batch large numbers of requests from a single client, generated while the client grows a write-back cache. They can also batch large numbers of requests generated by different clients, in the case of concurrency, when many clients are updating a single object.

**9.1.2. Contents of this Paper.** We first describe the CFS/MDS protocol as this provides the most insight into the operation of the system. We refer to this protocol as the meta-data protocol. It is driven by the requirements of the VFS methods found in the OS. An overview of the protocol is shown in figure 9.1.1. This part is split into a section about reading meta-data, a section describing updates, and one on the meta-data server implementation.

**9.1.3. Locking.** Clients need locks for file I/O and meta-data I/O, and meta-data servers need concurrency control for storage objects representing meta-data. We use a single lock service module, that can be incorporated in the client MDS protocols, in the object storage protocol, i.e. the protocol between clients and the OST, and in the protocols among the MDS systems.

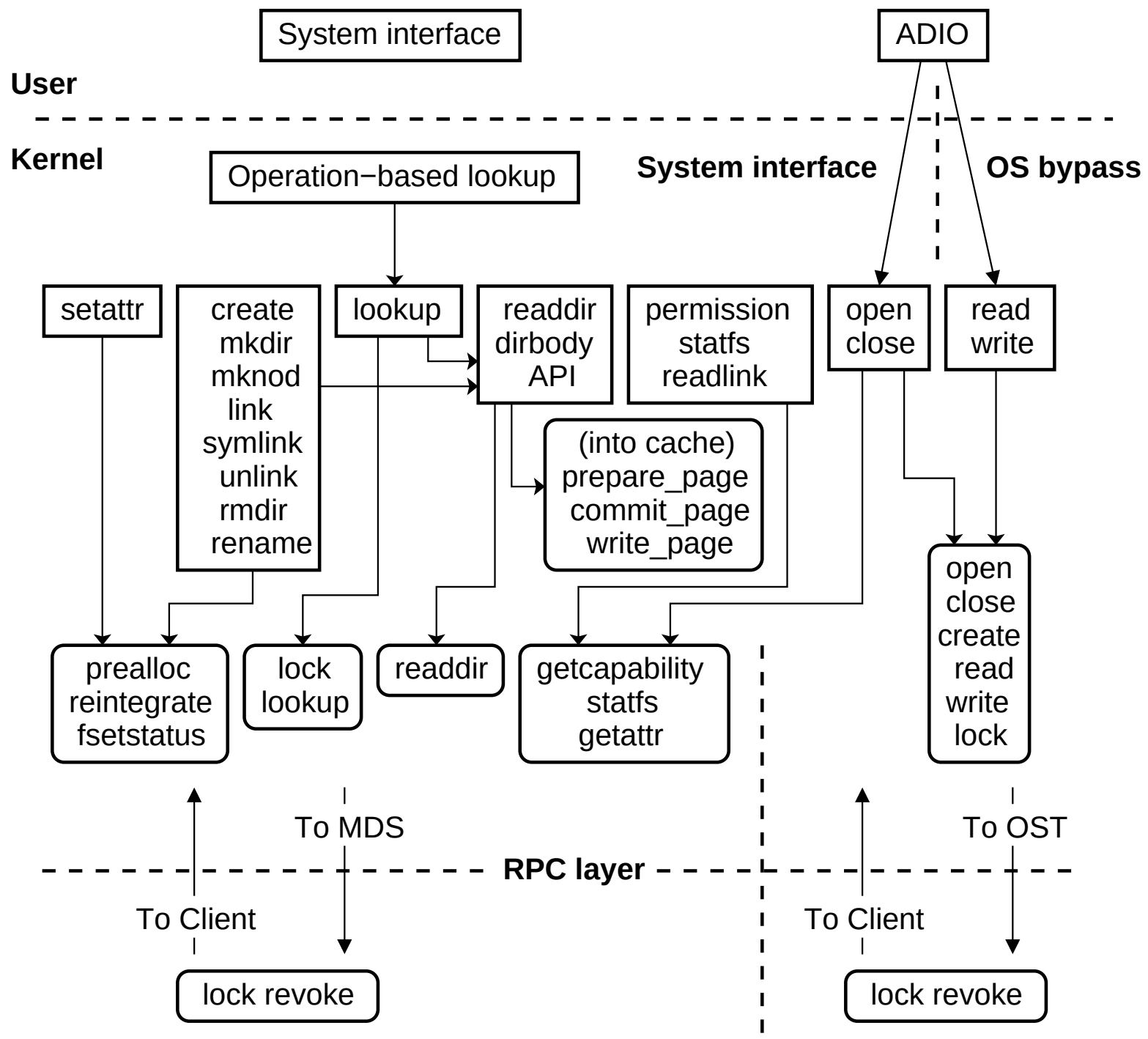

FIGURE 9.1.1. Meta-data

All file locking is served from OST systems. There are file locks to support Unix semantics and to provide **lock and flock** and **specialized locking** for collective I/O operations associated with parallel I/O support.

MDS servers offer two kinds of lock services to clients. First there is a resource location service that tracks which MDS systems are mastering resources. Clients contact this service to locate the MDS system handling a resource it wishes to update. The MDS then grants locks on filesystem objects to clients.

When MDS's update meta-data on persistent yet different resources, they require lock service. We will describe one possible implementation of this, which is to use an object-based meta-data filesystem at the core. This system is attractive because it has a large overlap with the file I/O system.

## 9.2. Mount Interfaces

Lustre will be managed by the global namespace filtering filesystem. This system is described in chapter 3.

Inside the filesystem, *read_super* needs to instantiate an inode for the root of the filesystem. This does not involve a request for this inode from the MDS - that happens only upon *lookup* operations inside the root inode or an *opendir* of the root inode. Lustre will instead build an artificial root directory, which merely contains a file named `mntinfo`. This can be used normally by the global namespace mechanisms to enter the root fileset of the Lustre filesystem. As a result, Lustre will mount, even in the absence of network connections.

The information is passed up to the automount daemon when the root directory is stored in the `mntinfo` file. Its default value is `fset::ROOT`. This can be modified in a mount-argument passed to the Lustre *read_super* method by the *mount* command.

A very similar process takes place when a mount-object elsewhere in the filesystem is entered (see 3). The Lustre mount infrastructure is minimal:

```
mount -t gnsfs -o
rootinfo=fset::SPECIALFSET@cellname,cache_type=lustre none /lustre
```

### 9.3. Reading Meta-data

There are three aspects to reading meta-data:

**Inode meta-data:** Acquired during *lookup* methods.
**Directory data:** Acquired during *readdir*.
**Dentry data:** Set up when a *lookup* for a name is executed.

**9.3.1. Inode Meta-data.** In Linux, inode meta-data is acquired upon lookup. There are two mechanisms in use to fetch inode meta-data. Most common is the *read_inode* super method, which is called during the instantiation of inodes with *iget*. However, in Lustre (as in some other filesystems) the attributes must be fetched before the *iget* call is made. The *read_inode* method is skipped and the attributes are copied into the inode after *iget* has delivered an empty inode.

Most other operating systems have a *getattr* method which is called in conjunction with using an object representing a file or directory.

Accordingly we need a request method for transporting inode meta-data from MDS to CFS systems:

```
struct req_desc {
fileid_t   fid;
attrmask   mask;
...
};

struct resp_desc {
...
attr *attr;
...
};
int l_getattr(IN req_desc, OUT resp_desc, OUT l_attr);
```

Lustre will use the attributes specified in the OBD protocol for the purposes of RPC's to give a relatively space efficient, platform independent, and flexible way of moving attributes from MDS to client systems.

Access control lists are considered meta-data by Lustre, and can be **acquired** with the *l_getattr* call if capabilities allow.

Other extended attributes may be present. The detailed protocol will follow the ideas of the EA specification `http://www.betbits.at/`.

In Lustre, inodes can have a large amount of meta-data to describe the striping of the inodes, and this meta-data is not fetched when a *l_getattr* request triggered by *lookup* takes place, but only upon *file_open*. Similarly, extended attributes may be stored with the objects, but are not normally required. The mask can indicate what needs to be fetched.

9.3.1.1. *Inode Numbers and File Identifiers.* User level programs can see a 32 bit inode number in the stat structures filled in by *stat* system calls. The VFS caches inodes based on this 32 bit inode number. For distributed filesystems, this is unfortunate as 32 bit inode numbers are hard to use for a global filesystem which typically uses a significantly larger file identifier. The difficulties are as follows.

The inodes need to be given a 32 bit inode number for hashing in the VFS. This inode number needs to have a high probability of being unique as false equality might lead a program like "tar" to mistake different files for (hard) links to one object. When clients purge inodes from the cache and re-instantiate them, the 32 bit inode numbers should preferably remain the same to avoid the opposite problem: user level tools may treat the same file as two different objects. This is perhaps best done on the server. Mathematically, there are no good solutions for injective maps from a big file identifier space to a 32 bit inode space.

If filesets introduce a separate mount-point each with its own superblock, this problem becomes significantly easier, since we gain extra bits in the VFS file identifier space through the identity of super-blocks. Hard links across fileset boundaries would not be recognized.

It is important to emphasize that the problems with inode numbers only affect user level applications that use the inode number through the *stat* system call. The Linux kernel (and BSD and Windows systems) can manage files with larger identifiers. The Linux virtual filesystem internally uses large file identifiers through the VFS method *iget4* and fills in attributes with the *read_inode2* filesystem method. The latter is responsible for filling in inode meta-data and, in effect, needs to contact a server for this information.

**9.3.2. Directory Data.** Directory data is used by client systems and authoritatively managed on MDS systems. Directory data is used solely to iterate over directory contents (the *readdir* or *getdents* system calls) and to associate file identifiers with names during a *namei* operation in the VFS. *namei* operations are executed very frequently and most systems cache the results in a name-cache, called the *dcache*.

Lustre allows clients to cache directory data. Clients use it to perform lookup operations, i.e. to find file *id*'s corresponding to names. When clients perform updates that affect a directory, they modify the cached copy of the directory data. Updates of directory data intended for persistent storage are solely performed by MDS systems. Clients notify MDS systems about updates through an update record which describes what transaction needs to be performed. This model of directory updates is similar to that used by AFS/Coda/DCE-DFS.

Lustre is designed to handle multiple directory formats. Interesting formats are the basic *ufs/ext2* style directories and directories **hashed** by name to allow directory resources to be subdivided over multiple systems. Linux Ext2 and GFS filesystems have directories with *name-hashes*.

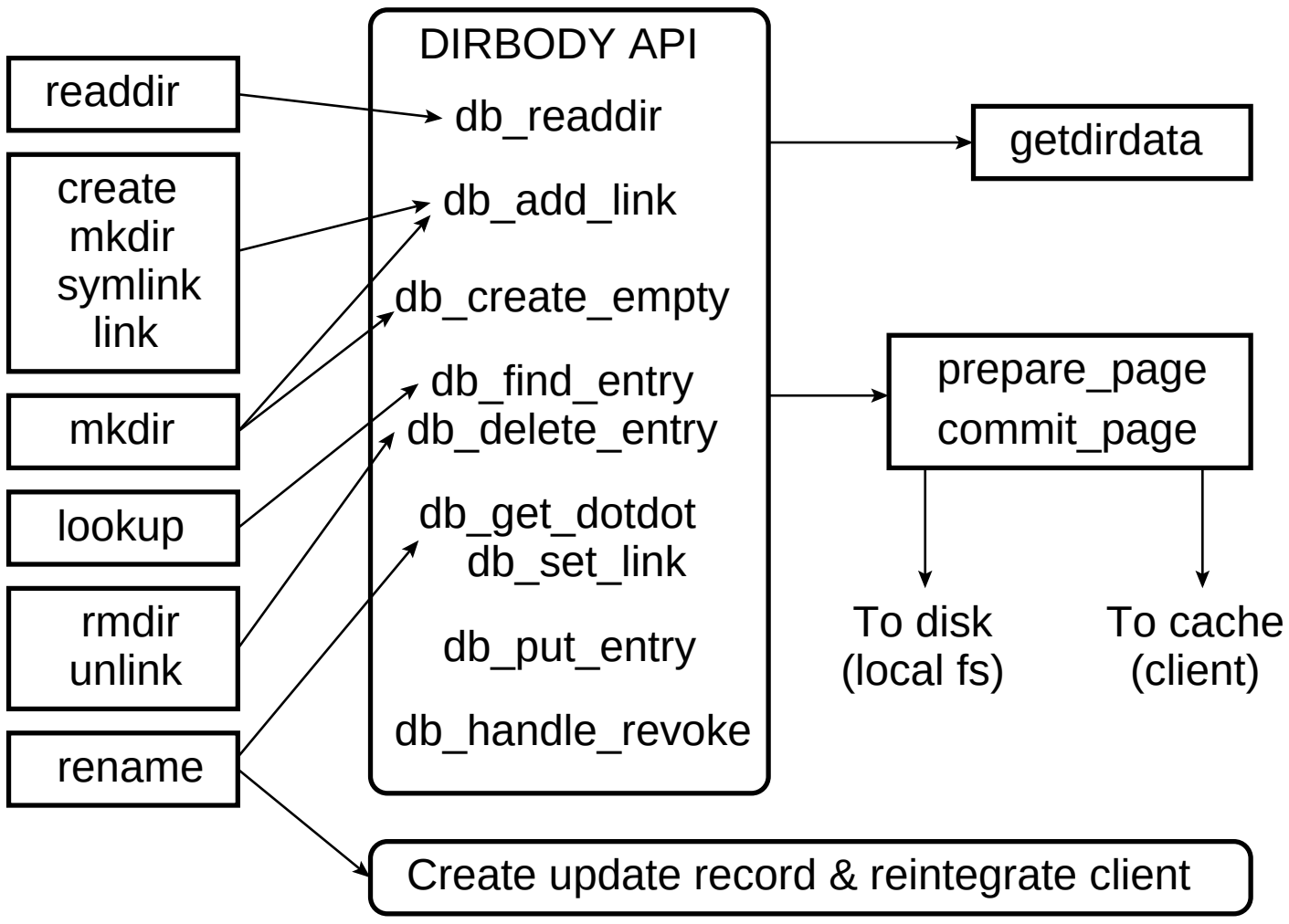

FIGURE 9.3.1. VFS

Handling of multiple directory formats is done through a **directory body API**. This API implements:

(1) The *readdir* VFS interface.
(2) An RPC client interface to fetch directory data.
(3) An RPC service to handle lock revocation.
(4) Several functions for internal use:
- ***db_add_link*:** To add an entry.
- ***db_create_empty*:** To create data for an empty directory.
- ***db_find_entry*:** To find an entry by name.
- ***db_delete_entry*:** To remove an entry.
- ***db_get_dotdot*:** To get the dotdot entry.
- ***db_set_link*:** To replace a file identifier.
- ***db_put*:** To put a locked directory entry away.

The component of the filesystem implementing the directory body handling must be able to fetch extents of directories, cache directory data, etc. We expect to require an RPC in the meta-data protocol to fetch directory data, which is implemented substantially differently for different directory formats. In the case of hashed directories, with directory data stored in pages, the parameters could have the following meaning:

```
int l_getdirdata(IN req_desc, OUT resp_desc, OUT vector_len, OUT *dirdata_bufs);
```

The request descriptor contains the file identifier of the directory and the requested hash-chain. The return descriptor contains the hashes that were actually sent to the client and is followed by an array of pages with directory contents.

**NOTE:** A *readdir+* variant, like in **NFS v3**, could be available to fetch hash-buckets of directories including the basic inode meta-data. It would be important to find good heuristics which would detect when *readdir+* is useful.

**NOTE:** A *readdir* must be restarted at a known good point when a directory changes while a *readdir* is traversing it. Usually this is implemented by versioning the file handle and the directory inode. *Readdir* is restarted if it is detected that the file handle version does not match the directory version. This leads to POSIX behavior of *readdir*/*getdents*.

9.3.2.1. *Page-based Directory Data.* The in-memory storage of the directory entries can take different forms, depending on the dirbody implementation. Typically we expect that directory pages are cached and consulted much as they are in disk and cluster filesystems.

The inodes will need to contain some information regarding the cached pages with directory information so that they can be released when all referencing dentries have been released. We can likely use the standard page cache of the directory inode.

9.3.2.2. *Lock Hierarchy for Page-based Directory Data.* There is a *read* lock associated with the hash-chain or directory pages which must be held for *readdir* or *lookup* operations to start.

As a result of cached directory data, the filesystem may instantiate dentries during *lookup*. The dentries are valid because a lock is held on the directory data. If the lock on the directory data is revoked, we have a number of possible outcomes:

(1) The dentries are purged.
(2) Before the lock on the page is released, the system locks the dentries that depend on it.
(3) The lock revocations on the directory data can include information on what fix to apply to avoid losing the lock. A client may surrender the lock on a hash-chain voluntarily and must then re-fetch the chain, or it may perform a **lock fixup** and make the required changes and retain the lock.

We will need to derive a policy which chooses the optimal method.

9.3.2.3. *Dentry-based Directory Data.* An alternative mechanism for storing directory data is to use the *namecache* directly. In this case, directories are stored as collections of **dentries**. The *d_data* file points to a structure which contains the device and object *id* of the corresponding inode.

- Dentries with *d_inode = NULL* and *d_data = NULL* are negative dentries, i.e. *cached* data indicating that a file does not exist. In a distributed system, negative dentries need attention. We avoid taking locks on negative dentries individually by purging negative dentries from a hash-chain when *read* locks are lost. When the locks are present, negative dentries can be placed and honored in the chain as long as the lock is present.

- Dentries with *d_inode = -1* and *d_data != NULL* are dentries for which inode attributes have not yet been fetched. They need to be intercepted by the *dentry_re-validate* method just before lookup takes a hold of the inode.

The hashing of dentries in the kernel dcache can be the same hash value as the on-disk representation when those hash values form a subset of the dcache hash values on the client. Otherwise, multiple disk hash values will get placed together in a single dcache hash-bucket. This is implemented through a dentry *d_hash* method.

With this use of dentries, the dentry *re-validate* call plays an important role in Lustre's lookup code. Since dcache entries may be present before inodes are attached to them, the kernel lookup mechanisms will indicate a cache hit for such an entry. When a cache hit happens, the filesystem is given the opportunity to re-validate the inode data in the dentry *d_re-validate* method.

The corresponding *readdir* directory file method is very similar to the *dcache_readdir* methods used in RAMFS.

**NOTE:** The dcache is a powerful component of the Linux VFS. In our work on Coda, we have implemented similar name caches for BSD, Windows 95, and NT. The considerations discussed here apply to those systems.

**9.3.3. Dentry Meta-data - *lookups*.** The purpose of a lookup call is to associate an inode with a pathname. In the Linux kernel, data structures named dentries are used for this purpose. This happens during all pathname traversals associated with execution of system calls.

The fundamental steps to perform a lookup are:

(1) Query the directory data to find the file identifier for a name.
(2) Fetch the attributes for that file identifier.
(3) Instantiate a **dentry** and **inode** with the correct attributes.

Observe that the attributes associated with the file identifier contain the inode number to be used on the client, as discussed above. For that reason the attributes, which contain the inode number, must be fetched before the inode is instantiated. We will not use the *read_inode* method and set the attributes in *lookup*. The *l_getattr* network method is used to fetch the attributes for the file identifier. This call was already discussed above for the read_inode interface.

```
lustre_lookup(struct inode *dir, struct dentry *de)
{
fid = db_find_name(dir, de);
l_getattr(fid, &attr);
inode = iget4(dir-i_sb, attr.ino, NULL, attr);
return 0;
}
```

```
lustre_read_inode2(struct super_block *sb, ino_t ino,
ino_finder, opaque)
{
attr = (lustre_attr *)opaque;
cpy_attr_to_inode(inode, attr);
set_methods(inode);
}
```

9.3.3.1. *Locking and Lookup.* Lookups are done with an **intent** to perform a further **operation**. A successful *lookup* will return with a guarantee that the name and inode information is correct for the intent. The operations fall into the following 4 classes:

(1) To continue a *namei* translation (for *lookups* on intermediate names in pathnames).
(2) To fetch the attributes of inodes associated with pathnames (for the *(f)stat* system calls).
(3) To modify the attributes of inodes associated with pathnames (for system calls like *chmod, chown, utimes*).
(4) To perform a modification on a directory (for system calls like *mkdir, rmdir, unlink, creat, mknod, rename, symlink*).

In each of these cases, different locking requirements govern the lookup. The first (continuing *namei* translations) requires a *read* lock on the directory. The second (*stat*-related) requires no lock, merely the attributes valid at the time of lookup. To modify inode attributes a *write* lock on the attributes of the inode is required. Finally, for directory modifications, the content of the directory data should be made available with a *write* lock.

In Lustre, *lookup* calls are serviced in two ways. By including the *nameidata* into the *lookup* argument it can be known what the intent of the *lookup* operation is. **Operation-based locking** exploits the knowledge of the intent on the server side of the locking call. The two methods of servicing the lookup calls are (1) to **not grant** a lock, but perform the operation in the intent, or (2) to **grant** a lock and not perform the intent.

The latter case enables writeback caching on the client, the former allows for a high concurrency on an object where server handling is more efficient. So for example, when *lookup* happens, it will:

(1) Utilize an existing lock on the directory hash-bucket and inode if available.
(2) Request a lock on the directory hash-bucket, presenting the intent of the lookup. Depending on the intent this could be a *read* or *write* lock.
(3) If the lock was granted, use method 1, otherwise return the result from the information included in the failed lock acquisition.

The lock request involves the meta-data protocol by including operation intents and results as in the *l_getattr* and *l_getdirdata* calls.

```
struct req_desc {
...
struct operation_buf *reqd_opbuf;
...
```

```
};

struct resp_desc {
...
struct opresult_buf *resp_opres;
...
};
```

We will regard this *lock/lookup* request as a part of the lock service protocol, where a callback function is invoked to try to execute the update when the locks are denied.

9.3.3.2. *Lock Conversion.* Finally, it may be necessary to convert a lock on directory data or attributes when the intent of a *lookup* operation becomes known. For this we will use a **lock conversion** protocol interface, which is again based on descriptors containing the intent.

### 9.4. Meta-data Updates - The Client Node Perspective

In this section we give an overview of how client nodes perform meta-data updates. There are several issues which need discussion:

**Locking:** Many distributed filesystems have had fairly sloppy locking semantics. We will discuss here what is required for cluster-wide Unix semantics and what issues need to be addressed in current Linux kernels.

**Describing updates:** Presently there is no standard way to encode a filesystem update. Implementations scatter various aspects of the updates over different regions of the kernel - some are done by the filesystem, some by the buffer cache, others by the VM. For a cluster filesystem it is useful to have a concise and complete description of an update for easy replay on other systems.

**Adaptation for concurrency:** Filesystems need to perform optimally when updating objects subject to high concurrency and also when concurrency is low. Depending on the concurrency of the object, the MDS may decide that the operation is most efficiently executed in a client/server model or alternatively grant a writeback lock to a client. We will describe how Lustre supports both models, and how it can provide improvements over existing network filesystems.

**Transactions and lock revocation:** We will describe how Lustre clients perform transactions and how groups of transactions are flushed to meta-data servers when locks are revoked.

This section gives an overview of these issues and it appears that they are intertwined at several points.

**9.4.1. Locking in Distributed and Cluster Filesystems.** Lustre will offer Unix semantics across multiple nodes. The concept of *Unix semantics* for network filesystems is not 100% precisely defined, but a good approximation is to say that it provides:

- Updates immediately visible to all processes in the cluster.
- Updates atomic with respect to other processes reading and updating filesystem information.
- Correctly implemented flags such as the *O_EXCL* option on open.

Disagreed upon areas concern file writes; large file writes tend to not be atomic. Secondly, memory mapped files rarely have Unix semantics across cluster nodes (this would imply that shared memory is implemented as part of the filesystem).

In practice, filesystems have implemented a variety of locking models:

1. **Strict distributed filesystem semantics:** Before a node is allowed to update an object, the nodes wait until all other nodes have surrendered their *read* locks on the object.
2. **Callback distributed filesystem semantics:** As soon as a node begins updating an object, other nodes are told they must re-acquire *read* locks. However, the update proceeds without waiting for processes that hold a *read* lock (AFS, Coda -connected).
3. **Optimistic distributed filesystem semantics:** Updates require a shared *write* lock on objects, i.e. one which can co-exist with read locks on the same object (InterMezzo, Coda - weakly connected).
4. **Lack of distributed filesystem semantics:** Multiple nodes can read and update objects (NFS).

It is vitally important here to understand that there are two notions related to locks:

**Pinning & releasing:** When a lock is released, it may be revoked; when it is pinned, revocations have to wait for the release. Locks cannot be pinned until they have been acquired.

**Acquiring & revoking:** When a lock has been acquired, it can be pinned for use. When a lock is revoked, it can no longer be pinned.

Many distributed filesystems have not been explicit about pinning and releasing locks and have relied on atomicity in the OS, usually known to the developers only.

Database transaction processing systems and a few exceptionally well implemented filesystems such as the VAXCluster Filesystem [7] have made these distinctions very clear since the early days, and usually have strict semantics. Distributed filesystems have had a history of weaker semantics, usually touted as a feature. These weaker semantics have led to endless complications in a correct implementation, particularly in NFS.

To implement strict semantics for a Linux cluster filesystem we will see that we face two obstacles which need to be addressed at the level of the VFS. For the discussion it is useful to consider the case of:

```
fd = open("filename", O_RDWR | O_CREATE | O_EXCL);
```

being *simultaneously* executed on more than a single node.

Linux update processing proceeds as follows (as of Linux 2.4.9):

(1) Copy strings from user memory to kernel memory.

(2) Perform a *lookup* on the parent(s).
(3) Perform a *lookup* on the affected entry/entries.
(4) Take **zombie** semaphores on parent inode.
(5) Perform permission checks, (non)existence checks; exit on failure.
(6) Take **kernel lock** .
(7) Perform the filesystem operation.
(8) Release kernel lock and semaphore locks.

As correct lookups are dependent on reading directory and inode information, a filesystem with strict semantics would want to acquire *read* locks during lookup. This is indeed what cluster filesystems like GFS do. The **first problem** is that at present:

> *There is no interface to hold read locks beyond the lookup and release them at the end of the operation. There is also no interface to promote read locks to exclusive locks during an update.*

As a result of this deficiency it is possible for another node to get a *write* lock between steps 3 and 7. The permission and existence checks performed by the VFS in step 5 are based on information which may have changed since the *lookup* completed. For example, node A may try *mkdir(foo)* while node B does *rmdir(foo)*. If node B completes its entire operation between step 4 and 5 of the operation on node A, then the operation on node A will fail, despite the fact that the directory was removed. Conversely, if both nodes performed the *mkdir* operation, all checks in step 5 would have to be repeated in step 7.

The typical method followed by filesystems is that locks are made exclusive when the FS-based update work starts. A second problem is that for issues like *O_EXCL*, the exclusive lock is most useful if it is acquired before step 5 so that the operation in step 7 could trust that the checks made in step 5 are correct. This leads to **problem two**:

> *There is no mechanism for a lookup to be aware of the* intent *of the operation.*

If a *lookup* were passed intent information, it would be the optimal place to acquire an exclusive lock for the purpose of an update. Note that the vast majority of *lookups* are for *read*-only information, so always acquiring a *write* lock is not an option that will lead to good performance.

There is a simple enhancement to the VFS which can provide the desired mechanisms, first publicly presented at a Red Hat Clustering get-together (see *Locking and the VFS* `http://www.lustre.org/docs/vfslocking.pdf`):

(1) When the first *lookup* is performed in stage 2, the FS is told about the entire intent of the operation.
(2) The VFS will pass this information to the filesystem layer in the *lookup* or *dentry_revalidate* methods. The FS methods can optionally acquire and pin locks and complete the the VFS *namei* translations.
(3) When dentries are put away after the operation, the FS will be notified again, to give it an opportunity to release locks (and possibly complete blocked revocations). A new step, step 9, is included in the update processing where the *nameidata* is released. A general filesystem needs notification so that it can release its locks.

Note that there is a reason that the new phase is a separate step and not part of the update operation. If after lock acquisition the checks in step 5 fail, the update method of step 7 will not be called. The filesystem must still be given the opportunity to release the locks.

The first major benefit of this method is already visible:

(1) The often delicate VFS coordinated semantic checks in step 5 can be preserved.

We will see below that knowledge of the intent can also lead to a highly optimized mechanism for making remote procedure calls.

**Other: platforms** Our focus here is on the detailed interaction between the meta-data updates and the VFS in Linux. On other platforms (Windows - excepting for file pages, BSD) we believe that the problems we are pointing out in the Linux VFS design are likely not present. However, the proposed implementation involving pinning, releasing, and using intents is valuable on all platforms.

**9.4.2. Update Records.** The update records we use are fundamentally quite similar to those used in InterMezzo. InterMezzo update records contain the complete information needed to replay operations on remote systems. Note that such replay involves meta-data updates which cannot normally be made from user space, such as setting the *ctime* of server inodes to that of the inodes on clients.

The key considerations are:

(1) The records will be packed in DAFS style packet format, but without the full DAFS request headers.
(2) Unlike InterMezzo version 1, the records will be based on *l_fileid* file-identifiers and not pathnames.
(3) Update records will not be written to permanent storage on the clients, but will be held in memory.
(4) Updates are grouped by dependencies on each other and dependencies on objects. Multiple update streams can be generated in the client systems in parallel.

The use of update records is most involved in the case of writeback caching which will be discussed below.

**9.4.3. Adaptation to Concurrency: Writeback and Client-server Mode.** As has already been indicated, Lustre envisages two paths to perform meta-data updates. In order to make an update, a CFS node must perform a *lookup* on the data involved, and acquire locks: this varies between a single inode when attributes are set, to as many as 4 inodes in the case of a *rename* operation.

The typical request sequence associated with a shared disk cluster filesystem processing is:

(1) Node sends lock request to lock manager for modification lock.
(2) Lock manager sends revocation messages.
(3) Lock manager receives revocation confirmation.
(4) Lock manager grants lock for modification.
(5) Node makes update.
(6) Other nodes re-acquire *read* locks, when accessing the modified objects.
(7) Update reaches stable storage, locks are granted.
(8) Modified object data is acquired by other nodes.

Lustre tries to follow a more efficient protocol. In cases of low concurrency, the MDS must be allowed to grant writeback privileges to clients. When the concurrency is high, it is more efficient if the MDS performs the operations and does not allow the clients to cache aggressively.

In cases of high concurrency it is considerably more efficient to ask the server to execute requests. The transition point between the two modes of operation is roughly where:

> Avg (remote execution) / [Avg (writeback processing) + Avg (lock establishment)] equals 1.

These two scenarios lead to the following methods for processing meta-data update requests, a processing also known as **reintegration**. If an exclusive lock is present, the operation can be performed with writeback caching. The client will build a record of the operation, much as in Intermezzo `http://www.inter-mezzo.org`, which is suitable for replay on a MDS. The records are organized in a variety of lists, so that all records that depended on a given lock can easily be found and such that record dependencies are implicit. Further grouping of records allows to reintegrate records to MDS when they have aged or when a device or file calls from the VFS. This is the case of **writeback caching**.

When MDS decides not to grant an exclusive lock to a CFS instance to perform an update, the MDS system will try to perform the operation as part of the lock acquisition. The lock status returned to the CFS will indicate if the lock was granted, and if not, whether the operation was performed on the MDS. If so, the client stores this in the *nameidata* associated with the operation and instantiates an in-memory copy of the modified object. Here we speak of **client-server processing**.

In this case of client-server processing, we have the following considerations:

(1) Passing the intent of the *lookup* to the FS. The mechanism for passing this information is to build an operation record, which contains the operation information, in the *nameidata*. The *nameidata* should also allow the filesystem to store pointers to lock structures.
(2) If locks are not present, the VFS builds an update record and transmits it to the MDS with the request for a lock.
(3) It remains to be determined if we only do this when the *lookup* hits the parent or if the entire filesystem pathname is passed to the MDS in case locks are not present. This is a special case of the usage of **intents**. Here the intent is to continue a *lookup*.
(4) The server will indicate that it performed the operation and that no lock was granted. This happens **before** the *lookup* in step 2 returns. The filesystem can now adjust the cached information in such a way that further *lookups* in step 2 and those in step 3 proceed without blocking, that the checks in step 5 pass without network traffic, and that the operation in step 7 sees in the dentries and inodes that all it needs to do is to update the VFS caches with the modification.

We now see further advantages of the intent-based locking:

- A single lock acquisition can suffice at the servers discretion. Multiple locks associated with lookups and conversion of locks associated with updates can be avoided completely.
- The RPC can be bundled with the *lookup* request processing. This will not only benefit Lustre but can improve an **SMBFS & NFS (version 3 and 4)** client as well by sending linked command chains.

**9.4.4. Writeback Caching and Lock Revocations.** The writeback caching protocol is more involved on the client:

(1) A lock request is fired off with the first *lookup*. All locks involved in the operation are granted, or already available.
(2) The update records which have been generated at lookup time are made part of a list of updates associated with the objects affected by the operation. Pointers to the request records are also associated with the dentries or inodes involved in the updates. For full concurrency it would be desirable that multiple updates can proceed in parallel.
(3) When the locks are taken, it is indicated that the dentries involved are using a reference to the lock by using a refcount.
(4) The *lookup* function now returns to the VFS, which subsequently performs tests to see if the *update* function can be invoked. In the case where it cannot, the operation fails, and a FS method is called which releases the update records (*release nameidata*).
(5) The client proceeds to make the updates to the dcache and cached inodes. It adds the update records to the writeback cache, and releases the lock reference and frees the *nameidata*.

In the writeback case the updates are propagated to the server when locks are revoked or when timeouts or memory pressure force caches to be shrunk. Additionally there is the case of clients requesting synchronous operations such as *sync*, *fsync* calls, or *updates* associated with files or directories opened with *O_SYNC*.

**Note:** Even when concurrency is high, the writeback locks will be useful. Imagine 10,000 clients creating subdirectories 1-10,000 of `/lustre/data`. Client `N` then proceeds to unpack kernel tarballs in `/lustre/data/N/`. It is important that the *mkdir* operation:

- Does not grant locks on `/lustre/data`.
- Does grant a lock on `/lustre/data/N` to the creating client.

The locks are acquired for objects. Objects may be involved in a transaction. Multiple transactions modifying a single meta-data object are possible and should be serialized (for the most part that happens automatically). When updates affect different objects, there is no need to serialize them from a transaction processing perspective (provided they are *isolated* in the strict sense of transaction processing), but it can be quite strange both to do so and not to do so.

Generating a single (hence ordered) update stream on a client is simplest. In case of a lock revocation, this would involve flushing all transactions preceding the ones associated with the lock.

Having multiple update streams is possible; it is slightly more complicated to recover from crashes in that case, but not impossible. The general arrangement of locks, objects, and updates is indicated in figure 9.4.1.

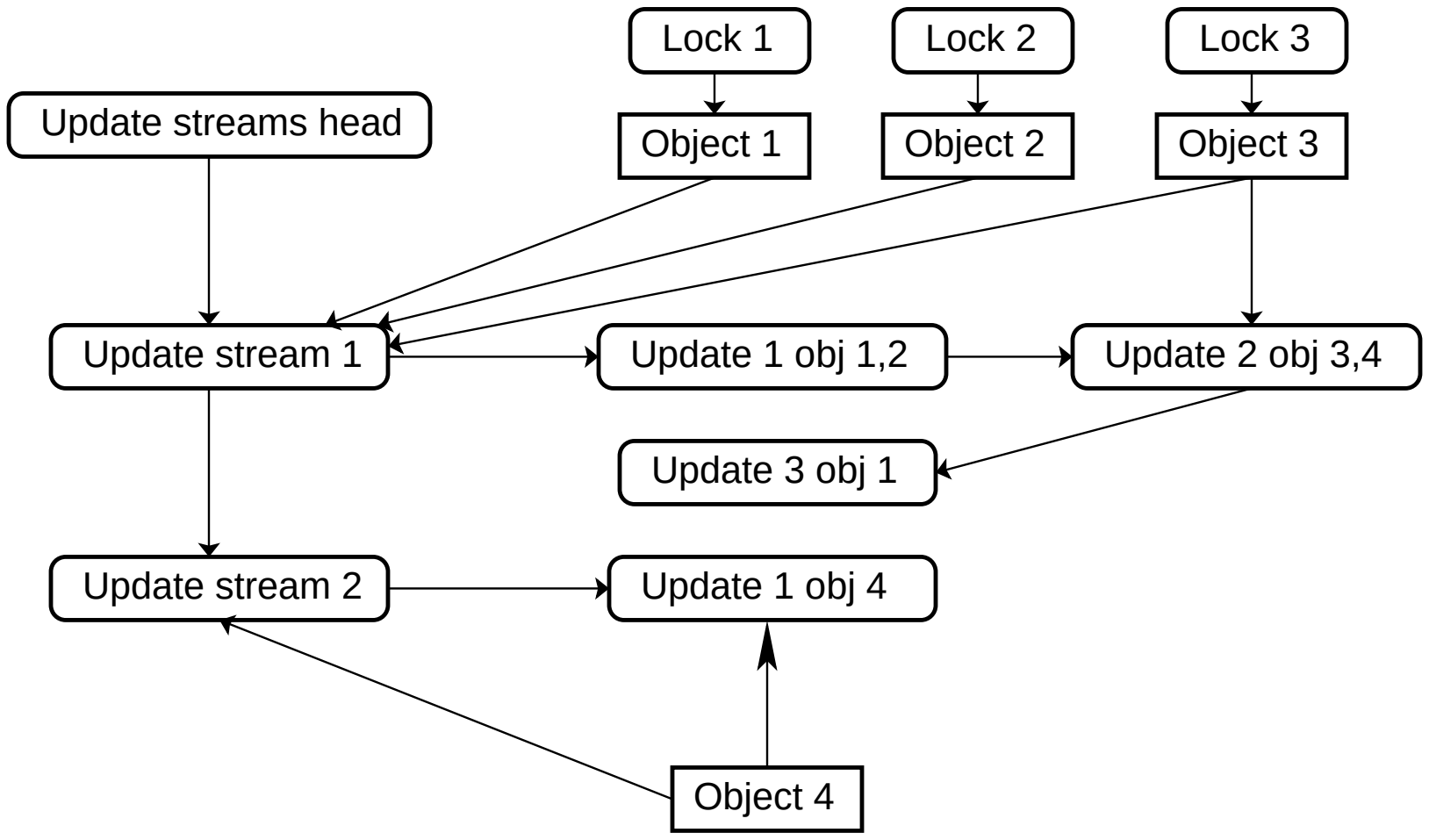

FIGURE 9.4.1. Overview of Meta-data Handling on a Client

Updates affecting related objects would still become part of one sequence (*tar zxf linux.tgz* would generate one sequence). If user U creates `dir1/f1` and then `dir2/f2` - would we accept, in the case of a crash, that possibly `dir2/f2` makes it to disk and `dir1/f1` did not? (I think we should allow this. Local FS's have different behavior for this - ext3 keeps order, I think, and ext2 doesn't.)

Another issue (recognized by Coda already) is that it is dangerous to have multiple users in a single update sequence. If a capability is refused by the server, the user owning that capability may block updates generated by another user. There are two reasons that capabilities might be refused: **token timeouts** and **administrative revocation authorization**. In the former case, update failures are needlessly annoying; in the latter, they are a beneficial feature and refusing all further updates from a client holding administratively revoked capabilities could be considered beneficial.

This rule doesn't have a big impact on the design, except in the case that an update is wedged behind another which might lead to a reboot of the client. A busy NFS/CIFS server working on a shared `file/directory` will see lots of updates on one object originating from different users. This is a case which merits further discussion.

#### 9.4.5. Meta-data Update Protocol Calls.

The core calls are:

**In l_reintegrate (IN int flags, IN int count, IN CHAR **records):** The first call sends a sequence of update records to the MDS for reintegration. There are in fact two completions of this call:

**In memory completion:** The server has updated its memory cache.

**Asynchronous commit notification:** The updates have reached stable storage.

### 9.5. MDS Handling of Client Meta-data Update Requests

As we have explained above, updates are performed in client-server or writeback mode. In both cases, updates are serialized in records, then transmitted to the MDS systems and processed there. Here we give more detail about the rationale behind this approach and the handling of the requests on the MDS.

We also pursue the implications for the server infrastructure arising from the use of file *id*'s. We explain why it is desirable to have multiple namespaces for filesystems, by file *id* and by name.

#### 9.5.1. Server-based Lock Management.

Standard lock sequences involve revocations of locks, and separate lock acquisition and conversion from filesystem updates. There are several optimizations in lock acquisition and revocation which Lustre utilizes:

**Lock versioning:** Lustre's lock will hold version information about objects. If locks are re-acquired, the system granting the lock will indicate what version of the data is still valid, so that it may not need refreshing. This is a fairly standard concept described in "VAXClusters "[7].

**Operation-based lock fixup:** When revoking a lock, the holder may be offered sufficient information to retain the lock provided a change is implemented.

**Operation-based lock acquisition:** When a lock is acquired the operation for which it is acquired is included in the lock request. The nodes can grant the lock and/or perform the operation to deal with situations of higher and lower concurrency.

(1) Node sends lock request which includes modification request to MDS.
(2) MDS validates that it has locks to perform modification and sends revocation messages including the update information to other nodes, and confirmation of operation to the requesting node.

This protocol is fundamentally quicker but has more complicated recovery properties in case the cluster sees network or system failures.

#### 9.5.2. Server Infrastructure.

The servers are going to see request sequences for directory and object meta-data, lock request traffic, and update records. Such a pattern is best served when frequently used objects are aggressively cached, but also receive messages from the VM system when shrinking is advisable. For this we intend to leverage existing caches in the OS, notably the inode and dentry caches.

It is of paramount importance to regard **file identifiers and names as two families of primary objects** in the server and client infrastructure. On the clients, usage of the filesystem will generate requests to the filesystem in terms of names. These are translated to requests for **file *id*'s** by the infrastructure sending requests to the server. During *read*-only filesystem traversal the MDS does not see names at all, except in directory pages which it doesn't parse.

Updates processed by the MDS are given in terms of pairs of file *id* and name. The file *id* is for the parent objects and the name is for the affected children. Similarly, revocations of locks are sent to clients in terms of file *id*'s.

The caches will rely on an object filesystem, or an object interface to existing filesystems. The names of the objects are the file *id*'s. The objects themselves, including their attributes, are stored in inodes. Directory inodes have pages for directory entries and file inodes have pages with I/O object information.

**9.5.3. The MDS Backend.** The MDS backend is the system which stores meta-data on persistent storage. As we have shown above, our model is that the MDS maintains a collection of persistent objects which has a similar layout as a filesystem. The major difference is that the file inodes contain references to other storage objects, not data for the files.

There are a number of possible choices for such meta-data storage:

(1) A filtering single node meta-data filesystem.
(2) An object-based meta-data filesystem.
(3) A shared block cluster filesystem.
(4) A clustered database.

We will make only one requirement on the MDS systems: after recovery they form a consistent meta-data filesystem. This means that all filesystem transactions were executed atomically with the transaction stream maintenance.

*Filtering MDS Systems.* A filtering MDS system is a filter layer over a filesystem that can implement the replay of update records, and the maintenance of the transactions in the update streams. Such a file server is very similar to the InterMezzo file server with a few minor differences:

- Journaled file writes: The meta-data regarding data objects associated with files will be stored as file data associated with file inodes on MDS systems. To give consistent updates, this data needs to be journaled.
- File identifier-based lookup: InterMezzo is a protocol based on pathnames; Lustre relies on file identifiers. Lookups in the MDS systems will have to find objects by file identifier.

*Clustered Object Meta-data Filesystem.* Perhaps our most natural approach is to use an object storage target for meta-data. The benefits are that the recover, lock, and object storage protocols used between clients and OST systems can be re-used to implement shared object filesystems. By using objects we also inherit the flexible API's for storing meta-data, e.g. those associated with security.

A possible drawback is that unless sophisticated logical modules are written that can execute transactions across multiple OST devices, the meta-data for a single fileset will be stored on a single OST. While such OST's can be made redundant and use heavyduty block storage backends, there is the potential for a bottleneck in cases of extremely heavy meta-data traffic on a single fileset. Filesystem transactions do not cross fileset boundaries: i.e. *link* and *rename* filesystem operations fail. We feel this limitation is acceptable.

*Shared Block Filesystem.* An alternative infrastructure would be an existing cluster filesystem, using file data for meta-data storage. While this might initially provide a head start, the addition of new features to the existing cluster filesystem might prove hard.

*Database-managed Meta-data.* A third proposal would store meta-data in a high end clustered database, much as SQL Server is rumored to underly the filesystem in future versions of Windows 2000. It is certainly true that databases can handle different formats of meta-data easily and have high transaction throughput capability; however, the glue between the meta-data server and the backend might involve some pretty unusual design. Given that our MDS is really intended as a transaction engine, this is worthy of further exploration.

### 9.6. The File Protocol

In this section we focus on file I/O. From the Lustre filesystem perspective, files are always stored on an object device. That object device can be a single direct drive, a remote object storage target which is addressed through an object storage client (OSC) device, or a logical object volume (LOV). The filesystem is not aware of the exact nature of the object storage device used, and handling these particulars is left to that device itself. The two extreme cases are large computer clusters where the files are layed out in 100's of OST's and at the other end the case of an existing filesystem exported as an MDS and OST, where the objects and their data are found among the inodes in the MDS.

We first introduce the descriptors for file inodes in Lustre. After this, we describe the protocols associated with file I/O. We review what opening a file entails. We follow this with I/O for filesystem use and parallel I/O support.

When files are opened, the MDS holding the meta-data is contacted and asked for the I/O meta-data. This I/O meta-data is contained in an (often extended) attribute of the file inode on the MDS. This meta-data contains a descripor of the object that holds the file data. The descriptor, except for the object identifier, is an opaque entity which is not interpreted on the MDS nor in the filesystem. It is handed to the object storage driver used by the filesystem for file I/O and interpreted only there. In the case of striping using an LOV, the meta-data may have a striping pattern and information about the stripes. Data can be striped over possibly 1000's of OST nodes and the I/O meta-data could be a few KiloBytes in size.

The MDS will also provide a capability, possible NASD style (see "NASD" [**10**]), to allow the OST to enforce access control.

The configuration database names the devices on which file inode meta-data names the logical device on which the data is stored. The configuration of a logical object volume associated with an MDS is somewhat involved. The LOV will be named in the resource database with a name and UUID. The setup of the device provides little more than a name for it, visible on the client. The true configuration happens when the client connects to the MDS and retrieves the LOV layout information. This is stored in a file on the MDS.

#### 9.6.1. Describing Devices.

**9.6.2. File Inodes.** Several invariants will be driving the definition of Lustre file inodes:

9.6.2.1. *File Inodes.* Lustre FS file inodes have data that is stored on a Logical Object Volume/device (LOV).

Naively, an LOV has a striping pattern and maps to a collection of OST devices. A UUID will be used to refer to a particular LOV. The cluster directory services contain a descriptor for the LOV that clients and MDS can cache. The descriptor tells the client/MDS what other logical drivers, client targets, and direct object volumes to use to access this LOV. For direct drivers, the bottom of the stack, this will come down to what disk partitions to use.

9.6.2.2. *Access Control.* All access to the file is governed by the meta-data cluster which will directly or indirectly authorize access. [I'm no longer sure if Kerberos or NASD style security is best for the OST's.] The MDS manages all resources (quota, allocations, etc.) of objects stored on OST's, including their use in LOV's.

9.6.2.3. *LOV: Logical Object Volumes.* The LOV descriptor can contain 100's and perhaps 1000's of references to direct OV's on OST's. It is important that the LOV is a manageable storage device (The direct OV's similarly are referenced by UUID through the resource directories).

Therefore, the objects stored in the LOV must be stored in a table to deal with operations that require iteration: migration, snapshot removal/rollback, failures and rebuilds of individual direct OV's, etc. This requires a table allowing (rapid) insertion/deletion/iteration of any/all objects on a LOV.

An important example is the following: a (redundant) disk array explodes. We need a fairly rapid way of marking the logical objects that reference the array as "bad objects", possibly even printing out a list (without asking files to get complicated errors in extents or views retrieved from the object).

For large collections of objects it is desirable that the table remains updatable during iterations, i.e. linked list structures may need to be embedded in the table.

Note that for direct OV's this directory is intrinsically part of the object device. Certain LOV's, such as the snapshot driver we built, have an object table that is a parasite of an underlying direct OV.

9.6.2.4. *File Inode Meta-data.* The LOV object may contain 1-1000's of subobjects; this configuration forms the meta-data of the LOV object. The MD, as far as I can see, consists of (a) the identity of these subobjects (what object on what OV), and (b) their persistent view descriptor (what part/stripe of the file do they describe).

The storage location of this LOV object meta-data can be data, meta-data, or extended attributes of the file inode in the MDS cluster.

There is a project called "Clusterfile" that has introduced neat data structures, called PITFALLS, for LOV object meta-data data. These descriptors are capable of describing pretty general MPI views, see the recent Monterey Cluster conference and upcoming Usenix FAST papers XXX.

9.6.2.5. *LOV Tables.* The table mentioning the inode and the pointer to the LOV object must have standard journal style FS recovery. Probably, therefore, the table of LOV objects (see 3 XXX) must be stored in the fileset that holds the meta-data for the fileset (otherwise consistency takes on an altogether new level of complexity).

So for these object volumes, the LOV object table is a parasite of the fileset meta-data that uses it.

9.6.2.6. *Consistency and Recovery of LOV Subobjects.* The subobjects reside on a different OV as the LOV and the consistency style between the LOV object stored in the MDS volume and subobjects stored all over is distributed; asynchronous replication consistency/recovery. These protocols are followed by InterMezzo and all database replication technology such as Oracle Snapshots. The InterMezzo implementation can be used in Lustre.

9.6.2.7. *LOV Descriptors and Storage Management.* The descriptors for OV's and LOV's held in the resource directory may change - this is why volumes must be described by reference. When the descriptor changes and the system (could be client, MDS, or OST) refreshes the data for it, it will install logical device drivers that match the new descriptor.

The classic example is somebody wanting to replace a disk array in an OST with a bigger one. The OST may be made aware of its bigger brother by updating its descriptor. It inserts a migrator and stores/retrieves data from the correct device depending on the state of the migration.

Another case is that we may replace an OST with a replicated OST (perhaps a remote encrypted one). In this case the client may see a new descriptor and send its write packets to both object drives, encrypted to the remote drive.

We need two fundamental operations: one is *callback* or lease-based updates of device descriptors with automatic system reconfiguration and second is a mechanism to prepare and introduce permissible device configuration changes in the resource directory. I suspect that the three fundamental pieces are (1) physical resources used, (2) descriptors, and (3) references by UUID to descriptors. References must change atomically and follow validation that the changes are permissible (sensible).

While Lustre is not going to be an exercise in object storage management, we want to leave infrastructure behind that can build on this.

**9.6.3. Opening and Creating Files.** The creation of a file is done between the client and meta-data server. The server creates an inode for the file, either immediately in an RPC case or at the clients in a writeback cache. Associated with the inode are a number of objects on storage controllers. The MDS has pre-allocated these with the OST's carrying the fileset. Clients can request the *id*'s of pre-allocated objects to perform full creations in writeback caching mode.

The MDS places pre-allocation calls to the OST (1) to reserve objects which the clients can use and (2) to reserve disk space on the OST, so that quota can be granted to the clients. Such pre-allocation calls are discussed in the Object Storage Protocol Specification found on the Lustre website `http://www.lustre.org`.

When existing files are opened by a CFS system, the CFS system makes a request to the MDS. The MDS transfers a certain amount of file inode meta-data to the client, including a token for security, containing:

(1) What objects on the storage controllers need to be used.
(2) What quota are available to clients before it needs to refresh the open handle on the MDS.

The meta-data belongs to a device which has a descriptor transferred to the CFS at fileset mount time. The device descriptor delivers:

(1) On which storage nodes data for the requested file resides.
(2) What striping information pertains to extents of the file.

The second phase of the *open* call consists of sending a *parallel* request from the CFS to the OST systems, to open the object. The object will also be created in the same call if it doesn't yet exist. This *open* call has not yet been documented for the Lustre object protocol, but is a necessary addition to introduce NASD-style security.

After the *open* call has completed, the Lustre file handle instantiated in the client kernel contains all the information needed to do I/O. This I/O can proceed in two different ways: (1) the system call interfaces for file I/O use the page cache for ordinary filesystem use, while (2) the other allows OS-bypass byte granular scatter-gather I/O from user space.

**9.6.4. Reading and Writing Files: System Interface.** For ordinary use of the filesystem, we will use the page cache. This will lead to a transfer of full pages between the OST and CFS systems. It is important to realize that on the CFS side, a relatively sophisticated logical object driver is layered over a collection of client drivers for OST devices. The logical driver activates client drivers to fetch appropriate extents of objects using the clients. This logical driver in turn can be driven by direct synchronous I/O or by the page cache.

A basic object storage-based page cache has been built as part of the Lustre project in the OBDFS (the XFS filesystem also has a page cache with similar capabilities as the OBDFS cache).

Unix semantics imply that the results of *writes* are immediately visible to readers. These semantics still have a number of variants about which the industry does not seem to agree. For example, some variants of Unix semantics state that *writes* will be atomic, while others only support this atomicity at a per page level.

In this setting of system interface file I/O, Lustre offers Unix semantics. *writes* to a file are preceded by the revocation of *read* locks on cached pages. Such revocation is managed through a lock service running on a proper subset of the OST I/O targets involved in accessing a LOV.

**9.6.5. Reading and Writing: OS Bypass.** In this section we cover so called parallel I/O usually associated with specialized applications. The I/O patterns seen in these applications can be summarized by a few key characteristics:

(1) The applications read and write small buffers.
(2) The applications do not require OS-supported synchronization of pages.
(3) The applications use fast networking and avoid user/kernel copies with OS-bypass interfaces.

In this section we cover the basic mechanisms supporting such I/O. We also outline the components of a more comprehensive parallel I/O implementation which exploits the object protocols.

9.6.5.1. *Parallel I/O.* When an application wants a handle to perform parallel I/O on a file, we will assume that the file will not be used through the system interfaces at the same time. Doing so will lead to undefined results.

Parallel I/O will take place over a user level handle to the object storage targets. Such handles are available to the kernel filesystem after a file has been opened. We will introduce an API that clones the kernel file handle to an authenticated user space file handle, which contains all the objects and storage target identifiers required to do I/O.

```
struct lustre_fh *lustre_clonefd(int fd)
```

A basic parallel I/O implementation now requires a vectored *read*/*write* interface:

```
struct buf {
char *addr;
uint64 len;
};
struct extent {
uint64 offset;
uint64 len;
};

struct lustre_fh *lustre_read(int fd, int count, struct extent **, struct buf **);
```

These calls map directly to the object storage *read*/*write* interfaces for objects. We expect that the semantics are that the *writes* are dispatched immediately and synchronously to the server, using OS bypass. There will be no locking of file extents when I/O is performed along this path.

9.6.5.2. *File Views and Collective Operations.* Increasingly the MPI-IO paradigm is used in scientific computations. The two core components are efficient ways to handle **file views** and mechanisms to perform **collective** operations. Using the object storage framework, an efficient and elegant implementation of these mechanisms is possible.

Applications can achieve better performance if they can coalesce the regions they need to *read* and *write* into larger buffers. On the client side, there are so-called **views** of files, which define a collection of extents through a pattern descriptor that often describes elements in a file associated with a phase of a scientific computation. Such views consist of many small buffers, at byte granularity.

On the storage side, it can be attractive to write files out in striping patterns that optimize some of the I/O accesses made to them. From this we see that a possibility for aggregation of buffers can exist:

(1) The application, or a **logical object module** on the client, translates between views and buffers.
(2) A logical module on the storage target translates between striping patterns and buffers.

Note that the input to 1. is the output to 2. and vice versa.

The Cluster file project has developed data structures and algorithms to perform these translations that seem eminently suitable for implementation in logical object modules.

9.6.5.3. *Collective Operations.* Collective operations are mechanisms where a collection of systems perform *writes* to a file and the systems need to synchronize when all these *writes* are done.

A logical module for parallel I/O purposes can easily implement the creation, tracking, and completion of collective operations. The details will be described elsewhere. A key feature to be included is to form a group of systems that participate in collective I/O. I/O requests from all other systems will be rejected.

**9.6.6. Miscellaneous File I/O issues.** A number of architectural issues have come to the foreground and require attention as Lustre develops.

9.6.6.1. *Adaptive Striping.* Instead of fixed striping patterns, it would be desirable if the filesystem could support striping descriptors that can make effective use of resources. For example, a policy might be to fill all storage devices evenly or to select the least busy storage nodes to get maximum throughput.

Going one step further, detecting the *write* patterns and building a descriptor dynamically that makes good use of the patterns, could help considerably. ROMIO (XXX insert link here) already adapts to many of the features that come with HDF5. It remains to be seen to what extent we need to go beyond that in terms of good support for HDF5.

9.6.6.2. *Special API's.* Mounting with no lock options. Disabling locking per file. A collective operation lock, blocking normal access until that lock is dropped. Mounting with *O_DIRECT* options. A battery of *fcntl* functions to change striping behavior.

9.6.6.3. *Allocation Policies.* OST's should get smarter about allocation policies. It is quite possible that the LOV striping information can provide very good hints to the OST's as to what to expect. Read ahead policy is crucial as our read performance is quite bad at present.

9.6.6.4. *Collaborative Cache.* A collaborative cache surrects a group of systems as proxy servers for file reads. Such servers should be published to clients trying to read, and the lock data available to OST's should show what file sections are already available on a certain proxy server.

In Labaratory environments, a dedicated group of systems can become the collaborative cache, while in commodity clusters, client nodes may also act as a cache node. This will be designed in detail for Lustre Lite Performance.

## 9.7. Changelog

**Version 3.0 (Dec. 2002)**

(1) P.D. Innes - updated and resized figures

**Version 2.0 (Nov. 2002)**

(1) P.D. Innes - updated text, URL's changed

**Version 1.5 (July 2002)**

(1) P.D. Innes - proofed, spell-checked, figures retooled for reference, Bibliography & Changelog added

**Version 1.0**

(1) P. Braam - rewrite

CHAPTER 10

# File striping configuration

## 10.1. Introduction

The Lustre filesystem may consist of a very large number of OSTs, and improved filesystem performance can be achieved by striping files across the various OSTs. File striping can help improve parallelism in the filesystem by having different clients access different parts of a single file concurrently. It is important to use a good striping pattern to achieve the maximum benefits from it. In this chapter we will first explain the various methods available with Lustre to set different striping patterns, we will also explain how per-file striping pattern can be set, we will conclude the discussion by explaining the effect of the various striping patterns on aggregate system performance.

## 10.2. Default LOV striping pattern under Lustre

There are several ways in which objects can be striped over multiple OSTs using the Logical Object Volume (LOV) driver.

There is a system-wide default striping pattern which is specified at filesystem creation time as part of the Lustre configuration. When a LOV is first specified to lmc, it takes the system-wide default striping configuration as parameters, and these are stored on the MDS.

```
$lmc -m <configfile> --lov lov1 mds1 <def_stripe_size> <def_stripe_count> 0
```

**configfile:** The name of the Lustre XML configuration file being created

**def_stripe_size:** The default number of bytes stored on each OST before file I/O moves on to the next OST.

**def_stripe_count:** The default number of OSTs that each file is striped across. Specifying def_number_stripes of 0 means stripe across all available OSTs. Specifying *def_number_stripes* of 1 means each file will be stored on only a single OST.

For example, if we specified *def_stripe_size* as 65536, and *def_stripe_count* as 2, then we bytes 0 through 65535 of each file would be in an object on the first OST for that file, and then bytes 65536-131071 would be in an object on the second OST for that file. Bytes 131072-196607 would again be in the first object on the first OST, and bytes 196608-262143 would again be in the second object on the second OST, etc.

Which OSTs are selected to hold the objects for a particular file are by default chosen based on the inode number returned from the MDS, so that objects are distributed evenly across all OSTs even if the default *def_stripe_count* is less than the total number of OSTs configured.

It is not possible to specify a striping pattern where the product of *stripe_size* and *stripe_count* is larger or equal to 4GB on 32-bit platforms such as i386.

### 10.3. Striping configuration per file

It is also possible to specify different striping configurations for each file created, at file creation time, via *ioctl(2)* from an application or by creating new files with the *lstripe* program before writing into them. It is not possible to change the striping pattern of a file after it has had objects allocated to it (which normally happens when the file is first opened, and even programs like *"touch"* will open a file to set the file modification time).

The *lstripe* program (part of the Lustre utility programs) allows one to create a new file and specify the stripe size in bytes, the starting OST, and the number of OSTs to stripe across. Normally, OSTs after the first one are chosen consecutively (returning to OST #0 when we hit the last OST). Usage is as follows:

```
$lstripe <filename> <stripe_size> <starting_ost> <stripe_count>
```

**filename:** The name of a file in the Lustre filesystem which does not yet exist.

**stripe_size:** The number of bytes stored on each OST before file I/O moves on to the next OST.

**starting_ost:** The number of the OST on which the file should start being written. OST numbers are assigned when the filesystem is first being created, in the order the OSTs were specified in the configuration file to lmc, and start with 0.

**stripe_count:** The number of OSTs over which to stripe the new file.

For example, if we specified *stripe_size* of 65536, *starting_ost* of 1, and *stripe_count* of 2 (and we had only 2 OSTs configured), then we would have the first 65536 bytes of the file on OST #1 and the second 65536 bytes of the file would be on OST #0, etc. If, instead, we specified *stripe_count* of 1, then all of the file data would be on OST #1, and the *stripe_size* is mostly irrelevant (although it may affect network I/O vector size).

To specify the striping pattern of a new file from within an application, one needs to open a new file with open(2), and then call ioctl(2) to set the striping pattern on the file before it is used. If another process opens the new file before the striping pattern has been set, it will use the default striping pattern for that file. If O_LOV_DELAY_CREATE is used, but then LL_IOC_LOV_SETSTRIPE is not called, the file descriptor returned from the initial open() is not usable.

```
#include <linux/lustre_lite.h> //for O_LOV_DELAY_CREATE, LL_IOC_LOV_SETSTRIPE
#include <linux/lustre_idl.h> //for struct lov_mds_md, LOV_MAGIC
struct lov_mds_md stripecfg;
int mode = 0644;
```

```
int fd;
fd = open("<filename>", O_CREATE | O_LOV_DELAY_CREATE, mode);
if (fd < 0) <failure>;
stripecfg.lmm_magic = LOV_MAGIC;
stripecfg.lmm_stripe_pattern = 0; // only 0 is available now
stripecfg.lmm_stripe_size = <stripe_size>;
stripecfg.lmm_stripe_offset = <starting_ost>;
stripecfg.lmm_stripe_count = <stripe_count>;
if (ioctl(fd, LL_IOC_LOV_SETSTRIPE, &stripecfg) < 0) <failure>;
```

### 10.4. LOV striping configuration and performance

It has been observed that while it is important to keep all OSTs in a LOV occupied for maximum aggregate performance, it is counter-productive to stripe all files across all OSTs to do so. This is caused by the fact that striping many files over many OSTs causes lots of disk contention on the OSTs. Also, an object is created on each OST on which a file is being striped over which slows down the initial create, and increases the number of RPCs needed for file size calculations, file data locking, etc.

For normal filesystem usage (i.e. many small files), it is probably optimal to "stripe" each file over only a single OST, since you will rarely have files larger than a single stripe and avoid overhead from doing per-file operations on multiple OSTs. Similarly, when many clients in a parallel application are each creating their own files, where the number of clients is significantly larger than the number of OSTs, the best aggregate performance is achieved when each object is put on only a single OST.

However, applications where multiple processes are all writing to one large file we need to stripe that single file over all of the available OSTs in order to achieve peak performance. Similarly, with few processes writing large files in large chunks, we need to stripe over enough OSTs so that we can keep several OSTs busy on both the write and the read path (e.g. 128kB stripes on 4 OSTs, and application writes of 512kB would be good).

### 10.5. Maximizing application read/write performance

Currently, applications which do file I/O using the O_DIRECT flag on open(2) (which also requires reads/writes be aligned on and multiples of 4096 bytes) have a significant performance benefit over regular file I/O. This is due to the fact that O_DIRECT writes allow multiple pages to be written with a single RPC over the network (up to 64kB at a time), and it also avoids using the page cache on the client nodes (which takes away from available memory for other applications).

For regular non-O_DIRECT read/write operations, the Linux VFS breaks requests into 4096-byte chunks, each of which is sent separately over the network. For a single thread and a large file, this can make a factor of 5-6x difference in the I/O speed. For smaller files, this will have less performance impact.

File I/O

This document describes the Lutre file I/O implementation.

## 10.6. Client

**10.6.1. Page Cache.** On the client the file system is using the Linux 2.4/2.5 page cache. This page cache periodically and under pressure pushes out data. This data is handed to the Logical Object Volume driver. This section contains a proposal how the LOV driver and the page cache interact for optimal transfers.

10.6.1.1. *Linux 2.4 VM Basics.* **prepare_write**, **commit_write**, and **writepage** are the most important *address space operations*, associated with the page cache of Linux inodes. These are involved in write caching.

*prepare_write* and *commit_write* are the two-phase write hooks and are called from generic_file_write. If *prepare_write* sees a partial write, it needs to ensure that the destination page is up to date, which can involve a read into the page, so that the page will be correct after the partial write. *commit_write* completes the write, marking the page as dirty in a write caching setup. *generic_file_write* holds the inode->i_sem over *prepare_write* and commit_write, which allows *commit_write* to update inode->i_size if the write extends the file.

*writepage* is handed dirty locked pages and is expected to write them to storage and unlock them. It is called from sync paths via *filemap_fdatasync*, from kupdate at regular intervals, from kswapd under memory pressure based on VM LRUs, and from just about any blocking allocation context when memory allocations would fail. Its primary job is to unlock the page it was given.

A locked page is marked by the ”PageLocked” atomic bit. Callers of ”lock_page” try to set this bit and block if its already set. More careful paths call ”TryLockPage” to set the lock and can fail if they find it already locked. ”unlock_page” clears the bit and wakes people waiting on it. Pages are locked before *writepage* is called and it’s *writepage*’s primary job to complete the io and unlock the page. Unlocking a page, then, is the only way to make progress for paths like *generic_file_write*, *filemap_fdatasync*, and *vmtruncate* that have gotten stuck trying to do work on a locked page.

The kernel makes heavy use of lists and flags to manage writeback state. Super blocks contain lists of dirty or locked inodes. Inodes have flags specifying if they’re dirty or are locked when someone is performing IO on them. Inodes contain lists of dirty, locked, and clean pages. Pages also contain state that marks them dirty (and up-to-date, which is different) or locked. A good place to start is with *set_page_dirty*, which marks pages dirty and puts their owner’s inode on the dirty list in its super block. The paths through *filemap_fdatasync* and *filemap_fdatawrite* do a pretty good job of documenting the migration from the dirty list, briefly through the locked list, and onto the clean list.

In 2.4, most filesystems implement write caching on top of buffer_heads. *commit_write* tends to call *generic_commit_write* and *block_write_full_page*. These associate pages with *buffer_heads* and the kernel keeps track of dirty buffer_heads.

10.6.1.2. *Lustre's 2.4 Caching Attempt.* Our caching in 2.4 is roughly as follows:

**prepare_write:** works as it did before; it only reads in a page if the target page isn't up to date and a partial write is coming.
**commit_write:** marks the page uptodate and calls "set_page_dirty" on the page which marks it and its inode dirty and puts them on its lists.
**writepage:** builds a "obd_brw_set" from the page it was given and also plucks dirty pages from the inode's dirty lists giving good batching on the wire.

When memory is getting tight on the system, "ll_file_write" throttles by trying to write back all the data under the superblock's dirty inodes and their dirty pages.

10.6.1.3. *Linux 2.6 VM.*

10.6.1.4. *Lustre Caching in Linux 2.6.*

**10.6.2. Memory mapping in the client file system.** 0. File Size: mmap should check the file size at the beginning of the mapping, that can probably be done with a simple getattr in the mmap method. I believe there is no requirement for mmap to track updates to the size, but there appears to be nothing saying that if a file grows a bit, and new data appears in a page, that we cannot show it.

1. What is the interaction between the flags passed to open and the options passed into mmap? From what you write it is pretty clear that a file opened for write can not be mmap'd with denywrite

(because decreasing the write count, already up'd by open, will fail with TXTBUSY) It does seem that DENYWRITE may require us to down the write count on the MDS if we want to follow a similar path as the local handling of this.

2. Faulting. The IBM approach seems to address just one issue. Clearly, the data in the file remains accessible to the application until the page is not mapped anymore. So we need a callback when the page leaves the page cache for the inode. I thought we just built such a callback, so i'm confused why you don't mention it. At that point locks associated with the mapping can be decreffed. This is also the answer on how to deal with cancellations, I think. You need to remove a page from the page cache and then cancel the lock. A progress rule can probably be constructed by allowing the job to run at least once before un-mapping the page.

3. Of course the hardware can generate a trap when a page is being modified for write. It seems that that trap should request a lock conversion from PR to PW. If things work satisfactorily at the moment for one client, I'm kind of inclined to do very little for 1.0. One thing that would be nice is to know that dirty data will be flushed within a few seconds, but that is probably already a property of our cache flush engine.

10.6.2.1. *Zach's notes on mmap & Lustre.* The initial mmap() syscall that sets up the syscall has the following steps that we care about:

(1) It checks permissions by comparing the requested modes for the mmap() operation with the modes stored in the file pointer from the time the file was opened. The mmap() code explicitly trusts the checks done at open time and doesn't care that they may not reflect reality anymore, as long as they once did.
(2) If the map was done with MAP_DENYWRITE, the mmap holds the inode->i_writecount for the duration of the mapping, forcing other writes to return -ETXTBUSY.
(3) the mmap() call does not seem to check the length of the mapping against the file size. some quick testing should verify this.
(4) the mmap() takes a reference on the file pointer, preventing ->release from being called until the mapping is destroyed

= mapping use – fault =

When a userspace address is referenced the filesystem is called in ->nopage to provide a struct page pointer to satisfy the mapping. Most do this with generic_file_nopage() which uses the page cache and ->readpage().

(1) notice that kernel accesses of userspace addresses, like the copy_from_user() in generic_file_read(), can cause a fault.
(2) ->nopage has to compare the faulting page with i_size to return NULL for accesses that are outside the file size.
(3) ->nopage isn't told if the fault is for a read or write access.
(4) ->nopage returns a page, it doesn't call a helper to instantiate the mapping. Notice that time may pass from when the page is returned to when the page appears in the address space.
(5) from some time after the mapping is in place, the userspace app can reference the page without running any kernel code at all

= sys_munmap() =

The syscall tears down the userspace mapping.

(1) it downs the count against inode->i_writecount, but without talking to the filesystem
(2) tearing down the mapping discovers dirty pages who make their way to->writepage()
(3) does an fput() on the file once the mapping is destroyed

10.6.2.2. *= implications for lustre =.*

sys_mmap. If we ignore the DENYWRITE mapping style, lustre seems to care very little about the creation of a mapping. The reference that is held by the mapping will shadow the ->open() ->release() handling alreeady present in lustre. For now it seems prudent to just return -EINVAL if a DENYWRITE mapping is requested, accompanied by a CERROR() with the name of the app. The permissions checking should just work as long as our ->open method makes sure that file->f_mode is properly initialized.

faulting. This is the exciting part. We need to take a DLM lock before allowing the faulter to proceed with their action. The complications are thus:

(1) We don't get called at some point in the future once the page satisfying the fault has really been unused. When do we drop our reference to the lock?
(2) The fault may occur in a context that already holds DLM locks. The app could be writing from an mmap()ed lustre file into an open fd on a lustre file. The lock for the write target might have already had its AST arrive, so the fault DLM match can't trivial match it.
(3) From 2.iv above, there is a delay between when the page is returned from nopage and when it is found in the mapping. If lock cancelation hits during this delay the page can be mapped into the mapping after cancelation and survive to return stale data.

It seems that IBM has cooked a patch to solve #3 at http://oss.software.ibm.com/linux/patches/?patch_id=923 . It adds a callback to the nopage caller that is called when the page has entered the mapping. I suspect this can be used to solve #1 as well. We will drop the lock reference after the page has entered the mapping. This doesn't guarantee progress if two nodes are just trading locks via dueling faults, but it seems less onerous than some manner of delaying refcounting via a timer.

Because ->nopage isn't told what kind of fault its satisfying, we'd initially have to get a PW lock on the page. This bodes ill for shared read mmaps, something that we may be interested in having work well. Should that day come, we can add a patch that adds a more reasonable ->nopage method which takes an argument to specify the type of fault.

Its not immediately clear to me how to solve the problem of nesting faults within other lock acquisitions. Maybe there is some lock flag magic we can use that I don't immediately understand? I hope we don't have to introduce some method that lets enqueues match locks that were issued by the same thread. That sounds horrible.

cancelation and invalidation. the truncate() system call path provides a cookbook for how to tear down mappings in the form of vmtruncate_list(). It is a small matter of programming to cook work-a-like function which compares the pages in the mappings to an extent instead of a single size that is being truncated to. Beyond that, and presuming we use IBM's fix to work around the race with instantiation, its not that interesting. It will need to transfer 'referenced' data from the page tables it is tearing down into the struct pages so that the rest of lock cancelation can issue writeback on the pages.

dirty mmap() writeback. writing through a mapping is interesting for the llite IO state machine because it results in ->writepage() being called on pages that ->commit_write() has never seen. It is relatively straight forward to teach ->writepage() to build up the obd_client_page in the page instead of relying on commit_write to have done so.

I'm not entirely sure how the fput() in unmapping is synchronized with the writeback of data that was dirtied by the mapping. I'll need to look into that.

**10.6.3. LOV.** the LOV treats I/O relatively simply. It simply hands down pages to the appropriate OSC, and offers a completion callback.

#### 10.6.4. OSC.

The OSC holds the important state machined for I/O. The state machine principally:

- plugs and unplugs the device, to stop and initiate I/O
- collects sufficiently many pages to warrant firing an I/O RPC to the OST
  **parameter::** max_pages_per_rpc
- manages a certain number of RPC's that is active at the same time called a flight group
  **parameter::** max_rpcs_inflight

Performance can be excellent (single threads reaching 300MB/sec), but depends on good tuning.

#### 10.6.5. OST I/O.

The OST now uses a variant of direct I/O. Key characteristics of this approach are:

- I/O is synchronous to the disk
- The page cache is (by default) not used
- A fixed maximum number of threads is doing I/O
- The I/O sizes are normally larger than 64K.
- Pages written can be *fliipped* into the OST read cache

We cannot use direct I/O as is, because the direct I/O api requires contiguous pages, but the lustre protocol may have discontiguous pages.

## 10.7. I/O Layering and API's

The page cache in llite now uses an asynchronous OBD api that gives the osc an opportunity to make intelligent rpc batching and issueing decisions. This page contains a description of this API.

The API relies on associating pages with objects. It is based on the long-term binding of IO state to a page in its object. When the caller knows it will be interested in performing IO on that page, it registers interest in the page and gets back an opaque token. From then on it uses that token to perform the IO operations on the page that the API offers: queueing IO on the page, increasing the priority of a previously queued IO, and destroying the association of the token and the page in the object.

The main goals of this improved api are to shield the layers from each other, allow better stacking between layers and concurrent IO, and more accurately expressing the difference between long-term arguments to the page binding and arguments to each IO performed during the binding.

### 10.7.1. OBD api layer.

10.7.1.1. *o_prep_async_page.*

Prototype.

```
void * obd_prep_async_page(exp, lsm, off, page, &callbacks, void *data)
```

Summary: Returns a cookie for a single page after registering the page in the OSC.

In the LOV this is pass through to the OSC.

The OSC sets up an osc_async_page with obd_async_page_ops, to be used as callbacks.

Description. This starts participation in the API. The caller provides details on the object and page with exp, lsm, and off. the presence of struct page here is a bit of a wart, but it is so because of the use of struct page deep down in ptlrpc to specify the memory. The callbacks allow the lower layers to get state about this io registration later. 'data' is an opaque argument returned to those callback functions.

the void *cookie returned is then used by the caller from then on to perform io from the memory pointed to by 'page' starting at 'off' in the object.

Then under the obd switch, the layers will be using this call to allocate state that they'll need during the IO. This also gives them, in particular the lov, the chance to cache state based on the offset in the file that will remain constant. For the typical striping model this lets all further calls through the lov dispatch immediately to the stripe in question, rather than calculating it based on the offset each and every call. (the divisions in that calculation showed up in profiles back when earlier versions of this api were client cpu bound).

each layer as it prepares the io stores the cookie of its lower layer in its state and passes its cookie to the caller. In practice the cookie will be a pointer to a struct that is guarded by a magic number at its head.

10.7.1.2. *o_queue_async_io.*

Prototype:

```
int obd_queue_async_io(exp, lsm, cookie, cmd, off, count, brw_flags, async_flags)
```

Summary: The cookies is used to identify the page which is going to see I/O. It is not safe to concurrently call multiple instances of methods or multiple methods with the same cookie.

This describes an I/O task for this page. Sets up state in the OSC, namely puts it on a list of objects ready for I/O. Each object has a list of pages that have I/O queued. Record a page under an object as having pending I/O.

Description: This queues io on the page/object binding specified by the cookie. cmd is read or write and off/count/brw_flags are the same arguments as in the brw_page api.

- async_flags are used to communicate properties of this io. Currently there are three:
  - ready: the io can be put in an rpc without having to call a caller's callback to make it ready
  - urgent: the io should be put in an rpc as soon as possible
  - fixed, or something: the io can be started without a callback to update whether or not the io should be performed. need a better name here.

The call primarily trickles down to the osc which puts the io in the queue to be put in an rpc based on the rpc issueing policy. the raid1 lov may take this chance to multiply the io over its active stripes if it so desires.

when the io completes the caller will get called via one of the callbacks specified at bind time. Due to races this completion could well arrive before this io queueing returns. The completion callback will not be called if the queueing api returns an error, the lov will have to be careful here if it issues multiple ios.

It will return -EBUSY if given a cookie that has already been queued for io which hasn't completed.

This call can block if the underlying layers decide that there isn't room for a new io to be queued. they'll sleep waiting for other io to complete to make room for this one.

exp/lsm are specified again only because it makes me nervous to cache them in the cookie at binding time.

10.7.1.3. *o_set_async_flags.*

Prototype.

```
int obd_set_io_flags(exp, lsm, cookie, async_flags)
```

Description. This lets the caller change the flags that an io currently has. Its only current use is to set an io urgent and ready that wasn't previously. being ready translates to a locked page in the llite use, and urgent is used when the filesystem thinks that this io is being waited on.

this call should return errors when the io hasn't been queued or has been completed, and perhaps should return errors for changes that don't make sense. the only example I can currently think of is that a page which is ready can not be made uh.. unready.. if its in an rpc.

The async flags can be one of:

**ASYNC_READY:** When I/O makes it towards an RPC, don't call make_ready operation on the osc_asyc_page. For Lustre lite, make_ready locks the page. Lets commit write queue write pages without being locked, typically when the page is already locked.

**ASYNC_URGENT:** Tells the RPC engine no more obd methods concerning this page will be made. Better put in an RPC and complete as soon as possible. In the file system writepage sets this flag and goes to sleep

**ASYNC_COUNT_STABLE:** When count is passed in in queue sync I/O. If this is set, refresh_count is not called. Used to avoid truncate racing with the file I/O.

10.7.1.4. *o_tear_down_async_page.*

Prototype.

```
int obd_teardown_async_pge(exp, lsm, void *cookie)
```

Description. Removes a page from the OSC if not active to stop I/O. It will fail with -EBUSY for an io that is currently in flight.

10.7.1.5. *o_queue_sync_io.*

Prototype.

```
int obd_queue_sync_io(exp, lsm, &container, void *cookie, off, count, brw, flags, etc)
```

Description. The key parameter is a list of pages. obd_sync_io containers make up this list.

The osc has a list of all sync io's per object. This list is headed off the loi.

There are callers (direct io, lock cancelation, liblustre?) who aren't really async but want good rpc concurrency. I'm tempted to have them use this api as well so that the rpc mechanics are the same underneath.

10.7.1.6. *o_trigger_sync_io.*

Prototype.

Description. Builds rpc's from the pages on the obdo's sync list.

10.7.1.7. *o_sync_io_wait.*

Prototype.

```
int obd_sync_io_wait(exp, lsm, &container)
```

Description. Waits for them to complete and returns an aggregate error code as decided by the under layers. If the layers did their 'preps' properly they'll be able tie together their per-page state, the callbacks, and the container struct that will probably contain things like lists, waitqueues, and counts remaining. the usual.

**10.7.2. Example use of the obd I/O methods.** Consider the Linux file system. The first file sytem I/O method is prepare_write, prepare write gets pages ready but does not interact with the OBD, except reading them in with obd_brw. The commit_write is called. Commit write calls prep_async_page, sets the page private to the a structure that includes the cookie. The commit_write calls queue_asyn_io. This is how a list of a dirty pages is built in the page cache and the OSC at the same time.

Suppose there is dirty list of pages. A daemon will appear and call writepage on each of them. Inside writepage set_async_flags. It sets the ASYNC_READY/URGENT flags on the async_page. This causes the OSC to put the page on the urgent list.

Lustre commit write will call the queue_sync_io and trigger_sync_io methods if it cannot queue the page in the cache. This can happen when the grants are at 0 or when the max_dirty_mb is 0 (in the osc /proc directory).

Now consider direct I/O. Direct I/O uses

#### 10.7.3. OSC state machine.

The OSC has lists of objects, objects can form RPC's. The loi are elements in the list of jects. Each loi contains list heads for lists of pages.

Each of the methods, prep, queue, setflags causes the page lists to be looked at. If the lists are ready for an RPC to be sent this formed and sent off. As part of preparing the RPC, the following callbacks are called:

**apcb_make_ready,:** Called if the READY flag is not set. This typically locks the page if it wasn't locked.
**apcb_refresh_count:** If the ASYNC_STABLE count is not set. If this returns 0, the I/O can complete immediately. This calls is per page.
**apcb_fill_obdo:** fills in the obdo associated with loi. This calls is per RPC.

#### 10.7.4. OST networking code.

#### 10.7.5. OBD Filter backend.

### 10.8. Read-ahead

In an effort to make parallel progress with the read business, I'd like some design-level input.

Currently reads in llite are not optimal. We've disabled the read-ahead in generic_file_read() and are trying to read-ahead in readpage(). But its not very clever – It doesn't issue pages well in the async API so batching isn't good. It also doesn't handle the page walking very well and stops reading ahead when it encounters pages that are under IO. That's very wrong – it should always try to maintain a window of reads in flight beyond the point where the app is currently blocked waiting for data. In realizing this, I hoped that we could punt to the kernel's read-ahead mechanics. It tracks read-ahead windows and such. A very nice patch later and we have less code and re-use the kernel's read-ahead code. But its not so hot. It only tries to maintain one read-ahead 'window' in flight at a time. It has the same sort of logic that resets the read-ahead window when a read blocks in a page that was previously issued by read-ahead. so we're back to doing it ourselves. I understand the desire to drive read-ahead decisions from the OSC so that llite doesn't have to base its guess on read-ahead on stripe sizing, or things of that sort. But the APIs aren't currently built for the api providers to initiate transfers. I worry about the OSC's knowledge of lock regions that are safe and its ability to have a place to store the read-ahead results while the app is off making concurrent progress. To that end:

(1) - are we OK with callbacks from the OSC up to ask if it should be initiating a read-ahead request for a page? This would get a page in the page cache and lock it in llite, possibly returning some error if the page was already uptodate or was beyond eof. This also seems the right place for llite to do the dlm match here as well to verify that the page is safe to be read from.
(2) depending on the striping an osc may need to do read-ahead without ever seeing the first read that triggers it. The lov would have to issue read-ahead in some way to hit these stripes.

(3) we'd have a brw_flag that specifies that a page should have read-ahead follow it? This would let prepare_write and o_direct stop the callbacks from happening.
(4) I don't know where liblustre will store the read-ahead results after the read() that generated the traffic has completed. My understanding of liblustre says that we operate as a blocking syscall context in the app. If it doesn't have a page-cache analog we'll need some place to store read results between syscalls so that later reads can be satisfied by the read-ahead data.
(5) I'm not sure how we'd maintain the read-ahead window. We can call into the OSC as reads are satisfied from the cache in llite, or we can call up to llite from the OSC as read-ahead requests are completed. Taking all this into account, it seems the path requiring the least code change is to track a read-ahead window in ll_inode_info and call into it near ll_file_read(). It requires the imperfection of having the read-ahead amount in llite, but it avoids all the code churn under the OBD API. I need to learn more about liblustre to see if it even has a place to store read-ahead. Whether it does or not might not even effect if its better to have it initiate the read-ahead instead of the OSC.

## 10.9. File I/O semantics

This page discusses how the clients manage acquisition and correct sharing of file size and file data.

**10.9.1. Definitions.** The following resources are associated with a file. The KMS is a Lustre specific resource which aids in the reduction of lock acquisitions and cancellations in the cluster.

**sz:** will denote the size of the file in the cluster

- For an inactive file, i.e. a file seeing no write or truncate activity, this is derived from the size of the inode(s) on the server nodes
- For an active file this is derived from max(KMS_i) and sizes on the server, where i ranges over all the nodes having cached stripes of the file.

**off:** will denote the offset associated with an open file
**[a,b]:** will denote the byte range in a file starting at a and including b as the last byte
**KMS:** denotes a lower bound on the file size known to a client, an integer maintained by each client for each file, satisfying sz = max(KMS's, server size).

**10.9.2. The known minimum size (KMS) attribute of an inode.** Lustre caches multiple extents, and protects extents with locks. File size can grow in multiple locations. To avoid too much concurrency on the file size lock, another approximate attribute of a file is maintained called the *known minimum size* ***(KMS).***

The KMS is a quantity associated with a Lustre cilent inode on a system, when the inode has at least one extent lock. Each client maintains this KMS integer, signifying the end of the data for an inode that the client is aware of. The KMS is easiest to maintain as an inode attribute on the client. Some clients can have a KMS that is higher than others. So when the file size is needed the KMS is not a sufficient piece of information to use, but often a lower bound on the file size is needed and this can be supplied by the KMS.

When the first lock is acquired for an OST object, the KMS is initialized. It is stored in the server side lock resource by retrieving the size of the object. It is then communicated to the lock holder upon acquisition of the lock. While the lock is valid, the quantity is known to be smaller than or equal to the file size. The precise changes made to the KMS are documented below.

A client changes the KMS in the following cases:

(1) It does a write to the file which appears to extend the file. Appears means that other clients may already have deposited data at a higher offset in the file, but the client is writing at an offset higher than all data it has cached.
(2) When a client truncates the file, it sets the KMS to the new size.
(3) When a client loses the lock with the highest extent it has cached, and the KMS lies in this extent, it moves the KMS down to the next highest extent.

When a lock revocation happens, the KMS needs to be updated on the client giving up the lock. It can only move downward, and the new value is the largest depends on the current value, the type of extent locks held. This may involve a query on the highest lock cached in a clients's DLM. Perhaps an interval tree is needed to find this lock. Read lock extents may be overlapping.

10.9.2.1. *Size probes.* The clients sometimes need to know the file size. They do this by doing a size glimpse/probe. The probe requests a lock to protect the size from the server, with a flag that the purpose of having the lock is merely to read the size of the inode. If the resource has no other (XXX recently used) locks, the server will grant the lock, but if it has locks, the server will make a special callback, called a **probe callback** to the client.

The probe callback does not cause the lock to be relinquished but instead the client returns the KMS to the OST. The OST moves the server resource resident copy of the KMS forward and sends the resulting size to the client. The probe callback is executed as an **intent** associated with the lock. Because the client side is not granted a lock, the size that has been retrieved is stored into the inode attributes by a callback handler.

These lock acquisitions are quite special, because the probing nature means that the reference on the lock can be released after the size has been taken from it. This eliminates the possibility of deadlock when the locks on multiple stripes are obtained out of order. As a result, the locks can be acquired in parallel.

**10.9.3. Resources requiring concurrency control.** We will use the following notation for resources protection:

**SZ-{R,W}:** size is protected for read / write while a lock is held
**SZ-*:** one read of the globally determined size is required
**[a,b]R:** data is protected for read
**[a,b]W:** data is protected for write

We will use corresponding notation for DLM resources, using { }, instead of [ ], to distinguish the extent lock implementation resources from the file metadata and data resources that are to be protected.

#### 10.9.4. Locks.

10.9.4.1. *Resources.* Lustre uses locks to control distributed consistency of file data. The following data structures need concurrency control:

(1) extents [a,b] of file data
(2) the file size attribute
(3) the KMS attribute

We make a distinction between the following types of locks:

**{a,b}R:** protect something for reading, implemented as a PR lock on the [a,b] extent. While this lock is held:
- No node shall have W lock
- The data in [a,b] will not change

**{a,b}W:** protect something for updating, implemented as a PW lock on the [a,b] extent While this lock is held:
- No data will be written in the [a,b] extent, except by the lock holder.
- The size can *jump over* [a,b], i.e. another node can write data behind [a,b].

10.9.4.2. *Semantics of locks on resources.*

**{p,-1}R:**

**semantics::**
- getattr returns correct sz
- data in [p, -1] read protected

**constraints::**

**acquisition::**
- KMS is set to sz, KMS and size cannot change

**client cancellation actions::**
- destroy cached pages

- adjust KMS
- mark size as not protected

**{p,q}R:**

**semantics::**
- signifies that if an extent [a,b] contained in [p,q] is read current data will be returned.
- a read lock on [p,q]
- guarantees: sz >= q
- constraints
- no {p,q}R locks will be granted that cross the size boundary, unless q == -1

**acquisition:**
- KMS is set upon acquisition.

**client cancellation action::**
- destroy cached pages
- client adjusts KMS value

**{p,-1}W:**

**semantics::**
- size is known
- data in [p, -1] is write protected
- setattr can be used to set the size to value in [p, -1]

**cancellation:** action:
- flush, then destroy cached pages
- send sz to server, perform update
- mark sz as not protected
- adjust KMS value

**{p,q}W:**

**semantics:**
- client can write new data in [p,q]
- data write protected in [p,q]
- KMS may be set in [p,q]
- under this lock the cluster inode size may change elsewhere

**acquisition::**
- initial KMS value for client is returned with lock

**client cancellation::**
- flush, then destroy cached pages
- send KMS to server, adjust size if necessary
- adjust KMS value

**Note::** without a lower bound on file size, this lock cannot be used for reading. To make it usable for reading, the initial KMS value is sent with the lock to the client. The client now knows how to handle reads, because the KMS can be moved up by the client and not down (only truncate does that and the {p,q}W lock would be revoked).

10.9.4.3. *Stripes and locks.* Conceptually it is clearest to think of a lock as something affecting an extent in the file as seen by the client side file system. However, the reality is that the file consists of several objects, and locks correspondingly are associated with the objects. This slightly complicates the description given above. The key points are:

**size known:** This happens when all objects associated with an inode have an extent lock including -1 in the extent.

**adjusting KMS:** Adjusting the KMS requires the system to scan all the locks associated with all objects and find the next lowest KMS. It is XXX possibly better to store a KMS with each object, because unallocated stripes could confuse the algorithm.

10.9.4.4. *The ldlm size value block.* The file size can be managed very efficiently the lock server. The lock server can easily keep **last known good** knowledge of file size in a lock value block associated with the resource. We call this variable **ldlm_size.** The rules for managing this are as follows:

- This is associated with the resource on the lock server
- Server updates happen upon cancellation and with the return of a glimpse callback and can be forced when writes happen.
- Clients receive the ldlm_size with the completion of an enqueue.
- Client propagate new ldlm_size to the server with cancallations and replies to glimpse callbacks.
- The ldlm_size is usually only increased, and is always bigger than or equal to the size of the object it is protecting.
- When a client truncates and cancels the truncation lock the size can be moved backwards

10.9.4.5. *Operations on ldlm_sz.*

**enqueue:**
(1) Initializes the ldlm_size. This is done with the lock value block initialization method (lvbo_init) and the implementation set the ldlm_size to the size of the file data object when the first lock is given out
(2) Returns the ldlm_size to the client enqueueing the lock

**completion of an enqueue:**
(1) of {p, -1} lock sends ldlm_sz to the client completing that lock

**cancel:**
(1) Canceling {p, -1}W sets ldlm_sz
(2) cancel of {p, q}W sends clients KMS to the server, ldlm_sz = max(ldlm_sz, KMS) .

**callbacks:**
(1) Client can enqueue with a GLIMPSE flag or without. When the flag is set, the lock server will not revoke other locks on the resource but perform a glimpse callback instead. The reply to the glimpse callback updates the ldlm_size as in a cancellation.
(2) The lock server will reply to the clients when all locks have been canceled (to complete the enqueue) or when the glimpse callbacks have all been made. The ldlm_size is returned to the client with that completion.

Notice that with multiple cancellations or glimpse callbacks of locks excluding -1 in the extent, the order in which the cancellations or glimpse callback replies are received does not affect the final value of ldlm_sz.

10.9.4.6. *Further Lock Optimizations.*

- Conversion from read to write lock is not handled yet
- Manipulating the extents in existing locks to avoid cancellations is not handled yet
- If multiple locks on one object protect a given region used by a system call, we do not handle this yet

- Write locks being called back for other write locks need only flush the data outside the extent requested by the new write lock, as the new writer will over-write the data.
- For one client read and write locks need not conflict.
- Using write locks for the purpose of read carries a problem that the KMS or size must be known or set for the read call to be useful.

**10.9.5. System calls and locks.** This section documents the system call return values and the distributed semantic behavior Lustre has chosen to implement. It is followed by a discussion how the DLM locks and lock value block negotiation can assist in doing this efficiently.

10.9.5.1. *System Call Semantics.*

**read(fd, buf, len):**

**returns::**

- min(sz, off + len) - min(sz, off)
- ie. reading behind the file returns 0 (this is EOF)
- ie. normal return code is len

**requires::**

- knowledge of SZ if off < sz and sz < off + len
- [off, off+len]R if sz > off + len
- [off, sz]R if sz <= off + len

**sets::**

- nothing

**write(fd, buf, len) O_APPEND:**

**returns::**

- len or ENOSPC if full amount cannot be written

**requires::**

- sz-W
- [sz, sz + len]W

**sets::**

- sz += len

**write(fd, buf, len):**

**returns::**

- len or ENOSPC

**requires::**

- [off, off+len]W

**sets:**

- KMS = max(off+len, KMS)

**stat("path",:** buf) fstat(fd, buf)”’

**requires:**

- one correct read of sz

**seek(fd, len, SEEK_END):**

**sets::**

- off = sz + len

**requires::**

- ***:** one correct read of sz
- ***:** ”’seek(fd, len, SEEK_CURR)”’

**sets::**

- ***:** off += len

**truncate("path", len) ftruncate(fd, len):**

**sets::**

- sz = len

**requires::**

- ***:** sz-W * [min(len,sz), max(len, sz)]

**mmap(start_addr, len, prot, flags, fd, offset):**

**semantics::**

- file size to be established at call time,
- there are no "extending mmap writes"
- if the size of the file changes, we do not have to notify the mmapped area

**requires:**

- read of sz if offset + len > sz
- [x,y]R when pages in extent are faulted in
- [x,y]W for pages in extent are dirty

**Questions about mmap**

- Is there are relationship between the permission with which the file was opened and the permissions requested in prot? If not, the MDS will have to get or deny write access in the PROT_WRITE and PROT_EXEC case. If yes, then this is inherited from the file open and nothing needs to be done for this.

10.9.5.2. *File System Method Semantics.*

**10.9.6. Locks associated with file system operations.** In this section we will prove that our lock protocol implementes sensible semantics.

10.9.6.1. *file read.* Read must have a lower bound on the file size or the know the file size to determine if EOF comes into play during the read. In older versions of Lustre read([a,b]) takes the {a, -1}R lock.

The return value for read involves the file size as can be seen above, so the following scheme for read is followed:

- For locking return: {a, b}R if there is concurrency on the lock, otherwise return {0, -1}R
- If a matching write lock {p,q}W is available that by itself does not carry a guarantee of the file size to execute the read.
- In that case probe for file size and use that as a lower bound during the read.

In the case where the file is striped, a successful match on the lock means that a collection of read and write locks is covering the interval to be read. If the KMS is calculated from these intervals an integer is found that lies above the highest read lock of the file. If this KMS is not above the interval we are reading a size probe must be made.

In pseudo code:

```
lustre_read(fd, buf, a, b)
{
        off_t top;
        lock = match_lock([a,b]);
        if (kms(fd) >= b)
               top = b
        else if (size_known(lock))
               top = size(fd);
        else
               top = glimpse_size(lock->resource);
        // now read
        do_read(fd, buf, a, top);
}
```

**Note:** A key issue to note is that even the KMS associated with a read or write lock [a,b] lies below b it is still necessary to validate the size. A write beyond b may change the file size. This affects reads, in that they wil not return short reads, but return 0's for unallocated space. However, the data cached in the [a,b] region is protected by the lock, so it does not need to be refreshed.

The reading performed by prepare write is not subject to this behavior. While it can only read data up to the end of the file, the return code of this read is not affected by the file size.

10.9.6.2. *write.* Getting the [a,b]W lock seems good. With this lock an initial value of the KMS must be given to the client, so that the lock can be used for reading.

10.9.6.3. *stat.* Probably it is best to get a lock only if no other locks are outstanding. All stat needs is a single read of the size file. Not recently used locks should probably be canceled.

10.9.6.4. *prepare_write.* Prepare write is always called with a lock on the range affected by the write.

(1) KMS is below that range - there is no data to read in prepare_write
(2) KMS is within that range – there is some data to read, but maybe not a whole page
(3) KMS is above that range – there is a full page of data to read

Prepare_write needs to update the KMS (or a local variable) but NOT update i_size. Prepare write updates this based on how much is written into the page. This updating can race with someone writing above our range, but it doesn't matter. we zero-fill our page in any case.

10.9.6.5. *Truncate.* In Lustre truncate() will be implemented as

(1) open
(2) truncate on OSTs
(3) close

This open will be an "MDS open by FID" because we have no parent or name. All network ops will be moved out of the inode truncation operation, ll_truncate, which cannot return an error, and into ll_setattr_raw. Also copy the small checks from vmtruncate(), so we can be sure that vmtruncate() will succeed if we get that far

### 10.9.7. Pseudo code.

10.9.7.1. *Psuedocode for ll_extent_lock.*

```
[ acquire a new lock, as before ]
down(&lli->lli_kms_sem);
/* If the KMS is already past the end of our lock, then do nothing.
* We could assert here that the KMS is covered by another lock, by
* doing a lock match! */
if (lli->lli_kms >= new_lock->extent.end)
    goto out;
/* If:
* the KMS < new_lock->extent.end, and
* the region [KMS, extent->end] is covered by an existing lock, then
* we know the existing KMS is valid for the purposes of prepare_write
* because nobody else could have changed the data covered by that lock. */
extent.start = lli->lli_kms;
extent.end = new_lock->extent.end - 1;
tmplock = lov_lock_match_a_different_lock(newlock, &extent);
if (tmplock != NULL) {
    unpin(lock);
    goto out;
}
update_kms(inode);
out:
up(&lli->lli_kms_sem);
```

10.9.7.2. *Pseudocode for update_kms.*

```
obdo = lov_getattr(lsm);
kms = obdo.o_size;
/* We must pin the lock, to prevent a race between setting the KMS and lock cancellati
lock = lov_pin_highest_lock(lsm);
/* This code assumes that it will only be called in the context of a
 * pinned lock, so we should match _something_. */
```

```
LASSERT(lock != NULL);
if (kms > lock->extent.end)
    kms = lock->extent.end;
down(&lli->lli_kms_sem);
lli->lli_kms = kms;
up(&lli->lli_kms_sem);
unpin(lock);
```

10.9.7.3. *Pseudocode for lock cancellation:* In the final cancel callback, after the lock has been marked as going away, all of the references are gone, and nobody can get a new one:

```
/* This function will not return the lock which is being cancelled, because it can't g
lock = ldlm_pin_highest_lock(resource);
down(&lli->lli_kms_sem);
if (lock == NULL) {
    lli->lli_kms = 0;
} else {
    lli->lli_kms = lock->extent.end;
    unpin(lock);
}
up(&lli->lli_kms_sem);
```

10.9.7.4. *Pseudocode for ll_file_write.*

```
[ setting the "extent" variable ... ]
lock = ll_extent_lock(inode, LCK_PW, &extent);
/* If we got a lock on the whole file, then we know that i_size ==
 * KMS, so we might as well record that fact. */
if (extent.start == 0 && extent.end == OBD_OBJECT_EOF)
    update_isize_from_kms(inode);
[ O_APPEND, maxbytes, generic_file_write, and the rest ... ]
```

10.9.7.5. *Pseudocode for ll_prepare_write.*

```
[ everything up to and including the "completely overwriting an existing page" code ..
down(&lli->lli_kms_sem);
if (lli->lli_kms <= offset) {
    memset(kmap(page), 0, PAGE_SIZE);
    kunmap(page);
   up(&lli->lli_kms_sem);
   GOTO(prepare_done, rc = 0);
}
up(&lli->lli_kms_sem);
```

```
ll_brw(OBD_BRW_READ, inode, page);
```

**10.9.8. MDS size management.** This section contains the details needed to let the MDS cache sizes. This has not been implemented and may turn out to be an unnecessary performance optimization.

By allowing the MDS to cache the file size and mtime we can significantly improve the performance of listing directories, because this avoid the necessity to make RPC's to the OST's to find the file sizes of the objects. However, the OST's are the nodes where the file size is changing. To do this we need to manage the consistency of the cached size on the MDS and the original values as they are encoded in the objects, in the face of crashes of the system. The mds should have ownership of the file size attribute while files are not open. This section contains more detail about this protocol.

10.9.8.1. *Client protocol during normal operation.* Under normal circumstances, when files are not open on the MDS - and therefore no i/o is active, see below - the MDS has the authoritative file size. Clients get this through mdc_getattr calls, and the *valid* attribute in the return indicates to the client that it has recieved a valid size from the MDS. Normally such getattr calls cause the client to have a lock on the inode attributes allowing them to re-use the attributes until the lock is canceled.

When a file is opened for write, the MDS will lose control over the file size and it will cancel such locks. At this point the inode revalidation routines will not acquire the file size from the MDS but instead they have to go through the LOV to the OST's to collect the sizes of the data objects. Just like for the MDS there are OST managed locks that control the validity of this attribute when it is cached. So fundamentally the getattr client algorithm which exploits MDS caching is:

```
int lustre_getattr(inode, iattr)
{
    rc = mdc_getattr(req, inode, iattr);
    if (rc || iattr->ia_valid & OBD_MD_FLSZ)
         return rc;
    rc = obd_getattr(req, inode, oa);
    return rc;
}
```

Upon mdc_close a client that knows the valid file size and has the recovery information associated with the I/O, it will communicate this to the MDS. If the MDS sees that this is the last closer, it can use that size to update its attributes and start giving out attribute locks again on the size. The client file close algorithm hears from the MDS if it needs to communicate valid file size and cookies to the MDS. If it has to, it registers this with a daemon that waits for completion of I/O to perform an getattr operation.

It is possible that no client has the authoritative size when it sends the mdc_close rpc to the MDS. The MDS may need to get the size from the client, if this is the last mdc_close it is expecting on that file. In that case the MDS will keep the file open and return an error to the client. The client will see this error and fetch the file size from the OST's and resend the close.

The following schematic code shows the client side close handling by the file system and close daemon:

```
void inode_begin_close(inode)
{
    page_cache_flush_register(perform_close_commit);
}
int ll_close(inode)
{
   // perhaps we can do it all now?
   if (!dirty(inode) && have_sz(inode) && have_recov_cookies(inode))
        pack_sz_cookie(req, inode);
    rc = mdc_close(req, inode);
    if (rc != GET_CLOSE_INFO) {
        // another client is responsible
        page_cache_flush_unregister(perform_close_commit);
    } else {
        // if size and cookies are here, make the rpc
        // rc can tell the client if it is the only user of the inode in which case
        // it likely has the size and certainly all the cookies
        page_cache_flush_activate(perform_close_commit, rc);
    }
}
void perform_close_commit(inode)
{
    // flushes caches in the cluster
    obd_get_sz_lock(inode);
    // get size and cookies from ost's
    rc = obd_getattr(inode, oa, OBD_MD_FLSZ|OBD_MD_FLCOOKIE);
    obd_drop_sz_lock(inode);
    // must register rc: it is treated as client failure
    rc = mdc_commit_close(req, rc, inode);
}
```

The daemon performing the closes just looks in the inode to see if the information required is already available, otherwise fetches it and communicates it to the MDS:

```
void close_daemon_main(struct super_block *sb)
{
     while (1) {
          sleep_on(daemon->work_waitq, inode_ready_for_close_commit());
          if (have_close_info(inode)) {
                page_cache_flush_unregister(inode);
                rc = mdc_commit_close(req, io_status(inode), inode);
```

```
            } else {
                  rc = perform_close_commit(inode);
            }
            if (rc)
                  recover_md(inode);
            continue;
      }
}
```

As part of the close, the MDS will perform a setattr command on the inode, to update the size stored on the MDS. We discuss this further down. We will wait for the reply from the MDS_CLOSE before sending a DONE_WRITING, (if necessary). This

(1) avoids extra RPCs, in the case where the close comes back and tells us that we are not the last closer
(2) more importantly, it avoids the race where the D_W arrives before the CLOSE

10.9.8.2. *OST Size Log records.* When the OST changes the file size this happens as part of a truncate or write call. Each of these calls passes in the file I/O epoch number. The epoch number is returned by the MDS to the client during file open and communicated to the I/O subsystem on the OST.

There are several cases:

- The object is already marked as having a changed size. This means that:
  - the object is pinned
  - the file I/O epoch number is stored, in memory, in an attribute of the object. This attribute is managed through the inode filterfs data field, a void * which points to Lustre generated filter data. The attribute is allocated if the pointer is still NULL and the epoch number is set to the value contained in the I/O or truncation request that is being serviced.
  - a size change log record was written with that epoch number, transactionally with the commit of the first size change of the object in the epoch
- If the object has no record yet of a size change, the OST will transactionally write a log record with the object size update. The log record contains the epoch and the object is now pinned in memory. The epoch in the object's attributes indicate which epoch is responsible for the pinning of the object.
- If the client sends an epoch different from the one that is seen in the object, it will determine if the size change is from an older or newer epoch. If it is from an older epoch or from the epoch currently pinning the object, it does nothing. If the client call contains a newer epoch, it updates the epoch number in the object to the new one, and schedules an atomic record addition with the next setattr commit on the object.

The OST algorithm for managing the epoch's pinned inodes are as follows. When I/O is started on an inode, together with the first file size and mtime changing transaction the OST writes a file change record. The following is called from within a transaction:

```
struct obdfilter_sz_rec {
    struct list_head   ofsr_hash_chain;
    struct list_head   ofsr_obdo_chain;
    __u64              ofsr_epoch;
    struct llog_cookie ofsr_cookie;
    struct llog_sz_    ofsr_rec;
    struct obdo       *ofsr_obdo
};
```

These are added to the OBDO's filterfsdata:

```
struct obdfilter_fsdata {
    struct list_head *ofsd_obdo_head;
    __u64             ofsd_epoch;
}
void obdfilter_write_szrec(obdo, req, rec)
{
    rec.sr_fid = req->data->mds_fid;
    rec.sr_epoch = req->data->mds_epoch;
    llog_add_rec(obdo->o_obd->obd_llog_ctxt, SZ_INDEX, &rec->ofsd_rec, &rec->ofsd_cook
}
```

The epoch lock controls concurrency on allocating filterdata and setting the epoch. We can use the obdfilter global lock:

```
int obdo_set_epoch(obdo, req)
{
    struct obdfilter_sz_rec *rec;
    struct obdo_cookie cookie;
    if (object_epoch(obdo) >= req->data->epoch)
        return;
    // start a new epoch
    if (!obdo_pinned(obdo)) {
        epoch_lock();
         // allocate obdfilter_fsdata, pin object
         obdo_pin(obdo);
         rec = obdo_alloc_filterdata(obdo);
         epoch_unlock();
    }
    obdo_set_epoch(obdo, req->data->epoch);
    obdfilter_write_szrec(obd, req, &rec);
    // places rec in a hash with all rec's and in d->ofsd_obdo_head
    hash_sz_rec(obdo, &rec);
}
```

Cancellations for these log records are only sent when the MDS has committed the file size change and terminated the epoch. When they arrive on the OST the llog cancel handler finds the record for the epoch base on the cookie. In all cases the record pointed to by the cookie will be canceled. The call will also un-pin the object if the epoch that is being canceled equals the epoch contained in the cookie. If the object is pinned for a newer epoch it is left in that state.

```
void obdfilter_cancel_szrec(cookie)
{
      struct obdfilter_sz_rec *rec;
      // use a hash to find the record from the cookie
      rec = find_rec(cookie);
      // the OST rebooted and old obdos are not pinned anymore
      if (rec) {
          epoch_lock();
          if (obdo->filterdata->epoch == rec->epoch) {
              obdo_free_cookies(obdo, rec);
              obdo_free_filterdata(obdo);
              obdo_unpin(obdo);
              unhash_szrec(rec);
          }
          epoch_unlock();
      }
      llog_cancel(cookie);
      return;
}
```

In many cases the OST will not be explicitly contacted by the client to fetch size and cookies, because the client already has the size and the cookies, for example if it is the only writer to the file. In the case where the client does make an explicit call to obtain cookies and size at the end of an epoch, it still should not cancel the epoch on the OST in its getattr RPC because the MDS may extend the epoch, unbeknownst to the client, in case a further open happens before the close commit completes. So the cookies are freed only when the records are finally canceled.

The I/O, truncation and obd_getattr calls which a client, or recovering MDS makes to the OST will retrieve not just size, atime, mtime from the obdo but also return the cookie to the client. For this the obdfilter walks the list:

```
void obdo_get_cookie(req, obdo)
{
      struct obdfilter_sz_rec *rec;
      rec = find_rec(obdo, req->data->io_epoch);
      LASSERT(rec);
      memcpy(cookie_repbuf(req), rec->ofsr_cookie);
}
```

10.9.8.3. *MDS getattr handling.* The MDS will only give out cached size of an inode if two conditions hold:

(1) A flag in the MDS device is set to indicate that recovery has completed the recovery of file sizes from OST's. This flag is cleared when recovery has retrieved all logs with pending size updates from the OST's.
(2) No I/O epoch is in progress for the inode.

The following pseudo code snippet shows this behavior in the mds handler for getattr requests. Note that such requests are often implicitly executed to add information to the replies of other requests such as opens.

```
int mds_getattr(req)
{
 ...
    if (mds->mds_sz_cache_active &&
        inode->filterfsdata->ioepoch == 0) {
         req->repdata->size= inode->i_size;
         req->repdata->valid |= OBD_MD_FLSZ;
 ....
}
```

10.9.8.4. *Definition of epochs, epoch numbering and open recovery.* An epoch on an MDS inode indicates that somewhere in the cluster file size changes can be initiated through interaction with OST's. An inode is not in an I/O epoch when no file size changes can possibly be made.

To assist the MDS in separating the epochs the client maintains the following invariant in their interaction with the objects:

> client has opportunity to change size of OST objects with writes truncates implies the inode on the MDS is open for write

This implies that Lustre will handle truncate in the kernel through ftruncate.

To execute the protocol it is beneficial if the epoch are numbered and the numbers increase. Epochs are an increasing sequences of integers, global to the MDS. If the MDS starts cleanly and sees no epoch continuations during the replay of open requests, the io-epochs will start at a random number selected at mds mount time, and increase with a small prime number. If during replay files are reopened for write, epoch are continued from the previous boot of the MDS.

During startup, when recovery has finished on the MDS and the processing of normal open calls resumes, the MDS executes the following algorithm:

```
void mds_initiate_io_epoch(mds)
{
    if (mds->mds_ioepoch == 0)
        get_random_bytes(&mds->mds_ioepoch, 8);
    return;
```

```
}
```

If during open replay a request for open contains an I/O epoch for an inode, the function *open_io_epoch* (see below) is called as follows:

```
open_io_epoch(mds, fid, req->req_replaydata->io_epoch)
```

This will pin the inode as usual but use the already allocated io_epoch.

When requests are re-sent, and replies were lost but are available from the MDS the epoch is part of the stored open reply data, and is communicated to the client during reply reconstruction.

10.9.8.5. *MDS I/O Epoch management.* The MDS initiates I/O epochs on a file when a file is opened for write, i.e. when O_RDWR or O_WRONLY is passed to open, and which the MDS mds_open command sees when FMODE_WRITE is set.

So whenever the are one or more openers for write for a file an epoch continues. The MDS epoch only ends when the file is not open for write anymore. Optionally we could extend the epoch on the MDS and end it when the setattr call associated with the size change has committed. Longer epochs expose us to longer periods without MDS authoritative file size information, shorter epochs cause size update records to be written more frequently on the OST's.

To manage the epoch's at during the processing of *mds_open* the MDS uses the following algorithm:

```
// if epoch is != 0, it is used, normally during replay, to set the epoch
void open_io_epoch(mds, fid, epoch)
{
     int res;
     inode = mds_fid2inode(fid);
     epoch_lock();
     if (inode->fsfiltdata && res = inode->fsfiltdata->ioepoch)
         goto out;
     pin_inode(inode);
     if (!inode->fsfiltdata)
         mds_alloc_fsfiltdata(inode);
     if (epoch)
         mds->mds_ioepoch = epoch;
     else
         mds->mds_ioepoch++;
     inode->fsfiltdata->ioepoch = mds->mds_ioepoch;
  out:
     epoch_unlock();
     return;
}
```

The epoch lock can probably be the same lock protecting the handling of open files on the MDS.

The client file system allows applications to close files at any time. When the last happens the I/O may not have been flushed, and the file may be re-opened before this I/O completes. The client goes through a two phase close protocol with the MDS. The file is closed normally, but then a protocol is engaged which transfers the authority of size management back to the MDS.

```
void epoch_end(inode)
{
     epoch_lock();
     inode->fsfilterdata->epoch = 0;
     free(ionde->fsfilterdata);
     epoch_unlock();
}
void mds_close_epoch_control(req, file)
{
     epoch_lock();
     if (openers(file->f_dentry->d_inode) > 1)
         goto out;
     if (mds_req_has_cookies_size(req)) {
         journal_start();
         mds_setattr(file, size, cookies, osc_size_commit_callback);
         journal_end();
         epoch_end(file->f_dentry->d_inode);
  out:
     epoch_unlock();
     return;
}
void osc_size_commit_callback(cookie)
{
     llog_replicator_cancel(cookie);
}
```

The Lustre lite inode info tracks a list of open files for this purpose and implements this as follows:

Note that this implies that when an MDS epoch on the file ends, the file has no openers and no pending I/O.

The MDS starts its epochs at some random non-zero integer and increases it each time a file increases. When no epoch is active on an MDS inode the epoch number is set to 0.

10.9.8.6. *MDS commit, failures and log recovery.* The only inode size changes occurring on the MDS are caused by mds_close_commit remote procedure calls. These initiate a settatr transaction with cancellation cookies for corresponding size log records on the OST's. Every file size change on the OST's has a log record. The log record must continue to exist until the MDS has committed its change. As we will see below this has important implications for recovery.

The MDS can only send out requests to cancel the size log records on the OST's when this commit has completed, and this is done as part of the MDS commit callback function. A key invariant to maintain is:

> The MDS will never execute a setattr transaction that changes the file size if no log record for that inode's file size change exists on the OST's.

A subtlety here is that a client may resend close and close_commit requests which cause the setattr transaction, which rolled back during a crash, to execute again. However, this is only done when the first transaction did not commit, so no cancellation record was sent to the OST for this record. Nevertheless, the OST already has a size change record for this file change, so our invariant is valid.

When the MDS recovers it will establish a connection using its LOV and/or OSC's to the OST's. For each OSC on the MDS this connection has a specific generation which together with a boot-count is used to determine a generation for log records on the OST.

The MDS OSC's will fetch the logs associated with the size changes from the OSTs and process all the records in the order they were entered in the log. It must only cancel records that will not be used by subsequent file change transactions on the MDS as processing the records includes their cancellation on the OST's. So the MDS's OSC's must process size logs ***after replay.***

```
osc_process_log_cb(loghdr, record, data)
{
     obd = (struct obd_device *data);
     if (loghdr->conn_generation == current_generation)
        return -EDONE;
     fid = record->fid;
     inode = fid2dentry(fid)->d_inode;
     get_stripe_ea(inode, &lsm);
     obd_getattr(obd, oa, lsm);
     mds_setattr(inode, oa->o_size, &cookie, osc_size_commit_callback);
     return 0;
}
```

Network failures cause the MDS to abort recovery, and run in degrated mode. The setattr transaction should not require space because no log records are written for it. Therefore failures in the execution could only be I/O errors and require manual intervention at this time.

The summary of catalog processing is then a new obd method, called o_process_logs. In order to implement this systematically the OSC during setup intializes a second set of llog_operations. It has already initialized the llog_operations to be used in conjunction with the llog_storage obd; it uses these operations as an originator of changes. The second set of operations is required when the OSC is acting as a replicator.

```
osc_process_logs( )
{
     // network call
```

```
        llog_repl_create(&handle, LLOG_OBD_SZ_LOG_HANDLE);
        llog_repl_init_handle(handle);
        // notice the new parameter to llog_cat_process, allowing us to select the operat
        rc = llog_cat_process(handle, osc->llog_repl_ops, osc_process_logs, osc->osc_llog
        if (rc == -EDONE)
            rc = 0;
        return rc;
}
```

The LOV collect error returns for these OSC catalog processors:

```
int lov_process_logs()
{
    foreach osc in lov->children {
        rc = obd_process_logs(osc);
        if (rc && !result)
              result = rc;
    }
    return result;
}
```

The MDS simply executes the process log method associated with the *mds->mds_osc_obd* device, but also makes a final decision if size caching can be re-enabled:

```
int mds_process_logs(obd)
{
     rc = OBP(mds->mds_osc_obd, process_logs)(mds->mds_osc_obd);
     if (rc == 0)
          obd->u.mds.mds_sz_cache_enabled = 1;
     return rc;
}
```

An error in log processing should cause events to initiate re-connect attempts and or to notify operations of I/O problems on the MDS.

If more than one system fails, including the MDS then the clients have to invalidate caches, and the MDS only processes logs. The MDS can start ordinary request processing while it is processin the logs.

10.9.8.7. *Client Failure.* If a client fails, and the MDS detects this failure it may as part of recovery close the files that the client had open. In some cases it will perform the last close on a file. Because of the client failure it has no opportunity to acquire the correct file size, from the client. The MDS will handle this with an LOV getattr to obtain the file size from the OST's.

The MDS may also not have obtained the log cookies required to cancel the size log records on the OST's. To supply this, the OST's will return the log cookies with the getattr command, as well as the epoch that are currently in use. The MDS then performs a setattr of this file size.

## 10.10. Group Locks

To support certain HPC installations, Lustre supports a group I/O lock. The semantics of the lock are as follows:

(1) All processes in a group of cooperating processes:
   (a) the processes share a group id, a 32 bit integer, which is generated in a way outside of the scope of this document.
   (b) mark the file as not requiring normal extent locks, and mark the file descriptor (as "usual") as blocking or non blocking.
   (c) take a concurrent write lock on a [0,-1] extent associated with a file. The concurrent write lock is passed the group id.
   (d) explicitly release this lock when done with their I/O, preceeded by a flush of cached data.
   (e) when the file is closed, deliberately or through exit, the group locks are dropped
(2) Readers on other nodes take [a,b]R locks which cannot be granted when group locks are present. Such readers can receive:
   (a) can be made to wait forever, interruptably. This is good for blocking file descriptors.
   (b) can get -EWOULDBLOCK, this is good for file descriptors that have been marked as non-blocking.
   (c) group enqueues with a different group id must wait for the current group and PR/PW locks to be released.

In case (a) this behavior causes further group locks to have to wait until the read is satisfied. This is not desirable, so we will let group locks jump over the waiting lists if other group locks have already been granted.

## 10.11. Lock server location and recovery

Lustre has chosen to implement the file locks per object, ie. on the OST that holds the object. If the locks for a file are held on a single OST to cover resource management for all stripes in the inode, new problems appear, associated with the triple "client, lock server, ost". The centrla issue is that failure to revoke a lock means a client needs to stop doing I/O to another system. This can be enforced in two ways:

**io fencing:** force the OST to stop accepting I/O or
**leases:** expect the client to renew its locks

**10.11.1. Leases.** The protocol for handling the leases is as follows. The lease is defined as the time that a lock is presumed to be valid.

a client knows the maximum amount of time that a write request should take: the amount of time that an OST is willing to let a request sit in the queue before discarding it, plus the longest time that a single PtlGet + write will take:

```
brwtime = (OST request queue timeout) + (PtlGet timeout) + (brw_kiovec time)
```

So

```
iotime = brwtime * # ios needed to flush the cache
```

This leads to the following:

(1) clients voluntarily stop writing at the end of the lease, the correctness issues are covered, as long as our estimates about write time are correct
(2) the DLM does not grant new locks, by disconnecting the failed client, until the iotime + the lock timeout have passed.
(3) To estimate and control the iotime:
  (a) the DLM should return with each lock an idea of how much data should be cached
  (b) each extra lock could reduce the amount by half; starts around the minimum amount for one client to keep the wire full (4-16 MB?)
  (c) the DLM can also scale its timeouts based on client count
  (d) clients could also send the stripe count with the lock request, which helps the timeout scaling

## 10.12. Grants

**10.12.1. Summary.** Clients need to cache write data to achieve maximum performance. Full sized RPCs to each OST (512kB), multiple RPCs in flight (at least 4 is best), can be a _large_ amount of data cached on clients if writing to many files.

To avoid client file caches containing data for which no space is available on the OST's *grants* of disk space are given to the clients, enabling them to use caches only when certainty exists about available space on the OST's.

The critical issue here is not to lose cached data when out of space on OSTs. To accomplish this OSTs grant available space to clients during each bulk RPC. Clients get minimum grant size initially, but grows/shrinks as client writes. The grant size also dynamically adapts to the number and available space. If clients are not writing we can't currently revoke their cache if OST is running short of space, but at most this is a few GB lost. But clients still subject to per-OSC cache limits.

A client will:

- Use the WB cache for a file write when there is space
- Receive updated grants piggybacked on replies to requests.
- Start with an initial 0 grant, which we might improve on by giving a first grant in obd_connect.
- Handle grants on a per-OST basis
- The client will not start using the WB cache until the grant exceeds the amount in the write request.

Space accounting on OST is difficult because of space consumed by metadata. Clients do synchronous IO if there is no grant available. The OST manages free space by keeping:

(1) 5% for the super user
(2) 1.5% for indirect blocks.

There is still considerable nervousness surrounding grants. If a lot of clients have a grant on an OST the amount of oustanding dirty data can be enormous in the face of a single OST throttling the flusing of this writeback data. This can cause lock timeouts. We expect to avoid timeouts only through giving progress indicators to waiting systems.

**10.12.2. Detailed Implementation.** Basically #974 patch is divided into two parts, OST granting logic and OSC logic to obey the granting. The OBD_MD_FLGRANT flag is introduced to let the client know OST is supporting granting.

10.12.2.1. *OST side.* Main part of OST code lives in obdfilter/filter_io.c

**filter_grant():** calculates amount of grant that we can give to this client,this is calculated based on amount of free space on the filesystem, andif there is no space shortage, the logic is to grant extra two megabytes of space in addition to what client have cached already (2 megabytes is totally madeup number that can be changed). This is done during read and write incoming obdo.

**filter_check_space():** checks the incoming write request pages if they are using the space from grant or not (this is usually the case with grant unaware client, or with sync requests (that are usually there when client cannot do any caching due to small grant or disabled client side caching (via sysctl))). Writtenamount of pages is not substracted from client's grant at this point, this is a job for fiter_grant() to adjust client's grant. If the page written is not from grant and there is enough free space on the filesystem, we just increase the grant and let operation to success, if there is not enough space, we see if this page is going to be written into already allocated area of a file, and if so, we still allow the page to be written. If the page writing request is not taking the space from grant and there is no free space, the entire incoming request is aborted with -ENOSPC.

**filter_grant_space_left():** just calculates how much space we have left on the OST filesystem, it tries to do some tricks to account for metadata too. Also there is a (totally made up) reserve of 10*PAGE_SIZE bytes for last_rcvd/llog data. This reserve will be eliminated later when #2059 is landed. We cannot live without this reserve, because when llog records cannot be written anymore, everything breaks and we cannot even delete files anymore. Also in addition to this, there is some statfs caching logic that is currently disabled, and I believe it cannot work in its current form (and I am not all that sure it is easy to implement statfs caching in this case at all).

**filter_grant_incoming():** is called on incoming write obdo. It checks for clients not using more cache than they were allowed to (warns if not). Also when the cache shrinking target was set for client, it checks for client meeting the target and only then OST-side accounting of client-cached data is changed.

**filter_inode_has_holes():** is just a helper function that checks if the underlying file region for given offset and lenght have all the blocks allocated or not.

10.12.2.2. *OSC side.*

**osc_announce_cached():** - preexisted code that reports to OST about size of client-side cache.

**osc_update_granted():** - preexisting code that gets granting info from OST replies and updates local structures. Now also calls osc_adjust_cache() when grant size is changed.

**osc_cache_cap():** is a helper function that returns amount of maximally allowed cache on client. If OST supports granting this is the minimum from granted size and max_dirty_mb sysctl value.

**osc_adjust_cache():** is called when cache size is changed. If cache size is requested to be increased, it wakes threads waiting for cache pages and allocates more pages for them until there is no waiters or new limit is hit. If cache is totally disabled, all the waiters are woken up and an error is returned which will lead to requests being resubmitted in sync mode. If cache size is requested to be decreased, nothing is done, as normal cache logic is able to deal with it.

**osc_enter_cache():** - preexisting code that manages client side cache. Now changed to take into account grant info as well.

**osc_exit_cache():** preexisting code. Now changed to check for possibly changed allowed cache size.

Client starts the connection with special (OST supports granting) flag set and grant size of zero, which leads to first write request to be synchronous on which the grant will be allocated on OST. Phil hints this is not all that good and that initial grant should be allocated and transferred to client at connect time. Also once the "OST supports granting" flag is reset (done in osc_invalidate_import), it is never reenabled back and client never listens to OST grant requests anymore. This is behavior wanted by zab for yet unknown reason.

Also There is no way for OST to tell if client listens to its grants or not (besides just watching how cache size grows past granted amount).

CHAPTER 11

# Recovery

As described in the earlier chapters, the distributed Lustre file system is made up of several components - the clients, MDS, OSTs and the network fabric, and is prone to be effected by failures in any of these. In such a cluster, either node failures could occur or transient network failures might be seen. A reliable and highly available file system like Lustre needs to provide support to recover in these different situations and return the file system to a consistent and performant state and also to make sure that a failure does not bring the whole filesystem down. There are a handful of different types of failures that can trigger recovery in Lustre:

(1) Client node failure
(2) MDS failure
(3) OST failure
(4) Transient network failure

The recovery is different for OST, MDS/clustered MDS systems, and for the clients. The state on a client is not persistent and the clients merely needs to recover the protocol, rejoin the cluster and be able to access the file system again. The MDS and OST targets have two additional complications: each must manage the consistency and recovery of its own persistent state relative to the other server's persistent state information. Also, these systems depend on the presence of other MDS and OST systems and an awareness of which of these are operating is necessary.

When OST or MDS systems fail and recover, they may offer a recovery period to clients that reconnect. During the recovery phase, clients are enabled to re-establish locks, re-open files, and replay updates which the server or OST may have lost. If a client misses this recovery window, it is evicted from the cluster.

The collaborative cache (COBD) is an upcoming feature in Lustre that would add another variable to the recovery equation. We will touch upon the failure modes for the COBD and the recovery issues involved there.

In the subsequent sections, we will decribe the basic recovery infrastructure made available in Lustre and then discuss how it is used in the various failure scenarios. For high availability, Lustre supports failover MDSs and OSTs, we will discuss this mechanism and state diagram for recovery in the various scenarios. The implementation details along with recovery related APIs will be discussed in the corresponding architecture section.

## 11.1. Recovery infrastructure

**11.1.1. Detecting failure.** In Lustre, *timeouts* are the mechanism used to detect failures. Every request is associated with a *timeout* value, the timer is started as soon as the request is sent. If the timer expires before a reply is received, it indicates some form of failure and will trigger recovery. The following situations will trigger recovery in Lustre:

- Lock requests timeout
- Requests timeout waiting for a reply
- Bulk I/O times out - A client might timeout waiting for the server to grab the data for *write* or complete a bulk *read* operation.
- Recovery is also triggered when we get a *-ENOTCONN* status in a reply after the initial attempt to connect failed. This typically happens during failback, when a device has been removed from one server but the server is still up handling requests for another server.

**11.1.2. Basic support.** In all our recovery discussions, we will assume that all the persistent storage resides on OST and MDS systems. If this is not the case, similar but different considerations will apply. This implies that client systems only have memory-based state.

Lustre recovery relies on some basic infrastructure - *epoch* number, *generation* number , the *incarnation* number and the *transaction* id to determine when recovery is needed, what level of recovery is needed, which requests qualify for recovery, these are all listed in table 1.

When recovery is ensued after a failure, a server should stop serving any new requests till the recovery is complete, this is called ***fencing.*** When connections between systems are used, the generation, incarnation, and epoch numbers are communicated to enable **fencing** between hosts in the cluster when they are not allowed to communicate.

**11.1.3. Client Request Management.** After recovery is completed, clients have to determine the actions to take for various requests. Clients keep lists of requests that belong to the different catagories - The client recovery protocol for the client will walk the list of all requests and take appropriate action. Some requests are merely in the list until the server commit confirmation is received. Other requests need to be replayed because the server lost them, or a reply needs to be reconstructed. If neither a reply was seen nor server processing the request, the client resends the request.

**Delayed:** Requests that can not be sent out until recovery is completed, this is I/O fencing.
**Sending:** Requests that have been sent but have not seen a reply come in.
**Replay:** Requests that have been sent but did not see a reply come back or any other confirmation that the changes were committed on the server.

## 11.2. Failure scenarios

In the following sections we discuss the various failure scenarios and the type of recovery that needs to be done in the various cases.

| Variable | Description | Purpose |
|---|---|---|
| Epoch number | Each storage controller maintains this counter on the persistent store and increments it upon each successful boot or recovery | The server maintains a file with information on received requests for every boot count. Based on current boot count, it can determine which requests qualify for recovery. |
| Incarnation number | The MDS cluster as a whole maintains an incarnation number, this increases everytime the cluster configuration changes. The initial value is the boot time of the MDS cluster leader. | This number can be used to assist recovery between the MDSs in the cluster. |
| Generation number | Connections between clients and other systems (OSTs, MDSs) have a generation number. | If a client is declared disconnected by an OST or MDS, the generation number changes. |
| Transaction ID(xid) | Monotonically increasing request number maintained by the client. This is unique with respect to each client, multiple clients might be using the same xid at the same time. | Each client request has a transaction id (xid) associated, the server keeps track of tracsactions received or committed using this number. This also helps the server determine which requests qualify for recovery and which have been committed. |
| Transaction number(transno) | A transaction number is maintained by the servers and is unique per server, but the clients might see the same transno from different servers. | Used to ensure that requests are replayed in exactly the order they happened. |
| Connection level | Every connection is associated with a level to indicate if normal processing is going on or if failure or recovery is in progress. | This information is used in I/O fencing. Each request tracks the connection level when it was created. Requests with level higher than the current connection level have to wait till recovery is completed. |
| Last_recvd | Transaction id of the last received request for a client - this information is kept on persistent store, in a *last_recvd* file on the servers. | This is used to determine which requests qualify for recovery. |
| Last_committed | Transaction id of the last_committed request for a client - also kept on persistent store on the servers. | This is used to determine which transactions were committed to the persistent store on the servers and do not need recovery. |

TABLE 1. Recovery support variable

**11.2.1. Persistent State Recovery.** As mentioned earlier, all persistent state in Lustre is on the MDS and the OST servers. The MDS stores the metadata for the objects on the OSTs. These servers are designed to be built on top of underlying journalled filesystems like ext3, ReiserFS, XFS, JFS. Currently *ext3* is the journalled file system used with Lustre, in future we will look into the other alternatives. All persistent state in the current design of Lustre is handled by journalling filesystems with asynchronous write-ahead logging. So, after a reboot we have the guarantee that disk state has been recovered by this journalling infrastructure. The only recovery that might be needed is to ensure that the information is consistent between the MDS and the OSTs. We need to make sure that in no situation does the OST have objects with no corresponding reference on the MDS, such objects are called *orphans*. We should also ensure that the MDS does not hold reference to non-existent objects. This could happen in the following cases -

(1) Create operation : A file create operation happens in 2 phases - object creation, recording the object information on the MDS. *Orphans* will be created if the MDS fails before the new object metadata is recorded on the persistent store. The same will happen if the client crashes before sending the object information to the MDS.
(2) Delete operation : OST fails to complete the delete/destroy operation leaving an object on the OST with no corresponding reference on the MDS inode. This could also happen if the client crashes before completing the delete.

The case of having MDS inodes with references to non-existent objects can be handled by dynamically creating new OST objects and storing them on the MDS inode as new extended attributes.

The MDS/OST coherency is maintained by maintaining a logical log of select operations being performed on the MDS and OST, which will be communicated in recovery scenarios. The OST will log the creation of all objects, the log can be cleanedup after the MDS commits and communicates this to the OST using a new *OST_SYNC_RPC*. At this point the OST can cleanup the log for committed objects. During the *unlink/delete* operation, the MDS will transactionally log the deletion records, once for each stripe of the file. A seperate log is maintained for each OST, this log can be removed after the OST destroy operation commits.

**11.2.2. Client failure.** A client which loses contact with a MDS or OST needs to recover. If the client lost contact due to a fatal event on the client and rebooted, this can be considered a normal start for the system. The only support needed for recovery would be for the revocation of locks and other resources held by the failed client, this will allow the surviving clients to continue operating. If a client fails to respond to lock cancellation callbacks from the lock server, or a bulk data operation times out, the client is assumed to have failed and is removed from the cluster. An active client will use a *pinger* in future to continuously announce its presence to the servers or to determine if a server is up after a failure, the server can use this to maintain a *last_heard* value for every client. If the server fails to hear from a client again within the expected time, it can pronounce the client as dead.

However, it is also possible that the network failed, or that the MDS/OST nodes failed due to hardware or software failures and through administrative events. In all these cases, a client has to perform some recovery to ensure that the requests that an application assumes to be committed were indeed completed. This might required a replay of locks, requests or reconstruction of replies. During MDS/OST failures, transactions for which replies were sent can be lost due to the asynchronous processing the MDS/OST undertakes. Similarly, the client may have to reconstruct replies when replies are lost for that were processed on the server. In case on MDS/OST failure, a client might also have to try to find failover/standby server to connect to before trying to replay requests/state. The path of recovery would depend on wether a failover server is available or not. If the client fails to reconnect and start recovery withing a specific window of time, it gets evicted from the cluster and will have to rejoin.

An MDS/OST may change its network address during recovery. This is managed by systems such as pensacola.

**11.2.3. MDS/OST failure.** An MDS or OST could fail due to hardware/software or administrative events. The recovery for persistent storage on the servers is ensured and taken care by the journalling filesystems they are built up on. Lustre supports a failover, each server can have a failback server that would takeover all services in the event the active server fails, this relies on the use of a shared storage for the MDS backing file system. In case of MDS, Lustre supports a failover server that is originally inactive and is made active only when the active server fails. On the other hand, in the case of OSTs, both the primary and the standby servers actively handle requests. When the primary server fails, the standy server is configured to take over the job done by the primary server along with its normal request processing. When the failed server comes back up, it takes over its original operations, the client requests failback onto this original server. In the case of OST failback, it is important to ensure that only a group of clients failback to the orginal server to ensure load balancing after failback.

The MDS server recovery functions in conjunction with some high availability software like *kimberlite* or *clumanager*. When a server fails, the *clumanager* detects this and indicates this to the standby/failover server. The failover server will do the required setup and start the required services, indicate to the LDAP server that it is the active server as shown in figure 11.2.1. The server will then handle any recovery associated with the client requests.

We define the recovery of the MDS service through the following requirements:

(1) The operations on the Lustre name-spaces as seen by clients and by the MDS in memory and on persistent storage form a strict and serializable system.
(2) The system will recover transparently from:
    (a) A crash of the MDS system in the absence of client failures
    (b) A temporary complete network failure followed by network restoration
(3) Failures of a single client or of the network between a single client and the MDS fall into two categories:
    **short failures:** are transparently recovered. The client sees no errors and MDS transaction processing proceeds normally. This involves handling:
        (a) The MDS protocol assists when a network or server failure causes **requests to be lost**.

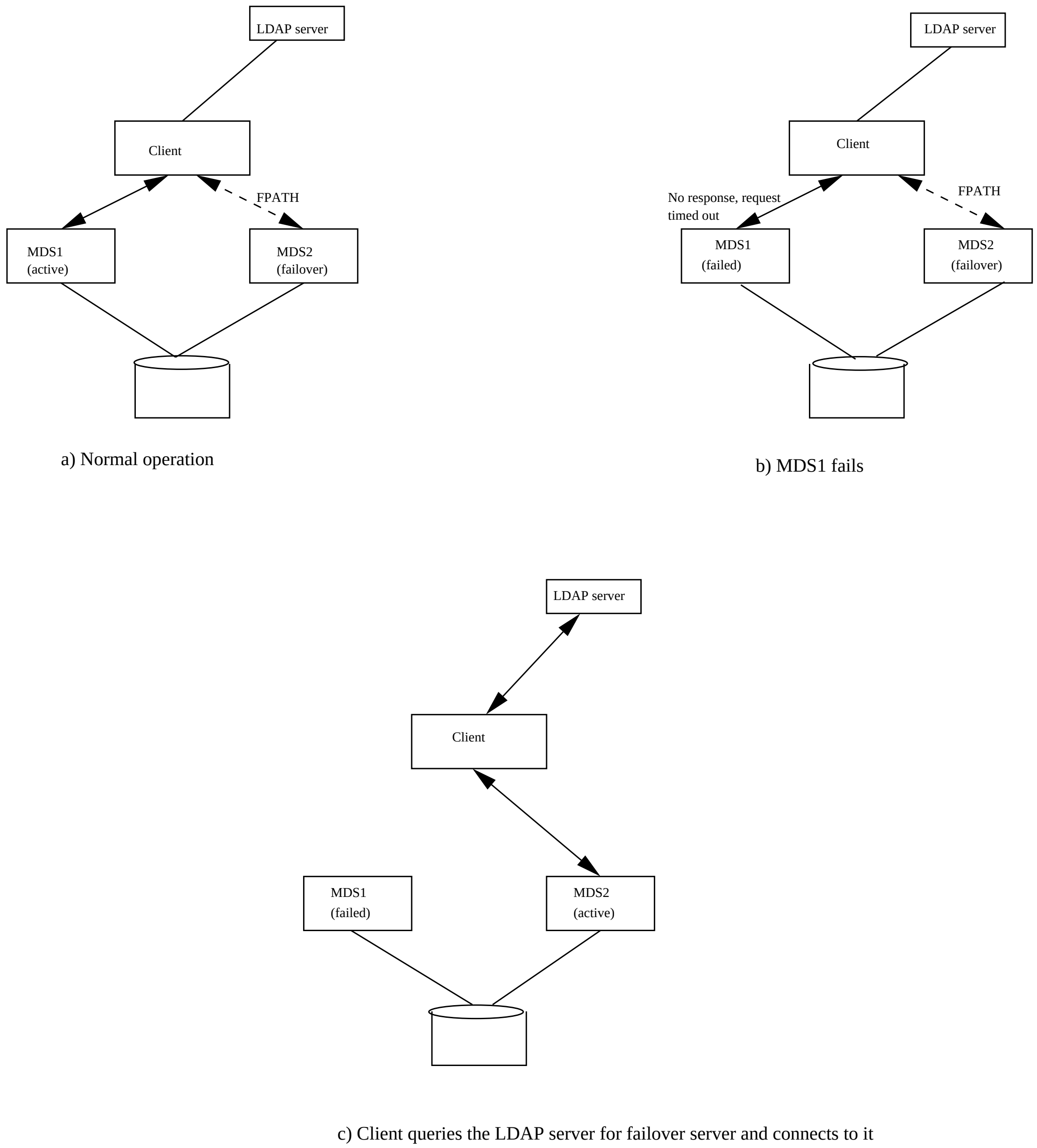

a) Normal operation

b) MDS1 fails

c) Client queries the LDAP server for failover server and connects to it

FIGURE 11.2.1. MDS failover

(b) The MDS provides a mechanism to re-establish lost replies for requests. This is done through the **retransmission** of requests.

**long failures:** lead to eviction of the client from the cluster. Applications on the client get error messages. Other clients continue processing after a short interval.

Recovery for the OSTs is defined by the following requirements:

(1) The clients always see the effect of complete atomic transactions on persistent storage, partial or incomplete transactions should not have any effect on the persistent store
(2) The OSTs should support transparent recovery in the following cases:
    (a) A failover OST is unavailable - This should not bring down the rest of the filesystem, the remaining system should continue to be available to all clients. Clients should see errors when trying to access the objects on the failed OST and prevented from trying to create new objects on this failed OST.
    (b) A failover OST is configured - In this case, a transparent failover to the failover OST should happen, any required requests should be replayed or resent.
    (c) Temporary network failures - The system should recovery transparently from temporary network failures.

The OST recovery involves several phases. First, the system must reboot and perform journal recovery on the logical volumes managed as object stores. After that, the MDS/OST protocol needs to be recovered. This protocol involves the following main components: file creation and removal, pre-allocation of objects, and object attribute updates. The protocol between MDS and OST's involve persistent data on both sides and we intend to use logging to do this as described in the section 11.2.1. If a failover/standby server is configured for OST, it will follow a path similar to that described for the MDS failover. On the other hand, if a failover server is not configured, the corresponding OSC will be marked as inactive and the clients will refrain from trying to create new objects on the failed server. The rest of the file system will remain accessible as usual.

11.2.3.1. *Design for failover OSTs.* The goal is to have failover OST server systems that uses redundant shared storage. The protocols should, transparent to the applications, recover read/write system calls if an OST fails.

Lustre already has general failover infrastructure, as described earlier, which can be adapted to OST's. In the lock, file system and request processing API's there is support for the protocol to transparently re-establish locks and open files after an OST failover. This support is not utilized for the OST failover, but is used for MDS failover; it is available for both Elan3 and IP networking. The key issue is that at present file writes are written into the OST's page cache, which may be lost in the case of an OST failure.

A necessary condition for correct failover of write calls is that a copy of the data continues to exist until it has reached persistent storage. We believe that writing twice, to multiple OST's, is not acceptable. That means that in our implementation the client needs to retain the data until the OST has flushed it to disk and that the OST needs to send a commit message back to the client.

The easiest way to do this is to explicitly wait for completion of writes on the OST. Initial profiling indicates that this does not negatively affect throughput of the OST (it in fact improves it) if the writes have reasonable sizes (like 64K). Such reasonable write sizes (as sent out by the client) can be accomplished in two ways: large synchronous writes, possibly with O_DIRECT and secondly page cache flushes on the client.

Having a write behind page cache for file writes will hide the file latencies from clients introduced by synchronous writes on the OST.

We now have a client-side write cache on Linux 2.4, so the writes on the client need not be synchronous, the writes on OSTs are now synchrnous. The most time consuming task will be to debug the general failover infrastructure for unexpected surprises.

This design proposal leaves us with one boundary case to consider which is that very many clients do very small writes to the OST's, which do not aggregate to large file writes automatically. To achieve maximum efficiency here, and make good use of the (redundant) caches in more expensive raid controllers behind the OST, it is very important that the I/O tasks are submitted and completed with events. So we do not issue these as blocking, synchronous I/O calls from OST service threads, because the threads would be unavailable to service other I/O requests. The completion events causes the OST to inform the client of completion. The submission process I/O tasks, allows the threads to push many small requests to the block device drivers on the OST. The SCSI/ATA block device infrastructure will aggregate these requests and avoid unnecessary waits on the client.

However, while this will work in theory, in practice it may well require tuning of the block device I/O request handling on Linux which would be a very time consuming proposition. We propose not to do such tuning now.

Unlike in the case of MDS failover, the standby OST server is also active and servicing requests before taking over the services for the failed server. When the primary server is restored, a group of clients need to fail back to it to ensure load balancing, OST failover is illustrated in figure 11.2.2.

**11.2.4. Network failure.** All network failure is detected because something **incurs a timeout**. The timeout doesn't mean that a message will be printed and the operation failed, the timeout merely activates recovery mechanisms. Unfortunately, there is no clear way to differentiate between a network failure and a server failure. A *timeout* could mean a server failure or a transient network failure, so an attemt is made to reconnect after a timeout is detected. If the reconnect fails, recovery is triggered.

Lustre provides a macro that can be used to wait for events to complete or certain conditions to be satisfied, it takes:

(1) A conditional expression to be satisfied
(2) A timeout after which an unconditional wakeup happens - This would be triggered unconditionally if the required event does not happen withing a specified time.
(3) A flag that indicates that "kill style" signals can cause wakeup.
(4) An optional timeout after which signals will be respected.

We have at present the following uses of *wait queue*'s in Lustre:

**"Client" request enqueues (*ptlrpc_queue_wait*):** ("client" is quoted because the MDS & OST make lock callbacks as a RPC client to the Lustre filesystem client systems as well).

**Bulk waits (hard-coded):** The server has waits for sending *bulkd* (ie. acting as a source) and as a *sink* (reading). The client only has a wait for acting as a source.

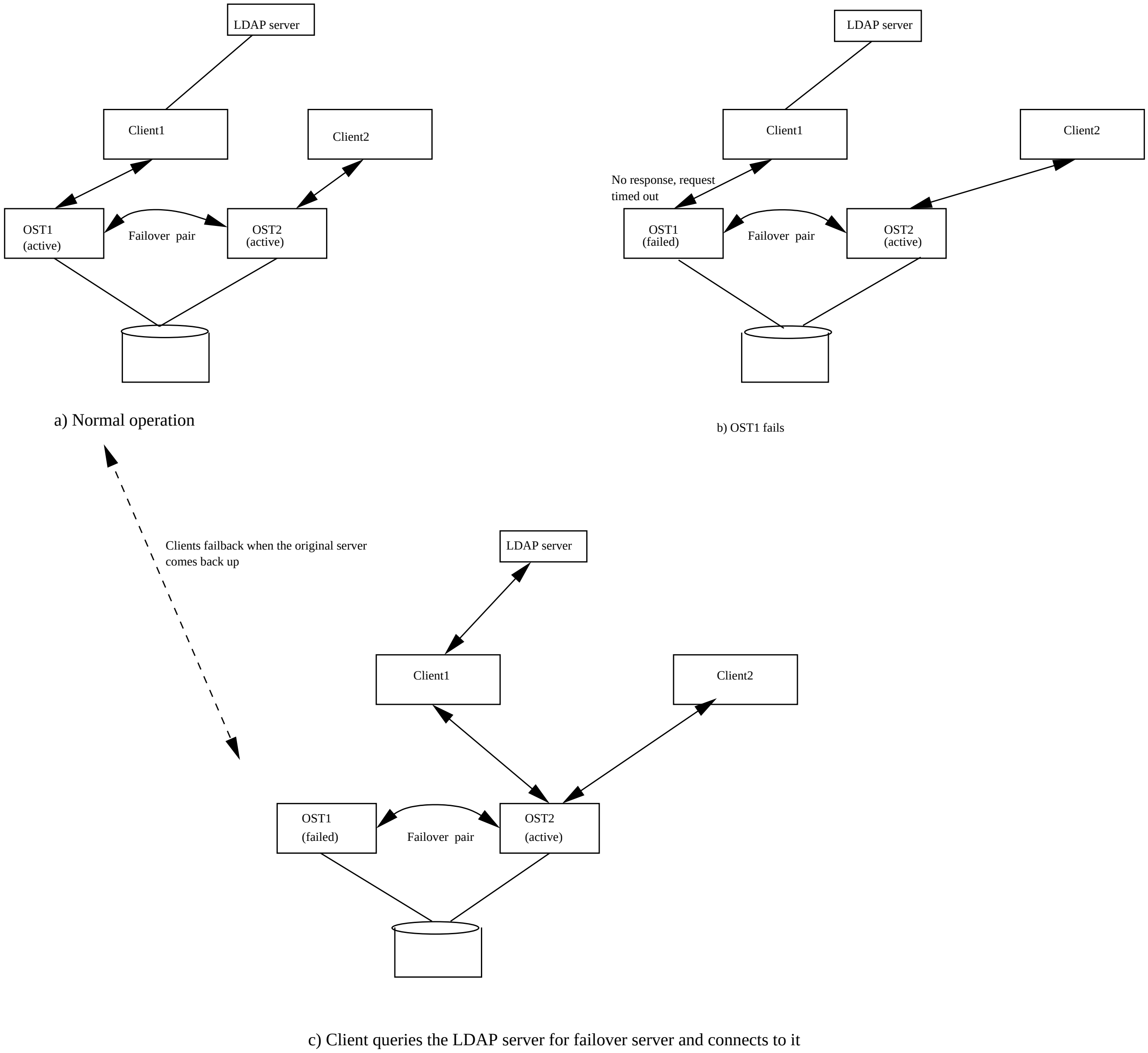

FIGURE 11.2.2. OST failover

**Waiting for locks to complete, ie. blocking AST waits:** Generally if a lock enqueue (or convert, which we don't use yet) returns a reply that the lock is taken, the system that is doing the enqueue calls a "blocking" callback. At the moment the blocking callback's are hard-coded waits (but they need to become functions). The completion callback's form the corresponding wakeup call, and these are fired off when a completion callback RPC is made by the lock server to the system enqueueing the lock. At present, the OST does not (yet) take locks, but the MDS takes locks and may sleep on the blocking callback, and so can clients, obviously.

**Waiting in request processing loops:** All waits will get timeouts. The triggering of these timeouts indicates a failure to receive the solicited response; this is the definition of a Lustre protocol failure. If any new reasons for waiting enter our system, we need to understand precisely what category they fall in, these timeouts will trigger recovery.

## 11.3. MDS Recovery

Lustre promises high availability for metadata service as well as the object storage targets and the ability to transparently handle single point failures. This requires a very well formed recovery infrastructure in place. In the following sections we will discuss how the basic infrastructure described earlier is used to provide sufficient recovery support.

**11.3.1. Introduction.** In this section we will use some terminology from the database literature, such as the concept of transaction processing and the read and write sets of transactions. See [2] for an excellent discussion of these topics.

The MDS manages storage for name-space objects. We require that when it starts this name-space is consistent, in the sense that it contains the image formed by a sequence of transactions, and that no effects of partial execution of transactions are visible. This is the normal *atomicity* requirement from the database literature.

The MDS requires operation very similar to a transaction engine. In practice Lustre implements this with a journaling file system, which provides transaction processing but not immediate durability. It is important to note that upon recovery the MDS storage can roll back, i.e. it can lose transactions whose requests were executed but not committed to the disk. This can happen both before and after a reply for a particular request was sent. The clients provide a mechanism of redundant availability of the request data which assists with guaranteeing durability in the absence of multiple failures. A case of multiple failures needs to be handled differently, the MDS supports a synchronous execution mode to handle such failure scenarios.

For a transaction engine *recoverability* and *strict execution* (see [2], 1.2) are important to simplify recovery and avoid cascading aborts. Lustre implements these features in the protocol through *acknowledging replies*, which provides ordering and replay semantics for transactions as we will see below.

*Serializability* (see [2] section 2.3) of the execution on the MDS is another important invariant to provide atomicity and isolation of transactions. Local file systems implement this with local locks on directories and inodes, which Lustre trivially replaces with DLM locks. The DLM locks have features that substantially enhance concurrency of transaction processing on the MDS.

The most basic design of the Lustre MDS with a single threaded metadata service based on a journaling file system leads to an even stronger set of guarantees than strict and serializable execution namely the property that all transaction histories are *serial* ([**2**] section 2.2.). A serial collection of transactions means that if T1 precedes T2 then all operations in T1 preceded all operations in T2. This has dramatic performance implications. In Lustre 1.0 the MDS is single threaded but does not ensure serializability, making transactions serializable as described here will definitely boost performance.

The *ordering* (see [**2**] section 1.3 & 1.4) of transactions is important from a client perspective. Lustre implements a scheme of *update streams* which allow clients to order transactions through a FIFO. Fifo's are often considered unnecessary, but having multiple streams combines the advantages of handshakes with the advantage of allowing to transfer the entire content of the FIFO to the MDS for execution without network transfers. This is particularly important when a client uses write behind caching of metadata updates. The reply acknowledgments and stream concepts in Lustre make up the *transaction scheduler* in database parlor.

A further issue in Lustre is that it is a distributed system in which network failure also need to be handled. Regardless whether a transaction rolled back or committed the reply to the client that issued it may be received by the client or lost. Lustre must recover from such situations.

**11.3.2. Request lifecycle.** A client software system that has cached (file system) state based on completed but not necessarily committed transactions can continue to use that state in the presence of MDS failures. To make this possible the MDS must provide a mechanism for the client to **replay** transactions that were not committed and lost.

In case of MDS recovery, the **replay** mechanism must ensure that all attributes related to the transaction that had been exposed to client system remain the same after replay and that execution is in exactly the same order as the original execution. Given that *fid's* are exposed to clients, replay must preserve *fids*. The reply acknowledgment system is important here.

In order to understand the details of recovery it is important to note the following states (see figure 11.3.1)that are normally achieved by all requests:

(1) **executed**: A request is executed when the server has completed the reads and updates and made them visible in the MDS name-space. This phase always occurs first, the following two stages can be reached in either order.
(2) **replied:** When a client is *in possession* of a reply to a request and that reply indicates where in the MDS transaction sequence the request was executed.
(3) **committed:** when the request is committed to disk, there is not need anymore for a client to replay the request. The client can then delete all cached information on this request.

The replied and committed phases can occur in *either order*, each of which brings its own complications to recovery.

**11.3.3. Tracking request execution and replies.** The first phase in a request's life is that it is executed. After that the request will be committed and replies will be received, but the order of these last two events is not always the same. Tracking the execution of requests would be required in different recovery scenarios, the recovery scenarios may involve the MDS failing or the network failing. In both cases it is possible that clients sent requests which were never received. The first requirement we need to fulfill is that the MDS knows what the last requests were that it executed.

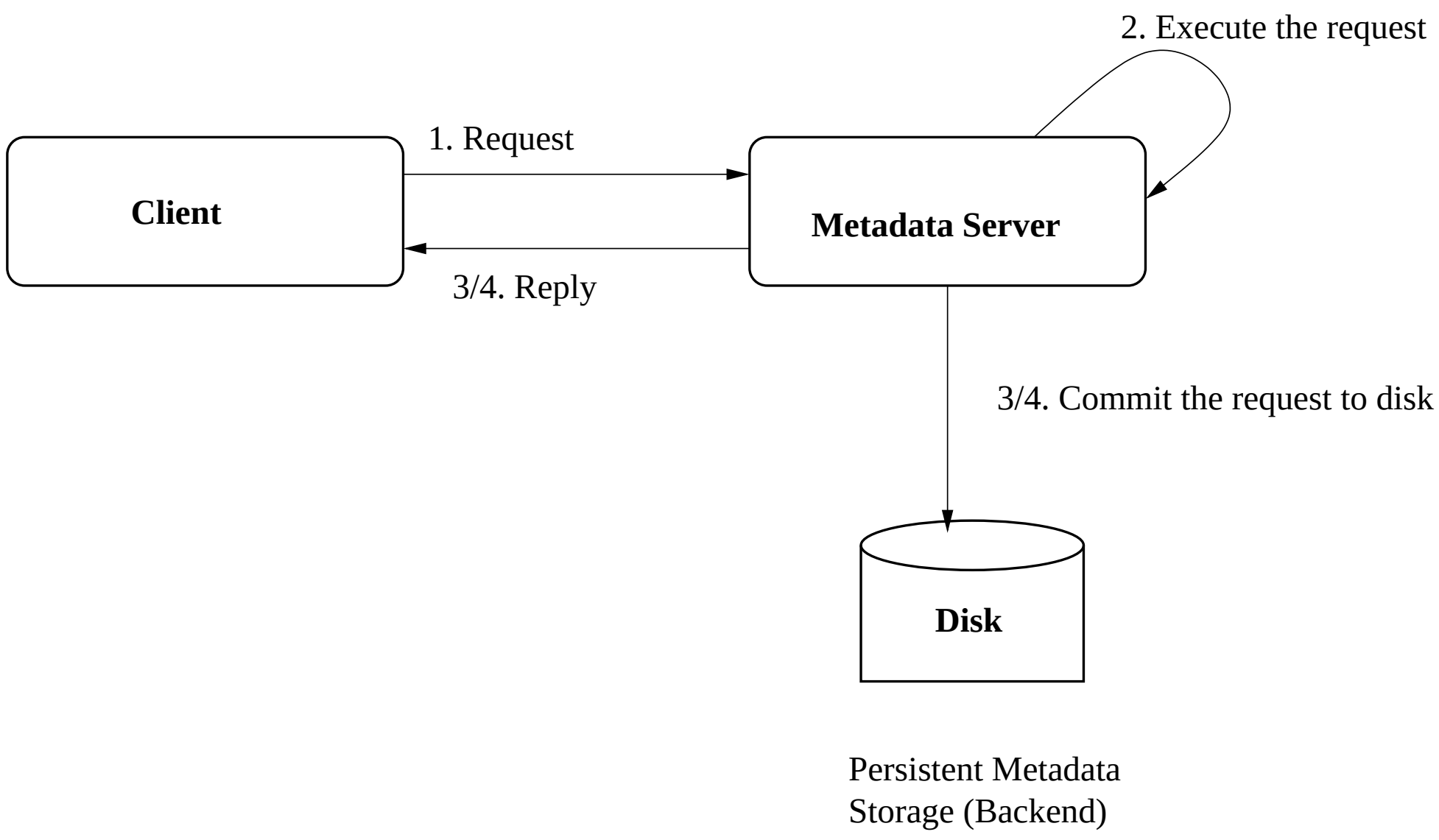

FIGURE 11.3.1. Path of a mds requests

The MDS processes incoming requests from all clients in parallel, and there is not necessarily a well defined ordering for incoming requests. To avoid possibly unbounded numbers of requests executing in parallel, each client has a number of update streams on which it can submit requests. The MDS executes requests from a single stream in order, and tracks the last executed transaction on each stream. Clients submit all requests with a *request id*, and this identifies the request that was last executed. If clients submit multiple requests that require ordered execution on the MDS, which is the case when performing write back caching, clients use a single stream for such requests.

A *last received* request number is maintained for each stream by the MDS in memory. It represents the last request it has processed.

During an MDS crash, a partially executed request or non-committed request will not leave a trace. After an MDS recovery the *last received* request equals the *last committed* request associated for each stream, but when the network stays up, the last received request can be different from the last committed request. The reconstruction of the transaction sequence has three parts:

(1) Requests that were not committed but executed and reply received need to be replayed in exactly the same order. We refer to this as **replay.**
(2) Requests that lost the reply need to be handled. Such requests may have been committed. This must happen after replay has completed. There is at most one such request per stream.

(3) Requests beyond last received can be retransmitted by clients and re-executed by the server, the ordering here can be determined by the client. For each stream there is at most one lost request. Lost requests can simply be resubmitted.

Tracking received replies is necessary because they guarantee that the MDS can rely on clients to reconstruct the transaction sequence. There is at most one outstanding request/reply pair per update stream and the next transaction can be started when the previous one is known to have its reply, we call this **reply-ack'ing**.

**11.3.4. Tracking committed requests.** Tracking committed requests is addressed by establishing a boundary between the transactions that are committed on disk and those that have been executed in memory. When replay is happening the clients send requests to the server which were executed previously in a certain order, and in order to reach recoverability the replay order must coincide with the original order of execution. It is obvious that a failure to do so would lead to an unreliable replay scenario.

This makes it necessary for the MDS to assign transaction numbers to transactions with the property that the set of transactions is ordered and that replay respects the order. Transaction numbers do not need to be unique as long as transactions with the same transaction number can be executed in any order. The database literature has described quite clearly when reordering transactions is permitted.

The metadata back-end generates transaction numbers that correspond to the order in which they are submitted for commit. The back-end is fitted with a callback mechanism to update the *last committed* transaction number.

A key issue now is how to correlate the last committed numbers with request numbers submitted by clients which are sometimes executed out of order. For this the metadata service enforces that requests coming from clients on a single stream are treated as FIFO's. Requests from a single stream are submitted sequentially to the back-end and will preserve ordering of the request and transaction numbering. To enable multi-threaded client systems, clients are allowed to open multiple update streams on the MDS and submit requests through any stream.

Lustre 1.0 features a single threaded metadata client, hence the complete list of requests is the update stream for that client. On the client side it is useful to extend the stream to include transactions that were executed and ack'd but not yet committed, to organize replay.

For each stream the MDS is required to produce a *last committed* transaction number upon recovery and report the *last committed* number for each stream to the client upon recovery. MDS servers store the *last committed* and corresponding request numbers for each stream on persistent storage as a part of every transaction executed. In the implementation this is referred to as the *last received file.*

MDS servers report the *last committed* numbers in every reply packet in a stream to ensure clients have knowledge of committed transactions so that they can free any replay data they may hold.

**11.3.5. Replay of the transaction sequence.** The client would have seen replies for requests on each stream except possibly the last request in the stream that was executed before a failure. Using the replies all requests that have a reply can be retransmitted to the server. The server orders the requests from multiple streams into the same transaction sequence as the original sequence for replay.

However, when multiple streams are executing concurrently there could be more outstanding reply packets and here the acknowledgments of replies become important. When the client has received a reply to the request it can make use of information in the request to extend the request with information that was not provided at the time of the original request, such as specifying a *fid* for a creation operation. So acknowledgment of replies allows server created data to be used during replay, which eliminates the need for pre-allocation (except during write back caching).

The recovery process on the MDS ensures that the transactions that clients replay are sorted and executed in the same order. Clearly a *double failure* of a client and the MDS can cause a gap in the recovery transaction sequence which may have cascading properties, because other clients cannot complete their replay. If this is undesirable a synchronous execution mode on the MDS server is available. Such synchronous execution may appear undesirable because of the negative impact on performance, but is in fact the norm on other cluster file systems.

We have now described how clients will replay the requests in streams from the last committed transaction up to the last replied transaction. Note that it is possible that this set of requests is empty. In each stream a single request may not have a reply. We discuss next how to handle the last requests in each stream that need to be recovered.

**11.3.6. Reply acknowledgment .** Reply acking is best implemented by defining resources in the MDS lock name-space that protect the file system objects that are read or written by transactions. No systems other than the MDS will obtain locks on these resources. When a transaction is started the MDS will take multiple reader single writer locks on these resources. When the MDS generates the reply packet, an acknowledgment structure is built and each of these locks is appended in this structure. To handle clients not supplying an *ACK* the structure is also listed in the client export for recovery purposes.

When the reply is ack'd all locks in the list are canceled. This requires the *Put* operation for the reply message to pass in a pointer to the *rep-ack* structure which is then freed when the event is handled. This implementation leads to good concurrent processing by only requiring the *ACK* when the transactions have conflicting read/write sets with any transaction for which an *ACK* for the reply is outstanding.

In the presence of the acknowledgments, lost replies are now possible only for transactions that are not followed by other ones that make conflicting updates or read modified data. A client can, upon retransmission to the MDS be certain that the server will know if the request was executed or not. If it was executed the server will retrieve, possibly from persistent storage, enough status information to assemble the reply corresponding to the request. Particularly for open the details are still somewhat involved.

It is perhaps surprising is that requests that fail or only read also need to be included in the acknowledgment processing. The reason is that their failure can be related to the position of the failed request in the sequence. An example is that two permission changes (setattr) requests on a single object may both succeed in one order and fail in the reverse order. The dependency of reading on previous updates makes it necessary for read requests (getattr, readdir) to be ordered.

If a client does not provide a timely acknowledgment to a reply the MDS can call sync and flush the locks since the transactions involved will then all be committed and never appear for replay.

While recovering, during replay reply acknowledgments need not be enabled, but from the moment that lost replies and reprocessing lost requests are handled, it must be enabled.

Notice that reply acknowledgments are unrelated to update streams, they are available for different purposes. Reply *ACKs* are used to enforce transaction sequence ordering for requests coming in from different clients. On the other hand, an update stream is used to provide a thread of MDS control to a client, a client could have multiple update streams allowing for parallelism.

**11.3.7. Lost reply boundary.** During a cluster failure, replies from the MDS to clients may be lost. Such replies will normally be associated with uncommitted transactions since the process of replying is typically much faster than a disk commit, but this is not guaranteed and replies may be lost for transactions that were committed.

If requests were sent but not committed there are cascading abort problems to be avoided. As an example, consider two clients, where the first client creates a directory. While the reply to the request is in transit a second client performs a lookup on the directory and obtains its *fid*. The problem which we see now is that unless the first client supplied a *fid* for the creation, it will not be able to replay the creation request correctly. These kind of issues provide the primary motivation for the reply barriers discussed above.

Reply reconstruction happens in two cases:

(1) MDS failed, the transaction was committed but the reply did not reach the client
(2) Network failed, the transaction was executed but the reply did not reach the client

In the first case the reply has to be reconstructed from the data stored in the last received log. In the second case, data in memory can be used as well, but we use a single mechanism for recovery, rather than two.

Reply reconstruction should be investigated call by call.

**getattr_lock, getattr, mds_readpage:** These calls can simply be re-executed. The barriers on replies assure us that the MDS will not have modified any data in the read set of these transactions.

**mds_setattr, mds_unlink, mds_rename, mds_create, mds_link:** These calls return simply an integer status. This status should be stored in the last received log and re-transmitted to the client. The *mds_unlink* also returns extended attributes . Because the cleanup of unlinked objects is part of the MDS-OST protocol, this should not be done during reply reconstruction.

**mds_open:** The open call needs the disposition and status in the last_rcvd file. If file handles are left open during a network failure and recovery then they can be retrieved from the list of open files by transaction id (*xid*). The lock that is passed back should be reconstructed during this search and returned to the client as is done by *mds_open*.

### 11.4. Overview of recovery algorithm

In these sections, we will discuss the details of recovery that entails in the various failure scenarious on the servers and clients.

**11.4.1. Client failure.** As mentioned earlier, if the client failure/crash occurs due to fatal events on the client, it will be considered as normal system startup. The only thing that needs to be done is to cleanup the state - locks, file handles associated with the failed client to ensure that the rest of the cluster continues to function normally. The server usually accomplishes this by sending out lock callbacks, if the client fails to respond within a certain time, the client is assumed to have failed and all the state for the failed client is cleaned up. The drawback of this is that it would take a very long time for the normal cluster functioning to start if it has to wait for several clients to timeout. An alternative to this is the new *pinger* operation, the client actively pings the servers at specified interval. The server uses this to keep track of when it last heard from a client, if the server does not hear from a client for a certain amount of time, the client is evicted from the cluster.

**11.4.2. MDS or OST failures.** On the clients, failure is detected through timeouts, this is followed by the following steps:

(1) **(I/O) fencing:** Once the failure is detected, further communication that changes the state of the system must be avoided. Lustre messaging uses/should use the following to control fencing:
    - Connection level is lowered to indicate failure or begining of recovery
    - Requests also have a level, request packets which are of a level higher than the connection, are not dispatched, these are hung of the *delayed requests* list in the imports. If they are sent out, they should get a reply, indicating that recovery is in progress.

(2) Check if the failed device(OST/MDS) is *replayable*, that is it has a failover server configured. Lustre always requires MDS failover to be configured to avoid bringing down the file system due to failure of a single MDS.
    - If it is not *replayable*, the corresponding OSC (in case of OST failure) is marked as *inactive*. This allows the client to refrain from trying to create new objects on the failed OSTs. Access to old objects on the failed server will result in an EIO error to indicate a failure.

(3) If the server is *replayable*, the client will run an *upcall* and use a *usermode* helper function to query the LDAP server for a failover/standby server.

(4) The clients connect to the failover server, the connection level is raised up to indicate successful re-establishment of the connection. It also obtains the *last_recvd* and the *last_committed* values from the server.
(5) Clients will go through all the requests in the *sending* and *replay* lists and determine the recovery action needed - resend sequest, resend request - ignore reply, cleanup up associated state for committed requests.
   (a) The client replays requests which were not committed on the server, but for which the client saw reply from server before it failed. This allows the server to replay the changes to the persistent store.
   (b) The client resends requests that were committed on the server, but the client did not see a reply for them, maybe due to server failure or network failure that caused the reply to be lost. This allows the server to reconstruct the reply and send it to the client.
   (c) The client resends requests that the server has not seen at all, these would be all requests with *transid* higher than the *last_rcvd* value from the server and the *last_committed* transid, and the reply seen flag is not set.
   (d) The client gets the *last_committed* transid information from the server and cleans up the state associated with requests that were committed on the server.
(6) After the recovery is complete, the requests on the *delayed* list are dispatched.

On the Server, the following steps are taken as a part of recovery:

(1) The failover server will be setup, all the required services will start
(2) As it sees the same shared storage as the failed server, it will retrieve the *last_rcvd* and the *last_committed* values from the *last_rcvd* file.
(3) It will expose a recovery window to all the clients to allow then to re-establish their connections and resend/replay requests. Clients that miss this window are evicted from the cluster, to avoid this in future the *pinger* operation will be used. A client will continuously ping the servers to make sure that they do not miss the window of recovery.
(4) During this phase clients can re-establish the state that they still retain. For example, the filesystem clients re-open files they have open, re-send transactions that they know the server has lost and that they know succeeded previously, and re-establish locks. All client state maintained for a connection is held in the import structures associated with the connection. It is very important that only clients that were connected before the current recovery cycle, ie. those that were present in the previous connection incarnation, may replay requests. Other clients will have to discard all state and return errors to all apps that use cached state. Subsequently there would be cleanup of the state associated with clients that were evicted.
(5) We have reached a normal state for the system, and processing can resume.

### 11.5. Persistent state recovery

As mentioned earlier, Lustre MDS and OSTs are built on underlying journalling file systems which transactionally ensure that the persistent state remains consistent all the time. But it is important to make sure that the metadata and data information are also consistent so that every object on each OST has a reference in some inode on the MDS and objects referred by all inodes indeed exist on the OSTs. The file creation or deletion in Lustre is a two-step process, creation requires objects to be created on the OSTs and this information stored in the extended attributes of the file inode on the MDS. Similarly, a file deletion involves deleting the objects as well as the corresponding reference on the MDS inode. This leaves room for inconsistencies to happen if an OST/MDS, client or network failure occurs before bothe the steps have been completed. Inconsistencies in the persistent state information might occur in the following scenarios:

(1) MDS crash on creates - the OST creates objects but the MDS crashes before recording the references on persistent store
(2) OST crash on delete - the OST fails to complete the destroy operation, while MDS removes its inode
(3) Client crash - *Orphans* (objects without corresponding reference in an inode on MDS) can also be created if the client crashes before sending out the object information to the MDS during a create operation
(4) OST crash on create - OST might crash before recording the object reference on persistent store resulting in an inode on MDS that holds reference to non-existent objects

The MDS/OST coherency will be maintained by keeping a logical log of select operations on the MDS and the OST, which will be communicated in recovery situations.

On create operations, the OST logs all object creations, on the other hand, during an unlink, the MDS will create and write log record of the operation. These logs will be used to detect and repair inconsistencies after a server failure and recovery. The logs will be cleaned up after the required recovery is complete.

## 11.6. Changelog

**Version 3.0 (May. 2003)**

(1) Radhika Vullikanti - Updated all the sections and reorganized to reflect the current state and plans for recovery.

**Version 2.0 (Dec. 2002)**

(1) P.D. Innes - updated text, Changelog added

**Version 1.5 (July 2002)**

(1) P.D. Innes - edited, proofed, spellchecked

**Version 1.0**

(1) P. Braam - original draft

CHAPTER 12

# Lustre Security

## 12.1. Introduction

This chapter outlines the security architecture for Lustre. Satyanarayanan gives an excellent treatment of the security in a distributed filesystem. Our approach seeks to follow the trail laid out in his discussion, although the implementation choices and details are quite different.

**12.1.1. Usability.** Only too often have security features led to a serious burden on administrators. Lustre tries to avoid this by using existing API's as much as possible, particularly in the area of integration with existing user and group databases. Lustre only uses standard Unix user API's for accessing such data for ordinary users. Special administrative accounts with un-usual privileges, to perform backups for example, require some extra configuration.

Lustre, unlike like AFS and DCE/DFS, does not mandate the use of a particular authorization engine or user and group database, but are happy to work with what is available. Lustre uses existing user & group databases and is happy to hook into LDAP, Active Directory, NIS, or more specialized databases through the standard NSS database switches. For example, in an environment where a small cluster wishes to use the */etc/passwd* and */etc/group* files as the basis of authentication and authorization, Lustre can easily be configured to use these files.

We follow in the footsteps of the Samba and NFS v4 projects in using existing ACL structures, avoiding the definitions, development, and maintenance of new access control schemes.

Lustre implements process authorizations groups as they provide more security from root setuid attacks, provided hardened kernels are used.

New features of Lustre are file encryption, careful analysis of cross realm authentication and authorization issues and file I/O authorization.

**12.1.2. Taxonomy.** The first question facing us is what the threats are. The threats are security violations and Lustre tries to avoid:

(1) Unauthorized release of information.
(2) Unauthorized modification of information.
(3) Denial of resource usage.

The latter topic is only very partly addressed. Alternative taxonomies of violations and threats exist and include concepts such as suspicion, modification, conservation, confinement, and initialization. We refer to Satya's discussion. The DOD categorization of security might fit Lustre in at broadly the C2 level, controlled access protection, which includes auditing.

**12.1.3. Layering.** On the whole Lustre server software is charged with maintaining persistent copies of data and should largely be trusted. Clients can take countermeasures to avoid too much trust of servers by optionally sending only encrypted file data to the servers. While clients are much less controlled than servers, they carry important obligations for trust. For example, a compromised client might steal users passwords and render strong security useless.

The security subsystem has many layers. Our security model, like much else in Lustre, leverages on existing efforts and tries to limit implementation to genuinely new components. The discussion in this chapter uses the following division of responsibilities.

**Trust model:** When the system activates network interfaces for the purpose of filesystem request processing or when it accepts connections from clients, the interfaces or connections are assigned a GSS-API security interface. Examples of these are Kerberos 5, LIPKEY, and OPEN.

**Authentication:** When a user of the Lustre filesystem first identifies herself to the system, credentials for the user need to be established. Based on the credentials, GSS will establish a security context between the client and server.

**Group & user management:** Files have owners, group owners, and access control lists which make reference to identities and membership relations for groups and users of the system. Slightly different models apply within the local realm, where user and group id's can be assumed to have global validity and outside that domain where a different user and group database would be used on client systems.

**Authorization:** Before the filesystem grants access to data for consumption or modification, it must do an authorization check based on identity and access control lists.

**Cross realm usage:** When users from one administrative domain requires access to the filesystem in a different domain, a few new problems arise for which we propose systematic solutions.

**File encryption:** Lustre uses the encryption and sharing model proposed by the StorageTek secure filesystem, SFS [13], but a variety of refinements and variants have been proposed by CFS.

**Auditing:** For secure environments, auditing file accesses can be a major deterrent for abuse and invaluable to find perpetrators. Lustre can audit clients, meta-data servers, object storage targets, and access to the key granting services for user credentials and file encryption keys.

## 12.2. Lustre Networks and Security Policy

Network trust is of particular importance to Lustre to balance the requirements of a high performance filesystem with those of a globally secure filesystem. On the whole Lustre makes few requirements of trust on the network and can handle insecure and secure networks with different policies which will seriously affect performance. The aim is to identify those networks for which cryptographic activities can be avoided, in cases where more trust exists. An initial observation is that there are two extreme cases that should be covered:

**Cluster network:** Lustre is likely to be used in compute clusters over networks where:
(1) Network traffic is private to sender and receiver, i.e. it will not be used by 3rd parties.
(2) Network traffic is unaltered between sender and receiver.

**Other network:** On other networks, the trust level is much lower. No assumptions are made.

We realize that there are a variety of cases different from these two extremes that might merit special treatment. Such special treatment will be left to mechanisms outside Lustre. Examples of special treatment might be to use a VPN to connect a trusted group of client systems to Lustre with relaxed assumptions.

Lustre uses Portals. We will **not** change the Portals API to include features to address security. Instead, we will use Portals network interfaces to assign GSS security mechanisms to different streams of incoming events. Lustre will associate a security policy with a Lustre network. The policy is one of:

(1) No security
(2) GSS security with integrity
(3) GSS security with encrypted RPC data
(4) GSS security with data encrypted.

**12.2.1. Binding GSS Security to Portals Network Interfaces.** Incoming and outgoing Portals traffic uses an instance of a network abstraction layer (NAL). Such an instance is called a Portals Network Interface. Lustre binds a Portals Network Interface to traffic from a group of network endpoints called Lustre Network, which is identified by a netid. Certain networks are connectionless and it is less easy to intercept the traffic such as UDP, QSW or Myrinet networks. Once Portals binds to the interface, packets may arrive and will face a default security policy associated with the event queue as described above.

In the case of TCP/IP, a client (socket file descriptor based) connection cannot be made available to the server subsystems (MDS and OST) until it has been accepted. The accept is handled by a small auxiliary program called the acceptor. The basic acceptor functionality, as shown in figure XXX [need to add] is to accept the socket, determine from what lustre netid the client is connecting, and give the accepted socket to the kernel level Portals NAL. The Portals NAL then starts listening for traffic on the socket, and interacts with the portals library for packet delivery.

To summarize policy selection:

**Connectionless networks:** The security policy is associated with the interface at startup time, through configuration information passed to the kernel at setup time. If no configuration information is passed, a strong security backend is selected.

**Connection-based networks:** When the network has connections, the *acceptor* of connections decides what Portals NI will handle the connection. It thereby affects security decisions and assign a security policy to a connection.

We invoke the selected security policies before sending traffic or after receiving Portals events for arriving traffic. As shown in figure 12.2.1 this is easily done by using different Portals network interfaces and event queues. The event queues ultimately trigger the appropriate GSS backend to invoke traffic on an NI. Outgoing traffic is handled similarly at the level of a connectionless network interface. For TCP connections are made in the kernel for outgoing traffice and the kernel needs awareness of what security policy applies to a certain network. At present this is a configuration option, if the need arises it could be negotiated with the acceptor on the server.

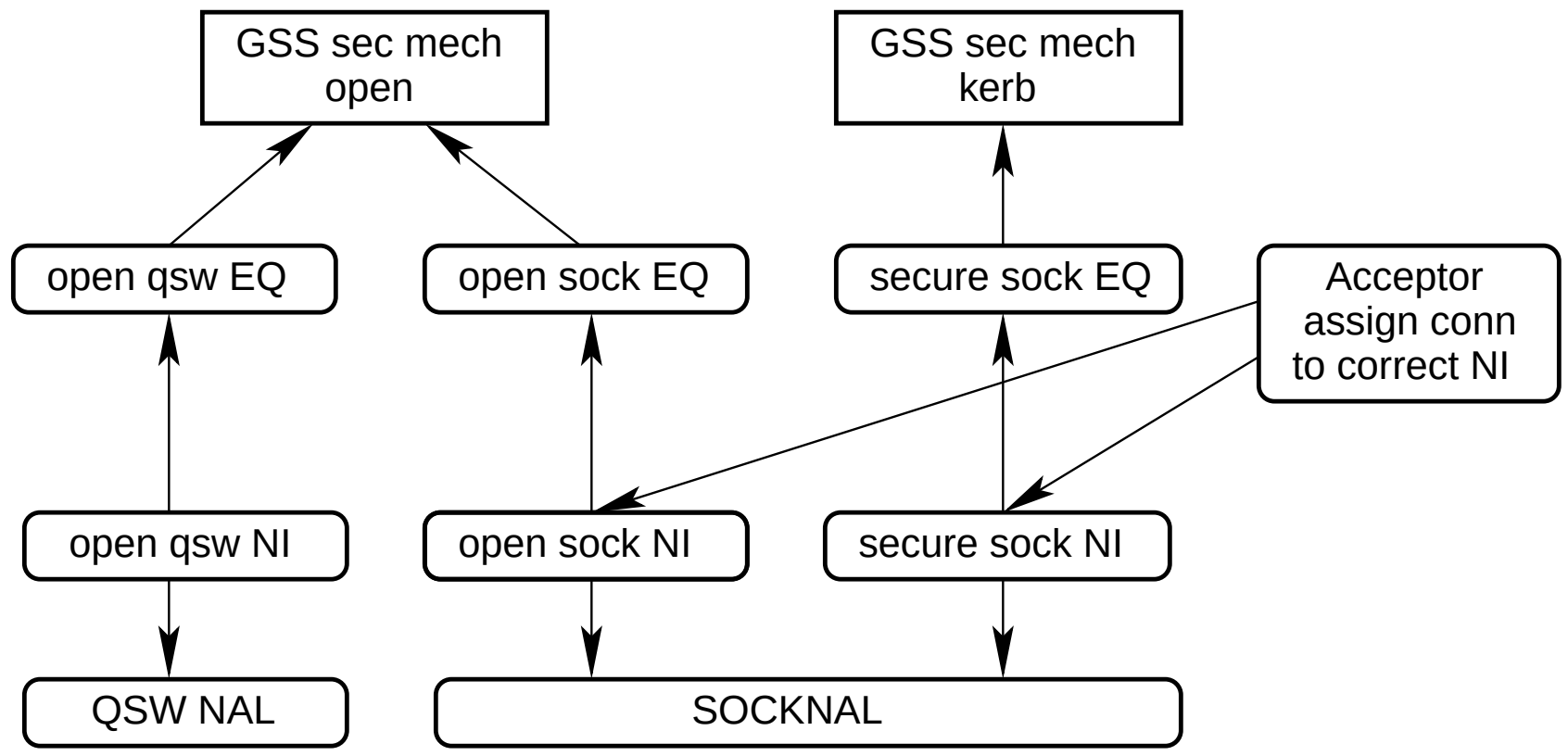

FIGURE 12.2.1. Interaction of Portals with GSS API

### 12.3. GSS Authentication & Trust Model

The critical question is to discuss what enforces security in Lustre and what is security enforcement? Lustre uses the GSS-API as a model for authentication and integrity of network traffic. Through the GSS API we can be sure that messages sent to server systems originate from users with proven identities, according to a GSS security policy installed at startup or connection acceptance time. The different levels of security arise from different GSS security backends. On trusted networks we need ones which are very efficient to avoid disturbing the high performance characteristics of the filesystem, but we also need to be prepared to run over insecure networks.

**12.3.1. The GSS-API.** The GSS-API provides for 3 important security features. Each of these mechanisms is used in a particular security context which the API establishes:

(1) The acquisition of credentials with a limited lifetime based on a principal or service name.
(2) Message integrity.
(3) Message privacy.

In order to do so, the GSSAPI is linked to a security mechanism. At present the GSSAPI offers the Kerberos 5 security mechanism as a backend. Typically the GSS-API is used as middle ware between the request processing layer and a backend security mechanism, as illustrated in figure 12.3.1.

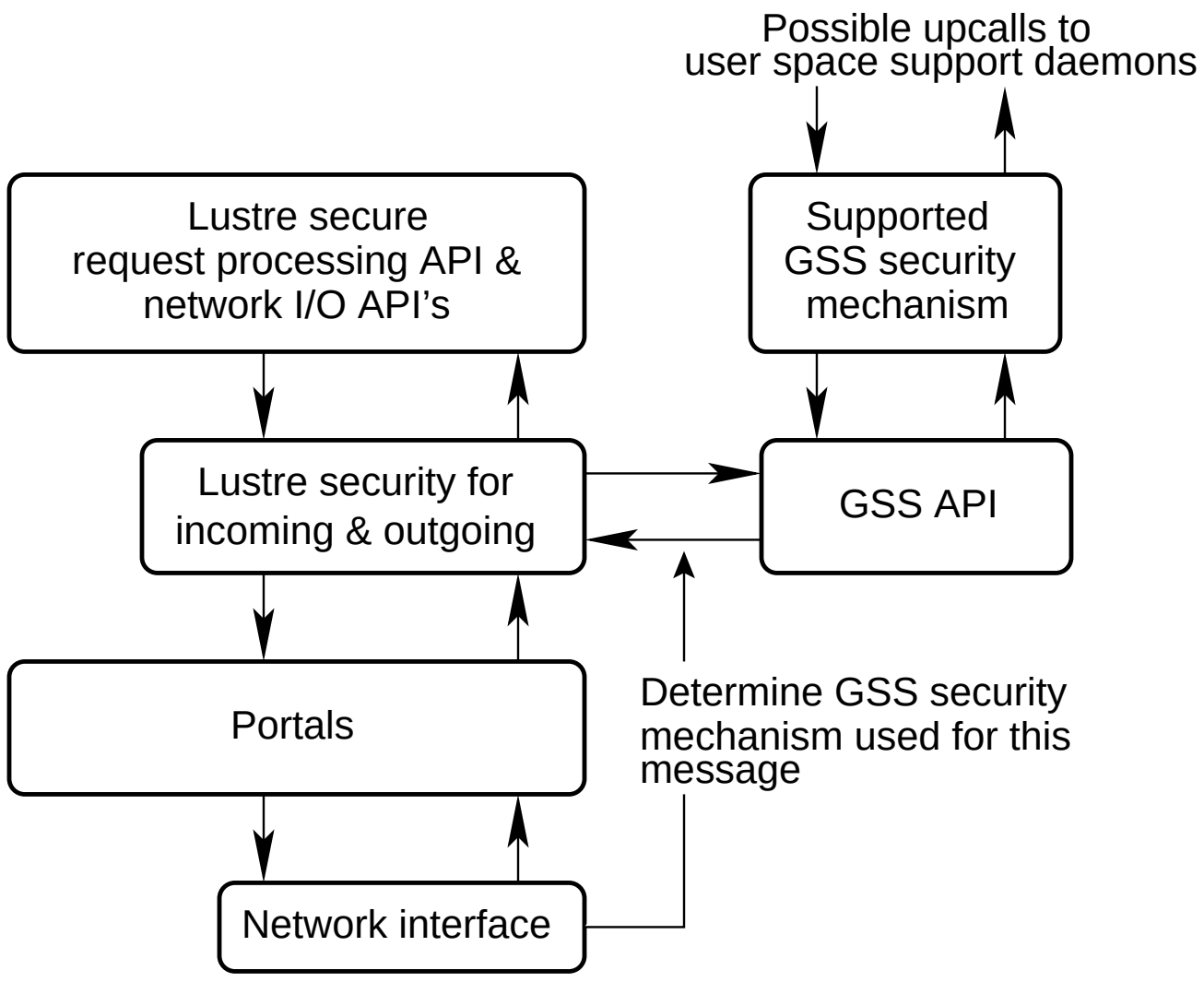

FIGURE 12.3.1. GSS API Security Protocol

For Lustre to use the GSS-API the following steps have been taken:

(1) Locate or build a kernel level implementation of the GSS-API with support for the TriLabs-required Kerberos 5 security mechanism. This can be obtained from the NFS v4 project. 0-copy properties of our networking API's cannot be preserved with that implementation, and changes have to be made to avoid the use of XDR and SUN RPC.
(2) Modify the Lustre request processing and network I/O API's to make use of the GSS API to provide their services. This will be original work requiring a fairly detailed design specification for peer scrutiny. The resulting API's will be similar to those provided by the RPCSEC_GSS-secure RPC API used in NFS v4. They will include various pieces of data returned by the GSS calls in the network packets.

**12.3.2. Removing credentials.** The kdestroy command will remove Kerberos credentials from the user level GSS daemon. However, we also need to provide a mechanism to flush the kernel cache of credentials. If this is not handled by the user level GSS daemon an lustre-unlog (ala kunlog for AFS) should be built.

**12.3.3. Special cases.** There are a few special connections that need to be maintained. The most important one is the family of MDS - OSS connections. The OSS should accept such connections and the MDS should have a permanently installed mechanism to provide GSS credentails to the authentication mechanisms. The OSS will treat the MDS principal as priviliged, just like some other utilities like backup, data migration and HSM software.

## 12.4. Process Identity and Authentication

Credentials should be acquired on the basis of a group of processes that can reasonably be expected to originate from the same authenticated principal. If that process group is determined by the user id of the process vulnerabilities can arise when unauthorized users can assume this uid.

One of the most critical security flaws of NFS is that a root user can setuid to any user and acquire the identity of this user for NFS authorization. In NFS v4 this is still the case - except that the uid for which su is performed, should have valid credentails.

The process authentication groups introduced by AFS can partly address this issue, however, it is only provides true protection on clients with hardened kernel software that make it difficult for the root user to change kernel memory. SELinux provides such capabilities. Without such, the extra security offered by PAGs is superficial and should not be provided.

PAGs may also help if processes under a single uid on a workstation arising from network logins may not be authenticated as a group. In environments where workstations provide strong authentication there may be no need for this, but pags can provide effective protection here.

**12.4.1. Process Authentication Groups.** Unix authorizes processes based on their *uid* - the *uid* defines a partition of the set of processes. Many distributed filesystems find this division of the processes too coarse to give effective protection; such systems introduce smaller Process Authentication Groups (PAG's).

A group of client processes can be tagged with a PAG. PAG's are organized to give processes that truly originate from a single authentication event the same PAG and all other processes a different PAG. This can separate processes into different PAG's even if the user *id* of the process is the same and it can bundle processes together that run under different user id's into the same PAG.

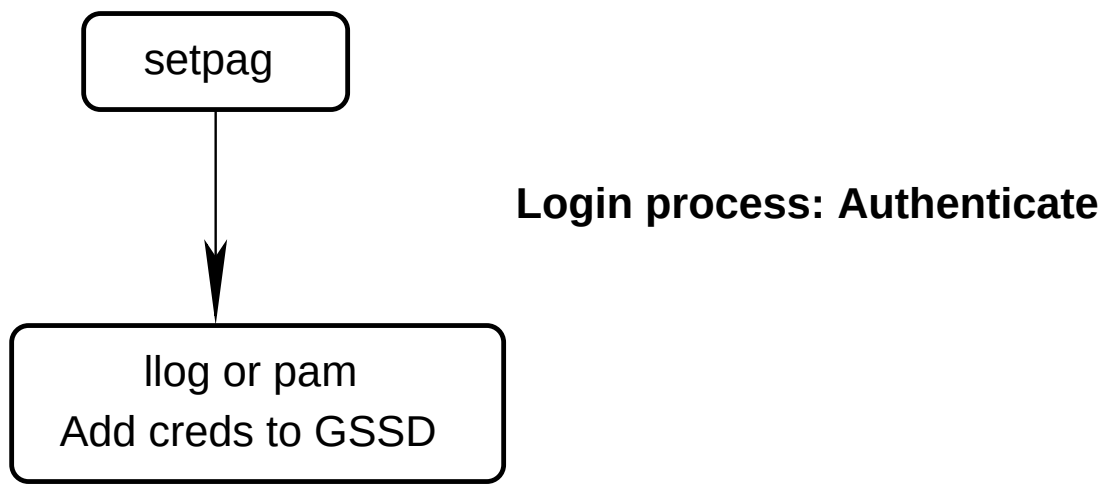

FIGURE 12.4.1. Login process

**12.4.2. Properties of a PAG.** The smaller group of processes for which authentication should give access is called a PAG, defined by the following:

(1) Every process should belong to a PAG.
(2) PAG's are inherited by fork.
(3) At boot time, init has a zero PAG.
(4) When a process executes a login-related operation (preferably through a PAM module), this login process would execute a "newpag" system call which places the process in a new PAG.
(5) Any process can execute *newpag* and thereby leave an authentication group of which it was a member.

**12.4.3. Implementation.** Lustre could implement a PAG as a 64 bit number associated with a process. Login operations will execute a setpag operation.

A Pluggable Authentication Module (PAM) associated with kinit and login procedures, or the *llog* program, can establish GSSAPI supported credentials with a user level GSS daemon during or after login. It is as this point that the PAG for these credentials should be well defined.

When the filesystem attempts to execute a filesystem operation for a PAG for which credentials are not yet known to the kernel, an upcall could be made to the GSS daemon to fetch credentials for the PAG. The Lustre system maintains a cache of security contexts hashed by PAG. A GSSAPI authentication handshake will provide credentials to the meta-data server and establish a security context for the session; this is illustrated in figure 12.4.2.

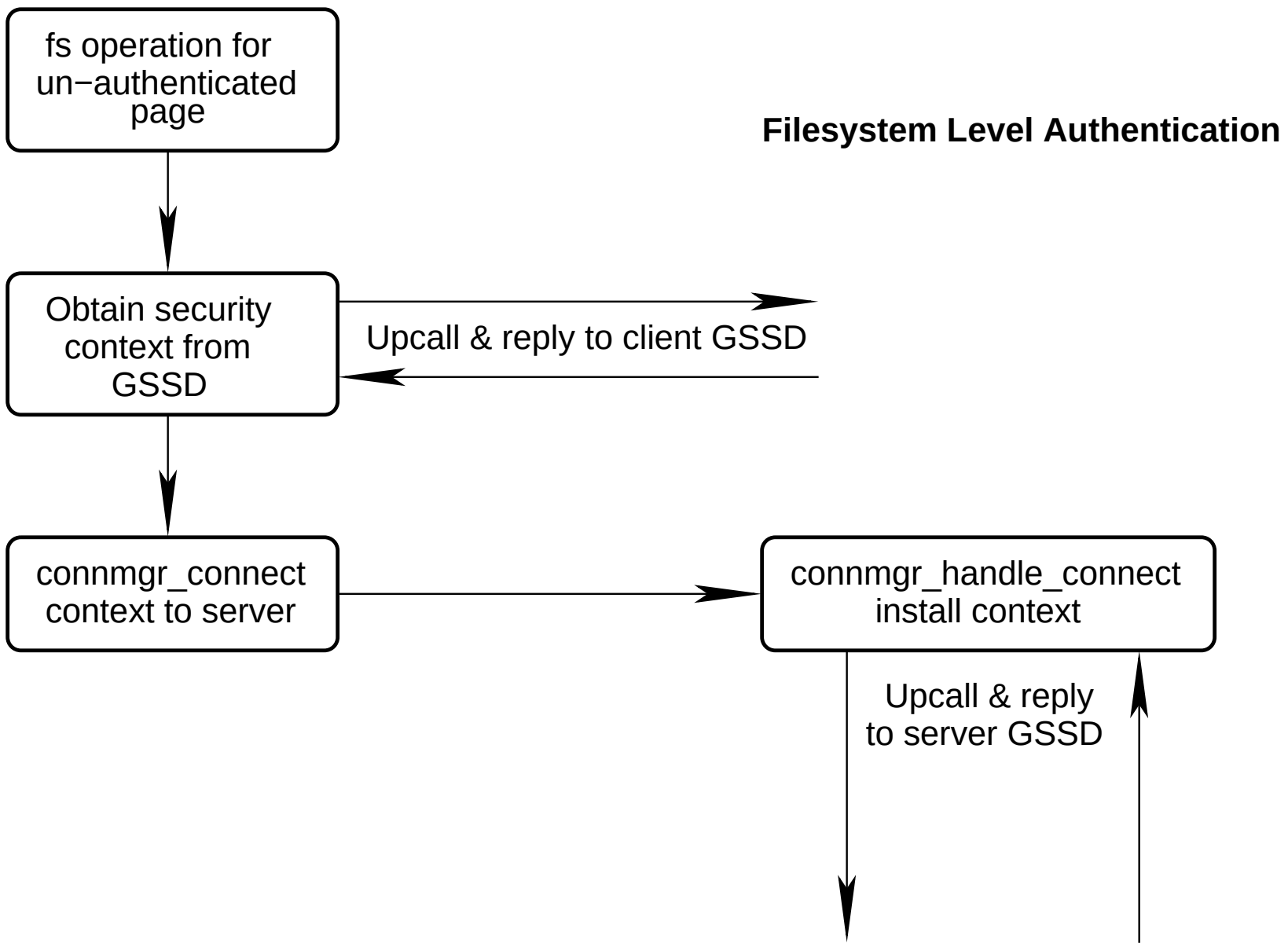

FIGURE 12.4.2. GSS API Authentication Handshake

Once the identity of the PAG has been established, both the client and the server will have user identities and group memberships associated with that identity. How those are handled will be discussed in the next section. Before authentication has taken place, a process only gets the credentials of the anonymous user.

#### 12.4.4. Alternatives.

12.4.4.1. *AFS implementation.* Design Note: The Andrew project used PAG's for AFS authentication. They were "hacked" in the sense that they used 2 fields in the groups array. Root can fairly easily change fields in the group array on some systems, but apart from that this implementation avoided changing the kernel.

The Andrew project called the system interface call "*setpag*", which was executed in terms of *getgroups* and *setgroups*.

12.4.4.2. *PAG and authentication and authorization data.* Probably a better way to proceed is to assign a data structure with security context and allow all processes in the same PAG to point to it and take a reference on the data structure. This authorization data would have room to store a list of credentials for use on different filesets and security operations. *newpag* will be a simple system call decreasing the refcount on the current PAG of a process and allocating a new one. We could use */proc/pags* to hold a list of PAG's.

## 12.5. The user and Group Databases

Lustre uses standard (default) user and group databases and interfaces to these databases, so that either enterprise scale LDAP NIS active directories can be queried or local */etc/passwd /etc/group* databases can be used.

Users and groups appear fundamentally in two forms to the filesystem:

(1) As identities of processes executing filesystem calls.
(2) As *user and group owners* of files, thereby influencing authorization.

Lustre assumes that within an administrative domain the results of querying for a user or group name or *id* will give consistent results. Lustre also assumes that some special groups and users are created in the authentication databases for use by the filesystem. These address the needs to deal with administrative users and to handle unknown remote users.

The user and group databases enter into the filesystem-related API's in just a few places:

**Client authorization:** The client filesystem will check group membership and identity of a process against the content of an ACL to enforce protection.

**Server authorization:** The server performs another authorization check. The server assumes the identity and group membership of client processes as determined by the security context. It sets the values of file and group owners before creation of new objects.

**Client filesystem utilities:** Utilities like ***ls*** require a means to translate user id's to names and query the user databases in the process. The filesystem has knowledge of the realm from which the inodes were obtained, but the system call interface provides no means to transfer this information to user level utilities.

As we will explain below when covering cross realm situations there is a fundamental mismatch between the two uses and the UNIX API's. Lustre's solution is presented in the next section.

Lustre security descriptor and the Current Protection Sub Domain

The fundamental question is **'Can agent X perform operation Y on object Z?**'. The protection domain is the collection of agents for which such a question can be asked. In Lustre, the protection domain consists of:

(1) Users and groups.
(2) Client, MDS, and OST systems.

For a particular user a current protection subdomain (CPSD) exists, which is the collection of all agents the user is a member of. This is shown in figure 12.5.1.

UNIX systems introduce a standard protection domain based on what the UNIX group membership and user identity are. These are obtained from the */etc/passwd* and */etc/group* files, or their network analogues through the NSS switch model. The UNIX task structure can embed this CPSD information in the task structure of a running process. A user process running with root permissions can use the *set(fs)uid, set(fs)gid,* and *setgroups* system calls, to change the CPSD information.

Things are more involved for a kernel level server system, to which a user has authenticated over the network. In that case, the kernel has to reach out to user space to fetch the membership information and cache it in the kernel to have knowledge of the CPSD. Such caches may need to be refreshed if the principal changes it's uid and is authorized to do so by the server systems. Lustre servers hold CPSD attributes associated with a principal in the Lustre Security Descriptor (LSD).

**12.5.1. Basic handling of users and groups in Lustre.** When a clients performs an authentication rpc with the server, the server will built a security descriptor for the principal. The security descriptor is obtained by an upcall. The upcall uses standard Unix API's to determine:

(1) the uid and principal group id associated with the username obtained from the principals name
(2) the group membership of this uid

This information is held and cached, with limited life time, in kerel server memory in the LSD structure associated with the security context for the principal. Other information that will be held in the LSD is information applicable in non-standard situations.

(1) The uid and principal group id of the principal on the client. If the client is not a local client with the same group and user database this is used as described in the next section
(2) Special server resident attributes of the principal for example:

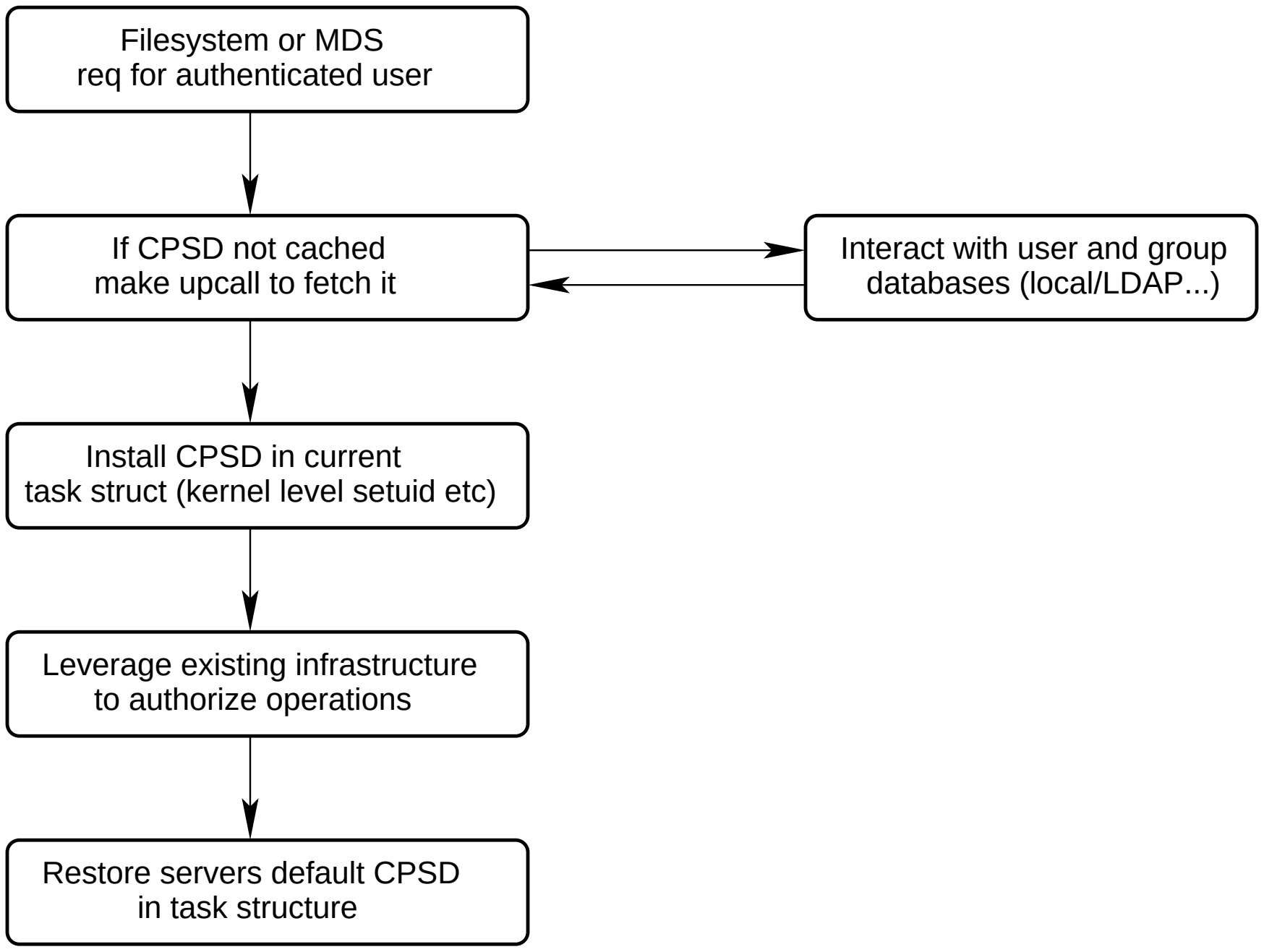

FIGURE 12.5.1. Obtaining the CPSD Information for a User

(a) Is the principal elegible for the server to respect setuid/setgid/setgroups information supplied by the client (these will only be honoured if the file system has an appropriate attribute also).
(b) Which group/uid values will be respected may be set?
(c) Is this principal able to access inodes by file identifier only (without a random cookie). This is needed for Lustre raid repair and certain client cache synchronizations.
(d) Should this principal get decryption keys for the files even when identies and ACL's would not provide these.
(e) Should this principal be able to restore backups (e.g. allow it to place encrypted files into the file system)

**12.5.2. Handling setuid, setgid and setgroups.** There are several alternative ways in which these issues can be handled in the context of network security for a file system.

12.5.2.1. *Priviliged principals.* A daemon offering GSS authenticated services can sometimes perform credential forwarding. Kerberos provides a way to forward credentials. This can provide excellent NFS v4 Lustre integration. This mechanism is external to Lustre.

When the service authenticates a user it can hand its credentials to the user level GSS daemon, which can use them to re-authenticate for furhter services. Therefore if Lustre requires a credential for a server process that has properly forwarded the credentials to the GSS daemon, it can transparently authenticate for this. Note that in this case the Lustre credential should be associated with the user id and the PAG (or optionally just with the user id).

If the server is a user level server, the setuid/setgid/setgroups calls can be intercepted to change the security descriptor associated with the process, in order for its credentials to be refreshed. If this is done the threat of root setuid, discussed above, is also eliminated.

12.5.2.2. *Forwarding credentials.* When an unmodified non GSS server is running on a Lustre client exporting file system information, there may be no facility for the server system to have access to credentials for the user. At the mimimum the principal would have to log in to the server system and provide authentication information, then the PAG system would have to be bypassed to allow the server to obtain the users credentials to become available to the servers PAG. AFS has recognized this a serious usability issue.

In order to not render Lustre unusable in this environment, a server resident capability can be associated with the triple: client, file system and principal. This capability will allow the client to forward user id, group id and setgroup arrays.

Extreme prejudice is required and by default no client, file system and principal has this capability.

## 12.6. Cross Realm Authentication and Authorization.

In global filesystems such as Lustre, filesets can be imported from different realms. The authentication problems associated with this are suitably solved by systems such as Kerberos.

A fundamental problem arises from the clash of the user *id* / group *id* namespaces used in the different realms. These problems are present in different forms on clients, where remote user id's need to be translated to sensible names in the absence of an UNIX API to do so. On servers adding a remote principal to an access control list or assigning ownership of a file object to a remote principal, the creation of a user *id* associated with that principal is required.

Lustre will address both problems transparently to users through the creation of local accounts. It will also have fileset options to not translate remote user id's, translate them lazily, or translate them synchronously to accommodate various use patterns of the filesystem.

**12.6.1. The Fundamental Problem in Cross Realm Authorization.** File ownership in UNIX is in terms of *uid*'s and *gid*'s. File ownership on UNIX in a cross realm environment has two fundamental issues:

(1) Clients need to find a textual representation of a user id.
(2) Servers need to store a *uid* as owner of an inode, even when they only have realm and remote user *id* available.

Utility invocation such as *ls -l file* issues a *stat* call to the kernel to retrieve the owner and user, and then use the C-library to issue a *getpwent* call to retrieve the textual representation of the user id.

The problem with this is that while the Lustre filesystem may have knowledge that the user name should be retrieved from a user database in a remote realm, the UNIX API has no mechanism to transfer this information to the application.

This is in contrast with the Windows API where files and users are identified by **SID's** which lie in a much larger namespace and which are endowed with a lookup function that can cross Windows domains (the function name to do so is *lookupaccountsid*).

When the filesystem spans multiple administrative domains, the Unix API's are not suitable to correctly identify a user.

A server cannot really make a remote user an owner/group owner of a file nor can it make ACL entries for such users, unless it can represent the remote user correctly in terms of the available user and group databases.

**12.6.2. Lustre handling of remote user id's.** When a connection is made to a Lustre metadata server the key question that arises is:

> Is the user / group database on the client identical to that on the server?

We call such a client *local* with respect to the servers. Lustre makes that decision as follows:

(1) The acceptor, used to accept TCP clients has a list of local networks. Clients initiating connection from a local network will be marked as local.
(2) There is a per fileset default that can overrides when the tcp decision is not present. This decision may not be present when clients on other networks connect.

Each lustre system, client and server, should have an account *nllu* "Non local Lustre user" installed. On the client this is made known to the file system as a mount option, on the server it is a similar startup option, part of the configuration log. On the client is is important that there is a name associated with the nllu user id, to make listings look attractive.

When the client connects and authenticates a user, it presents the client's uid of this user to the server. The client uid also presents kerberos identity of the user to the server, and this is used by the server to establish the server uid of the principal. For each client the server has a list of authenticated principals.

When the server handles a non-local client, it proceeds as follows for each uid that the server wants to transfer to the client or vice versa:

(1) If uid is handled by the server, and it is among the list of authenticated user id's translate it.
(2) All other uids are translated to the server or clients *nllu* user id.

**12.6.3. Limited manupilation of access control lists on non-local clients.** In order to provide an interface to ACL's from non local clients, group and user names must be given as text, for processing on the server. Lustre's lfs command will provide an interface to list and set ACL's. However, the normal system calls to change ACL's are not available for remote manipulation of ACL's.

**12.6.4. Solutions in Other Filesystems.**

**AFS:** We believe there is no work around for the *getpwent* issues in the AFS client filesystem.
The Andrew filesystem has a work-around for the fundamental problem on the server side. When users gain privileges in remote cells that require them to appear as owners of files or in access control lists, the *cklog* program can be used and creates an entry in the FileSystem Protection DataBase (**FSPDB**) recording user id's and group membership of the remote realm. The file server can now set owner, group owner, and ACL entries for the remote user correctly.
This creation is required only once, but allows the remote cell to treat a cross-realm user in an identical fashion as a local user. For details, see the AFS documentation [**12**].

**Windows:** The Windows filesystem stores user identities in a much larger field than a 32 bit integer and the fundamental problem does not exist in Windows. The Win32 function *lookupaccountsid* maps a security *id* to full information about the user, including the domain from where the *sid* originated. File owners are stored as *sid*'s on the disk.

**NFS v4:** This filesystem appears not to explicitly address this problem. NFS v4 transfers the file and group owners of inodes to the clients in terms of a string. On the whole this is a bad idea for scalability, as it forces the server to make numerous lookups of such names from userid's, even when such data is not necessarily going to be used.
If it is desirable to give clients textual information about users, they should probably interact with the user databases themselves to avoid generating a server bottleneck.

## 12.7. MDS Authorization: Access Control Lists

Our desire is to implement authorization through access control lists. The lists must give Linux Lustre users POSIX ACL semantics. Given that we handle cross realm users through the creation of local accounts for those users, we can rely on the POSIX ACL mechanisms. Lustre will use existing ACL mechanisms available in the Linux kernel and filesystems to authorize access.

This is the same mechanism used by Samba and NFSv4.

Good but not perfect compatibility has been established between CIFS, NFSv4, and POSIX ACL's. The subtle semantic differences between Windows, NFSv4, and POSIX ACL's can be further refined by adding such ACL handling to the filesystems supported by Lustre.

A secondary and separate "access" control list may be added to filesets that have enabled file encryption. This ACL will be handled separately after the POSIX ACL has granted access to the inode.

**12.7.1. Fid guessing.** During pathname traversal the client goes to the parent on the MDS, going through ACL's mode bits etc, to get its lookup authorized. When complete, the client is given the FID of the object it is looking for. If permisions on a parent of the fid change, a client may not be able to repeat this directory traversal. A well behaved client will drop the cached fid it obtained when it sees permission changes on any parents. To do so it uses the directory cache on the client.

A fid guessing attack consists of a rogue client re-using a fid previously obtained or obtained through guessing in order to start pathname traversal halfway through a pathname, at the location of the guessed fid. Protecting against access to MDS inodes through "fid guessing" is important in the case of restrictive permissions on a parent, and less strict permissions underneath.

To prevent this, the Lustre MDS generates a capability during lookup which allows the fid to be re-used for a short time upon presentation of the capability. Any fid based operation would fail unless the fid cookie is provided. This limites the exposure to rogue clients to a short interval, of which users should preferrably be aware.

12.7.1.1. *Alternatives:* NFS has made file handles "obscure" to achieve the same.

**12.7.2. Implementation Details.** A fundamental observation about access control lists is that typically there are a few access control lists per file owner, but thousands of files and directories with that owner. As a result it is not efficient, though widespread practice, to store a copy of the ACL's with each inode.

The Ext3 filesystem has implemented ACL's with an indirection scheme. We leverage that scheme on the server, but not yet on the client.

## 12.8. Auditing

Lustre uses a filter layer called smfs which can intercept all events happening in a filesystem and on OST's.

Auditing happens on all systems. Auditing on clients is necessary to record access to cached information which only the client filesystem can intercept at reasonable granularity; operations that result in RPC's are not cached for efficiency reasons. On the MDS systems, audit logs are perhaps the most important since they contain the first point of access to the file and directories. On the OSS's a summary audit log can be written, with a reference to the entry on the MDS that needs to be looked at in conjunction with this. For this the objects on the OSS carry a copy of the FID of the MDS inode.

Lustre will send this information to the syslog daemon. The granularity of the information logged will be tunable. A tool is available to combine the information obtained from servers and clients and to scan for anomalies.

A critical piece of information that needs to be logged on the OSS is the full file identitier of the MDS inode beloning with an object. Moreover, file inodes on the MDS should contain a pointer to parent directories to produce traceable pathnames.

**12.8.1. Alternatives.** Such mechanisms are described in Howard Gobioff's thesis [XXX] section 4.4.3.

## 12.9. SFS Style Encryption of File Data

The StorageTek SFS filesystem provides a very interesting way to store file data encrypted on disks, while enabling sharing of the data between organizations. SFS is briefly described in [**13**] and [**14**]. In this subsection, we review some of the SFS design a proposed integration with Lustre. We also provide a more light weight cryptographic file system capability that is much easier to implement.

**12.9.1. Encrypted File Data.** In SFS, file data can be encrypted. Each file has a unique random key, which is created at the time the file is created. It is stored with the file, but it is encrypted and a third party agent, called the group server must be access to provide the unencrypted file key. The key never changes, and remains attached with the file until the inode of the file is freed.

The file is encrypted with a technique called *countermode,* see [**15**], *section 2.4.5.* Countermode is a simple mechanism to encrypt an arbitrary extent in a file without overhead related to the offset at which the extent is located.

Ultimately this cryptographic information leads to a bit stream wich is used to x-or'd with the file data. Patches probably exist for Linux kernels to introduce counter mode encryption of files relatively easily.

**12.9.2. Creating a New File.** An information producer creates a new file and can define who can share this file. At the time of file creation the file is encrypted with a random key, and an access control list for the file is generated, granting access to the file. The group server is involved for two reasons:

(1) It encrypts the file key with a group server key.
(2) It signs the access control list, including the key, so that its integrity remains known.

The encrypted file key and the signed access control list are stored with the file.

**12.9.3. The SFS Access Control List.** SFS defines an **access control list**, which is perhaps an unfortunate term because it is more a sharing control description. We call the SFS access control list the SFS control list.

The SFS control list contains identity descriptors which contain a name of a **group** *(*confusingly called a **project** in the SFS literature) and the file key encrypted with a public key of a group server. Once an application has access to the inode, it can scan the SFS control list and present an identity to a group server, which then returns the key to the file. This description, taken from the SFS papers fails to address the issue of integrity of the ACL for which some measures must be taken.

A variety of more complicated identities can be added to the SFS control list. Escrow can be defined by entries that state that any K of N identities must be presented to the group server before the key will be released. There is also a mechanism for an identity to be recursive with respect to group servers and require more than one group server to decrypt before the key is presented.

In principle anyone who can modify the SFS control list of the file can add further entries defining groups managed by other group servers, by encrypting such entities with the public key of the group server, provided the group server permits this operation.

12.9.3.1. *The Group Services.* The user, or the filesystem on behalf of the user, presents an identity found in the access control list and the user credentials to a *group server.* Group server checks that the user is a member of the group and returns the un-encrypted key to the filesystem to allow it to decrypt the file.

The group server can build an audit trail of access to files.

The group server must be trusted since it can generate keys to all files that have an ACL entry encrypted with the public key of the group server.

The group acts a bit like a KDC, but it distributes file keys, not session keys.

Some aspects of the group service are the subject of a patent application filed by StorageTek.

12.9.3.2. *Weaknesses Noted.*

**Counter mode encryption:** This technique has some weaknesses, called *malleability*, but adding **mixing** can fix this. Mixing algorithms are worked on but will be patented. [see Rogaway as UCSC.]

**Access control:** The SFS access control lists have, at least theoretically, a weakness. While it is debatable if the system actually gives the key to a user, once the key has been given out to a user the user may retain access to the file data permanently. For a database file which remains in existence permanently, this is not an optimal situation.
Ordinary access control lists need to supplement the authorization. This will prevent unauthorized access to the file. However, a user with a key remains a more risky individual with respect to theft of the encrypted data.

**12.9.4. Lustre SFS.** Lustre provides hooks for a client node to invoke the services of the group key service as proposed by SFS. The SFS access control list will be stored in an extended attribute, in **addition to normal ACL's** discussed above. A key feature of this group server is that principals can manipulate the database, in contrast with system group databases, which usually allow only root to make any modifications.

Lustre also implements a simpler encryption scheme where the group key service runs in the MDS nodes. This scheme uses the normal ACL with an extended attribute to store the encrypted file encryption key. The MDS has access to the group server key, and provides the client with the unencrypted key after authorization for file read or write succeeds, based on the normal POSIX ACL. Lustre also has a server option on principals that allow decryption on certain client nodes, regardless of the ACL contents. It is recommend that the acquisition of credentials for such operations follows extremely secure authentication, such as multiple principals using specially crafted frontends to the GSS security daemons.

**12.9.5. Controlling encryption.** An MDS target can have a setting to have none, all or part of the files encrypted. When part of the files is encrypted, the user *lfs* can mark a directory subtree for encryption.

**12.9.6. Encrypting Directory Operations.** Encrypting directory data is a major challenge for filesystems. It appears possible to use a scheme like the SFS scheme to encrypt directory names. MDS directory inodes can hold an encrypted data encryption key that is used to encrypt & decrypt each entry in the directory.

Clients encrypt names so that the server can perform lookup on encrypted entries. The client receives encrypted directory entries and for directorly listings, the client performs decryption of the content of the directory.

## 12.10. File I/O authorization

**12.10.1. Capabilities to access objects.** The clients request the OSS to perform create/read/write/truncate/delete operations on objects. Truncate can probably be treated as write, particularly because Lustre already has append only inode flags to protect files from truncation. The goal is to efficiently authorize these operations, securily. This section contains the design for this functionality.

When a client wishes to perform an operation on an object it has to present a capability. The capability is prepared by the MDS when a file is opened and sent to a client. Properties of the capabilities are:

(1) They are signed with a key shared by the MDS and OSS
(2) Possibly specify a version of an object for which the capability is valid
(3) They specify the fid for which objects may be accessed
(4) They specify what operations are authorized.
(5) A validity time period is specified, assuming coarsely synchronous clocks between the MDS and OSS.
(6) The kerberos principal for which the capability is specified is included in the capability.

**12.10.2. Network is secure/insecure.** If the network is secure, capabilities cannot cannot be snooped off the wire so no network encryption is needed. However, normally capabilities have to be transmitted in an encrypted form between the MDS and the client and between the client and the OSS to avoid stealing the capability off the wire.

GSS can be used for that. If GSS authenticates each user to the OSS a particularly strong scenario is reached.

**12.10.3. Multiple principals.** If a single client perform I/O for multiple users, the client Lustre software establishes capabilities for each principal through MDS open. Ultimately the I/O hinges on a single capability still being valid.

**12.10.4. Revocability and trusted software on client.** If a malicious user is detected, all OSS's can refuse access through a "blacklist". This leads to immediate revocability.

If client software is trusted, clients will discard cached capabilities associated with files when permissions change, for example. Cached capabilities only exist if cache of open file handles is used. If software on clients cannot be trusted, a client may regain access to the file data as long as his credentials are valid.

This could be refined by immediately expiring capabilities on the OSS's, by propagating an object version number to the OSS's and including it in the capability. This would slow down setattr operations, but increase security.

**12.10.5. Corner cases.**

12.10.5.1. *Cache flushes.* Cache flushes can happend after a file is closed. If file inode capability cookies are replicated to objects, this can lead to problems, because a cache flush could encounter a -EWRONGCOOKIE error, but no open file handle is available to re-authorize the I/O. If cookies are replicated, when the file is closed the data needs to be flushed, postponing closes has proven to be very hard.

12.10.5.2. *Replay of open after recovery.* If files saw permission reduction changes while open, replay of open involves trusting clients to replay honestly, or including a signed capability to the client to replay open with pre-authorized access on the MDS. At present Lustre checks permission on replay again, so open replay may not be transparent and may cause client eviction.

12.10.5.3. *Client open file cache.* With the client open cache reauthorization after the initial open is possible but somewhat pointless. If the client software cannot be trusted data could be shared between processes on the client anyway. Lustre uses the client to re-authorize opens from the open cache.

12.10.5.4. *Write back cache.* With the write back cache, a client should be authorized to create inodes with objects and set initial cookies on the objects it creates. For the master OSS where the objects will finally go, such authorization should involve an MDS granted capability, for the cache OSS, the client can manage security.

12.10.5.5. *Pools.* Pools have a security parameter attached to them to authorize clients in a certain network to perform read, modify, create, delete operations on objects on a certain OST. This authorization is done as part of file open, create and unlink. The MDS will not grant capabilities to perform operations on objects not allowed by the pool descriptor.

## 12.11. Odds and Ends

**12.11.1. Recovery and the security server.** The security server provides GSS/kerberos (or other GSS services) and networked user/group database services to Lustre. This is 3rd party software and Lustre has not planned modifications to it to become failover proof. The following details the situation further:

The software will consists of:

(1) LDAP services, and here the client is the C library queries to that database partly through PAM modules and utilities.
(2) kerberos KDC, the client is the client and server GSS daemons and kerberos utilities like kinit, their library equivalents and PAM modules

The server parts of 1,2 can easily be made redundant as standard IP services. For 1 & 2, client server protocol failure recovery would consist of retries and transaction recovery code for the services. This recovery, for the protocols in 1 & 2, would be completely outside the scope of Lustre. It just _may_ exist already, but I doubt these protocols have good retry capabilities.

However, if Lustre components, MDS, OSS and clients fail and recover they will re-use these services appropriately to recover. In some cases Lustre's retry mechanisms may, by coincidence invoke appropriate rety on protocol 1 & 2.

**12.11.2. Renewing Session Keys.** Long running jobs need to renew their session keys. Lustre will contain sufficient error handling to refresh credentials from the user level GSS daemons transparently.

**12.11.3. Portability to Less Powerful Operating Systems.** When Lustre is running as a library on a system which may not have access to IP services, some restrictions in the security model are required. For example a GSS security backend running on a service node operating the job dispatcher should supply a context that can be used by all client systems. [XXX: Is this the Red Storm model? It does not fit BG/L.]

Every effort will be made to implement a single security infrastructure and treat such special cases as policies.

## 12.12. Summary Tables

| attribute/structure | system | containing data structure | notes |
|---|---|---|---|
| pag | client | current process & user level GSS daemon | a number belonging to a process gro authenticated process groups |
| ignore pag | client | super block | do not use pags, but use uid's to get |
| client remote | client | super block | treat this client as one in a remote do |
| gss context | client/MDS/OSS | associated with a principal and service export | a list of GSS supplied contexts assoc the export/import on servers/clients |
| LSD | MDS | associated with a principal and service export | describes attributes and policies for a principal using a file system target. |
| security policy handler | MDS/OSS | lustre network interface | methods describing the server securi on this network |
| root squash | MDS | MDS target descriptor | root identity is mapped one-way on t |
| target allow setid | MDS | MDS target descriptor | grant certain client/principals setid c |
| modify cookie | MDS | MDS target descriptor | when this flag is set, operations char mode bits or owners will cause the v modified on MDS and OSS's associ inodes. |
| principal allow setid | MDS | LSD | allow principal setid |
| principal setids | MDS | LSD | which id values can be set by a princ a permitted value) |
| client allow setid | MDS | LSD | when a principal is found that has se the client list is given to the MDS |
| local or remote domain | MDS | lustre connection | is this export to a local or remote cli the kernel by the acceptor upon conn set as a configured per server default with the network. A client may over request to be remote during connecti |
| client uid-gid/server uid-gid | MDS | LSD | client and server uids/gids associated security context, used for remote cli |
| Lustre unknown uid/gid | client/MDS | superblock/LDS | unknown user id to be used with this when translating server inode owner client owner/groups, used only with clients. |
| groups array | MDS | LSD | server cached group membership arr |
| mds directory inode cookie | MDS | mds inodes | random number to authorize fid base MDS inodes |
| inode crypt key | MDS | mds inodes | random file data encruption key, enc the group server key |
| parent fid | MDS | mds inodes | the fid of a parent inode of a file inod pathname reconstruction in audit log |
| file inode cookies | MDS/OSS | mds inodes/oss objects | random numbers enabling authoriza operations |
| file crypt master key | LGKS | lkgs memory | key to decrypt file encryption keys |
| MDS audit mask | MDS | target smfs filter file system | mask to describe what should be log |
| Client audit mask | client filter | superblock of filter | mask to describe what should be log |
| OSS audit mask | | target smfs filter file system | |
| | | | |

### 12.12.1. Data structures and variables.

### 12.12.2. Client configuration options.

**remote lustre user id::** *mount option rluid=<int>*
**remote lustre group id::** mount *option rlgid=<int>*
**client is remote::** mount option *remote*

**don’t use pag::** mount option *nopag*

#### 12.12.3. MDS configuration options.

| **feature** | **description** |
|---|---|
| **network secure** | **configuration option: a network with a given <*netid*> is secure: <neti secpol=<open\|GSS\|Integrity\|Encrypt>** |
| **principal db:** | **allow setid <principal> <allowed setuids> <allowed setgids> <allowed targets <allowed clients>** |
| | **allow decrypt <principal> <allowed targets> <allowed clients>** |
| | **allow fid access <principal> <allowed targets> <allowed clients>** |
| **target security parameters** | **no root squash <allowed targets> < allowed clients>** |
| **pool security parameters** | **<pool name> <client netid> <c\|m\|r\|u>** |
| **audit** | **audit <audit options> <what target> <what clients>** |
| **encrypt** | **encrypt <all\|none\|partial> <what target>** |
| **file encryption master key** | **<key>** |

What client descriptors are lists of *netid* and lists of *netid,nid,pid* triples.

**12.12.4. OSS configuration options.** The OSS needs a principal DB to grant the MDS and certain administrative users raw object access without cookies.

**princpal db:** allow raw <principal> <allowed targets> <allowed clients/mds>

#### 12.12.5. Extended attributes.

| **what** | **system** | **data structure** | **description** |
|---|---|---|---|
| dir cookie | MDS | MDS dir inode | random 64 numb |
| file read cookie | MDS/OSS | MDS file inodes & objects | Part of the capab |
| crypt key | MDS | encrypted MDS inodes | encrypted inode e |
| parent inode pointer | MDS | all MDS inodes | pointer to an dire |
| crypt subtree | MDS | MDS directory inodes | all file inodes und |
| | | | |

### 12.13. Changelog

**Version 4.0 (Sep 2005)** Peter J. Braam, update for security deliverable.

**Version 3.0 (Aug 2003)** Peter J. Braam, rewritten after security CDR

**Version 2.0 (Dec. 2002)** P.D. Innes - updated figures and text, added Changelog

**Version 1.0** P. Braam - original draft

CHAPTER 13

# Configuration Management

## 13.1. Introduction

This section gives an overview of Lustre configuration management; we will describe the various steps, the tools available, and explain how to use these tools with some examples. Lustre configuration involves a couple of steps: first is to create a configuration file to describe the system, second is to do the actual low level device configuration based on the configuration file created. We will discuss the layering of the drivers on the relevant systems: the client, OST, and MDS. On each of these a sequence of low level configuration commands are issued at startup time; this will be described. We will next describe the XML configuration files used, the LDAP schema for Lustre, and the tools available to create the XML files and do the Lustre device configuration.

The configuration file can be fed to a utility, ***lconf***, that does the low level device configuration using the ***lctl*** utility. The *lctl* utility can set up the various modules, start the various services required and do device specific configuration. This process is illustrated in figure 13.1.1.

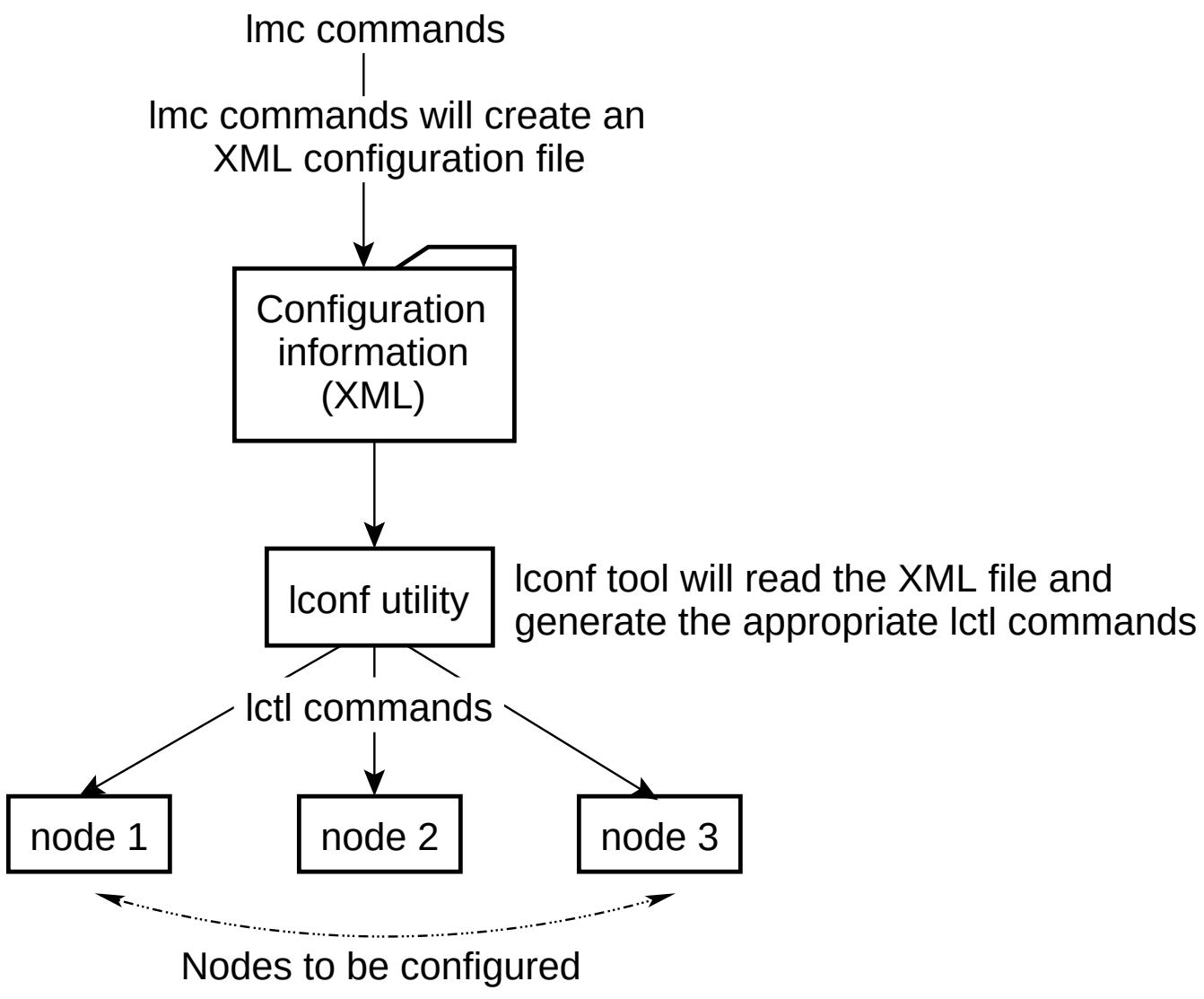

FIGURE 13.1.1. Lustre Configuration

In the subsequent sections, we will also describe SNMP and the use of Lustre configuration management infrastructure for monitoring and debugging.

## 13.2. Driver Organization

Drivers are introduced into the Linux kernel as loadable modules on client, MDS and OST systems. All systems run the networking stack, class driver and lock manager. Common modules are loaded in the order they appear below, using the Linux *modprobe* or *insmod* commands:

(1) **Portals**
(2) **ksocknal/rkqswnal:** Except on routing systems, one of these modules is loaded.
(3) **OBDclass**
(4) **ptlrpc**
(5) **ldlm**

The OST systems at the lowest level have the ExtN filesystem, a version of Ext3 which may have minor extensions. The filter drivers offer an object API to partitions which the kernel mounts (invisibly) as ExtN filesystems and the OST driver offers these as network accessible targets. The drivers are loaded in the following order:

(1) **ExtN**
(2) **OBDfilter**
(3) **OST**

The situation on the MDS is similar to that of the OST, but there is not yet a separate meta-data device driver (analogous to the OBDfilter driver) and meta-data target; they are integrated in the MDS driver. For recovery the MDS driver utilizes a small module, *fsfilt_extN*, with somewhat specialized functions using ExtN:

(1) **ExtN**
(2) **fsfilt_extN**
(3) **MDS**

On clients there is obviously the filesystem module, client drivers for the MDS and OST, and a logical volume driver for striping over OSC's:

(1) **OSC**
(2) **MDC**
(3) **LOV**
(4) **llite**

**13.2.1. Normal System Startup.** Lustre startup is relatively straightforward, involving the various drivers described above. Startup is organized by *lconf* which works its way through a variety of **levels**. The following diagram shows what subsystems are activated at what level and figure 13.2.1 illustrates the steps in a normal startup.

**10:** Network
**20:** Device, ldlm
**30:** OBD, MDD
**40:** MDS, OST
**50:** MDC, OSC
**60:** LOV
**70:** Mount-points

Upon startup:

**Clients:** Will contact MDS and OST systems.
**MDS:** Will contact OST systems.

Because any network failures, either during early connection establishment or during mount, will trigger retries and are not sensitive to order, cluster startup can proceed in parallel.

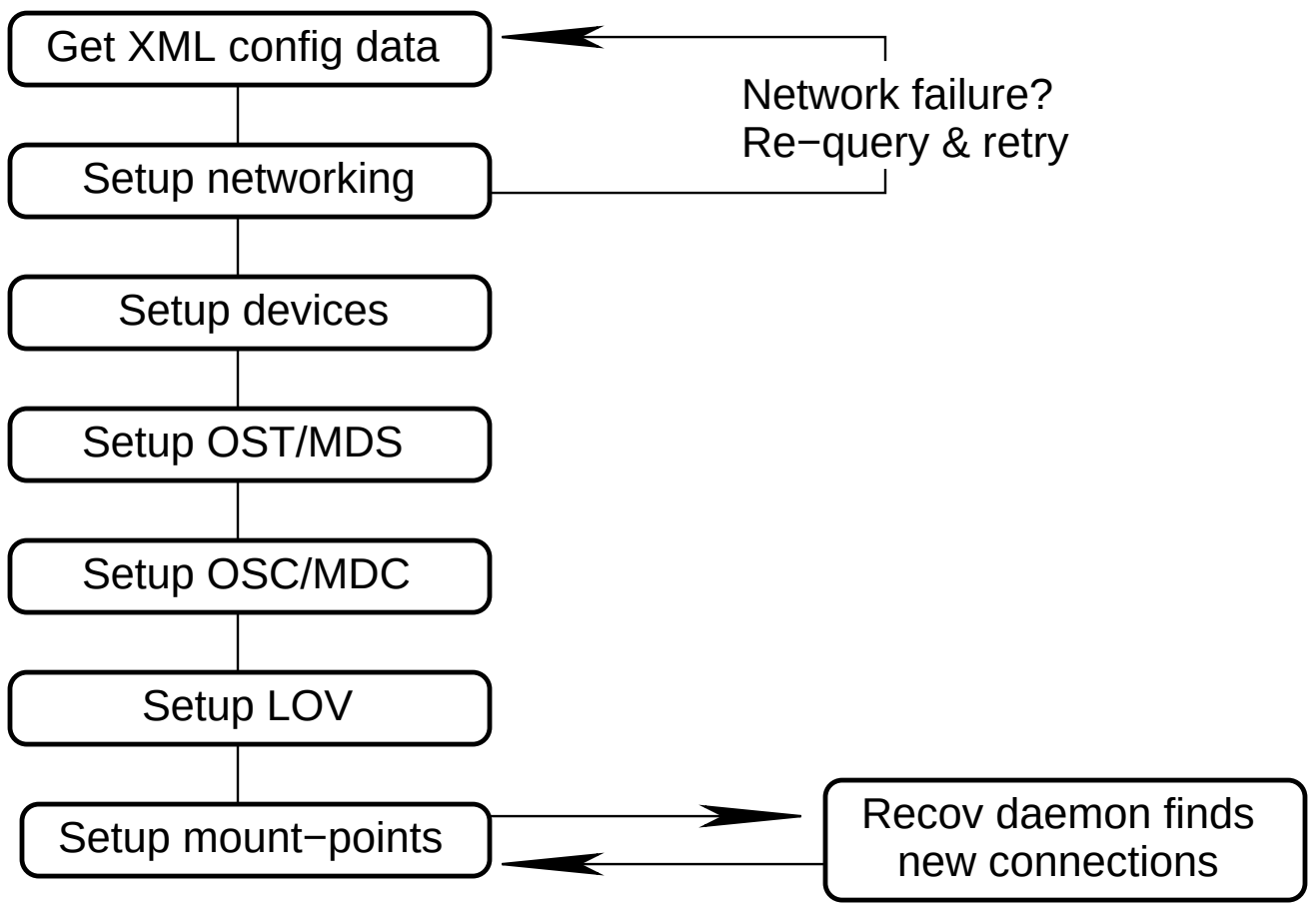

FIGURE 13.2.1. Normal System Startup

## 13.3. XML Configuration Files

To ease the process of configuring, we use eXtended Markup Language (XML) to describe the lustre configurations. Such XML configuration files can be easily generated using *lmc* and implemented using *lconf* . The use of XML configuration files can eliminate misconfiguration errors by checking the syntax against the Lustre configuration Data Type Definition (DTD). The configuration of a Lustre system proceeds by level:

**Level 1:** Network, all NAL's, and the systems own identity and routing tables.
**Level 2:** ldlm
**Level_3:** OBD's and a soon to come MDD
**Level 4:** OSC, OST
**Level_5:** MDC, MDS
**Level_6:** LOV
**Level 7:** Filesystems (mount-points)

Configuration parses the file and works down the elements by level when carrying out the configuration. An example of the DTD for Lustre can be found in chapter 34.

The top level entities in the XML file are usually compound structures, which follow one by one. For example, one of the entries is for an **OBD device**. It defines the name and UUID of the object device and the filesystem type required to mount it:

```
<!ELEMENT obd (obdtype|device|autoformat)>
<!ATTLIST obd
name #CDATA #REQUIRED
uuid #CDATA #REQUIRED
type #CDATA #REQUIRED>
```

So, an example OBD device configuration may read:

```
<obd name="obd-srv" uuid="obd-srv-UUID" type="obdfilter">
<fstype>extN</fstype>
<device size="10000">/dev/loop1</device>
<autoformat>no</autoformat>
</obd>
```

When the XML parser (*lconf* for Lustre) has seen the OBD elements in the XML configuration file, it invokes *obdcontrol* commands equivalent to:

```
lctl „ EOF
newdev
attach obdfilter obd-srv obd-srv-UUID
setup /dev/loop1 extN
quit
EOF
```

Another element in the XML file is an OSC; the OSC entry is as shown below:

```
<!ELEMENT osc (ost_ref obd_ref)>
<!ATTLIST osc
name #CDATA #REQUIRED
uuid #CDATA #REQUIRED>
```

An example OSC is the following:

```
<osc name="osc-bluearc1"
uuid="osc-bluearc1-UUID">
<obd_ref uuidref="bluearc1-OBD-UUID"/>
<ost_ref uuidref="bluearc1-OST-UUID"/>
</osc>
<ost name="bluearc1-OST"
uuid="bluearc1-OST-UUID">
<net_ref uuidref="bluearc1-NET-UUID"/>
<obd_ref uuidref="bluearc1-OBD-UUID"/>
</ost>
```

To set this up, the equivalent of the following control program is invoked:

```
lctl << EOF
newdev
attach osc osc-bluearc1 osc-bluearc1-UUID
setup bluearc1-OST-UUID bluearc1-network-UUID
quit
EOF
```

The *obd_connect* call now ships the service UUID (`BLUEARC1_OST_UUID`) to the node (`BLUEARC1_NETWORK_UUID`).

The network address for the node can be queried in the kernel with the *lustre_uuid2peer* function, that allows us to send the packet to the right station, then the handler for *obd_connect* will search the list of managed OST services to find the one with the `BLUEARC1_OST_UUID` UUID (which was given to it at attach time on the OST). The return for *obd_connect* will contain a "handle" (a pointer that can be sent secure over the net) for the "export" information of the target to the client. All further calls avoid any lookups on the server.

### 13.4. Lustre LDAP Schema

Configuration data can conveniently be stored in a LDAP repository, following a schema that matches that of the XML configuration descriptors. For small installations XML *config* file-based management is sufficient, but for larger systems we support LDAP-based configuration management. The LDAP-based configuration centralizes the configuration data for large clusters in one place. It serves three primary purposes:

(1) Centralize the management: Every system finds its configuration information in one place.
(2) Default profiles: Systems configure themselves according to the profile that is assigned to them. Particularly the client profile can be in use by very many systems, but does not lead to entries for every single system.
(3) Avoiding errors: Mis-configuring a storage system can have disastrous consequences. The record structure has been made robust with a variety of checks.

This scheme makes it a pleasant and doable task to configure a 10,000 node cluster and has different levels of abstractions so that on systems without disks, etc. a proxy/remote agent can translate high level *config* stuff into a stream of low level commands that can be executed without XML/LDAP, etc. There are several good references on LDAP:

(1) Some generalities `http://ldap-project.berkeley.edu/`.
(2) The most up to date and easiest to follow instructions for getting things working`http://www.openldap.org`.
(3) An excellent IBM book on LDAP`http://publib.boulder.ibm.com/pubs/pdfs/redbooks/sg244986.pdf`.

The root name of a Lustre LDAP schema is augmented by distinguished names for the objects. In some cases a directory structure is appropriate. The common names are guaranteed to be unique, but to limit the chance of accidental misconfiguration, easily achieved when renaming entries, we have included UUID's wherever entries are related to each other. The LDAP schema for Lustre is illustrated in chapter 34.

**13.4.1. Putting Together LDAP & *lctl*.** The two configuration systems are easily put together. A system queries the resource database to find the profiles that match its identity. With the profiles in hand, it retrieves the setup information of whatever needs configuration, converts it into an *lctl* script and configures as described in 13.5.

On every OST/MDS system, we store the node UUID of the system in a file. When the system starts up, it registers its network addresses in the NODE entry with the RDB and retrieves its profile. Any system can name itself like this. A named system has an entry in the resource database and can be configured with non-default profiles.

A Lustre client that reboots has no remaining Lustre state; we give the client a new Lustre client UUID each time it reboots.

MDS's and OST's that have lost a client remove its state when the client has no chance of recovering state with the MDS/OST anymore, ie. when the client has to start a completely new session with the servers. When a client reconnects to the cluster it can present its UUID. If the server system (MDS/OST) no longer has this UUID, it will enter the new one in the relevant databases (some of the MDS and OST databases related to transaction sequences and pre-allocated objects are persistent).

## 13.5. Configuration Management Tools

We provide several tools that make Lustre configuration easy.

**13.5.1. *LMC*.** *LMC* (Lustre Make Configuration) is a utility written in Python; it produces or merges XML descriptors. Each node in the system is identified using a **nodename**. The first step is to describe the networking at each node. After this, we need to describe each of the components of Lustre (MDS, OST and CFS) - the location of the component using the nodename, a name for the component and its size. There could be some component specific information required; for instance, for OST configuration we need to specify the LOV (if any), and for the client we need to specify the mount-point for the Lustre filesystem. LMC generates a XML-based configuration file based on these inputs.

More details about this utility can be found in the management pages.

**13.5.2. *LConf*.** *LConf* (Lustre Configuration) utility is another tool that helps in doing the low level configuration based upon the XML configuration file generated using *lmc*. The configuration information can be fed to *lconf* as an XML file or in the form of a URL that indicates the location of all the configuration information. If a system consists of multiple nodes, the nodename can be specified in *lconf*. Based on the nodename specified or using the default nodename, *lconf* determines the configuration information for the node being configured. Then it invokes *lctl* to do the low level device specific configuration.This utility automatically loads all the modules needed for the Lustre system. This utility is described in further details in the management page.

**13.5.3. *LCtl*.** *LCtl* (Lustre Control) is a low level *config* tool that can be used for configuration as well as troubleshooting/debugging. This utility eats low level text instructions, installs kernel configuration for everything, and can mount a Lustre filesystem. It is invoked by the *lconf* tool described previously. The management page describes this utility in detail along with several examples of its usage.

**13.5.4. *LFS* (Following the AFS/Coda fs/Cfs Tradition).** *LFS* is a tool to do Lustre-specific FileSystem operations, e.g.:

***lfs_showobj <file>*:** Give information about data object in a file.
***lfs_setacl*:** Set access control list.

## 13.6. SNMP

We see at least two interesting ways in which this can interact with SNMP (Simple Network Management Protocol):

(1) ***/proc* entries for agents:** When *lctl* enables a device, it will make some default *proc* entries, which reflect its configuration. A device specific callback then adds further files/fields to the /proc/lustre/devices/<devname>/ directory that can be monitored through the agent of an SNMP monitor.
(2) **Consistency check:** SNMP monitoring can use a "dump" produced by *lconf* to watch the cluster and generate traps when desired *config* information is no longer visible according to */proc*.

## 13.7. Dynamic Reconfiguration

We don't plan on implementing this immediately, but the following features will become available:

(1) **Kernel Traps:** Something is timing out or no longer working. The trap causes a re-query of LDAP for a new MDS server, for example.
(2) **Suspend/Replace Device:** *lctl* will have a *suspend* command that devices can optionally honor. This will allow a device to "empty itself", and queue up new requests while being reconfigured. Example: install a migrator.
(3) **Client Force:** We will be able to force a client to re-fetch its configuration and execute "explicit changes" in it. Example: remove a dead array from the cluster.

## 13.8. Configuration and High Availability

Lustre clients can transparently perform a failover from one server to another. The actions that take place at the level of configuration management are described here.

The failover service is responsible for designating an active server. When the service fails this can be detected by a client or by the failover cluster service itself and, either way, a transition is forced to failover the service to a new active server. When the failover completes the migration of the service to the new node, it registers the identity in LDAP. Further queries for the service UUID in LDAP will give the new server.

A client affected by a failed service will time out in one of its requests and the connection manager thread makes an upcall to a user level helper application. The helper application will query for a the name of the new active server, find its network address, and register it with the kernel code. From here forward the filesystem can recover pending transactions, against the new failover service.

## 13.9. Using Lustre for Monitoring and Configuration

Filesystems such as AFS, Coda, and InterMezzo are essentially self-configuring. Lustre could go beyond that by not only importing configuration data almost invisibly through the global namespace mechanism, as discussed in the Global Namespace Chapter, but also export monitoring and debugging information through Lustre.

An ultimate goal could be that:

(1) A cluster resource database exports a filesystem, over a default network, like IP that contains the information required to set up other networks and filesets. The schema currently implemented in XML could easily be exported systematically with a file tree of configuration information. After obtaining this filesystem as a default mount (perhaps this should be the root fileset), the client could proceed to configure other mounts.
(2) Each client exports monitoring, debugging, and performance information as a Lustre export of a filesystem. In order to do this efficiently, we will want a *mds_ost_filter* module that exports an existing filesystem (or a subtree thereof) through a filter driver.

(3) A management station can build a namespace of Lustre exports of all clients using the usual mechanisms.
(4) The client can export a part of */proc/sys* as a *read-write* filesystem through Lustre and allow centrally-managed tuning by writing to files in this area from the management station.

We find this idea very attractive, but it will require a great deal of confidence in Lustre as well as some new infrastructure.

## 13.10. Changelog

**Version 4.0 (Dec. 2002)**

(1) P.D. Innes - updated text & figure, Changelog formatted

**Version 3.0 (Oct. 2002)**

(1) P. Braam - chapter rewrite

**Version 2.0 (Jul. 2002)**

(1) P.D. Innes - proofed, edited, & spell-checked, added figure float

**Version 1.5 (Jun. 2002)**

(1) P. Braam - beginning of details on XML configuration and relationship to networking

**Version 1.0 (Apr. 2002)**

(1) P. Braam - original draft, introductory notes on LDAP

CHAPTER 14

# Configuration Management

## 14.1. Introduction

This section describes the design Lustre configuration management. We will first discuss the layering of the drivers on the relevant systems, the client, OST, MDS. On each of these a sequence of low level configuration commands is issued at startup time and this is covered next. We describe how the configuration is specified in XML and what low level API is used to configure the drivers and devices. Finally there is a basic API between the systems and the resource database which we address in the final section.

## 14.2. Driver Organization

Drivers are introduced into the Linux kernel as loadable modules on client, MDS and OST systems. All systems run the networking stack, class driver and lock manager. Common modules are loaded in the order they appear below, using the Linux *modprobe* or *insmod* commands.

**portals:**
**ksocknal/rkqswnal:** except on routing systems, one of these modules is loaded
**obdclass:**
**ptlrpc:**
**ldlm:**

The OST systems at the lowest level have the ExtN file system, a version of Ext3 which may have minor extensions. The filter drivers offer and object api to partitions which the kernel mounts (invisibly) as ExtN file systems and finally the ost (object storage target) driver offers these as network accessible targets. The drivers are loaded in the following order:

**extN:**
**obdfilter:**
**ost:**

The situation on the MDS is similar to that of the OST, but there is not yet a separate metadata device driver (analogous to the obdfilter driver) and metadata target, they are integrated in the mds driver. For recovery the mds driver utilizes a small module fsfilt_extN with somewhat specialized functions using extN.

**extN:**

**fsfilt_extN:**
**mds:**

On clients there is obviously the file system module, client drivers for the MDS and OST and a logival volume driver for striping over OSC's.

**osc:**
**mdc:**
**lov:**
**llite:**

## 14.3. Low Level Configuration Command API

**14.3.1. Block device preparation.** Each OST and MDS requires a block device to export a target. The block device must be formatted as an *ext3* file system. There are no further constraints on the device, e.g. using IDE/SCSI disk partitions, logical volumes, RAID'd devices, loop devices or ramdisks is possible. The result of this operation is that a certain device name is available for use by the MDS or OST. An OST device requires no further configuration at this phase.

To configure an MDS to use a logical volume, the logical volume configuration must be stored on the MDS device. This is done in an *lctl* command session as follows:

Start lctl from the shell:

(1) start an lctl session
**lctl:**
(2) A free device number is found.
**newdev:** find the first free device number
**device <number>:** use the given number as the device
(3) Attache the device to the MDS:
**attach mds <anyname> <anyuuid>:**
(4) Setup the device to use the formatted block device described above.
**setup <mds block device pathname> extN:**
(5) Connect
**connect:**
(6) Store the LOV target list on the MDS device, for later use by the file system:
**lov_setconfig lov-uuid stripe-count stripe-size offset pattern UUID1 [UUID2...]:**
This writes the UUID of the lov to be used to the disk, as well as thedefault number of stripes to be used, the stripesize in bytes and theoffset number of the first stripe. Pattern is currently ignored and leads to striping over all devices. If a stripe width of 1 is given, then the object is written to a single OST. The list of UUID's following the initial parameters are the UUID's of the OSC devices to be used. At least one must be passed.
(7) Disconnect
**disconnect:**
(8) Cleanup

**cleanup:**

(9) Detach

**detach:**

(10) Quit

**quit:**

**14.3.2. Lustre Network Configuration.** Network configuration can proceed when the

(1) portals
(2) rqswnal / ksocknal

modules are loaded.

If the socknal is used an OST and MDS must start the *acceptor* program, which accepts hands TCP connections, established at user level to the the kernel. Options to the acceptor are:

If one system is required to initiate commands to another, for example a client to an MDS and OST, or the MDS to the OST, then this system must establish a connection. The system needs to do this for every system it connects with. The configuration of connectsion proceeds in an *lctl sessoin.*

(1) Start lctl

**lctl:**

**Network configuration::**

**network [tcp/elan/myrinet]:** The commands that follow apply to net.
**connect<hostname port> | <id>:** For tcp/elan connect to a remote NID.
**disconnect <hostnid>:** Disconnect from a remote NID.
**add_uuid <name> <uuid> <hostnid>:** Associate a name/UUID with a NID.
**close_uuid uuid:** Disconnect a UUID.
**del_uuid uuid:** Remove a UUID association.
**mynid NID:** Inform the socknal of the local NID.
**add_route:** TargetNID gatewayNID niID add an entry to the routing table.
**del_route:** TargetNID delete an entry from the routing table.
**print_routes:** Print the routing table recv_mem size Set socket receive buffer size.
**send_mem size:** Set socket send buffer size.
**nagle:** Nagle [on/off] enable/Disable.

**14.3.3. Lustre Device Configuration.** Configuration of drivers proceeds generally in two steps. Lustre offers *devices* each device has a *type* controlled by the *driver*

The modules must be loaded (future versions may employ autoloading of modules). For each lustre device configuration is done with the *lctl* utility. The utility is started from the shell and executes 3 commands.

(1) Start lctl from the shell: start an lctl session

**lctl:**

(2) A free device number is found.

**newdev:** find the first free device number
**device <number>:** use the given number as the device

(3) In the lctl session a driver is attached to a new device. A name and UUID is given to identify and uniquify a device:
**attach <type> <device name> <device uuid>:**
The device name and uuid are entered as text. There is a maximum of 36 characters available. The device type is one of the following:
(a) ptlrpc
(b) ldlm
(c) obdfilter
(d) osc
(e) ost
(f) mdc
(g) mds
(h) lov

(4) A setup command is issued for the device:
**setup <arguments...>:**
The arguments given to setup differ per device type and are as follows.
**ptlrpc:** no arguments
**ldlm:** no arguments. The setup commands starts service threads for lock callback service, and is only required on client systems
**obdfilter: <device> <filesystem type>:** Currently the only file system supported is extN.
**osc: <OBDFILTER-UUID> <OST-server-UUID>:**
**mdc: <MDS-UUID> <MDS-server-UUID>:**
**mds: <device> <fstype>:** only extN is fully supported at present.
**ost: <obdfilter-UUID>:**
**lov::** no arguments

(5) Quit the lctl session
**quit:**

## 14.4. XML Configuration

The number of Lustre network connections and devices that need configuration is oftne prohibitively large to allow low level configuration. The low level configuration can also not directly benefit from the relational structure among the systems to be configured. To ease the process of configuring Lustre clusters configurations can be described in XML. Such configurations can be generated with the *lmc* utility and such configuration files can be implemented on a system by lconf.

**14.4.1. The XML configuration DTD.** A simple DTD describes entries that can be made in the XML configuration file. The DTD describes valid configuration entries in a Lustre configuration file, and they are easily seen to match up precisely with the low level configuration.

```
<!-- Lustre Management DTD -->
<!-- basic entities -->
<!ENTITY % tag.content "(#PCDATA)">
<!ENTITY % tag.ref "num CDATA #IMPLIED name CDATA #IMPLIED uuidref CDATA #REQUIRED">
<!ENTITY % tag.attr "   name CDATA #REQUIRED   uuid CDATA #REQUIRED">
<!-- main elements -->
<!ELEMENT lustre (node | profile | mountpoint | ldlm | mds | mdc | obd | ost | osc | lov|router)*>
<!ELEMENT profile (service_ref)*>
<!ATTLIST profile %tag.attr;>
<!ELEMENT mountpoint (path | fileset | mdc_ref | osc_ref)*>
<!ATTLIST mountpoint %tag.attr;>
<!ELEMENT node (network*, profile_ref)>
<!ATTLIST node %tag.attr;>
<!ELEMENT ldlm EMPTY>
<!ATTLIST ldlm %tag.attr;>
<!ELEMENT obd (fstype | device | autoformat)*>
<!ATTLIST obd %tag.attr; type (obdfilter | obdext2 | obdecho) 'obdfilter'>
<!ELEMENT ost (network_ref | obd_ref | failover_ref)*>
<!ATTLIST ost %tag.attr;>
<!ELEMENT mds (network_ref | fstype | device | autoformat | server_ref | failover_ref)*>
<!ATTLIST mds %tag.attr;>
<!ELEMENT osc (network_ref | obd_ref)*>
<!ATTLIST osc %tag.attr;>
<!ELEMENT mdc (network_ref | mds_ref)*>
<!ATTLIST mdc %tag.attr;>
<!ELEMENT lov (devices | mdc_ref)*>
<!ATTLIST lov %tag.attr;>
<!ELEMENT devices (osc_ref)+>
<!ATTLIST devices stripesize CDATA #REQUIRED stripeoffset CDATA #REQUIRED                   pattern    CD
<!ELEMENT router (misc)*>
<!ATTLIST router %tag.attr;>
<!-- basic elements -->
<!ELEMENT network (server | port)*>
<!ATTLIST network type (tcp | elan | myrinet) 'tcp' %tag.attr;>
<!ELEMENT fstype       %tag.content;>
<!ELEMENT device       %tag.content;>
<!ATTLIST device       size CDATA #IMPLIED>
<!ELEMENT server       %tag.content;>
<!ELEMENT port         %tag.content;>
<!ELEMENT autoformat   %tag.content;>
<!ELEMENT path         %tag.content;>
<!ELEMENT fileset      %tag.content;>
<!-- id tag elements -->
<!ELEMENT network_ref   %tag.content;>
<!ATTLIST network_ref   %tag.ref;>
<!ELEMENT profile_ref   %tag.content;>
<!ATTLIST profile_ref   %tag.ref;>
<!ELEMENT obd_ref       %tag.content;>
<!ATTLIST obd_ref       %tag.ref;>
<!ELEMENT mds_ref       %tag.content;>
```

```
<!ATTLIST mds_ref        %tag.ref;>
<!ELEMENT osc_ref        %tag.content;>
<!ATTLIST osc_ref        %tag.ref;>
<!ELEMENT lov_ref        %tag.content;>
<!ATTLIST lov_ref        %tag.ref;>
<!ELEMENT mdc_ref        %tag.content;>
<!ATTLIST mdc_ref        %tag.ref;>
<!ELEMENT mountpoint_ref %tag.content;>
<!ATTLIST mountpoint_ref %tag.ref;>
<!ELEMENT service_ref    %tag.content;>
<!ATTLIST service_ref    %tag.ref;>
<!ELEMENT server_ref     %tag.content;>
<!ATTLIST server_ref     %tag.ref;>
<!ELEMENT failover_ref   %tag.content;>
<!ATTLIST failover_ref   %tag.ref;>
```

**14.4.2. lmc.** LMC (Lustre Make Configuration) produces lustre configurations in XML format. It is intended to be from a script generated by system adminstrator to create a lustre desrcriptor for a cluster. Each element of a configuration requires a seperate invocation of LMC, there for a large cluster may require a hundred or more calls to LMC.

14.4.2.1. *Commands:*

(1) **–node ”node_name”** By itself, this will create a new node. This is also used to specify a specific node for other elements, and the
(2) **–net host_name net_type [port]** Adds a network interface to a node. If host_name is ’*’, then the local address while be substituted when the node is configured with lconf. This requires a **–node** argument.
(3) **–mds mds_name** Adds an MDS to the specified node. This requires a **–node** argument.
(4) **–lov lov_name mds_name stripe_sz stripe_off pattern** Creates an LOV with the specified parameters. The mds_name must already exist in the descriptor. When used as an argument to other commands, only the lov_name is specified.
(5) **–ost device [size]** Creates an OBD, OST, and OSC. The OST and OBD are created on the specified node. If an lov option is used, then the OSC will be added to the LOV. Requires: **–node.**
(6) **–mtpt /mnt/path mds_name lov_name|osc_name** Creates a mount point on the specified node. Either an LOV or OSC name can be used. Requires: **–node**

**Options**:

(1) **–output filename** Sends output to the file. If the file exists, it will be overwritten.
(2) **–merge filename** Add the new element to an existing file.
(3) **–format** Turns on autoformat. The block devices for MDS and OBD will be formatted if necessary.

14.4.2.2. *Example, lov.sh.* This example script (found in the tests directory in the lustre source distribution) creates a simple one node LOV configuration. All the lustre devices are configured on one node, and the LOV has two OBDs attached.

```
#!/bin/bash
config=lov.xml
LMC=../utils/lmc
# create nodes
${LMC} -o $config --node localhost --net localhost tcp
# configure mds server
${LMC} -m $config --format --node localhost --mds mds1 /tmp/mds1 50000
# configure ost
${LMC} -m $config --lov lov1 mds1 4096 0 0 ${LMC} -m $config --format --node localhost
${LMC} -m $config --format --node localhost --lov lov1 --ost /tmp/ost2 100000
# create client config
${LMC} -m $config  --node localhost --mtpt /mnt/lustre mds1 lov1
```

**14.4.3. lconf.** LCONF configures one node using the XML file created by LMC. It perfoms all the required steps to configure a lustre node, including loading kernel modules and preparing block devices, if necesary. LCONF has one required argument, the xml file.

14.4.3.1. *Options.*

(1) **–node node_name** Specifiy a specific node to configure. By default, LCONF will search for nodes with the local hostname and 'localhost'. When –node is used, only node_name is searched for. If a matching node is not found in the config, then LCONF exits with an error.
(2) **–cleanup** Unconfigure a node. The same config and –node argument used for configuration needs to be used for cleanup as well. This will attempt to undo all of the configuration steps done by LCONF, including unloading the kernel modules.
(3) **–noexec** Print, but don't execute, the steps lconf will perform. This is useful for debugging a configuration, and when used with –node, can be run on any host.
(4) **–gdb** Causes LCONF to print a message and pause for 5 seconds after creating a gdb module script and before doing any lustre configuration. (The gdb module script is always created, however.)

## 14.5. Normal System Startup

Lustre startup is relatively straightforward, but does involve many drivers. Startup is organized by *lconf* which works its way through a variety of *levels*. The following diagram shows what subsystems are activated at what level.

**10:** network
**20:** device, ldlm
**30:** obd, mdd

**40:** mds, ost
**50:** mdc, osc
**60:** lov
**70:** mountpoints

Upon startup:

**clients:** will contact MDS and OST systems,
**mds:** will contact OST systems

Because any network failures, either during early connection establisment or during mount, will trigger retries and are not sensitive to order, cluster startup can proceed in parallel.

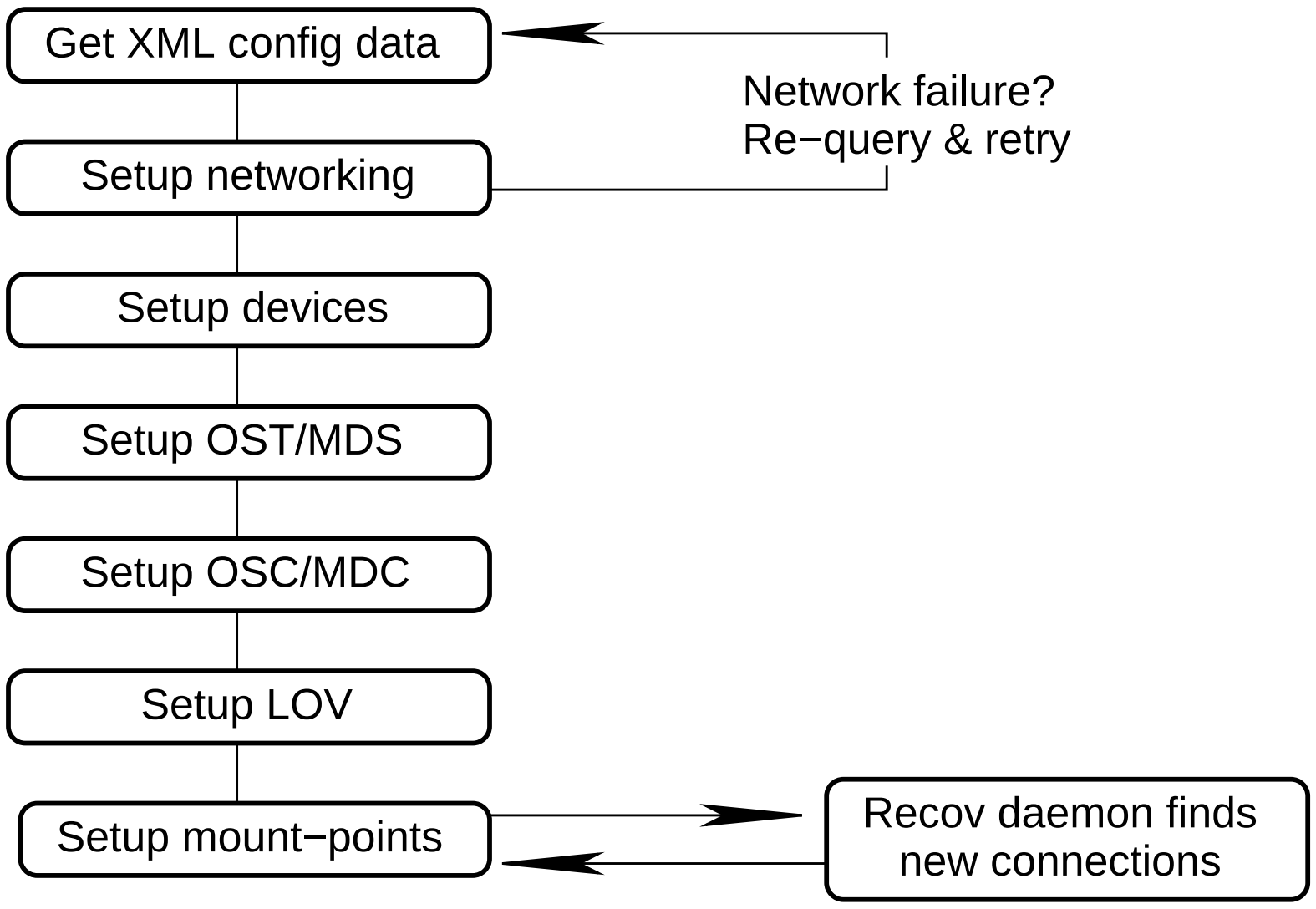

## 14.6. LDAP

Configuration data is conveniently stored in an LDAP repository, following a schema that matches that of the XML configuration descriptors.

**14.6.1. Access API's.** The MDS and OST server failover mechanism execute queries and updates to LDAP. When the failover system, such as RedHat's cluman has reactivated the service on the failover node, it updates the *active server* field in the configuration.

**lustre-modify:** -d dn -a attr=val [-h host | -p port |\n" " -u user | -w passwd | -v?]

- -h host LDAP hostname (default %s)
- -p port LDAP port (default %d)
- -u user LDAP user (default none)
- -w passwd LDAP port (default none)
- -d dn LDAP dn\n" " -a attr=val Modify attribs, attr1=val1+attr2=val2,val3+...

- -v Verbose output
- -? Show this help

The client system will initiate recovery after a timeout event takes place. It will query ldap to find the currently active MDS or OST node. This is done with the lustre-query command.

**lustre-query:** [-h host | -p port | -u user | -w passwd | -s search | -vimdneaf?]

- -h host LDAP hostname (default %s)
- -p port LDAP port (default %d)
- -u user LDAP user (default none)
- -w passwd LDAP port (default none)
- -s search LDAP filter (default %s)
- -v Verbose output
- -i Output the lustreCommonID attribure
- -m Output the lustreUUID attribute
- -d Output the lustreDesc attribute
- -n Output the lustreIPAddress attribute
- -e Output the lustreElanAddress attribute
- -a Output the lustreActiveUUID attribute
- -f Output the lustreFailoverUUID attribute
- -? Show this help

**14.6.2. LDAP schema.** The following LDAP schema closely matches the XML DTD given above and is used for storage in LDAP.

```
# Schema for Lustre configuration
#
# Peter J. Braam <braam@clusterfs.com>
# Brian Behlendorf <behlendorf1@llnl.gov>
# 1.3.1.4.0.1.1 - LDAP Attribute types
# 1.3.1.4.0.1.1.0 - Name Attribute types
# 1.3.1.4.0.1.1.0.1 - Mutliple ASCII strings
# 1.3.1.4.0.1.1.0.2 - Single ASCII strings
# 1.3.1.4.0.1.1.1 - Lustre Identifiers
# 1.3.1.4.0.1.1.1.1 - UUID
# 1.3.1.4.0.1.1.1.2 - Common Name
# 1.3.1.4.0.1.1.1.3 - Short Description
# 1.3.1.4.0.1.1.1.4 - Filesystem Name
# 1.3.1.4.0.1.1.2 - LDAP Network Attribute types
# 1.3.1.4.0.1.1.2.1 - Network Address
# 1.3.1.4.0.1.1.2.1.1 - IP
# 1.3.1.4.0.1.1.2.1.2 - Elan
# 1.3.1.4.0.1.1.2.2 - Active UUID
# 1.3.1.4.0.1.1.2.3 - Failover UUID
# 1.3.1.4.0.1.2 - LDAP Object classes
# 1.3.1.4.0.1.2.1 - lustre
# 1.3.1.4.0.1.2.2 - lustreNode
# 1.3.1.4.0.1.2.3 - lustreMDS
```

```
# 1.3.1.4.0.2 - SNMP
# Multi-Name Attribute Type Definitions
attributetype ( 1.3.1.4.0.1.1.0.1 NAME 'mname'
        EQUALITY caseIgnoreIA5Match
        SUBSTR caseIgnoreSubstringsMatch
        SYNTAX 1.3.6.1.4.1.1466.115.121.1.26 )
# Single-Name Attribute Type Definitions
attributetype (1.3.1.4.0.1.1.0.2 NAME 'sname' SUP mname SINGLE-VALUE)
# A UUID to identify something exactly
attributetype (1.3.1.4.0.1.1.1.1 NAME ('lustreUUID' 'uuid') SUP sname)
# A common name to identify something
attributetype (1.3.1.4.0.1.1.1.2 NAME ('lustreCommonId' 'id') SUP sname)
# A short description
attributetype (1.3.1.4.0.1.1.1.3 NAME ('lustreDesc' 'desc') SUP sname)
# A common name to identify something
attributetype (1.3.1.4.0.1.1.1.4 NAME ('lustreFileSystem' 'fs') SUP sname)
# Lustre IP address
attributetype (1.3.1.4.0.1.1.2.1.1 NAME ('lustreIPAddress' 'ip') SUP mname)
# Lustre Elan address
attributetype (1.3.1.4.0.1.1.2.1.2 NAME ('lustreElanAddress' 'elan') SUP mname)
# Active UUID
attributetype (1.3.1.4.0.1.1.2.2 NAME ('lustreActiveUUID' 'auuid') SUP sname)
# Failover UUIDs
attributetype (1.3.1.4.0.1.1.2.3 NAME ('lustreFailoverUUID' 'fuuid') SUP mname)
# Object Class Definitions
objectclass ( 1.3.1.4.0.1.2.1 NAME 'lustre' SUP top STRUCTURAL
        DESC 'Identify a Lustre filesystem'
        MUST fs
        MAY desc )
objectclass ( 1.3.1.4.0.1.2.2 NAME 'lustreNode' SUP top STRUCTURAL
        DESC 'Identify a Lustre node'
        MUST ( uuid $ id )
        MAY ( lustreElanAddress $ lustreIPAddress ) )
objectclass ( 1.3.1.4.0.1.2.3 NAME 'lustreMDS' SUP top STRUCTURAL
        DESC 'Identify a Lustre MDS'
        MUST ( uuid $ id $ auuid $ fuuid ) )
```

## 14.7. Design for OST Migration, Addition and Removal

**14.7.1. Introduction.** We should make are using the same terminology.

Although we informally use the term OST to refer to a node that exports OSTs, the real meaning of OST is a specific object storage target, and an OSD is a physical access path to an OST. Although the OSDs have ”uuids” in the configuration data, they are only used by the config tools, and the actual OST ”uuid” is used when setting up an OSD. The ”LOVTGTS” file is actually a list of OST ”uuids”, not OSD ”uuids”.

Almost every reference in the design document to OSD should really be OST. It is important to make this distinction, because adding an OSD to a cluster can mean something completely different from adding an OST. Infact, there can be multiple OSTs above one OSD just like there can be multiple NFS exports for a single directory. For instance, one export can be world accessible but read-only, the other could be read-write.

A single OST cannot be shared by multiple file systems at the moment, and possibly never. The purpose for OST's is really to share the same data with different access controls like available in NFS (e.g. root_squash, read-only, nosuid etc).

So one can add both OST's to existing OSD's or add new OSD's with OST's to a cluster.

When migrating data the OSD is involved and the OST's will all migrate to the target of the migration.

**14.7.2. Configuration Issues.** These are not part of this design but the primitives that are to be used are discussed in the DogFoodConfig page.

**14.7.3. Migration.** We can build a migrating driver on the following principles:

- Configuration management: as in dogfood config *
- Run the migrator on the target system, for normal operation the migrator should be very similar to the cobd:
  - if an object is requested, create it and migrate data on the fly
- The iterator thread:
  - acquires with an ioctl what the last object number is on the source
  - makes an ioctl to the (empty) target to start new creations at that number
  - starts walking the object directory on the source (see Michael Bishop's notes about the iterator)
  - pull data

**14.7.4. Addition and Removal.** We can already handle a OST becoming inactive, which is essentially the same as a temporary deletion, as far as the client is concerned. There is probably some code that can be shared here.

Under the issue of OST addition, you need to deal both with adding OST's, i.e. export descriptors or OSD's and with OSD's.

- I'd like to see quite detailed explanations how the two interact etc.
- Also you say you cannot perform delete and add requests simultaneously. How, in some detail, do you assure that this results in an error condition?

On the race conditions, I'm not so enamoured with the idea of using a lock. It is possible better to have a version match between client, MDS and OST configurations. Servers return errors when a client presents a wrong configuration version in the requests and clients fetch new configurations after such errors and then retry their requests. This avoids reconfiguration storms in large clusters which would be provoked by a callback. Also, I'm not so hot on having 8000 configuration locks for 8000 clients (Red Storm).

Unfortunately, you left out the hard part, namely how to manage in flight requests AND in flight recovery when it is detected that the configuration has changed. We need to know how you plan to do this both for the case of OST failover recovery OST's and for the other case where failing OST's has a completely different effect, they are ignored. Can you provide us with detailed thoughts about this? Quite likely we should assist you with this, unless you are familiar withthe recovery subsystem.

In the current architecture, UUID's are random, reserved UUID's don't exist. Can you come up with a better proposal for the deleted OST's and OSD's?

Clearly the addition/removal needs to be transparent, so the questions in the preceeding paragraph should focus on that.

I object to any proposal that walks the MDS file system because it is not scalable in the face of trillions of files on the MDS (in particular I consider the fsck tool a useful debugging utility but not a tool useful for production). There are better solutions.

In due course (see the Lustre book chapter on security) the OST objects will contain the referencing MDS fid, and we can implement that in the LLP timeframe, because all it needs is an (embedded) EA on the OST. It makes a lot of sense then to make the deletion routine client driven, optimally from a client file system mounted on the OST. That client file system has all the infrastructure to run an iterator on the OST that is being deleted to update only those objects on the MDS that are affected, based on the MDS fid found in the object. It also has the infrastructure to then make updates to MDS inodes and adjust the striping information to make the file sparse, without an object (a questionable policy, but requested for Hendrix) or to rename the objects thus found.

So all in all this is a useful start, but quite a bit more is needed before we can call this a design that we agree with.

CHAPTER 15

# Redundant Object Storage Targets

## 15.1. Introduction

Lustre will be used for data archives. In these situations, data is usually written once and a high degree of redundancy and reliability of the archives is desired. This document proposes a design for Lustre to provide this reliability.

The first line of attack against losing data will be to use a mirrored OST (RAID1). Essentially, a replica of all objects would be kept on a different OST. So, if the first OST fails, the objects will still be available on another OST. Another impact of mirroring is load balancing, in the presence of these multiple copies of the same data, read load can be shared by the two OSTs holding the replicas.

## 15.2. Implementation

Lustre clients use the logical object volume (LOV) to hide multiple OSTs beneath it, we will enhance the LOV to do mirroring. Each client subsystem will use the LOV driver to mirror data writes to two OST's. The LOV will store a striping configuration on the Metadata Server (MDS) in the LOV configuration data.

The object ids and location information for a file is stored in the extended attribute portion of an inode, as with striped inodes.

**15.2.1. OST choices.** The application can use the file system running on the OST node that holds the objects. The file system then has a local and remote OST. We will attempt (in due course) to eliminate the client-server components sitting between the local OSC and OBD.

The second object will be on a remote OST. It has been argued that it is better to not pair up the OST's but chose a random second OST for striping. This limits the loss of data in case of a double failure.

**15.2.2. Defining failures.** When writes to an OST happen, the implementation has the choice to declare a write successful if is successful to one, all or a subset of the replicas (If a subset is chosen it is typically chosen sufficiently large to allow quorum.).

The chosen strategy is to declare the write to be successful when the write succeeds on a single OST. The LOV driver will gracefully handle a failure to one of the replica OST's.

However, this still constitutes a failure of the replicated OST pair for which a re-synchronization is needed. This can be done lazily with the help of a replication log.

**15.2.3. Recovery.** There are pretty major issues involved in the correct recovery handling of the raid1 ost.

The problem is the usual one that if the power goes off we need consistent disk state or a way to reach such a state. Without constraints, the writes can have happened on both, one or neither of the nodes.

Handling this involves a replication algoritm of some kind - I studied several over the weekend. Without a very expensive sequence number per write (as in intermezzo's kml sequence), it is probably hard to avoid chosing a "primary" during recovery. It is important to note that raid1 disks also chose a primary for recovery.

Here are a few options...

I looked at the algorithms used by DRBD (the distributed replicating block device) and that has an interesting status vector to indicate who is the primary (it is set, for example, when a nodes promotes to primary, by the operator, to force failover, or at initialization time). Here writes are propagated from the primary block device to another, it is not really raided. The secondary can take over when the primary fails. Recovery in many cases can be very quick but if the power goes off, we need to handle "orphaned" writes.

Another approach is to try to build a device where clients write to both OST's. The client (by 1 failure point assumption) can control the situation well when one of the OST's goes down, but the OST's still need an "orphaned write" recovery, a write never reported to a client, which possibly made it to only one OST. This can be done with our default open/close orphan infrastructure, or with an improved scheme, writing extents that are modified. Normal raid 1 disks need to copy the entire disk in the case of failures of this kind, hence I think that a "recover per object" is a good step forward to a finer granularity, and the extents are not needed now.

Let me elaborate on this one more, since this is what Jacob has started. So each node writes a log record when the first write to an opened file happens (we can make the open part of the first write). We need an explicit close/cancel call. When the file is closed, this cancels the log records. [This btw, is extremely similar to what we need for letting the MDS handle the size.] In case of a failure, a choice of primary determines in which direction the sync goes to work to synchronize the data if orphan records exist on both sides (such a choice is also made for RAID 1 block devices). If one node fails, orphan records on the other one will not be removed. [This one looks doable.]

The third approach is to build lazy synchronization, intermezzo style. In this case the primary (local) OST is more or less a rw-cobd for the second one. I'd like to re-consider that in the fall when we do the MD WB caching.

To enable rebuilds, the first write to an object following an open needs to be recorded in the replication log, as an LML record (see the InterMezzo literature). The LML records will become replication records upon close, these will enable lazy rebuilds, or allow replicas to be built.

When the replica has reached stable storage the record can be cleared. This can be done by calling a sync call before closing the object or by awaiting a log cancellation from the replica's OST node.

**15.2.4. Sync on close.** To improve reliability, data can be synced to disk on close - the application knows in this way that the data is on stable storage when the file has been closed.

If this sync is also applied to the remote object the replication log records can be cleared quickly.

**15.2.5. Disable partially failed pairs.** When one of the two OST's in a mirrored pair fails, there is a higher risk of losing data on the other OST since the remaining OST will get the entire load, instead of sharing it with the other OST in the pair. By disabling further updates in a pair as soon as one of the nodes fails, we eliminate the risks associated with the double load. We also greatly improve re-synchronization time in case the failed OST can be resurrected, and we eliminate the risk associated with new objects being stored on just a single OST.

The LOV would no longer honour creation of new objects on this pair. It would mask errors by re-fetching the configuration data from the OST allocating and determine which RAID pair to use

This would proceed through a recovery upcall, made by the OST that detects the failure. The configuration data would be updated to eliminate the OST pair from the policy for new object allocation.

**15.2.6. Re-synchronization.** The re-synchronization can use the OST resident replication records derived from first write records.

This is entirely analogous to recoverability between MDS and OST of the mtime and size attribute updates which the OST performs and propagates to the MDS on close.

## 15.3. Other Mechanisms for use in an archive

**15.3.1. Load Balancing.** We envisage that the incoming service will be able to determine what pathname in the file system will be required.

It is attractive for Lustre to provide a module that enables the load balancing component to look at the pathname in the protocol, determine an optimal server for delivering the data and redirect or forward the request to that node.

We envisage that an LVS module can do this quite easily and that it is best if the load balancer runs on the metadata server to eliminate the network traffic associated with the frequent lookups it may have to do.

The mechanism used by this redirection module would be to use the file system API to perform a lookup (or stat system call) on the pathname. Then obtaining the extended attributes associated with striping will give the information required by the load balancer. The load balancer can then use the information about object replicas to direct reads to the appropriate servers.

#### 15.3.2. Immutable Files.

To further limit the possibilities for accidental damage to files, we propose to use an immutable bit on the inodes. The immutable bit would be transferred to the MDS inode and OST objects upon close.

In order to remove a file from the archive an operator will have to use a chattr command to clear the immutable attribute, and then issue the rm command to perform the unlink associated with the object.

## 15.4. Implementation Plan

#### 15.4.1. Initial Targets.

(1) LOV mirrored writes
(2) Graceful write failure masking
(3) Recovery upcall generation
(4) Create disabling on failed pairs
(5) Sync on close option in file system
(6) Immutable attributes
(7) Mechanisms for re-synchronization of one object
(8) Scripts to re-synchronize a failed OST
(9) Scripts to restore a recovered pair in cluster
(10) Changes to lconf to support setup of OST pairs.

#### 15.4.2. Second set of targets.

(1) LVS load balancing module
(2) Replication / re-synchronization log
(3) Use log for fast re-synchronization
(4) Eliminate OST client - server for local OST
(5) Ability to add OSTs to a live cluster.

CHAPTER 16

# Caching OBD

## 16.1. Introduction

An important aspect of a filesystem is read scalability. Many client file systems may try to contact the file servers to access data at the same time, e.g. when a cluster boots that has shared root and system file systems. If every client has to read data from the same single server, the network load at the server would be exteremly high and the available network bandwidth would pose a bottleneck for the system. A collaborative cache (COBD) distributes this read load over multiple nodes which can be dedicated cache servers or clients. In a filesystem using COBD, there could be copies of the same data at multiple nodes, so read for the same data could be serviced by these various nodes. This can result in an un-precedented improvement in scalability for reads. Instead of trying to read data from a single server everytime, a client can be redirected to a peer node that has already cached the data.

The Lustre cluster file system has three main components, the *client file system* (CFS), *the metadata server* (MDS) and the *object storage targets* (OSTs). In Lustre file system there can be either read or write requests going to the OSTs. The *writes are* flushed down to the OSTs as soon as possible so as to give a consistent view of the data, and the client file system will use the standard mechanisms to cache writes in the page cache and flush them in cases of direct I/O or sync system calls. A *read* access is made through the page cache from the OST's. But such read requests could be made from other *caching* at other nodes as long as it is known that they have or can acquire the correct version of the data. The node would get a local copy of the data, provided the data does not get changed subsequent *read* request for the same data can be handled without contacting the storage device. This would improve the *read* performance considerably.

Caching nodes use existing client (OSC) and server (OST) infrastructure but introduce a new object driver, the Caching Object Driver (COBD). The COBD can be introduced on client nodes or run on dedicated caching server and such nodes can now service the client read requests. This plan would considerably reduce the number of requests that need to be serviced by the OSTs.

In this chapter we will explore the various issues pertaining to the design of such collaborative caches, some of the important issues are failure recovery, collaboration management and cache coherency. In this document we will examine the current architecture for lustre and propose an addition to the design to incorporate a distributed collaborative read cache that would improve the performance.

## 16.2. Current Architecture

The various components in the current design are shown in figure 16.2.1.

Anytime a client wants to read a page, they have to determine which OST to contact and then send the read request to that OST. In a situation with several thousand clients, sending the read requests to the same OST would seriously limit the scalability of the system.

FIGURE 16.2.1. Current architecture for Lustre

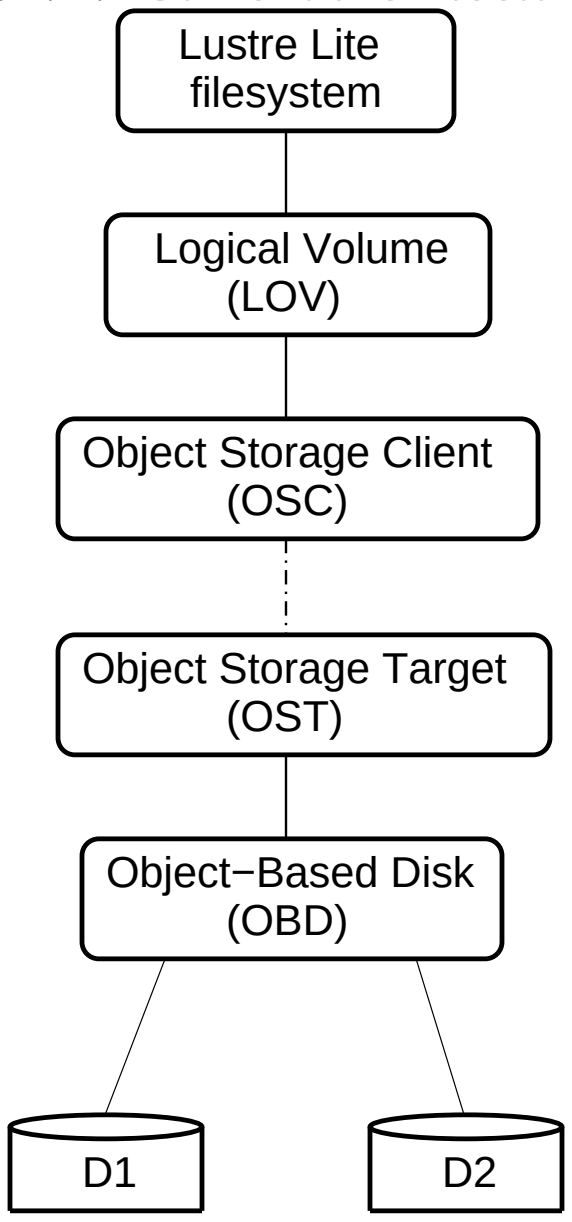

## 16.3. New Architecture

We propose a design for a collaborative cache to distribute the *read request* load among several nodes that cache the same data pages. The new request path is shown in figure 16.3.1.

This architecture introduces secondary *caching* OST's into the picture, which supports the *target OST* in the service of read requests. The target OST must refer read requests to suitably chosen caching OST's. The two OST's will be identical code modules, with a different backend. In case of the *target* OST the backend would be the actual disks that store the data with the OBD layer in between. In the case of the *caching* OST, it would be the in-memory cached pages of various objects with a layer of the new COBD between them.

The figure also shows a new OBD driver, used by the caching OST's known as COBD, this is the read cache layer.A read request would be first sent to the OST which has the requested object. The OST checks to see if it knows of caching OST's running on other client nodes or on dedicated cache servers that might be already caching the requested file extent. The client then gets a refferal for the COBD and the read request is forwarded to COBD. The subsequent subsections highlight the various aspects of COBD and the changes needed to implement it.

FIGURE 16.3.1. Proposed design for collaborative cache

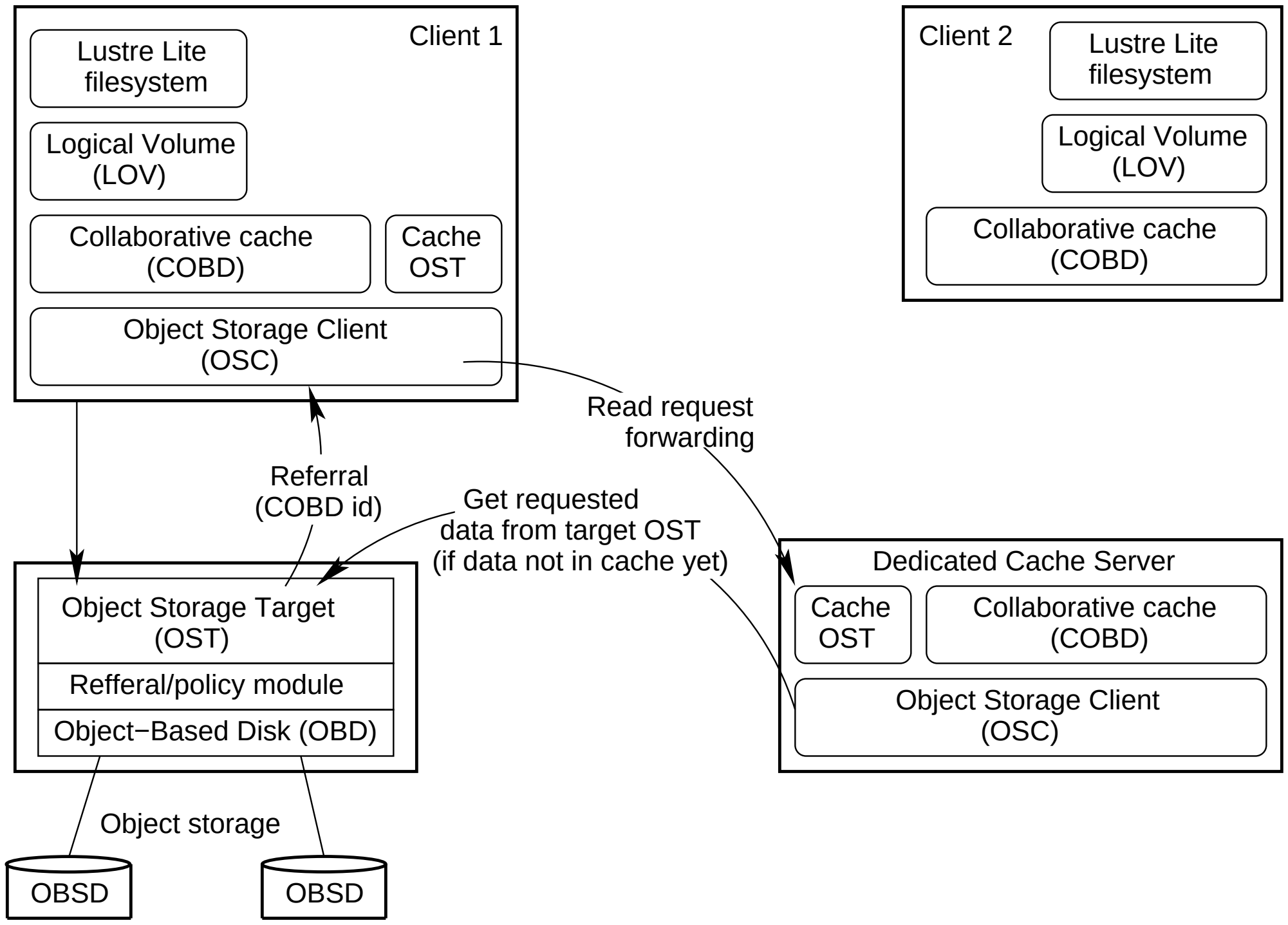

**16.3.1. Read Cache (COBD).** This could be running on a dedicated cache server or a peer node (Client1 in figure 16.3.1) that would export an OST so that it can receive and process *read* requests. If COBD was on a peer, i.e. client node, it will be used by the file system to service requests. Lock requests still are made to the OST, so we will disable any lock granting ability at the OST on top of COBD. The COBD should simply service the read request. When accessed by the client file system, other requests are passed through to the OST by the COBD, but when accessed by the OST, the COBD only honours read requests.

The COBD needs to maintain a cache of mapped pages of objects. This cache needs to shrink when kernel memory comes under pressure. We will implement this by associating cached objects with (in memory) inodes. Inodes have an associated address space to manage a page cache on which the kernel can exert pressure and which include suitable callbacks for cleanup of related data (such as locks) in the COBD. If this page cache is backed by swap it will be possible to use persistent storage for large caches.

Even more attractivily, but not planned immediately, we might try to make the persistent cache use a file system so that it might be usable after a reboot. In this case a direct OBD would be used for storage.

The COBD needs to maintain read locks on the data it caches to ensure its consitency. It fetches such read locks from the target OST, as the file system is presently doing, but there is one modification. The target OST when faced with a client lock request will grant as big a lock as possible. For a request initiated by the COBD the target OST will grant a lock that reflects the actual data that the COBD caches. This puts the target OST in a position to have reliable information about which referrals of read requests can be served from cached data.

16.3.1.1. *Referral Policy and API for lock query.* The target OST will receive a read request and must select a caching OST to service the request. It will use the lock manager on the target OST to search for caching OST's that have the file extent cached in the COBD.

As shown in figure 16.3.1, the referral decision is made in the new Referral/Policy module between the OST and OBD. This is a new API that will take as input a resource name(file & extents) and determine the nodes that have a lock on it and the type of locks held. The request should be reffered to COBD only if *'PR'* compatible locks are in the granted list for the resource. The API should return a pointer to the list of nodes that have a *'PR'* compatible lock on the resource and 0 as the return value to indicate the success of referral. Such nodes can then be used in a round-robin fashion to service the request.

If no nodes are found, the target OST makes a cache population policy decision and may still direct a client to a caching OST. When the caching OST receives the read request, the COBD will not find it in the cache, take out a lock and use the standard OSC to populate its cache.

**16.3.2. Referral.** The referral simply consists of the host name of the caching OST. Based on the list of nodes holding the 'PR' lock on the resource, one of them is selected as a referral. The UUID of the caching OST is returned as a refferal to the client. The client can send a request to the caching OST where the COBD will service the read request. An further optimization would be to return a referral at the same time a read lock is given to the client. This would reduce the number of messages required. To grant the read lock to the client, the *target* OST would have to check the lock information for the resource, the same information could be used to determine the *caching* OST as well. The figure 16.3.2 shows the path followed by a read request in the current architecture.

In the proposed design, the locks would be taken from the *target* OST, but the read request would be directed to a *caching* OST as shown in figure 16.3.3.

A further issue here is that of *populating the caching OST's.* The target OST must have an algorithm to decide which caching OST's should be responsible for what data. Schemes of interest are a tree of caching OST's, the entire cluster forming a cooperative cache and probably a variety of others.

**16.3.3. No Lock Service.** The caching node will also have an OST layer to enable communication of requests to it, but we need to ensure that it will only provide read service. Since the source code for the OST module is the same as the one used on top of OBD, we could add a flag to indicate that these are reffered requests.

FIGURE 16.3.2. The path for a read request in current architecture

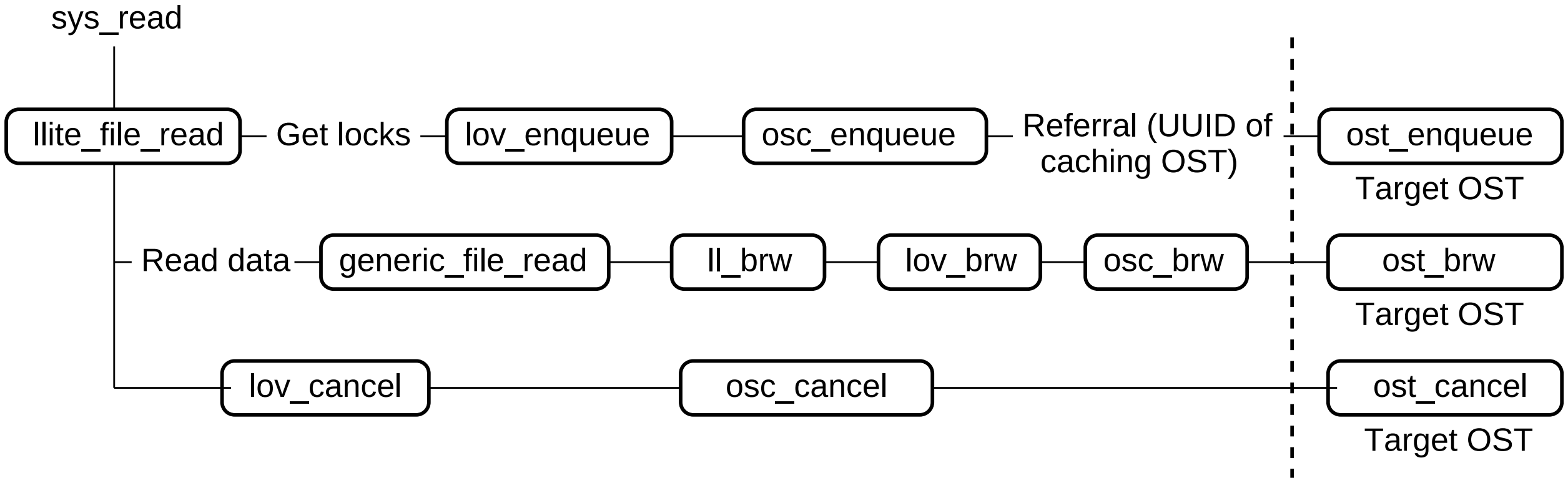

FIGURE 16.3.3. Path for a read request with COBD

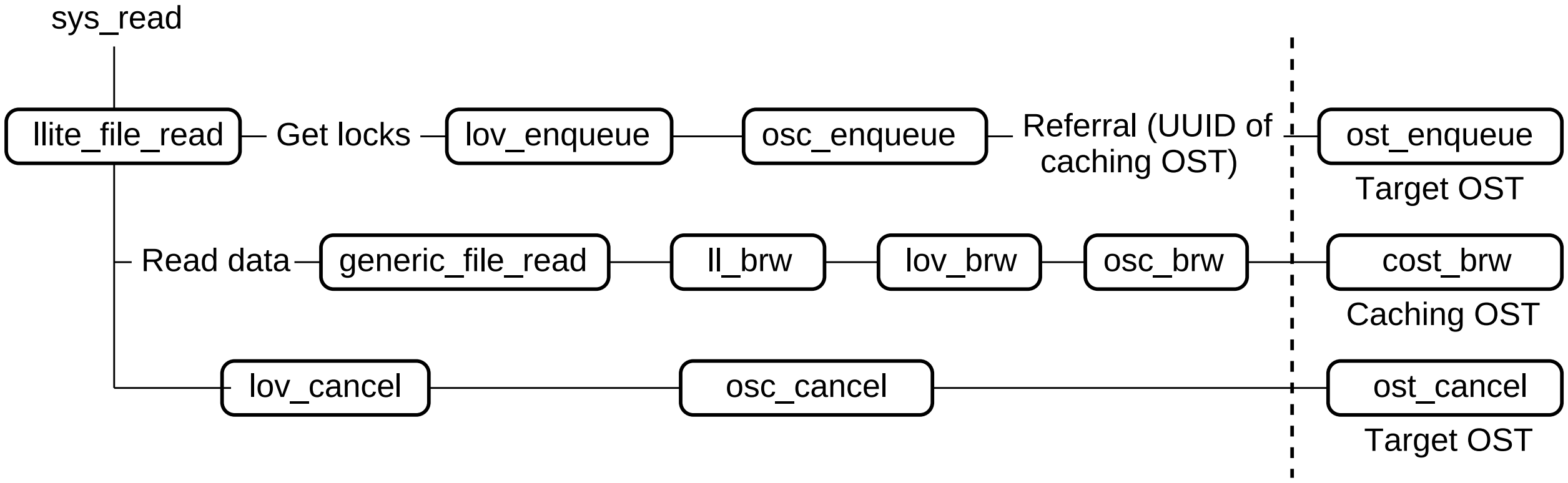

**16.3.4. Changes to Locking Policy.** We need tighter locking policy, at present Lustre gives out optimistic locks, so if there are no contentions on a resource a client might get lock for the complete file rather than just the extents they are interested in. This would hinder the performance of Collaborative cache. We need to change the locking scheme so that COBD's can make requests that do not return optimistic locks but instead return locks only for the extents to be accessed are locked.

We already have the logic to lock extents of a file, we merely need to allow a non-optimistic policy for caching OST's in addition to the existing optimistic policy for clients.

**16.3.5. Cache Consistency.** Another aspect to caching is maintaining the consistency, in Lustre this is already taken care of. When a request for a write is received by an target OST it will revoke all the read locks that were held on that resource and invalidate any cached copies. There would be no subsequent referrals to the same caching OST until it's COBD again obains the updated resource under *'PR'* lock. It is possible that the lock revocation happens just after a referral is returned to the client OSC and before the forwarded read request is serviced by the COBD. In this case we have decided to let COBD handle the read by trying to obtain a lock on the resource and get the updated copy of the resource.

**16.3.6. The client file system page cache and the COBD cache.** The client file system has a page cache and we will integrate that page cache with the COBD cache. This is straightforward in the case of the client file system interacting directly with a COBD, but slightly more involved when an LOV sits between the file system and the COBD.

This will require a small change to the object API which is to return a pointer to a cached page instead of reading a cached page.

**16.3.7. Policy.** In the COBD implementation, besides the basic mechanism, we would have to make several policy decisions. In this subsection we list some of the decisions that we will have to make.

(1) Based on the list returned from the lock query API, the OST has to use some policy to determine which COBD should be selected as the cache to process the new read request. At this point this could be a simple round-robin algorithm.
(2) When returning a referral with the UUID of the caching OST, we could also add a field that would indicate the the file extent range available at that caching node.
(3) In case the *target* OST has the request data in its own cache (saving it a disk I/O), should it service the request there instead of returning a refferal? This could be done, it would help by reducing the number of network messages that would be required to reply with a referral and for the client to send a read request to the *caching* OST. But servicing the read at the *target* OST would still consume network bandwidth there.
(4) It is possible that no single caching OST has the required data, but more than one might collectively have the required data. In such a case, we could return multiple referrals along with the extent range available at each of them.

CHAPTER 17

# Lustre Metadata Write Back Caching

This page documents the proposed implementation of the write back cache. Our starting point were the following requirements:

- Minimal changes to Lustre Lite
- The design should make a persistent writeback cache possible
- Exploit existing api's whereever possible

We believe that the proposal here satisfies all these requirements.

## 17.1. Mechanisms

**17.1.1. Layering.** File system requests are directed by the VFS to the Lustre Lite layer. Lustre Lite checks what locks are required, if any and gives out cached data that is known to be good. It also forwards update requests with lock requests through our intent mechanism. When a writeback cache is enabled several new issues enter the picture:

(1) a decision is made if a writeback lock is available for the objects the operation is accessing
(2) the cache is queried for the object
(3) cache misses are handled
(4) updates are made locally
(5) log records of the updates are created

We introduce two layers below the client file system. The lowest is the local metadata cache. The pivot of our construction is to construct the cache as a locally running metadata "server". We call this the local MDS. The second one is the write back metadata client: WBC. This is a logical metadata driver to assist handling cache misses in the local cache. The WBC accepts commands from Lustre Lite and interacts with the local and remote MDS to handle cache misses.

**17.1.2. Cache Queries.** If a writeback lock is found or acquired, lookup and attribute requests are serviced through the local MDS. The WBC driver first determines if the cache has the object. To do so, it uses two commands:

- lookups in the local MDS
- analyzing extended attributes in directory inodes in the local MDS. These extended attributes encode what parts of directories have been cached locally.

**17.1.3. Cache Misses.** If a cache miss is encountered the WBC will use the usual MDC API to the remote MDS to acquire the object or data it needs. In case it fetches a partial directory is will set extended attributes accordingly to indicate what part of the directory is available in cache.

**17.1.4. Local updates.** The clustered metadata driver will include a local MDS that runs over a file system. This file system can be a memory based file system like shmfs or a disk based one. The local MDS has the same API as a remote MDS and the updates to the local cache can be made by dispatching the command as usual, but to the local MDS instead of the remote MDS. Local updates require preallocated inodes. These will be regarded as cached objects and their availability will be handled by the WBC similar to the cache miss handling.

**17.1.5. Log Records.** Lustre Lite has a logging API, carefully contructed for recovery with foresight for us in the writeback caching. As part of the local update a log record, either in memory or transactionally consistent on disk through nested transactions (as in recovery) is created. A daemon, sensitive to memory pressure propagages update records to the remote MDS and handles cancellations of the records when they commit remotely (another already existing api for the logging records).

**17.1.6. Cache Purges.** The local MDS, like an AFS cache, will use a prescribed amount of storage space, in memory or on disk depending on the file system underneath the local MDS. The cache will run one or more flush daemons that operate like the well known ext2 ”’balance dirty”’ code. If the cache is empty the daemon will run periodically, if the cache is moderately full it is activated. When the cache is very full, the daemon synchronously removes cached data.

Removal of cached data is dependent on any associated log entries being flushed to the remote MDS first. The cache daemon will interact with the WBC for this purpose and to establish that objects are not busy when they are being purged.

**17.1.7. Readdir.** Lustre Lite will implement a new directory content management scheme. This directory layer is present in two subsystems:

- Lustre Client File System. Key aspects are:
- cached directory entries will be extremely similar to shmemfs dentry based directory implementation.
- inode attribute attributes to describe what parts of the directory have been cached in memory.
- MDS, both the local and remote MDS.
- These will offer a directory fetch API.
- Locks will control cache validity
- Readdir cookies will assist with re-reading after directory updates.
- When the write back driver is present the api will be executed against the local cache MDS
- Without the write back driver, against the remote MDS.

## 17.2. Lock Management

**17.2.1. Write back locks.** When a client accesses a remote MDS directory the MDS will give out a subtree lock. Subtree locks can be read or write locks following the single writer multiple reader model among clients. If a subtree has not seen recent activity and directories are small enough to reasonably fit into a client cache, subtree locks will be given out. If the directory is already locked or has seen very recent activity a normal intent lock execution is performed.

Subtree locks are associated with the namespace.

**17.2.2. Sub Tree Change time (stctime).** To measure if a sub tree has seen changes, the MDS will update not just the ctime of particular inodes involved in the transaction, but it will also modify an extended attribute holding this change time on all ancestors of the modified object or objects. This stc time has nanosecond granularity and can validate the version of an entire subtree.

This update can be made in a persistent manner or in a memory only way. The stctime associated with files open on multiple nodes will change only when all nodes not holding a writeback lock close it.

**17.2.3. Filling and revalidating the cache.** When a client with a subtreee lock fills its cache it fetches objects from the MDS. During this process the MDS may revoke other locks on such objects. If open objects, files or directories, are encountered, the object is be given to the client with a flag indicating the object is subject to remote operations. No operations may be performed in WB mode unless they commute with the remote operations that may be performed on such objects. This can happen for renames only, we think, in which case the writeback lock will be yanked away by the MDS before the operation can complete.

**17.2.4. Lock revocation.** When other client acquire new locks underneath a subtree lock held by another client, then the write back caching locks need to be revoked. It may be that a lock is requested associated with a fid on the directory. The fid2dentry code will need to be modified to return a connected dentry for this purpose (NFS shows how to do this trivially). The cases where a non-connected dentry for a fid could be found appear to be somewhat exotic but can happen.

After the connected dentry is found the MDS can establish trivially if there is an active subtree lock on the tree and revoke it. Revocation has the following effects:

- WB metadate update log flush
- Convert open mds and OST objects under the subtree lock to remotely opened locks

**17.2.5. Open objects.** If a client with a WB lock fetches an object into its cache that is associated with an MDS open (ie. an open file, an open directory, a current working directory or a mount point) then the cache fill will return that object with a flag that it is subject to remote operations. (This is precisely what Sprite did in the early 80's!) Open objects can get involved in operations that do not commute with the write back log. In that case the writeback lock will be canceled.

**17.2.6. Expiration and nested subtree locks.** Subtree locks can be nested. When a subtree lock is found on an ancestor the client will continue to see if an older ancestor has a subtree lock also. If so, the first one found is canceled after the update log associated with it is flushed.

Subtree locks are subject to expiry, like all unused locks. If a subtree lock is not active, the expiry callback function is called, which will flush any remaining log entries and cancel the lock.

## 17.3. Preallocation

We will follow a simple scheme to create preallocated objects in a directory pointed to by the export on the MDS, and transfer the attributes (fid) to the client. This does mean that some create operations will turn into rename operations on the MDS when new fids arise from pre-allocation.

## 17.4. Recovery

The protocol is remains recoverable in the same way, because it has not changed. A pinger is needed to notify a client sufficiently quickly of an MDS failure. In this way the client can replay it's WB lock and reintegrate pending changes after failover.

CHAPTER 18

# Caching and XDSM

Lustre will implement an XDSM api as part of its writeback cache. This document describes how this is implemented.

## 18.1. Introduction

**18.1.1. What does a writeback cache require?** A cache is a secondary storage space for data. Secondary can have multiple meanings and some of them are:

- **partial:** The cache holds a part of the entire data collection, typically on media that provide faster access.
- **disconnected:** A cache is made available when the primary storage space is offline.

Typical tasks associated with cache management are:

- **validation:** Data was found in the cache. Is it valid, i.e. is it the correct version of the data?
- **write back_updates:** Data is added to the cache or changed in the cache
- **reintegration:** Data in the cache was changed, it needs to be propagated to the primary storage location.
- **cache miss handling:** Data is accessed, but not found in the cache.
- **space management:** The cache is used to add new data, but space may not be available.

A key problem encountered with caches is that the cache may need to be modified during the process of accessing it. This makes it often akward to integrate the cache with a file system. A better approach is a wrapping file system that uses an underlying file system as the cache, and can operate on that file system freely, because VFS locks are taken on different inodes.

## 18.2. Redo, Undo records

Our cache sometimes needs to have redo or undo records available.

## 18.3. Space Management

The cache should maintain a list similar to an LRU list to purge data that was not recently used. Our design is as follows.

A cache replacement algorithm is usually based on an LRU. We may use a modified algorithm called the ARC cace replacement which appears to offer serious advantages. There are only minor mechanical differences between the two mechanisms so we will simply refer to the replacement database as the LRU.

The key difficulty that needs to be addressed is that insertions and removals from the database should have

- good locality of reference, if actions happen closely together in time, the updates to the database cover the same disk blocks as much as possible
- efficient removals

We believe the scheme presented here addresses the needs.

**18.3.1. Mechanisms.** Each time space is needed and LRU list is queried to find objects that can be removed. Space is freed by removing sufficiently many objects from the cache to satisfy a request. The objects that are eligible for removal must be:

(1) leaf nodes in the file system, others cannot be removed
(2) objects that are not busy or dirty
(3) objects that have not been pinned in the cache by the user

To satisfy these requirements, the inodes in the file system underlying the metadata cache will be given extended attributes. Some of these have already been discussed above, for space management we add some additional fields:

```
__u32 hoard_priority
struct llog_cookie lru_cookie
```

The hoard priority gives an indication to the cache manager if the object is eligible for eviction at all. Users may wish to keep certain files cached at all times.

The LRU is a list that is persistent, and transactionally updated, as follows. An entry in the LRU is a log record, that contains

- a fid for the inode that is cached

A cache purge simply walks the log and removes objects until goals have been met. The removal of the objects uses the lvfs api.

Changes in the file system cause the cache to be updated. When an object is accessed and it is a leaf node that is eligible for purging it is

- If present in the database, deleted from the llog, using the cookie in the EA and then re-appended. The EA is updated. Note that this leads to the inode being re-written, possibly for atime updates and to write out the new EA. Deletion from the log is simply clearing a bit in the llog bitmap.
- If not present yet, appended to the llog.
- if a directory becomes a leaf node, through a removal of its children then it is inserted in the LRU.

A more refined version would store in the LRU what part of the file was accessed most recently and have a cookie for every segment of say XMB size, that was recently accessed. This is much more difficult to handle. Perhaps files that cannot be cached in their entirety should not be cached at all.

These updates can all be made transactional quite simply by wrapping file system updates in nested transactions that can add a small amount of data to the llog or flip the bits in the header.

**18.3.2. Interaction with smfs.** The SMFS pre-actions related to cache space management are to check for space, and if necessary remove objects as listed in the LRU.

The post actions are to update the LRU if previous operations were successful.

**18.3.3. Cache Purges.** The cache, like an AFS cache, will use a prescribed amount of storage space, in memory or on disk depending on the file system underneath the local MDS. The cache will run one or more flush daemons that operate like the well known ext2 ”’balance dirty”’ code. If the cache is empty the daemon will run periodically, if the cache is moderately full it is activated. When the cache is very full, the daemon synchronously removes cached data.

CHAPTER 19

# Lib Lustre - a Lustre Library Implementation

This chapter will describe the design and implementation of a Lustre file access library, this library provides access to Lustre file systems to an application linked to it. The library is designed in a highly portable fashion and shares almost all code with the kernel implementation of Lustre. The primary goals for the library are to provide a portable mechanism to access Lustre from different POSIX compliant operating systems - from microkernel based systems to the Windows operating system.

The Lustre library version, known as *liblustre*, uses a thin virtual file system like layer to provide the interface between the application and the file system as illustrated in figure 19.0.1. The *liblustre* implementation for the Cray-RedStorm project exploits the *libsysio* module from Sandia National Laboratory for this purpose, library support for Windows OS will be provided using the *wine* interface.

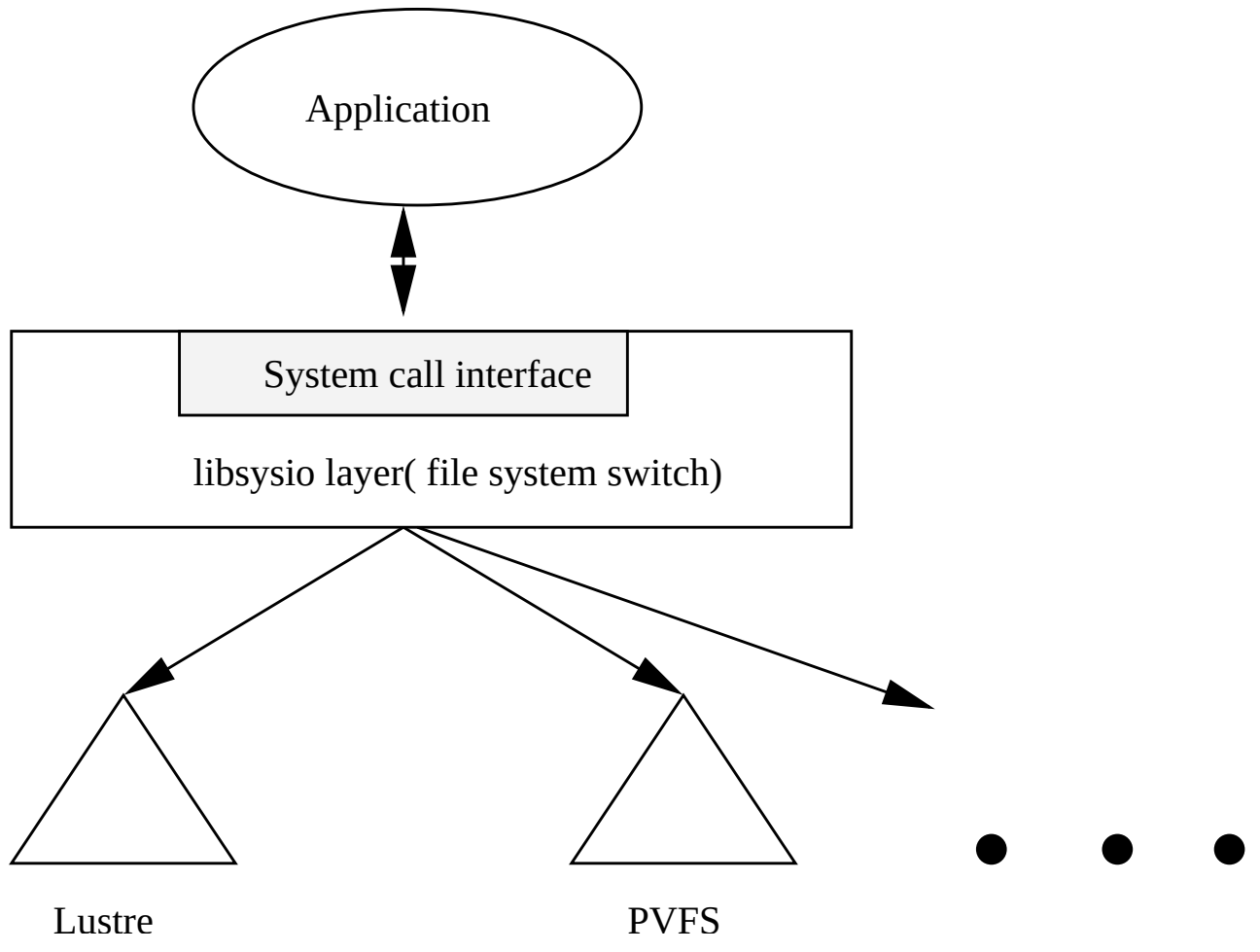

FIGURE 19.0.1. Overview of library version

## 19.1. Introduction

The Lustre file system in general consists of three main layers - Lustre client, metadata server (MDS) for handling namespace operations and the object storage targets (OSTs) for I/O operations.

The system calls from applications virtual file system layer in the Linux kernel will redirect all the requests for Lustre file system to the Lustre client which then communicates with the MDS and OSTs to serve the request. The library version provides I/O access with the help of an additional layer - the *libsysio* layer. *Libsysio* exports a POSIX compliant API interface to the applications and acts like a file system switch and directs requests to the appropriate file system. For Lustre, these requests are directed to a library version of the client - *liblustre*. The contrast between the two methodologies is shown in the figure 19.1.1.

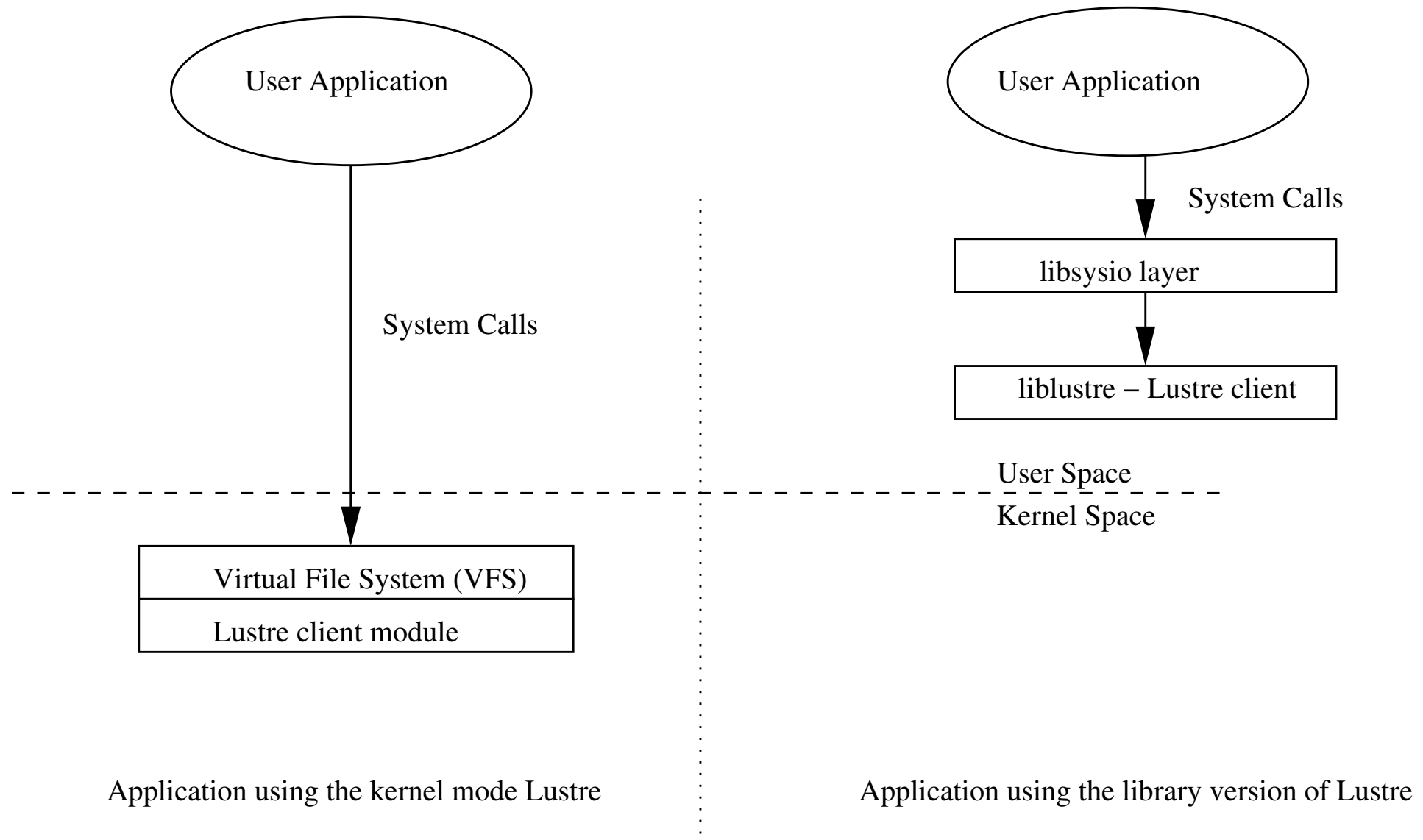

FIGURE 19.1.1. Comparison of the kernel mode and library versions of Lustre

In order to ensure complete portability for Lustre, focus also needs to be on the following areas:

**Build system:** The build system for Lustre has evolved into a cross compilation environment. A lot was learnt from the Coda project in which cross compilation was used for all Windows platforms.

**Configuration system:** The configuration management system initially depended on specific Linux interfaces, such as a character device and to a lesser extent the *proc* and *sysctl* file systems. These interfaces were replaced with more generic ones.

**API calls:** Lustre initially depended on calls to the Linux kernel API's, for memory management, event handling (scheduling, sleeping, notification) and to some file system interfaces. These interfaces where wrapped into a library which is linked in to the Lib Lustre code module to make it depend only on interfaces in C libraries available in the runtime system.

**Threading:** The Linux kernel implementation of Lustre is dependent on kernel threads for multiple reasons. Clients need to handle lock revocation and completion callbacks, servers offer many other services. Additonally the configuration system uses the thread in which the configuration tools run to configure the state of the system. These interfaces were reviewed and changed to accomodate both multithreaded user level environments in which Lustre may be available as well as the single threaded environments.

## 19.2. Build System

The GNU autoconf system is used to configure the Lustre source tree for building. For a file system the appropriate build flags are:

**build:** The system on which the software will be built. Linux is presently the only supported type.
**host:** The system on which the software will be run.

The **target** variable is not used, since that only applies to cross compiler setup. The initially host target systems that Lustre will support are:

- Linux - the Linux kernel implementation. This is the default and can be omitted from the configuration commands.
- POSIX - a Lib Lustre library for Unix systems
- ST - a single threaded Lib Lustre implementation, suitable for microkernels like Catamount
- Windows - a win32 file system API library or explorer plugin to provide Lustre access to Windows applications.

To set up Lustre for cross compilation, it needs to be compiled with the correct *host* flag specified in the *configure* command. Portals has now been made a part of the Lustre tree, so it will automatically be compiled with the same flags.

## 19.3. Configuration Management

**19.3.1. Configuration interface.** Under Linux, Lustre is configured through ioctl's to a character device, */dev/lustre.*

In a multithreaded environment, like that of a Windows or Unix workstation gaining access to Lustre through a library, it remains desirable to preserve much of this infrastructure. We will adapt the lustre configuration tools to dynamically configure the library by sending commands through a Unix domain socket.

For a single threaded client it is mandatory that a startup function can be called and perform all initialization.

The configuration mechanism for the Lustre library makes use of a lot of the existing infrastructure, the servers are configured as usual. On the clients, *lconf* utility is used to dump all the configuration information to a *dump_file*, this file will be read by *liblustre* for Lustre configuration.

The figure 19.3.1below shows the path followed by *liblustre* initialization and setup.

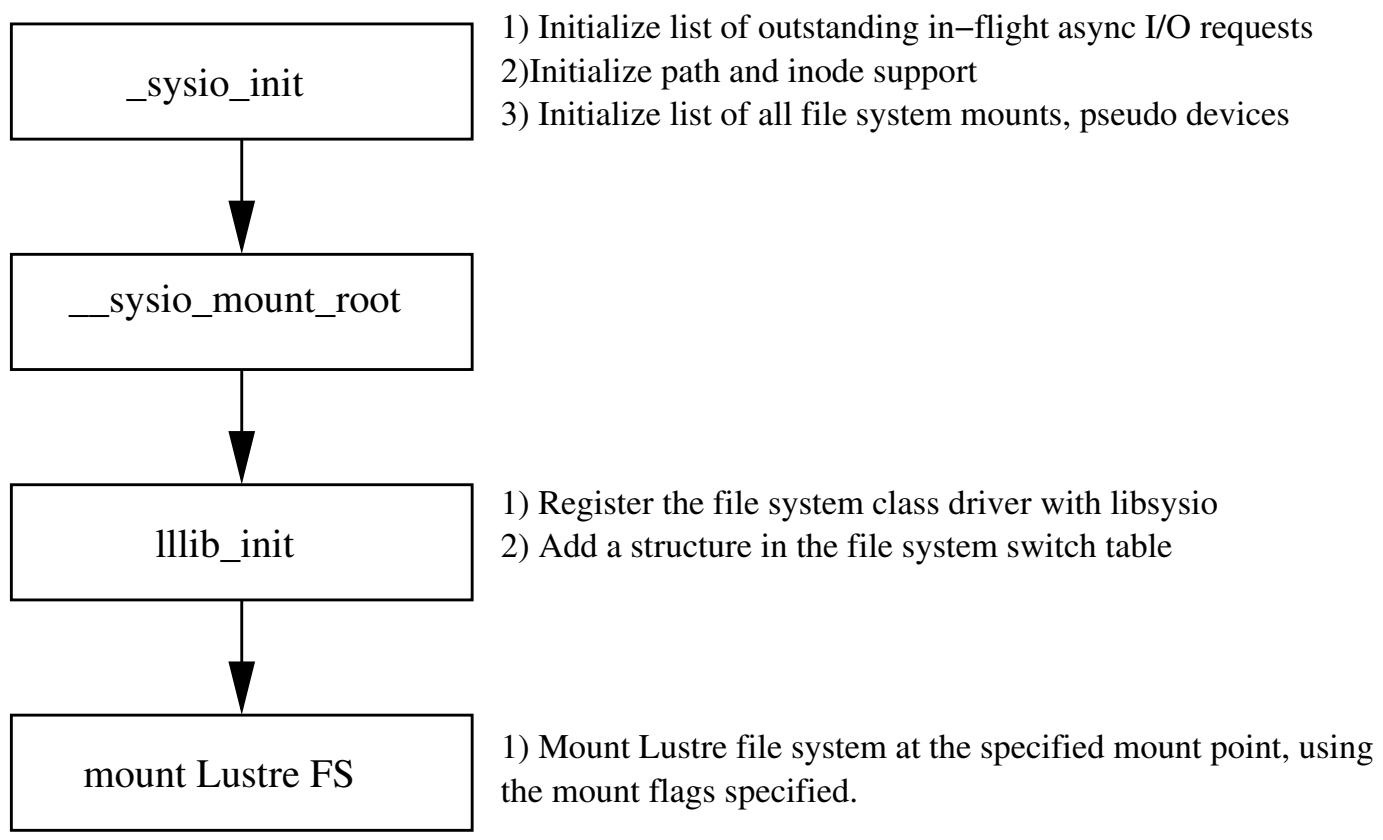

FIGURE 19.3.1. Lustre mount

**19.3.2. OBD devices.** The obd devices are the cornerstone in Lustre linking drivers to devices. The existing kenel code will be made to work in user space.

## 19.4. PtlRpc module

The portals RPC module is the layer built upon the Portals stack from Sandia, this is used by Lustre for message passing. In the kernel version of the Lustre file system, there is a fairly heavy use of kernel infrastructure to :

(1) Provide service - Callback service is offered for the client lock.
(2) Handle waiting, timing out and handling signals for processes executing remote procedure calls (RPCs)
(3) Handle recovery and replay of failed or reverted RPCs

This module has seen a number of changes to make it portable and work in the user space. We will change this module to make use of support modules for 3 environments:

(1) Linux kernel
(2) A multi-threaded user level client environment
(3) A single threaded client environment

A key issue here, that might turn out to be helpful (although it can probably be avoided), is a PtlEQWait that can wait on multiple event queues. The case where this is most likely to help us is where we have mutliple network interfaces on client systems - a situation not so relevant for Red Storm.

In the following section, we will discuss the changes made to this module in more details. This is still in the process of evolution and is likely to see a few more changes.

**19.4.1. Modifications for *liblustre*.** In the current implementation of *ptltpc*, each NAL (socknal, qswnal etc.) have two kind of queues:

(1) ***Serving queue*** **-** This is used by all the service threads (OST, MDS, LDLM) which rely on event to trigger request processing. In this case, it is important that we do not lose any event information here.
(2) ***Other 8 queues*** **-** These are not event triggered, we only use callbacks for these and so we can let the events wrap around when overflow occurs.

In *liblustre*, the main thread can only wait on one event queue using *PtlEQWait()* function. We can not lose an event here also, but the event queue could be short since we save the event at anothre place if needed in the callback. Once it returns from callback, the event is no longer useful and can be overwritten.

In the userspace, *l_wait_event* was redefined to use *PtlEQWait()* to wait for an event to happen. This implies that only some portals event could change the condition status we are waiting for. Furthermore, user thread will also check for event using *PtlEQGet* to check the event at many execution points because we might need to handle some urgent request such as ldlm revoke request etc.

There are some other details, such as reqring the service buffer offset management, some trick to avoid memory leak etc. Soon all this would be written in a more elegant fashion.

## 19.5. OSC & MDC

The object storage client is probably easily ported as the design here has carefully avoided using any types and structures except those defined by Lustre itself.

The MDC is unfortunately not in good shape for porting. Only late in the project did we realize how important metadata abstractions would become and we will have to seriously rework the interfaces the MDC currently offers and their use in the file system.

## 19.6. LDLM

The distributed lock service will be easy to port to a client environment.

## 19.7. File I/O

The LibSysio interface has rich features including asynchronous I/O. To implement iowait and iodone is straightforward as their functionality is embedded in the standard kernel level I/O completion callback handlers (see lustre/llite/rw.c and osc/osc_request.c). These routines collect portals events and signal completion when the last event associated with an I/O operation has completed.

The subsequent sections show the path for some of the system calls in the presence of libsysio.

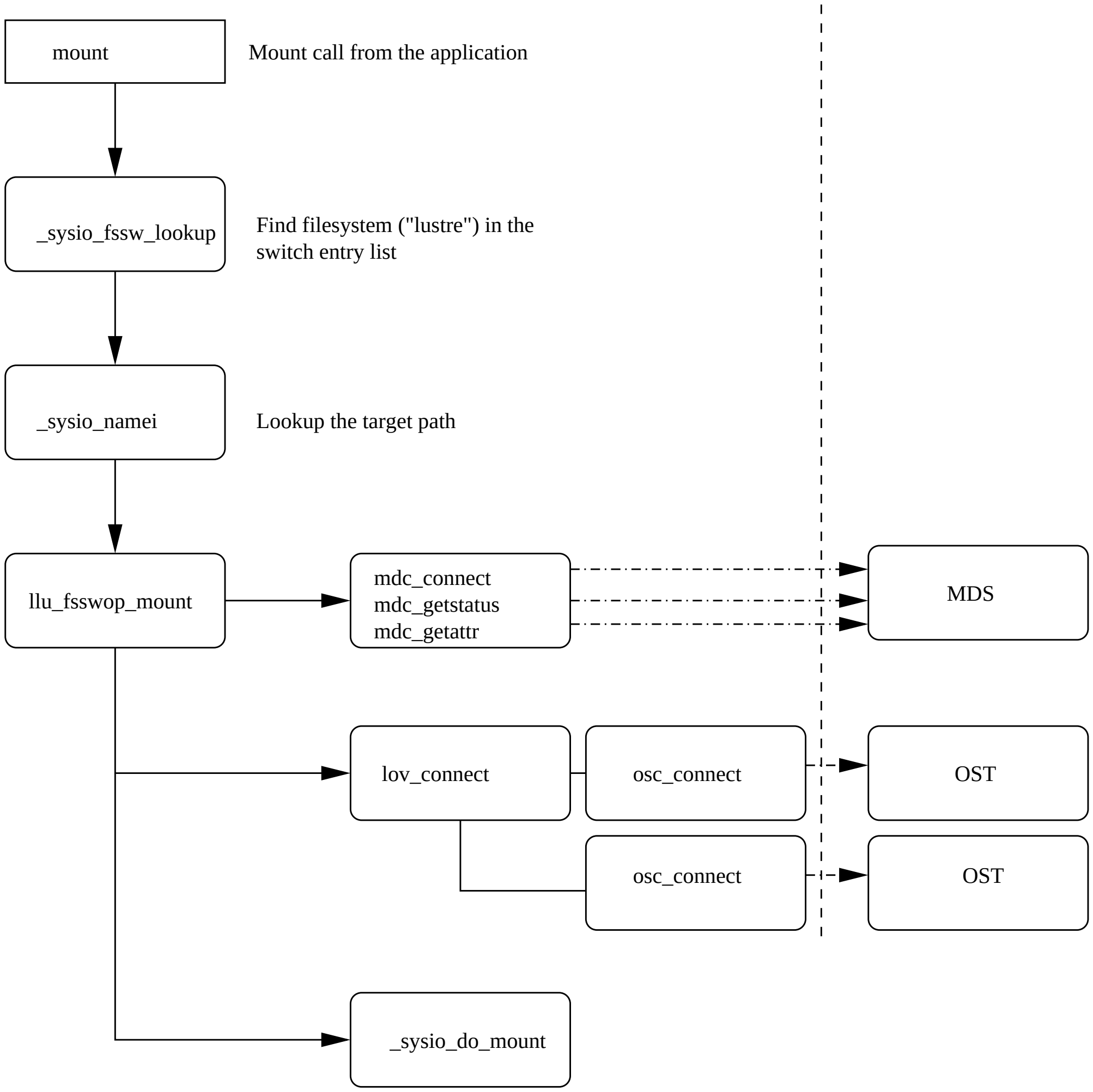

FIGURE 19.7.1. Path followed by mount

**19.7.1. Mount.** The application's call for mount would result in the *libsysio* layer calling the appropriate filesystem function, this is shown in figure 19.7.1.

CHAPTER 20

# Quality Of Service - Implementation, APIs

This chapter gives an overview of the Quality Of Service (QOS) protocol implemented in Lustre. The QOS can be defined in a variety of ways - ensure that all OSTs are uniformly utilized, provide preferential treatment to certain applications, provide quality guarantees to some applications. Our initial QOS implementation will answer the first requirement - mainly it will ensure that the OSTs are uniformly utilized and that an OST that might be filling up is intelligently avoided. An outcome of this protocol would be to help balance the I/O load across all the OSTs.

We will start with a description of the QOS implementation in Lustre and then describe the various APIs developed to provide this support.

## 20.1. Quality of service - Overview

**20.1.1. Overview.** In the absence of QOS implementation, Lustre randomly determined a starting OST for every file and then chose consecutive OSTs starting with the randomly chosen OST to place the objects on (depending on the striping pattern). This implementation might result in unequal utilization of the OSTs, some OSTs might be nearly full while some others might have a lot of space available.

The QOS protocol was intended to reduce this disparity - the outcome would be a more uniform utilization of all the OSTs, it might also help balance the I/O load more uniformly across all the OSTs.

The implementation of QOS requires a few important pieces:

- A feedback mechanism - this is required to provide information about how full the various OSTs are
- *QOS intelligence* - Based on the continuous feedback recieved, the clients need to make intelligent decisions on where to put the file objects on. This requires the client to also maintain some lists with this feedback and continuosly/periodically update the feedback.
- Performance metrics used - this determines the information that needs to be tracked for our QOS purposes. The initial implementation that deals only with the utilization of all OSTs requires - number of free blocks available, number of objects allocated.

In order to avoid extra RPC load resulting from clients explicitly querying the OSTs for these performance metrics, we require this information to be piggy-backed on any reply from the OSTs to the clients. Ofcourse, this implies that we run a small risk of making decisions based on stale information (this is possible when the I/O load on the clients was very low for a period), but the savings in terms of not requiring extra RPCs out-weighs the risks introduced by the stale information.

## 20.2. Implementation details

**20.2.1. QOS implementation.** As described earlier, one of the key requirements of the QOS design was to add as little overhead as possible. So, we avoid the need to send out any extra RPCs from the clients to OSTs to gather the metrics information, instead the OST piggy-backs this information to the client on every reply. On the *create*, *delete*, *getattr*, *setattr*, *open*, *close*, and *punch* replies, the OST inlines the current performance metrics into the RPC reply. The performance metrics that are measured are:

- number of blocks available.
- number of objects allocated.

The wire protocol has also fields for operations per second, kilobytes per second, lock requests per second and latency. These fields are reserved for future usage and are not currently used.

The client uses round robin object allocation in *lov_create* unless a QOS threshold is violated. The thresholds are (in order for priority):

- number of blocks available on OST
- number of objects allocated on OST
- percentage of storage available on OST - this is calculated based on the *blocks available* and *total blocks* information

On reception of a status update in an RPC reply, the client checks whether is has updated its priority lists in the last *QoS-update-interval*. If not, it adjusts the the priority lists with the new information. This update is performed only every *QoS-update-interval* to reduce the compute overhead on the client.

During object creation, the client checks whether the priority lists have been rescanned in the last *QoS-rescan-interval*. If so, it uses the list as-is, if not it walks through all OSCs and sees whether the status information is older than the *QoS-refresh interval*. If it is older, the client can do a *statfs* call to update the information and readjust the list. If the information is newer, but the entry has not been sorted in the last interval, it will be sorted .

The client checks the thresholds to determine if QoS allocation policy must be used or the normal round robin allocation. If QoS allocation must be used, the starting OST is choosen from the list in priority order. The best OST is used as first, the others in ascending order of their index

The update intervals and the thresholds are configurable through */proc* on each client on the fly.

**20.2.2. Data structures and tunable parameters.** The QOS implementation required some new data-structures and modifications to a few existing ones. It also introduces some new variables (tunable parameters) as described below:

(1) *qos_stats* (*include/linux/lustre_idl.h*) - This structure holds the QOS information which is piggy-backed on the RPC replies. The inline field is currently 60 Bytes in size and limits the size of this structure. The current structure fits on IA32 and IA64.
(2) *obd_inline_data* - This is the new structure which is piggy-backed on the replies from OSTs to clients. It holds the *qos_stats* structure.
(3) *lov_tgt_desc* (*include/linux/obd.h)* - This is an existing structure used to hold the list of all OSTs, it has been modified to have some new fields.
   (a) index: number of this OSC in *lov_obd*, for faster retrival
   (b) tgt_lock: spinlock protecting updates on *tgt_qos_stats.* The locks should be taken in the following order - lov_lock (in struct lov_obd), tgt_lock. The lock order between tgt_locks of multiple lov_tgt_desc instances is undefined since there is no need to hold more than one lock and any time.
   (c) tgt_qos_stats: QOS information as received from OST with RPC reply is stored in this structure on the clients.
   (d) tgt_btotal: total blocks on OST divided by 100, used for percentage free calculation. To avoid a 64 bit division on IA32 we also calculate tgt_shift to simplify the division.
   (e) qos_stats_time_sorted: last time the entry was sorted into the lists
   (f) qos_stats_time_stored: Last time the entry was stored/updated. Whenever we get information from an RPC we store it in the *tgt_qos_stats* field. If the lists have not been reordered in the last *QOS_UPDATE_INT* (adjustable through /proc), the lists get resorted and the sorted time set to the stored time. This mechanism prevents too much overhead on resorting the lists on each update. The sorting effort is little - the lists are always kept sorted, if a value changes we only need to move the element who's value has changed. This is performed by *lov_adjust_lists* function.
   (g) tgt_bfree: free blocks on OST. This value is updated on list re-sorts. It is protected by the *lov_lock* in *lov_obd* struct.
   (h) tgt_nobjects: number of objects allocated on ths target OSTs.
   (i) tgt_freeblock_percent: percentage free.
   (j) tgt_bfree_list: linked list of targets, sorted by blocks free
   (k) tgt_nobjects_list: linked list of targets, sorted reverse by number of objects allocated
   (l) tgt_percent_list: linked list of targets, sorted by percentage free. All lists are protected by the *lov_lock.*
(4) *lov_obd* (*include/linux/obd.h)* - This is also an existing structure that has been modified to contain some new fields -
   (a) qos_list_time_stamp: last time the whole lists were scaned for outdated updates. Scan intervals are *QOS_RESCAN_INT*. On scans it detects all targets older than *QOS_REFRESH_INT* and forces a read of current QOS information with a statfs call to the OST.

(b) *tgt_*_list*: list heads for lists as described above *max*, *min*: maximum and minimum values found in targets. These values are used to determine whether QOS policies must be applied or whether normal allocation done. Storing these values allows us a fast decision without scanning the lists.

(5) *qos_statfs_int* (*include/linux/obd_support.h*) - interval to cache the metrics information on the server for inlining it to RPC replies. Benchmarking showed that too many statfs slow down the OST. The interval is in seconds and can be adjusted through */proc*.
(6) *qos_update_int* - This is in seconds while *QOS_UPDATE_INT* is in jiffies.
(7) *qos_rescan_int* - This is again in seconds, but *QOS_RESCAN_INT* is in jiffies.
(8) *qos_refresh_int* - This is in seconds, but *QOS_REFRESH_INT* is in jiffies.
(9) *qos_nobjects_imbalance* - If minimum and maximum of objects allocated on targets exceed qos_nobjects_imbalance, QOS allocation policy sorted by nobjects is used.
(10) *qos_freeblock_imbalance* - same, but for free blocks qos_freeblock_percent: same, but for percentage free Priorities are: blocks free, number of objects, percentage free.
(11) *aa_lov* (*include/linux/obd_lov.h*) - We also collect QOS information on asynchronous getattr calls, the *aa_lov* field is used to pass the information between the phases.

## 20.3. QOS APIs

This section describes the modifications made to some existing APIs to support QOS and also describes the new functions added.

### 20.3.1. lov_connect.

20.3.1.1. *Prototype.*

```
static int lov_connect(struct lustre_handle *conn, struct obd_device *obd,
struct obd_uuid *cluuid)
```

20.3.1.2. *Parameters.*

**input: conn:** Existing lustre handle.
**input: obd:** Target device.
**input: cluuid:** UUID of the osc

20.3.1.3. *Return Values.*

20.3.1.4. *Description.* This function has been modified to now initialize the new fields added for QOS. The *obd_statfs* call initializes the statfs cache and the stored QOS information.

### 20.3.2. lov_setup.

20.3.2.1. *Prototype.*

```
static int lov_setup(struct obd_device *obd, obd_count len, void *buf)
```

20.3.2.2. *Parameters.*

**input: obd:** Device structure for the LOV to be setup.
**input: len:** number of OSTs.
**input: buf:** input parameters for setup (UUID, MDC name etc)

20.3.2.3. *Return Values.*

20.3.2.4. *Description.* This function has been modified to initialize the LOV list heads.

### 20.3.3. lov_adjust_lists.

20.3.3.1. *Prototype.*

```
void lov_adjust_lists(struct lov_tgt_desc *tgt, struct lov_obd *lov,
int force)
```

20.3.3.2. *Parameters.*

**input: tgt:** The structure that has list of all OST targets.
**input: lov:** Device structure for the LOV.
**input: force:** Flag to indicate if the list sort should be forced irrespective of when it last happened

20.3.3.3. *Return Values.*

20.3.3.4. *Description.* This is a helper function used to sort the target list based on the metrics information collected. The *force* variable indicates if this should be done only when the associated threshold is exceeded, or if the update and sort should be forced.

### 20.3.4. lov_update_qstats.

20.3.4.1. *Prototype.*

```
static void lov_update_qstats(struct obdo *oa, struct lov_tgt_desc *tgt,
struct lov_obd *lov)
```

20.3.4.2. *Parameters.*

**input: oa:** Structure used to inline the metrics information from OSTs to clients as a part of the RPC replies.
**input: lov:** Device structure for the LOV.
**input: tgt:** List of targets stored in the LOV

20.3.4.3. *Return Values.*

20.3.4.4. *Description.* This is called on receiving an RPC reply to copy the new metrics information from OST into the LOV structure.

### 20.3.5. lov_get_qos_list.

20.3.5.1. *Prototype.*

```
static int *lov_get_qos_lists(struct obd_device *obd, struct lov_obd *lov,
int startset, int ost_idx, unsigned stripe_count, int *idxs_idx)
```

20.3.5.2. *Parameters.*

**input: obd:** Device structure for the LOV device.
**input: lov:** Structure with LOV specific information.
**input: startset:** User preffered starting OST
**input: ost_idx:**
**input: stripe_count:** Number of stripes in the file that dictates the number of OSTs to stripe across

20.3.5.3. *Return Values.*

20.3.5.4. *Description.* This function returns indexes of targets in QOS prefered order if QOS policies should be used. It honors user requests for number of OSTs, prefered first OST etc.

## 20.4. Changelog

**Version 1.0 (10/06/2003)**

(1) Radhika Vullikanti - Added this new chapter on QOS implementation.

## 20.5. File I/O

**20.5.1. Read Ahead.** The current read performance of Lustre Lite is poor, but we believe this is for well understood reasons. The improvements we intend to make for LLP are as follows:

**Batching reads:** Presently Lustre reads page by page. A simple change will initiate all the reads associated with a network ost_brw_read command, and do a single wait. This involves a simple change to the function ost_preprw to initiate I/O on the individual pages and then wait for all of it to complete. At present, for historical reasons, it initiates I/O on a page and waits for that to complete before issuing another read.

**OST read-ahead:** In Linux 2.4 OST read ahead was fairly hard to arrange. With 2.5 the generic read-ahead mechanisms are page cache based. The file structure is responsible for keeping track of optimal read-ahead windows and the abstractions are separated out from the generic file write routines.

The osc_read function is implemented on the OST in terms of obd_preprw and obd_commitrw, which prepare and finalize the pages respectively. Obd_preprw is responsible for reading the pages in, when it returns the DMA will take place and, for reading obd_commitrw simply releases the pages.

We will introduce a read-ahead in obd_preprw which is extremely similar to the read-ahead for normal file I/O to a disk file system. The design for this will use the open file structure on the OST, which the client established for each data object associated with the inode at open time. The OST read function will become quite similar to do_generic_file_read with the exception that it is passed a list of pages, not a buffer, namely the pages which need to be DMA'd to the client.

The read-ahead mechanism will initiate I/O on pages beyond the ones passed so that these will be enter the page cache to speed up further reads.

Read ahead can consume a significant portion of the available I/O bandwidth. The OST call to obd_preprw has the opportunity to adjust read-ahead windows in the face of traffic conditions on the OST, to ensure it does not exceed maximum load requirements based on the current traffic at the OST, or does not fall below a level where QOS guarantees might not be met.

**Client filesystem read-ahead:**

We will ensure that the client side read-ahead behavior matches what can be done on the OST side. In particular, the client should not initiate read-ahead until the OST has initiated the data stream to the client. [The new Portals EVENT_START_PUT may make this particularly simple to implement.]

**Quality assurance:** We will make it possible to trace the read-ahead behavior and show that it matches expectations. At present this could be done with tcpdump but is cumbersome.

**20.5.2. Block Allocation Optimization.** A key consideration in OST performance is the block allocation mechanisms. Fundamentally there are *extent-based* allocation schemes where blocks are allocated in ranges, and *block* allocation schemes where blocks are allocated one-by-one. The complexity of the extent mechanisms is considerable, and it has not been established with rigorous analysis if the extent approach pays off (for example, bench marks have almost universally placed ext2 ahead of XFS in write performance).

20.5.2.1. *Approach.* We will perform a detailed analysis of the overhead associated with block allocation. We expect that this will point in the direction of optimizations, but if evidence shows that extents are needed to get really high performance, then we will add extent features to ext3. Possible design and implementation approaches of extent features have been discussed in the ext2 community and will not be discussed here.

20.5.2.2. *Known optimizations.* There is certainly low hanging fruit in block allocation optimization. First, OST's are typically performing large writes. However block allocation is done one by one. This has at least three negative effects:

(1) In memory journal transactions are started and stopped more than once. Note that because memory transactions are aggregated into large disk transactions this does not necessarily mean that more disk transactions take place.
(2) Allocation data is searched more than once.
(3) There is a risk of fragmentation. When two writes to the same place are taking place simultaneously the blocks allocated my easily not be contiguous.

The solution is to enhance the disk file system with a bmap function that allocates a range of blocks in one fell swoop. The ext3 team has already welcomed this small change to ext3.

It is expected from CPU utilization studies done by Andrew Morton that with this algorithm a typical server workstation will be able to write many 100's of MB/sec, probably well beyond what has currently been tested.

### 20.5.3. Optimizations of file size and change times.

20.5.3.1. *File Size.* The in memory management of file sizes for files which are open is done with distributed locks as described in the LL CDR. When files are not open the size may well be queried by the stat call, e.g. in the *ls -l* scenario.

Even in this case Lustre Lite manages file sizes by querying the the data objects. This means that the file system makes a *obd_getattr* call to the LOV which queries all the OST's it is managing. This has an unpleasant side effect, namely that each stat call gets all but one attribute from a single RPC to the MDS but then must query all the OST's to get the file size.

Fixing this is easy in the absence of crashes: simply store the file size on the MDS. When the client closes the file it first closes it on the OST's. This gives the client up to date file size information, which it can transfer to the MDS. The MDS is aware when the last process is closing the file and can then store the correct file size.

Fundamentally the issue is one of propagating state information about an object from the OST, which modifies the file size to the MDS. In order for such information not to be lost a replay log needs to be maintained. The mechanism we describe here is almost identical to the InterMezzo replication algorithm and it is also used, in a slightly more complicated form, to handle file creation recovery in Lustre [see updates to the LL CDR documentation].

The replay log record is atomically created in persistent form *before* the file data is written. Fortunately *ext3* has the ordered write semantics which make it possible to guarantee this. As a result the OST system is aware of the update of the inode. Under normal operation the replay record is invalidated when the the server file size has been updated.

(1) A file is opened by one or more processes
(2) The first write to the file causes an intent log to be written
(3) With the last close the intent record is transferred to the client which updates the MDS.
(4) The client updates the MDS file size
(5) When commit notification comes back from the MDS, the client asks the OST to invalidate the record.

To make this a recoverable operation a client failure is treated as a failure in communications. The recovery from this failure is that the MDS must ask the OST for the replay log and replay any setattr commands on the MDS inodes that it obtains from the OST's (through the LOV).

The MDS must block further requests for file sizes while this operation is in progress, or instruct the client to fetch file sizes through the LOV.

**Further considerations**

(1) For small files the overhead of the replay log may be significant. In this case it may be better to retain the current algorithm, particularly since the number of data objects is likely equal to one.

20.5.3.2. *mtime, ctime and such.* The above methodology can also be used to recover updated mtime's and ctime's from the OST. These might not be exactly equal to those that the client would have reported to the MDS but would likely be very accurate as the OST can update the mtime when it does file writes.

The above scheme goes through without modification.

**20.5.4. Reading of sparse files.** File system scalability is affected a lot by the available network bandwidth. So, to improve the scalability of a file system, we try to reduce the amount of network usage in performing various file I/O operations. File *reads* are one of the most common file operations, so our intention here is to reduce the network usage for reads. Several of the large reads might actually be to sparse files. Sparse files contain *holes*, or several pages that are not mapped. Currently, such pages would also be sent on the wire as zero-filled pages, contributing to the network overhead. The idea that we present here is that in case of a read to a sparse file, it is not necessary to tranfer the holes as zero filled pages, instead we could return a flag corresponding to the holes which would indicate to the client that the page did not have any actual data. This allows us to reduce the number of pages that we would need to transmit in bulk and thereby also reduce the network usage. We propose to make changes to some APIs that would allow us to use a flag to indicate the holes in a file. The changes would be limited to the *Object Storage Targets(OST)* and *Object Storage Client(OSC)* layers. We will use try to use functionality already available in the underlying file system to avoid any kernel changes.

The next section briefly describes the current algorithm to service read requests for sparse files. The subsequent section will describe the changes we propose to enable a more efficient *read* for sparse files.

20.5.4.1. *Sparse file reads in Lustre Lite.* A file *read* request travels from the client to the *OST* through the *OSC* layer. At the *OST* all the *read* request follows the path as isslustrated in 20.5.4.1. As this figure shows, if the page is unmapped (i.e it is actually a hole in the file), the *block_read_full_page* function would zero-fill it and return this page. So, the upper layers are unaware that the page was zero-filled and did not contain any data.

(One of the initial options we initially thought of was to add a new flag to the cache, when the function does a memset for the page it could set the flag to indicate the presence of a *hole* there, but this would mean changing some underlying kernel data structures. Moreover, once a *write* is done to the page, we have to have a way to reset the flag. So , we chose the scheme described in the next section instead.)

Figure 4.4.1: Code Path for file reads in Lustre

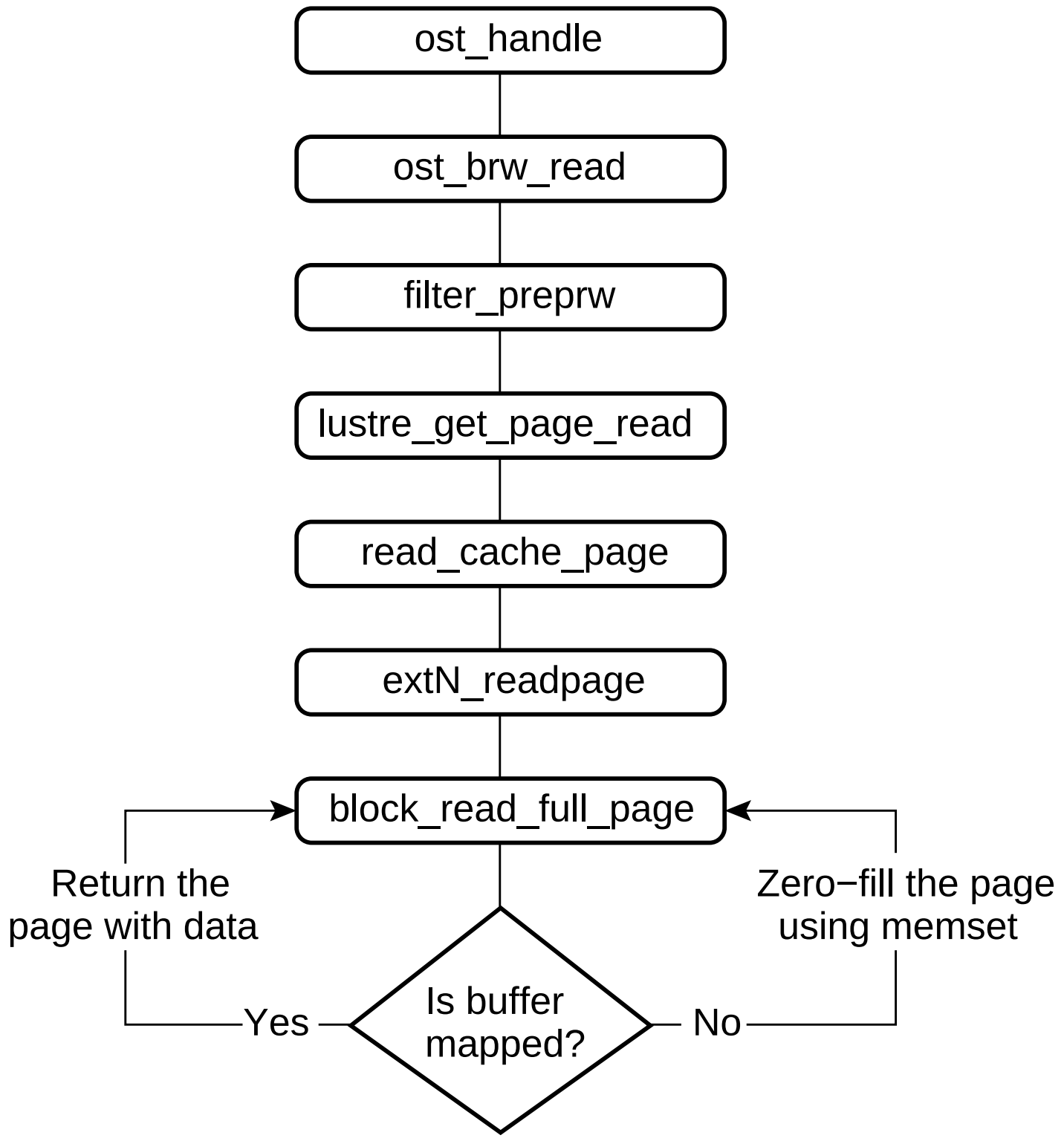

20.5.4.2. *Proposed changes to the read algorithm.* Sparse file reads can be optimized if we do not send the holes in the file as zero-filled pages over the wire. To make this possible, we need some way to determine if a page is mapped or not. We also need to be able to inform the *OSC* about this. We propose very minor changes to some functions and a new data structure to hold the status information for every page in each read request. This would be sent back to the *OSC* layer either by tagging it onto the bulk send or by sending a separate reply with this status information.

The algorithm that we propose is shown in 20.5.4.2. We use a function *bmap* already supported in most file systems to determine if a page is mapped or not. For every page that needs to be read, the function *filter_preprw* will first call *bmap* to check if the page is mapped or not. If the page is mapped, we go ahead in the normal path. We check if the page is in cache, otherwise bring it into cache. On the otherhand, if the page is not mapped, we set the *flag* field in the new *niobuf_status* structure to indicate that the page was not mapped. The *status* field can be set to -1 in case of unmapped pages or the length in other cases.This prevents the request for an unmapped page from even going to the *lustre_get_page_read* function which would unneccessarily bring in a zero-filled page for any unmapped page.

The new structure we suggest should have the following fields:

```
struct niobuf_status{
__u64 offset
__u32 status
__u32 flag
}
```

We would use the *flag* field to indicate if the page was mapped or not. On the receiving end, the *OSC* can read this status information and then create zero-filled pages for the unmapped pages and pass the whole set of pages to the client. This keeps the scheme transparent to the client at the same time saving usage on the wire.

The next issue that we need to handle is to decide a way to return this information to the *OSC*. We also need to make changes such that *ptlrpc_prep_bulk* would now only send the pages that were read in. We can return the status information either by tagging it onto the bulk send or by a separate reply. The next section describes some of the issues related to Portals, these need to be clarified to ensure that Portals support the new scheme.

20.5.4.3. *Portals issues.* In the previous section, we have assumed that Portals would not have any problems with the new scheme. We need to confirm that, so I have listed all the possible things that Portals might need to handle:

- We need to be able change the bulk send RPC to now send only the pages that were mapped.
- It is possible that in a request for 6 pages of a file, only page 1 and page 6 were mapped and the data for them is sent back. For the other, a status would be sent indicating that they were unmapped. Some sink pages might have been set aside by Portals to receive these pages, would it be able to determine which pages have arrived and fill the correct buffers? How would it handle such a situation?
- The *OSC* will be expecting say 6 pages in read reply, but it only gets 2. It should be able to determine which pages are available and which are not by reading the status information (But the status would arrive after the bulk does) and zero fill the other pages?

Figure 4.4.2: New algorithm for file reads

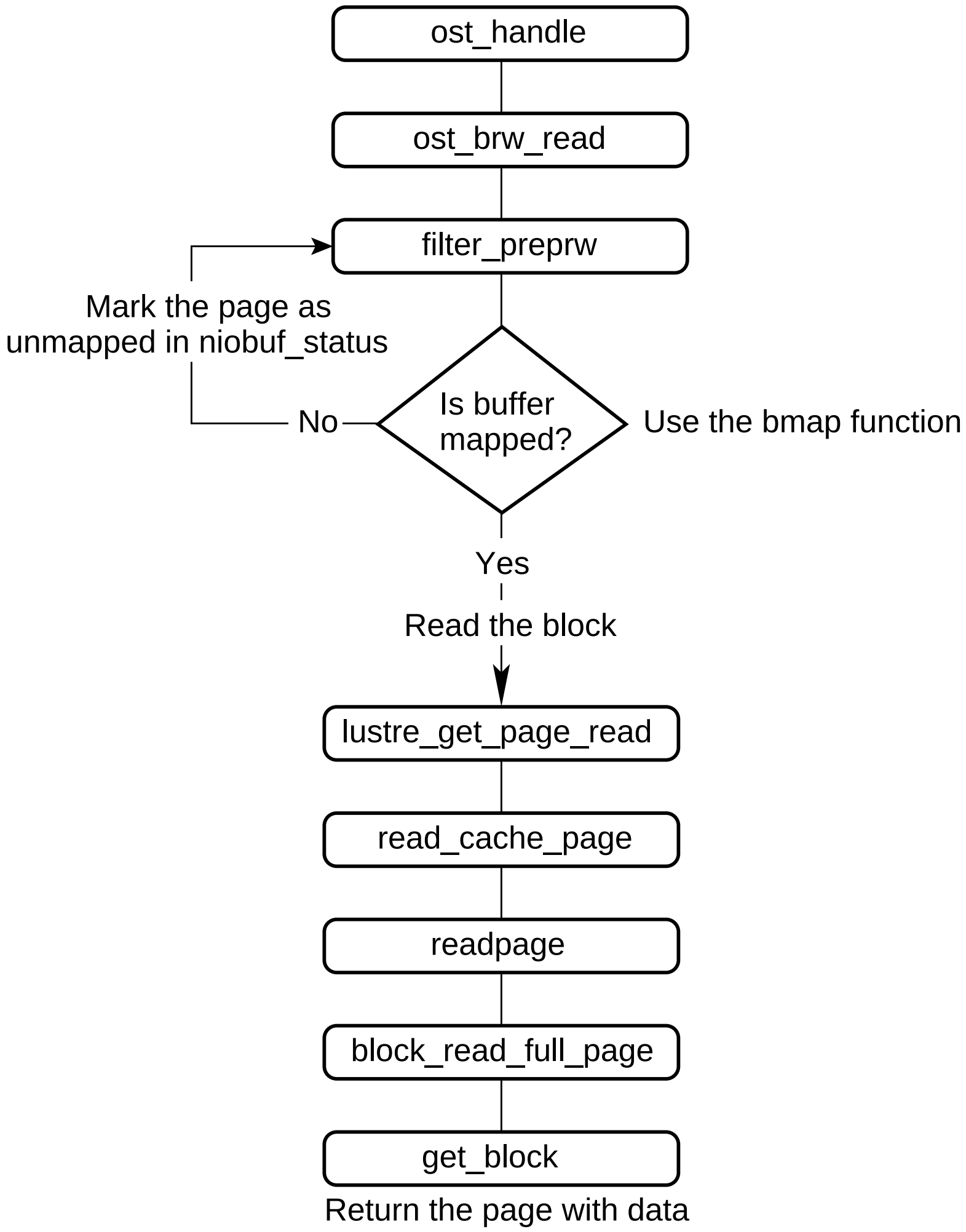

- What is the best way to send back the status information to the *OSC* ? Could we send it along with the bulk? Or should we send it as a seperate reply from the *OST* to the *OSC ?*

——— We have decided to send this status information as a part of the reply to the rpc sent from the *OST* to the *OSC*.

**20.5.5. Using PtlGet for file writes.** The Portals package is used with Lustre to provide a good networking stack. Portals message passing API allows two asynchronous I/O functions PtlPut and PtlGet. The PtlPut function is invoked by the initiator and enables it to transfer bulk data to the target process. On the other hand PtlGet invokes a remote *read* request, in this case the target should send the PtlGet request to indicate the data it wants to read. The issue here is to change from the use of PtlPut function to PtlGet function to give the receiver more control for writes. The next section describes the current code path for *writes* using PtlPut function. Section 4.5.2 describes changes that would be needed to replace the PtlPut function with PtlGet function for *writes*.

20.5.5.1. *Current Code Path for sending buffers using portals.* The write request is handled at the OSC by the osc_brw_write function. This function prepares the bulk pages needed for the write operation, and uses ptlrpc_send_bulk function to send the data to OST. The ptlrpc_send_bulk invokes the PtlPut function to trasmit the data. PtlPut function initiates an asynchronous I/O to the remote node. It will put the data from the source pages into the sink pages indicated for the remote end. The function would return PTL_OK if successful. This is illustrated in figure20.5.5.1 20.5.5.2. *Changes to replace PtlPut with PtlGet.* Now we want to use PtlGet for write I/O instead of PtlPut. The new path is shown in figure 20.5.5.2. Here instead of the osc_brw_write handler calling PtlPut to put the data for write into the sink pages at the OST, we will now have the ost_brw_write handler call the PtlGet function to read the data from OSC. The PtlGet request would be sent as soon as the OST prepares the sink pages based on the information it receives from the OSC in the *write* request. The osc_handler will wait on the PTL_EVENT_GET event that will indicate the completion of the *get* event and allow the OSC to free the related memory buffers.

Figure 4.5.1: Portals code path for file writes

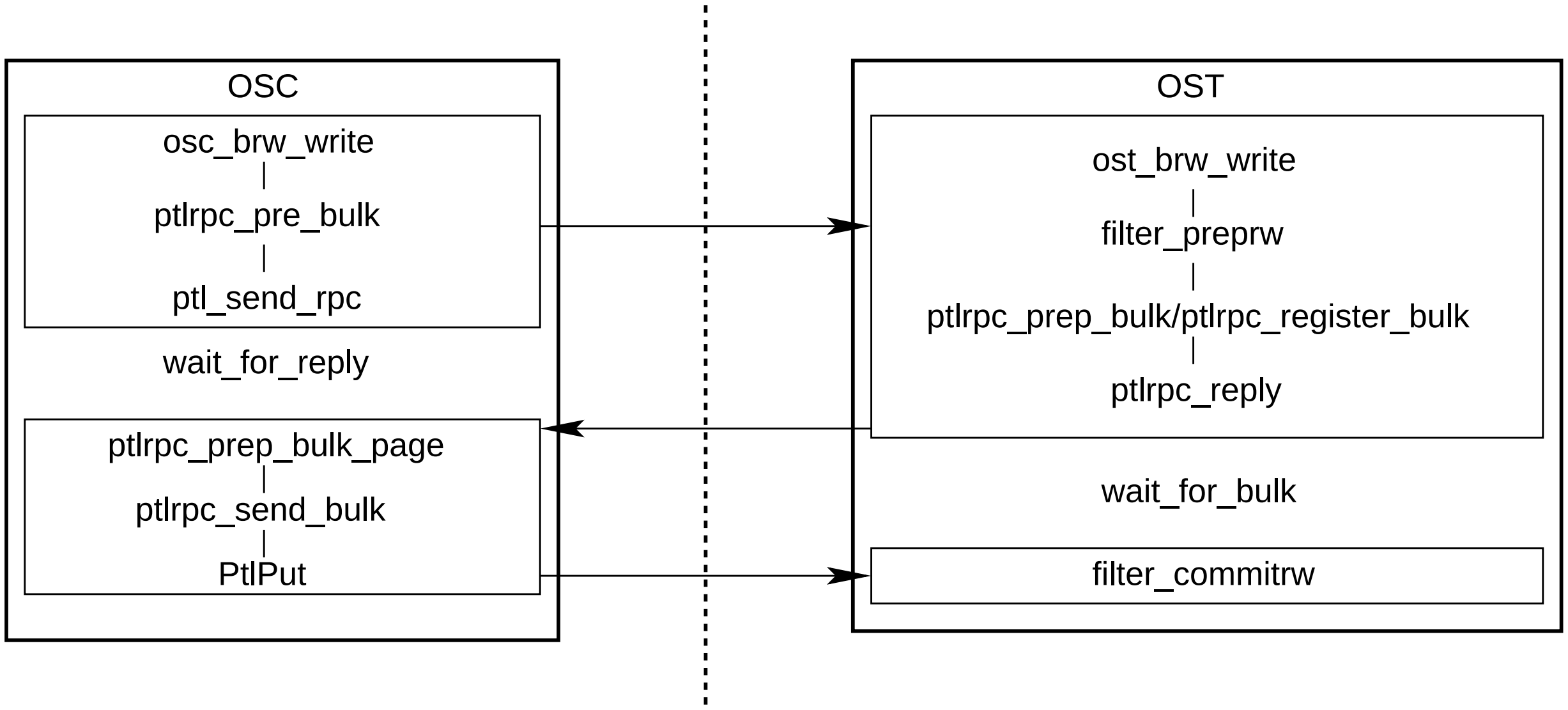

**20.5.6. OST Status return code.** We propose that the OSTs would return some status information to the clients. The applications can use this information in several ways. Maybe the application can decide to schedule certain jobs when the load on the OSTs it needs is below a certain threshold. The I/O access patterns at the OSTs change very quickly, so it would be useful if this information were sent back with every request from the client.

a) We also need to determine the kind of information that would be useful, this is shown in the new structure given below:

Figure 4.5.2: Write request using PtlGet function

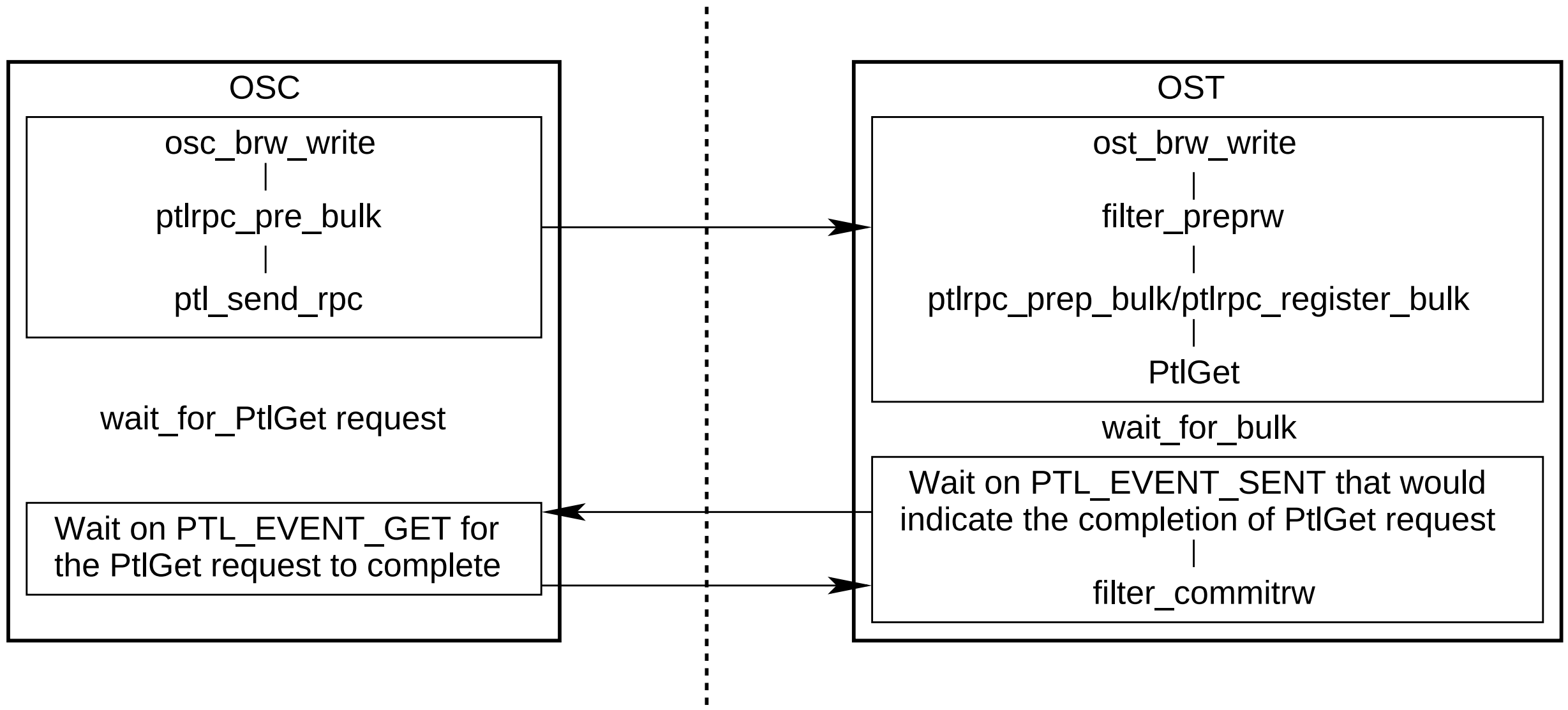

```
struct ost_statbuf{
  __u32 os_nops;  - This can be used to determine how                      busy the OS
__u32 os_ucpu;    - This gives the % CPU utilization
__u64 os_fbavl;   - available free blocks at the OST
__u32 os_netbw;   - available network bandwidth at the OST
}
```

The ost_handler.c is the function that sees all the I/O requests arriving at the OST. So, this could keep track of the ops/sec field. This information has to be returned with every request to the OST. So, we can tag it to the rpc reply that will be sent by the OST to OSC for every rpc received. This can be done in ptlrpc_reply function. We will have to expand the ptlrpc_request structure to add this new struct in it.

## 20.6. LOV Module

**20.6.1. Delayed object allocation.** We are proposing that objects for larger files are only allocated when required. This is a simple change to the LOV write method. If no object is yet allocated in a stripe, it will do so. The it will return a value to the caller which indicates that it has updates the extended attribute for the LOV object, which the caller will then store on the MDS.

Due to acceptance tests requiring the immediate creation of objects that behavior must remain an optional feature.

As a result collections of small files will always use a single object.

**20.6.2. Object allocation and striping policies.** We will, at least for the purpose of profiling, want to understand the performance implications of:

(1) The stripe sizes change
(2) The number of striped objects

We propose mechanisms to change these from the defaults on a per directory or per file basis as detailed below.

With respect to profiling, we are solliciting suggestions from TriLabs. Fundamentally there are two interesting things to study:

(1) OST dependence on stripe size
(2) OST dependence on the number of clients writing stripes.

**20.6.3. Changing object allocation and striping policies.** After discussions with many parties opinions on policies for object allocations appear to differ widely. CFS will implement a few simple mechanisms to change behavior of the file system:

**system interface:** There will be operations which can be executed on a directory or on an open file to change the behavior of every file in that directory or the file.
**lfs:** the lfs utility can make such settings on directories from the command line.
**environment variables:** GPFS has introduced a mechanism where environment variables can be used to influence stripe sizes etc.

**20.6.4. OST status information at MDS.** The MDS also needs to have some status information about all the OST to be able to do some intelligent space management within the cluster or to provide QOS gaurantees. The following information would be useful to the MDS:

- Available space on the OST(%) - Based on this MDS could use some intelligence to do better space management in the cluster. It can try to distribute the data equally among all OSTs, if some OST group is getting full it can use that information to decide on using some other available group of OSTs.
- Aggregate load on an OST in terms of ops/sec - This determines how busy an OST is. So maybe we can take some QOS decisions based on this information. Maybe for some performance critical applications with minimum bandwidth requirements we could decide to use less busy OSTs store and get data from.
- Available network bandwidth at the OST - This can be used to grant QOS gaurantees to applications

The other issue here would be to decide how to get this information to the MDS. Since the MDS already has all the information about all the OSTs in the cluster, we would allow the MDS to periodically contact all the OSTs and obtain this information. The frequency can be decided maybe based on the type of access pattern expected.

**20.6.5. QOS from OST.** This issue is related to the previous issue we discussed. In the real world, different applications have different requirements from the file servers. QOS is a gaurantee that the application would never receive less than what it was promised, it could always get more if possible. The QOS is especially very important for multimedia applications - video, audio. These applications need some gaurantees from the underlying file system, otherwise the content delivery would not be very good. So, the QOS requirements could depend on the kind of applications that access the OSTs, some examples are listed below:

a) In some cases, the clients would need a preset minimum bandwidth from the OSTs. This gaurantee can be given to the client by the MDS. The MDS would take the decision based on the information it has about the OSTs. The client will present QOS gaurantee to the OST with every request. This would allow the OST to service requests based on the QOS gaurantees a client has been promised. The OST could maintain a priority queue to handle the requests.

b) In other situations the clients could be promised only upto certain maximum bandwidth. For example upto 5% to this client, rest 95% to everyone else.

c)If there is only a single client who has been gauranteed 5% of the bandwidth, but at this point there is no other client that the OST is handling, then this client can be given as much as is needed to service their request. But as soon as there is contention with some other client's QOS gaurantees, the first client should be bumped down.

We now need some structure that will store information to indicate the QOS promised to the client. This can be given to the client on an open, and then the client would present this to the OST with every subsequent request. The OST can use this information to determine how to service requests from that client.

```
struct lustre_qos{
   __qos_min_bw;
   __qos_max_bw;
}
```

This would be valid only on that open, subsequent could give different QOS gaurantee depending on the status information.

The various components that we would need for implemeting this:

(1) A new API that would allow the MDS to query the OST status
```
mds_getinfo(ost_conn, struct ost_statbuf);
```
(2) an interface at the MDS to allow it to send out these queries? We would need an interface similar to the OSC at the client for the communication between the MDS and the OSTs.
(3) a new API that would handle this new request from the MDS and reply to it.

```
ost_handle(new switch option here for the new request type - OST_STATQ)
|
|
v
```

```
ost_get_info(new function)
|
|
v
filter_get_info
```

(4) filter_get_info - This is a function currently available that accepts a key and returns its value. This is not currently used, we can try to make use of it to get the status information.

## 20.7. Special API's

**20.7.1. Implementation mechanisms.** There are a variety of mechanisms to implement file system specific behavior:

**ioctls:** can be made on an open file (including directories and special files). Ioctls are passed on to the file system by the VFS with little interference.
**fcntl:** is not passed on to the file system. This call is suitable for setting flags that are part of POSIX.
**mount options:** mount options can be passed that affect all files.
**environment variables:** GPFS has introduced environment variables to set stripe sizes etc. We believe this is a fundamental violation of the normal usage of environment variables. Given that this will not be acceptable to the kernel community we propose not to support this.

**20.7.2. Special API's.** The following special API's are proposed:

**invalidate page range:** drop pages from the cache (after flushing them if they are dirty).
**no locks:** Already present: an ioctl to disable all lock fetches associated with an open file. One might argue that this is only sensible in conjunction with O_DIRECT but fsync and the above call provide another mechanism to manage file data from user space. This option is also available as a mount option. [TODO: exempt the file size lock from this policy.]
**lazy attribute handling:** What exactly do we want to be lazy about? File sizes could be reported without a file size lock, that is a big savings when files are open. Attributes that are cached but for which there is no lock could be reported [that's somewhat doubtful].

# Part 2

# Design

CHAPTER 21

# Introduction

The first part of the book gave a very high level view of the Lustre filesystem. In this design part, we will concentrate on the implementation level details. Lustre has been designed as a POSIX compliant filesystem. Lustre leverages heavily on the open standards of Linux operating system and some underlying journalling filesystems - Ext3, JFS, ReiserFS, and XFS.

Lustre uses the open source Portals stack for networking, but has its own request processing layer on top of this stack. This layer provides all the methods for establishing new connections, setting up buffers for IOV, support recovery, and error handling. The Portals stack sits on top of the Network Abstraction Layer (NAL) that can support a variety of devices. In Lustre, connections are associated with the *export* and *import* structures on the server and client. The *export* is a datastructure used at the server/target to manage all the connections it handles. The *import* is a local data structure with a corresponding *export* on the target.

We will also describe the details of data sent over the network. The Portals stack is used for networking, so Lustre messages are encapsulated within the Portals packets. A Lustre message (*request*/*reply*) is message buffers and often has multiple buffers in a single message.

In this part of the book we will also describe all the API's used for various object drivers. Each lustre device is associated with a device structure that determines the various methods the device supports. The methods for object devices are presented as simple stub API's that would invoke the appropriate method on the appropriate driver using the device structure associated with it. This design makes adding new drivers very easy. For example, if it is the COBD driver, the stub API *obd_brw* should invoke the *cobd_brw* method. So, there could be different implementations of the same API for different drivers like *obdfilter*, *lovobd*, *osc*, *snapobd*, or *cobd*.

Lustre implements distributed locking mechanism; each Object Storage Target (OST) is allowed to manage locks for the object it stores. The meta-data server manages the locks on file meta-data. Lustre associates several data-structures with handles; a handle would consist of the address of the data-structure location in memory along with a randomly generated cookie. These can be used by the remote end to get to the corresponding data-structure in memory easily. The presence of the cookie allows the validation of handles to be done, this protects against any possibility of **fake/old** handles being sent. For example, before a client sends a lock request to the server, it will create a local lock structure to hold this lock. It will pass a lock handle (address of the local lock structure plus a random cookie) in the request buffers sent to the server. The server will return a handle to its own lock structure and also include the local handle in its reply. So, when the client gets the reply, it can use the local handle to quickly access the lock information. We will also explain the extent-based locking mechanism provided in Lustre.

Lustre leverages extensively on the underlying journaling filesystems for persistent state recovery. The recovery protocol in Lustre is based upon an **epoch number** (boot count for each storage controller), an **incarnation number** (bootcount for the meta-data cluster), and a **generation number** (unique for every connection between the client and other systems), making it resilient to failures of OST's or MDS's and allowing it to quickly recover from such failures or from network failures. We describe this protocol, along with the API's used, in detail in the chapter on recovery. All other Lustre modules leverage on this recovery module; for example, if a request from OSC to OST incures a time-out, it will trigger the recovery module into action.

The rest of the design part of the book will describe each of these issues more indepth. We will list and explain all of the API's exported by the various modules. We also include the design and implementation details for several upcoming features of Lustre - meta-data writeback caching, file writeback caching, and collaborative caching.

**Version 2.0 (Dec. 2002)**

(1) P.D. Innes - edited, proofed, & spell-checked text, added Changelog

**Version 1.0**

(1) R. Vullikanti - original draft

CHAPTER 22

# Networking API's

## 22.1. Introduction

The networking API's span multiple layers, most of which are internal to the network stack itself. Here we document the API's used to perform remote procedure calls, bulk movement, and network buffer manipulations. This gives the information required to extend the Lustre filesystem. We also document an extension of Portals which we have implemented.

## 22.2. Multiple Portal Interfaces

A recent addition to the networking infrastructure in Lustre was to support multiple portals interfaces, earlier a *ptlrpc* service could only serve on the first interface it found. In the present architecture, we support only one interface of each type, this list is given in the *ni_name* structure.. In the following subsections, we describe the procedure required to initialize and the support needed for multiple interface scenario.

**22.2.1. Initialization of Portals interfaces.** Any node could have multiple portals interfaces, of different types. In this section we describe the initialization procedure for these interfaces. A list of all known portals network interfaces is maintained in the *ni_names* structure, any new interface should be added here to be initialized. During the portals initialization routine (*ptlrpc_init_portals*), each of these interfaces is initialized, this initialization involves allocating the required event queues. These event queues are allocated for general purpose callbacks, such as notification when bulk data has been received/sent or a request/reply has been sent or received. These callbacks are for solicited network operations to indicate the completion of an operation requested by the node. For example, a *reply_in* callback event will be generated only if the node had sent out a *request* earlier. So these can be short queues and we can simply let them wrap when they fill up.

If the NAL is loaded (i.e. inter_module_get() of the NAL's network interface succeeds), a struct ptlrpc_ni is created. This stores the NAL's name, its *ptl_ni_handle_t* and the handles of a set of event queues allocated on the NAL. It should be noted that the per NAL event queues are only used for their callbacks, they can be as small as we like. By default, these are allocated to handle 1024 events. Any NALs loaded after the ptlrpc module has initialised are ignored.

**22.2.2. Service Initialization.** The services required on a node depend on how it is configured, for example, it could just be a client, or it could be an OST or MDS server. A service initialization is done in *ptlrpc_init_svc* routine. The following things happen during a service initialization:

(1) When a service is initialized (see ptlrpc_init_svc()), it allocates an incoming request event queue for each NAL recorded. These are usually *unsolicited events*, a node will not be able to control how many or how often they happen. Due to this, the size of these event queues is very critical. If a queue fills up, all subsequent requests will be dropped. This might cause the sender (of the dropped requests) to think that the target node has failed and force it to go into recovery.
(2) It also allocates and posts request buffers for each NAL (i.e. request buffers are not shared between NALs). The current request buffer allocation and incoming request event queue sizing is not very intelligent, it treats all NALs the same. This may not be true at all, and may waste memory as a consequence.

When the callback of one of the service's incoming request event queues fires, it wakes one of the service's threads. These threads call *ptlrpc_check_event()*, which checks each of the service's event queues in turn, until it finds an event. It should be noted that this check always starts from a different NAL to prevent starvation. If this check always started on the same NAL, it is possible that events on another NAL might have to wait for an unbounded time before being serviced, this is not acceptable. The request buffer descriptor passed as the event's user pointer tells which interface the request was received on (and therefore, which interface the client may be reached on).

The *ptlrpc_uuid_to_peer()* replaces *kportal_uuid_to_peer()*, and *struct ptlrpc_peer* replaces *struct lustre_peer* within lustre. The *ptlrpc_uuid_to_peer()* initialises a struct ptlrpc_peer to contain the NID of the peer and a pointer to the struct ptlrpc_ni that contains the correct network interface and event queue handles for communicating with the peer.

## 22.3. Buffer Manipulation

Lustre messages consist of one or more buffers depending on the message type. As these buffers are sent and received across networks, it is important that these buffers are interpreted in the correct byte-order. Moreover, some error checking might also be needed to make sure that the message buffer lengths and contents are verified just like they were system call parameters - (a) this finds problems early, (b) it prevents buggy nodes crashing their peers and (c) we will be communicating directly with userspace programs when we're running the library implementation of lustre, this makes this error checking all the more important.

In order to make things efficient, senders transmit in host byte order. As a result, messages can be retransmitted without worrying about conversions. On the other hand, receivers convert to their own byte order by inspecting the *magic* value set in the lustre message, thie is done using the *lustre_msg_swabbed* function. If the swabbed value of *PTLRPC_MSG_MAGIC* is equal to the value in the received message, it implies that the sender and receiver have different byte ordering schemes. This requires that replies have their *magic* set properly. If the sender's byte-order is different from the receiver's, a in-place byte order conversion is done using type specific *swabbers*. The intention is that the incoming messages undergo basic validation and byte-order conversion as early as possible, therafter they can be interpreted and re-interpreted at leisure.

The *extended attribute data* is *opaque* to the network, the *lov_mds_md* is little-endian on disk, and stays that way (the MDS doesn't look inside it) until it arrives on the client, here it gets converted to host byte order. In the following subsections, we describe some of the functions that are used in Lustre buffer manipulation, packing and un-packing messages. We also describe some of the macros that are available to help in debugging problems.

### 22.3.1. *ptlrpc_prep_req*.

22.3.1.1. *Prototype.*

```
struct ptlrpc_request * ptlrpc_prep_req(
struct obd_import *imp,
int opcode,
int count,
int *lengths,
char **bufs)
```

22.3.1.2. *Parameters.*

***input: imp*** This import structure contains the handle for existing connection to the server.
***input: opcode*** The operation which should be invoked on the server.
***input: count*** The number of buffers in the Lustre request message.
***input: lengths*** An array containing the sizes of all the buffers in the message.
***input: bufs*** An array of the message buffers.

22.3.1.3. *Return Values.* The function will return a pointer to the newly allocated request structure if successful, or else NULL is returned when there is no memory available to allocate the request or the buffer.

22.3.1.4. *Description.* This function will allocate a new request structure and the required message buffers for it based on the input parameters count and lengths. It will also pack the input buffers into the newly allocated request.

### 22.3.2. *lustre_msg_size*.

22.3.2.1. *Prototype.*

```
int lustre_msg_size(int count, int *lengths)
```

22.3.2.2. *Parameters.*

***input: count*** The number of buffers in the message.
***input: lengths*** An array of sizes of all the buffers in the message.

22.3.2.3. *Return Values.* The function will return an integer value.

22.3.2.4. *Description.* This is a helper function that calculates the total length of a Lustre message containing an array of buffers as indicated by count and lengths parameters and return the total length.

### 22.3.3. *lustre_msg_buf*.

22.3.3.1. *Prototype.*

```
void *lustre_msg_buf(
struct lustre_msg *m,
int n,
int min_size
)
```

22.3.3.2. *Parameters.*

***input: m*** The message in which to find the pointer to the n-th buffer.
***input: n*** The buffer number for which a pointer is requested.
***input: min_size*** The minimum possible buffer length

22.3.3.3. *Return Values.* If successful the function will return a pointer to the requested message buffer; it will return NULL if the number of buffers available is less than the buffer number whose pointer is requested, or if the buffer length is zero or less than the minimum size specified.

22.3.3.4. *Description.* This is a helper function that will return a pointer to the n-th buffer in the Lustre message *m*. The minimum size parameter can be specified irrespective of whether a message is being packed or unpacked. This function can be used to check if the size vector on the outgoing messages is correctly set. A minimum size of zero can be passed to this routine only when there is no idea about how big the data might be.

### 22.3.4. *lustre_unpack_msg*.

22.3.4.1. *Prototype.*

```
int lustre_unpack_msg(
struct lustre_msg *m,
int len
)
```

22.3.4.2. *Parameters.*

***input: m*** The message to be analyzed.
***input: len*** The known length of the message passed. This is used during unpacking to avoid buffer overflows and establish the validity of the buffers contained in the message.

22.3.4.3. *Return Values.* 0 on success, *EINVAL* if the *len* is smaller than the required length to contain a *lustre_msg* and the buffers described thereby.

22.3.4.4. *Description.* This converts the basic lustre message to host byte order. This is done in handle_incoming_request() on servers and in ptlrpc_queue_wait(), ptlrpc_replay_request() on clients.

### 22.3.5. *lustre_pack_msg*.

22.3.5.1. *Prototype.*

```
int lustre_pack_msg(int count,
int *lens,
char **bufs,
int *len,
struct lustre_msg **msg
)
```

22.3.5.2. *Parameters.*

***input: count*** Number of buffers in the Lustre message.
***input: lens*** Lengths of the buffers in the Lustre message.
***input: bufs*** Contents of buffers in the message, when not NULL.
***output: len*** Pointer to the length of the message that will be returned.
***output: msg*** Pointer to the message buffer that will be constructed.

22.3.5.3. *Return Values.* 0 upon success, -*ENOMEM* when the required buffers could not be allocated.

22.3.5.4. *Description.* Allocates the memory required to hold a Lustre message with count buffers of lengths given in *lens*. Copies in the data of length *len* from the bufs array.

### 22.3.6. *lustre_swab_typename*.

22.3.6.1. *Prototype.*

```
int lustre_swab_typename(struct typename *t)
```

22.3.6.2. *Parameters.*

***input: typename*** structure that needs to be swabbed

22.3.6.3. *Return Values.* No return value.

22.3.6.4. *Description.* There would be a routine for every structure that is sent over the wire, some examples are - l*ustre_swab_ost_body*, *lustre_swab_mds_body*, *lustre_swab_lov_desc*, *lustre_swab_ll_fid*. All of these are defined in *ptlrpc/pack_generic.c* and exported for use as needed. So, whenever the wire protocol changes, the corresponding swab routine needs to be modified.

If the structure that you need to swab is variable length, multiple swabbers are needed - one for the fixed length part and the other for the variable length part.

### 22.3.7. *lustre_msg_string*.

22.3.7.1. *Prototype.*

```
char *lustre_msg_string(struct lustre_msg *m, int index, int max_len)
```

22.3.7.2. *Parameters.*

***input: message*** a Lustre message
***input: index*** Index of the buffer within the Lustre message
***input: max_len*** maximum length of the buffer at *index*

22.3.7.3. *Return Values.* This will return the string if successful, else it will return NULL.

22.3.7.4. *Description.* This function is useful only for unpacking. It will first extract the buffer at the specified index. If the max_len is specified as 0, then it verifies that there is a NULL terminated string that fills the specified buffer. If a max_len value is specfied, the function will verify that a string of length less than or equal to max_len is present in the specified buffer, it also checks if the string is NULL terminated.

### 22.3.8. *lustre_swab_reqbuf, lustre_swab_repbuf*.

22.3.8.1. *Prototype.*

```
void *lustre_swab_reqbuf(struct ptlrpc_request *req, int index, int min_size, void *swabber)
void *lustre_swab_repbuf(struct ptlrpc_request *req, int index, int min_size, void *swabber)
```

22.3.8.2. *Parameters.*

***input: req*** The portals request
***input: index*** Index of the request/reply buffer within the message
***input: min_size*** minimum length of the buffer at *index*
***input: swabber*** swabber function

22.3.8.3. *Return Values.* Returns the swabbed (i.e byte-order corrected) buffer if successful, else returns NULL to indicate some errors.

22.3.8.4. *Description.* These two functions are used to swab Lustre requests or replies. The first step would be to extract the Lustre request/reply buffer using the *lustre_msg_buf* function. The specified swab function is then applied on this buffer, the swabbed buffer is returned. Before calling the swabber function, a quick check is done using the *lustre_msg_swabbed* function to see that the buffer is not already in the required order. If the buffer to be swabbed is of fixed size, it can be swabbed by calling the type specific function. For e.g, in the requests to OSTs, the first buffer (index 0) has the fixed size *struct ost_body*, this can be swabbed using the function *lustre_swab_ost_body*. This function will use *lustre_msg_buf* to extract the buffer, swab it in place and return the swabbed buffer.

In cases where an array of fixed size things is passed, the first can be swabbed using the *lustre_swab_reqbuf*. The total number of things can be determined using the buffer length, then each of them can be swabbed using the structure specific swab function. For e.g, in *ost_brw_read*, the first *struct obd_ioobj* is swabbed using the *lustre_swab_reqbuf*, the remaining are swabbed by calling the *lustre_swab_obd_ioobj* for each of them.

In the case of variable sized buffer, we need to have a swab function for the fixed part of the buffer and a seperate swab function for the variable part. We should first swab the fixed part, inspect it to make sure that the the expected amount of data is available, then call the swabber for the variable length buffer.

## 22.4. Client Setup & Cleanup

Setting up a client to use a service is done by setting up a client device, finding the connection matching a UUID of the remote system, and finally allocating a ptlrpc client structure containing control information for the API and recovery to be used.

### 22.4.1. *ptlrpc_uuid_to_connection*.

22.4.1.1. *Prototype.*

```
struct ptlrpc_connection *ptlrpc_uuid_to_connection(
obd_uuid_t *uuid
)
```

22.4.1.2. *Parameters.*

***input: uuid*** The UUID of the system to connect to.

22.4.1.3. *Return Values.* NULL upon failure, a valid connection with refcount incremented upon success.

22.4.1.4. *Description.* This function looks for a match, based upon the UUID given as input parameter, in the global list of connections being used as well as unused connections. If there is no existing structure found, a new one is allocated and added to the list of connections. Upon success in either case, the refcount for the connection corresponding to the given UUID is incremented. The function will return a pointer to the connection structure when successful.

### 22.4.2. *ptlrpc_init_client*.

22.4.2.1. *Prototype.*

```
void *ptlrpc_init_client(
int req_portal,
int rep_portal,
char *name,
struct ptlrpc_client *cl)
```

22.4.2.2. *Parameters.*

***input: req_portal*** Request portal to use.
***input: rep_portal*** Reply portal to use.
***input: name*** Name of the client .

22.4.2.3. *Return Values.* Void.

22.4.2.4. *Description.* This call initializes a client structure with the input parameters passed for further use.

### 22.4.3. *ptlrpc_cleanup_client*.

22.4.3.1. *Prototype.*

```
void ptlrpc_cleanup_client(struct obd_import *imp)
```

22.4.3.2. *Parameters.*

***input: imp*** The import structure to be cleaned up, this structure is associated with an existing connection to a server.

22.4.3.3. *Return Values.* Void.

22.4.3.4. *Description.* Cleans up a client, decrements the refcount, and frees up the memory allocated for the client structure and name string.

### 22.4.4. *ptlrpc_put_connection*.

22.4.4.1. *Prototype.*

```
int ptlrpc_put_connection(struct ptlrpc_connection *c)
```

22.4.4.2. *Parameters.*

***input: c*** Connection to put away.

22.4.4.3. *Return Values.* 1 if this was the last use of this connection, 0 otherwise.

22.4.4.4. *Description.* This function reduces the refcount on a connection by one. If the refcount falls to zero, the connection is put back on the unused list.

## 22.5. Server Setup & Cleanup

### 22.5.1. Description of Server Request Processing.

22.5.1.1. *Memory Management.* The ptlrpc layer installs a ring of buffers into which the requests are delivered by Portals. When the buffer is full, it is unlinked, no requests are delivered into unlinked buffers. Portals uses a next match entry with a similar buffer. A reference count is incremented and remains on the buffer for every request that is being processed until the handler completes, after which the refcount goes down. When the refcount of an unlinked buffer falls to 0 the buffer is added to the ring again. It should be noted that no memory is allocated for requests except at startup. In order to handle about 6,000 client threads, a few Megabytes of pre-allocated memory are used.

Reply buffers are allocated by the request handlers and are freed by a Portals event handler when the reply leaves the system. In the following subsections, we describe the APIs used to start and initialize network services and finally the cleanup associated with it.

### 22.5.2. *ptlrpc_init_svc*.

22.5.2.1. *Prototype.*

```
struct ptlrpc_service * ptlrpc_init_svc(
__u32 nevents,
__u32 nbufs,
__u32 bufsize,
__u32 max_req_size,
int req_portal,
int rep_portal,
obd_uuid_t uuid,
svc_handler_t handler,
char *name)
```

22.5.2.2. *Parameters.*

***input: nevents*** The number of events to be stored in the event queue.
***input: nbufs*** The number of buffers in the ring of request buffers.
***input: bufsize*** Size of the buffers to use in the ring of request buffers.
***input: max_req_size*** Maximum request size for the specified service.
***input: req_portal*** Portal number for incoming requests.
***input: rep_portal*** Portal number where replies should be sent.
***input: uuid*** UUID of the system on which the service is running (typically set to “self”).
***input: handler*** Service handler function.
***input: name*** Name of the service, set for sanity checks to *mds, ost, ldlm.*

22.5.2.3. *Return Values.* If successful, the function returns a pointer to the newly allocated service structure, or else *NULL* is returned if allocation or Portals errors take place.

22.5.2.4. *Description.* This allocates a service structure, which contains lists of requests, buffers, portal numbers, and locks that are used during request processing and the handler function for the requests. It registers a ring of incoming request buffers (the ring length is presently statically set to *RPC_RING_LENGTH* [value 10]), and is known to have to be larger than the number of threads handling the requests. The incoming request buffers are allocated.

A Portals event queue with the specified number of events, and match entries for all buffers (which are of length *bufsize)* is allocated, and an event handler will wake up a default service thread which calls the *handler* function to process the request.

### 22.5.3. *ptlrpc_start_thread*.

22.5.3.1. *Prototype.*

```
int ptlrpc_start_thread(struct obd_device *dev,
struct ptlrpc_service *svc, char *name)
```

22.5.3.2. *Parameters.*

***input: dev*** The device on for which a new kernel thread has to be started.
***input: svc*** The service to which a service thread is added.
***input: name*** Name of the new thread.

22.5.3.3. *Return Values.* If successful, the function will return 0, or else -ENOMEM can be returned if no memory can be allocated. Any error from kernel_thread may be returned to the caller.

22.5.3.4. *Description.* Attaches a worker thread to the service processing requests. This is added to the list of threads in the service structure for the given service on the specified device.

### 22.5.4. *ptlrpc_stop_all_threads*.

22.5.4.1. *Prototype.*

```
void ptlrpc_stop_all_threads(
struct ptlrpc_service *svc
)
```

22.5.4.2. *Parameters.*

***input: svc*** The service to be stopped.

22.5.4.3. *Return Values.* Void.

22.5.4.4. *Description.* Stops all service threads running for a service

### 22.5.5. *ptlrpc_unregister_service*.

22.5.5.1. *Prototype.*

```
int ptlrpc_unregister_service(
struct ptlrpc_service *service)
```

22.5.5.2. *Parameters.*

***input: service*** The service to cleanup.

22.5.5.3. *Return Values.* This service can return a number of errors that arise when the lists of requests are not empty or, when *PtlMEUnlink* fails, it could return the following error code : *PTL_NOINIT*.

22.5.5.4. *Description.* Frees all buffers and matches entries and the service structure associated with the service.

## 22.6. Bulk Movement

**22.6.1. Bulk Movement.** Bulk movement is initiated as a read or write call by clients. It is convenient to speak of the *source* and *sink* for the transfer.

The calls start by requesting the *sink* to prepare buffers. In the case of reading, the client can send the Portals match information for the registered sink buffers and the server can send the buffers back using the portals **PtlPut** operation. In the case of writes, the path has been changed recently. The client initiates the request by registering some memory, the server will then initiate a transfer using the portals **PtlGet** operation.

During the preparation phase, the client and server register the pages as an IOV memory descriptor, i.e. as a single contiguous buffer in the Elan address space, although the Linux kernel addresses of the pages may not be contiguous.

When the pages are ready, the *source,* i.e. client for writing or server for reading issue a PtlPut command to send the buffer across.

**22.6.2. Portals API Extension: IOV's.** We have made a backward compatible API extension to Portals by adding an option to the MD, called *PTL_MD_IOV*. When this is set, the start address will be interpreted as a pointer to an IOV. The lib will retrieve this from user memory and use it to send the IOV. At the moment the IOV's that will be transferred correctly are up to 16 elements long. A new field that was introduced in *ptl_md_t* to store the IOV count.

**22.6.3. *ptlrpc_bulk_put*.**

22.6.3.1. *Prototype.*

```
int ptlrpc_bulk_put(
struct ptlrpc_bulk_desc * );
```

22.6.3.2. *Parameters.*

```
struct ptlrpc_bulk_desc{
struct list_head bd_set_chain; /*Entry in obd_brw_set*/
struct obd_brw_set *bd_brw_set;
int bd_flags;
struct ptlrpc_connection *bd_connection;
struct ptlrpc_client *bd_client;
__u32 bd_portal;
struct lustre_handle bd_conn;
void (*bd_ptl_ev_handler) (struct ptlrpc_bulk_desc *);
wait_queue_head_t bd_waitq;
struct list_head bd_page_list;
__u32 bd_page_count;
atomic_t bd_refcount;
void *b_desc_private;
#if (LINUX_VERSION_CODE >= KERNEL_VERSION(2,5,0))
struct work_struct bd_queue;
#else
struct tq_struct bd_queue;
#endif
ptl_md_t bd_md;
ptl_handle_md_t bd_md_h;
ptl_handle_me_t bd_me_h;
atomic_t bd_source_callback_count;
struct iovec bd_iov[16];
}
```

***input: desc*** The bulk descriptor for the transfer.

***bd_page_list, bd_page_count, bd_connection, bd_portal*:** Must be initialized with a list of mapped bulk pages, the total count, the ptlrpc_connection and portal number to use for the match entries.

22.6.3.3. *Return Values.* This function will return 0 if successful, else it will return one of the following error codes:

***ENOMEM*:** No memory available to allocate the required IOVEC.
***PTL_NOINIT*:** PtlMDAttach/Bind/Update not initialized, PtlPut not initialized.
***PTL_SEGV*:**

22.6.3.4. *Description.* This initiates the movement of local source data into remote sink buffers via PtlPut, typically performed on the server when the client requests a read operation. The bulk descriptor is the source descriptor. All pages are currently sent as a single IOV.

### 22.6.4. *ptlrpc_bulk_get*.

22.6.4.1. *Prototype.*

```
int ptlrpc_bulk_get(
struct ptlrpc_bulk_desc * );
```

22.6.4.2. *Parameters.*

```
struct ptlrpc_bulk_desc{
struct list_head bd_set_chain; /*Entry in obd_brw_set*/
struct obd_brw_set *bd_brw_set;
int bd_flags;
struct ptlrpc_connection *bd_connection;
struct ptlrpc_client *bd_client;
__u32 bd_portal;
struct lustre_handle bd_conn;
void (*bd_ptl_ev_handler) (struct ptlrpc_bulk_desc *);
wait_queue_head_t bd_waitq;
struct list_head bd_page_list;
__u32 bd_page_count;
atomic_t bd_refcount;
void *b_desc_private;
#if (LINUX_VERSION_CODE >= KERNEL_VERSION(2,5,0))
struct work_struct bd_queue;
#else
struct tq_struct bd_queue;
#endif
ptl_md_t bd_md;
ptl_handle_md_t bd_md_h;
ptl_handle_me_t bd_me_h;
atomic_t bd_source_callback_count;
struct iovec bd_iov[16];
}
```

***input: desc*** The bulk descriptor for the transfer.

***bd_page_list, bd_page_count, bd_connection, bd_portal*:** Must be initialized with a list of mapped bulk pages, the total count, the ptlrpc_connection and portal number to use for the match entries.

22.6.4.3. *Return Values.* This function will return 0 if successful, else it will return one of the following error codes:

***ENOMEM*:** No memory available to allocate the required IOVEC.
***PTL_NOINIT*:** PtlMDAttach/Bind/Update not initialized, PtlPut not initialized.
***PTL_SEGV*:**

22.6.4.4. *Description.* This initiates the movement of remote source data into local sink buffers via PtlGet, typically performed on the server when the client requests a write operation. The bulk descriptor is the source descriptor. All pages are currently sent as a single IOV.

### 22.6.5. *ptlrpc_register_bulk_put*.

22.6.5.1. *Prototype.*

```
int ptlrpc_register_bulk_put(struct ptlrpc_bulk_desc *);
```

22.6.5.2. *Parameters.*

***input: desc*** The bulk descriptor to be used for transfer.

***bd_page_list, bd_page_count, bd_connection, bd_portal*:** Must be initialized with a list of mapped bulk pages, the total count, the ptlrpc_connection and portal number to use for the match entries.

22.6.5.3. *Return Values.* If successful the function will return 0, else one of the following errors will be returned:

***EINVAL*:** IOV size longer than the maximum allowed (PTL_MD_MAX_IOV).
***ENOMEM*:** No memory available to allocate the required IOVEC.
***PTL_NOINIT*:** PtlMDAttach not initialized, PtlGetId not initialized.
***PTL_SEGV*:**

22.6.5.4. *Description.* This is used by the client to register the bulk sink buffers for a read operation. The server will write into the sink buffers by performing a PtlPut operation via ptlrpc_bulk_put.

### 22.6.6. *ptlrpc_register_bulk_get*.

22.6.6.1. *Prototype.*

```
int ptlrpc_register_bulk_get(struct ptlrpc_bulk_desc *);
```

22.6.6.2. *Parameters.*

***input: desc*** The bulk descriptor to be used for transfer.

***bd_page_list, bd_page_count, bd_connection, bd_portal*:** Must be initialized with a list of mapped bulk pages, the total count, the ptlrpc_connection and portal number to use for the match entries.

22.6.6.3. *Return Values.* If successful the function will return 0, else one of the following errors will be returned:

***EINVAL*:** IOV size longer than the maximum allowed (PTL_MD_MAX_IOV).
***ENOMEM*:** No memory available to allocate the required IOVEC.
***PTL_NOINIT*:** PtlMDAttach not initialized, PtlGetId not initialized.
***PTL_SEGV*:**

22.6.6.4. *Description.* This is used in case of a write request, the client will register a bulk descriptor that the server can use to transer the writes from using **PtlGet** operation.

#### 22.6.7. *ptlrpc_abort_bulk.*

22.6.7.1. *Prototype.*

```
int ptlrpc_abort_bulk(
struct ptlrpc_bulk_desc *bulk
);
```

22.6.7.2. *Parameters.*

***input: bulk*** An allocated and initialized bulk descriptor.

22.6.7.3. *Return Values.* If successful the function will return 0, else one of the following error codes will be returned:

***PTL_NOINIT*:**
***PTL_SEGV*:**

22.6.7.4. *Description.* This cleans up a *sink* bulk descriptor. Cleaning up means unlinking the ME's and MD's registered with Portals.

### 22.7. Flowchart for Bulk Data Movement

Bulk movement is initiated as a read or write operation requested of the server by the clients. It is convinient to also realize the *source* and *sink* of the transfer – the data always flows from the *source* to the *sink*. The operations start by the client requesting the operation of the server.

**22.7.1. Bulk data movement for reads.** In the case of reading, the client registers its bulk sink buffers with portals and sends the resulting match information to the server. The server will also register its source memory buffers with portals. The server performs the read operation by pushing the bulk data to the client's sink buffers with a PtlPut operation. The server message consists of the *PTL_MSG_PUT* header along with the payload, the header contains information about the receiver memory buffers where the data should land. Depending on the information in the Portals header, the receiver will send back a *PTL_MSG_ACK* to the sender specified buffers when the receive is successfully completed.

Figure 22.7.1 indicates the events taking place during a bulk transfer for a read operation. We have deliberately not included the rendez-vous handshakes which are normally accomplished through RPC's.

The key aspect of *ptlrpc_bulk_put* is to prepare a Portals MD with a bulk IOV and perform a *PtlPut* operation on that memory descriptor. The event handler for the bulk source is responsible for cleaning up the descriptor. Although the API allows users to specify their own event handlers, Lustre only uses *brw_finish* at present.

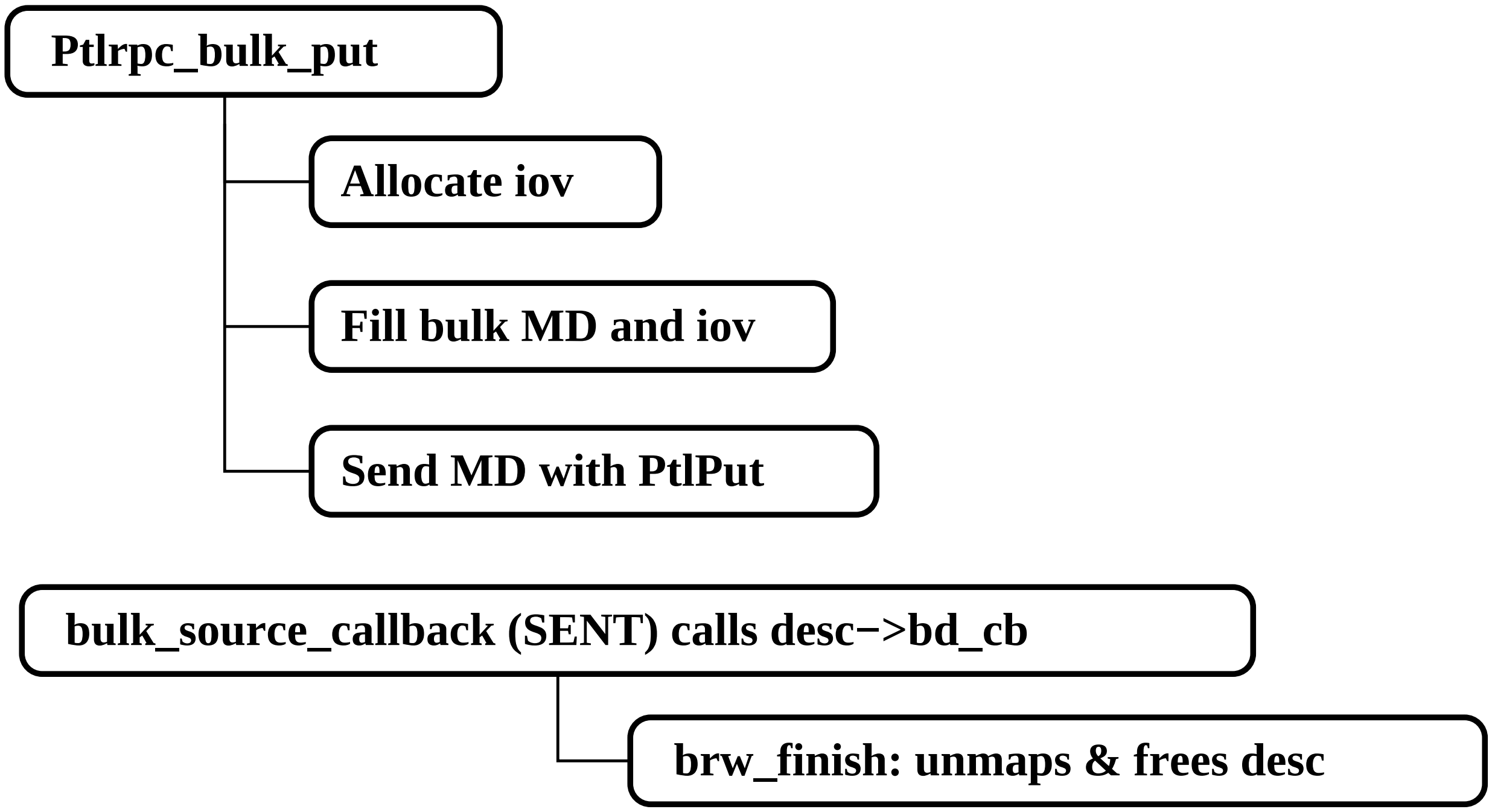

FIGURE 22.7.1. Bulk Data Movement at the Source for read operation

The sink situation is more passive, as shown in figure 22.7.2.

The sink is passive until the sink callback handler is called, which frees the descriptor and invokes secondary callbacks to wake up processing sleeping on the bulk.

Not shown in the figures are the mechanisms to generate a timeout event, if bulk transfers are somehow failing.

**22.7.2. Bulk data movement for writes.** In the case of a write operation, the client is the *source* and the server is the *sink*. The client registers the bulk source buffers with portals, and sends their match information to the server. The receiver (server) sends a *PTL_MSG_GET* with the source and sink buffer information. The source/client sends a *PTL_MSG_REPLY* message with the payload, the message header specifies the sink (server) buffers where the payload should land.. The path for a bulk write operation at the source is illustrated in 22.7.3.

The key aspect of *ptlrpc_bulk_get* is to prepare a Portals MD with a bulk IOV, the server will then perform a **PtlGet** initiate the write transfer. So, in this case the *source* situation is more passive, it simply waits for the *sink* to initiate and complete the data transfer from the MD it indicates.

The sink situation is more active, as shown in figure 22.7.4.

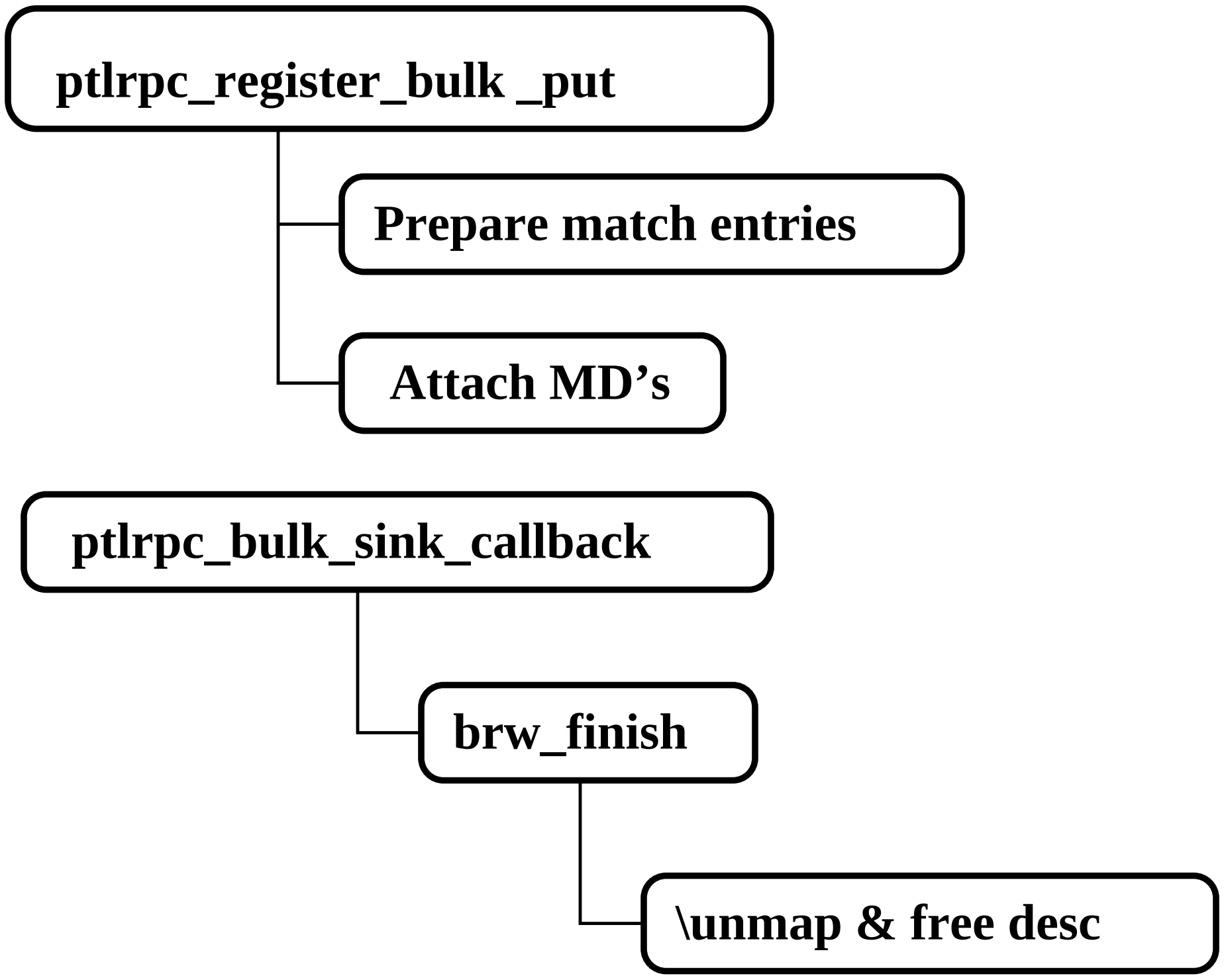

FIGURE 22.7.2. Bulk Data Movement at the Sink for read operation

In the case of write operations, the *sink* is more active, it is the *sink* that initiates the transfer. The callback handler is called when the transfer is completed, this frees the descriptor and invokes secondary callbacks to wake up processing sleeping on the bulk.

In an earlier implementation of bulk writes, *PtlPut* was used. The error handling was more difficult in that case. The *PTL_EVENT_ACK* callback involved waking up threads, the wake-up of the OSC thread could happen even before the OST had completed the write. The OSC could then proceed to perform other operations that depend on the completion of the write, they could find out later that the OST did not finish the write.

## 22.8. RPC's

RPC's are used to communicate across the network. In this section, we will give an overview of the Portals RPC's used to transmit messages in Lustre.

### 22.8.1. *ptlrpc_reply*.

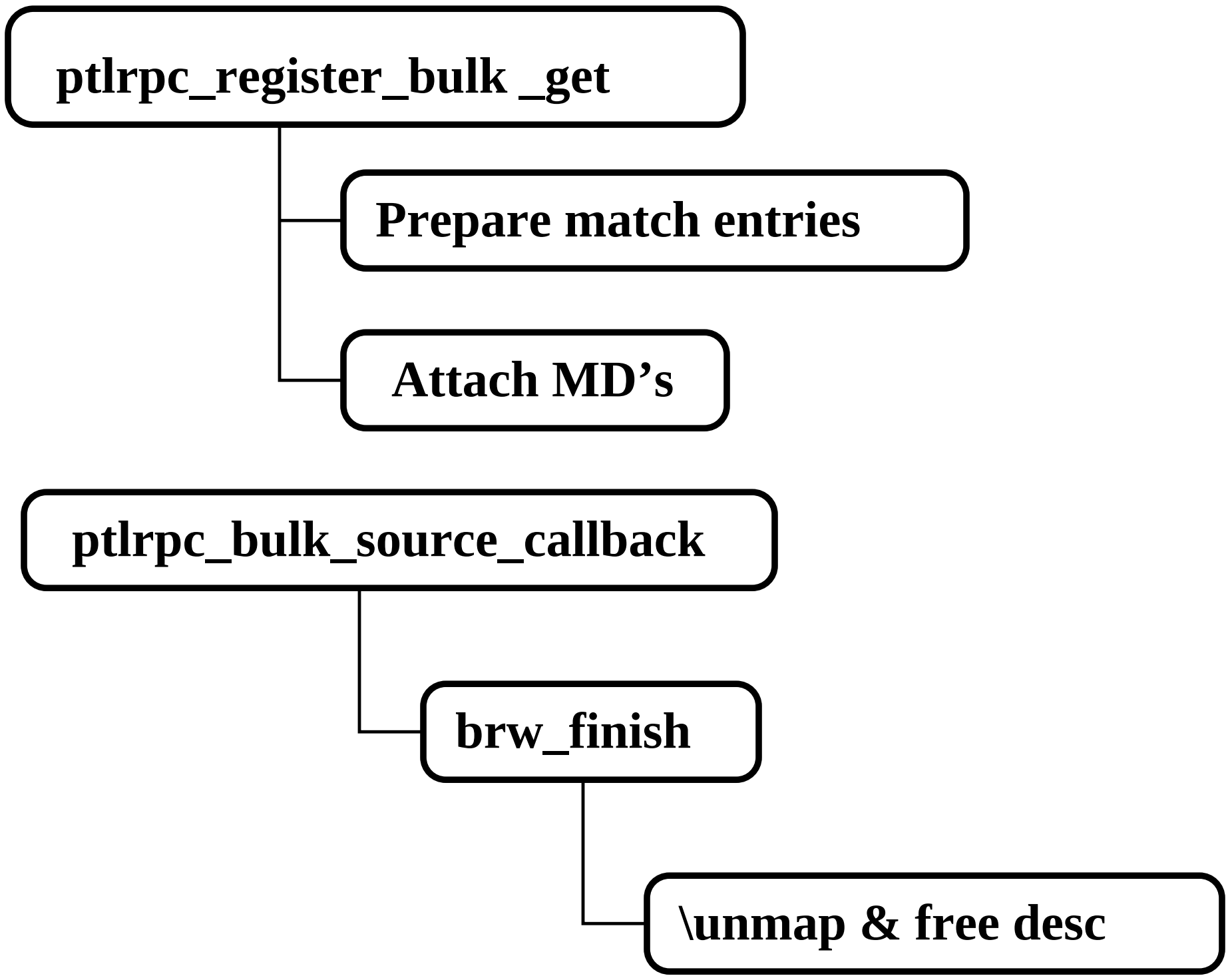

FIGURE 22.7.3. Bulk Data Movement at the Source for write operation

22.8.1.1. *Prototype.*

```
int ptlrpc_reply(struct ptlrpc_service *svc, struct ptlrpc_request *req);
```

22.8.1.2. *Parameters.*

***input: svc*** The service structure which is used to get the reply_portal number.
***input: req*** The request for which the reply is to be sent.

22.8.1.3. *Return Values.* If successful the function returns 0, else it will return one of the following errors:

***EINVAL*:** The *repmsg* structure in the request is not allocated.
***PTL_NOINIT*:**
***PTL_SEGV*:**

22.8.1.4. *Description.* This function sends out the reply contained in the request structure to the reply-portal specified in the service structure.

**22.8.2. *ptlrpc_error*.**

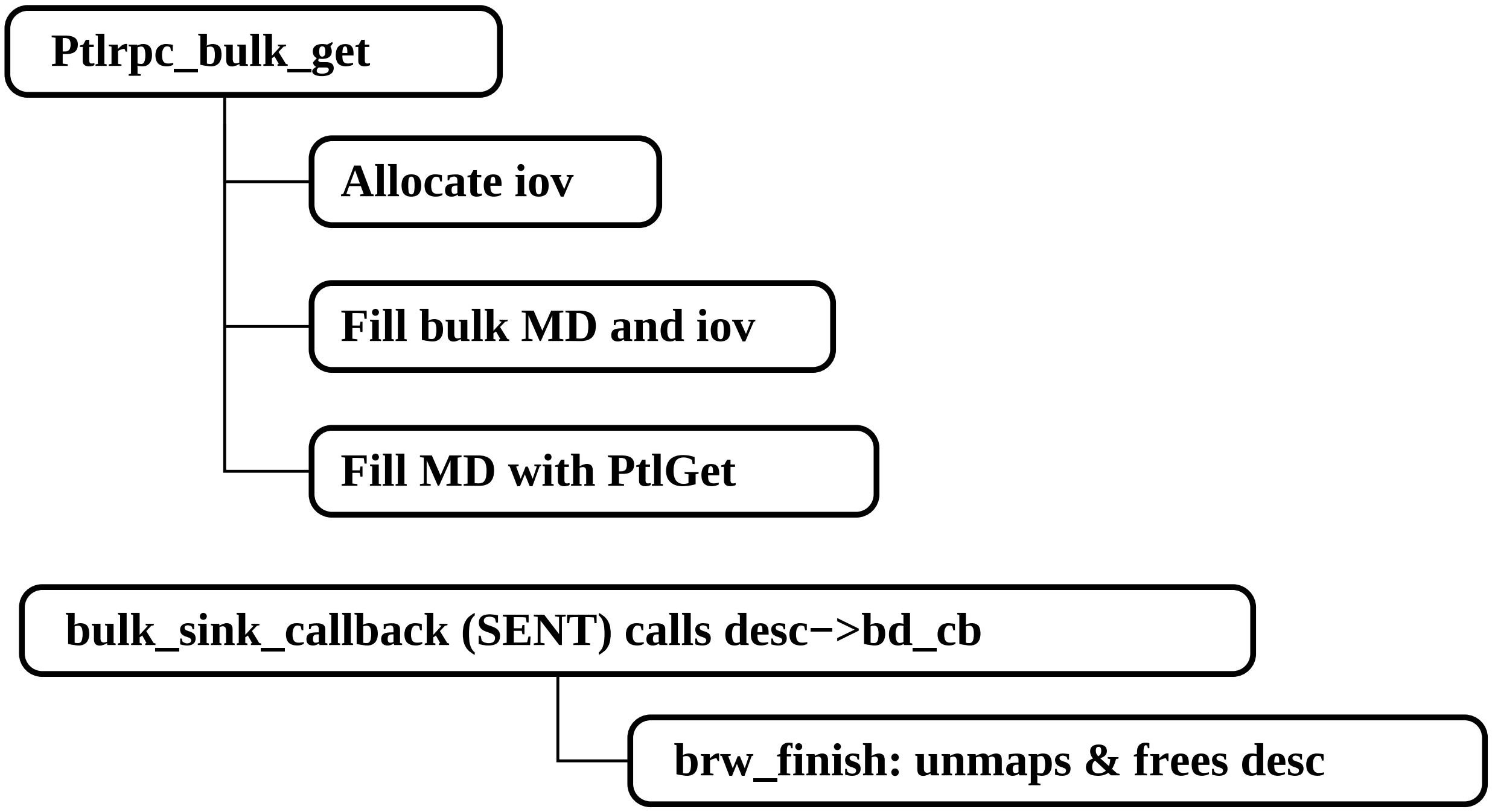

FIGURE 22.7.4. Bulk Data Movement at the Sink for write operation

22.8.2.1. *Prototype.*

```
int ptlrpc_error(struct ptlrpc_service *svc, struct ptlrpc_request *req);
```

22.8.2.2. *Parameters.*

***input: svc*** The service structure which contains the reply_portal number.
***input: req*** The request for which reply is to be sent.

22.8.2.3. *Return Values.* This function has the same return values as for function *ptlrpc_reply* above. It also throws the following error:

***ENOMEM*:** No memory available to allocate the buffer required for the reply.

22.8.2.4. *Description.* This function is used to send errors back to the client. It uses the *ptlrpc_reply* function to send this message.

**22.8.3. *ptlrpc_resend*.**

22.8.3.1. *Prototype.*

```
void ptlrpc_resend_req(struct ptlrpc_request *request);
```

22.8.3.2. *Parameters.*

***input: request*** The request to be resent.

22.8.3.3. *Return Values.*

22.8.3.4. *Description.* This retransmits a request, used in protocol recovery.

### 22.8.4. *ptl_send_rpc.*

22.8.4.1. *Prototype.*

```
int ptl_send_rpc(struct ptlrpc_request *request);
```

22.8.4.2. *Parameters.*

***input: request*** The request to be transmitted.

22.8.4.3. *Return Values.* If successful the function returns 0, else it returns one of the following error codes:

***EINVAL*:** Wrong request type.
***ENOMEM*:** No memory available to allocate the reply buffer.
***PTL_NOINIT*:**
***PTL_SEGV*:**

22.8.4.4. *Description.* This function sends out the given message after setting up the reply buffer for it.

## 22.9. Local RPC

Lustre will frequently run services on the node that is local. A simple construction has been made to make this possible.

We have added a local_rpc in the *client_obd* structure to make client device can get the service handler function directly, and this field of client_obd would be setup via lctl probe.

We also add a PROBE iocontrol for mdt/ost, this iocontrol operation would probe all mdc/osc device on local machine and setup their local_rpc field properly. (lctl probe has been disabled since self export be added, see bug 2353)

Modify some ptlrpc layer functions for local RPC * ptl_send_rpc: if client obd has local_rpc, call local_send_rpc. * ptlrpc_reply/error: if request type indicating a local request, call local_reply/error. * ptlrpc_bulk_get/put: if request type indicating a local request, call local_bulk_move.

Add some local rpc functions. In general case, local_send_rpc will just call service handler to process this request then set all the request flags/status correctly, if current thread has an uncompleted journal, local_send_rpc will start another thread to process this request and wait it exit. for bulk I/O request, I just simply copy the memory form source page list to sink page list.

## 22.10. Changelog

**Version 4.0 (Apr. 2003)**

(1) Radhika Vullikanti (04/03/2003) - Add networking snippets provided by Eric - about multiple portals interface support, APIs for the new swab support.

**Version 3.5 (Jan. 2003)**

(1) Radhika Vullikanti (01/31/2003) - Updated section 12.5 to reflect the changes made to replace PTLPUT with PTLGET for bulk writes.

**Version 3.0 (Dec. 2002)**

(1) P.D. Innes - updated text and figures, added Changelog

**Version 2.0 (Nov. 2002)**

(1) P. D. Innes - edited, proofed, spell-checked, added figures

**Version 1.0 (Oct. 2002)**

(1) R. Vullikanti - original draft

CHAPTER 23

# Wire Level Protocol Description

## 23.1. Introduction

This chapter gives an overview of the wire formats used by Lustre. Lustre embeds its messages in Portals packets. Lustre employs a *Lustre message* structure to encapsulate requests and replies which is extremely efficient for the purpose of packing and unpacking and shares many properties of the DAFS wire format. The Lustre systems embed their requests and replies in the Lustre messages as *buffers.* A message frequently contains multiple buffers, such as a request header and a pathname, or includes multiple requests, as is used in the intent locking case. Every Lustre message is encapsulated within a Portals message header, the figure 23.1.1 illustrates this. A lustre message could have message body with multiple message buffers.

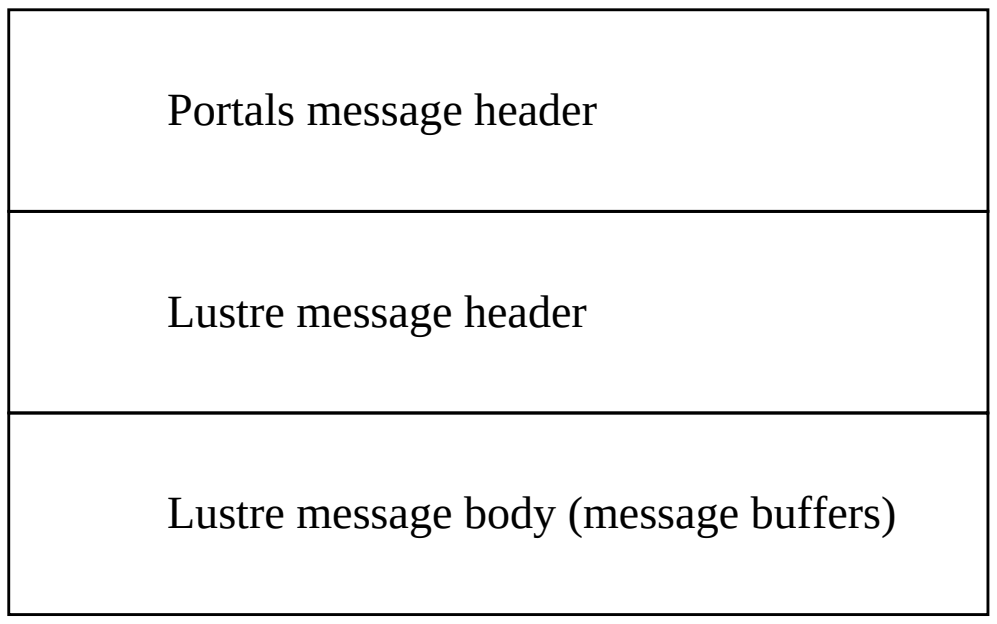

FIGURE 23.1.1. Message structure

In this chapter we will describe the structure of the various message headers and the formats of the message buffers that are sent/received in Lustre for the different type of operations.

## 23.2. Portals message headers

As previously discussed, all Lustre network traffic is handled by Portals. Portals introduces its own headers. The vast majority of all Lustre packets are put on the wire with PtlPut and PtlGet, and are simple in structure. The various Portals packet types we use are PtlPut packets, PtlGet packets and ACK packets and reply. Lustre sends Lustre request, reply, and bulk read as PtlPut packets, this gets translated into PTL_MSG_PUT and a PTL_MSG_ACK packets. The bulk write messages are sent as PtlGet packets and this is translated into PTL_MSG_GET and a PTL_MSG_REPLY.

Each Lustre packet is wrapped in a portals header. The portals header can be visualized as consisting of two part, one portion is common to all packets and a second part that depends on the type of packet (PtlGet, PtlPut, PtlAck,PtlReply). The first part has a fixed sized, the size of the second part again depends on the type of packet and is stored within the structure for the various packet types.

The fields in the common portals header have the semantic meaning in Lustre illustrated in table 1.

Some Lustre clusters use routing, but this shall not affect the content of any Portals header packet that is routed.

| Bytes | Description (*ptl_hdr_t*) | Lustre Semantics incoming packets | Outgoing packet semantics |
|---|---|---|---|
| 8 | Destination nid | This equals the Lustre nid of the receiver. The nid can be an IP address or Elan node-id. | This field is set to the final destination Lu work id: IP addr or Elan Id. When a reply packet is sent, this field will be set to the nid of the request packet that was received nection with this reply. |
| 8 | Source nid | Source Lustre network id, this as many does not equal the network address of the node sending the packet when the packet is routed. | This field is set to the Lustre node-id fror the packet originates. |
| 4 | Destination pid | 0 for Lustre | 0 for Lustre |
| 4 | Source pid | 0 for Lustre | 0 for Lustre |
| 4 | Message type | *PTL_MSG_PUT* or *PTL_MSG_ACK* *PTL_MSG_GET* *PTL_MSG_REPLY* | *PTL_MSG_PUT* or *PTL_MSG_AC* *PTL_MSG_GET or PTL_MSG_REPLY* |
| size depend on the message type | msg | depends on message type | depends on message type |
| size depend on the message type | msg_filler | The ptl_hdr_t structure size is 72, after the initial common header and the message specific structure, the rest of the structure is padded for the sieze to be 72 | The ptl_hdr_t structure size is 72, after th common header and the message specifi ture, the rest of the structure is padded for t to be 72 |

TABLE 1. Portals Header Structure

**Notes:**

(1) A Portals version field needs to be included in this header.
(2) All 32 bit integers will be little endian. 16 byte integers have the most significant bytes preceding the least significant.
(3) The nid will be the Lustre network id. In a routed network this may not coincide with the origin of the packet.
(4) ACK packets, which are generated when a PtlPut arrives with a nonzero ACK MD, are similar. But the fields following the ACK MD token are not used and the header is padded out to 88 bytes, and any padding must be ignored by the receiver.

The *msg* field is determined by the message type, each of the message types has a different structure. The table 2 illustrates the header for a PtlPut message

| **Bytes** | **Description (*ptl_put*)** | **Lustre Semantics incoming packets** | **Outgoing packet semantics** |
|---|---|---|---|
| 4 | Portal index | See section 4.1 | See section 4.1 |
| 16 | ACK MD index | The sending node of a PtlPut packet for which an ACK is requested includes this as a cookie for the packet for which an ACK is to be sent. On incoming ACK packets, this field will be used to generate an event for the corresponding packet (in portals lingo, for the memory descriptor for it). When set on incoming PltPut packets this field will be copied into outgoing ACK packets, but except for its' presence this field is not interpreted by the receiver of a PtlPut packet. | Memory descriptor used for ACK event. This field is set for PtlPut packets for Lustre bulk messages to indicate that an ACK packet is requested. On outgoing ACK packets this will equal the MD handle on the associated incoming PtlPut packet. Unless the first 64 bits of this field are set to ~0 an ACK will be generated. |
| 8 | Match bits | In an incoming Lustre request PtlPut message, this is set to an integer that the recipient must include in (i) Lustre reply PtlPut packets for that request, and (ii) incoming or outgoing Lustre bulk PtlPut packets. On incoming replies this field must equal the xid of the request to which the packet belongs. For incoming, i.e. sink side, bulk packets, this field must equal the xid's the sink sent to the source in the bulk handshake. | On outgoing request packets this field must be set to a unique integer. On outgoing reply packets this field must be set to the xid of the transaction associated with the request. For bulk packets, the source must set this field to match the xid's sent by the sink during the preparatory bulk handshake. |
| 4 | Length | Packet body length, not including the Portals header. | Packet body length, not including the Portals header. |
| 4 | Offset | Sender managed offset: 0 in Lustre. | Sender managed offset: 0 in Lustre. |
| 8 | Header data | Reserved for Lustre use. | Reserved for Lustre use. |

TABLE 2. Portals PTL_MSG_PUT message Structure

Here, the xid is a request identifier that in normal use of the system increases by 1 for every request sent from a particular node to a service. During recovery of systems requests may be re-transmitted, hence the xid is not totally unique.

The table 3 shows the structure for a PtlGet message.

The table 5 shows the structure for an Ack packet.

Finally, the table shows the PtlReply packet:

| Bytes | Description (*ptl_get*) | Lustre Semantics incoming packets | Outgoing packet semantics |
|---|---|---|---|
| 4 | Portal index | The portals index on the osc that the write source buffers have been posted under | |
| 16 | return MD index | This is the OST trying to provide a handle to the write target buffers that the PTL_MSG_REPLY should be sent to | The OSC will use to send the PTL_MSG_REPLY |
| 8 | Match bits | In an incoming Lustre request PtlGet message, this is set to an integer that the recipient must include in (i) Lustre reply PtlGet packets for that request, and (ii) incoming or outgoing Lustre bulk PtlGet packets. On incoming replies this field must equal the xid of the request to which the packet belongs. For incoming, i.e. sink side, bulk packets, this field must equal the xid's the sink sent to the source in the bulk handshake. | On outgoing request packets this field must be set to a unique integer. On outgoing reply packets this field must be set to the xid of the transaction associated with the request. For bulk packets, the source must set this field to match the xid's sent by the sink during the preparatory bulk handshake. |
| 4 | Length | The portals agent that receives the get msg uses the length to find a local match entry (list of memory regions) that will provide source of the bulk flow. The length of the source buffer is specified here. | The length of the data region from which the get originated. |
| 4 | source offset | Offset within the memory region that we get from the match bits | Sender managed offset: 0 in Lustre. |
| 4 | return offset | This offset is set by the OST and simply copied by the OSC | The offset within the return wmd that the data should be send to by the source. |
| 4 | sink length | length of the server buffer where the writes would go- set by the server | sink length - used to verify that the matching packet is of correct length and not larger. |

TABLE 3. Portals PTL_MSG_GET message Structure

For more information about the precise meaning of these headers, see the Portals specification [http://www.cs.sandia.gov/~bright/papers/portals3/portals3.html].

## 23.3. Lustre Messages: RPC's

Lustre uses Portals to send *Lustre messages* across the net. All request and reply messages are packaged as *Lustre messages* and embedded within the Portals header. Lustre messages have fields in the body that describe the contents of the message and assist in recovery. Bulk packets are sent as raw Portals packets.

| Bytes | Description (*ptl_ack*) | Lustre Semantics incoming packets | Outgoing packet semantics |
|---|---|---|---|
| 4 | m_length | | |
| 16 | DST MD index | This the receiver's MD as copied from the PtlPut packet | This is the destination MD copied from the corresponding PtlPut packet |
| 8 | Match bits | | |
| 4 | Length | Packet body length, not including the Portals header. | Packet body length, not including the Portals header. |

TABLE 4. Portals PTL_MSG_ACK message structure Structure

| Bytes | Description (*ptl_reply*) | Lustre Semantics incoming packets | Outgoing packet semantics |
|---|---|---|---|
| 4 | unused | | |
| 16 | DST MD index | This the receiver's MD as copied from the PtlPut packet | This is the destination MD copied from the corresponding PtlPut packet |
| 4 | dst_offset | | |
| 4 | unused | | |
| 4 | Length | Packet body length, not including the Portals header. | Packet body length, not including the Portals header. |

TABLE 5. Portals PTL_MSG_REPLY message Structure

Lustre requests and replies (but not bulk data) fill the "packet data" field with at least a *lustre_msg* structure explained in table 6. Some requests/replies might have additional buffers, this would be indicated by the *buffer count* field in the *lustre_msg* structure. Every *Lustre message* requires a corresponding reply.

Each buffer and the message itself is padded to an 8-byte boundary, but this padding is not included in the buffer lengths.

The structure of the data in the buffers in *lustre_msg* structure would depend on the type of operation. In the following sections we would describe the buffer data formats for all the possible Lustre request/reply types.

The structure described above has a field to contain the last received transaction number, this might not be required in every reply. The *last committed* transaction number is required for recovery purposes. We might also need to add the *generation* number of the connection on every outgoing request to enable the target to IO fence old requests. It might also be useful to add protocol version information in every outgoing request packet. In future, we might include a *subsystem* field similar to that in SUN RPC and an *authenticator*.

| Bytes | Name (*struct lustre_msg*) | Use on incoming packets | Use on outgoing packets |
|---|---|---|---|
| 8 | Export/import handle cookie | The first 8 bytes provide a *Lustre handle* to the export or import data associated with the message. It is used by the receiver to locate the object. The export data handle is used by services for incoming request packets; the import handle is used by clients for incoming ASTs. | The handle is a copy of the handle excha the peer during the subsystem connec shake. On outgoing request packets, handle of the target service is include going ASTs, included is the import han client. |
| 4 | Magic | Magic constant 0x0BD00BD0. | Magic constant 0x0BD00BD0. |
| 4 | Type | *PTL_RPC_MSG_REQUEST* | *PTL_RPC_MSG_REPLY* *PTL_RPC_MSG_ERR* |
| 4 | Lustre msg version and protocol version:version Current value 0x00040001 in little endian | Most significant 16 bits: Lustre msg protocol version; least significant 16 bits: subsystem protocol version. This is checked by the service for request packets against the available protocols offered by the receiver. Not used by clients. | This field is set by the client to indica sions used. |
| 4 | Protocol and Opcode: opc | Most significant 16 bits: Lustre subsystem protocol number; least significant 16 bits: opcode for request in protocol. Used by the service to locate the request handler. | Set by the client to indicate what reque sent. Not set by client |
| 8 | Last received counter | In replies: last transaction no for MDS/OST. | |
| 8 | Last committed counter | In replies: last committed transaction. | |
| 8 | Transaction number | In replies: transaction no for request. | |
| 4 | Status | Return value of handler. | |
| 4 | Buffer count: bufcount | How many buffers are included. | |
| 4 | flag | Operation specifics flags use the top 16 bits (eg. MSG_CONNECT_RECONNECT)and common flags used bottom 16 bits (eg. MSG_LAST_REPLAY). | |
| “buffer count“ * 4 | Buffer lengths buflens[] | What is the length of each of these. | |
| total of “buffer lengths“ | Message data | | |

TABLE 6. Lustre Message Structure

## 23.4. OSC - OST Network Protocol

In this section we will describe the buffer data formats for all the requests/replies exchanged between the object storage clients (OSC) and the targets (OSTs). There are eleven OST opcodes and each of them has different information in the buffers:

***OST_CONNECT*:** This is sent as a Lustre message with a three buffers. The first buffer will contain only the UUID of the target to be connected to. On the server side this UUID is translated into the target device. The second buffer is a UUID identifying the connecting client. The target instantiates a unique export structure for the client which is passed in on every further request. The third buffer holds the import handle which is sent back to the client for lock callbacks.

***OST_DISCONNECT*:** This is a Lustre message without any additional buffers. The server tears down the export specified in the *lustre_msg* structure.

***OST_GETATTR, OST_SETATTR, OST_OPEN, OST_CLOSE, OST_CREATE, OST_DESTROY,*:**

***OST_PUNCH*:** These OST requests have the structure shown in table 7 in a single message buffer (hereafter referred to as the *OST body*):

| Bytes | Description (*struct ost_body*) |
|---|---|
| | OBD Object (OBDO) |

TABLE 7. OST request structure

An OBD object, as illustrated in table 8, is similar to a cross-platform inode. It describes the attributes of a given object. The *valid* flag is used to indicate the fields that are relevant for a request. As an example, the **OST_SETATTR** will use the *valid* flag to indicate which attributes need to be written to the device.

***OST_READ, OST_WRITE*:** These requests have different structure. The first buffer in the lustre message is a network ioobject. This is followed by an array of *remote niobufs* in buffer 2. There is one IO object (see table 9) and one niobuf (see table 11) per file extent. In case of reads, each niobuf includes a return code that indicates success/failure/errors on a per-page basis as shown in

When an ***OST_READ*** request is received, the data is sent to portals match entries equal to the xid given in the niobuf_remote structure. In case of reads, each niobuf includes a return code that indicates success/failure/errors on a per-page basis as shown in 10.

For an ***OST_WRITE*** request, buffers with such match bits are prepared by the client so that the server can get data from the buffers. The bulk data described in those structures is sent as a standard Portals packet, without any Lustre RPC header.

***OST_STATFS*:** This function has one reply message buffer which contains a *struct obd_statfs*. The contents are shown in table 12.

The server should fill in the critical fields at the minimum, relating to the number of free/total file objects and blocks and zero-fill the unused fields.

| Bytes | Description (*struct obdo*) |
|---|---|
| 8 | id |
| 8 | Group |
| 8 | atime |
| 8 | mtime |
| 8 | ctime |
| 8 | Size |
| 8 | Blocks |
| 8 | rdev |
| 4 | Block size |
| 4 | Mode |
| 4 | uid |
| 4 | gid |
| 4 | Flags |
| 4 | Link count |
| 4 | Generation |
| 4 | Valid |
| 4 | OBDflags |
| 4 | o_easize |
| 60 | o_inline |

TABLE 8. OBD object

| Bytes | Description (*struct obd_ioobj*) |
|---|---|
| 8 | id |
| 8 | Group |
| 4 | Type |
| 4 | Buffer count |

TABLE 9. IO object

## 23.5. Lustre DLM Network Protocol

The Lustre lock manager has 3 regular calls and 2 callbacks. The regular calls are sent on the same portal as the affected service; for example, meta-data lock requests are sent to the MDS portal. The callbacks are sent to the portal reserved for DLM RPC's. Every request to the lock manager has at least a single buffer with the *ldlm_request* structure as shown in table 13 in it or the *ldlm_reply* structure (see table 17).

**23.5.1. Lustre lock request structure.** Any lock request in lustre consists of atleast the *ldlm_request* structure (see table 13).

| Bytes | Description (*struct niobuf_remote*) |
|---|---|
| 8 | Offset |
| 8 | xid |
| 4 | Length |
| 4 | Flags |
| 4 | return code |
| | addr |
| sizeof page struct | Flags |
| | target_private |
| sizeof dentry struct | dentry |

TABLE 10. Niobuf_local

| Bytes | Description (*struct niobuf_remote*) |
|---|---|
| 8 | Offset |
| 8 | xid |
| 4 | Length |
| 4 | Flags |

TABLE 11. Niobuf

| Bytes | Field name | Meaning |
|---|---|---|
| 8 | os_type | Magic constant describing the type of OBD (not defined yet). |
| 8 | os_blocks | Total number of blocks on OST. |
| 8 | os_bfree | Free blocks. |
| 8 | os_bavail | Available blocks (free minus reserved). |
| 8 | os_files | Total number of objects. |
| 8 | os_ffree | Number of unallocated objects. |
| 40 | os_fsid | UUID of OST. |
| 4 | os_bsize | Block size. |
| 4 | os_namelen | Length of OST name. |
| 48 | os_spare | Reserved. |

TABLE 12. OBD Status structure

| Bytes | Name | Description |
|---|---|---|
| 4 | lock_flags | flag filled by the server to indicate statusof the lock |
| 92 | lock_desc | Lock descriptor is filled with requested type, name, and extent. |
| 8 | lock_handle | |
| 8 | lock_handle2 | |

TABLE 13. The lock request structure

As shown in table 13, every lock request would contain a *lock description* structure as shown in 16. This structure has several sub-components. It contains a *struct ldlm_extent* (see table 14) structure that describes the file extent covered by the lock.

| Bytes | Name | Description |
|---|---|---|
| 8 | start | Start of extent. |
| 8 | end | End of the extent. |

TABLE 14. Lock extent descriptor

Secondly, we have resource descriptors, *struct ldlm_resource_desc* (see table 15), this is used to describe the resource for which a lock is requested. This is an unaligned structure, its allright as long as this is used only in *ldlm_request* structure.

| Bytes | Name | Description |
|---|---|---|
| 4 | lr_type | Resource type: one of *LDLM_PLAIN*, *LDLM_INTENT*, *LDLM_EXTENT*. |
| 8*3 | lr_name | Resource name. |
| 4*4 | lr_version | Version of the resource (not yet used). |

TABLE 15. Lock resource descriptor

| Bytes | Name | Description |
|---|---|---|
| 44 | l_resource | description of the resource for the lock (see 15) |
| 4 | l_req_mode | Requested lock mode, one of *LCK_EX* (=1), *LCK_PW*, *LCK_PR*, *LCK_CW*, *LCK_CR*, LCK_NL (=6) File I/O uses *PR* and *PW* locks. |
| 4 | l_granted_mode | Lock mode that is granted on this lock. |
| 16 | l_extent | Extent required for this lock (see 14) |
| 4*4 | l_version | Version of this lock. |

TABLE 16. Lock descriptor

**23.5.2. Lustre lock reply structure.** The reply message contains a reply (see table 17).

**23.5.3. Message structures for the various locking operations.** In the following sections we will describe the message structures for the various locking operations supported in Lustre.

23.5.3.1. LDLM_ENQUEUE. This message is used to obtain a new lock. The Lustre message contains a single buffer with a *struct ldlm_request* .

23.5.3.2. LDLM_CANCEL. This message cancels an existing lock. It places a *struct ldlm_request* in the Lustre message, but only uses the *lock_handle1* part of the request (we will shrink this in the future). The reply contains just a *lustre_msg*.

| Bytes | Name | Description |
|---|---|---|
| 4 | lock_flags | Flags set during *enqueue*. |
| 4 | lock_mode | The server may change the lock mode; if this quantity is non-zero, the client should update its lock structure accordingly. |
| 8 * 3 | lock_resource_name | Resource actually given to the requester. |
| 8 | lock_handle | Handle for the lock that was granted. |
| 16 | lock_extent | Extent that was granted (will move to policy results). |
| 8 | lock_policy_res1 | Field one for policy results. |
| 8 | lock_policy_res2 | Field two for policy results. |

TABLE 17. Reply for a lock request

23.5.3.3. LDLM_CONVERT. This message converts the lock type of an existing lock. The request contains an *ldlm request* structure, as in *enqueue*. The requested mode field contains the mode requested after conversion. An *ldlm_reply* message is returned to the client.

23.5.3.4. LDLM_BL_CALLBACK. This message is sent by the lock server to the client to indicate that a lock held by the client is blocking another lock request. This sends a *struct ldlm_request* with the attributes of the blocked lock in *lock_desc*.

23.5.3.5. LDLM_CP_CALLBACK. This message is sent by the lock server to the client to indicate that a prior unfulfilled lock request is now being granted. This too sends a *struct ldlm_request* with the attributes of the granted lock in *lock_desc*. Note that these attributes may differ from those that the client originally requested, in particular the resource name and lock mode.

### 23.6. Client / Meta-data Server

The client meta-data network protocol consists of just a few calls. Again, we first explain the components that make up the Lustre messages and then turn to the network structure of the individual requests. The MDC-MDS protocol has significant similarity with the OSC-OST protocol.

Messages have the following Portals related attributes:

(1) Destination portal for requests: *MDS_REQUEST_PORTAL*
(2) Reply packets go to: *MDC_REPLY_PORTAL*
(3) Readdir bulk packets travel to: *MDC_BULK_PORTAL*

A few other constants are important. We have a sequence of call numbers:

```
   #define MDS_GETATTR    1
#define MDS_OPEN       2
#define MDS_CLOSE      3
#define MDS_REINT      4
#define MDS_READPAGE   6
#define MDS_CONNECT    7
#define MDS_DISCONNECT 8
```

```
#define MDS_GETSTATUS  9
#define MDS_STATFS     10
#define MDS_GETLOVINFO 11
```

The update records are numbered too, to indicate their type:

```
   #define REINT_SETATTR 1
#define REINT_CREATE 2
#define REINT_LINK 3
#define REINT_UNLINK 4
#define REINT_RENAME 5
#define REINT_RECREATE 6
```

**23.6.1. Meta-data Related Wire Structures.** As indicated in table 1, many messages to MDS contain an *mds_body* (see table 18).

| Bytes | Name | Description |
|---|---|---|
| 16 | fid1 | First fid. |
| 16 | fid2 | Second fid. |
| 16 | handle | Lustre handle |
| 8 | size | File size. |
| 8 | blocks | blocks |
| 4 | ino | inode number |
| 4 | valid | Bitmap of valid fields sent in / returned. |
| 4 | fsuid | effective user id for file access. |
| 4 | fsgid | effective group id for file access. |
| 4 | capability | Not currently used |
| 4 | mode | Mode of file |
| 4 | uid | real user id. |
| 4 | gid | real group id. |
| 4 | mtime | Last modification time. |
| 4 | ctime | Last inode change time. |
| 4 | atime | Last access time. |
| 4 | flags | Flags. |
| 4 | rdev | device |
| 4 | nlink | Linkcount. |
| 4 | generation | Generation. |
| 4 | suppgid | |
| 4 | eadatasize | Size of extended attributes. |

TABLE 18. MDS Body

In the *mds_body* structure a file identifier is used to identify a file ( see table 19).

The file type is a platform independent enumeration:

| Bytes | Name | Description |
|---|---|---|
| 8 | id | Inode id |
| 4 | generation | Inode generation |
| 4 | f_type | Inode type |

TABLE 19. File Identifier Structure

```
#define S_IFSOCK 0140000
#define S_IFLNK  0120000
#define S_IFREG  0100000
#define S_IFBLK  0060000
#define S_IFDIR  0040000
#define S_IFCHR  0020000
#define S_IFIFO  0010000
```

The MDS stores the file striping information, which includes the object information, as extended atttributes. It might be required to send this information across the wire for certain operations. This can be done using the variable length data structure shown in described in table 20.

| Bytes | Name | Description |
|---|---|---|
| 4 | lmm_magic | 0x0BD00BD0, the striping magic (read, obdo-obdo). |
| 8 | lmm_object_id | The id of the object as seen by the LOV. |
| 4 | lmm_stripe_size | Stripe size. |
| 4 | lmm_stripe_offset | Stripe offset. |
| 2 | lmm_stripe_count | How many stripes are used for the file. |
| 2 | lmm_ost_count | Total number of OSTs in the cluster (determines the maximum stripe count) |
| 8*n | lmm_objects | An array of object id, in the order that they appear in the LOV descriptor. |

TABLE 20. Variable Length Structure

**23.6.2. MDS Update Record Packet Structure.** In this section we will describe the message structures for all the metadata operations that result in update of the file metadata on the MDS. The structure of the update record will depend on the operation type, all update records contain a 32 bit opcode at the begining for identification.

23.6.2.1. *REINT_SETATTR.* The *setattr* message contains a structure containing the attributes that will be set, in a format commonly used across Unix systems as shown in table 21.

23.6.2.2. *REINT_CREATE.* The *create* record is used to build files, symbolic links, directories, and special files. In all cases the record shown in figure 22 is included, and a second buffer in the Lustre message contains the name to be created. For files this is followed by a further buffer containing striping meta-data. For symbolic link a third buffer is also present, containing the null terminated name of the link.

The reply contains only an *mds_body* structure along with the *lustre_msg* structure.

| Bytes | Name | Description |
|---|---|---|
| 4 | sa_opcode | opcode of the update record that follows. |
| 4 | sa_fsuid | effective user id for file access. |
| 4 | sa_fsgid | effective group id for file access. |
| 4 | sa_cap | Not currently used |
| 4 | sa_reserved | Not currently used |
| 4 | sa_valid | Bitmap of valid fields. |
| 16 | sa_fid | fid of object to update. |
| 4 | sa_mode | Mode |
| 4 | sa_uid | uid |
| 4 | sa_gid | gid |
| 4 | sa_attr_flags | Flags |
| 8 | sa_size | Inode size. |
| 8 | sa_atime | atime |
| 8 | sa_mtime | mtime |
| 8 | sa_ctime | ctime |
| 4 | sa_suppgid | Not currently used |

TABLE 21. *setattr* Message Structure

| Bytes | Name | Description |
|---|---|---|
| 4 | cr_opcode | opcode |
| 4 | cr_fsuid | effective user id for file access. |
| 4 | cr_fsgid | effective group id for file access. |
| 4 | cr_cap | Not currently used |
| 4 | sa_flags | for use with open |
| 4 | cr_mode | Mode |
| 16 | cr_fid | fid of parent. |
| 16 | cr_replayfid | fid of parent used to replay request. |
| 4 | cr_uid | uid |
| 4 | cr_gid | gid |
| 8 | cr_time | Time |
| 8 | cr_rdev | Raw device. |
| 4 | cr_suppgid | Not currently used |

TABLE 22. Create record

23.6.2.3. *REINT_LINK*. The *link* Lustre message contains 2 fields: an *mds_rec_link* record described in table 23 followed by a null terminated name to be used in the target directory.

The reply consists of an *mds_body*.

| Bytes | Name | Description |
|---|---|---|
| 4 | lk_opcode | |
| 4 | lk_fsuid | effective user id for file access. |
| 4 | lk_fsgid | effective group id for file access. |
| 4 | lk_cap | Not currently used |
| 4 | lk_suppgid | Not currently used |
| 16 | lk_fid1 | fid of source. |
| 16 | lk_fid2 | fid of target parent. |

TABLE 23. File link Records

23.6.2.4. *REINT_UNLINK.* The *unlink* Lustre message contains 2 fields: an *mds_rec_unlink* record described in table 24 followed by a null terminated name to be used in the target directory.

| Bytes | Name | Description |
|---|---|---|
| 4 | ul_opcode | |
| 4 | ul_fsuid | effective user id for file access. |
| 4 | ul_fsgid | effective group id for file access. |
| 4 | ul_cap | Not currently used |
| 4 | ul_reserved | Not currently used |
| 4 | ul_mode | Mode |
| 4 | ul_suppgid | Not currently used |
| 16 | ul_fid1 | fid of source. |
| 16 | ul_fid2 | fid of file to be removed. |

TABLE 24. File unlink Records

The reply consists of an *mds_body*. Notice that one that the *lk_fid2* is superfluous, but useful for checking correctness of the protocol.

23.6.2.5. *REINT_RENAME.* The rename lustre message contains 2 fields: an *mds_rec_rename* record (see table 25) followed by two null terminated names, indicating the source and destination name.

The reply consists of an *mds_body*.

23.6.2.6. *REINT_RECREATE.* This request is present for recovery purposes and identically formatted to that of REINT_CREATE, except for the value of the opcode.

### 23.6.3. MDS Request record packet structure.

23.6.3.1. *MDS_GETATTR.* The getattr request contains a mds_body as request. The parameters that are relevant in the request are the fid and valid fields. In WB caching mode, the attributes are received by using the fid in the *mds_body*, but in CS mode the fid is that of the parent directory and the attributes are retrieved by a name included as a buffer in the lustre message following the *mds_body.*

| Bytes | Name | Description |
|---|---|---|
| 4 | rn_opcode | |
| 4 | rn_fsuid | effective user id for file access. |
| 4 | rnl_fsgid | effective group id for file access. |
| 4 | rn_cap | Not currently used |
| 4 | rn_suppgid1 | Not currently used |
| 4 | rn_suppgid2 | Mode |
| 16 | rn_lk_fid1 | fid of source. |
| 16 | rn_lk_fid2 | fid of target parent. |

TABLE 25. File rename Records

The reply may be followed by mds striping data in the case of a fid of type S_IFREG, or can be followed by a linkname. This happens when in the *valid* field the *OBD_MD_LINKNAME* bit is set.

23.6.3.2. *MDS_OPEN.* The open request contains and *mds_fileh_body* (see figure 26), followed by an optional *lov_stripe_md.* The stripe meta-data is used to store the object identities on the MDS, in case the objects were created only at open time on the OST's. The fid indicates what object is opened. The handle in the request is a local file handle, to deal with re-opening files, during cluster recovery.

| Bytes | Name | Description |
|---|---|---|
| 16 | fid | File id of object to open / close. |
| 16 | file handle | File handle passed or returned. |

TABLE 26. File handler structure for open/close requests

The reply contains the same structure. The Lustre handle contains the remote file handle and a security token; the body has the attributes of the inode.

23.6.3.3. *MDS_CLOSE.* The structure of this call is equal to that of open, although *lov_stripe_md* can currently not be passed.

23.6.3.4. *MDS_REINT.* The message structure contains an empty Lustre message with an update record. The update records are described above and appended as the first buffer in the Lustre message. The reply is an *mds_body.*

23.6.3.5. *MDS_READPAGE.* The request structure contains an *mds_body.* The *fid1* field contains the identifier of the file; the size field gives the offset of the page to be read. The reply is simply a Lustre message.

23.6.3.6. *MDS_CONNECT.* See *OST_CONNECT.*

23.6.3.7. *MDS_DISCONNECT.* See *OST_DISCONNECT.*

23.6.3.8. *MDS_GETSTATUS.* This call will retrieve the root fid from the MDS. The request message is a *lustre_msg*; the reply contains a *lustre_fid*, the root fid of the file system.

23.6.3.9. *MDS_STATFS.*

23.6.3.10. *MDS_GETLOVINFO.*

## 23.7. Client - MDS/OST recovery protocol

We have introduced a new operation in which clients ping all the servers periodically. When a server (MDS/OST) fails, all the connected clients need to participate in recovery within a given time, if they miss the recovery window, they are removed from the cluster. The client will then lose all the cached updates. The *ping* operation can be used by the clients to continuously check if the servers are up or not. If a failover server is available, the clients need to find and connect to them. A new opcode, ***OBD_PING*** has been introduced for this purpose, this is understood by both OST and MDS nodes. This new opcode has a value of 400, and no request or reply body (both have length 0), the figure 27 illustrates this message.

| **Bytes** | **Name (*struct lustre_msg*)** | ***Value*** |
|---|---|---|
| 8 | Export/import handle cookie | Contains the import/export handle for the request |
| 4 | Magic constant | 0x0BD00BD0. |
| 4 | Type | *PTL_RPC_MSG_REQUEST / PTL_RPC_MSG_REPLY* |
| 4 | Lustre-msg version and protocol version | Current value 0x00040001 in little endian |
| 4 | Opcode | 400 |
| 8 | Last received counter | In replies: last transaction no for MDS/OST. |
| 8 | Last committed counter | In replies: last committed transaction. |
| 8 | Transaction number | In replies: transaction no for request. |
| 4 | Status | Return value of handler. |
| 4 | Buffer count: bufcount | 0 |
| “buffer count“ * 4 | Buffer lengths buflens[] | No buffers |

TABLE 27. Lustre message for the new OBD_PING operation

On the server side, zero-to-minimal processing should be done for this new type of Lustre message. In addition, OBD_PING can be sent with a request message containing an addr and cookie of zero (no export), and should be processed without any change in that case. Specifically, it should not return ***-ENOTCONN*** for a mis-matched export handle, if the addr and cookie are both zero.

Another scenario in which the pinger plays an important role is during cleanup, in a cluster if a clients are shutdown while they hold locks, the OSTs will have to wait for a long time for timeouts to occur for all the clients. At this point, the server can assume that the client died and cleanup the locks held by it. On the other hand, in the presence of the *ping* operation, the OST will keep track of *time_last_heard* parameter for every client. The server can use this variable to track when it last heard from a client, if the time exceeds a certain threshold value, the OSTs can mark the clients as dead.

## 23.8. Changelog

**Version 2.2 (Apr. 2003)**

(1) Radhika Vullikanti (28 Apr 2003) - Updated the structures to reflect the current protocol.

**Version 2.1 (Apr. 2003)**

(1) Phil Schwan (26 Apr 2003) - Updated wire protocols to reflect changes made between Lustre versions 0.6.0.3 and 0.6.0.4 (bugs 593, 1154, 1175, and 1178). All sizes are now in bytes.

**Version 2.0 (Apr. 2003)**

(1) Radhika Vullikanti (04/01/2003) - Added a new section describing the wire protocol changes made for recovery purposes.

**Version 1.5 (Jan. 2003)**

(1) Radhika Vullikanti (01/31/2003) - Updated section 13.2 to reflect the changes that were made to the wire protocol for using ptlget for bulk writes.

CHAPTER 24

# NAL's: Network Abstraction Layers for Portals

Portals interacts with network interfaces through a *Network Abstraction Layer (NAL)* allowing easy implementation on different networks. The NAL implementations can exploit networking capabilities such as RDMA.

When a NAL is initialized it does whatever is required to receive notifications about incoming data. For the TCP NAL, this means using a hook in the TCP stack; for other networks it may involve other libraries or a handshake with the NIC. When those incoming data notifications occur, the NAL is responsible for reading data off the wire and giving it to Portals as requested. Once Portals has composed an outgoing packet, the NAL is asked to put the data on the network. Both of these are asynchronous mechanisms; Portals is notified via lib_finalize upon completion.

So far three types of NALs have emerged:

**Library NALs:** Alternatively, NALs can layer on sophisticated libraries such as Quadrics's QSW kernel or Myricom's GM library and exploit remote DMA mechanisms with low overhead. Cluster File Systems and LLNL have developed such NALs.

**NIC resident NALs:** Specialized NALs can use programmable network interface cards to complete offload of the Portal protocol to the NIC. Sandia has developed such NALs.

**Basic NALs:** NALs that are relatively dumb and send packets down a device or IP socket from within the kernel.

This chapter documents the design of the Portals NAL's that are used by Lustre.

## 24.1. Interactions Between Portals and the NAL

The figure 24.1.1 is crucial in understanding the interaction between Portals and the NAL's.

## 24.2. KDAPL NAL

This section describes how a KDAPL NAL can be constructed using the v1.0 API. We make frequent reference to the KDAPL v1.0 specification and adopt terminology from this document.

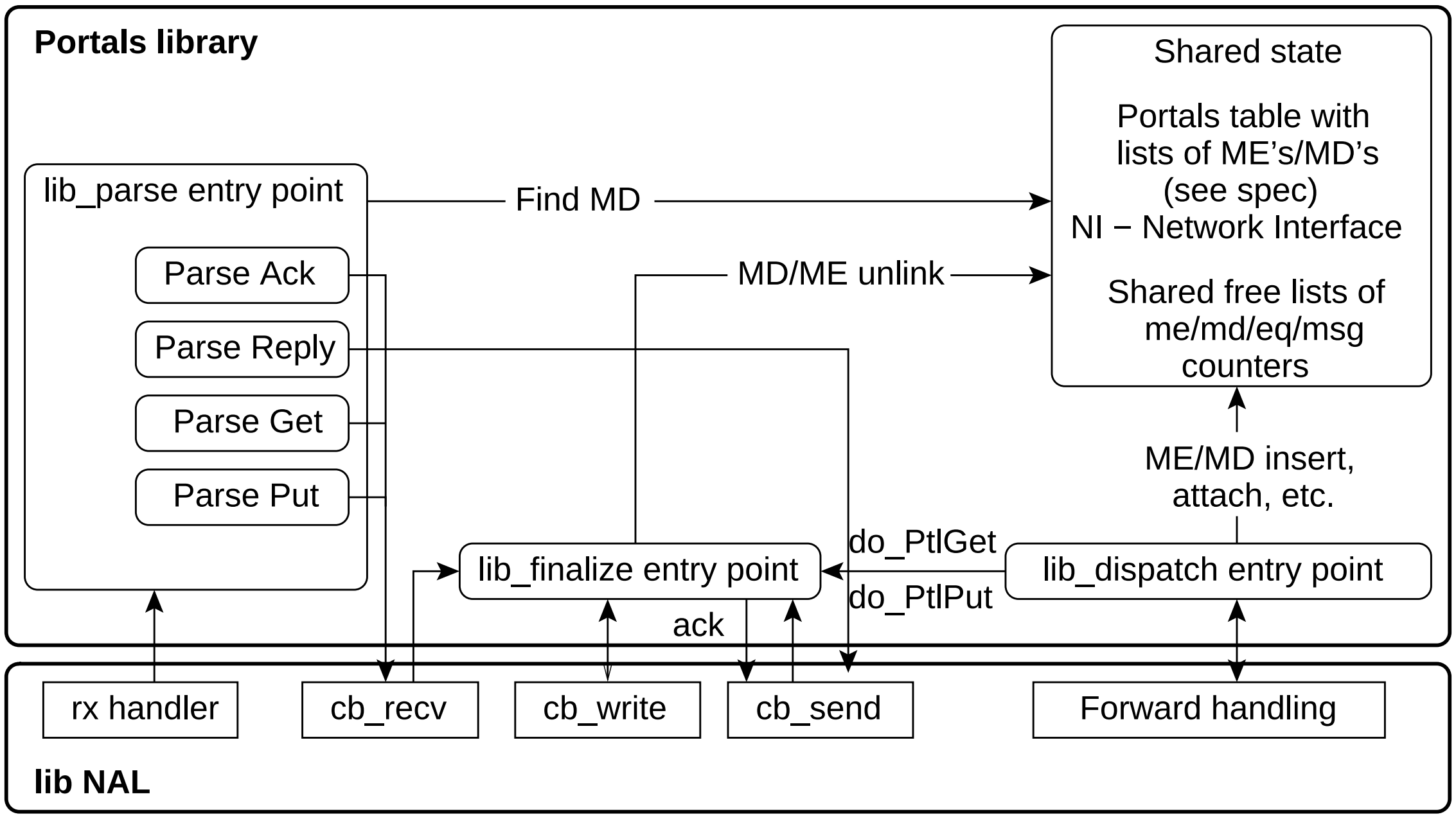

FIGURE 24.1.1. Interaction between Portals and NALs

**24.2.1. Initialization.** The NAL can be loaded into the kernel following the registration of a provider library and the generic portals code. The provider library registers itself with the KDAPL registry (DAT_REGISTRY_ADD_PROVIDER), or is contained in the static registry, in which case it can be loaded automatically.

The first thing the NAL will do is to open an interface (DAT_IA_OPEN) ultimately based on a directive in a Lustre configuration file, but invoked by the Portals NAL initialization function.

At this point the NAL should get ready for service, making threads available to handle events. This will be immediately followed by creating an event dispatcher with DAT_EVD_KCREATE.

All state changes managed by DAT and exported to the provider are event driven. This means that an event handler is run and an event queue is available to be polled. Each interface that is opened has an associated queue, the length of which is passed in at DAT_IA_OPEN time.

The event handling function will enable the threads pick up instances of events that need service. Such events include:

- connection requests
- connection completions
- RDMA completion and error events.

The event handling might involve actions that are different for passive and active nodes. Passive nodes will handle connection request events, the active side will connect to a service point offered by an acceptor and handle connecion completion.

**24.2.2. Event Handling.** A kernel level service thread should wait on events from the event dispatcher and handle connection requests.

(1) Create a service thread, which sleeps on a condition

```
    void dapl_handle_event_main(void *data)
{
     cb_nal_t *nal = (cb_nal_t *) data;
     struct dapl_ni *ni = (struct dapl_ni *)
                                  nal->data;
     while (1) {
         wait_event(ni->dni_event_waitq,
                       event_waiting(ni));
            if (ni->time_to_quit)
                 break;
            event = event_dequeue(ni);
            dapl_handle_event(event);
}
```

(2) Create an event dispatcher

The Upcall parameter should be a simple function that wakes up the service thread.

```
    typedef struct dat_upcall_object {
    DAT_PVOID        instance_data;
    DAT_UPCALL_FUNC  upcall_func;
} DAT_UPCALL_OBJECT;

typedef void(*DAT_UPCALL_FUNC)(
    DAT_PVOID instance_data,
    const DAT_EVENT event,
    DATA_BOOLEAN more_events
);
```

The instance data should probably be the ni managed by the NAL, in which appropriate pointers can be stored.

```
    void dapl_nal_handle_event(
    void *instance,
    DAT_EVENT event,
    DAT_BOOLEAN more)
{
      struct dapl_ni *ni = (struct dapl_ni *)instance;
      event_enque(ni, event);
      wake_up(ni->dni_event_waitq);
}
```

**24.2.3. Connection management.** The KDAPL NAL can provide connection management somewhat analogous to the Lustre socknal.

**24.2.4. Receive handler.** The receive handler should post a READ on each connection and wait for traffic. An event will come in when the read completes. In the simplest NAL thinkable the model is now identical to the socknal in the Lustre code. However, this NAL may suffer from buffer overflows easily.

A small refinement of this would be for larger packets to not send a body but instead to send the LMR of the sender, which creates a corresponding RMR for the sender. Now an RDMA write and read pair can be sent up to complete the transfer. (This is conceptually similar to the RTS / CTS handshake found in the Sandia NAL's.)

The receiver maintains a state machine which indicates:

**header:** the Portals header is being read.

When the header has been read, it can call lib_parse. The Portals lib will return the memory descriptor where the data needs to be copied, and the receive handler on the connection can move the state to body. While in the body state the receiver can post more receives until the whole packet is read.

In can continue to read or register a memory region for DMA, send a CTS packet back on the connection and initiate an RDMA read.

A header which is not a Portals header but which is important is the CTS header which indicates that an DAT_EP_Post_RDMA_Write.

**body:** the packet body is being moved.

In this state a small packet can be received with a DAT_EP_Post_Recv by the handler.

In case the header packet included a DAT_RMR_CONTEXT the receive handler must register the packet for DMA and post an RDMA read. To get data moving it should DAT_EP_Post_Send message to the sender to indicate that it is free to do it's send. This can be used on the remote side to register initiate a remote DMA_READ.

When it has the whole packet, it calls lib_finalize Portals generates the appropriate event.

The mechanisms to implement the receive handler are very similar to the connection acceptor. Everything generates an event and the event can wake up one (or more) threads to look at the event, analyze data correspondinly and move the state machine one step forward.

**24.2.5. Portals methods.** Perhaps the simplest part of the implementation of a KDAPL NAL is defining the Portals methods. There are two sets of methods, the api and the lib methods. The api methods are not relevant for the current discussion and can be the same as those used by the socknal and qswnal.

**cb_recv:** All cb_recv does is kick the receive handler into the body state, as in the socknal.

**cb_send:** For small packets, which were registerd without a bind of a remote region to the connection (always in the initial implementation) cb_send should simply post a send on the connection with DAT_EP_POST_send. It should possibly loop until all of the packet has been copied over, depending on credits on the connection (this is why the initial implementation may have difficulties scaling, whereas one with an RTS/CTS style mechasism will scale better).

In case this is for a large buffer, cb_send would send a small RTS packet which could simply be a Portals packet with a flag set to indicate it is an RTS and include the DAT_RMR_CONTEXT in the packet.

The lib_finalize call associated with the completion of this request must be posted later by the thread, when it wakes up on the corresponding event.

**cb_validate:** Called when building up a memory descriptor to be used by Put or Get operations.

The NAL is responsible for pinning the memory. Next it should create a local memory region DAT_LMR_kcreate. Now there are two options: one is to stop here and let the cb_send operation post a send.

The other one is to create a corresponding remote region with DAT_RMR_kcreate, register the LMR/RMR with the connection through DAT_RMR_Bind. If the latter is done, a DAT_RMR_CONTEXT is returned that can be used by cb_send to do remote DMA operations.

**cb_invalidate:** Should call DAT_LMR_free to free the memor region. It may also have to undo the binding and destroy the RMR with DAT_RMR_Free.

**cb_read:** as in other nals

**cb_write:** as in other nals

**cb_malloc:** as in other nals

**cb_free:** as in other nals

**cb_sti:** as in other nals

**cb_cli:** as in other nals

**cb_printf:** as in other nals

**cb_dist:** no idea what this is, we don't use it.

## 24.3. SockNAL

In this section we will discuss the working of the SockNAL (version 0.5.14.2). We will describe its initialization, details of its working and the SockNAL scheduler. This will provide a very good understanding of sockNALs.

The Portals stack has been explained in details in the Lustre book (`http://www.lustre.org/docs/lustre.pdf` ). As mentioned there the portals stack consists of the following layers : Portals API, NAL API, Portal Library and the NAL library. The NAL layer provides independence from the underlying network. The separation between the API and the library domain allows the flexibility to place the API either in the kernel space or in the user space. The figure shows the routing of any calls made to the Portals API

### 24.3.1. Initialization.

The initialization starts with portals initialization which is handled by *init_kportals_module*. This tests the validity of *kmem_cache* using *test_kmem_cache_validate* and then registers a miscellaneous device called *’portals’*. It then calls *PtlInit* to initialize the portals library, an entry for the *’portals’* device is also inserted into the proc file system.

The initialization for sockNAL is handled by *ksocknal_module_init* and follows the path shown below:

- Initializes the *ksocknal_data(nal_t)* data structure with function pointers for the ksocknal API functions
- Initializes *ksocknal_data(ksocknat_data_t)* - this data structure is needed globally for maintaining different connections along with their transmission and reception status and other details.
- It calls *PtlInit* to obtain the handle for the network interface (NI). This creates a portals table for the use of kernel processes. It is registered as *ksocknal_ni* allowing other kernel processes to use the handle whenever required.
- It creates a number of scheduler threads equal to the number of CPUs available
- It starts a *reaper* thread
- Calls *kpr_register*

**24.3.2. Interaction with the user space NAL library component.** At present, some parts of the NAL library reside in the user space. The NAL library exchanges information with this component at several places.

SockNAL does not create sockets in the kernel space, instead it creates them in the user space. Then it establishes the connection and hands it over to the kernal space NAL, this is done using *ioctl* that is provided newly created device *’portals’* (IOC_PORTALS_REGISTER_CLIENT_FD and IOC_PORTAL_REGISTER_SERVER_FD) . The *ioctl* IOC_PORTAL_REGISTER_MYNID passes the network id of the local host to the kernel.

**24.3.3. SockNAL Scheduler.** As described in section 24.3.1, the socknal initialization spawns several schedular threads (equal in number to the number of CPUs). These threads are in-charge of the actual data movement into and out of the sockets. the scheduler designed such that it maintains fairness between multiple senders and receivers.

24.3.3.1. *Data transmission by a process.* Data will be transmitted in two cases - when a process calls *PtlPut* and when the data that is received needs to be forwarded. In the first case, the call eventuallt reaches the send routine in *ksocknal_lib* structure, which is the *ksocknal_send* call. The *ksocknal_send* creates a packet (with the header attached) in the form of the *ksock_ltx* structure. This is handed over to the *ksocknal_launch_packet* function. This function appends the packet to the *ksnc_tx_queue* in the *ksock_conn_t* structure for the process. Further, it adds the *ksock_conn_t* structure to the *ksnd_tx_conns* list that is maintained in the *ksocknal_data_t* structure. The scheduler operates on the connections stored in this list. The *ksocknal_launch_packet* function also wakes up the scheduler if it is waiting for work to be done.

24.3.3.2. *Data transmission by the scheduler*. The scheduler takes *ksock_conn_t* structures from the *ksocknal_data_t* and passes them to the *ksocknal_process_transmit* function one at a time. This function passes the data to the sendmsg function with the MSG_DONT_WAIT flag. If all the data is passed, the packet is removed from the list in the connection, if some data is pending, the connection is added to the *ksock_tx_conns* list, otherwise it is removed. In case overflowing of the socket buffers occurs, the connection is not added to the *ksock_tx_conns*, it is handled by the *ksock_write_space* function when space is available in the socket buffers.

24.3.3.3. *Receiving data.* Whenever data arrives on a connection, the connection is added to the *ksnd_rx_conns* list in the *ksocknal_data* structure. This is done by the *ksocknal_data_ready* function, this function is registered by the sockna; in place of the original *data_ready* function used by the corresponding TCP protocol. The scheduler will call the *rcvmesg* function of the corresponding protocol with the MSG_DONTWAIT flag for all the connections enqueued in this queue (to ensure a non-blocking receive call). Depending upon the data that is received, the data in the socknal goes through a finite state machine dealing with the data. The received data goes through the following 6 states, this is also illustrated in figure 24.3.1 :

(1) SOCKNAL_RX_HEADER : indicates that the connection is supposed to accept *portal* header of the packet.
(2) SOCKNAL_RX_BODY : indicates that the connection is supposed to accept the rest of the body of the packet.
(3) SOCKNAL_RX_SLOP : indicates that the connection is supposed to neglect the rest of the data in the packet.
(4) SOCKNAL_RX_BODY_FWD : indicates that the connection is reading body of the packet that is to be forwarded.
(5) SOCKNAL_RX_GET_FMB : indicates that the connection is scheduled for forwarding of the packet.
(6) SOCKNAL_RX_FMB_SLEEP : indicates that the connection is blocked for a forward descriptor.

**24.3.4. The reaper thread.** In the case of calls to NAL shutdown API (*api_shutdown*), the socknal calls *ksocknal_close_sock* which internally calls *ksocknal_put_conn* function. This decrements the reference count for the socket, if the count becomes NULL, it closes the socket connections. The actual socket close is done by *ksocknal_close_conn* function. Closing the socket in an interrupt handler is too much work for the handler, so in that case the *ksocknal_put_conn* just appends the *ksock_conn_t* structure to a reaper list. The reaper thread, which is in-charge of this list, calls the *ksocknal_close_conn* for these connections. This thread sleeps if there are no connections on this list.

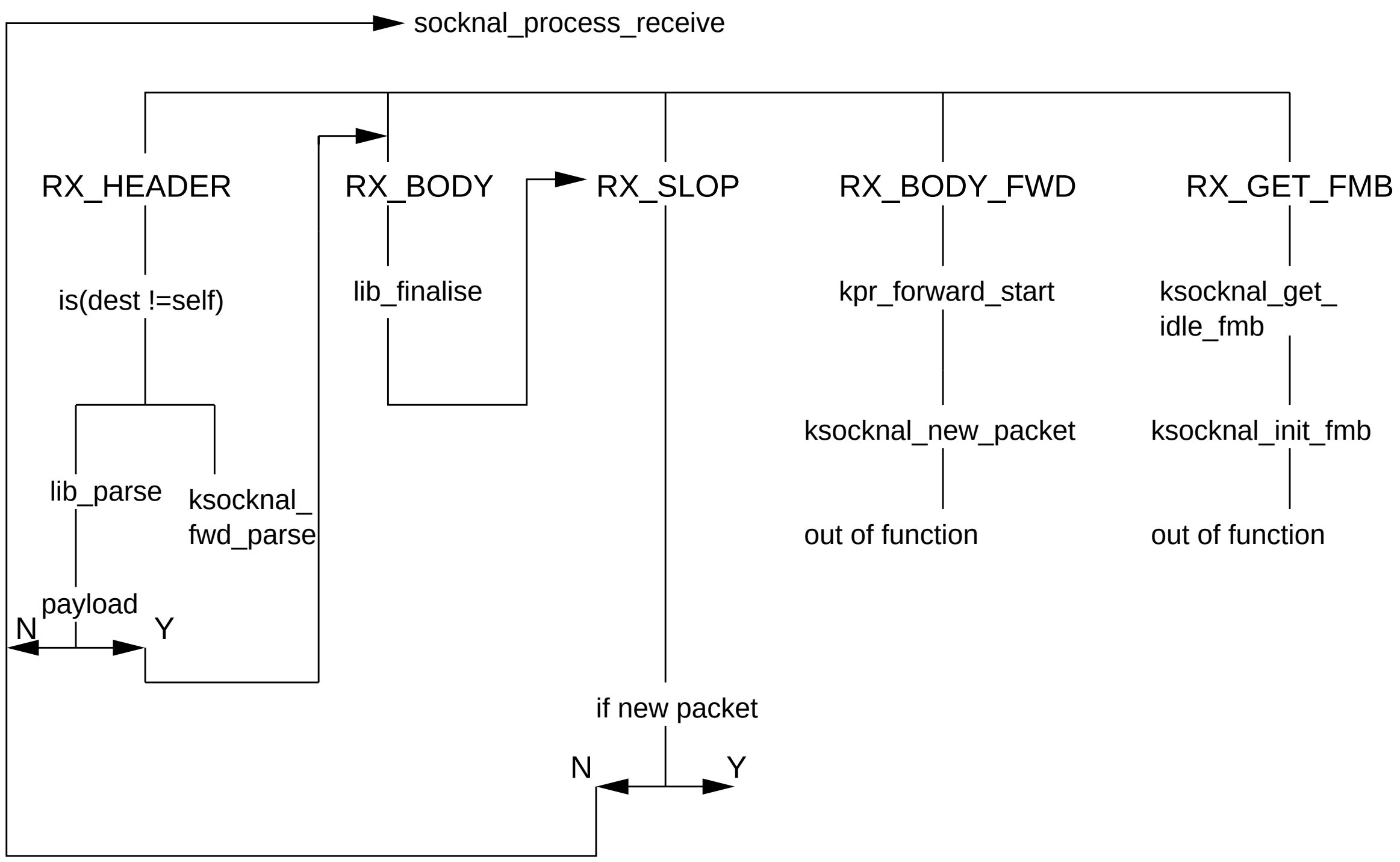

FIGURE 24.3.1. Finite state machine for received data processing in socknal

CHAPTER 25

# Object Storage API

## 25.1. Introduction

As described in the architecture section, object based storage consists of a stack of Object Based Drivers (OBD's) layered on top of the storage devices. There could be a variety of OBD's - filter, snapshot , logical volume (LOV), and MDS OBD's. We have also introduced a new Caching OBD (COBD) driver that can be used to improve read scalability as detailed in the architecture section. A serious effort has been put to make the API's stackable. All methods on object devices are placed through a simple stub API which invokes the method on the appropriate driver. The following object API descriptions are incomplete in that their semantics depend on the object driver and are possibly not correct for all conceivable future drivers.

All API calls will return an integer as their status. We define return values as a subset of the Linux i386 error numbers. When heterogeneous systems use OBD's error numbers, they need perform appropriate translation and network encapsulation.

Some of the methods are passed a *struct OBDO* as input parameter. In such cases, care must be taken to correctly set the *o_valid* field for the *OBDO*. Only those fields in the *OBDO* whose *o_valid* flag is set are used by the method. The caller sets flags in *o_valid* to indicate which fields are to be passed to the method (for *getattr*, *setattr*, etc.). The method sets flags in *o_valid* to indicate fields that it has changed when returning the *OBDO* to the caller (e.g. blocks for the *write* method). It is up to the caller to save *o_valid* flags across method calls if needed. Strict adherence to this guideline ensures consistent metadata is passed between class drivers.

In this chapter we will first describe the object API and then the utility functions and special cases offered by the class driver. We will then describe how the new caching OBD will leverage on these API's.

## 25.2. Object Interface

### 25.2.1. *iocontrol*.

25.2.1.1. *Prototype.*

```
int iocontrol (int cmd, struct lustre_handle *conn, obd_count len,
void *karg, void *uarg);
```

25.2.1.2. *Parameters.* The arguments are made available as a kernel and user buffer. All parameters are ***input/output*** parameters, i.e. they may be modified by the driver.

***input: cmd*** This is the command passed to the iocontrol request; some of the possible values are listed below:

- IOC_ LDLM_TEST, IOC_LDLM_DUMP, IOC_LDLM_REGRESS_START, IOC_LDLM_REGRESS_STOP
- IOC_OSC_REGISTER_LOV
- IOC_REQUEST_GETATTR, IOC_REQUEST_READPAGE
- IOC_LOV_SET_OSC_ACTIVE

***input: conn*** Existing Lustre handle to a device.
***input: len***
***input/output: karg*** These parameters can be input/output.

25.2.1.3. *Return Values.* The return value will depend on the *cmd* passed in to the *iocontrol* API. Below, we list some of the errors returned for the *cmd* IOC_LOV_SET_OSC_ACTIVE:

**ENOTCONN:** There is no existing connection to the OSC device specified.
**EBADF:** The type passed in the *struct OBD* is incorrect.
**EALREADY:** The OSC is already active.

25.2.1.4. *Description.* This function allows a specific OBD device to implement iocontrol commands which are not handled by the class driver's generic *ioctls*. Examples of these are commands to format devices, or partition them, or to manipulate snapshot or RAID devices.

This function may well be activated by user space administrative programs, and not by kernel code. But it is also likely to be activated through a client/target interface, in which case it is called from kernel mode.

25.2.1.5. *Examples.* The SCSI OBD driver has *ioctls* for *formatting* and *partitioning* the drive. The Ext2 OBD driver should also implement these, but it may be preferable for the Ext2 driver to do this in user space, through an upcall mechanism.

The snapshot driver has *ioctls* to *remove* and *restore* snapshots.

The *ptlrpc* driver has an *ioctl* to initiate recovery after a connection has been re-established

The MDS driver has an *ioctl* to store LOV striping information on the MDS device, as shown in the figure 25.2.1.

25.2.1.6. *Outstanding Issues.*

- Should there be user space address parameters at all?
- Should the *ioctls* handled by all kinds of drivers be specified?
- Should there be a generic upcall system for user level handling of methods?

#### 25.2.2. *get_info.*

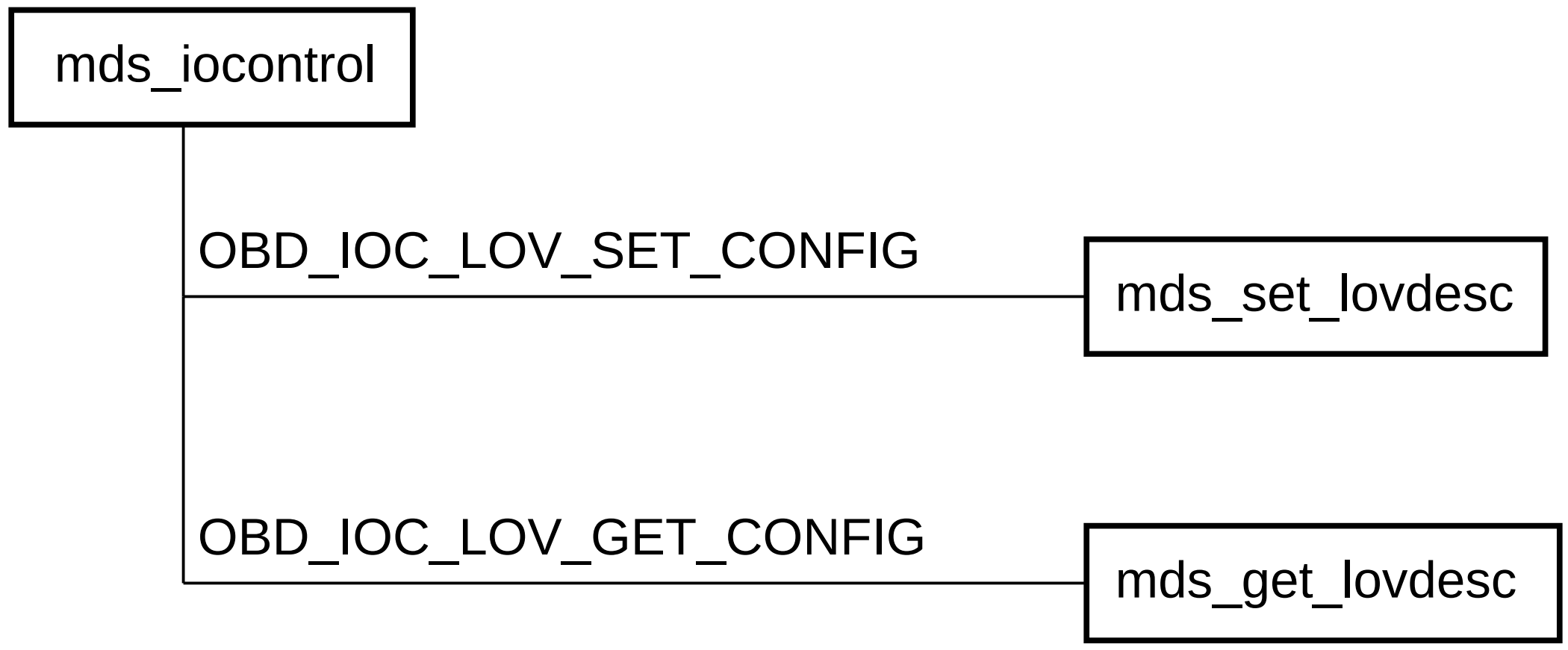

FIGURE 25.2.1. IOControl Path for MDS Driver

25.2.2.1. *Prototype.*

```
int get_info (struct lustre_handle *conn, obd_count keylen, void *key,
 obd_count *vallen, void *val);
```

25.2.2.2. *Parameters.*

***input: conn*** Existing client connection.
***input: keylen*** Length in bytes of the parameter name and other input values.
***input: key*** Null terminated name of value to be retrieved.
***input/output: vallen*** Length in bytes of the return data array, and the length of the returned data.
***output: val*** Caller-allocated array to hold data value retrieved from device.

25.2.2.3. *Return Values.* Upon successful completion *get_info* will return 0, otherwise one of the following error codes will be returned:

**EINVAL:** An invalid connection or key is specified.
**EFAULT:** The val pointer is invalid.
**EFBIG:** The return data value is too large to fit in the supplied buffer (larger than vallen).

25.2.2.4. *Description.* The *get_info* method retrieves device parameters, specified by name and possibly other input data (e.g. OST number or similar).

25.2.2.5. *Examples.* To retrieve the device blocksize, use the "blocksize" key.

25.2.2.6. *Outstanding Issues.* It would be good if direct OBD drivers provided a simple database interface for this purpose. An agreement on mandatory information which needs to be supplied by the driver is desirable too.

### 25.2.3. *set_info* /* not presently used */.

25.2.3.1. *Prototype.*

```
int set_info (struct lustre_handle *conn, obd_count keylen, void *key,
obd_count vallen, void *val);
```

25.2.3.2. *Parameters.*

***input: conn*** Existing client connection.
***input: keylen*** Length in bytes of the parameter name.
***input: key*** Null terminated name of value to be set.
***input: vallen*** Length in bytes of the input data value.
***input: val*** Data value to be stored into the parameter.

25.2.3.3. *Return Values.* Upon successful completion *set_info* will return 0, otherwise one of the following error codes will be returned:

**EINVAL:** An invalid connection or parameter is specified.
**EOPNOTSUPP:** It is not possible to set this parameter.

25.2.3.4. *Description.* The *set_info* method will set the device parameter given by *key* to *val*.

25.2.3.5. *Examples.*

25.2.3.6. *Outstanding Issues.*

### 25.2.4. *attach* /* not presently used */.

25.2.4.1. *Prototype.*

```
int attach (struct obd_device *dev, obd_count len, void *data);
```

25.2.4.2. *Parameters.*

***input: dev*** Device being attached to OBD driver, with type. The list of clients is initialized.
***input: len*** Length in bytes of type specific attachment data.
***input: data*** Attach data that is private to the particular type.

25.2.4.3. *Return Values.* Upon successful completion *attach* will return 0, otherwise one of the following error codes will be returned:

**EBUSY:** The device is busy or already in use.
**ENODEV:** An invalid or unconfigured device is specified.
**EFAULT:** Data pointer is invalid or inaccessible.
**EINVAL:** Invalid data is specified.
**ENOMEM:** No memory is available.

25.2.4.4. *Description.* The *attach* function performs device specific attach operations beyond the generic setting of the OBD type of the *dev*, which also associates the appropriate OBD methods for accessing the device.

25.2.4.5. *Examples.* An *attach* function is not required for an OBD driver. For the SCSI driver the *data* field holds SCSI connection information - adapter, bus, ID, LUN.

25.2.4.6. *Outstanding Issues.* Is this perhaps a service only offered by the class driver?

The naming of devices should perhaps not be associated with a character device.

The locking of modules should be more closely interacting with attach.

### 25.2.5. *detach* /* not used */.

25.2.5.1. *Prototype.*

```
int detach (struct obd_device *dev);
```

25.2.5.2. *Parameters.*

***input: dev*** Specifies the device to be unconfigured.

25.2.5.3. *Return Values.* Upon successful completion *detach* will return 0, otherwise one of the following error codes will be returned:

**EBUSY:** The device is still set up or has existing connections.
**ENODEV:** An invalid or unconfigured device is specified.

25.2.5.4. *Description.* The *detach* function removes the OBD methods and unregisters its OBD type. Prior to a successful *detach* command, all existing connections must be closed and the *cleanup* routine for the device been called.

25.2.5.5. *Examples.*

25.2.5.6. *Outstanding Issues.* Is this method necessary?

### 25.2.6. *setup*.

25.2.6.1. *Prototype.*

```
int setup (struct obd_device *dev, obd_count len, void *data);
```

25.2.6.2. *Parameters.*

***input: dev*** Specifies the device to be setup.
***input: len*** Length in bytes of type specific attachment data.
***input: data*** Setup data that is private to the particular type.

25.2.6.3. *Return Values.* Upon successful completion *setup* will return 0, otherwise one of the following error codes will be returned:

**EBUSY:** The device is busy or already in use.
**ENODEV:** An invalid or unattached device is specified.
**EINVAL:** Invalid data is specified.

25.2.6.4. *Description.* The *setup* performs device configuration for the driver. It must be called after the *attach* method for the device. What is done by *setup* is unwound by *cleanup*.

25.2.6.5. *Examples.* The snapshot driver uses *setup* to establish connections to its multiple child devices. The *OBDfilter* driver uses the *setup* to mount the underlying filesystem and perform driver initialization tasks.

25.2.6.6. *Outstanding Issues.* We have not reached totally firm conclusion as to what is to be undertaken at attach time and what should be left to the setup routine.

### 25.2.7. *cleanup.*

25.2.7.1. *Prototype.*

```
int cleanup (struct obd_device *dev);
```

25.2.7.2. *Parameters.*

***input: dev*** Device being cleaned up/ the list of clients is cleared.

25.2.7.3. *Return Values.* Upon successful completion *cleanup* will return 0, otherwise one of the following error codes will be returned:

**EBUSY:** The device still has existing clients.

25.2.7.4. *Description.* The *cleanup* method unconfigures the device. All client connections must have been *disconnected* from this device prior to the call. It must be called before the device can *detach*. After the function returns, no data associated with the client driver for this minor should remain.

25.2.7.5. *Examples.* The *OBDfilter* driver uses *cleanup* to unmount the underlying filesystem and flush all pending I/O to disk.

25.2.7.6. *Outstanding Issues.*

### 25.2.8. *connect.*

25.2.8.1. *Prototype.*

```
int connect (struct lustre_handle *conn, struct obd_device *obd,  obd_uuid_t
client_uuid);
```

25.2.8.2. *Parameters.*

***input/output: conn*** Holds the import handle the target of the connect call will store. Upon return it will contain the export handle to be used by the client for further calls.
***input: obd*** Device to connect to.
***input: client_uuid*** UUID of the client. Can be used by the target of the connect call to re-establish a connection that was already open, and not dropped yet.

25.2.8.3. *Return Values.* Upon successful completion *connect* will return 0, otherwise one of the following error codes will be returned:

**ENODEV:** An invalid or unconfigured, or unsetup device is specified.
**EINVAL:** Invalid data is specified.
**ENOMEM:** No memory is available.

25.2.8.4. *Description.* This function initializes a connection or session to a device. This allows for connection specific data to be managed by the device and cleaned up when a connection is severed. Typical examples of such data are preallocated object IDs or reserved drive space. Connection IDs are not reused until the driver is restarted.

25.2.8.5. *Examples.* The filesystem (Lustre Lite) connects to the driver during mount. A logical driver connects to its children drivers (think e.g. of RAID1 over multiple devices) during its initialization.

25.2.8.6. *Outstanding Issues.*

### 25.2.9. *disconnect.*

25.2.9.1. *Prototype.*

```
int disconnect (struct lustre_handle *conn);
```

25.2.9.2. *Parameters.*

***input: conn*** Connection to be stopped.

25.2.9.3. *Return Values.* Upon successful completion *disconnect* will return 0, otherwise one of the following error codes will be returned:

**ENODEV:** An invalid or unconfigured device is specified.
**EINVAL:** Invalid data is specified.

25.2.9.4. *Description.* The *disconnect* method frees any resources associated with a client connection and closes the connection to the driver.

25.2.9.5. *Examples.* The *OBDfilter disconnect* method will drop open file references when it is closed.

25.2.9.6. *Outstanding Issues.*

### 25.2.10. *statfs.*

25.2.10.1. *Prototype.*

```
int statfs (struct lustre_handle *conn, struct obd_statfs *osfs);
```

25.2.10.2. *Parameters.*

***input: conn*** Existing client connection.
***output: osfs*** Structure to hold device status information.

25.2.10.3. *Return Values.* Upon successful completion *statfs* will return 0, otherwise one of the following error codes will be returned:

**EINVAL:** An invalid connection is specified.

25.2.10.4. *Description.* The *statfs* method returns status information about the device, such as the device type, block size, the number of total and free blocks and objects.

25.2.10.5. *Examples.*

25.2.10.6. *Outstanding Issues.* For snapshot devices, the number of free blocks and objects will decrease even if new objects are not created because of object replication within the driver.

### 25.2.11. *preallocate* /* not used yet */.

25.2.11.1. *Prototype.*

```
int preallocate (struct lustre_handle *conn, obd_size *req, obd_off *IDs);
```

25.2.11.2. *Parameters.*

***input: conn*** Existing client connection.
***input/output: req*** Number of objects requested/returns number allocated.
***output: IDs*** Array of object IDs allocated.

25.2.11.3. *Return Values.* Upon successful completion *preallocate* will return 0, otherwise one of the following error codes will be returned:

**EBUSY:** The device is busy or already in use.
**ENODEV:** An invalid or unconfigured device is specified.
**EFAULT:** Data pointer is invalid or inaccessible.
**EINVAL:** Invalid data is specified.
**ENOMEM:** No memory is available.

25.2.11.4. *Description.* The *preallocate* method will mark a number of objects as reserved for a particular connection (i.e. they are guaranteed to be usable by the client).

The number of requested objects can be at most 32. The number of objects actually preallocated will be returned in *req*, and is not guaranteed to equal the requested number.

25.2.11.5. *Examples.* Preallocation of objects can be used to mitigate latency when round-trip communication between the client and target imposes significant overhead. This API can also be used in the implementation of *writeback cache* at the clients. In the presence of the *writeback cache*, dirty pages will not be flushed to the *target OST's* immediately. Instead they would be kept in the *writeback cache* and marked *dirty*; the *write* call would then return indicating to the client that the *write* was successful (even though the write request did not go to the target OST's). In this case, when the data is flushed out to the OST's at a later time, we need to ensure that there is space available for it. To ensure this, we can invoke a *preallocate* call on the target OST's to reserve space for the *unflushed writes* before indicating to the client that the *write* request was completed.

25.2.11.6. *Outstanding Issues.* Preallocation of objects is implemented, but not yet exploited.

### 25.2.12. *create.*

25.2.12.1. *Prototype.*

```
int create (struct lustre_handle *conn, struct obdo *oa, struct
lov_stripe_md **md);
```

25.2.12.2. *Parameters.*

***input: conn*** Existing client connection.
***input/output: oa*** *OBDO* to be created. Fields in the *OBDO* can be set to give hints to the driver when allocating the object.
***input/output: md*** Contains the object IDs when the *OBD_create* call is placed against a striped device.

25.2.12.3. *Return Values.* Upon successful completion *create* will return 0, otherwise one of the following error codes will be returned:

**EINVAL:** Invalid data is specified.
An invalid connection was given.
**ENOMEM:** No memory is available.
**EIO:** A read or write error occurred while accessing the device.
**ENOSPC:** No space left on device.

25.2.12.4. *Description.* The *create* method will create a new object on the device. Fields in the *OBDO* can be specified when calling *create* (e.g. object ID, mode, or size), along with the corresponding *valid* flag, in order to give hints to the driver when creating the object. The driver is not constrained by these hints. Object IDs can be preallocated for this connection with the *preallocate* method.

We hope that the chosen form of the parameters for *create* are flexible enough for future extensions of this kind.

25.2.12.5. *Examples.* The *OBDflags* field can interact with the parameters to control the behavior of the disk during *create*, for example in the following ways:

(1) If the ID is set to 0 or the ID valid flag is not set, the drive will find an available ID to use.
(2) The *OBDflags* may contain hints to pre-allocate space for the object, in conjunction with the size field. It can make sure the object is contiguous or make sure the object is layed out such that various bandwidth requirements can be met.
(3) If the ID is set, the *OBDflags* can indicate that the ID was preallocated, and use the given ID. Otherwise, a valid ID flag indicates an attempt to create an object with a given ID without certainty of success.
(4) The *OBDOflags* can be used in conjunction with the *OBDmd* to transfer a crypto method and key to be applied to the file.
(5) The *OBDO* metadata can also be used to specify a nearby object which the drive may use in conjunction with the *OBDflags* field to, for example, locate files near the directory they were first created in.

25.2.12.6. *Outstanding Issues.* None of the *OBDO* creation hints are used at this time.

### 25.2.13. *destroy*.

25.2.13.1. *Prototype.*

```
int destroy (struct lustre_handle *conn, struct obdo *oa, struct
lov_stripe_md *md);
```

25.2.13.2. *Parameters.*

***input: conn*** Existing client connection.
***input/output: oa*** *OBDO* to be deleted. At least the ID field must be set.
***input: md*** The *md* contains the IDs of the objects scattered over a striped device when the obd_destroy call is placed against a striped device.

25.2.13.3. *Return Values.* Upon successful completion *destroy* will return 0, otherwise one of the following error codes will be returned:

**EINVAL:** An invalid connection is specified.
**ENOENT:** An invalid object ID is specified.

25.2.13.4. *Description.* The *destroy* method will delete the specified object from the device.

25.2.13.5. *Examples.*

25.2.13.6. *Outstanding Issues.* *Destroy* can make good use of the flags in the *OBDO*. For example, it can erase an object for secure deletion, or initiate backups for objects tagged as needing one upon deletion.

### 25.2.14. *setattr*.

25.2.14.1. *Prototype.*

```
int setattr (struct lustre_handle *conn, struct obdo *oa, struct
lov_stripe_md *ea);
```

25.2.14.2. *Parameters.*

***input: conn*** Existing connection to OBD driver.
***input: oa*** Attributes of an *OBDO*. The valid field specifies which attributes contain meaningful data and should be written to the device.
***input: ea*** The ea will contain the object IDs when the call is placed against a striped device.

25.2.14.3. *Return Values.* Upon successful completion *setattr* will return 0, otherwise one of the following error codes will be returned:

**ENOENT:** An invalid object was specified.
**EINVAL:** Invalid data is specified.
**ENOMEM:** No memory is available.

25.2.14.4. *Description.* The *setattr* method writes the object attributes such as atime, mtime, size, etc to the device. Only attributes specified by *valid* flags will be stored to the device.

25.2.14.5. *Examples.* The *setattr* method is called by *notify_change* method to set specific attributes. The caller is responsible for setting the valid flags.

25.2.14.6. *Outstanding Issues.* Caution should be taken when specifying which attributes are to be copied. OBD metadata is normally private to a particular OBD driver and not to be shared among multiple drivers, since this introduces false sharing.

The definition of sufficiently rich metadata will continue to be an issue.

### 25.2.15. *getattr.*

25.2.15.1. *Prototype.*

```
int getattr (struct lustre_handle *conn, struct obdo *oa, struct
lov_stripe_md *md);
```

25.2.15.2. *Parameters.*

***input: conn*** Existing connection to OBD driver.
***input/output: oa*** Attributes of an *OBDO*. The ID and valid fields must be set prior to calling *getattr*.
***input/output: md*** When placing the call, the *md* will contain the object IDs of the objects scattered over a striped device. The *md* will return the attributes stored in a striped device when the call is made against a logical volume.

25.2.15.3. *Return Values.* Upon successful completion, *getattr* will return 0, otherwise one of the following error codes will be returned:

**ENOENT:** An invalid object ID was specified.
**EINVAL:** Invalid data is specified.
**ENOMEM:** No memory is available.

25.2.15.4. *Description.* The *getattr* retrieves metadata such as atime, mtime, size, etc... for the given object ID. Only those fields specified by *valid* will be retrieved from the device and stored in the *OBDO*.

25.2.15.5. *Examples.*

25.2.15.6. *Outstanding Issues.*

### 25.2.16. *open*.

25.2.16.1. *Prototype.*

```
int open (struct lustre_handle *conn, struct obdo *oa, struct
lov_stripe_md *md);
```

25.2.16.2. *Parameters.*

***input: conn*** Existing client connection.
***input/output: oa*** Attributes of an *OBDO*. The ID, mode, and valid fields must be set prior to calling open.
***input/output: md*** When placing the call, the *md* will contain the object IDs of the objects scattered over a striped device. The *md* will return the attributes stored in a striped device, when the call is made against a logical volume.

25.2.16.3. *Return Values.* Upon successful completion *open* will return 0, otherwise one of the following error codes will be returned:

**EINVAL:** Invalid data is specified.
**ENOENT:** The object ID specified does not exist.
**EIO:** An error was encountered accessing the device.

25.2.16.4. *Description.* The *open* method will open a handle for the given object ID on the device. This allows the device to track usage of an object and avoid destroying the object if it is still in use.

25.2.16.5. *Examples.* The *open* call would be used in case of an open request from the client for an object on the *target OST.* The client will first obtain the metadata information from the MDS; this will contain the object IDs. The client request would contain the object ID for the object to be opened; the *OST_open* call is traced in figure 25.2.2.

25.2.16.6. *Outstanding Issues.*

FIGURE 25.2.2. OST open request

### 25.2.17. *close.*

25.2.17.1. *Prototype.*

```
int close (struct lustre_handle *conn, struct obdo *oa, struct
 lov_stripe_md  *md);
```

25.2.17.2. *Parameters.*

***input: conn*** Existing client connection.
***input: oa*** Attributes of an *OBDO*. The ID, mode, and valid fields must be set prior to calling open.
***input: md*** Contains the object IDs of the objects scattered over a striped device.

25.2.17.3. *Return Values.* Upon successful completion *close* will return 0, otherwise one of the following error codes will be returned:

**EINVAL:** Invalid data is specified.
**ENOENT:** The object ID specified does not exist.
**EIO:** An error was encountered accessing the device.

25.2.17.4. *Description.* The *close* method will close a previously opened handle for the given object ID on the device. This method will verify if there was a valid connection to the client who sent the close request. It will then check if a valid *handle* exists for the object requested to be closed. If everything is alright then the object is closed.

25.2.17.5. *Examples.*

25.2.17.6. *Outstanding Issues.*

### 25.2.18. *brw.*

25.2.18.1. *Prototype.*

```
int brw(int rw, struct lustre_handle *conn, struct lov_stripe_md *md,
 obd_count bufs, struct brw_page *pgarray, brw_callback_t *callback, void
 *data);
```

25.2.18.2. *Parameters.*

***input: rw*** Flag indicating if a read or write operation is done, set to either OBD_BRW_READ or OBD_BRW_WRITE.
***input: conn*** Existing client connection.
***input: md*** Contains a descriptor of a striped object to which the I/O will be done.
***input: bufs*** The number of pages to transfer.
***input: pgarray*** Array of buffer descriptors.

```
struct brw_page {
  struct page *pg;
  obd_size count;
obd_off offset;
  obd_flag flag;
};
```

The pages must be aligned on a page boundary.
***input: callback*** Callback function to invoke when the I/O has been initiated and completed.

```
typedef int (*brw_callback_t)(void *, int err, int phase);
```

**phase** will be passed in by the function calling the callback and will CB_PHASE_START, CB_PHASE_FINISH.
**data** will equal the data parameter passed in.
**err** will contain the error value of the operation.

25.2.18.3. *Return Values.* Upon successful completion *brw* will return 0, otherwise one of the following error codes will be returned:

**EFAULT:** Data pointer is invalid or inaccessible.
**EINVAL:** Invalid data is specified.
**ENOMEM:** No memory is available.
**EIO:** An error was encountered accessing the device.

25.2.18.4. *Description.* The *brw* method will do asynchronous I/O on an object; it will perform page aligned I/Os to the device. Based upon the value of the *rw* flag, it will invoke *brw_read* or *brw_write*. The callback function would be invoked after the required work has been done at each driver level. So, for an LOV device the callback sleeps until the start phase has happened and the I/O on all the OSC devices below the LOV have completed. On the other hand, the callback for an OSC device simply sleeps until both the start and finish phases for the request have passed.

25.2.18.5. *Examples.* An instance of this method can be found in the OSC driver. The *OSC_brw* method is invoked when a read/write request has to be performed on a *target OST*. Depending on the value of *rw* flag, it invokes either *OSC_brw_read* or *OSC_brw_write*. The method performs vector I/O by reading/writing multiple pages at one time. A limit to the IOV (I/O Vector) size is set by the PTL_MD_MAX_IOV (presently set to 16).

In case of *OSC_brw_read*, we have to first prepare a bulk descriptor into which the pages would be read and these have to be registered with portals. The call is blocked till this read request completes.

For reads we always read a full page, but this could be optimized to NULL out the page section that lies beyond the end of the file.

For a *write* request, *OSC_brw_write* function is called. In this case, an *OST_WRITE* request is first sent to the *target OST* with information about the size of the write; the *target OST* will prepare and register the required sink for the *write*. The actual *write* would be done using a *ptlrpc_send_bulk* method. There were a couple of issues with the *OSC_brw_write* method. First *OSC_brw_write* sent small buffers for small writes, but these were accidentally dumped into pages that were expecting a full overwrite. That was fixed by calling *filter_get_page_write* with a correct length for the preparation of the page. Secondly, mapped writes were always sending whole pages across which leads to file size extensions. We now only send the data up to the end of the file.

25.2.18.6. *Outstanding Issues.* Extensions for raw, asynchronous, parallel, and collective I/O are possible with this interface and can be added. An issue that remains open in *OSC_brw_write* is how to handle the bounce pages we introduce when the writes hit locked pages (this is not possible for single page writes but it is for IOV's). At present, we fill the page with byte 0xBA for badness to alert developers to this problem.

Another issue that remains is of sparse file reads. A read request for sparse file could include a request for unmapped pages of the file. At present, we send back zero-filled pages from the *target OST's* to the *OSC's* in case of an unmapped page; this is a waste of network bandwidth. We need to modify the *filter_preprw* on the *OBDfilter* device so that the unmapped pages are not sent.

#### 25.2.19. *preprw.*

25.2.19.1. *Prototype.*

```
int preprw (int cmd, struct lustre_handle *conn, int objcount, struct
 obd_ioobj *obj, int niocount, struct niobuf_remote *remote, struct
 niobuf_local *local, void **desc_private);
```

25.2.19.2. *Parameters.*

***input: cmd*** Flag indicating if a read or write operation is done, set to either OBD_BRW_READ or OBD_BRW_WRITE.
***input: conn*** Existing client connection.
***input: objcount*** The number of objects being passed in the *obj* parameter.
***input: obj*** Array of *objcount* structures identifying the object(s) to read from or write to and the number of buffers in *remote* and *local* which belong to each object.

```
struct obd_ioobj {
  obd_id  ioo_id;
  obd_gr  ioo_gr;
  __u32   ioo_type;
  __u32   ioo_bufcnt;
};
```

***input: niocount*** The total number of buffers in the *remote* and *local* arrays.
***input: remote*** Array of *niocount* structures identifying the read or write buffers to be prepared on the device. The *xid* field holds the caller's transaction ID.

```
struct niobuf_remote {
  __u64  offset;
  __u32  len;
  __u32  xid;
  __u32  flags;
};
```

***output: local*** Array of *niocount* structures identifying the buffers prepared by this call. The array is allocated by the caller and filled in by *preprw*.

```
struct niobuf_local {
  __u64  offset;
  __u32  len;
  __u32  xid;
  __u32  flags;
  void *addr;
  struct page *page;
  void *target_private;
  struct dentry *dentry;
};
```

***output: desc_private*** A private pointer that can be used to pass data between *preprw* and *commitrw.*

25.2.19.3. *Return Values.* Upon successful completion *preprw* will return 0, otherwise one of the following error codes will be returned:

**EINVAL:** Invalid data is specified.
**ENOENT:** The object ID specified does not exist.
**ENOSPC:** No space was available to write to the disk.
**ENOMEM:** No memory was available to allocate buffers.
**EIO:** An error was encountered accessing the device.

25.2.19.4. *Description.* For a data write, the *preprw* method will prepare receive buffers in memory and allocate space in the filesystem to write into for the objects and at the offsets and lengths specified. After the *preprw* call completes, the data will be written into the buffer addresses returned in the *local* struct.

For a data read, the *preprw* method will read the data for the objects at the offsets and lengths specified and return the buffer addresses in the *local* struct.

The *local* and *remote* buffers are packed into a single array in the same order as the objects in the *obj* array.

25.2.19.5. *Examples.* To write 2 pages with offsets A1, A2, B1, and B2 to each of 2 objects with objIDs A and B, we would have:

```
objcount = 2;
obj = {{ .ioo_id = A, .ioo_bufcnt = 2 },
       { .ioo_id = B, .ioo_bufctn = 2 }};
niocount = 4;
niobuf_remote = {{ .offset = A1, .len = 4096 },
                 { .offset = A2, .len = 4096 },
                 { .offset = B1, .len = 4096 },
                 { .offset = B2, .len = 4096 }};
```

The preprw call will prepare the 4 buffers and allocate space in the filesystem for these buffers.

25.2.19.6. *Outstanding Issues.* The specified offsets must be page aligned, and the lengths must be a multiple of the page size. The *filter_preprw* method needs to be modified for sparse file reads.

**25.2.20. *commitrw*.**

25.2.20.1. *Prototype.*

```
int commitrw (int cmd, struct lustre_handle *conn, int objcount, struct
 obd_ioobj *obj, int niocount, struct  niobuf_local *local, void
*desc_private);
```

25.2.20.2. *Parameters.*

***input: cmd*** Flag indicating if a read or write operation is done, set to either OBD_BRW_READ or OBD_BRW_WRITE.
***input: conn*** Existing client connection.
***input: objcount*** The number of objects being passed in the *obj* parameter.
***input: obj*** Array of *objcount* structures identifying the object(s) to read from or write to and the number of buffers in *local* which belong to each object.

```
struct obd_ioobj {
  obd_id  ioo_id;
  obd_gr  ioo_gr;
  __u32   ioo_type;
  __u32   ioo_bufcnt;
};
```

***input: niocount*** The total number of buffers in the *local* array.
***output: local*** Array of *niocount* structures identifying the buffers prepared by this call. The array is allocated by the caller and filled in by *preprw*.

```
struct niobuf_local {
  __u64  offset;
  __u32  len;
  __u32  xid;
  __u32  flags;
  void *addr;
  struct page *page;
  void *target_private;
  struct dentry *dentry;
};
```

***output: desc_private*** A private pointer that can be used to pass data between *preprw* and *commitrw.*

25.2.20.3. *Return Values.* Upon successful completion *commitrw* will return 0, otherwise one of the following error codes will be returned:

**EINVAL:** Invalid data is specified.
**EIO:** An error was encountered accessing the device.

25.2.20.4. *Description.* For a data write, the *commitrw* method will write the buffers allocated in *preprw* from memory into the filesystem.

It then frees any buffers allocated and transactions started in *preprw*. The *commitrw* call will always be called after a *preprw* call completed successfully.

25.2.20.5. *Examples.*

25.2.20.6. *Outstanding Issues.*

**25.2.21. *enqueue.***

25.2.21.1. *Prototype.*

```
int enqueue (struct lustre_handle *connh, struct lov_stripe_md *md, struct
 lustre_handle *parent_lock, __u32 type, void *extentp, int extent_len,
__u32  mode, int *flags, void *callback, void *data, int datalen, struct
 lustre_handle *lockh);
```

25.2.21.2. *Parameters.*

***input: connh*** Existing client connection.
***input: md*** LOV descriptor of the object to be locked.
***input: parent_lock*** Valid lock handle of a parent lock, or NULL.
***input: type*** Lock type (one of LDLM_PLAIN, LDLM_EXTENT, or LDLM_MDSINTENT).
***input: extentp*** If type is LDLM_EXTENT, this is a pointer to struct ldlm_extent.
***input: extent_len*** If type is LDLM_EXTENT, this is the size of eventp.
***input: mode*** Lock mode of the new lock.
***input: callback*** A callback function of the type *ldlm_blocking_callback*.

***input: data*** Opaque data stored in the lock.
***input: datalen*** Length of the opaque data.
***input/output: flags*** Used to pass information to and receive information from ldlm_lock_enqueue.
***output: lockh*** The new lock handle(s).

25.2.21.3. *Return Values.*

**ENOMEM:** Unable to allocate locks, buffers, or temporary structures.
**EINTR:** The user aborted the lock request after a timeout.
**EINVAL:** One of the many input parameters is invalid.

25.2.21.4. *Description.* This method is used to acquire referenced locks. Exactly how is left to the implementation, and may involve matches, enqueues, or conversions. *enqueue* is not required to return exactly the lock that is requested, but rather can return any lock that the underlying driver knows will protect the operation.

*lockh* may be an array of lock handles, and multiple locks may be enqueued, if *md* describes many objects.

No callers currently pass a non-NULL parent_lock.

25.2.21.5. *Examples.* `osc/osc_request.c:osc_enqueue()` is an excellent example, because it goes through a number of steps to acquire a suitable lock. First it tries to match existing locks of the mode (PW or PR) requested. If unsuccessful, it tries to match the other of the two modes. If the caller needs a PR lock and we already have a PW lock, then we trust the VFS to protect the reader/writer conflicts and return a reference to the PW lock.

If the caller requests a PW lock and we already have a PR lock, we attempt to convert the lock. Finally, if we have no locks that match, it enqueues a new lock at the requested mode.

25.2.21.6. *Outstanding Issues.* *extentp* and *extent_len* are to become a generic cookie for the lock enqueue function. *callback* would be of the type ldlm_blocking_callback, but some header dependency issues need to be resolved.

### 25.2.22. *cancel.*

25.2.22.1. *Prototype.*

```
int cancel (struct lustre_handle *connh, struct lov_stripe_md *md, __u32
 mode, struct lustre_handle *lockh);
```

25.2.22.2. *Parameters.*

***input: connh*** Existing client connection.
***input: md*** The LOV_stripe_md which was passed to enqueue.
***input: mode*** The lock mode which was passed to enqueue.
***input: lockh*** Lock handle to cancel.

25.2.22.3. *Return Values.*

**EINVAL:** The lock handle is invalid.

25.2.22.4. *Description.* This method releases the caller's hold on the given lock(s)–*lockh* may be an array of lock handles, if *md* refers to multiple objects.

25.2.22.5. *Examples.*

25.2.22.6. *Outstanding Issues.* This method is poorly named, as it never actually cancels the lock, just decreases the reference count.

### 25.2.23. *sync* /* not implemented yet */.

25.2.23.1. *Prototype.*

```
int sync (struct lustre_handle *conn, struct obdo *tgt, struct
 lov_stripe_md *md, obd_off offset, obd_size size);
```

25.2.23.2. *Parameters.*

***input: conn*** Existing client connection.
***input: oa*** Existing *OBDO* to sync data.
***input: offset*** Offset in bytes from start of file to begin flush.
***input: count*** Number of bytes to flushed to device.

25.2.23.3. *Return Values.* Upon successful completion *sync* will return 0, otherwise one of the following error codes will be returned:

**EINVAL:** Invalid data is specified.
**EIO:** An error was encountered accessing the device.

25.2.23.4. *Description.* The *sync* method will flush data from buffers to the underlying device. It is possible to flush only a portion of the buffers by specifying the *count* and *offset*. An offset of 0 and count equal to the object size indicate that all of the outstanding I/O on the object should be completed.

The *sync* method will either return after pending I/O is scheduled for writing to the device, or it will wait for all pending device I/O to complete, as specified in the *OBDflags*.

25.2.23.5. *Examples.*

25.2.23.6. *Outstanding Issues.* Implemented but not used.

### 25.2.24. *punch*.

25.2.24.1. *Prototype.*

```
int punch (struct lustre_handle *conn, struct obdo *tgt, obd_size count,
 obd_count offset);
```

25.2.24.2. *Parameters.*

***input: conn*** Existing client connection.
***input: oa*** Existing *OBDO*.
***input: count*** Number of bytes to punch from object.
***input: offset*** Offset in bytes from start of file to do punch.

25.2.24.3. *Return Values.* Upon successful completion *punch* will return 0, otherwise one of the following error codes will be returned:

**EINVAL:** Invalid data is specified.
**EIO:** An error was encountered accessing the device.

25.2.24.4. *Description.* The *punch* method will create a hole *count* bytes long starting at *offset* bytes from the start of the file. For holes that are not block aligned and of block size, the partial block(s) will be zero-filled at the specified location. The object will be truncated to be *offset* bytes long by using *offset+count* larger than *o_size*.

25.2.24.5. *Examples.* For OBDfs truncate is mapped onto punch. This function is also useful for HSM and other arenas where truncate isn't flexible enough.

25.2.24.6. *Outstanding Issues.* Prototype code is included, but not fully implemented. The semantics of mapping truncate onto punch are not fully specified.

### 25.2.25. *migrate.*

25.2.25.1. *Prototype.*

```
int migrate (struct lustre_handle *conn, struct obdo *dst, struct obdo *src,
 obd_size *size, obd_off *offset);
```

25.2.25.2. *Parameters.*

***input: conn*** Existing client connection.
***input/output: dst*** Existing *OBDO* to migrate to.
***input: src*** Existing *OBDO* to migrate from.
***input: size*** Number of bytes to migrate.
***input: offset*** Offset in bytes from start of file to migrate.

25.2.25.3. *Return Values.* Upon successful completion *migrate* will return 0, otherwise one of the following error codes will be returned:

**EINVAL:** Invalid data is specified.
**ENOMEM:** No memory is available.

25.2.25.4. *Description.* The *migrate* method will move data blocks from one object to another object at the specified *offset*. The entire object will be migrated with an offset of 0, and a size of *o_size*.

Depending on the implementation, there may be a requirement that the *offset* and *size* be an even multiple of the page size.

25.2.25.5. *Examples.* The snapshot driver uses the *migrate* method to do copy-on-write migration of data from the primary object to an indirect object.

25.2.25.6. *Outstanding Issues.* Currently all the object data is migrated from src to dst. It would be interesting to only migrate certain segments of the data.

#### 25.2.26. *copy*.

25.2.26.1. *Prototype.*

```
int copy (struct lustre_handle *dstconn, struct obdo *dst, struct
     lustre_handle *srcconn, struct obdo *src, obd_size *size, obd_off *offset);
```

25.2.26.2. *Parameters.*

***input: dstconn*** Existing client connection for destination *OBDO*.
***input/output: dst*** Existing *OBDO* to copy data to.
***input: srcconn*** Existing client connection for source *OBDO*.
***input: src*** Existing *OBDO* to copy data from.
***input: size*** Number of bytes to copy.
*input: offset* Offset in bytes from start of file to copy.

25.2.26.3. *Return Values.* Upon successful completion *cop*y will return 0, otherwise one of the following error codes will be returned:

**ENOENT:** An invalid object was specified.
**EINVAL:** Invalid data is specified.
**ENOMEM:** No memory is available.

25.2.26.4. *Description.* The *copy* method will copy data from one object to another object at the specified *offset*. The offset and *size* must be multiples of the page size.

The *dst OBDO* will have the size and blocks fields set from *src*.

25.2.26.5. *Examples.*

25.2.26.6. *Outstanding Issues.* Offset and *count* are not implemented. The entire object is copied, and it would be helpful to only copy parts of it.

#### 25.2.27. *iterate*.

25.2.27.1. *Prototype.*

```
int iterate (struct lustre_handle *conn, int (*repeat) (obd_id, obd_gr, void
 *), obd_id *startid, obd_gr group, void *data);
int repeat (obd_id startid, obd_gr group, void *data);
```

25.2.27.2. *Parameters.*

***input: conn*** Existing client connection.
***input: repeat*** Function to run on each existing object.
***input/output: startid*** Object ID where iterator starts running .
Returns object ID for next object to be iterated.
***input: group*** Object group.
***input/output: data*** Private data for the repeat function.

25.2.27.3. *Return Values.* Upon successful completion *iterate* will return 0, otherwise one of the following error codes will be returned:

**EINVAL:** Invalid data is specified.
**EOPNOTSUPP:** The iterator function couldn't be run.
**ENOMEM:** No memory is available.

25.2.27.4. *Description.* The *iterate* method will call the *repeat* function once for each existing object on the device with the object ID. It is up to the repeat function to determine what action, if any, should be taken on a specific object. If the repeat function returns a non-zero value, the iterator will stop running and return this value to the caller.

If the iterator did not successfully iterate through all of the objects on the device, either because the repeat function returned non-zero, or because of an error (such as not being able to run the repeat function on a remote device), then the *startid* will be set with the ID of the object on which the iterator stopped. This makes it possible for the caller to locally run the repeat function on the returned object ID, increment the *startid*, and then call the iterate function with a new *startid* to retrieve the next existing object ID.

Their iterator is not guaranteed to operate atomically on the existing objects at the time the method is called. It is up to the caller to ensure that no objects are created that will place the device in an inconsistent state.

25.2.27.5. *Examples.* The *iterate* method is used in the snapshot driver when removing or restoring a snapshot. The snapshot repeat function determines whether the object belongs to the snapshot being worked on, and ignores others.

25.2.27.6. *Outstanding Issues.* The iterator repeat function currently needs to be able to execute on the device.

## 25.3. Caching Nodes

Caching nodes can be used to improve read *scalability* as described in the architecture section; these can be dedicated cache servers or clients. Caching nodes use the existing client (OSC) and server (OST) infrastructure, but add a new caching driver (COBD). The COBD will export several of the API's described in the previous section. The COBD layer can only honor *read* requests. We should not provide any *lock granting* ability on the COBD; *read* locks will still be granted by the *target OST's*. Figure 25.3.1 traces a *read* request in the presence of caching nodes. The COBD layer is stacked on top of an OSC, so if the requested data is not cached, COBD can use the underlying OSC layer to request the *target OST* for the data.

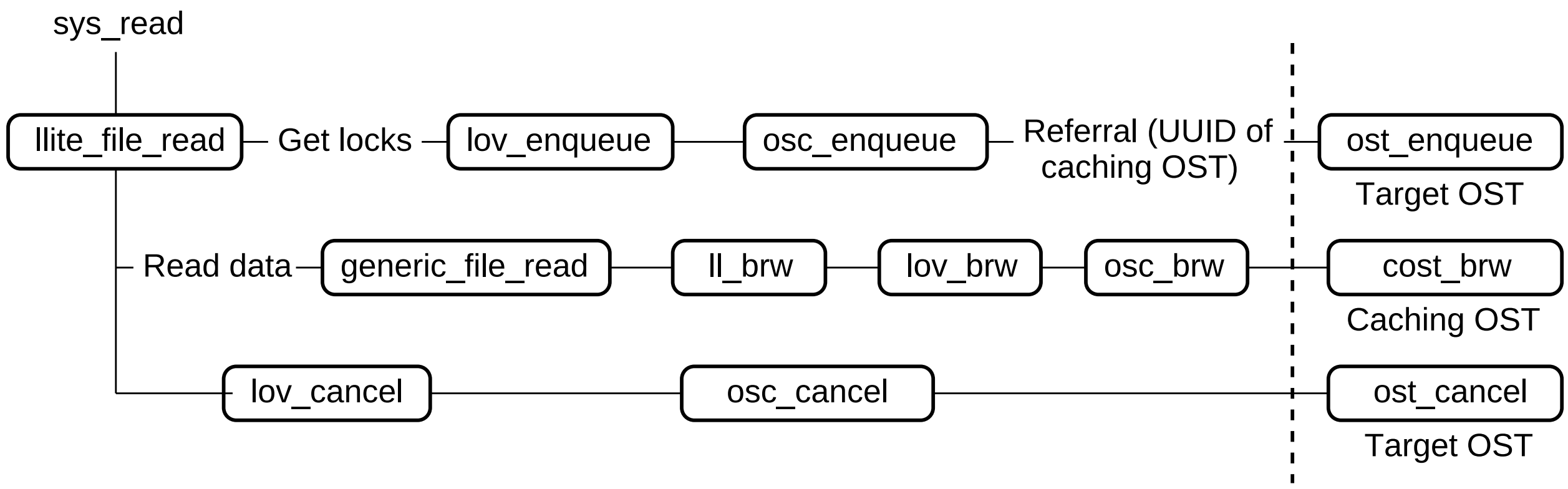

FIGURE 25.3.1. *read* requests serviced by caching node

As shown in figure 25.3.1, the client first obtains a *read lock* from the *target OST*; a referral for a caching node is returned in the same reply. The caching node has a caching OST to service the read requests. The caching OST will export a *cost_brw* method that should only honor *read* requests.

## 25.4. A Diagram of Bulk Data Movement Between OSC and OST.

Figure 25.4.1 shows how the OSC and OST negotiate a bulk data transfer; at present the write is done using PtlPut by the source.

Error handling here is quite involved. On client side, the RPC to obtain the descriptor can time out. These timeouts are fairly normal recovery issues. If a time out happens during the sending of the bulk pages, the memory buffer and descriptors given to Portals must be cleaned up.

On the server, the error handling similarly has a new aspect. The wait on the bulk transfer can time out. In this case it is fairly easy to unlink the match entries and release the buffers.

Instead of using PtlPut, an alternative is to use PtlGet at the sink. This will significantly simplify the error handling and the handling of synchronous I/O. On the source, the error handling in case of network problems is not optimal. We need an event indicating a failure, at which point the buffers can be released. The possible use of PtlGet is illustrated in figure 25.4.2.

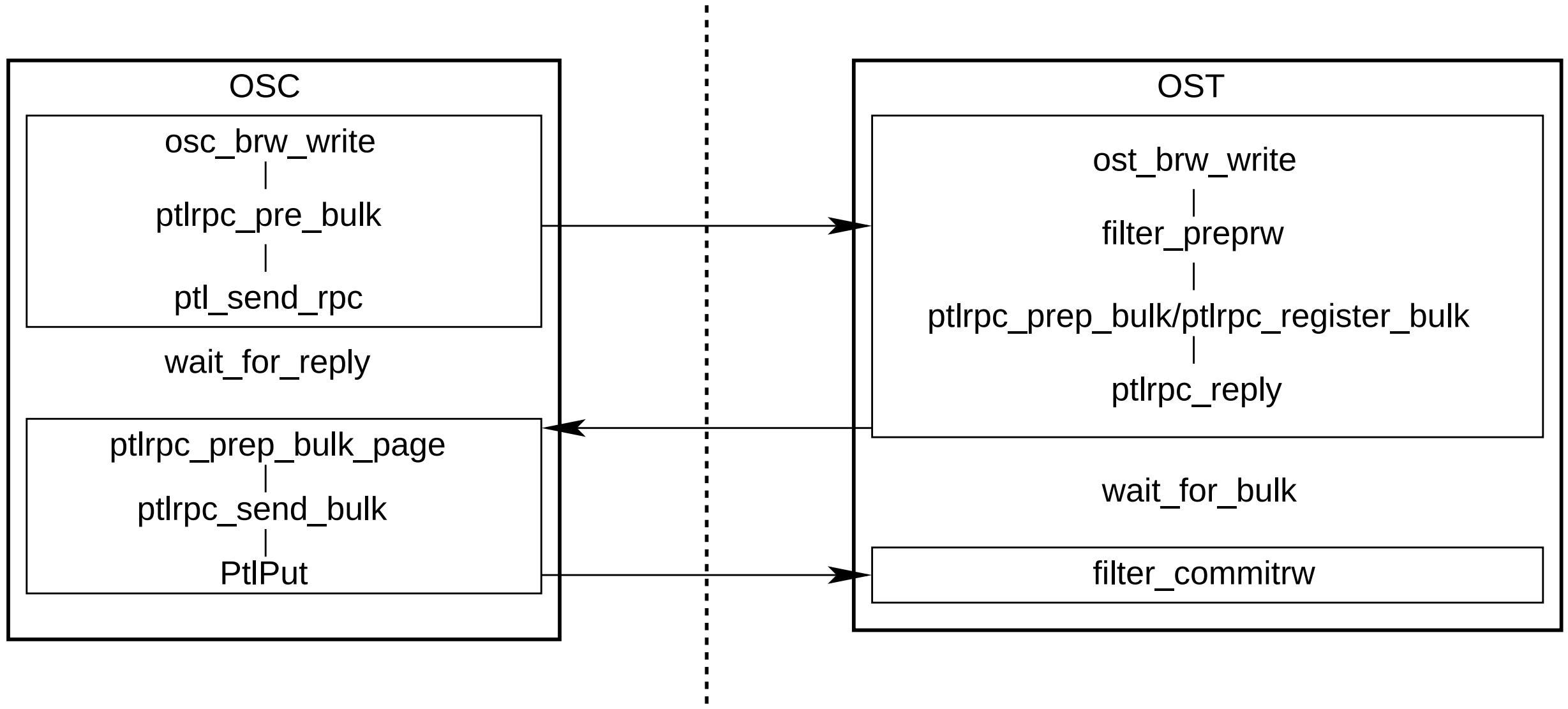

FIGURE 25.4.1. Bulk data transfer between OSC and OST

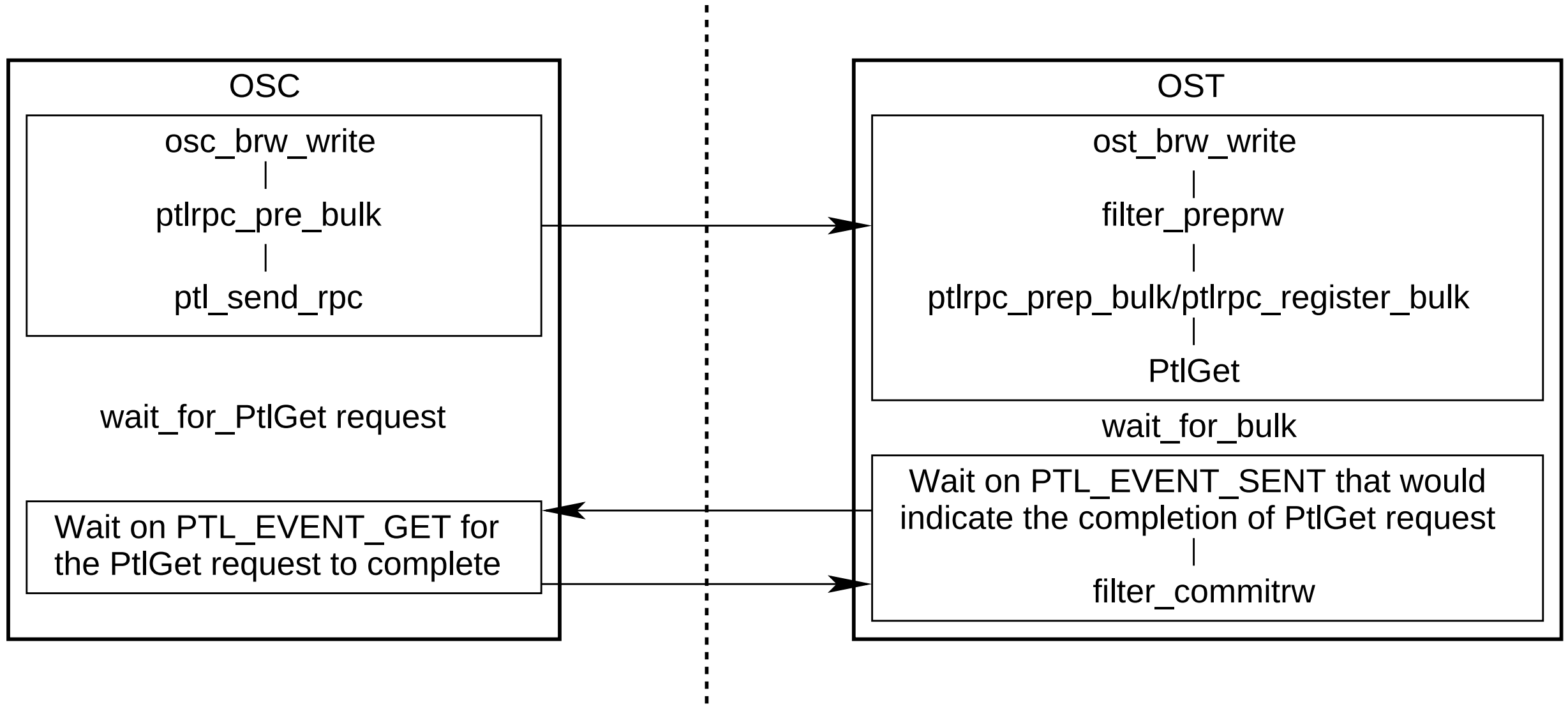

FIGURE 25.4.2. OSC-OST bulk movement using PtlGet

## 25.5. Class Driver Interfaces

The class driver can be opened and closed for the purpose of making ioctls to it. There are no read/write methods on the class driver, and the driver is a character device (to avoid any confusion with block devices).

Certain OBD devices may find it useful to enrich the class interface with the read/write commands, for example to pass upcalls to a daemon. For such devices, a naming scheme that allows the read/write commands to take place correctly is important.

The ioctl interface to the class driver corresponds to the generic OBD interface, and forms the basis of user level libraries exploiting OBD storage. The class driver exports a few functions callable from other modules; these are described below.

### 25.5.1. *class_nm_to_type*.

25.5.1.1. *Prototype.*

```
struct obd_type * class_nm_to_type(char *nm)
```

25.5.1.2. *Parameters.*

***input: nm*** OBD type name.

25.5.1.3. *Return Values.* This function will return the OBD_type structure corresponding to the given name if the name is found in the list of OBD_types, otherwise it will return NULL to indicate failure.

25.5.1.4. *Description.* This function will take as input an OBD type value and search the list of OBD_types to find the OBD_type structure corresponding to the specified OBD type.

### 25.5.2. *class_register_type*.

25.5.2.1. *Prototype.*

```
int class_register_type(struct obd_ops *ops, char *nm)
```

25.5.2.2. *Parameters.*

***input: nm*** Name of the new OBD type.
***input: ops*** The operations exported by the new OBD type.

25.5.2.3. *Return Values.* If successful, the function would register the new OBD type and return 0, otherwise it will return one of the following errors:

**EEXISTS:** The type specified already exists.
**ENOMEM:** No memory available to allocate the space for the new OBD type structure.

25.5.2.4. *Description.* This function takes in a new OBD type value and checks if that type is already registered. If it is not, the new OBD device type is registered, otherwise the appropriate error code is returned. If successfully registered, the usage count for the module is incremented.

### 25.5.3. *class_unregister_type*.

25.5.3.1. *Prototype.*

```
int class_unregister_type(char *nm)
```

25.5.3.2. *Parameters.*

***input: nm*** Name of the OBD type to be unregistered.

25.5.3.3. *Return Values.* If successfully unregistered the function returns 0, otherwise it will return one of the following error codes:

**EINVAL:** Unknown OBD type.
**EBUSY:** The module is still in use and can not be freed.

25.5.3.4. *Description.* This function gets the OBD_type structure corresponding to the specified name from the list of OBD_types. If found, it frees the related buffers and decrements the usage count for the module.

**25.5.4. *class_name2dev.***

25.5.4.1. *Prototype.*

```
int class_name2dev(char *nm)
```

25.5.4.2. *Parameters.*

***input: nm*** Name of the device.

25.5.4.3. *Return Values.* If successful, the function would return the index corresponding to the location of the name in the array of OBD devices. If the name is not found, -1 is returned to indicate the failure.

25.5.4.4. *Description.* The function compares the given name to the name in each entry of the array of OBD devices (OBD_dev). If a match is found, the corresponding index is returned.

**25.5.5. *class_uuid2dev.***

25.5.5.1. *Prototype.*

```
int class_uuid2dev(char *uuid)
```

25.5.5.2. *Parameters.*

***input: uuid*** UUID of a device.

25.5.5.3. *Return Values.* The function returns the index of the given UUID in the array of OBD devices if successful, else it returns -1.

25.5.5.4. *Description.* Searches the OBD_dev list (list of all OBD devices) for the given UUID. If a match is found, return the array index, else return -1.

**25.5.6. *class_uuid2obd.***

25.5.6.1. *Prototype.*

```
struct obd_device *class_uuid2obd(char *uuid)
```

25.5.6.2. *Parameters.*

***input: uuid*** UUID of an OBD.

25.5.6.3. *Return value.* Returns a pointer to the OBD_device structure in the list of OBD devices if a match is found, otherwise returns NULL.

25.5.6.4. *Description.* This function searches the list of OBD devices (OBD_dev) for the given UUID; if a match is found it returns a pointer to that OBD_device struct.

#### 25.5.7. *obd_cleandup_caches.*

25.5.7.1. *Prototype.*

```
void obd_cleanup_caches(void)
```

25.5.7.2. *Parameters.*

25.5.7.3. *Return Values.*

25.5.7.4. *Description.* This function is called to cleanup/free all the OBD device related caches.

#### 25.5.8. *obd_init_caches.*

25.5.8.1. *Prototype.*

```
int obd_cleanup_caches(void)
```

25.5.8.2. *Parameters.*

25.5.8.3. *Return Values.* If successful, the function returns 0, otherwise it returns:

**ENOMEM:** No memory available to initialize the various caches.

25.5.8.4. *Description.* This function is all the OBD device related caches.

#### 25.5.9. *class_conn2export.*

25.5.9.1. *Prototype.*

```
struct obd_export * class_conn2export(struct lustre_handle *conn)
```

25.5.9.2. *Parameters.*

***input: conn*** Handle to an existing connection .

25.5.9.3. *Return Values.* If successful, the function returns a pointer to an OBD_export structure, else it returns NULL.

25.5.9.4. *Description.* The export structure describes the state held for a given connection. The function accepts a connection handle and returns the corresponding OBD_export structure found in the export_cachep cache. The ’cookie’ information is used for validation of the handle given.

#### 25.5.10. *class_conn2obd.*

25.5.10.1. *Prototype.*

```
struct obd_device * class_conn2obd(struct lustre_handle *conn)
```

25.5.10.2. *Parameters.*

***input: conn*** An existing connection handle.

25.5.10.3. *Return Values.* Returns NULL if it does not find a device corresponding to the given Lustre_handle, otherwise returns a pointer to that device.

25.5.10.4. *Description.* This function is used to determine the OBD device that the given connection is for. It first locates the export struct corresponding to the connection handle from the export_cachp cache; this structure contains a pointer to the OBD_device.

#### 25.5.11. *class_new_export.*

25.5.11.1. *Prototype.*

```
struct obd_export * class_new_export(struct obd_device *obddev)
```

25.5.11.2. *Parameters.*

***input: obddev*** The OBD device for which the export structure needs to be allocated.

25.5.11.3. *Return Values.* This function returns a pointer to the newly allocated export structure if successful, else it returns NULL if there was no memory available to allocate a new struct.

25.5.11.4. *Description.* This function allocates a new export structure in the export_cachep cache and initializes it. It also obtains a random number for the cookie in the export structure. If there was no memory available in the cache, the function returns NULL. Set the pointer in the device structure, as well, to point to the new export struct.

#### 25.5.12. *class_destroy_export.*

25.5.12.1. *Prototype.*

```
void class_destroy_export(struct obd_export *exp)
```

25.5.12.2. *Parameters.*

***input:exp*** The export structure to be freed.

25.5.12.3. *Return Values.*

25.5.12.4. *Description.* This is a cleanup routine to destroy the specified export structure and free the memory used in cache for it.

#### 25.5.13. *class_connect.*

25.5.13.1. *Prototype.*

```
int class_connect(struct lustre_handle *conn, struct obd_device *obd,
 obd_uuid_t cluuid)
```

25.5.13.2. *Parameters.*

***input/output: conn*** The new connection to be initialized.
***input: obd*** OBD device for the connection.
***input: cluuid***

25.5.13.3. *Return Values.* Return 0 if successful, or one of the following error codes:

**EINVAL:** Conn/OBD is NULL.
**ENOMEM:** No memory available to allocate the export struct for the new connection.

25.5.13.4. *Description.* This function creates and initializes a new connection for the specified device. It allocates the required export struct and initializes the conn structure.

**25.5.14. *class_disconnect.***

25.5.14.1. *Prototype.*

```
int class_disconnect(struct lustre_handle *conn)
```

25.5.14.2. *Parameters.*

***input: conn*** The connection to be disconnected.

25.5.14.3. *Return Values.* Return 0 if successful, else:

**EINVAL:** Attempting to free a non-existent client.

25.5.14.4. *Description.* This function will free all the buffers associated with the given connection.

**25.5.15. *class_conn2cliimp.***

25.5.15.1. *Prototype.*

```
struct obd_import class_conn2cliimp(struct lustre_handle *conn)
```

25.5.15.2. *Parameters.*

***input: conn*** An existing client connection.

25.5.15.3. *Return Values.* Return a pointer to the OBD_imports struct if successful or else returns NULL.

25.5.15.4. *Description.* This function will return the OBD_import structure associated with the given connection handle.

#### 25.5.16. *class_disconnect_all*.

25.5.16.1. *Prototype.*

```
void class_disconnect_all(struct obd_device *obddev)
```

25.5.16.2. *Parameters.*

***input: obddev*** The OBD device for which all the connections are to be disconnected.

25.5.16.3. *Return Values.*

25.5.16.4. *Description.* This function will disconnect all the connections associated with the given OBD device. It will loop through the exports list for the device, disconnecting each connection.

### 25.6. Basic Support Functions

We provide several support functions and macros that have been used all over Lustre source code. This section describes some of these functions.

#### 25.6.1. NTOH__u32.

25.6.1.1. *Prototype.*

```
NTOH__u32(var)
```

25.6.1.2. *Description.* This is a macro that would convert the given 32 bit variable from network byteorder to the host byteorder using the function le32_to_cpu.

#### 25.6.2. NTOH__u64.

25.6.2.1. *Prototype.*

```
NTOH__u64(var)
```

25.6.2.2. *Description.* This is a macro that would convert the given 64 bit variable from network byteorder to the host byteorder using the function le64_to_cpu.

#### 25.6.3. HTON__u32.

25.6.3.1. *Prototype.*

```
HTON__u32(var)
```

25.6.3.2. *Description.* This is a macro that would convert the given 32 bit variable from host byteorder to the network byteorder using the function cpu_to_le32.

### 25.6.4. HTON__u64.

25.6.4.1. *Prototype.*

```
HTON__u64(var)
```

25.6.4.2. *Description.* This is a macro that would convert the given 64 bit variable from host byteorder to the network byteorder using the function cpu_to_le64.

### 25.6.5. UNLOGV.

25.6.5.1. *Prototype.*

```
UNLOGV(var, type, ptr, end)
```

25.6.5.2. *Description.* This macro copies sizeof(type) from ptr to var; moves ptr by sizeof(type). When ptr goes beyond end, the macro throws an error EFAULT.

### 25.6.6. LUNLOGV.

25.6.6.1. *Prototype.*

```
LUNLOGV(var, type, ptr, end)
```

25.6.6.2. *Description.* This macro copies sizeof(type) from ptr to var after converting it to host byteorder using NTOH__u32/NTOH__u64 (depending on the type field); moves ptr by sizeof(type). When ptr goes beyond end, the macro throws an error EFAULT.

### 25.6.7. LOGV.

25.6.7.1. *Prototype.*

```
LOGV(var, type, ptr)
```

25.6.7.2. *Description.* This macro copies new value from var to ptr and increments ptr by sizeof(type).

### 25.6.8. LLOGV.

25.6.8.1. *Prototype.*

```
LLOGV(var, type, ptr)
```

25.6.8.2. *Description.* This macro converts var to network byteorder using HTON__u32/HTON__u64, as appropriate based on the value of type, and copies the result into ptr. It increments otr by sizeof(type) everytime.

### 25.6.9. UNLOGP.

25.6.9.1. *Prototype.*

```
UNLOGP(var, type, ptr, end)
```

25.6.9.2. *Description.* This macro makes var point to (type *)ptr and increments ptr by sizeof(type). If ptr goes beyond end, the macro throws an error EFAULT.

### 25.6.10. LOGP.

25.6.10.1. *Prototype.*

```
LOGP(var, type, ptr)
```

25.6.10.2. *Description.* Copies the contents at location var into the location ptr using memcpy and increments ptr by sizeof(type).

### 25.6.11. LOGL.

25.6.11.1. *Prototype.*

```
LOGL(var, len, ptr)
```

25.6.11.2. *Description.* Copies 'len' amount of data from memory location var to ptr. Increments ptr by size_round(len).

### 25.6.12. LOG0.

25.6.12.1. *Prototype.*

```
LOG0(var, len, ptr)
```

25.6.12.2. *Description.* Copies 'len' amount of data from the location var to ptr, delimits ptr by 0, and then increments ptr by size_round(len+1). Essentially, the macro copies values into ptr separated by 0.

## 25.7. Changelog

1. Radhika Vullikanti (10/03/2002) : Added descriptions to the class driver interface section

CHAPTER 26

# Metadata API

## 26.1. Introduction

The Lustre clients communicate with the metadata server (MDS) for all the operations that involve inodes and directories. The MDS also handles locking for the metadata operations. Every client has a metadata client (MDC) module to support this interaction. The metadata server has a *metadata target (MDT)* and an *MDD* similar to the sub-division done for the object store (OST, OBD). The *MDT* will act as the server component for a corresponding *MDC* on the client and recieve client requests, package replies. The *MDT* on the other hand will act as an interface to the underlying filesystem that acts as the persistent store for the Lustre file system metadata. The MDS will use the same locking APIs as used by the OSC to handle locking, it will just use it based on a different policy.

The MDS stores the file inodes for all Lustre files, the file attributes - size, mtime, ctime, atime are maintained here. The information about the file striping pattern, objects of the file, location of the objects is kept as a part of the *extended attribute* on the file inode. When a file is closed, the MDS is authoritative on the *size*, *mtime*, *ctime* and *atime*. But when a file is open by one or more clients, the OSTs would be better informed on the correct values for all these attributes. The MDS will be updated only after the last close on the file. We will discuss the protocol used to manage the file attributes and synchronize the information between the MDS and the OSTs in the presence of single failures.

Some of the other features that will be added to Lustre metadata handling are - Writeback metadata cache and the clustered metadata server. These would be quite useful in improving the metadata performance of Lustre. We will discuss the implementaion planned for them.

Finally, in this chapter, we describe the APIs used between the clients and MDS. The metadata APIs will evolve to become a first class citizen in the Lustre project and will support stacking, much as the object API's already do. The description here presents the current state of these APIs, these might evolve and be updated in future if needed.

## 26.2. Attribute Management

In this sectione, we discuss how Lustre inode attributes are managed, the focus is on the shared client-server case. In the metadata write back case, as we will see later, the whole subtree is owned by a single client, this makes things much simpler.

**26.2.1. File size.** File size management is very critical, the correctness of several operations can depend on having the correct file size - like write or truncate. So, it is important that clients have an accurate way to obtain file size. On closed files, the MDS should be able to provide this information.

On the other hand, for open files and for files for which the normal protocol failed due to system our network outages, the correct file size is computed by adding up all the object sizes for the file objects. The correct size will be computed by the LOV driver and returned to the client. Similarly, during MDS recovery, all OST's holding a stripe of the file need to be queried to obtain the size.

The clients requesting for a file size will first contact all the OSTs holding the file stripes to obtain a file size lock on the file, this ensures that no other client will extend the file while the size is being determined. If another client is holding a write lock on the file, it could be doing either a write or a truncate operation, both of which effect the file size. Truncates are internally handled as ftruncate

Currently, a request for file size lock would prompt the DLM on the target OSTs to send out lock callbacks in case there are other clients holding write locks on any file extents. Once all the other clients have given up their write locks, the OSTs will grant the required lock to the client, the client will then use *getattr* to query all involved OSTs for the object size, all these sizes will be added up to determine the file size. This algorithm could lead to really bad performance for operations that involve obtaining the file size information. We have planned for some optimizations for this protocol:

(1) Instead of revoking locks from other clients and granting a size lock to the client querying the file size, the OSTs can send intent based callbacks to the clients with write locks, indicate that the callback is only to get the file size. The clients can then return the size in callback replies, servers can send these to the client requesting file size in the reply to the size lock request.
(2) Caching the file sizes on the MDS - this allows the MDS to be authoritative for all closed files.
(3) Have a single lock manager for each file, this could be the OST holding the first stripe of the file

**26.2.2. mtime management.** The *mtime* is stored the same way as size. The different handling arises from the following reasons:

(1) multiple clients can modify the mtime
(2) the mtime is slightly less critical to be absolutely correct

However the following are done to update *mtime*:

- *utime* calls are handled as truncate calls.
- *write* calls propagate client mtime to OST's where it is normally written to disk, upon close it is propagated to the MDS
- the default is that client clocks are used when setting mtime for writes, truncations or utime operations. Optionally an approximate server time is used.

The *mtime* is acquired roughly along the same lines as the size. However the following refinements are planned or implemented: in the following D is proposed to be 900(ms).

- The client wrote to the file itself during the last **D**ms, has a write lock: the client returns the locally stored *mtime*.
- he OST holding the locks can use or return to the client a list of clients that may have modified the mtime and a list of ost's that may have the modified mtime. This list can be used to limit the number RPC.
- The flush daemon can write every Dms, making OST based *mtime* reliable up to POSIX compliance test boundaries. The lock manager OST may have seen a write or a lock operation for an operation during the last **D**ms: the OST will return its *mtime*.
- If during queries of the the *mtime*, the current time is found this can be used as the *mtime*.

The *ctime* management is done very similar to *mtime* management.

26.2.2.1. *Options for mtime.*

- Server time can be used
- **D** may be adjusted - in fact **D** will be much larger than 900ms when writeback caching is operational.

**26.2.3. atime.** The *atime* management can be even more relaxed, the preferred mode of operation:

- the protocol negotiates *atime* in memory updates, i.e. always locally on a single client
- network updates will be only piggybacked on other requests associated with the same object or inode.
- no disk updates are made which only set atime

Motivation for this goes back to the Sprite file system of 1982.

26.2.3.1. *atime options.*

- Such *atimes* can optionally be written to disk on the MDS after close
- *Atimes* could be be stored on the OST's as well if customer demand warrants this.
- *Atimes* could be made recoverable if customer demands warrants this.
- *Atimes* could be negotiated over the network as *mtimes*.

In an almost-write-only environment running HSM's opinions appear to be that this is of significant importance. The omnipresent web server farms would incur a huge overhead through any of these mechanisms

## 26.3. Writeback metadata cache implementation

In this section we describe the implementation details of the proposed write back metadata cache. The desire for this implementation is:

- Minimal changes to Lustre Lite
- The design should make a persistent writeback cache possible
- Exploit existing API's where ever possible

We believe that the proposal here satisfies all these requirements.

**26.3.1. WB Cache organization.** File system requests are directed by the VFS to the Lustre Lite layer. The first step is to request for required locks for a given operation, Lustre Lite checks what locks are required. If it already has the required lock available, its granted locally and if the cached data is known to be valid, that is given out. Otherwise, it also forwards update requests with lock requests through our intent mechanism to the remote MDS.

When a writeback cache is enabled several new issues enter the picture:

- a decision is made if a writeback lock is available for the objects the operation is accessing - writeback lock is usually taken on a directory sub-tree
- the cache is queried for the object
- cache misses are handled
- updates are made locally
- log records of the updates are created
- the updates are sent to the MDS at a later time driven by memory pressures or some other requirement

In the non-writeback mode, a metadata client driver is used to interact with the remote MDS. But to satisfy the issues listed about in the writeback mode, we introduce two layers below the client file system. The lowest is the local metadata cache, this could be in-memory or on persistent store. This acts as a locally running *metadata server* and we call it the *local* MDS.

The second layer one is the *write back metadata client* - WBC. This is a logical metadata driver to interact with the *local* MDS as well as the remote MDS and assist handling cache misses in the local cache. The WBC accepts commands from Lustre Lite and interacts with the local and remote MDS to handle cache misses. This organization or layering for the metadata writeback cache is illustrated in figure 26.3.1.

26.3.1.1. *Cache Queries.* If a writeback lock is found or acquired, lookup and attribute requests are serviced through the local MDS. The WBC driver first determines if the cache has the object. To do so, it uses two commands:

- lookups in the local MDS
- analyzing extended attributes in directory inodes in the local MDS. These extended attributes encode what parts of directories have been cached locally.

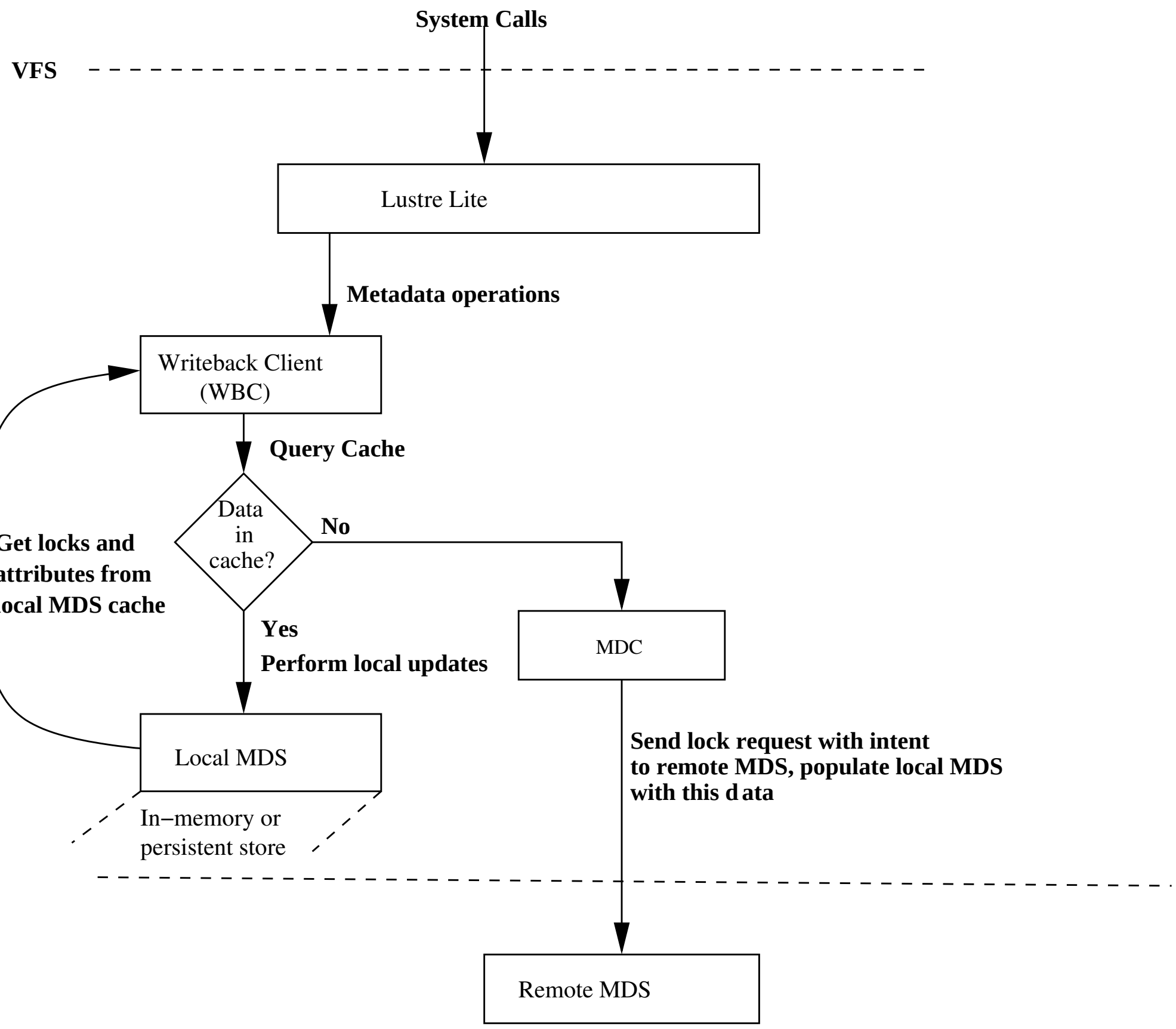

FIGURE 26.3.1. Metadata writeback cache organization

26.3.1.2. *Cache Misses.* If a cache miss is encountered the WBC will use the existing MDC APIs to the remote MDS to acquire the object or data it needs. In case it fetches a partial directory it will set extended attributes accordingly to indicate what part of the directory is available in cache.

26.3.1.3. *Local updates.* The clustered metadata driver will include a local MDS that runs over a file system. This file system can be a memory based file system like shmfs or a disk based one.

The local MDS exposes the same APIs as a remote MDS and the updates to the local cache can be made by dispatching the command as usual, but to the local MDS instead of the remote MDS. The only difference is that the local updates require preallocated inodes to ensure that the operation does not receive an error when the cache is flushed at a later point of time. Once local updates are made, the application making the system call gets a completion return code. The local MDS content will be regarded as cached objects and their availability will be handled by the WBC similar to the cache miss handling.

**26.3.2. Log Records.** Lustre Lite has a logging API, carefully constructed for recovery with foresight for us in the writeback caching. As part of the local update a log record is created, either in memory or transactionally consistent on disk through nested transactions (as in recovery).

A daemon, sensitive to memory pressure propagates update records to the remote MDS and handles cancellations of the records when they have been committed remotely (another already existing API for the logging records).

**26.3.3. Cache Purges.** The local MDS, like an AFS cache, will use a prescribed amount of storage space, in memory or on disk depending on the file system underneath the local MDS.

The cache will run one or more flush daemons that operate like the well known *ext2* balance dirty code. If the cache is empty the daemon will run periodically, if the cache is moderately full it is activated. When the cache is very full, the daemon synchronously removes cached data.

Removal of cached data is dependent on any associated log entries being flushed to the remote MDS and committed to persistent store first.

The daemon also has to ensure that its not trying to yank and flush inodes in use, this can be done by maintaning reference counts for the inodes.

**26.3.4. Directory content management.** Lustre Lite will implement a new directory content management scheme. This directory layer is present in two subsystems:

- Lustre Client File System. Key aspects are:
  - cached directory entries will be extremely similar to *shmemfs* dentry based directory implementation.
  - inode extended attributes to describe what parts of the directory have been cached in memory.
- MDS, both the local and remote MDS These will offer a directory fetch API.
  - Locks will control cache validity
  - Readdir cookies will assist with re-reading after directory updates.
  - When the write back lock is present the API will be executed against the local cache MDS
  - Without the write back lock, against the remote MDS.

**26.3.5. Lock Management.**

26.3.5.1. *Write back locks.* When a client accesses a remote MDS directory the MDS will give out a subtree lock. Subtree locks can be read or write locks following the single writer multiple reader model among clients. If a subtree has not seen recent activity and directories are small enough to reasonably fit into a client cache, subtree locks will be given out.

If the directory is already locked or has seen very recent activity a normal intent lock execution is performed.

Subtree locks are associated with the namespace.

26.3.5.2. *Sub Tree Change time (stctime).* To measure if a sub tree has seen changes, the MDS will update not just the *ctime* of particular inodes involved in the transaction, but it will also modify an extended attribute holding this change time on all ancestors of the modified object or objects. This *stctime* has nanosecond granularity and can validate the version of an entire subtree.

This update can be made in a persistent manner or in a memory only way.

The *stctime* associated with files open on multiple nodes will change only when all nodes not holding a writeback lock close it.

26.3.5.3. *Filling and revalidating the cache.* When a client with a subtree lock fills its cache it fetches objects from the remote MDS. During this process the MDS may revoke other locks on such objects.

If open objects, files or directories, are encountered, the object is be given to the client with a flag indicating the object is subject to remote operations. No operations may be performed in WB mode unless they commute with the remote operations that may be performed on such objects. This can happen for renames only, we think, in which case the writeback lock will be yanked away by the MDS before the operation can complete.

26.3.5.4. *Lock revocation.* When other client acquire new locks underneath a subtree lock held by another client, then the write back caching locks need to be revoked.

It may be that a lock is requested associated with a *fid* on the directory. The *fid2dentry* code will need to be modified to return a connected dentry for this purpose (NFS shows how to do this trivially). The cases where a non-connected dentry for a *fid* could be found appear to be somewhat exotic but can happen.

After the connected dentry is found the MDS can establish trivially if there is an active subtree lock on the tree and revoke it.

Revocation has the following effects:

- WB metadate update log flush
- Convert open mds and OST objects under the subtree lock to remotely opened locks

26.3.5.5. *Open objects.* If a client with a WB lock fetches an object into its cache that is associated with an MDS open (ie. an open file, an open directory, a current working directory or a mount point) then the cache fill will return that object with a flag that it is subject to remote operations. (This is precisely what Sprite did in the early 80's!)

Open objects can get involved in operations that do not commute with the write back log. In that case the writeback lock will be canceled.

26.3.5.6. *Expiration and nested subtree locks.* Subtree locks can be nested. When a subtree lock is found on an ancestor the client will continue to see if an older ancestor has a subtree lock also. If so, the first one found is canceled after the update log associated with it is flushed.

Subtree locks are subject to expiry, like all unused locks. If a subtree lock is not active, the expiry callback function is called, which will flush any remaining log entries and cancel the lock.

**26.3.6. Preallocation.** We will follow a simple scheme to create preallocated objects in a directory pointed to by the export on the MDS, and transfer the attributes (fid) to the client. This does mean that some create operations will turn into rename operations on the MDS when new fids arise from pre-allocation.

**26.3.7. Writeback cache recovery.** The recovery in general is achieved using transaction numbers, connection levels, replies and reply acknowledgements as described in the recovery chapter. The clients will also have a *pinger* that periodically pings the servers and announces its presence as well as ensures that the server is still available. The presence of a metadata writeback cache adds another issue, it is possible that a client node has a writeback lock on a partial directory tree and is servicing the client requests locally. Meanwhile the remote MDS has a failure, the failover MDS takes over its job. In such a scenario, the *pinger* has to detect when the new server is active, re-establish its writeback lock and reintegrate its changes.

## 26.4. Clustered MDS implementation

This document describes the design of the clustered metadata handling for Lustre. This material depends on other Lustre design, such as:

- General recovery
- Orphan Recovery
- Metadata Write Back caching

Overall the clustered metadata handling is structured as follows.

- A cluster of metadata servers manage a collection of inode groups. Each inode group is a Lustre device exporting the usual metadata APIs, augmented with a few operations specifically crafted for metadata clustering. We call these collections of inodes groups.
- Directory formats for file systems used on the MDS devices are changed to introduce and allow directory entries to contain an inode group and identifier of the inode.
- A logical clustered metadata driver is introduced below the client Lustre file system write back cache driver that maintains connections with the MDS servers.
- There is a single metadata protocol that is used by the client file system to make updates on the MDS's and by the MDS's to make updates involving other MDS's.
- There is a single recovery protocol that is used by the clients - MDS and MDS-MDS service.
- Directories can be split across multiple MDS nodes. In that case a primary MDS directory inode contains an extended attribute that points at other MDS inodes which we call directory objects.

**26.4.1. Configuration management and Startup.** The configuration will name an MDS server, and optionally a failover node, which hold the root inode for a fileset. Clients will contact that MDS for the root inode during mount, as they do already.

They will also fetch from it a clustering descriptor. The clustering descriptor contains a header and an array lists what inode groups are served by what server.

Through normal mechanisms clients will wait and probe for available metadata servers, during startup and cluster transitions. When new servers are found or configurations have changed they can update their clustering descriptor as they update the LOV striping descriptor for OST's.

**26.4.2. Data Structures.** The information key to clustered MDS functioning is the enhanced *fid,* the fid will contain a new 32 bit integer to name the inode group. This new information will be used to determine which MDS should service a certain.

The directory entries (*dentry)* will also contain a new 32 bit integer to name the inode group.

Directory inodes on the MDS, when large, contain a new extended attribute (EA) which is a descriptor of how the directory is split over directory objects, residing on other MDS's. This EA is subject to ordinary concurrency control by the MDS holding the inode. The EA is virtually identical to the LOV EA.

**26.4.3. The Logical Metadata Volume.** Client systems will use the cache obd or client file system directly to communicate with the LMV driver. The LMV driver offers the metadata API to the file system and uses the metadata API offered by a collection of MDC drivers. Each MDC driver manages the metadata traffic to one MDS.

The function of the LMV is very simple: it figures out from the command issued what MDC to use. This is based on:

- the inode groups in the request
- a hash value of names used in the request, combined with the EA of a primary inode involved in the request.
- for readdir the directory offset combined with the EA of the primary inode
- the clustering descriptor

In any case every command is dispatched to a single metadata server, the clients will not engage more than one metadata server for a single request.

The API changes here are minimal and the client part of the implementation is very trivial. The figure 26.4.1 illustrates the layering used for clustered MDS.

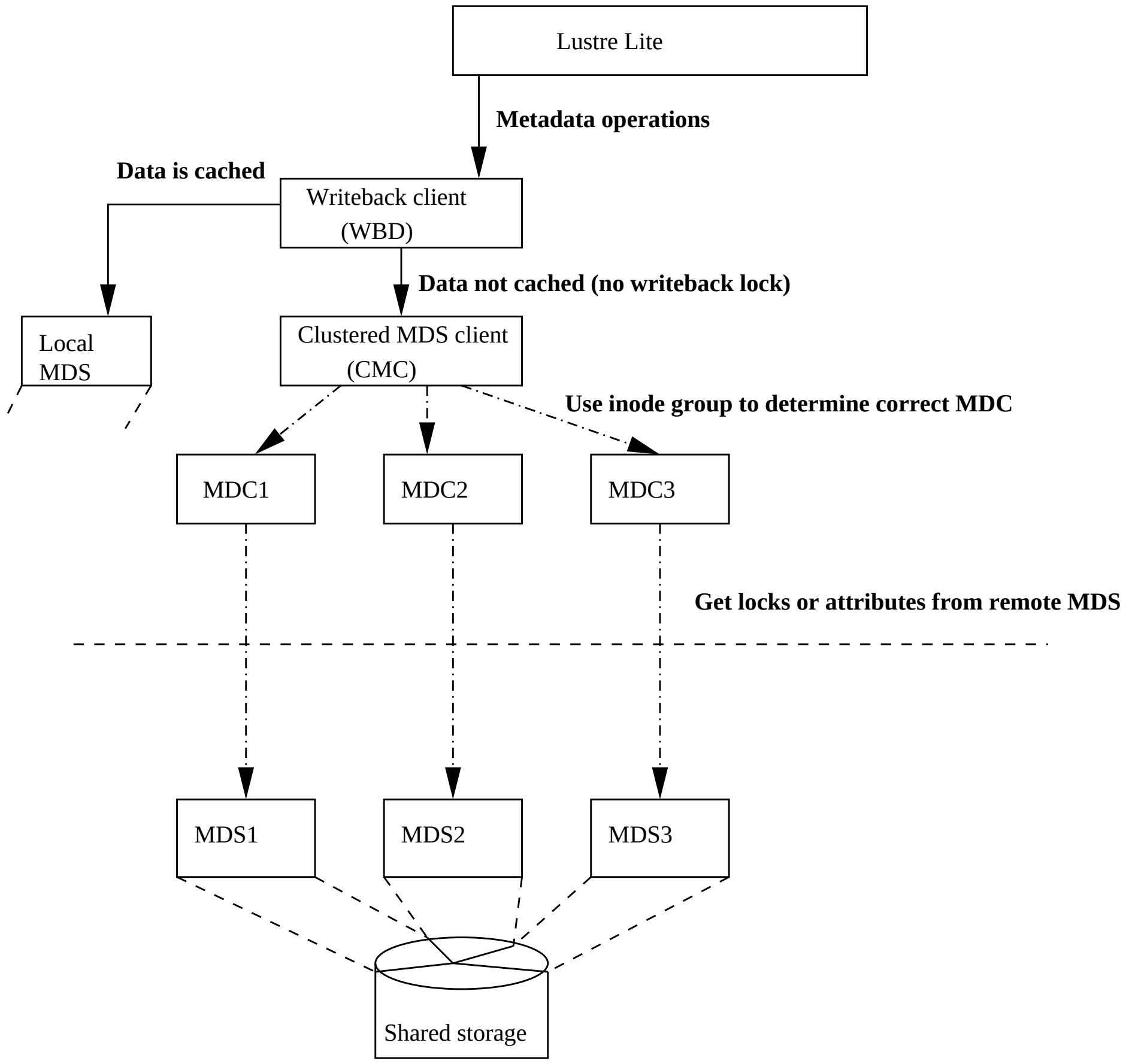

FIGURE 26.4.1. Layering for Clustered MDS

**26.4.4. MDS implementation.** For the most part, operations in the presence of clustered MDS are extremely similar or identical to what they were before. The only difference, instead of directed all operations to a single MDS, now operations are directed towards one of the MDSs in the cluster. In some cases, multiple mds servers are involved in updates due to a directory split, or operations spanning more than one directory. The *getattr*, *open*, *readdir*, *setattr* and *lookup* methods are unaffected.

Methods adding entries to directories are modified in some cases:

- *mkdir* always create the new directory on another MDS - This is to allow maximum spreading of the data across all MDSs.
- *unlink*, *rmdir*, *rename* - these may involve more than one MDS as the two directories might exist on different MDSs
- large directories - all operations making updates to directories can cause a directory split. The directory split is discussed below.

- other operations - If no large directories are encountered and splits are not required, all other operations proceed as they are executed on one MDS.

26.4.4.1. *Directory Split.* A directory that is growing larger will be split. There is a fairly heavy penalty associated with splitting the directory and also with renames within split directories. Moreover, at the point of splitting, inodes become remote and will incur a penalty upon unlink.

Probably it is best to delay the split until the directory is fairly large, and then to split over several nodes, to avoid further splits being necessary soon afterwards.

**26.4.5. Recovery.** The general recovery algorithms and infrastructure is described in much details in the recovery chapters. In the presence of clustered MDS, multiple MDS nodes may be involved in some operations, in such cases it is essential to coordinate the recovery efforts between them. So, here we will discuss the additional issues for a clustered MDS scenario.

26.4.5.1. *Transaction Replay.* The MDS - MDS interaction is managed as follows. The node approached with a request change is made the coordinator of the transaction.

26.4.5.2. *Mechanisms.* The coordinator will first establish that the transaction can commit on all nodes, by acquiring locks on directories and checking for available space existing entries with the same name etc. It may also first perform a directory split if the size is becoming too large, and more MDS nodes are still available.

All nodes involved in the transaction need to have a transaction sequence number to place the transaction into their sequence and allow correctly replay.

At this point the coordinator will:

- start a transaction locally.
- It will then report the transaction sequence number to all other nodes involved in the transaction.
- These nodes will commit (in memory as usual), write a journal record for replay and reply to the coordinator.
- The coordinator will then commit its own transaction.
- The replay log records are subject to normal log commit cancelation messages, but on the coordinator commit messages must be received from all other nodes before the record will be canceled.

In this way if the results of the transaction survive on any of the nodes, they can be replayed on all.

26.4.5.3. *Cluster crashes and the transaction sequence.* If the cluster crashes abruptly, there is the opportunity for transactions to be in progress affecting multiple nodes. Dependencies between the transactions must be managed to ensure serializibility of the protocol. The mechanism was discussed in the MDS architecture chapter.

For example, in transaction T1, a node X creates directories. Then in transaction T2 a cross MDS node rename moves a file with a directory entry on node Y into this directory. It is now possible for this file to lose its directory entry on Y and for the transaction on X not to commit. More complex examples exist.

26.4.5.4. *Replay.* To order transaction sequences Lustre uses reply acknowledgements, the ACKs serve only one purpose; to release a lock that enforces ordering of the transaction sequence. In the case where MDS operations involve more than server, the reply *ACK* from the primary to secondary servers should only be sent after the client has sent the *ACK* to the first server. This MDS-MDS reply ACK is now not really an ACK anymore, but a simple lock cancelation review.

Clients will replay lost transactions to the MDS which they originally engaged for the request.

Orphaned children will be cleaned up only after replay completes to allow orphaned objects to be re-used during replay.

26.4.5.5. *Failover.* The configuration data can designate a standby MDS that will take over from a failed MDS. By organizing the servers in one or more rings, the nearest working left neighbor MDS can be the failover node. This leads to a simple scheme with multiple failover nodes, avoiding quorum and other complications beyond what is needed for two node clusters.

**26.4.6. Locking.** The locking for metadata operations in the clustered metadata implementation is very similar to that employed by the Linux VFS. The Linux VFS locking is described in detail in the linux kernel documentation, *Documentation/filesystems/directory-locking.*

When more than one element needs to be locked, the locking order we follow has three components:

(1) ancestor first
(2) ordered by mds
(3) order by inode number

The only difference with normal VFS locking is that instead of inode number ordering of locks we order by the pair (mds no, inode number).

As an example, consider a cross directory rename. The client can determine the locking order of the source and destination parent directories, because it has the fids and directory tree relationship for the parents. In one possible implementation, the rename operation is sent by the client to the MDS node which is holds the source directory. This node acquires the two locks on the source and destination in the correct order. The source lock will be local, the destination lock will be remote. If the target of the rename exists, it needs to be locked also, and this is always done last. Now the rename operation can proceed.

In an alternative implementation the client places its rename RPC to the source or destination directory with the highest lock order. This node takes the lock first and then performs an RPC to the other node which locks the lower order of the two parents. The operation completes as above. In this second implementation, the MDS locks are always taken locally.

**Note:** In cmd1 code, by accident, lock ordering forgot to take into account MDS order.

A further complication is addressed by the Linux global file-system lock. This lock is taken for cross-directory renames (before any other locks, so it is non-contended), to prevent the ancestry relationships changing during a rename. Lustre similarly will take a global-file system lock during cross directory renames. Note that while this lock is taken, all operations (lookups, creations, deletions etc) can proceed except other cross directory renames. Global locks should be avoided as much as possible, and the following can be taken into account:

(1) The only operation that takes a global lock is a cross directory rename, where the object being renamed is also a directory (the latter optimization is not made in the Linux VFS). No other operations take this lock, in particular, it is not taken by renames within one directory or hard link operations.
(2) If the source and destination directory have a common ancestor, the global file system lock can be replaced by a lock on that ancestor. Directory placement policies will attempt to keep subtrees like users home directories on one MDS node, so a common ancestor can rapidly be found on this MDS node for all renames within a home directory. This optimization is also not present in Linux.

The above reference contains proofs that the locking operations among all file system operations are deadlock free and do not introduce cycles into the directory tree. The proofs can be applied *mutatis-mutandum* to the case at hand: the locks taken by the MDS are not semaphores but instead DLM PW locks to protect directory content and PW locks on leaf objects being processed.

## 26.5. Metadata Client APIs

In the earlier implementations of the metedata API's for Lustre, there was possibility of race conditions happening between various operations. A very big change that was made to these API's was to define a ***mdc_rpc_lock*** semaphore, every metadata operation is protected by this. In every operation, this semaphore is grabbed before sending the request over to the server and it is released only after a reply is received from the server. The structure for thei new lock is shown below:

```
Struct mdc_rpc_lock{
     struct semaphore rpcl_sem;
     struct lookup_intent *rpcl_it;
}
```

Most of the intent operations were removed and replaced by ***RPC mode*** that does a single RPC to the server and returns the rc immediately, bypassing most of the VFS. Below we have described the various metadata APIs that are currently used. There is no *mdc_open* call in this list because we do intent-based open. An *open* intent is specified in the lock request sent to the server during lookup. The server then performs all the open related operations and returns this information in the intent structure. This is sent back to the client, the VFS will invoke the create and open operation, at this time the open data is pulled out of the intent structure completing the *open* operation without another RPC to the server.

### 26.5.1. mdc_enqueue.

26.5.1.1. *Prototype.*

```
int mdc_enqueue(
  struct obd_export *exp,
  int lock_type,
  struct lookup_intent *it,
  int lock_mode,
  struct mdc_op_data *data,
  struct lustre_handle *lockh,
  void *lmm,
  int lmmlen,
  ldlm_completion_callback cb_completion,
  ldlm_blocking_callback cb_blocking,
  void *cb_data);
```

26.5.1.2. *Parameters.*

***input: exp*** Existing MDC export.
***input: lock_type*** Lock type (always LDLM_MDSINTENT).
***input: it*** Lookup-intent of this locking operation.
***input: lock_mode*** DLM lock mode (LCK_PR, LCK_PW, LCK_EX, etc.).
***input: data*** MD operation hint, which contains object ids, names, etc.
***input: lmm*** Object additional metadata information (optional).
***input: lmmlen*** Object additional metadata information size.
***input: cb_completion*** Lock completion call back function pointer. Usualy ldlm_completion_ast() is used.
***input: cb_blocking*** Lock activities notification call back function. For instance, if lock is going to be canced, this function will take control and have a chance to perform something here. Local caches related to locked resource may be invalidated here, as they become not up-to-date.
***input: cb_data*** Useful opaque pointer, which is passed to both call back functions above and can be used for bringing context related info to lock activities handlers. For instance, if lookup lock on a directory is going to be canceled due to directory modifications, this pointer can contain some information about the directory and can be used during directory pages truncating. That could be directory inode or whatever else.
***output: lockh*** Lock handle of the newly-acquired lock.

26.5.1.3. *Return Values.* Upon successful completion *mdc_enqueue* will return 0, otherwise one of the following error codes will be returned:

**EINVAL:** An invalid or corrupt intent was passed in; or an invalid reply packet was sent from the server.
**ENOMEM:** Not enough memory to allocate locks, request/reply buffers, etc.

In either error case, the handle in ***lockh*** cannot be trusted.

26.5.1.4. *Description.* Execute an intent lock request. It will contact the MDS to get a lock on an object in order to do some operation (as given by the lookup_intent parameter). The MDS is free to perform the intended operation on behalf of the client and return a lock on a different object (if you are creating a new file), or not return a lock at all (if you are unlinking a file).

### 26.5.2. mdc_getstatus.

26.5.2.1. *Prototype.*

```
int mdc_getstatus(
   struct obd_export *exp,
   struct ll_fid *rootfid
}
```

26.5.2.2. *Parameters.*

***input: exp*** Existing MDC export.
***output: rootfid*** Object identifier for the MDS root directory.

26.5.2.3. *Return Values.* Upon successful completion *mdc_getstatus* will return 0, otherwise one of the following error codes will be returned:

**EINVAL:** An invalid export is specified.
**ENOMEM:** Not enough memory to allocate the request structure.

26.5.2.4. *Description.* This function invokes the helper function ***send_getstatus*** to obtain the *rootfid* of a mount point. In the ***send_getstatus*** function, we first grab the ***mdc_rpc_lock*** semaphore, send a request to the server for the *rootfid*. When a reply is received, release the semaphore and copy the *fid* from the reply body to the *rootfid* variable. The most recent transaction number that the MDS has processed for the client is also returned as a part of this reply to help the client in recovery.

### 26.5.3. mdc_getattr.

26.5.3.1. *Prototype.*

```
int mdc_getattr(
  struct obd_export *exp,
  struct ll_fid *fid,
  unsigned long valid,
  size_t ea_size,
  struct ptlrpc_request **request);
```

26.5.3.2. *Parameters.*

***input: exp*** Existing MDC export.
***input: fid*** The object identifier from which to get the attributes.
***input: valid*** The mask of which file attributes to retrieve.
***input: ea_size*** The size of the buffer to allocate to recieve LOV attributes.
***output: request*** Request struct for the issued RPC, which holds the returned attribute data in the reply buffers.

26.5.3.3. *Return Values.* Upon successful completion *mdc_getattr* will return 0, otherwise one of the following error codes will be returned:

**EINVAL:** An invalid export is specified.
**ENOMEM:** Not enough memory to recieve the configuration data.
**ENOENT:** The object does not exist.
**EIO:** Error reading the LOV configuration from disk.

26.5.3.4. *Description.* Obtain file attributes for a Lustre inode. The first reply buffer will hold the *mds_body* struct with the attributes. Only those attributes marked in the valid field hold correct data. If the file is a regular file, then the second reply buffer will hold the LOV attribute data. If the *OBD_MD_LINKNAME* flag was set and the file is a symbolic link, then the second buffer will hold the symbolic link target. This operation will also first try to grab the ***mdc_rpc_lock*** before sending the request to the server.

### 26.5.4. mdc_getattr_lock.

26.5.4.1. *Prototype.*

```
int mdc_getattr_lock(
  struct obd_export *exp,
  struct ll_fid *fid,
  char *filename,
  int namelen,
  unsigned long valid,
  size_t ea_size,
  struct ptlrpc_request **request);
```

26.5.4.2. *Parameters.*

***input: exp*** Existing MDC export.
***input: fid*** The object identifier for the parent directory which has the file that we want to get attributes of.
***input: filename*** The name of file for which we need to get the attributes.
***input: namelen*** The length of filename.
***input: valid*** The mask of which file attributes to retrieve.
***input: easize*** The size of the buffer to allocate to recieve LOV attributes.

***output: request*** Request struct for the issued RPC, which holds the returned attribute data in the reply buffers.

26.5.4.3. *Return Values.* Upon successful completion *mdc_getattr_lock* will return 0, otherwise one of the following error codes will be returned:

**EINVAL:** An invalid connection or key is specified.
**ENOMEM:** Not enough memory to recieve the configuration data.
**ENOENT:** The object does not exist.
**EIO:** Error reading the LOV configuration from disk.

26.5.4.4. *Description.* This is a new call that was added to get attributes of a file using the filename. This operation will also first try to grab the ***mdc_rpc_lock*** before sending the request to the server. Only the fields flagged in the *valid* field will be returned with correct values. The server will lookup the filename and take the required locks.The first reply buffer will hold the *mds_body* struct with the attributes. If the file is a regular file, then the second reply buffer will hold the LOV attribute data. If the *OBD_MD_LINKNAME* flag was set and the file is a symbolic link, then the second buffer will hold the symbolic link target.

#### 26.5.5. mdc_statfs.

26.5.5.1. *Prototype.*

```
int mdc_statfs(
   struct obd_device *obd,
   struct obd_statfs *osfs,
 );
```

26.5.5.2. *Parameters.*

***input: obd*** MDC obd device.
***output: osfs*** The structure used to receive the filesystem status.

26.5.5.3. *Return Values.* Upon successful completion *mdc_statfs* will return 0, otherwise it will return one of the following error codes :

**ENOMEM:** Not enough memory to allocate the buffer for RPC request.
**EINTR:** If request was interrupted.
**ETIMEDOUT:** If the request timed out.

26.5.5.4. *Description.* Obtain file system status information from the MDS. A request is prepared and sent to obtain the filesystem details, the returned status is unpacked into the *osfs* structure.

#### 26.5.6. mdc_setattr.

26.5.6.1. *Prototype.*

```
int mdc_setattr(
  struct obd_export *exp,
  struct mdc_op_data *data,
  struct iattr *iattr,
  void *ea,
  int ealen,
  void *ea2,
  int ea2len,
  struct ptlrpc_request **request);
```

26.5.6.2. *Parameters.*

***input: exp*** An existing MDC export.
***input: data*** MD hint data, which also contains object whose attributes need to be set.
***input: iattr*** The new values for the attributes.
***input: ea*** The striping information for this file
***input: ealen*** Length of the extended attribute field
***input: ea2*** Additional info needed for creating log cancel data on MDS.
***input: ea2len*** Length of the additional info.
***output: request*** Request struct for the issued RPC.

26.5.6.3. *Return Values.* Upon successful completion *mdc_setattr* will return 0, otherwise it will return one of the following error codes :

**ENOMEM:** Not enough memory to allocate the buffer for RPC request.

26.5.6.4. *Description.* Change attributes of a Lustre inode on the MDS. Create and send a request to the MDS with the new attributes for a specific inode. The *setattr* calls that are invoked from the open path are required to take a different lock, the ***mdc_setattr_lock***, and be directed to the *setattr* portal. All other *setattr* calls would take the normal ***mdc_rpc_lock*** and go to the normal portal.

### 26.5.7. mdc_close.

26.5.7.1. *Prototype.*

```
int mdc_close(
   struct obd_export *exp,
   struct mds_body *body,
   struct obd_client_handle *och,
   struct ptlrpc_request **req);
```

26.5.7.2. *Parameters.*

***input: exp*** An existing MDC export.
***input: body*** The object hint for the file to be closed.
***input: och*** Client handle of the object. It contains MD hint etc.
***output: request*** Request buffer for the issued RPC request to MDS.

26.5.7.3. *Return Values.* Upon successful completion *mdc_close* will return 0, otherwise it will return one of the following error codes :

**ENOMEM:** Not enough memory to allocate the buffer for RPC request.
**EINVAL:** Wrong packet type for the message to MDS.
**ENOENT:** Object to be closed does not exist.

26.5.7.4. *Description.* Close a file. This API creates and sends a file close request to the MDS. The file to be closed is identified using the inode number and the handle provided as input parameters.

### 26.5.8. mdc_readpage.

26.5.8.1. *Prototype.*

```
int mdc_readpage(
  struct obd_export *exp,
  struct ll_fid *fid,
  __u64 offset,
  struct page *page,
  struct ptlrpc_request **request);
```

26.5.8.2. *Parameters.*

***input: exp*** An existing MDC export.
***input: fid*** The object identifier for the file to be read.
***input: offset*** The position to start reading from.
***output: page*** A page to be filled up by data.
***output: request*** Request buffer for the issued RPC request to MDS.

26.5.8.3. *Return Values.* Upon successful completion *mdc_readpage* will return 0, otherwise it will return one of the following error codes :

**ENOMEM:** Not enough memory to allocate the buffer for RPC request.
**EINVAL:** Wrong packet type.

26.5.8.4. *Description.* Read a directory page from the MDS. This API creates and sends a request to MDS for a page from a file. A sink for the page is allocated; error is thrown if this allocation fails. The API specifies the file to read from and an offset within the file. This request is sent to the ***MDS_READPAGE_PORTAL***.

### 26.5.9. mdc_create.

26.5.9.1. *Prototype.*

```
int mdc_create(
  struct obd_export *exp,
  struct mdc_op_data *data,
  const char *data,
  int datalen,
  int mode,
  __u32 uid,
  __u32 gid,
  __u64 rdev,
  struct ptlrpc_request **request);
```

26.5.9.2. *Parameters.*

***input: exp*** Existing MDC export.
***input: data*** MD hint with object identifier info in which to create the new file inode.
***input: data*** Additional info related to creating an object. This may be symlink target if we create a symlink.
***input: namelen*** The length of additional info.
***input: mode*** The mode information for the new file, e.g type, access mode.
***input: uid*** *userid* of the file's owner.
***input: gid*** *groupid* of the file's owner.
***input: rdev*** The device identifier for the file.
***output: request*** Request buffer for the issued RPC request to MDS.

26.5.9.3. *Return Values.* Upon successful completion *mdc_create* will return 0, otherwise it will return one of the following error code :

**ENOMEM:** Not enough memory to allocate the buffer for RPC request.
**EINVAL:** Bad file type.
**ENOSPC:** Error creating the new object.

26.5.9.4. *Description.* Create an inode for a file, directory, symlink, or special file on the MDS. This API creates and sends a request to the MDS to create a new inode on MDS for a new file. The *name* field has the name of the new inode to be created. The *target* specifies the name of the object for which a symbolic link should be created.

### 26.5.10. mdc_unlink.

26.5.10.1. *Prototype.*

```
int mdc_unlink(
  struct obd_export *exp,
  struct mdc_op_data *data,
  struct ptlrpc_request **);
```

26.5.10.2. *Parameters.*

***input: exp*** Existing MDC export.
***input: data*** The MD operation hint, which contain also object for objects, which be touched during this MD operation.
***output: request*** Request buffer for the issued RPC request to MDS.

26.5.10.3. *Return Values.* On successful completion *mdc_unlink* will return 0, otherwise it will return one of the following error code :

**ENOMEM:** Not enough memory to allocate the buffer for RPC request.
**ENOENT:** The child entry that the request wants to unlink does not exist.
**EINVAL:** Bad file type for unlink operation.

26.5.10.4. *Description.* Unlink an inode on the MDS. Create and send a request to the MDS to remove link. The inode of the link is specified if known; it is used to verify that the correct file is being removed. The *dir* field gives the inode of the parent directory from which the specified file is to be unlinked. The mode bits are used to choose between *rmdir* and *unlink*, if the request is to unlink a directory, the server calls *vfs_rmdir*. On the other hand, if the unlink is for a regular file, the server will verify if this is the last reference to the inode. If it is a last reference, the server will pack the object attributes in the reply to be passed back to the client, the client can use this information to delete the objects from the OSTs.

### 26.5.11. mdc_link.

26.5.11.1. *Prototype.*

```
int mdc_link(
  struct obd_export *exp,
  struct mdc_op_data *data,
  struct ptlrpc_request **request);
```

26.5.11.2. *Parameters.*

***input: exp*** Existing MDC export.
***input: src*** MD operation hint, which contains also object info for file/directory which should be unlinked.
***output: request*** Request buffer for the issued RPC.

26.5.11.3. *Return Values.* Upon successful completion *mdc_link* will return 0, otherwise it will return one of the following error code :

**ENOMEM:** Not enough memory to allocate the buffer for RPC request.
**ESTALE:** The information about the *src* or *dir* fields is not valid.

26.5.11.4. *Description.* Perform a link operation on MDS inodes. Create and send a request to the MDS to create a link to the file identified by the *src* inode.

#### 26.5.12. mdc_rename.

26.5.12.1. *Prototype.*

```
int mdc_rename(
  struct obd_export *exp,
  struct mdc_op_data *data,
  const char *old,
  int oldlen,
  const char *new,
  int newlen,
  struct ptlrpc_request **request);
```

26.5.12.2. *Parameters.*

***input: exp*** Existing MDC export.
***input: data*** MD operation hint. Here it contains also object info about old and new parents.
***input: old*** Old name of the file.
***input: oldlen*** The length of the old file name.
***input: new*** The new name for the file.
***input: newlen*** The length of the new name.
***output: request*** Request buffer for the issued RPC request to MDS.

26.5.12.3. *Return Values.* On successful completion *mdc_rename* will return 0, otherwise it will return the following error code :

**ENOMEM:** Not enough memory to allocate the buffer for RPC request.
**ESTALE:** Information about the *src* or *tgt* directories is no longer valid.
**ENOENT:** The object we want to rename does not exist.

26.5.12.4. *Description.* Change a file name in a directory to another name. This API takes the name in the source directory and the target directory, as well as the new name, creating a rename request which is sent to the MDS. At the MDS, the old file is given the new name and moved to the target specified.

#### 26.5.13. mdc_change_cbdata.

26.5.13.1. *Prototype.*

```
int mdc_change_cbdata(
  struct obd_export *exp,
  struct ll_fid *fid,
  ldlm_iterator_t it,
  void *data);
```

26.5.13.2. *Parameters.*

***input: exp*** Existing MDC export.
***input: fid*** Object identifier to be used for forming/getting resource id to be used.
***input: it*** The ldml resource iterator.
***input: data*** Data used for iterating over resource.

26.5.13.3. *Return Values.* Always returns zero, as ldlm_change_cbdata() functions does not return any error code.

26.5.13.4. *Description.* This function is used for releasing ldlm resource under passed fid, which denotes object identifier. Thus, it is called in inode clear time.

## 26.6. Logical Medata Volume API

This section details the metadata api of the LMV driver

### 26.6.1. lmv_getstatus.

26.6.1.1. *Prototype.*

```
static int lmv_getstatus(
    struct obd_export *exp,
    struct ll_fid *fid)
```

26.6.1.2. *Parameters.*

***input: exp*:** handle for this object device

***output: fid*** pointer to allocated ll_fid

26.6.1.3. *Return values.*

**0:** success
**non-zero:** an error occurred which is returned to the caller.

26.6.1.4. *Description.* This retrieves the fid of the root inode of the file system. This is only called from the fill super functions used by the mount related functions. It makes an RPC to the primary MDS, stored in the 0-slot of the targets array of the LMV.

### 26.6.2. lmv_getattr.

26.6.2.1. *Prototype.*

```
static int lmv_getattr(
    struct obd_export *exp,
    struct ll_fid *fid,
    unsigned long valid, unsigned int ea_size,
    struct ptlrpc_request **request)
```

26.6.2.2. *Parameters.*

**input: exp:** handle for this LMV device
**input: fid:** file identifier for which attributes should be retrieved
**input: valid:** contains flags dictating which attributes need to be retrieved
**input: ea_size:** size of the preallocated EA buffer that may be returned
**output: request:** place holder for the request that will be returned

26.6.2.3. *Return values.*

**0:** on success
**non-zero:** errors, passed to the caller

26.6.2.4. *Description.* This call retrieves attributes for the fid by calling down to the underlying metadata driver in the LMV targets. If the fid is split RPC's are made to all the MDS's holding objects associated with the fid.

### 26.6.3. lmv_change_cbdata.

26.6.3.1. *Prototype.*

```
static int lmv_change_cbdata_name(
    struct obd_export *exp,
    struct ll_fid *pfid,
    char *name,
    int len,
    struct ll_fid *cfid,
    ldlm_iterator_t it,
    void *data)
```

26.6.3.2. *Parameters.*

**input: exp:** handle for this LMV device
**input: fid:** file identifier for which callback data should be changed
**input: name:** NULL or name in the pfid directory
**input: len:** length of name
**output: cfid:** fid of the name for which cbdata was changed
**input: it:** iterator to retrieve locks associated with the fid
**input: data:** data to place in the locks for callback purposes

26.6.3.3. *Return values.*

26.6.3.4. *Description.* If name is NULL, this changes the cbdata on the pfid. Otherwise it finds the name in the pfid directory and changes the cb data associated with that fid. It finds the target MD driver on which to execute this by taking the using the hash of the name and the object structure of the directory fid passed in. It calls the MD driver below the LMV and executes the same method in that driver.

### 26.6.4. lmv_close.

26.6.4.1. *Prototype.*

```
int lmv_close(
    struct obd_export *exp,
    struct mds_body *body,
    struct obd_client_handle *och,
    struct ptlrpc_request **request)
```

26.6.4.2. *Parameters.*

**input: exp:** handle for this LMV device
**input: body:** file identifier for which callback data should be changed
**input: och:** file handle that should be closed
**output: request:** placeholder for the request that will return results

26.6.4.3. *Return values.*

**0:** for success

26.6.4.4. *Description.* This call will find the MDS that is holding the file handle and invoke the MDC driver for that MDS to close the handle.

### 26.6.5. lmv_create.

26.6.5.1. *Prototype.*

```
int lmv_create(struct obd_export *exp,
    struct mdc_op_data *op_data,
    const void *data,
    int datalen,
    int mode,
    __u32 uid,
    __u32 gid,
    __u64 rdev,
  struct ptlrpc_request **request)
```

26.6.5.2. *Parameters.*

**input: exp:** handle for this LMV device
**input: data:** holds symlink target
**intput: datalen:** length of data
**input: uid:** owner
**input: gid:** group owner
**input: mode:** creation mode
**input: rdev:** device numbers to be used when mode shows a device file
**input: op_data:** security context and fids and metadata extended attributes of the operation
**output: request:** result placeholder

26.6.5.3. *Return codes.*

**0:** on success
**nonzero:** error will be returned to the caller

26.6.5.4. *Description.* This function will find which target to approach and dispatch the create call accordingly to the underlying MD driver.

**26.6.6. lmv_done_writing.** Will not be implemented, to be removed.

**26.6.7. lmv_enqueue.**

26.6.7.1. *Prototype.*

```
int lmv_enqueue(struct obd_export *exp,
    int lock_type,
    struct lookup_intent *it,
    int lock_mode,
    struct mdc_op_data *data,
    struct lustre_handle *lockh,
    void *lmm,
    int lmmsize,
    ldlm_completion_callback cb_completion,
    ldlm_blocking_callback cb_blocking,
    void *cb_data)
```

26.6.7.2. *Parameters.*

**input: exp:** handle for this LMV device
**input: lock_type:** type of the lock to request
**input: it:** intention description structure for the request
**input: lock_mode:** mode of the lock to request
**intput: data:** holds symlink target
**output: lockh:** lock handle f
**input: lmm:** message buffer for special object creation **input: lmmsize:** length of lmm buffer
**input: cb_completion:** callback function to be called when enqueue operations completed
**input: cb_blocking:** callback function to be called when enqueue request is blocked
**input: cb_data:** ast data in the lock for callback functions

26.6.7.3. *Return codes.*

**0:** on success
**nonzero:** error will be returned to the caller

26.6.7.4. *Description.* The function dispatches an enqueue request to the low-level MD driver.

### 26.6.8. lmv_getattr_lock.

26.6.8.1. *Prototype.*

```
int lmv_getattr_lock(
   struct obd_export *exp,
   struct ll_fid *fid,
   char *filename,
   int namelen,
   unsigned long valid,
   unsigned int ea_size,
   struct ptlrpc_request **request)
```

26.6.8.2. *Parameters.*

**input: exp:** handle for this LMV device
**input: fid:** fid of the parent directory
**input: filename:** the child filename
**input: namelen:** filename length
**input: valid:** validation flag for the operation
**input: ea_size:** size of EA
**output: request:** placeholder for the request that will return results

26.6.8.3. *Return Codes.*

**0:** on success
**nonzero:** error will be returned to the caller

26.6.8.4. *Description.* Use the filename and parent directory fid to retrieve the attribution of the file.

### 26.6.9. lmv_intent_lock.

26.6.9.1. *Prototype.*

```
int lmv_intent_lock(struct obd_export *exp,
    struct ll_uctxt *uctxt,
    struct ll_fid *pfid,
    const char *name,
    int len,
    void *lmm,
    int lmmsize,
    struct ll_fid *cfid,
    struct lookup_intent *it,
    int flags,
    struct ptlrpc_request **reqp,
    ldlm_blocking_callback cb_blocking)
```

26.6.9.2. *Parameters.*

**input: exp:** handle for this LMV device
**input: uctxt:** structure including parent & child inode group id information
**input: pfid:** fid of the parent directory
**input: name:** the name of file to handle
**input: len:** length of the filename
**input: lmm:** message buffer for special object creation
**input: lmmsize:** the size of lmm message
**input: cfid:** fid of the file whose name is input ’filename’
**input: it:** intent description structure
**input: flags:** lookup flags
**output: reqp:** placeholder for the request that will return results
**input: cb_blocking:** callback function to be called while request is blocked

26.6.9.3. *Return Codes.*

**0:** on success
**nonzero:** error will be returned to the caller

26.6.9.4. *Description.* Dispatch an intent lock request to low-level MD driver.

### 26.6.10. lmv_link.

26.6.10.1. *Prototype.*

```
int lmv_link(struct obd_export *exp,
     struct mdc_op_data *data,
     struct ptlrpc_request **request)
```

26.6.10.2. *Parameters.*

**input: exp:** handle for this LMV device
**input: data:** holds symlink target
**output: reqp:** placeholder for the request that will return results

26.6.10.3. *Return Codes.*

**0:** on success
**nonzero:** error will be returned to the caller

26.6.10.4. *Description.* Dispatch a file link request to low-level MD driver.

### 26.6.11. lmv_rename.

26.6.11.1. *Prototype.*

```
int lmv_rename(
   struct obd_export *exp,
   struct mdc_op_data *data,
   const char *old,
   int oldlen,
   const char *new,
   int newlen,
   struct ptlrpc_request **request)
```

26.6.11.2. *Parameters.*

**input: exp:** handle for this LMV device
**input: data:** holds symlink target
**input: old:** the old filename
**input: oldlen:** length of the old filename
**input: new:** the new filename
**input: newlen:** length of the new filename
**output: request:** placeholder for the request that will return results

26.6.11.3. *Return Codes.*

**0:** on success
**nonzero:** error will be returned to the caller

26.6.11.4. *Description.* Dispatch a rename request to low-level MD driver.

### 26.6.12. lmv_setattr.

26.6.12.1. *Prototype.*

```
int lmv_setattr(
   struct obd_export *exp,
   struct mdc_op_data *data,
   struct iattr *iattr,
   void *ea,
   int ealen,
   void *ea2,
   int ea2len,
   struct ptlrpc_request **request)
```

26.6.12.2. *Parameters.*

**input: exp:** handle for this LMV device
**input: data:** holds the symlink target
**input: iattr:** attribution to update
**input: ea:** the striping information for this file
**input: ealen:** length of ea structure
**output: ea2:** extened attribution in the reply
**input: ea2len:** length of ea2 structure
**output: request:** placeholder for the request that will return results

26.6.12.3. *Return Codes.*

**0:** on success
**nonzero:** error will be returned to the caller

26.6.12.4. *Description.* Dispatch a setattr request to low-level MD driver.

### 26.6.13. lmv_sync.

26.6.13.1. *Prototype.*

```
int lmv_sync(
   struct obd_export *exp,
   struct ll_fid *fid,
   struct ptlrpc_request **request)
```

26.6.13.2. *Parameters.*

**input: exp:** handle for this LMV device
**input: fid:** the fid of the file to synced
**output: request:** placeholder for the request that will return results

26.6.13.3. *Return Codes.*

**0:** on success
**nonzero:** error will be returned to the caller

26.6.13.4. *Description.* Dispatch a sync request to low-level MD driver.

### 26.6.14. lmv_readpage.

26.6.14.1. *Prototype.*

```
int lmv_readpage(
   struct obd_export *exp,
   struct ll_fid *mdc_fid,
   __u64 offset,
   struct page *page,
   struct ptlrpc_request **request)
```

26.6.14.2. *Parameters.*

**input: exp:** handle for this LMV device
**input: mdc_fid:** fid of the file to be read
**input: offset:** offset to the start of the file
**input: page:** the page descriptor of the page to hold the data
**output: request:** placeholder for the request that will return results

26.6.14.3. *Return Codes.*

**0:** on success
**nonzero:** error will be returned to the caller

26.6.14.4. *Description.* Dispatch a readpage request to low-level MD driver.

### 26.6.15. lmv_unlink.

26.6.15.1. *Prototype.*

```
int lmv_unlink(
   struct obd_export *exp,
   struct mdc_op_data *data,
   struct ptlrpc_request **request)
```

26.6.15.2. *Parameters.*

**input: exp:** handle for this LMV device
**input: data:** holds the symlink target
**output: request:** placeholder for the request that will return results

26.6.15.3. *Return Codes.*

**0:** on success
**nonzero:** error will be returned to the caller

26.6.15.4. *Description.* Dispatch an unlink request to low-level MD driver.

### 26.6.16. lmv_get_real_obd.

26.6.16.1. *Prototype.*

```
struct obd_device *lmv_get_real_obd(
   struct obd_export *exp,
   char *name,
   int len)
```

26.6.16.2. *Parameters.*

26.6.16.3. *Return Codes.*

26.6.16.4. *Description.*

### 26.6.17. lmv_valid_attrs.

26.6.17.1. *Prototype.*

```
static int lmv_valid_attrs(
   struct obd_export *exp,
   struct ll_fid *fid)
```

26.6.17.2. *Parameters.*

**input: exp:** handle for this LMV device
**input: fid:** the fid to be checked

26.6.17.3. *Return Codes.*

**0:** maybe invalid
**1:** valid

26.6.17.4. *Description.* Check if the cached attribution of file is still valid.

### 26.6.18. lmv_delete_object.

26.6.18.1. *Prototype.*

```
int lmv_delete_object(
   struct obd_export *exp,
   struct ll_fid *fid)
```

26.6.18.2. *Parameters.*

26.6.18.3. *Return Codes.*

26.6.18.4. *Description.*

## 26.7. Changelog

**Version 3.0 (Aug 2004)** Peter Braam, Yury Umanets, Jackson He: update after CMD1

**Version 2.0 (Mar. 2003)** Radhika Vullikanti : Updated all the metadata APIs to reflect the changes that were made, removed the mdc_open function that is not used.

**Version 1.5 (Sept. 2002)** Radhika Vullikanti : Added descriptions for mdc_setattr, mdc_close, mdc_open, mdc_link, mdc_unlink

CHAPTER 27

# Lock Management API and Internals

## 27.1. Introduction

An overview of the Lustre lock manager was provided in an earlier chapter in the architecture section. In this chapter we discuss the implementation details of the lock management in Lustre. In Lustre the locking is of two kinds - locking for meta-data operations and locking for file I/O. The meta-data locks are handled only by the meta-data server which stores all the meta-data. The locks for file I/O are handled by the target OST's that store the file objects. In the following sections we will first describe the various data structures that are fundamental to the lock management and other details of lock manager implementation, then we will describe the various API's that make this possible.

## 27.2. Lock Manager Internals

**27.2.1. Data Structures.** There are some data structures that form a central part of the lock manager:

***ldlm_namespace*:** This structure describes a collection of resources and locks on the resources as shown in figure 27.2.1. This can exist both on the client space as well as the server space, the distinction is made using a flag in the structure which would be set as *LDLM_NAMESPACE_CLIENT* or *LDLM_NAMESPACE _SERVER* respectively to indicate that. On the server node all locks that clients have enqueued are managed in the namespace; on the *local* namespace the client copies of such enqueues are held. The namespace is organized as a hash table, hashed by resource name. Each namespace is identified by a *name*; this is usually the common name of a device. For example, if a client has an OSC with a common name *OSC_name*, all the locks managed by it will be under the namespace *OSC_name*. The creation of a namespace can be done using the API *ldlm_namespace_new.*

***ldlm_resource*:** This is a data structure that defines a resource. A resource is always defined within an existing namespace and should have an associated *parent* resource. The *parent* for a root resource would be set as 0. A new resource can only be created using the *ldlm_resource_get* method.This structure also contains an entry for the type of lock being held on the resource - *PLAIN, EXTENT, MDSINTENT.* A count is kept to keep track of the number of locks being held on the resource. Each resource is also associated with the following queues:

***lr_granted*:** This is the queue of granted locks.
***lr_converting*:** This is a queue of blocked converting locks on the resource.
***lr_waiting*:** This is the queue of locks waiting on other converting and granted locks.
***lr_tmp*:** A list of RPC requests associated with processing the queues

***ldlm_lock*:** This is the data structure that describes the lock being held on a resource. It contains a pointer to the structure for the resource on which the lock is being held. It also keeps note of the associated completion and blocking callback functions. The *l_readers* and *l_writers* help to keep track of the number of readers/writers on the resource. Lustre allows extent-based locking and the *l_extent* structure is used to store the information (start and end offsets) for the extent on which the lock is being held. The *l_remote_handle* is a handle to the associated information on the same lock at the remote end (the remote could be server or client). On the lock server, the information about the connection with which the lock is associated is kept in the *l_export* structure. Every lock is also associated with a wait queue to allow processes to wait on the lock to be granted or to be intimated when the lock is no longer in use.

***ldlm_res_id*:** This data structure is used to store the resource identifier.

The Lustre lock manager is formed using these primary data structures as shown in figure 27.2.1.

**27.2.2. Data Structures: the Evaporation/Refcounting Problem.** In Lustre DLM, effectively we have one compound data structure, namely the namespace with resources and locks as shown in figure 27.2.1.

There is a variety of ways to get to objects. For example, we can get to a lock structure either directly if a pointer to that structure is available or from the resource structure. Each should return the object in such a way that it can't be taken away from us. We use reference counts to ensure that an object is not taken away from us before we are done with it. In this refcounting business there are two special crucial events that increase and decrease the refcount: one is creation and one is killing. Creation obviously happens once and killing happens once because the dying flag prevents a second killer from doing it again. Every subsequent reference to the object will also increase the refcount.

**27.2.3. Locks.** Every lock is represented by a lock structure and is associated with a handle; the handle is simply the address of the lock structure that is passed from server to client along with a random cookie. The handle is included by the clients in any request on the lock and allows the server to quickly get to the lock information. A random cookie in the lock handle allows validation of any handle received by the server.

As described earlier, we use refcounts even for the lock objects to ensure safe access to them. The initial creation of a lock causes the refcount to be incremented twice - once for lock allocation (this will be decremented only when the lock is destroyed in *ldlm_lock_destroy*), and again to indicate that the lock is in use (this prevents the lock from being taken away).

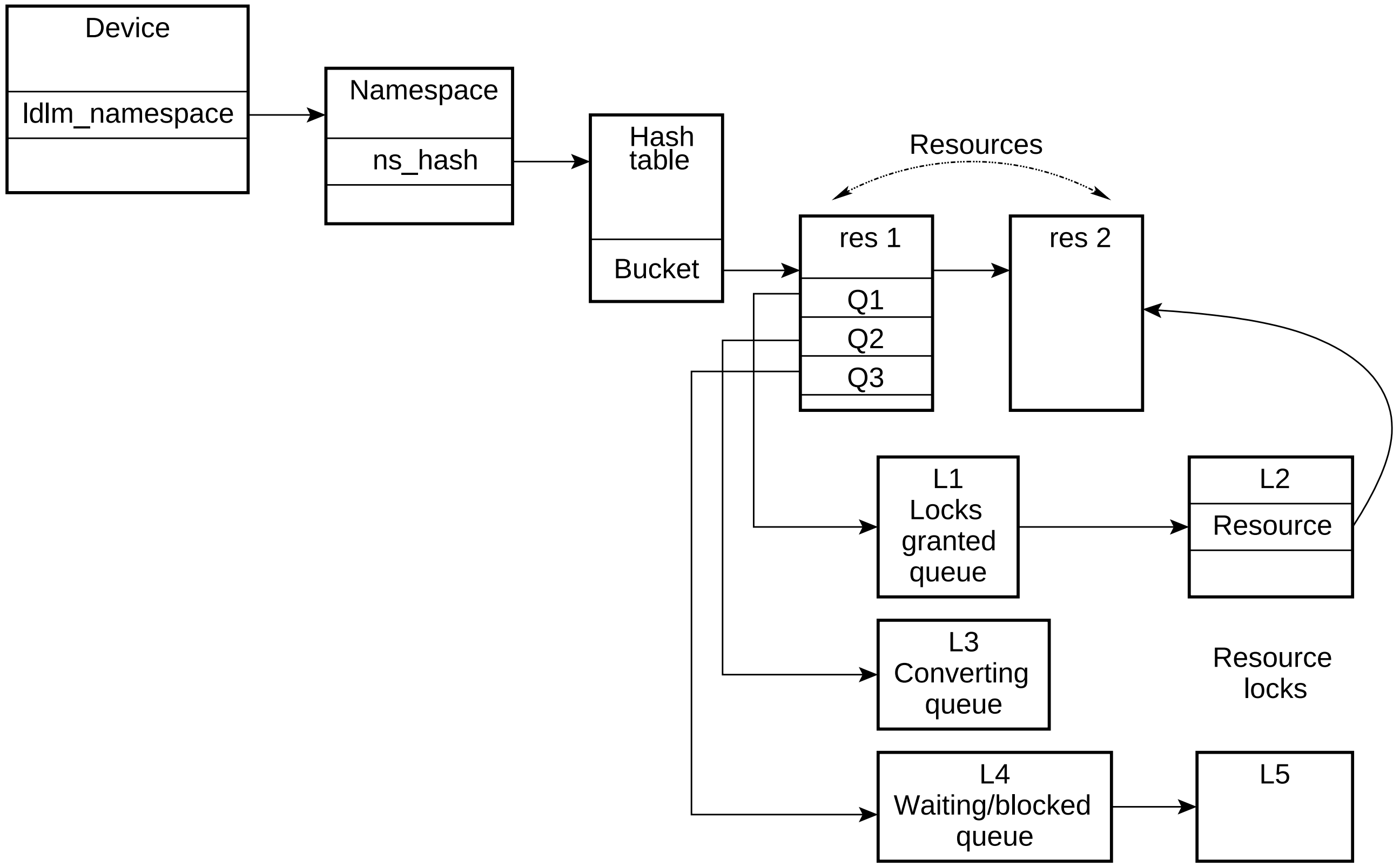

FIGURE 27.2.1. View of Lustre DLM

We have a complementary *ldlm_lock_put* operation to free the lock structure and associated resource, but this can be done only when the refcount falls to 0 indicating that no one is using the lock. Every call to *ldlm_lock_addref* will increase the locks reference count.

27.2.3.1. *References on Locks.* Every lock is associated with a reference count, and it also keeps a separate count for readers and writers. When an application (filesystem) uses a lock it adds a reference to the lock for the type of usage, e.g. a reference for reading/writing (we will have references for every lock mode). These references allow applications ensure that their locks will not be taken away while they are using them. It is important that the applications decrease the references when they are done with the lock.

Such references increase the refcount on the lock, and are only added for *local locks.*

Internally, adding a reader/writer reference increases the refcount on the lock and can postpone the execution of callback handlers. A lock can be canceled only when the reference count goes down to zero.

**27.2.4. Resources.** Resources exist within a namespace and are associated with a *parent*; the *ldlm_resource* structure is used to describe a resource as mentioned earlier. Locks are stored in the namespace within a hash table. The hash table is indexed by the hash value generated from the resource name and its parent. The *ldlm_resource_get* function is used to add a new resource to an existing namespace. A resource is associated with queues of locks on the resources - granted locks, blocked locks, and locks waiting to be converted.

These lists are protected by the namespace lock which can be taken multiple times safely by the same thread.

A resource is also associated with a refcount used to pin it down; the refcount for every lock on a resource's lists contributes to the resource refcount. A resource can also be destroyed only when the refcount goes down to 0.

Some important points to note:

(1) Having access to a referenced lock/resource means the lock/resource won't be freed underneath us.
(2) Lock resources can change in only 3 locations: in the policy, in the completion, and in the final phase of the enqueue.
(3) A careful understanding when things are dying in relation to gaining access to the object. This involves the ordering of put and the locks on resources in some cases (as you do in your *ldlm_local_lock_match* very nicely).
(4) A precise understanding when a lock enters/leaves a list; when it has been killed it should perhaps go on a dying list if it is still referenced.

**27.2.5. Callbacks.** The Lustre DLM exploits the completion and blocking call back infrastructure for both server lock trees and for local lock trees. The server lock trees have callbacks on the locks that make an RPC to the client to invoke the callbacks on the local locks.

Callbacks associated with server locks are prepared as work items while walking the namespace queues and handled afterward.

***local_lock_callbacks*:**
- ***lock_decref*:** When the reference count falls to 0.
- ***handle_callback*:** If there are no users of the lock.

***server_lock_callbacks*:** Generated from:
- ***ldlm_lock_compat_list*:** Checks a list of granted locks to see what compatibility exists. Prepares a blocking AST work item if an incompatible lock is found.
- ***grant_lock*:** In each case the callback function is the *ldlm_cli_callback* function.

On the server side, to avoid concurrency during lengthy callbacks, work items are built that dispatch the callbacks after the global namespace lock is dropped.

**27.2.6. Blocking.** The lock services see a huge amount of re-entrant calls. While one may argue that efficient lock management will eliminate most of these, the lock manager should be able to handle such situations.

While the data structures described above are being modified, no blocking routines should be called. This means that:

(1) Policy must take a refcounted lock away from enqueue, but not hold the namespace lock as policies generally may use the lock manager for enqueues
(2) All RPC's should be added to the *lr_tmp* list.
(3) I suspect it is easier to handle replies with callbacks than by blocking (we can easily add these to *ptlrpc_queue_wait*, or even to the handling functions).

**27.2.7. Request Ordering.** What we have already seen is that sequences of related requests/replies can arrive in just about any reasonable order. For example, a client originating enqueue request may block. The following messages can arrive in any order after the request has been sent:

(1) The reply to enqueue which indicates that the lock is blocked
(2) The blocking AST's can come from the same namespace and remove other locks held on the resource.
(3) The completion AST's may reach the client before the blocking or reply messages come.

## 27.3. Locking API

### 27.3.1. *ldlm_cli_enqueue*.

27.3.1.1. *Prototype.*

```
int ldlm_cli_enqueue
{
  struct lustre_handle *conn,
  struct ptlrpc_request *req,
  struct ldlm_namespace *ns,
  struct lustre_handle *parent_lock_handle,
  struct ldlm_res_id *res_id,
  __u32 type,
  void *cookie,
  int cookielen,
  ldlm_mode_t mode,
  int *flags,
  ldlm_completion_callback completion,
  ldlm_blocking_callback callback,
  void *data,
  void * cp_data,
  struct lustre_handle *lockh
}
```

27.3.1.2. *Parameters.*

***input: conn*** Connection handle to use to reach the server.
***input: req*** Request to place request in.
***input: ns*** Namespace to search for the lock.
***input: parent_lock_handle*** Parent lock to enqueue for.
***input: res_id*** The resource id for which the lock request has to be enqueued.
***input: type*** Type of lock request - *LDLM_PLAI*N, *LDLM_EXTENT*, *LDLM_MDSINTENT.*
***input: cookie*** Pointer to the extent requested.
***input: cookielen*** The length of the extent structure.
***input: mode*** The requested lock mode.
***output: flags*** The flags set upon return by the lock server.
***input: completion*** Lock completion callback function.
***input: callback*** Lock blocking callback function.
***input: data*** Data passed to the completion handlers (e.g. an inode).
***input: cp_data*** Not used at present.
***input: lockh*** The handle.

27.3.1.3. *Return Values.*

**EINVAL:** The lock handle passed in was not valid.
**ENOMEM:** The request or reply, lock or resource could not be allocated.

27.3.1.4. *Description.* This call enqueues a new lock request for the resource passed in, in the namespace given. The request is packaged as the first message in the requests if that pointer is not NULL. The cookie is currently used to pass in an extent, in case the lock is an extent lock. The completion and blocking callbacks are passed the data pointer. The result of the call consists of an error code and upon success a filled in lock handle.

The *flags* field is used to return the status of the lock request and can have one of the following values:

***LDLM_FL_BLOCK_CONV*:** The enqueue blocked on a converting lock.
***LDLM_FL_BLOCK_GRANTED*:** The enqueue blocked on a granted lock.
***LDLM_FL_BLOCK_WAITING*:** The enqueue blocked on a waiting lock.
***LDLM_FL_LOCK_CHANGED*:** The lock manager granted a lock on a resource different from the one requested in the call. This is typical in the case of intent locks. For example, if a lock was requested on a parent directory with an intent to create a new file under that directory, the MDS will create the new file and return a lock on this new file instead.

In figure 27.3.1, we have traced the path of a lock request from client to server starting at the *ldlm_cli_enqueue* function.

#### 27.3.2. *ldlm_match_or_enqueue*.

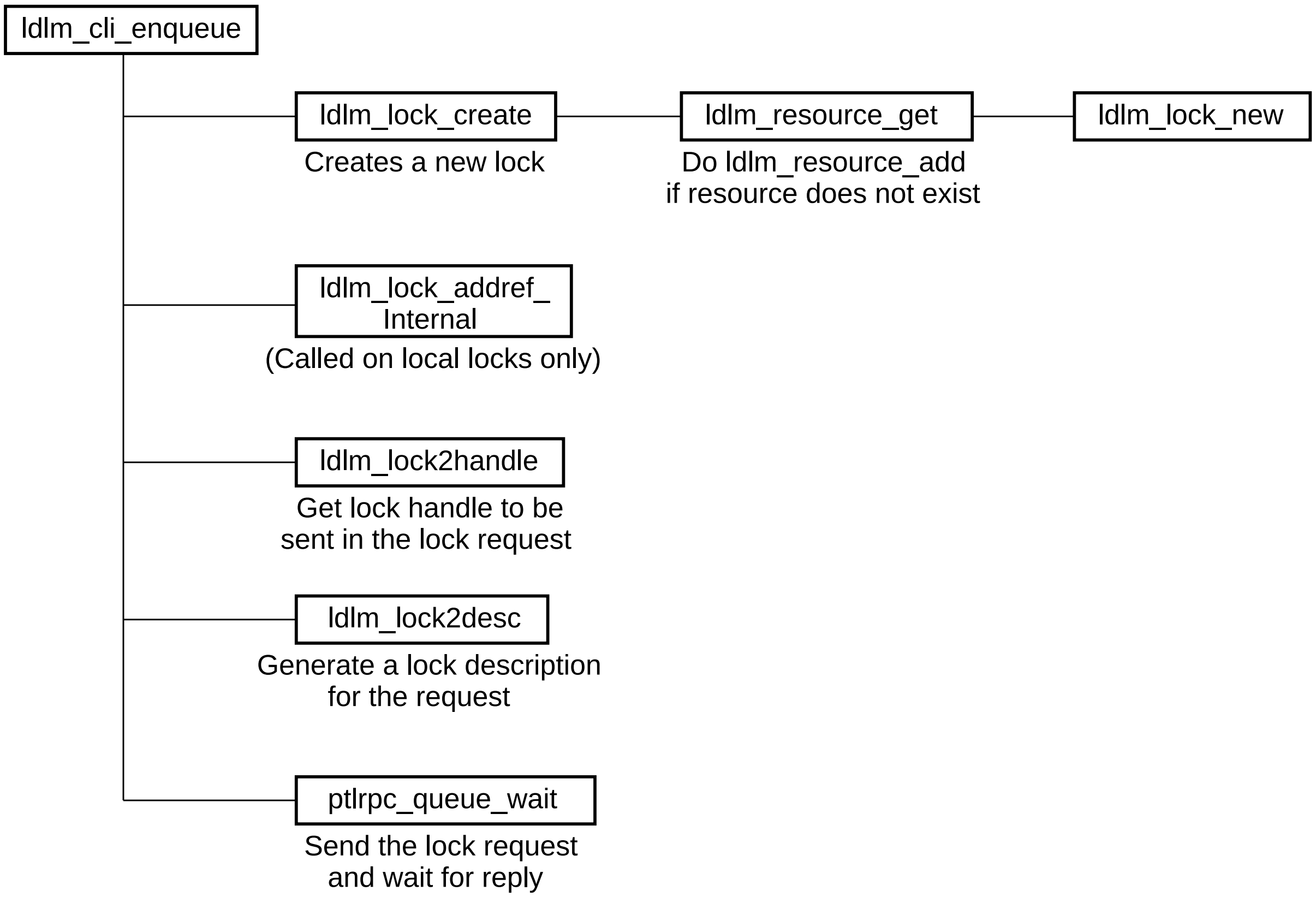

FIGURE 27.3.1. Path trace for a Lock Request to Server

27.3.2.1. *Prototype.*

```
int ldlm_match_or_enqueue
{
  struct lustre_handle *conn,
  struct ptlrpc_request *req,
  struct ldlm_namespace *ns,
  struct lustre_handle *parent_lock_handle,
  struct ldlm_res_id *res_id,
  __u32 type,
  void *cookie,
  int cookielen,
  ldlm_mode_t mode,
  int *flags,
  ldlm_completion_callback completion,
  ldlm_blocking_callback callback,
```

```
    void *data,
    void * cp_data,
    struct lustre_handle *lockh
  }
```

27.3.2.2. *Parameters - see ldlm_cli_enqueue.*

***input: conn*** Connection handle to use to reach the server.
***input: req*** Request to place request in.
***input: ns*** Namespace to search for the lock.
***input: parent_lock_handle*** Parent lock to enqueue for.
***input: res_id*** The resource id for which the lock request has to be enqueued.
***input: type*** Type of lock request - *LDLM_PLAIN*, *LDLM_EXTENT*, *LDLM_MDSINTENT*.
***input: cookie*** Pointer to the extent requested.
***input: cookielen*** The length of the extent structure.
***input: mode*** The requested lock mode.
***output: flags*** Flags set upon return by the lock server.
***input: completion*** Lock completion callback function.
***input: callback*** Lock blocking callback function.
***input: data*** Data passed to the completion handlers (e.g. an inode).
***input: cp_data*** Not used at present.
***output: lockh*** The handle.

27.3.2.3. *Return Values.* The function will return 0 if successful, or else it will return one of the error codes returned by *ldlm_cli_enqueue* function.

27.3.2.4. *Description.* This function is used by a server to try to find a lock that matches the mode (for the new lock request) passed in, which is already held by the client. If a match is found, it fills in the lock handle and places a refcount on the lock, otherwise it tries to obtain a new lock to satisfy the client request.

### 27.3.3. *ldlm_lock_match.*

27.3.3.1. *Prototype.*

```
int ldlm_lock_match(struct ldlm_namespace *ns, struct ldlm_res_id * res_id, __u32 type
   void *cookie, int cookielen, ldlm_mode_t mode, struct lustre_handle *lockh)
```

27.3.3.2. *Parameters.*

***input: ns*** The namespace in which to search for this lock.
***input: res_id*** The resource identifier for which a lock is requested.
***input: type*** The type of lock requested for.
***input: cookie*** This structure is currently used to hold the extent in the lock request.
***input: cookielen*** Length of the cookie structure.
***input: mode*** The requested lock mode.

***output: lockh*** The handle of the matched lock.

27.3.3.3. *Return Values.* This function will return 1 if a matching lock is found, otherwise it will return 0.

27.3.3.4. *Description.* When a server gets a lock request for a particular resource, it can use this function to determine if a matching lock already exists. The function searches the granted/converting/blocked queues for the resource specified in the namespace indicated. If a matching lock is found, it will increase the refcount of the lock (to make sure that the lock manager is aware that the lock is being used) and return the handle of this existing lock. Another usage for this function is to find dplicate locks, in this case ***ns*** is passed as NULL.

#### 27.3.4. *ldlm_cli_cancel*.

27.3.4.1. *Prototype.*

```
int ldlm_cli_cancel
{
   struct lustre_handle *lockh
 }
```

27.3.4.2. *Parameters.*

***input: lockh*** Handle to the lock that has to be canceled.

27.3.4.3. *Return Values.*

**EINVAL:** Client side lock handle was not valid.
**ESTALE:** Server side lock handle is not valid.
**ENOMEM:** Request or reply could not be allocated.

27.3.4.4. *Description.* This call cancels the lock on the server node and frees the local copy of the lock. If the resource of the lock is no longer used, the resource is freed also. It is an error to call this on a lock that still has references.

#### 27.3.5. *ldlm_cli_cancel_unused*.

27.3.5.1. *Prototype.*

```
int ldlm_cli_cancel_unused
{
 struct ldlm_namespace *ns,
 struct ldlm_res_id *res_id,
 int flags
}
```

27.3.5.2. *Parameters.*

**ns** The namespace which 'res_id' belongs to.
**res_id** The resource on which to cancel unused locks.
**flags** If non-zero, cancel these locks locally without notifying the server.

27.3.5.3. *Return Values.*

**EINVAL:** Invalid resource name.
**ENOMEM:** Insufficient memory for temporary buffers.

27.3.5.4. *Description.* This call cancels all locks on the given resource which have a reader/writer refcount of zero. If the flag is set to *local_only*, the client can discard the locks without informing the server.

#### 27.3.6. *ldlm_cli_convert.*

27.3.6.1. *Prototype.*

```
int ldlm_cli_convert
{
 struct lustre_handle *lockh,
 int new_mode,
 int *flags
}
```

27.3.6.2. *Parameters.*

***input: lockh*** The handle for an existing lock which needs to be converted to a new mode.
***input: new_mode*** The new mode to convert the lock to.
***output: flags*** This field is used to return the status of the operation.

27.3.6.3. *Return Values.*

**EINVAL:** The lock handle passed in was not valid.
**ENOMEM:** The request or reply could not be allocated.
**Flags:** Can be set to:
***LDLM_FL_BLOCK_CONV*:** The conversion blocked on a converting lock.
***LDLM_FL_BLOCK_GRANTED*:** The conversion blocked on a granted lock.

27.3.6.4. *Description.* This call converts a lock to a new mode. The call may succeed or reply that it is blocked. The lock will be returned with a reference for the new mode, when granted.

#### 27.3.7. *ldlm_lock_decref.*

27.3.7.1. *Prototype.*

```
void ldlm_lock_decref
{
   struct lustre_handle *lockh,
   __u32 mode
 }
```

27.3.7.2. *Parameters.*

***input: lockh*** Handle for the lock to decref.
***input: mode*** The usage mode of the lock handle that is decremented.

27.3.7.3. *Return Values.* Void.

27.3.7.4. *Description.* This call is issued by a program that has ceased to use a lock in a certain mode. When this call is issued and callbacks are pending on this lock, it will trigger their AST's, which in normal circumstances cancel or revoke the lock.

#### 27.3.8. *ldlm_lock2handle.*

27.3.8.1. *Prototype.*

```
void ldlm_lock2handle
{
  struct ldlm_lock *lock,
  struct lustre_handle *handle
}
```

27.3.8.2. *Parameters.*

***input: lock*** A non-zero pointer to an existing lock.
***output: handle*** The handle that is obtained for this lock.

27.3.8.3. *Return Values.* Void.

27.3.8.4. *Description.* Converts a lock to handle.

#### 27.3.9. *ldlm_lock_dump.*

27.3.9.1. *Prototype.*

```
ldlm_lock_dump
{
  struct ldlm_lock *lock
 }
```

27.3.9.2. *Parameters.*

***input: lock*** A non-null lock pointer for an existing lock.

27.3.9.3. *Return Values.* Void.

27.3.9.4. *Description.* This is a helper function used for debugging purpose. It can be used to dump out all the information associated with the specified lock in a debugging log.

### 27.3.10. *ldlm_namespace_new.*

27.3.10.1. *Prototype.*

```
struct ldlm_namespace *ldlm_namespace_new
{
  char *name,
  __u32 client
}
```

27.3.10.2. *Parameters.*

***input: name*** The name of the new name space to be created.
***input: client*** This field is used to indicate if the namespace is on the client side or on the server side, it can have two possible values - *LDLM_NAMESPACE_SERVER*, *LDLM_NAMESPACE_CLIENT*.

27.3.10.3. *Return Values.* If successful, the function will return a pointer to the newly created namespace, or it will return NULL if an error occurred.

27.3.10.4. *Description.* This call instantiates a new namespace for locks. If the ldlm module is loaded, attached, and set up (as is customary for all modules), this is all that is needed to start acquiring locks, or to start offering lock service (if client = 0).

### 27.3.11. *ldlm_namespace_free.*

27.3.11.1. *Prototype.*

```
int ldlm_namespace_free
{
  struct ldlm_namespace *namespace
 }
```

27.3.11.2. *Parameters.*

***input: namespace*** The namespace to be freed.

27.3.11.3. *Return Values.* This function will return *DLM_OK* (success) if the namespace did not exist, or if the namespace existed and all it was cleaned up.

27.3.11.4. *Description.* Walks all the resources in the namespace. It cancels all locks if the resource is a client resource , i.e. the resource is mastered on a different node. It frees the local copy of such locks and when the resource is no longer used, frees the resource. All resources, locks, and the namespace itself will be freed after this call completes. If this is called on the server, the function will log a warning when a lock is freed that is referenced by a client.

### 27.3.12. *ldlm_resource_get.*

27.3.12.1. *Prototype.*

```
struct ldlm_resource* ldlm_resource_get(struct ldlm_namespace *ns, struct
 ldlm_resource* parent, __u64 *name, __u32 type, int create)
```

27.3.12.2. *Parameters.*

***input: ns*** Namespace in which the new resource should be created.
***input: parent*** The parent of this new resource.
***input: name*** Name of the new resource.
***input: type*** Type of resource.
***input: create*** Flag to indicate if the resource should be created or not.

27.3.12.3. *Return Values.* This function will return a pointer to the newly created resource if successful, otherwise it will return NULL.

27.3.12.4. *Description.* This function is used to create a new resource in the specified namespace. The function is passed a pointer to the parent resource. Since the namespace is organized as a hash table, the function will first determine the correct bucket for this resource by calculating a hash based on the parent resource and the new resource. The function will first search for the new resource in this bucket; if the resource is already there, it would simply increase the refcount for the resource and return a pointer to the match found. If the resource does not exist and the create flag is set, a new resource structure will be created and added to the hash bucket; its reference count will be initialized to 1 to indicate that it is in use.

### 27.3.13. *ldlm_resource_getref.*

27.3.13.1. *Prototype.*

```
struct ldlm_resource *ldlm_resource_getref(struct ldlm_resource *res)
```

27.3.13.2. *Parameters.*

***input: res*** The resource whose refcount needs to be incremented.

27.3.13.3. *Return Values.* This function will return a pointer to the resource that was modified.

27.3.13.4. *Description.* This function can be called to atomically increment the refcount of the specified resource.

### 27.3.14. *ldlm_resource_put.*

27.3.14.1. *Prototype.*

```
int ldlm_resource_put(struct ldlm_resource *res)
```

27.3.14.2. *Parameters.*

***input: res*** The resource to be freed up.

27.3.14.3. *Return Values.* This function will return 1 if it successfully freed up the resource, otherwise it will return 0.

27.3.14.4. *Description.* The function is called to free up a resource when it is no longer used, i.e. its reference count is down to 0. It will first decrement and test the refcount for the resource. This will be double checked to ensure that no one grabbed the resource before we got a lock on it. If someone did, then the freeing operation is canceled. If we successfully obtain a lock on the resource, then it checks to make sure that all the queues associated with it are empty (its a bug if they are not since the refcount is 0), decrement the namespace refcount, and free the data structure for the resource.

### 27.3.15. *ldlm_resource_add_lock*.

27.3.15.1. *Prototype.*

```
void ldlm_resource_add_lock(struct ldlm_resource *res, struct list_head
 *head, struct ldlm_lock *lock)
```

27.3.15.2. *Parameters.*

***input: res*** The resource on which the new lock is to be added.
***input: head***
***input: lock*** The new lock on the specified resource.

27.3.15.3. *Return Values.*

27.3.15.4. *Description.* This function is used to add a new lock to the granted/converting/blocked queues of a resource.

### 27.3.16. *ldlm_resource_unlink_lock*.

27.3.16.1. *Prototype.*

```
void ldlm_resource_unlink_lock(struct ldlm_lock *lock)
```

27.3.16.2. *Parameters.*

***input: lock*** The lock to be removed.

27.3.16.3. *Return Values.*

27.3.16.4. *Description.* This is a helper function. This function is called to remove a lock from the resource lock queue. It will first take a lock on the namespace.

### 27.3.17. *ldlm_res2desc*.

27.3.17.1. *Prototype.*

```
void ldlm_res2desc(struct ldlm_resource *res, struct ldlm_resource_desc
 *desc)
```

27.3.17.2. *Parameters.*

***input: res*** An existing resource.
***output: desc*** The description of the resource.

27.3.17.3. *Return Values.*

27.3.17.4. *Description.* This function can be used to obtain the resource description (type, name, version) from the resource structure.

### 27.3.18. *ldlm_completion_ast*.

27.3.18.1. *Prototype.*

```
int ldlm_completion_ast(struct ldlm_lock *lock, int flags)
```

27.3.18.2. *Parameters.*

***input: lock*** An existing lock.
***input: flags*** Flag indicating the status of a lock request.

27.3.18.3. *Return Values.*

27.3.18.4. *Description.*

### 27.3.19. *ldlm_lock_get*.

27.3.19.1. *Prototype.*

```
struct ldlm_lock *ldlm_lock_get(struct ldlm_lock *lock)
```

27.3.19.2. *Parameters.*

***input: lock*** An existing lock on a resource.

27.3.19.3. *Return Values.* This function will return a pointer to the lock.

27.3.19.4. *Description.* This is a helper function that may be called to increment the reference count associated with the lock and with the resource associated with the lock.

### 27.3.20. *ldlm_lock_put*.

27.3.20.1. *Prototype.*

```
void ldlm_lock_put(struct ldlm_lock *lock)
```

27.3.20.2. *Parameters.*

***input: lock*** An existing lock.

27.3.20.3. *Return Values.*

27.3.20.4. *Description.* This function is used to delete a lock. It will first decrement the lock's refcount and then free the resource structure for the lock resource if its refcount is 0. If the lock refcount is 0 and its flag has been set to *LDLM_FL_DESTROYED*, then the lock structure is free too and the lock is destroyed. If there is a connection associated with the lock, that is also cleaned up.

### 27.3.21. *ldlm_lock_destroy*.

27.3.21.1. *Prototype.*

```
void ldlm_lock_destroy(struct ldlm_lock *lock)
```

27.3.21.2. *Parameters.*

***input: lock*** An existing lock to be destroyed.

27.3.21.3. *Return Values.*

27.3.21.4. *Description.*

### 27.3.22. *ldlm_lock_change_resource*.

27.3.22.1. *Prototype.*

```
int ldlm_lock_change_resource(struct ldlm_lock *lock, __u64 new_resid[3])
```

27.3.22.2. *Parameters.*

***input/output: lock*** An existing lock on some resource.
***input: new_resid*** A new resource id.

27.3.22.3. *Return Values.* This function will return 0 if successful, or else it will return:

**ENOMEM:** No memory available to allocated the structure for the new resource.

27.3.22.4. *Description.* This function is called to change the resource in an existing lock structure. It is useful in case of *intent-based locking*. For example, the client might send a request to the MDS to get a lock on the parent directory with an *intent* of later creating a new file in it. The MDS will decide to execute the intent instead; it will create the new directory for the client. It will then change the resource in the lock to show the new directory, and return a lock on the new directory to the client.

### 27.3.23. *ldlm_lock2desc*.

27.3.23.1. *Prototype.*

```
void ldlm_lock2desc(struct ldlm_lock *lock, struct ldlm_lock_desc *desc)
```

27.3.23.2. *Parameters.*

***input: lock*** An existing lock.
***input: desc*** A description structure for the lock.

27.3.23.3. *Return Values.*

27.3.23.4. *Description.* This function helps to fill the lock description structure with information about the lock - *requested_mode*, *granted_mode*, *extent*, *version*.

### 27.3.24. *ldlm_lock_addref*.

27.3.24.1. *Prototype.*

```
void ldlm_lock_addref(struct lustre_handle *lockh, __u32 mode)
```

27.3.24.2. *Parameters.*

***input: lockh*** A handle for an existing lock.
***input: mode*** The locking mode requested.

27.3.24.3. *Return Values.*

27.3.24.4. *Description.* This function is used to increase the reader/writer count in a lock depending on the requested locking mode. The lock modes *LK_NL*, *LK_PR*, and *LK_CR* will increase the reader count; all other locking modes will cause the writer count to be incremented. This is done only on local locks.

## 27.4. Changelog

**Version 1.5 (Oct 2002)**

(1) Radhika Vullikanti (11/09/02) : Added more detailed explainations and some pictures to section 17.2.

**Version 1.5 (Oct 2002)**

(1) Radhika Vullikanti (10/03/2002) - Made some modifications to some details of lock management

CHAPTER 28

# File System Design

## 28.1. Introduction

The Lustre file system consists of three subsystems - the *client filesystem*, the *Meta-Data Server(MDS)*, and *Object Storage Targets (OST's)*. An overview of the system can be found in the architecture section of the book. All of these subsystems leverage on an existing filesystem - for example the meta-data server uses an underlying extN (ext3 filesystem with some extensions) to store the object meta-data in persistent storage.

The Lustre filesystem exposes a POSIX compliant VFS layer to the client applications. It leverages extensively on the Linux operating system and the underlying journaling file systems like ext3, reiserFS. Lustre uses existing API's to Linux page caches, for example for reading a page, the *lustre_get_page_read* invokes the Linux API *read_cache_page*. The *filter_get_page_write* Lustre API calls the Linux API *grab_cache_page* to grab a page in cache. Several optimizations are also done on top of this, for example if the page to be written is locked by another write, Lustre would write to a temporary page till it is able to grab the lock on the correct page.

In this chapter we will describe some of the implementation details for the Lustre filesystem. We start with a brief overview of the various filesystem operations exported by Lustre, then describe the data structures used and their location. We have a discussion of file I/O locking and meta-data locking. We will also consider one of the issues that would help improve the file system scalability - *writeback cache* for file I/O. We then proceed to flowcharts of some important system calls.

## 28.2. File System Data Structures

As described earlier, the Lustre filesystem has two subsystems the clients interact with - the MDS and OST. All the meta-data operations - creating new directories, files, symbolic links, or acquiring and updating inodes, are handled by the MDS. All the file I/O related operations are directed to the OST's. The following tables give a summary overview of how filesystem related data is managed in a persistent and transient form.

**28.2.1. Persistent Data Storage.** Table 1 details persistent storage of filesystem objects, as well as which functions move the data between systems.

**28.2.2. Cached & Persistent Data Updates.** Table 2 illustrates where the caches are used.

| | Stored here | Protocol used to move data to client | FS operations | Purpose |
|---|---|---|---|---|
| inode meta-data | MDS | *mds_getattr_lock*<br>*mds_getattr* | *lookup2* (CS)<br>*read_super* | Inodes building up the file system. |
| directory data | MDS | *mds_readpage* | *readdir*<br>*lookup* (WB) | directory pages maintain the filesystem namespace |
| file data | OBD | *obd_brw* | *file_read*<br>*file_write* | As usual. Stored in data objects, implemented as OBD inodes. |
| file block allocation | OBD | Never sent to client. | *file_read/*<br>*write* | Spread allocation data evenly over the cluster. |
| symbolic links | MDS | *mds_getattr*<br>*mds_getattr_lock* | *readlink*<br>*followlink* | Usual |
| LOV striping data | MDS inode, as EA | *mds_getattr*<br>*mds_getattr_lock* | *lookup*<br>*file_open* | striping information used to decide where and how many objects should be created |
| access control lists | MDS inode, as EA | *mds_getattr*<br>*mds_getattr_lock* | N/A | will be used for security purposes |
| LOV configuration | MDS | *mds_get_lovinfo* | *read_super* | |

TABLE 1. Persistent Data Storage

| | | Updated by... | Notes |
|---|---|---|---|
| inodes | MDS | *mds_reint*<br>(all reint records update inodes) | store file metadata |
| directory data | MDS | *mds_reint* | *mds_reint* can be called from *ldlm_enqueue* to execute intents. |
| directory data | client | Refreshed upon readpage, changed when client does updates. | Directory data is NEVER propagated from client to MDS. All updates are propagated through update records. |
| striping meta-data | MDS | *mds_reint* (create (WB))<br>*mds_open* (delayed object allocation (CS) ) | |
| LOV configuration | MDS | *lctl* | |
| symbolic links | MDS | *mds_reint* (symlink) | |
| dcache | MDS | *mds_reint*, *mds_fid2dentry*, *mds_getattr_lock* | MDS dcache caches named dentries and dentries by file id. |
| dcache | client | *lookup2*, file system updates | |

TABLE 2. Cached & Persistent Data Updates

### 28.3. File Locking Strategy

The locking strategy for Lustre Lite is basic, and many refinements are expected for Lustre Lite Performance. In this section we discuss locking for semantic purposes, ***i.e. this section does not cover flock and lockf system calls.***

**28.3.1. File I/O Locks.** In the Lustre locking scheme, locks for file I/O are handled by the OST's which store the file data. The OST's are capable of granting locks on byte range extents; this allows increased concurrent access to a file, i.e. different clients could have locks on different extents at the same time. Below we describe how the extent-based locking scheme is used in different cases:

(1) Locks on *a byte range extent* [a,b] - a single *writer*(client) can hold a *write* lock on this extent, there could be multiple *readers* accessing different extents.
(2) Locks that *prevent the file size from changing* - a file operation like *append* or *truncate* would cause the file length to change by a node; this client should have a lock such that no other client can change the file length while this operation is being done. This can be achieved by granting an extent lock to the first node with the upper extent equal to -1 (i.e. 2^64 -1). This prevents any other client from changing the file length.
(3) Locks that require *the file size to be known* on a client - these use an extent [0,-1]. When this lock request is made to an OST it will return the current size of the object to avoid another *getattr* call.

A lock is *granted* only if the request does not *conflict* with the other locks being held on the same resource. A *conflict* is a different mode that is not compatible with a granted mode for overlapping of extents. At present, Lustre follows a very optimistic locking scheme; when locks for file extents are requested the lock manager will grant the largest available file extent that does not conflict with other locks. As an optimization, the lock manager returns the file size with any lock granted.

Locks are acquired only when the client doesn't already own a sufficiently strong lock. When the *file size known* lock is first acquired, the file size is updated on the client; when it is re-acquired, the return code -EALREADY from *obd_enqueue* indicates that it should not be updated.

File I/O locking can be eliminated if applications are known to behave well, this can be done using the ioctl *LL_FILE_IGNORE_LOCK*.

A key consideration is that the two type of locks may interact when an extent crosses the end of file. The extent conflicts handle this automatically.

**28.3.2. System Calls Requiring File Locks.** The following system calls require protection of an extent in the file or the lock on the file size:

***sys_write*:** This call invokes the *ll_file_write* filesystem method. It would always need locks in *'PW'* mode. If the file is opened with *O_APPEND* flag, this will need the lock on the file size [0, -1]. In other cases, an extent lock suffices.

***sys_read*:** This system call invokes a *ll_file_read* method from the Lustre file system methods. It requires a *'PR'* mode lock on the extent that has to be read.

***sys_lseek*:** This call invokes the *ll_seek* method on Lustre filesystems. It is essential that the correct file size is know and it does not change while this operation is being performed, especially when the *whence* parameter equals *SEEK_END* . So, it requires a *'PR'* lock on the file size using [0, -1] as the extent. Any other locks that may possibly affect the file size should be revoked before this lock is granted.

***sys_truncate*:** This system call invokes the *ll_truncate* call on Lustre. Since it involves reducing the file length to specified *offset* , it requests a *'PW '* lock on extent [*offset*, -1]. This blocks any other operation from changing the file size, but at the same time allows operations that affect the file before *offset* to proceed by acquiring a lock on the required extent. For example, while a 'PW' lock has been granted for file extent [*offset*, -1], a *reader* could request for a read lock on file extent [a, b] if a, b < *offset*.

***sys_stat*:** This invokes the *ll_lookup2* call on Lustre. If the intent is *IT_GETATTR* or *IT_SETATTR*, a *'PR'* lock is requested on the file size using the extent [0, -1].

***sys_close*:** This call invokes the *ll_close* method on Lustre filesystems. The close method involves updating the file size on the meta-data server as well, so it requires the file size knowledge across the *mds_close* call. This means no one else should be able to change the file size till the *mds_close* is done, so we take an extent lock on [0,-1].

**28.3.3. Client Lock Namespace.** The locks on files are handled by the OST's which store the file objects. On the clients, its is the OSC layer that communicates with OST's. So, on the clients, the OSC's are used as the namespace for locks that are granted to the client. When a client requests for a lock on a particular resource, the OSC tries to find a matching lock (a lock that would be at least as strong as the requested lock). If a matching lock is found, that is used, or else the lock request is sent to the OST.

**28.3.4. LOV Locking.** The Logical Object Volume (LOV) can be used to virtualize the storage devices underneath. The LOV might be managing several OST's, but will appear as a single device to all the upper layers, shown in figure 28.3.1a. The LOV communicates with the OST's using the OSC driver. A file can be striped across all the OST's that are a part of an LOV, as shown in figure 28.3.1b. The round-robin algorithm would be a simple striping scheme in which we can stripe across all the OST's in the LOV; there could be other schemes as well.

In an LOV, a particular file extent might consist of stripes managed by multiple OSC's. So when the LOV receives a lock request for a file extent, it will break it up into lock requests for extents managed by a particular OSC. For example, in figure 28.3.1b, if the lock request is for a file extent consisting of stripes 1 & 2, the LOV will break the lock request into two - one for stripe 1 managed by OSC1, the other for stripe 2 managed by OSC2. The general pattern of the striping is that the extent functions are quasilinear, increasing functions that map extents to other extents; this is shown in figure 28.3.2.

When an LOV sees a lock request for file extent [0, -1], it breaks this into lock requests for every OSC in the striping pool.

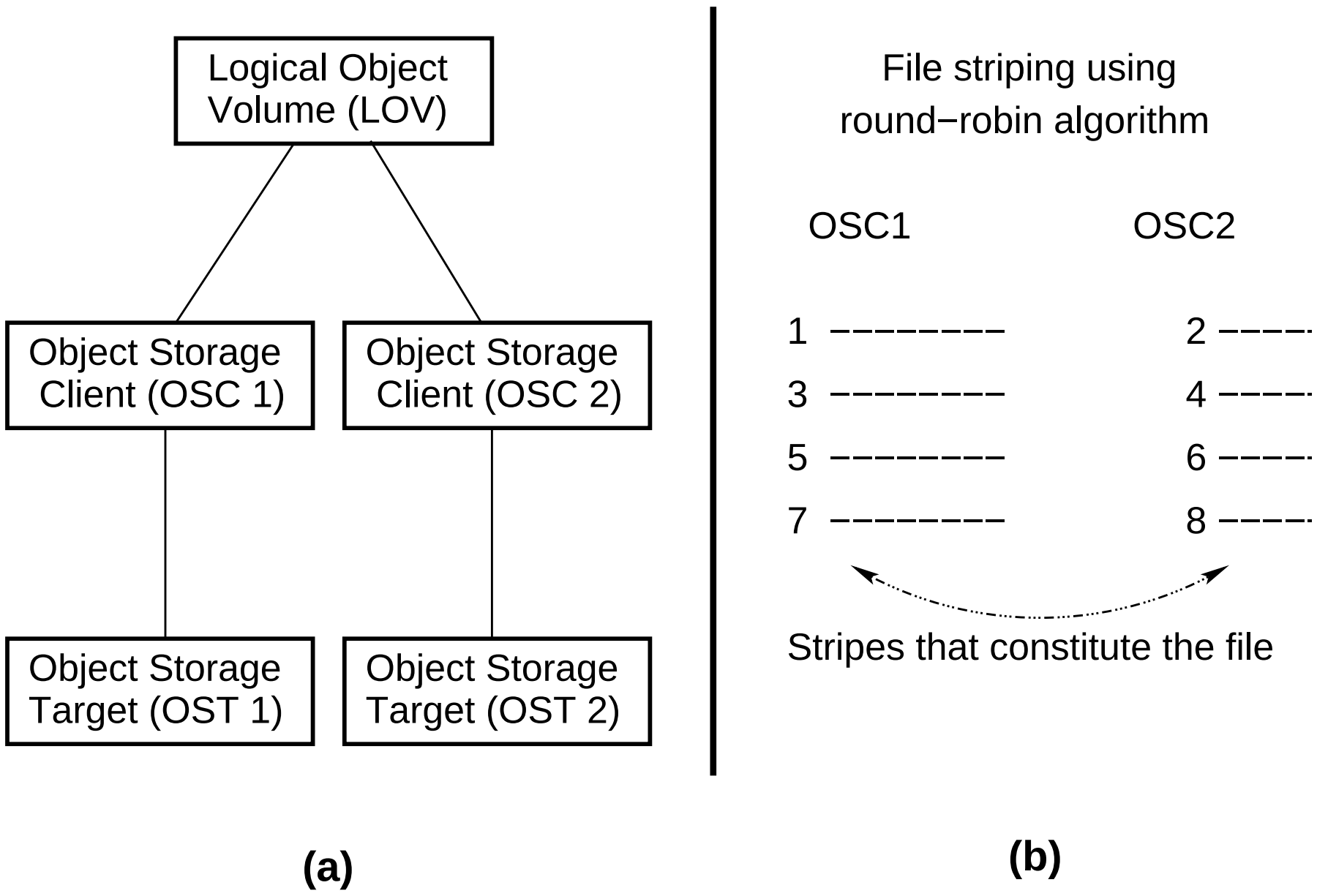

FIGURE 28.3.1. File Striping in LOV

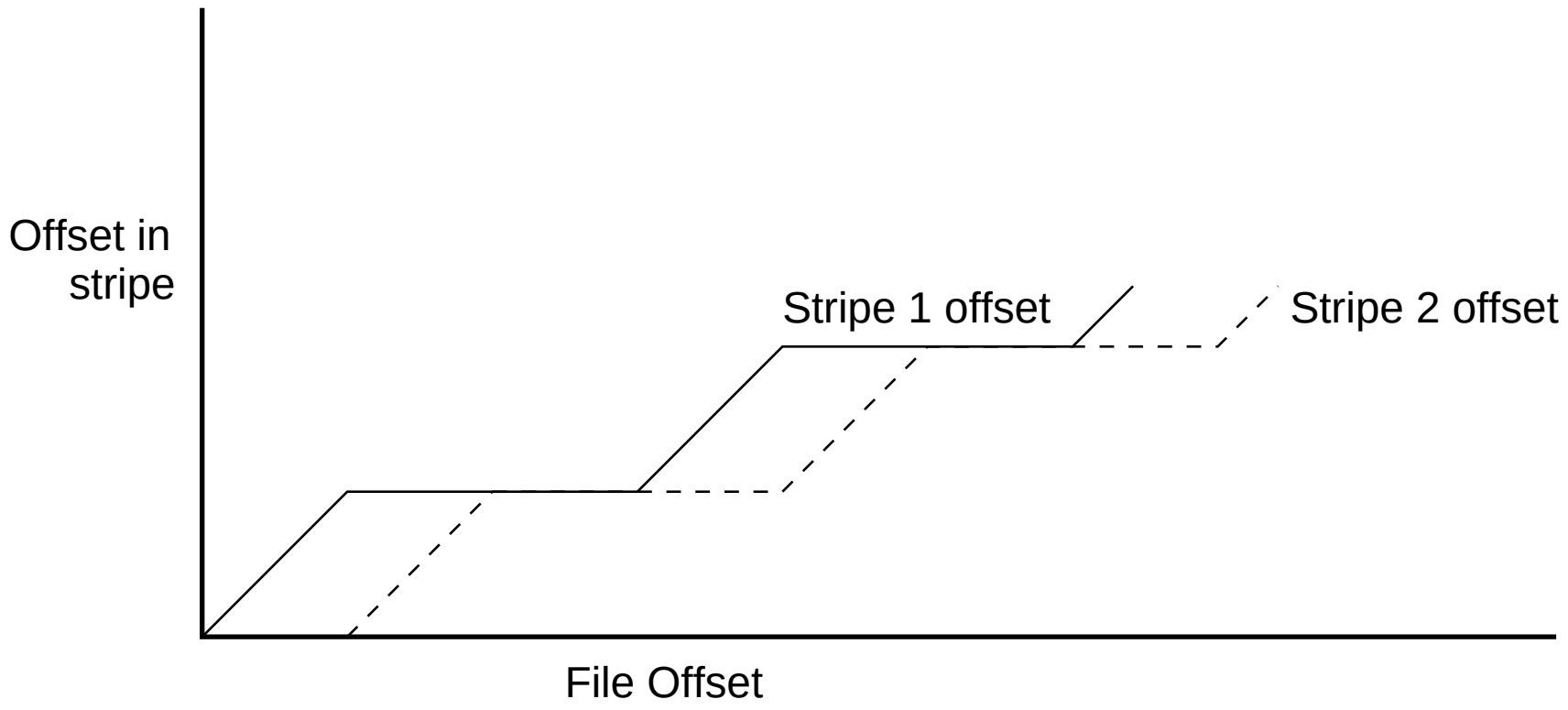

FIGURE 28.3.2. LOV Extents

**28.3.5. Lock refinements.** In order to improve performance, some lock refinements have been proposed and are now being implemented. Earlier, each OST managed the locks for the file extents it held. This meant that no single OST had a global view of the changes being done on a file or the various locks being held. This proved to be sub-optimal in several cases. The major refinement that is now in progress is to have a single lock manager for the complete file. The choice of the manager could be random, or a simple algorithm would be to choose the OST that holds the first stripe of a file to be its lock manager. The selection of starting OST is done randomly, so this simple policy should distribute the locking load quite uniformly.

Another optimization proposed is to improve very simple but probably very common operations lik *ls -l*. We currently require a read lock on the file size for any client wanting to query the file size, this means yanking all other locks held on the file. A few improvements have been suggested to solve this problem - return object size with every lock request, intent based OST callbacks and callback replies. When an OST gets a request for file size, it can send an *intent-based-callback* to a small sub-set of clients that are known to have write locks started from the highest file offset. The callback is tagged with an intent that indicates that the reason for the callback is to grant a size lock and return file size to client. Clients can then simply return the updated object sizes in the callback replies and continue to hang on to their write locks. The helps to reduce tremendous amount of RPC traffic and unneccesary lock yanking.

## 28.4. Meta-data Locking

In this section we give an overview of the meta-data locking. The fundamental problem is to protect data that may be cached on a client, MDS, or OST. This data consists of cached dentries, cached inode attributes, and cached directory data. All of these are used for multiple purposes by the file system.

### 28.4.1. Cached Data Structures Requiring Locks.

**dentries:** A cached dentry represents a node in the file namespace. These dentries tend to remain valid for a very long time, namely until they are removed or renamed.

- **validity:** Must be guaranteed with a lock.
- **current working directory:** Process with its current working directory in a directory. Does this require a lock?
- **writeback subtree:** A dentry may be marked as the root of a subtree that has a meta-data writeback caching enabled. In this case a write lock is needed.

The locks on the dentries protect the dentry caches on the clients from incoherency. The resource names of the locks protecting these objects will be composed of the *fid* of the inode they are protecting together with a flag to indicate if they are a writeback lock or a single dentry lock.

**file inodes:** A cached inode has several attributes for which validity is important:

- **regular file size, file data:** For regular files this was discussed in 28.3.1, and is managed by OST based locks.

**directory inodes:**

- **directory file data:**
  1. Read lock is needed for *readdir* (only on the client) and *lookups* (which can happen on the MDS and client) .
  2. Write lock is needed for updates to the directory, made on the client in write-back mode and on MDS in intent mode.
  3. An exclusive lock is needed for *rmdir* to invalidate any other outstanding locks.

We believe that directory data requires a separate lock resource name from the resource name used for attributes.

**all inodes:**

**extended attributes:** Contain the striping information for files.

**other attributes:**

(1) Read locks for stat, permission checks, and the like.

(2) Write locks are needed for *setattr* calls.

**files:** These data structures represent open files.

**28.4.2. Writeback and Intent Based Updates.** Lustre supports two modes of meta-data updates. The fundamental mechanisms are described in the architecture section.

In both writeback and intent mode, read-only access to the filesystem can be based on data cached in the client. In intent mode, updates are not done with client-based write locks, but exclusively with MDS-based write locks. In writeback mode, the client caches directory data and updates it locally. It batches groups of update transactions and a client based meta-data flush daemon sends these to the MDS.

The locking model of intent based updates is very simple. Only the MDS obtains write locks on the content of directories, while the clients typically only get attributes on attributes of inodes and locks on dentries. The locks on the attributes are revoked when other systems update the inodes, and the dentry locks are only revoked when files are removed or renamed.

Writeback caching mechanisms can be introduced. These work fundamentally like a file I/O write-back cache, which is flushed when other systems need to see the data, or under memory pressure or after a timeout. Fundamentally a writeback lock on a directory grants the holder to modify that directory and any objects beneath it in the directory tree.

The implementation of writeback caching requires us to have a tree structure in the locks that reflect the directory tree. This will made part of the lock policy function. The lock policy function will build a tree of locks and insert new locks in the tree (figure 28.4.1).

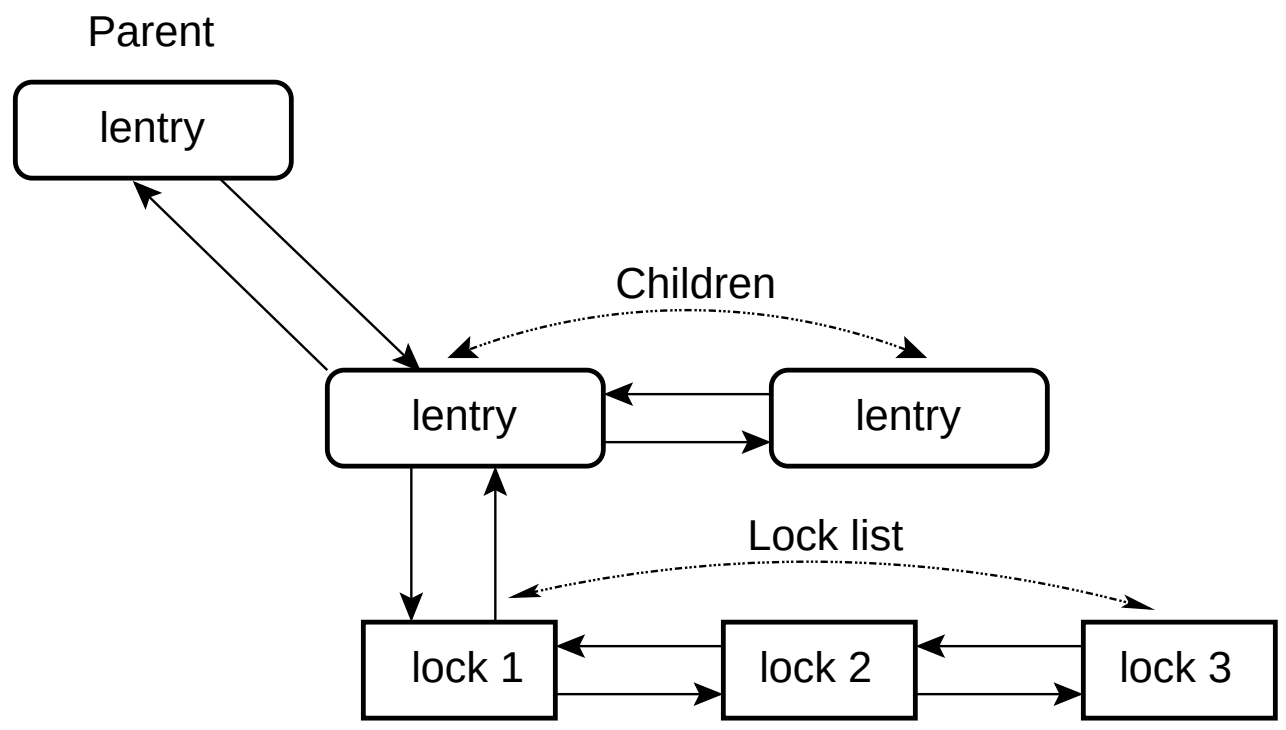

FIGURE 28.4.1. Lock Tree

In order to build this tree we have two options.

First we can introduce a data structure similar to a dentry:

```
struct lentry {
    struct lentry   *l_parent;
    struct list_head l_child_of;
    struct list_head l_children;
    struct list_head l_locks;
}
```

and inside the Lustre lock structure we similarly require:

```
struct ldlm_lock {
    ...
    struct lentry   *ll_lentry;
    struct list_head ll_lentry_list;
    ...
}
```

The code performing lookups is now responsible for linking the locks into the parents entry appropriately, on both the client and the server.

The second possibility is to let the meta-data intent policy functions link the locks into the dentry tree. This means that dentries for objects with active locks will remain cached. While this cache may be larger than the cache of lentries required, the virtual memory subsystem will be able to exert pressure on this cache, something it cannot do on the lentry cache all that easily.

The basis for granting client writeback locks are frequent server-based operations that come from few clients.

**28.4.3. Lock Optimizations.** Lustre Lite Performance will try to improve performance of the system with the following locking refinements and optimizations:

- The extent policy of always allocating as much as possible is crude, and sometimes sub-optimal. In fact, it may be almost the worst possible policy for usual IOR configurations.
- We never down-convert a lock when we get a blocking AST, only cancel. If we down-converted, we could handle it right away instead of waiting for a reader/writer to finish.
- We could send this conversion (or a cancellation) in the reply to the AST.
- It's a little more complicated (and I must have been feeling behind) so the server side convert handler always just puts it on the conversion queue, even if it can be granted immediately. This could be fixed to save having to issue an AST in this case (although we currently never convert because of conversion deadlocks).
- When we lose a lock–any lock–on a file or directory, we currently flush all inode pages from memory instead of just the ones covered by that lock. Once we have lock versioning, we can mark these pages as out of date and possibly mark them clean again after re-acquiring an unchanged lock.
- We plan on eventually sending update records along with lock revocations that are the results of an intent lock, so that clients can update their caches and keep their read locks.

- In addition, as we talk about optimizations, we may well want to consider whether we really want to send the AST's serially for multiple locks help by the same client. Maybe it's not a big deal once we get to a more state-machine model, and we don't risk having a lot of clients stuck behind a pile of round trips to one other client.

## 28.5. Writeback Cache for File I/O

Another very important optimization that has been added to Lustre is the writeback cache for file I/O. Earlier file *writes* were write through, i.e *writes* would not be complete till the changes are flushed to the target OST's. This could impose a considerable delay on each *write* operation, especially in a very busy cluster. In the presence of a writeback cache, the *writes* would first go to this cache and then flushed to the target OST at a later time. This could occur either periodically, when memory pressure increases and the cache size has to be reduced, or when the particular cache page is needed for another operation. The periodic *write* flush could be batched, further improving the performance. The Linux kernel has changed considerably from 2.4 to 2.5. In this section we will describe the writeback cache implementation for Linux 2.5 as well as Linux 2.4 to enable backporting of this feature.

**28.5.1. File write operation.** The write operation in Lustre without the writeback cache is illustrated in figure 28.5.2. As shown here, the two functions - *prepare_write* and *commit_write* are the two-phase write hooks and are called from *generic_file_write*. If *prepare_write* sees a partial write, it needs to ensure that the destination page is up to date so that the page will be correct after the partial write. So it firsts reads in the page from disk. In case of complete page writes, this step does not do anything. The *commit_write* function completes the write, marking the page as dirty in a write caching setup. The *generic_file_write* holds the semaphore to lock the file inode (*inode->i_sem*) over *prepare_write* and *commit_write* phases, this allows *commit_write* to update the file size (*inode->i_size*) if the write extends the file. The writes have to be flushed to persistent store synchronously in the absence of writeback cache.

The *writepage* routine is handed dirty locked pages and is expected to write them to persistent store and unlock them. Its primary job is to unlock the page it was given.

A locked page is marked by the *PageLocked* atomic bit. Callers of *lock_page* try to set this bit and block if its already set. More careful paths call *TryLockPage* to set the lock and can fail if they find it already locked. The *unlock_page* routine clears the bit and wakes people waiting on it. Pages are locked before writepage is called and it's the primary job of *writepage* routine to complete the I/O and unlock the page. Unlocking a page, then, is the only way to make progress for paths like *generic_file_write*, *filemap_fdatasync*, and *vmtruncate* that have gotten stuck trying to do work on a locked page.

**28.5.2. Writeback Cache Design (Based on Linux 2.4).** The Linux 2.4 kernel provides some support to manage writeback state. The filesystem super block contain lists of dirty or locked inodes. The inodes have flags specifying if they are dirty or are locked when someone is performing I/O on them. The inodes contain lists of dirty, locked, and clean pages. Pages also contain state that

wbcache_24lin.eps not found!

FIGURE 28.5.1. Write path in the presence of writeback cache

marks them dirty (and up-to-date, which is different) or locked. A writeback cache implementation in Linux 2.4 makes use of all this infrastructure.

The *prepare_write* and *commit_write* functions perform the operations as decribed in the previous section. The page is then market dirty and the inode is also marked dirty. Earlier, the *writepage* routine simply tried to complete the I/O and unlock the page. Instead, now it will build a set of dirty pages, these can be batched and sent out in a single update to the server, this is shown in figure 28.5.1. This is called from sync paths via *filemap_fdatasync*, from *kupdate* at regular intervals, from *kswapd* under memory pressure based on VM LRUs, and from just about any blocking allocation context when memory allocations would fail. The application would receive a write completion signal as soon as the write is in the cache. In the present design, it is possible taht when the cache is flushed at a later time, it might be discovered that the OSTs do not have space for its data. The application might have already completed and taken decisions assuming the write completion, so it is important to prevent such situations. One way is to use some preallocation mechanism to decide how big the cache can be and how much data can be cached there without the risk of an ENOSPC error on cache flush. With this change in place, the code path might look as shown in fugure 28.5.3.

The introduction of this cache has also changed the lock revocation path. Earlier a lock revocation did not require any dirty pages to be written out to the disk since writes were always synchronous. But now since writes are asynchronous, writes have to written out to disk either when a lock revocation is called on that page or when pages have to be released under memory pressure. So the *ll_extent_lock_callback* function now calls *ll_pgcache_remove_extent* to complete the I/O to disk and release the lock.

**28.5.3. Writeback Cache Design (Based on Linux 2.5).** As described in the previous section, the *file write* operation in Lustre is started with the *ll_file_write* method. The *ll_prepare_write* method prepares the pages to be written; for example if there is a partial write to a page, it will first invoke *ll_brw* to read the old data for the page and then the synchronous *write* for the page is completed by sending the data for the page over the network to the target OST. This process is repeated for every page to be written, and is illustrated in figure 28.5.2. We will now describe the changes that are needed to implement the writeback cache on Linux 2.5.

The changes to *ll_prepare_write* and *ll_commit_write* would be similar to those described for Linux 2.4. Here again, preallocation or reservation of space for cache flush is important and can be done with a new function before *ll_commit_write* as shown in 28.5.3.

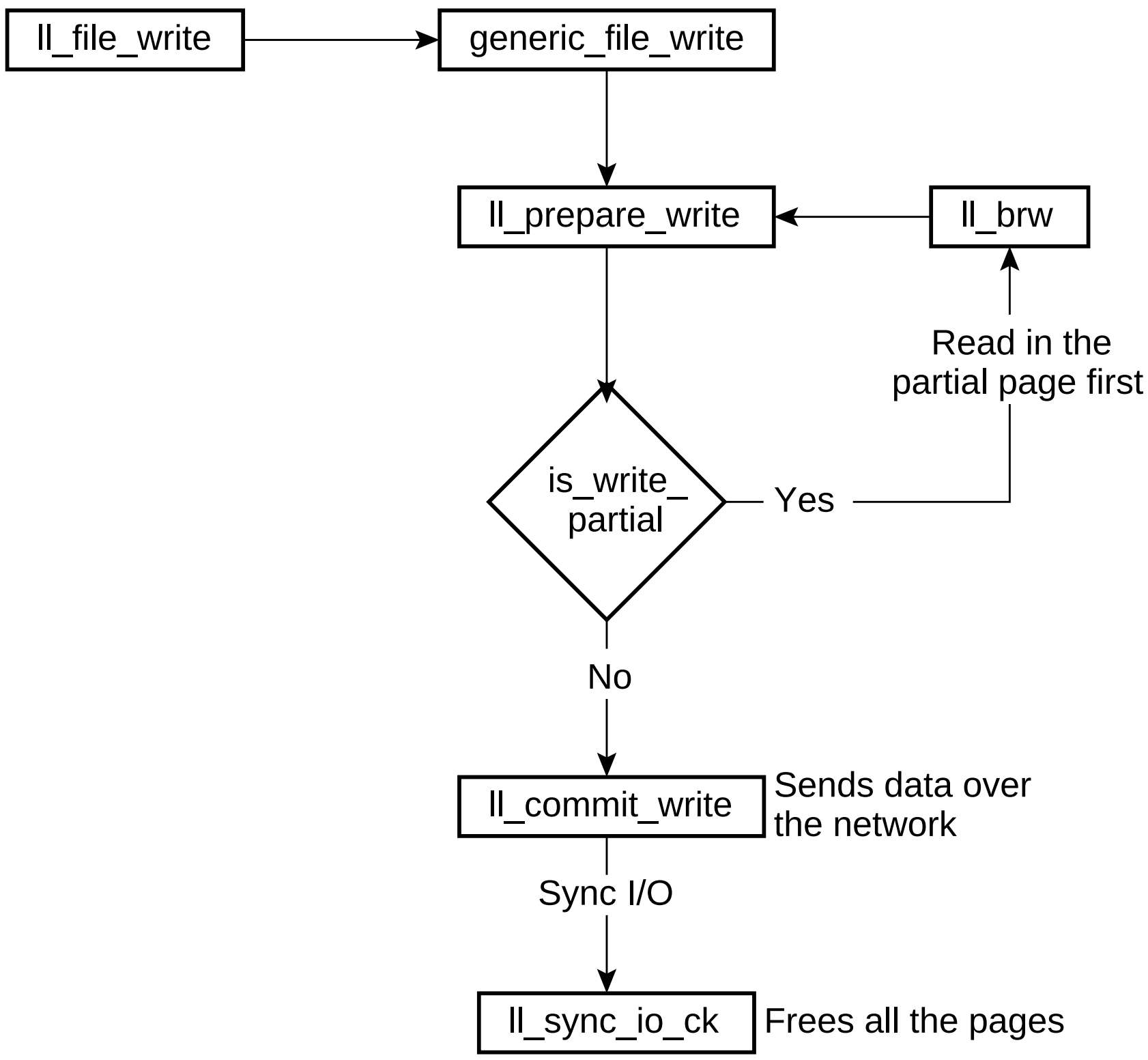

FIGURE 28.5.2. Synchronous File Writes in Lustre

We would also have to modify the *ll_commit_write* function to simply mark the pages dirty by hanging them off the correct list in the *address_space* structure for the file inode (the *dirty_pages* list) and then returning to the client to indicate a successful write. The new code path is shown in figure 28.5.3. The writes are first completed in the writeback cache, and flushed to OST at a later time either under memory pressure or when a lock revocation is done using *ll_lock_callback* function. It invalidates all the pages and, it will have to call *ll_writepages_internal* (or a similar a new function) and pass it the file extent that needs to be released. Only that extent will have to be written back and the corresponding pages invalidated.

The same *ll_writepages_internal* function would be called in the *pdflush* code path as well as in the *lock callback* path. The *pdflush* routine does not have to choose a set of pages, it can just pick a certain number of pages at random for writeback or just writeback all the pages. But in the case of *lock callback* routine, we need some way to indicate the file extent that needs to be written back. We can add the *ll_writepages_internal* as a helper function. In the *pdflush* path, *ll_writepages* would call *ll_writepages_internal* and specify the extent as the start and end of the file. On the other hand, *ll_lock_callback* would pass the exact file extent to be written out to *ll_writepages_internal* as illustrated in figure 28.5.4.

We will leave the actual writeback of dirty data to the *pdflush daemon* as described in the next section.

28.5.3.1. *Periodic Flush of Dirty Data.* The Linux 2.5 kernel provides a *pdflush daemon* which allows multiple worker threads for writing back dirty file system data periodically. Usually this is done when the system is experiencing memory pressure and the number of free pages drops below a certain threshold. This path is shown in figure 28.5.4.

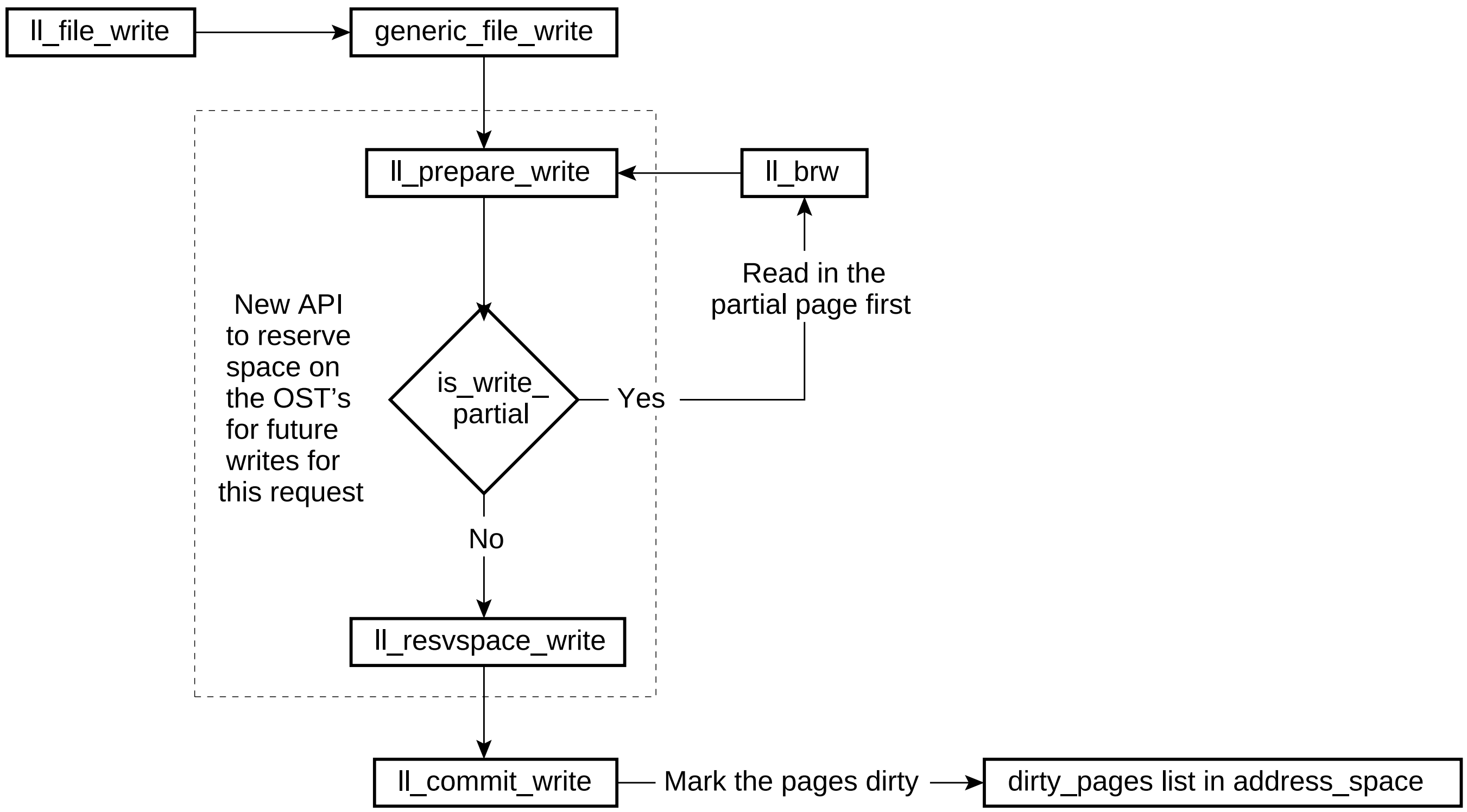

FIGURE 28.5.3. Writeback Cache Path

We need to have some way to pass to *ll_writepages* the actual extent of pages to be written back along with the number of pages that is already allowed in *write_back_control* structure. This would be required in the case of trying to revoke lock on a file extent.

## 28.6. System Call Implementation

In this section we will trace the path for several of the basic system calls. In all cases we discuss a cluster where a logical volume is used for striping.

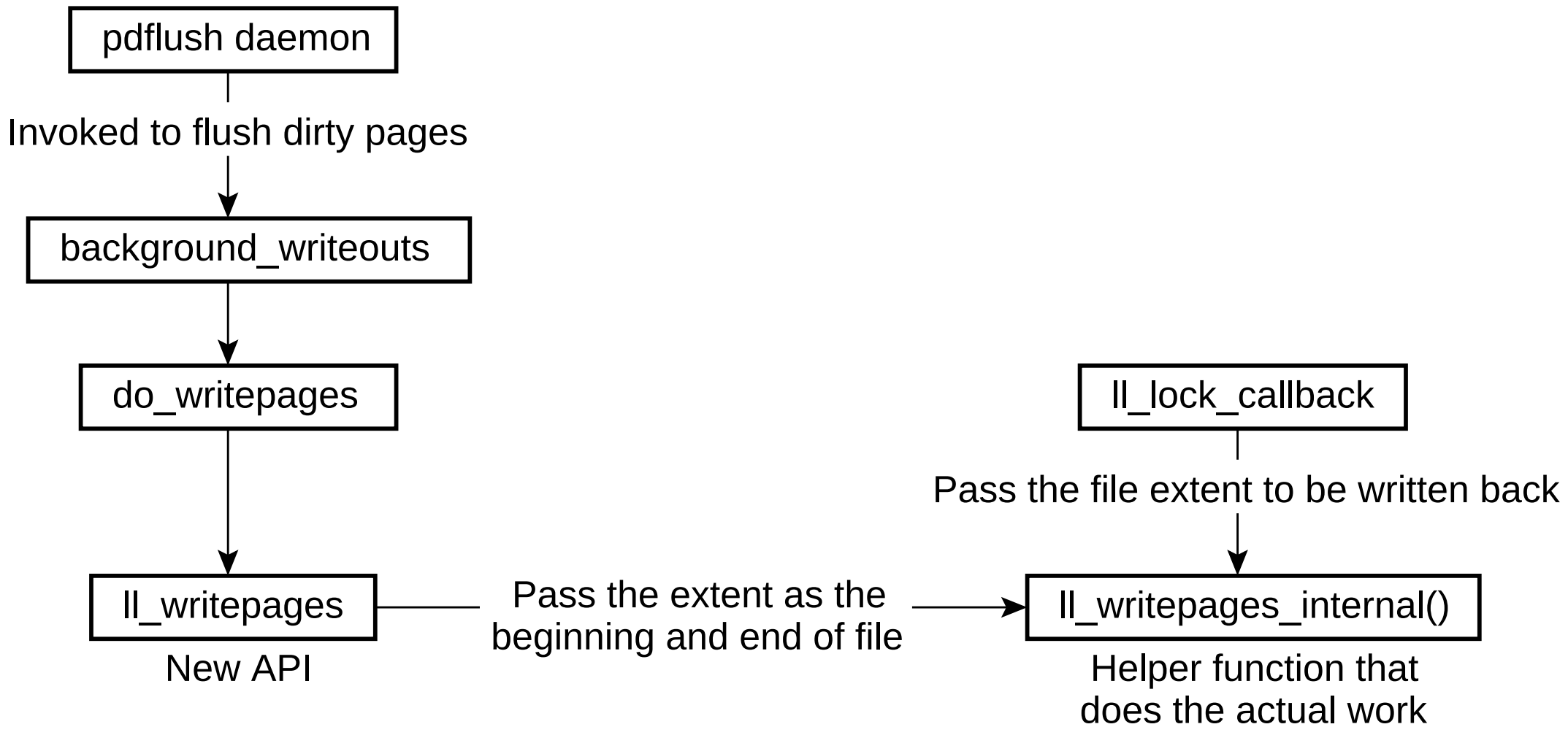

FIGURE 28.5.4. New API's for Lustre Writeback

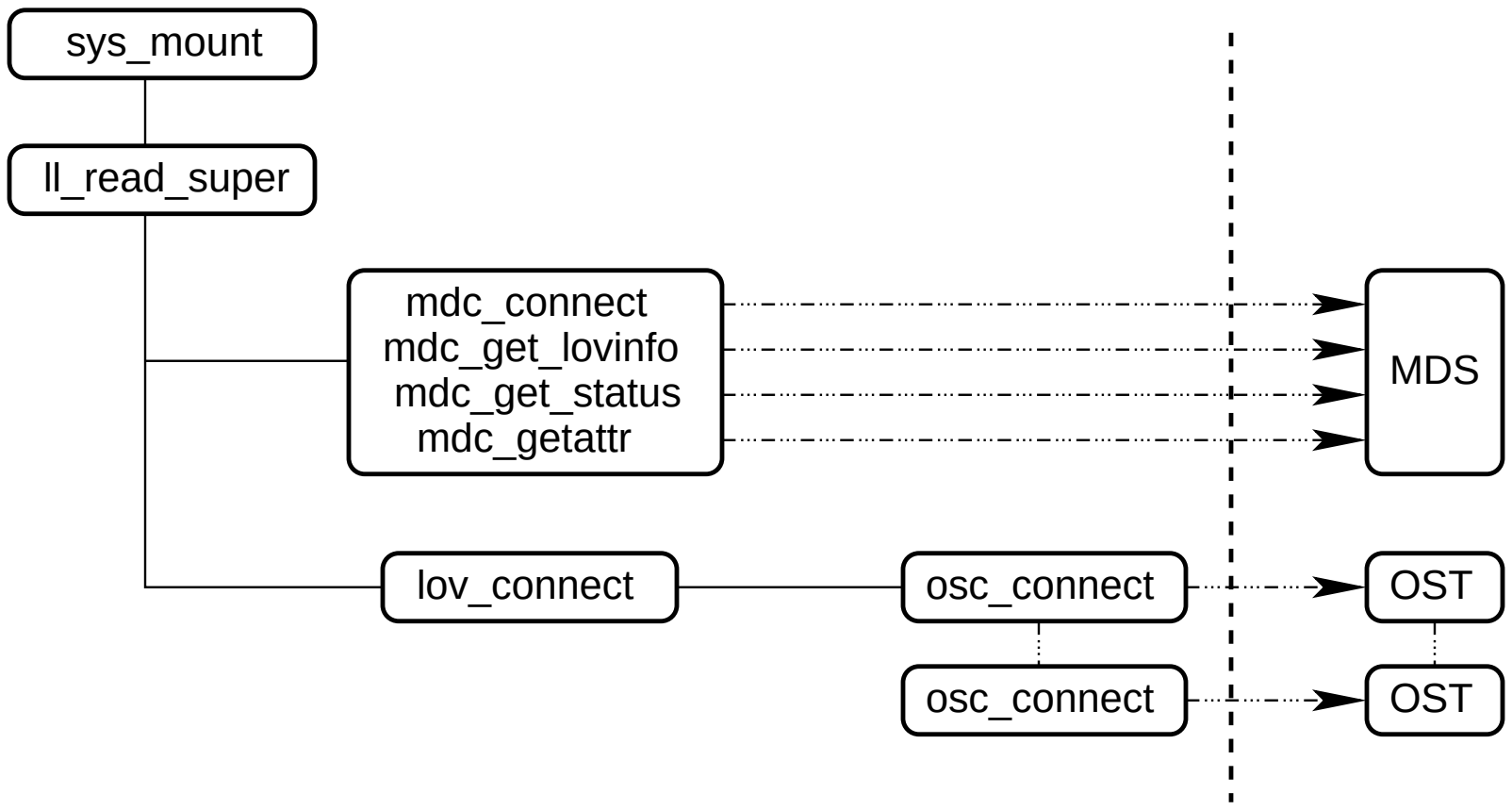

FIGURE 28.6.1. *sys_mount* Tree

**28.6.1. *sys_mount*.** The first call we discuss is the *sys_mount* call (figure 28.6.1), which is relatively un-involved.

The server side calls operate as follows.

***mds_connect*:** For each client the MDS maintains a persistent structure called the *mds_client_data*.

```
struct mds_client_data {
   uuid_t mcd_uuid[37]; /* Client UUID */
   uuid_t uuid_padding[3]; /* Unused */
   __u64  mcd_mount_count; /* MDS incarnation number */
   __u64  mcd_last_transno; /* MDS last completed trans no - for reply after recove
```

```
    __u64  mcd_last_xid; /* xid for the last transaction */
    __u32  mcd_last_result; /* Result from the last RPC */
    __u32  mcd_last_data; /* Per op data - disposition for open & c.*/
    __u8 padding [MDS_LR_SIZE - 74];
}
```

This structure is supplemented by volatile data, a *struct mds_export_data*, maintained for each connected client in the export structure, which consists of a list of open files. There is also a list of granted locks for a client which is held in the *struct ldlm_export_data*.

This call will search among the cached structures of *mds_client_data* if the client was recently connected. If so it will use that client data in the export data, otherwise a new structure is initialized and used.

***mds_get_lovinfo*:** Obtains the structure data of the LOV device which the client uses to stripe files over.

***mds_get_status*:** A general type of call which retrieves the root *fid* for the file set which is the MDS target.

***mds_getattr*:** Transfers the inode attributes of the root inode of the file set to the client.

***osc_connect*:** Operates almost identically to the *mds_connect* call.

**28.6.2. *sys_stat*.** In the *stat* system call in figure 28.6.2 we have called out the details of the lock interactions.

This call requires almost no error handling as it changes little state.

**28.6.3. *sys_write*.** This system call (figure 28.6.3) interacts heavily with the OST's for data movement.

**28.6.4. *sys_write* (with *O_DIRECT*).** This direct I/O situation (figure 28.6.4) is similar to that of the generic write system call, but cheaper since no reading of pages can happen.

**28.6.5. *sys_open*.** Next we deal with *sys_open* when called with the *O_CREAT* flag, to create a new file (figure 28.6.5).

The server handling in this function is typical of MDS service:

***ldlm_handle_enqueue*:** The main entry point. The client tries to get a lock on the directory containing the object it wishes to open. The type of lock is an *MDS_INTENT* lock and the lock manager calls the *mds_lock_policy* function. The lock policy decides that an intent *LOOKUP* or *CREATE* is not worth a lock and performs the *lookup* or *create* operations on the MDS.

Now the client can instantiate a valid inode for the object it wishes to open.

***lov_create* & *ost_create*:** If the object did not yet exist, *ll_file_open* will start by creating it. An LOV creation simply creates objects according to a striping pattern on multiple OST's. The object id's are returned to the client.

***lov_open* & *osc_open*:** These functions open the file. This is mostly done to protect the inode from unlinks (I/O to open unlinked files must go through).

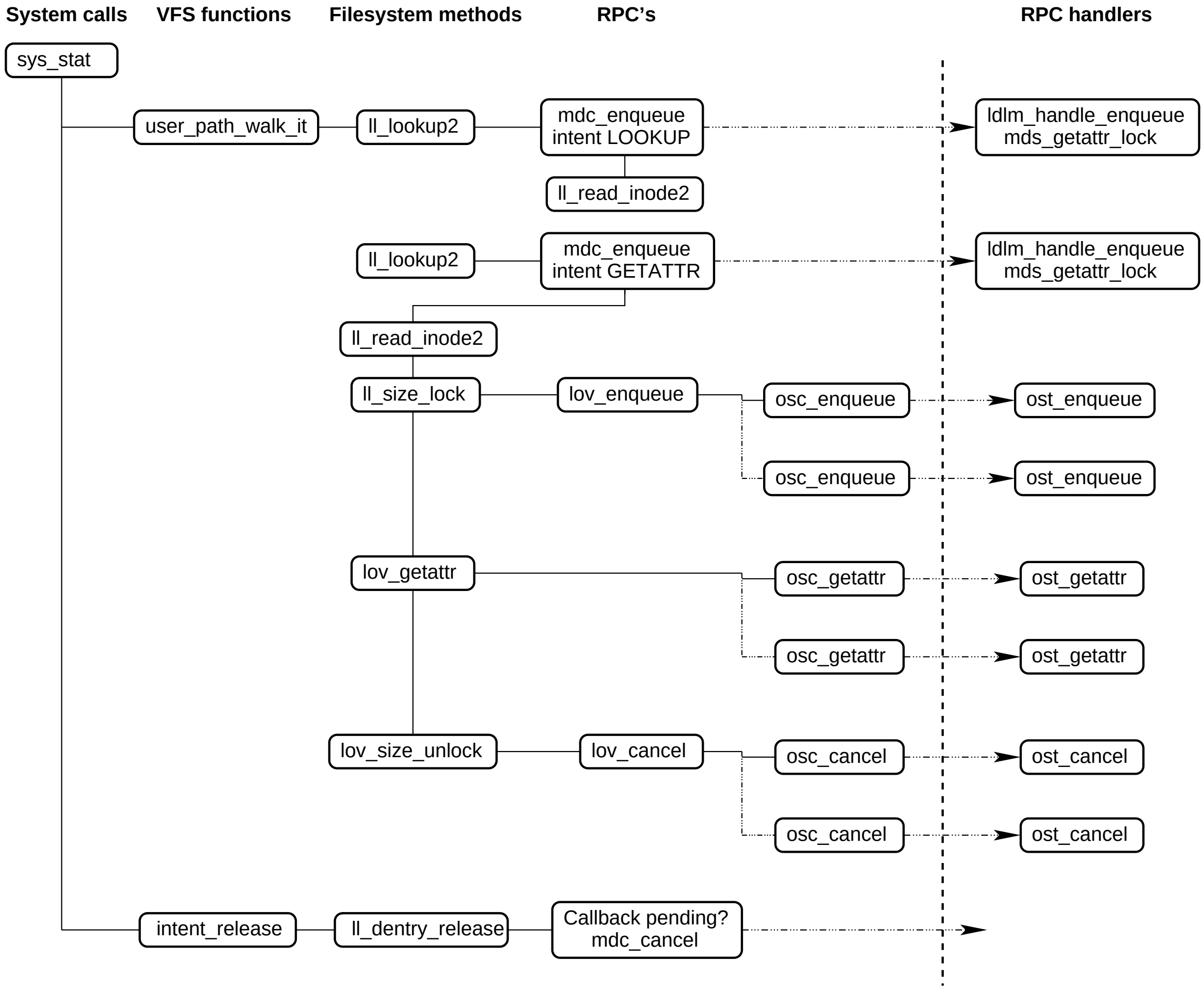

FIGURE 28.6.2. *sys_stat* Tree

***mdc_open***: Analogous to *osc_open* and keeps a file handle open on the MDS. Included in the open request are the object id's if the object was just created. The open call on the MDS stores the striping information (object id's, pattern, stripe size and stripe offset) as inode meta-data on the MDS.

**28.6.6. *mds_open*.** The *mds_open* call is schematically shown in figure 28.6.6. Notice that the call makes various efforts to retain and use per-client data stored in the export to be usable during the replay stage of file system recovery.

**28.6.7. *sys_unlink*.** The *unlink* system call is straightforward, as shown in figure 28.6.7. Noteworthy are that the MDS revokes all locks on the object it is removing (to avoid getting orphased lock resources) and that the removal of data objects is only done when the inode link count hits zero.

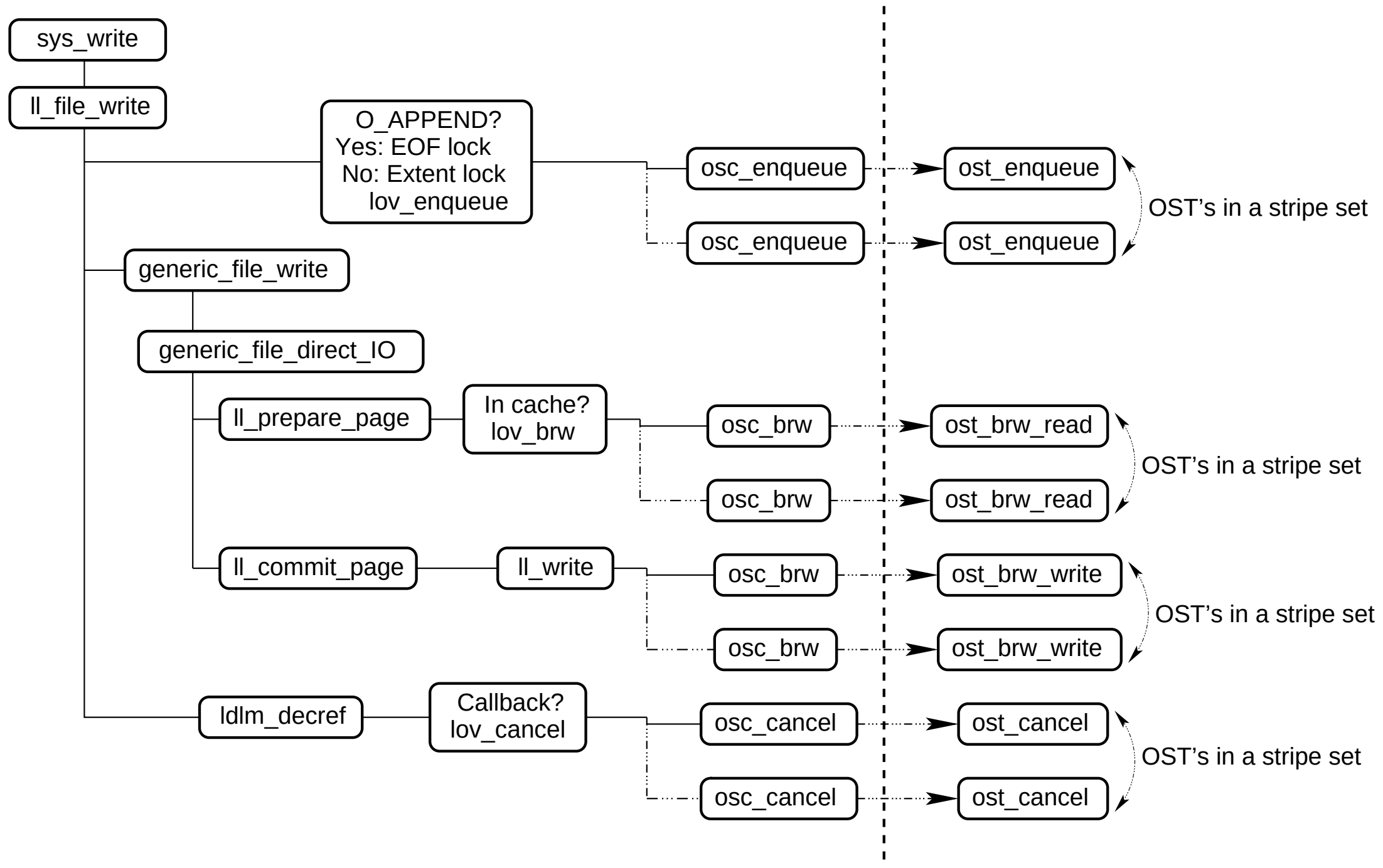

FIGURE 28.6.3. *sys_write* Tree

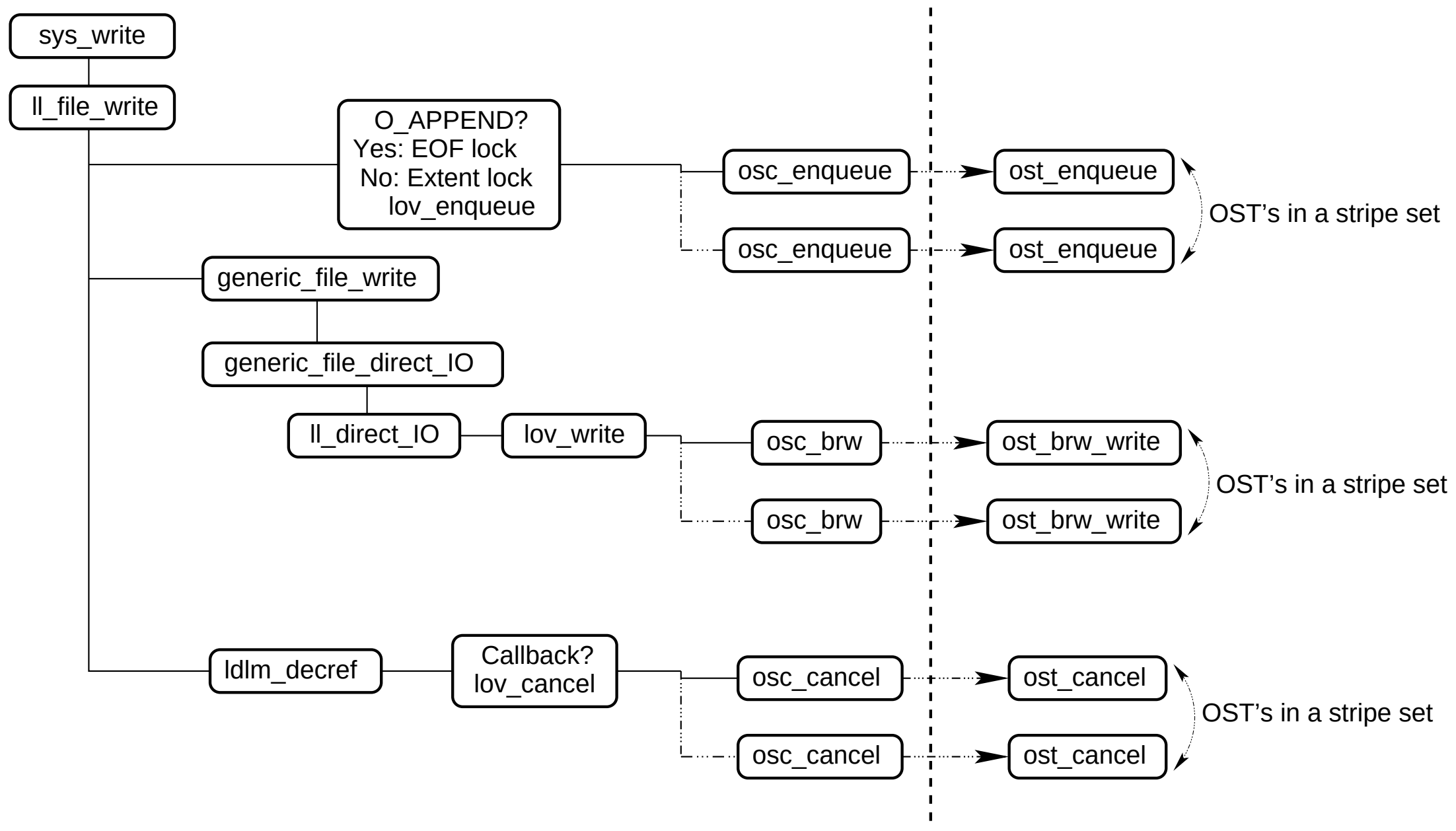

FIGURE 28.6.4. *sys_write O_Direct* Tree

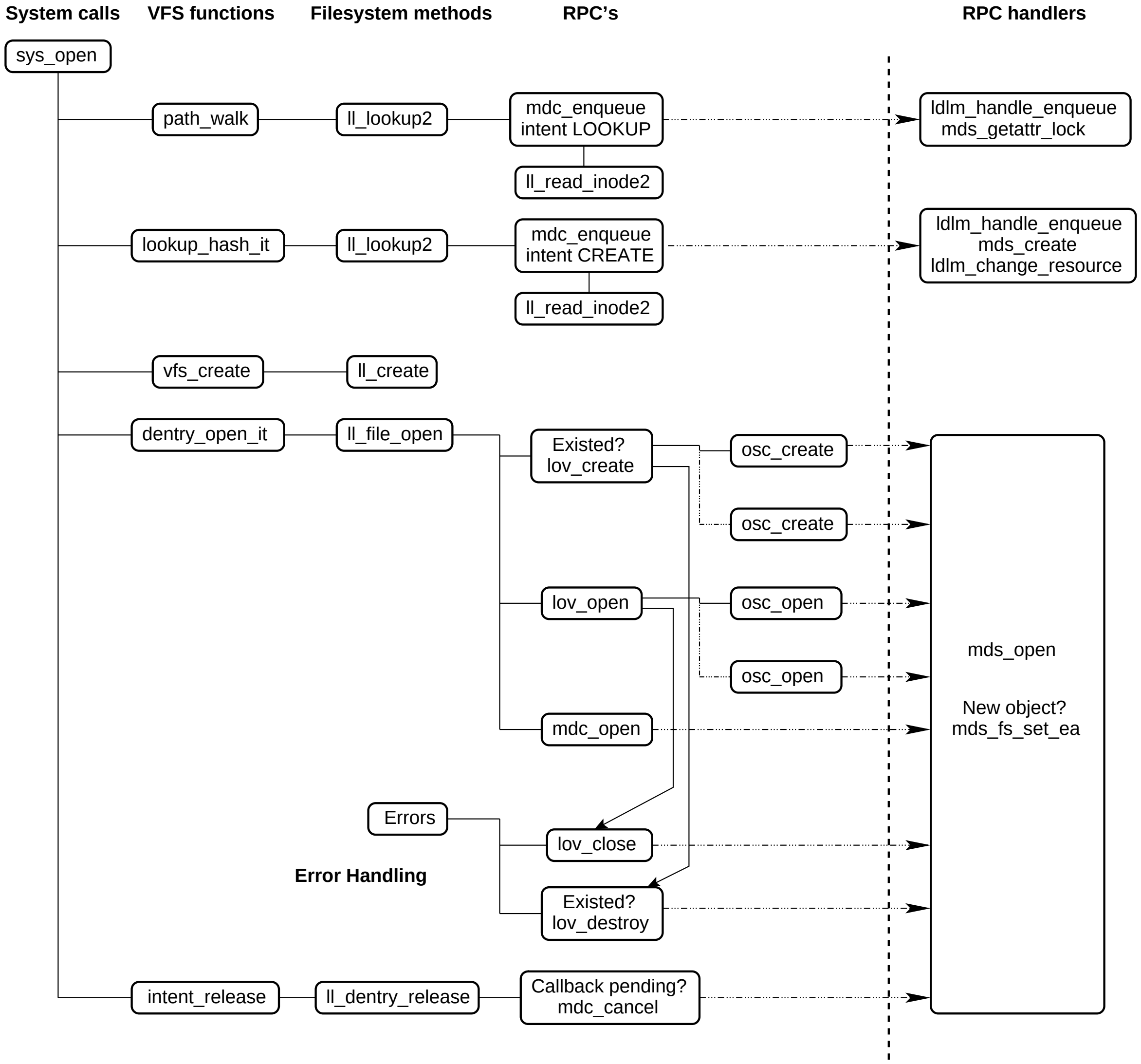

FIGURE 28.6.5. *sys_open* Tree

The client will update cached directory pages itself to avoid having to fetch them another time.

**28.6.8. *sys_mkdir*.** The *mkdir* system call (figure 28.6.8) is very similar to the unlink case discussed above.

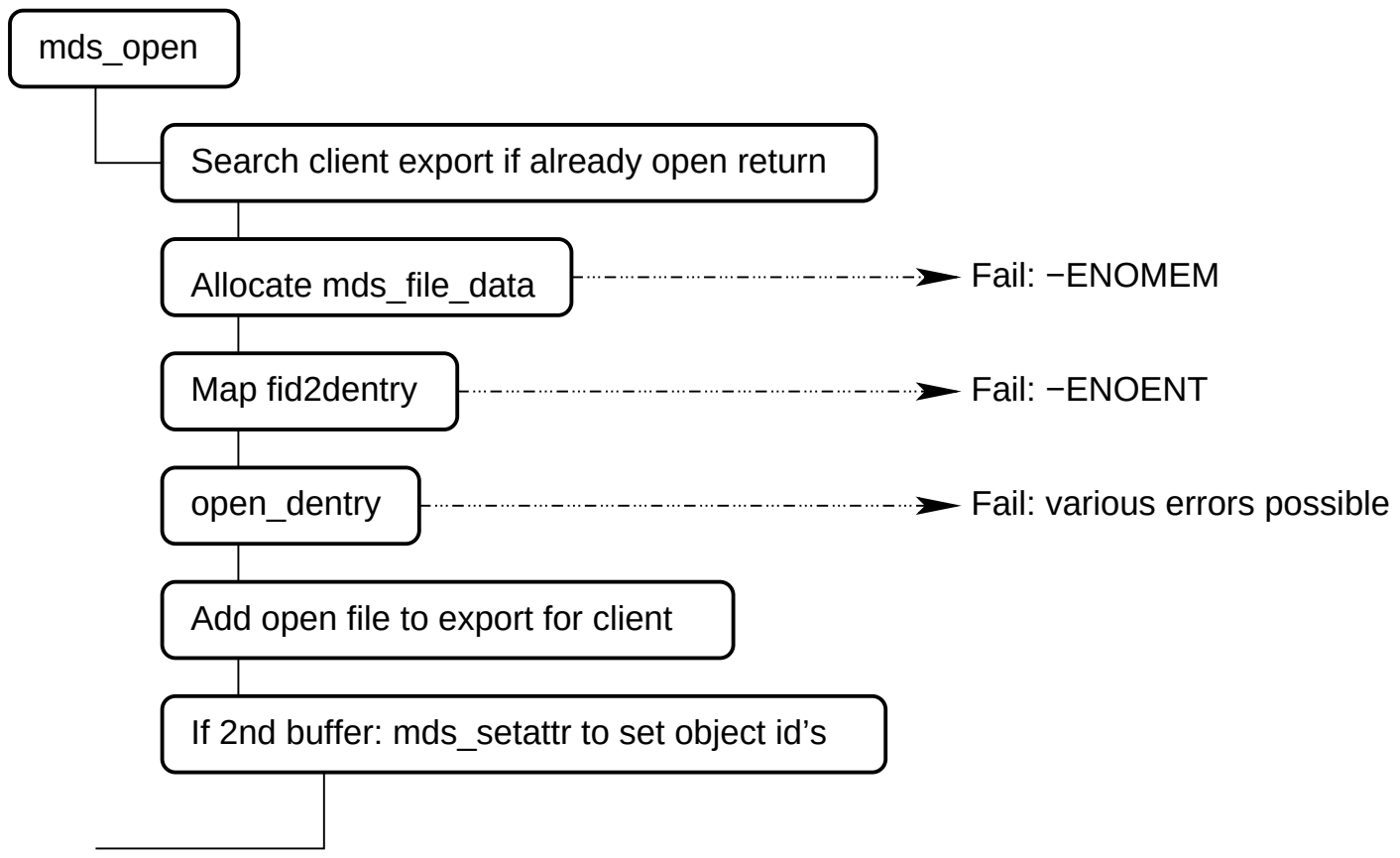

FIGURE 28.6.6. *mds_open* Tree

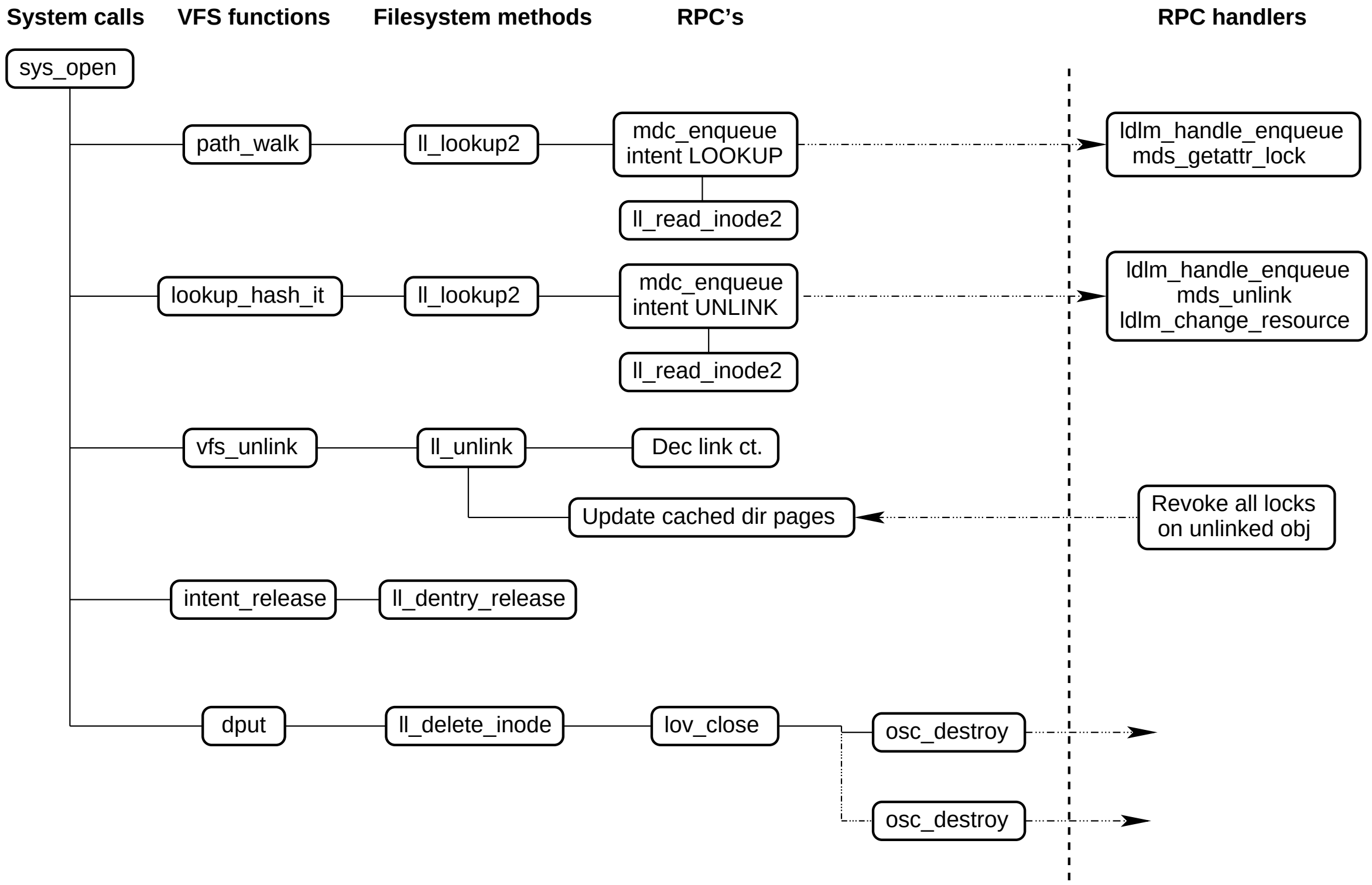

FIGURE 28.6.7. *sys_unlink* Tree

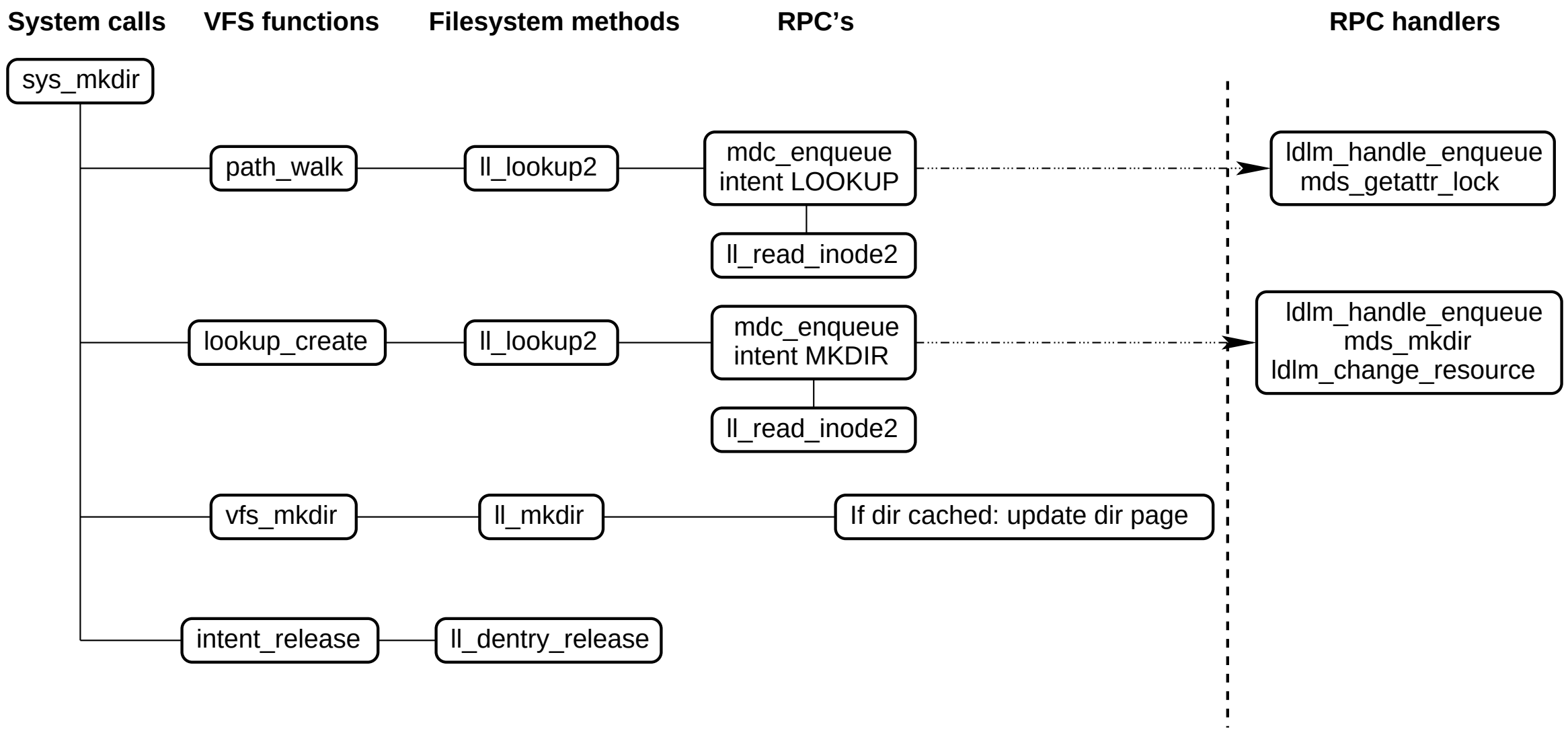

FIGURE 28.6.8. *sys_mkdir* Tree

## 28.7. Change history

(1) Version 1.0
(2) Version 1.1 Radhika Vullikanti - Reorganized the chapter, added section for writeback caching.
(3) Version 1.2 D. Innes - Edited and spellchecked Nov 15/02
(4) Version 1.5 Radhika Vullikanti - Removed the section on writeback metadata cache, this is already avalable in detail in the metadata chapter. Made some corrections and updates.

CHAPTER 29

# Recovery API's

This chapter gives an overview of the recovery mechanisms implemented in Lustre. We start with a description of the design of the recovery daemon in Lustre, we will describe the implementation for recovery protocols between the various subsystems in detail, and then we will describe some details about the recovery API's used by the system.In the Lustre filesystem, there are various subsystems that interact and the communication might be over a network. There could be various failure situation s - clients could fail, MDS could fail (in which case a transition has to occur to a failover MDS if available), OST s could fail (this might also require transition to new OST) and more importantly network failures could occur or network packets could be lost requiring replay/resending of requests/replies. In all these cases, recovery would be needed to ensure that the filesystem remains in a consistent state and data/meta-data is always made available to the clients. As described in details in the architecture section, Lustre relies on three variables for recovery. The first is an boot count; maintained by every storage controller, it is increased on every boot. Then there is an epoch number, a monotonously increasing number that changes everytime the cluster configuration changes. And finally, a unique generation number maintained for every connection from a client to other subsystems. The persistent state recovery in Lustre is very nicely taken care of by the underlying journaling file system. Lustre support both MDS and OST failover with the help of an LDAP server to store and retrieve information about replacement servers, this can work in conjunction with availability software like kimberlite or clumanager.

This chapter gives an overview of the recovery mechanisms implemented in Lustre. We start with a description of the recovery infrastructure in Lustre, we will describe the implementation for recovery protocols between the various subsystems in detail. We will also describe the failover mechanisms in scenarios where a failover/standby server is available. Finally, we will describe some details about the recovery API s used by the system

## 29.1. Lustre Recovery Management

**29.1.1. The Lustre Recovery Infrastructure.** Lustre currently relies on timeouts to determine if servers are up or not - for example is a client request times-out waiting for a reply to come back, the client will assume that either the server has failed or some transient network failure has occured. Moreover, even if the client has seen a reply, its possible that the serevr fails before the change is pushed into persistent state on the server, in either of these cases the client needs to be informed and asked to take appropriate action. Similarly, an MDS might fail before the creation of objects is recorded on persistent store resulting in orphan objects on the OSTs that need to be cleaned up. In this section, we will describe some of the infrastructure available and used in Lustre to help in recovery.

A client keeps a ptlrpc_connection for every server it has a connection to. It also uses the obd_import structure to maintain, track and monitor per-connection information and also to hang the per-request status information. This associates an imp_level with a connection to indicate the connection status, a level is also associated with every request to indicate when it was sent out, for example all requests generated when the connection is active would have a request level of LUSTRE_CONN_FULL, but the server might fail before the request is sent out, at this time the import associated with the connection to the failed server will be marked to have a import level of LUSTRE_CONN_RECOVD to indicate that recovery needs to be completed. Requests at a higher level than the import level can not be sent out, this allows us to do I/O fencing and hold back from sending requests when recovery is underway.

(1) LUSTRE_CONN_NEW - This is the request level for the request sent out to the server to setup a new connection.
(2) LUSTRE_CONN_CON - This is the import level when a new connection is requested.
(3) LUSTRE_CONN_NOTCONN - Import level when a client has been evicted
(4) LUSTRE_CONN_RECOVD - Level of an import associated when recovery is going on
(5) LUSTRE_CONN_FULL - Level of an import associated with an active connection

The obd_import structure has three lists of requests:

(1) **sending_list** - requests that have been sent once and are waiting for reply are hung off this list.
(2) **replay_list** - requests that have been sent but did not seen a reply come back or confirmation that the changes were committed on the server are tracked using this list. If a request incurs a timeout due to server failure or network problems and a recovery has already been started, this request will be deleted from the sending_list and moved into the replay_list. After recovery is completed, these requests need to be handled before allowing new requests (those on the delayed_list or completely new requests) to be processed.
(3) **delayed_list** - requests at an import level higher than the import level of the connection are held back on this list till recovery is completed.

Some of the other variables that for the basic framework of Lustre recovery framework are:

(1) *imp_replayable* : This field in the *obd_import* structure indicates if failover is configured for a specific MDS/OST. The variable is set if failover is supported.
(2) *imp_invalid* : This indicates the status of a server, it is set if the server fails and a failover/standy server does not exist.
(3) *trans_no* : Each server maintains transaction numbers to enable ordering of client requests when replay is needed.

The server uses the export structure to track the various clients connected to the server.

**29.1.2. Persistent State Recovery in Lustre.** Recovery of persistent state is another aspect of recovery in any filesystem. In Lustre, all persistent state in the current design is handled by journaling file systems with asynchronous write ahead logging. After a reboot we have the guarantee that disk state has been recovered by the journaling infrastructure. The only recovery that might be needed is to maintain consistent between the OST and MDS persistent state information to make sure there are no orphan objects and no reference in inode to non-existent objects. This is termed as orphan recovery and will be discussed in a later section.

## 29.2. Client Recovery

**29.2.1. Basic Recovery Scheme.** A Lustre client interacts with both the MDS and OST's. It might lose contact for several reasons. A fatal event on the client might cause it to be rebooted; this can be considered as a normal start for the client system. However, it is also possible that the network failed or that the meta-data nodes failed, due to hardware or software failures and through administrative events. Any of these events could trigger recovery process in the clients. The client recovery scheme is shown in figure 29.2.1.

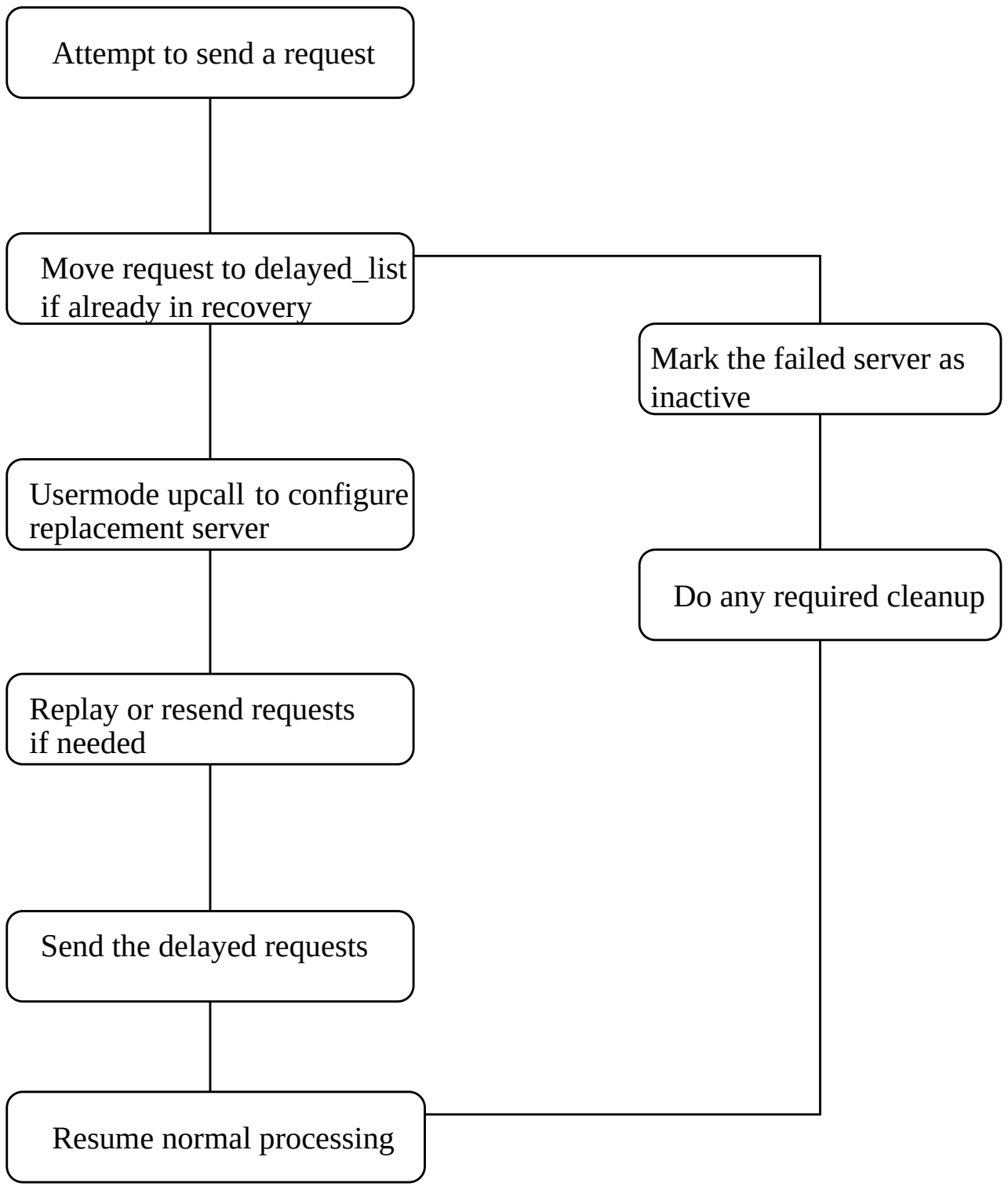

FIGURE 29.2.1. Client Recovery Overview

In the case of MDS/OST failures, a client will lose replies for transactions that were processed asynchronously at the server. In such a situation, the client will first have to determine a new MDS/OST address using an upcall to a user level program. Usually, this information is stored in the LDAP server. The client will then attempt to reconnect using the new address; it has to suspend new I/O request processing during this recovery phase. If the client connects to a new MDS/OST, it will examine all the retained requests and replay or resend if required.

After a client reconnects and the server has indicated that replay is required, the client enters the **REPLAY** state (see ImportStates) and starts sending its uncomitted requests. If it has no uncommitted requests, it sends a **LAST_REPLAY** message and waits for replay to finish. Transactions with a *transno* <= *last_committed* (which must be an OPEN) are replayed by the server as soon as they are recieved, in any order. Transactions with *transno* > *last_commited* are queued and processed strictly in order.

If the client times out during replay, it will attempt to reconnect, and if unable to reconnect, **REPLAY** will fail and the client returns to **DISCON** state. It is possible that clients will timeout frequently during REPLAY, so reconnection should not delay an already slow process more than necessary. We can mitagate this by increasing the timeout during replay. We may also need a special case for resending the *LAST_REPLAY* message if that times out.

It is extremely important that requests that should be re-played are properly separated from requests that should be resent. For that reason it is important to insert a barrier in the recovery process where the client stops interpreting replies. Because unregistering a buffer can be a blocking operation (the buffer may be busy receiving data at a very low rate), a special unlinking api is provided to assist with this, which is called when pending requests are timed out.

This whole process is illustrated in figure 29.2.1. Once the recovery process is complete, normal processing will be resumed.

In figures 29.2.2 we have traced the recovery path in the Lustre source showing the various functions that would be executed for clients in case of an OSC failure.

**29.2.2. Open request recovery handling.** In order to ensure that applications are insulated from the effects of failover, we must ensure that any open filehandles are restored during the failover process. In addition, opens and closes must be carefully replayed to ensure that the semantics of open-unlink and O_CREAT are preserved, as well as keeping a one-to-one balance of open and close requests for a given filehandle. This section describes only the MDC/MDS version of open-replay.

29.2.2.1. *Open Requests.* Open requests are always issued transaction numbers, so they will be retained (at least) until the transaction stream has been committed to that point. This preserves the open in the sequence of reintegrated operations, so that open-unlink and O_CREAT effects are always well-ordered.

A reference is also kept by the file descriptor (*fd->fd_mds_och.och_req*), to ensure that the open request is replayed appropriately to restore/update the file handle (see mdc_replay_open, called via rq_replay_cb in ptlrpc_replay_req). The transaction number ensures that the correct ordering is preserved with respect to a possibly *committed* open operation and subsequent unlinks. Until the file descriptor's reference is dropped (which will be performed by the close-commit callback) the *rq_replay* flag is set, so that the open request is kept on the replay list even after the peer's last-committed value would indicate that it could be released.

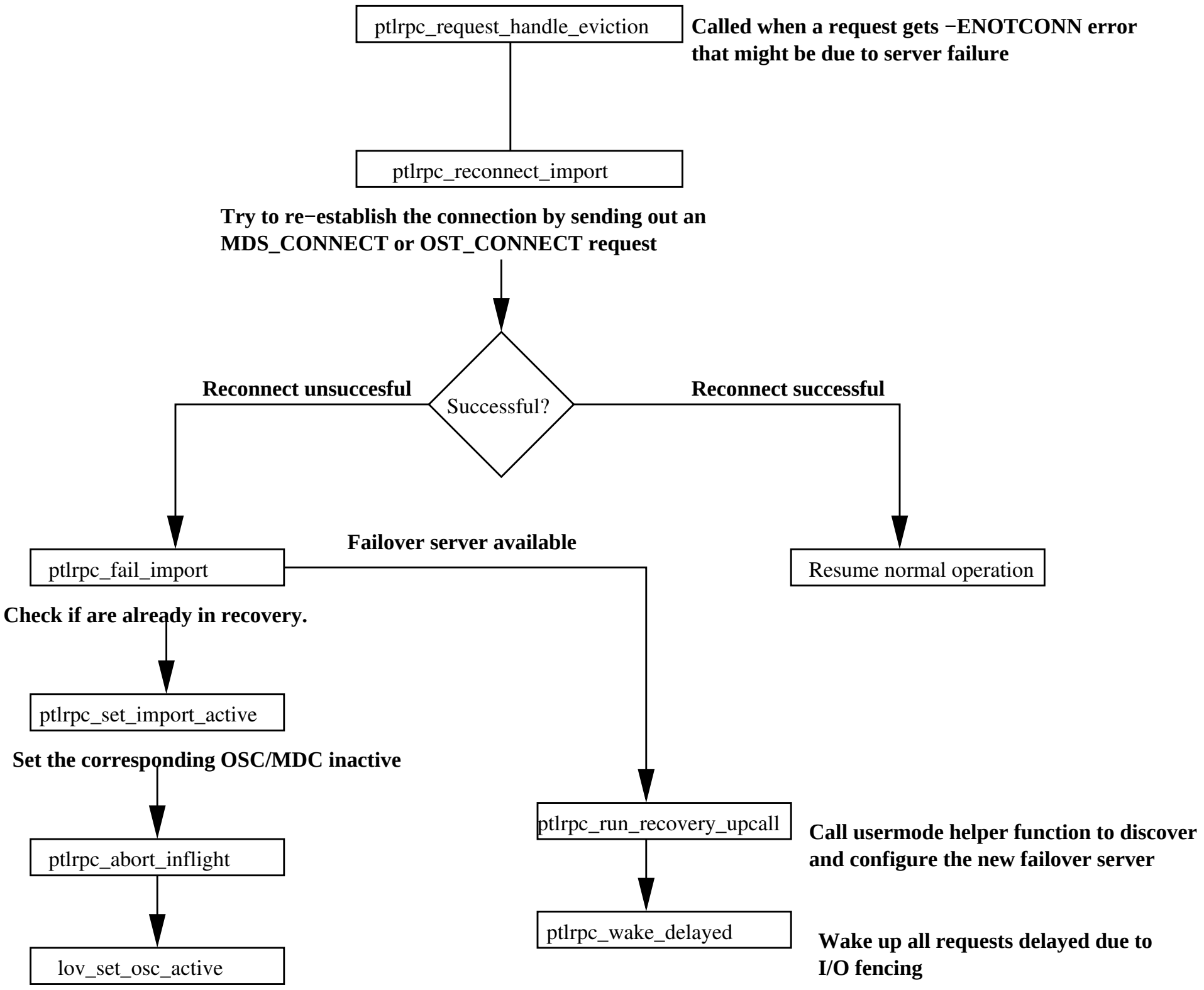

FIGURE 29.2.2. Tracing the Client Recovery Path

29.2.2.2. *Close requests.* Close requests are also always issued transaction numbers, chiefly so that the effects of open-unlink-close are preserved across failover.

When a close request has been executed, the file descriptor's open-reference is transferred to the close request, and a commit handler is installed (mdc_commit_close, called via rq_commit_cb from ptlrpc_free_committed). The commit handler removes the *rq_replay* flag and drops the close request's reference, so that the request is freed by the next ptlrpc_free_committed pass.

(It's perhaps useful to think of the file descriptor's reference and close request's reference as being two phases of the same "open file handle" reference. In the first phase, we have to store the pointer to open_req in the file descriptor, because we haven't created any close request yet. In the second phase, the client's file descriptor has been destroyed, so the pointer is moved to the close request's *rq_cb_data*.)

## 29.3. Server Recovery

In Lustre, an MDS or OST failure can be termed as server failure. A reliable Lustre operation requires that a peer node is configured to take over the services that were being performed by the failed node. This relies on a shared storage behind the server node and the corresponding failover node.

In MDS, we support active-standby failover, if the active MDS fails, a standby MDS takes over its operations till the active MDS comes back up. On the other hand, on OSTs we support an active-active failover. A peer active OST can be configured to take over the operations of another OST in case that OST fails. In both cases, if the original server comes back up, load balancing will take place or in case of MDS, the original server will take over.

**29.3.1. Basic Server Recovery Scheme.** A server here could be an MDS or an OST. A server will initiate recovery in two cases:

(1) A timeout on a bulk transfer, expected lock cancellation, or lock revocation.
(2) A reconnect from a client has been discovered.

The target/server handles a reconnect request from clients as shown in figure 29.3.1. The server will check if it is in recovery, if it is, a flag in the reply to client will be set to inform the client that the server is in recovery. If the client is the first one to try to reconnect after a server failure and recovery, the server will start a timer to expose a recovery window during which all other clients can also try to reconnect. Clients that fail to reconnect during this period will be evicted from the cluster. If the client is in recovery, it will deny connect requests from new clients. The server will first determine if the client is new; in this case a new export will be initiated for the client. If the server finds an export corresponding to the client, it knows that the client is trying to reconnect.

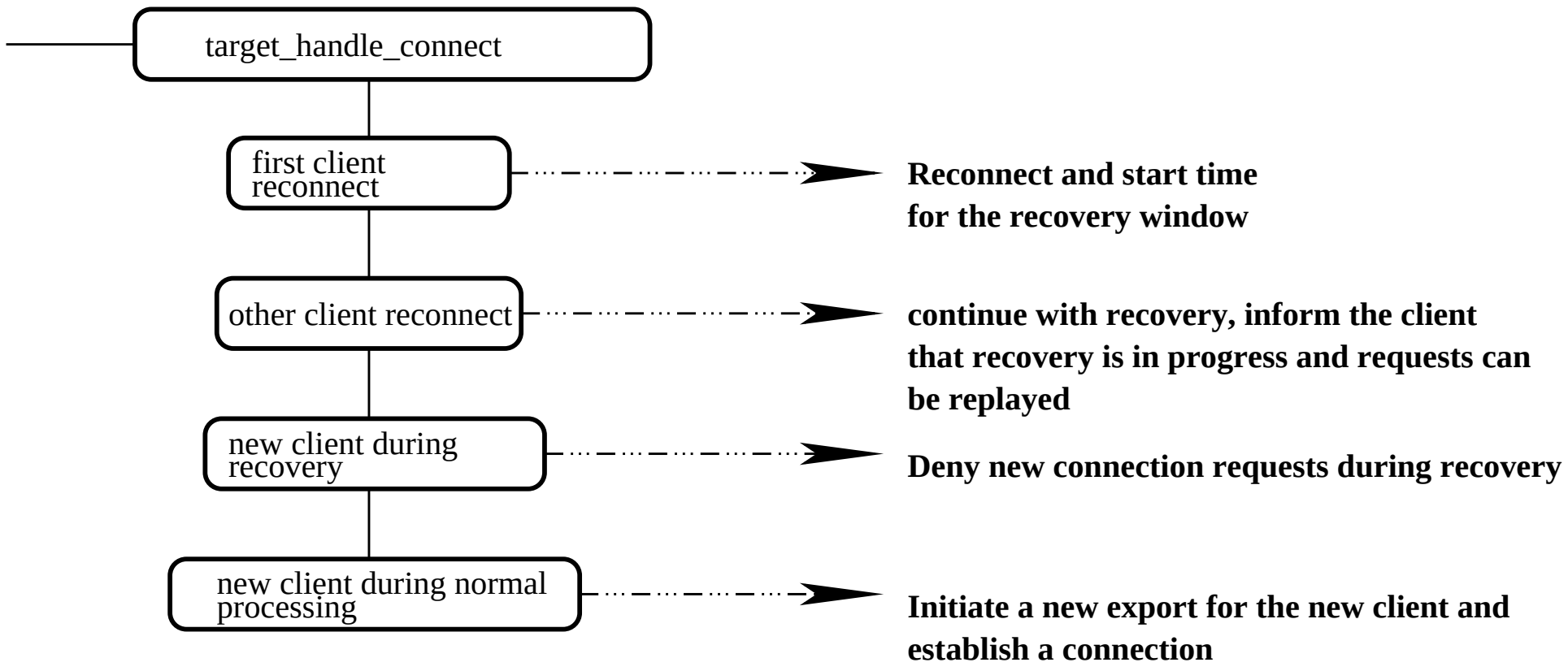

FIGURE 29.3.1. Server Reconnect Handler

**29.3.2. Replay mode on the MDS server.** When the server recovers or a failover server starts up, it determines the number of clients that need to reconnect and recover, this is indicated by the *export->nbr_recovering_clients* variable. The server starts a timer with an extended time-out period to allow for enough time for all the clients to recover. If timeout occurs, the server will destroy the remaining shadow exports to prevent the clients from reconnecting, and reduce *nbr_recovering_clients* which will allow replay to begin. At the end of replay, however, the server will evict all the clients (even if no transaction number gaps were found).

The details of recovery on the client during this phase is described in section 29.2.1.

The server will not start processing queued transactions until after all the clients have connected and either queued a transaction for replay, or sent a **LAST_REPLAY** message which removes the client from replay processing and reduces *nbr_recovering_clients*. Processing can begin when *nbr_recovering_clients* == *nbr_replay_requests_in_queue*.

Requests are replayed in the order of the transaction number. If a gap is found in the transaction numbers during replay, then recovery stops immediately and all the clients are evicted. It is possible that the client with the missing transno might be busy responding to a reply and hasn't sent it's next request yet, so the better design would be to allow the server to wait for *obd_timeout* before failing recovery. The server will have to recreate the state that was exposed to the clients, these are requests that completed successfully, but not committed to the disk before server failure. The server does not use any lock verification, it just trusts the clients to replay the state correctly. It is also important for the server to produce identical fids (inode number + generation) for creating objects, this is required to avoid mismatches for file handles that are already open on the creating clients.

The server ends the *replay phase* when it has received a **LAST_REPLAY** message from every client. If all the clients that were originally expected reconnected, then the server sends a successful reply to all the clients. Otherwise, all clients are evicted, and an error reply is sent. The reply to this *LAST_REPLAY message* is held back until all clients have completed replay and the sent a *MDS_GETSTATUS* request with the *LAST_REPLAY* set. This ensures that the clients don't send non-recovery requests before recovery is complete.

## 29.4. OST/MDS Persistent State Recovery

**29.4.1. Introduction.** The OST and MDS manage related persistent state. The meta-data server holds inodes for files which contain information about objects that are located on different OST's and constitute the file. It is essential that these two remain consistent with each other at any time. If an OST or MDS fail and recover, either the inode or the objects or both may have disappeared because they were not written to persistent storage.

Fundamentally the problem here is one where one system (the OST) creates persistent information, namely the object id's, and this needs to be replayed to another system (the MDS) which stores the object id's in the inodes. This is a variant of filesystem replication, the recovery of which was extensively studied in the InterMezzo project. In this section we will explore the following failure scenarios :

(1) File Creation: Data objects can be orphaned due to MDS failure - File creation in Lustre is done in two phases, creation of the objects on the OST's and then informing the MDS of all these new objects. It is possible that the MDS fails before the extended attributes (this is where the object information is kept within a file inode) are written down to the disk or the client fails to properly report the objects created to the MDS. So, this would leave us with objects with no corresponding information in the inodes, hence called orphaned objects.
(2) File Removal: Data objects can be orphaned due to client or OST failure - When a file is removed/unlinked, we need to delete the objects from the OST's as well as their information in the inodes. Objects might become orphaned if the extended attributes were modified but the OST/client failed before the objects could be removed as well.
(3) Object Lost, Inode Present: The object will be re-created if found missing during client RPC's and the MDS information will be updated

In this section, we will describe the infrastructure and algorithm to recover and repair persistent state inconsistencies.

**29.4.2. Logging infrastructure.** The persistent state recovery is implemented using logging. The MDS will keep logs for unlink operations and the OST will create logs for create operations. Log entries are variable length entries with a length stored at the begining and end of each log entry to facilitate efficient forward and backward walking of the logs. In normal operation, log entries are generated sequentially and cancelled without searching. In recovery operations, the log will be walked sequentially and cancellation cookies will be generated for each handled log entry, and then destroyed when the log is empty as normal (this allows several different users of the log to share in the recovery without the danger of removing the recovery log prematurely, at the expense of not having opaque log cookies).

In order to efficiently track recovery log entries, a 128-bit log cookie is generated in-memory for each logging event and passed over the network. This will be the 64-bit objid of the logfile a 32-bit log index and a 32-bit offset into the log file. These cookies are passed from the logging target to the complementary target, and then returned only to the logging target for log entry cancellation (i.e. for creates the OST does the logging, the cookies are passed to the MDS, and then returned back to the originating OST). They are not inspected anywhere other than at the logging host under normal operation.

We propose to create many small log objects (up to 64kB in size) as usage demands, to avoid the case of large log objects being kept when a small number of transactions therein were not committed. Having larger log objects means more wasted space in that case, and more (useless) log data transferred/walked at recovery time.

Two new methods and one RPC is added for MDS/OST recovery, namely the obd_log_add() and obd_log_cancel() calls, and the OST_LOG_CANCEL RPC. Calling obd_log_add() writes an entry to a local log object (on disk) and returns a log cookie for that entry. Calling obd_log_cancel() takes the number of cookies, and an arrray of log cookies as parameters, and batches cookies into a page in RAM until the page is full or it is called with a "flush log" flag, at which point it sends the cookies to the logging target for cancellation. At the logging target, the cancel cookies result in clearing of an in-use flag in the log bitmap header and destroying the log object if it no longer has in-use log entries.

**29.4.3. Create Logging.** The create sequence is as follows, the client performs:

(1) *mds_open*
(2) *obd_create* (typically an LOV call) - This would create objects on one or more OSTs, and generates a LOV EA record to store on the MDS.
(3) *mds_setattr* - Send the LOV EA record to the MDS, this would be stored in the extended attribute of the inode.

Each OST will need to write a record (transactionally, or first) with the creation of the object. The record would contain:

(1) the MDS inode number/generation (fid)
(2) the OST object id, which will be unique (currently it is not)

The logging of this information on each OST generates an OST log cookie that is returned to the client, combined into a LOV array if applicable (like the LOV EA but in a separate lustre message of the setattr RPC), and passed on to the MDS with the setattr call.

If the log is not written transactionally (but was committed to disk, while the object create was not), then this does not affect OST crash handling. In the case of a simultaneous MDS and client crash, then either the setattr committed (which is no different than if we had logged the create transactionally - the MDS references a non-existent object), or it did not commit and the MDS will get an -ENOENT during the object destroy step, which can be ignored.

29.4.3.1. *Normal operation.* During a normal operation (that is absence of any failures), the following steps are followed:

(1) The MDS keeps the LOV cookie array as opaque data in-memory until the setattr that is saving the LOV EA data has its commit callback called (telling us that the LOV EA data is now safely on disk).
(2) The MDS calls the *obd_log_cancel* method, which passes the log cookie array along with the LOV EA.
(3) If applicable, the LOV *log_cancel* code will split up the LOV cookie array into per-OST cookies (based on the LOV EA data) and call the *obd_log_cancel* call.
(4) Each OSC will keep a page of log cookies to be cancelled until the page is full, at which time it will send these cookies to the OST with a new *ost_log_cancel* RPC.
(5) The OST will take each cookie and cancel the corresponding log entry by marking the clearing that index in the log bitmap, and when the log has no more in-use indicies the log object is destroyed.

29.4.3.2. *OST Crash.* If the OST crashes, there exists the possibility that objects were not created on disk, and that log cancellation cookies were lost from the MDS.

(1) If the creating client is still running [failover OSCs only –phil], client/OST RPC replay is done. This will replay the OST create if it was not committed.
(2) If the creating client is no longer running, the MDS will refer to a non-existent object in the LOV EA.
(3) If the object was not created permanently, then it will be created lazily for this file the next time that this stripe is opened. Any data written to that stripe would have been lost if the client didn't do a sync on the file, if the OST was not doing synchronous writes for OST failover (this is normal even for local filesystems).

29.4.3.3. *MDS Crash.* If the MDS crashes, there exists the possibility that the setattr was not committed to disk. Also, unsent OST log cookies for which a commit happened will have been lost.

(1) If the creating client is still running, client/MDS RPC replay is done. This will re-send the MDS setattr if it was not committed.
(2) At some suitable time after client/MDS recovery is complete, the MDS first does an OST_SYNC RPC (to ensure all created objects are permanent) and tells the OST to start new logs (via a special OST_LOG_CANCEL RPC).
(3) The MDS does an *obd_get_info*(*recovery_log*, *last_log_objid*) RPC to get the next OST log objid.
(4) For each OST log object and performing a bulk read of that objid to get the log contents.
(5) For each uncommitted record in the log, check if the MDS fid exists and if the LOV EA contains that objid. If not, perform an OST destroy object RPC (ignoring -ENOENT if returned). In all cases, do an *obd_log_cancel()* for that record. This will clean up any old OST log records for which the cancellation was lost.
(6) It asks the OST to sync the destroys and remove the each orphan log when complete, and does another *get_info*(*recovery_log*) RPC to get the next log objid and continues OST/MDS recovery processing until no more recovery logs are available on any OSTs.

Alternately (via iterators and preserving layering/LOV EA opaqueness, after the *ost_sync_rpc*, and *ost_log_cancel* RPC with flags to start new logs steps above):

(1) The MDS calls *obd_iterate()* with a flag/value indicating it wants to do OST recovery log processing/iteration, passing *mds_remote_recovery_callback()*.
(2) If applicable, the *lov_iterate()* calls *obd_iterate()* on each connection passing *lov_remote_recovery_callback()* along with the OST recovery log processing flag/value.
(3) The *osc_iterate()* sees the log recovery flag/value and calls *osc_remote_recovery_iterator()* which does an *obd_get_info*(*recovery_log*, *last_log_objid*) RPC to get the log objid, and does a bulk read of the log object data.
(4) For each in-use log entry, the OSC calls the passed iterator callback (passing in the OSC connection and log data).

(5) If applicable, the *lov_remote_recovery_callback()* calls *mds_ost_recovery_callback()* with the OSC connection and log data.
(6) *mds_remote_recovery_callback()* checks if the MDS fid exists and if so returns the LOV EA data to the caller.
(7) If applicable, *lov_remote_recovery_callback()* extracts the LOV EA data for this OSC and returns it to the caller.
(8) *osc_remote_recovery_iterator()* examines the LOV EA to see if the logged objid exists therein, and if not it calls the *obd_destroy()* passing in the returned LOV EA data.

29.4.3.4. *Client Crash.* If the client crashes after objects were created, but before the MDS setattr was sent, then there exists the possibility of orphaned objects on the OSTs. Since the client does the MDS setattr call before any writes are done to an OST, no data that was sent to the OST can be lost due to client-only failure. Also, no other clients could have had access to this stripe data until after the setattr had completed, so there is no chance that other clients had references to these orphaned objects.

(1) The MDS will delete the orphan objects in the same fashion as in MDS Crash the next time the MDS performs recovery.

It would also be desirable to do this log recovery more often than when the MDS crashes. It could be possible to do this when the MDS detects a client timeout (or some fixed time thereafter like in full MDS recovery processing, to batch multi-client disconnection handling after any flood of client disconnects has passed).

(1) The MDS syncs the local filesystem to force commit all setattr operations (triggering OST cookie cleanup as normal) and then flushes any unsent log cancel messages to the OSTs. The latter would be an extension to the obd_log_cancel() API/RPC which flushes the local log cookies to the OST and causes the OST to start logging to a new log object (instead of having the mere act of requesting a log object cause a new log object to be started, which would loop indefinitely in this case).
(2) The MDS performs the same MDS Crash steps to clean up any OST orphans caused by the loss of one or more clients via timeout. This will also clean up any old OST log entries so that full MDS recovery processing would be sped up because it doesn't have to look at old recovery records.

**29.4.4. Unlink Logging.** When a client unlinks a file the sequence of calls is:

(1) mds_unlink - returns LOV EA data to the client if this is the last link for this file
(2) ost_destroy

During unlink the MDS will (transactionally) write a log record of the deletion, once for each stripe in the file. There is a separate log for each OST, and the record contains:

(1) OST object id

The record can be removed when the OST destroy operation commits. The logging of this information on the MDS generates a MDS log cookies (one for each stripe in the file) that are returned to the client, split into constituent cookies by the LOV driver (if applicable) and then a unique MDS log cookie is passed on to the each OST with the destroy calls.

29.4.4.1. *Normal Operation.* When the OST has committed the unlink of that objid, it calls the log_cancel() method, which batches up MDS cookies until a page has been filled, and then calls the OST_LOG_CANCEL RPC to send these cookies to the MDS for cleanup. The MDS processes the log cookies and clears the entries in its log object by clearing the in-use bit in the log header. When that log has no more uncommitted entries that object is destroyed.

29.4.4.2. *OST Crash.* If the OST crashed, it is possible that objects will not have been destroyed, or that log cancellations for committed unlinks are lost.

(1) If the client is still available , client/OST replay will redo the OST destroy (if uncommitted), which will resend the log_cancel cookie from the OST to the MDS.
(2) If the client is unavailable, the OST needs to do a *get_info(recovery_log)* RPC to the MDS to get its list of objects for which unlinks have been committed.
(3) For each objid in the log, perform a local destroy call, and send log_cancel cookies when they are committed or if -ENOENT is returned.

29.4.4.3. *MDS Crash.* If the MDS crashed, it is possible that the MDS unlink was reverted and the OST destroys succeeded. This will result in holes in the file (if one or more OSTs or the client crashed), or an empty file on the OST (if only the MDS crashed).

(1) The MDS executes *osc_destroy* for each record in the per-OST recovery logs for previous boot cycles, which will either return -ENOENT (MDS does a local log cancellation), or an unlink (and the OST will send a log cancellation cookie when the unlink has committed).

Alternately, with iterators (to preserve LOV stacking):

(1) The MDS executes *obd_iterate()* with a flag for MDS recovery log processing, passing *mds_local_recovery_callback()*.
(2) If applicable, *lov_iterate()* calls *osc_iterate()* for each OSC with *mds_local_recovery_callback()*.
(3) The OSC reads its recovery log and calls the passed callback with the OSC connection and LOV EA data as parameters.
(4) mds_local_recovery_callback() calls obd_destroy() with the passed connection and LOV EA.

29.4.4.4. *Client Crash.* If the client crashed during unlink, it is possible that it did not issue destroy RPCs to the OSTs.

(1) If the OST is still running, it flushes its local log cancel cookies to the MDS and tells it to start a new log.
(2) It waits for some period of time to allow clients with in-progress unlinks to finish operations.
(3) It proceeds with OST Crash recovery for the client-unavailable case.

(4) If more clients crash while recovery is pending

**29.4.5. Client crash on create/unlink.** A client crash in the middle of a create/unlink opration could also leave behind orphans. As an example, the client could die right after the objects are created on the OSTs, but before the mds_setattr which records the existence of these new objects on the MDS. Similarly, during an unlink operation, a client could complete an mds_unlink, deleting the inode, but might fail before ost_destroy is called to delete the objects. Both of these scenarios can be taken care by the recovery protocol between MDS and OSTs during the course of normal operation as described in the previous sections.

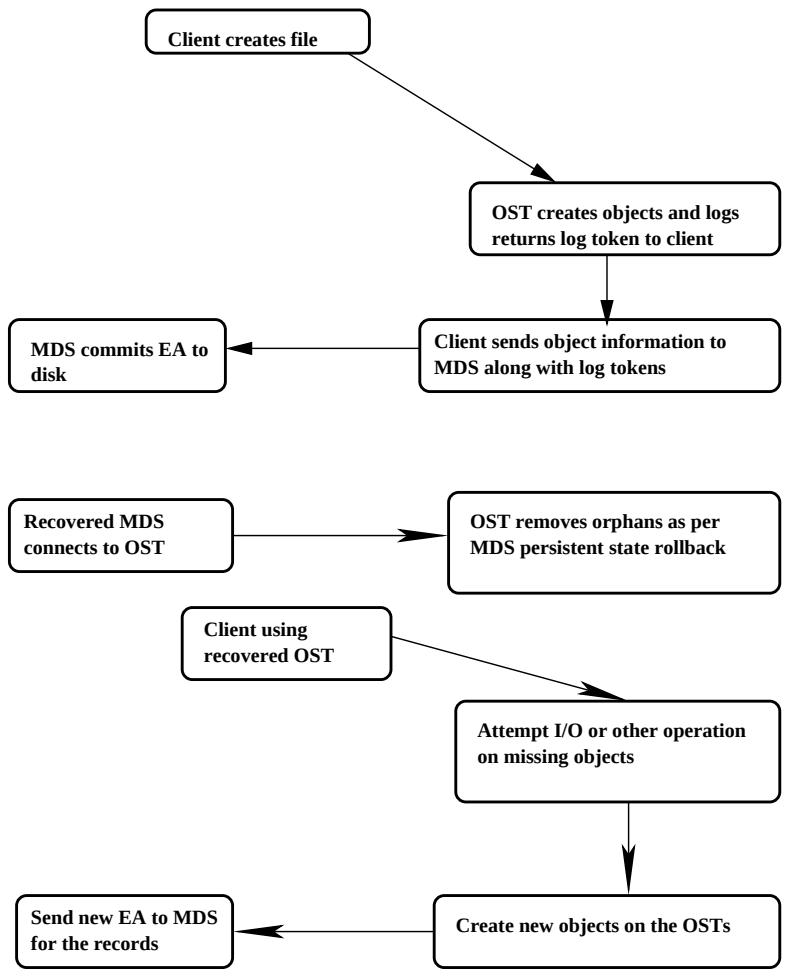

FIGURE 29.4.1. MDS/OST Recovery

## 29.5. Gateway Node Failover

Lustre has router or gateway nodes that can route Portals packets between different networks. These gateway nodes have a failover or redundancy model that we describe here.

(1) The portals router spreads messages over all gateways that can reach a particular destination. There is a new command...

lctl –net <nal> set_gw <nid> {up|down}

... to enables and disables individual gateways. When a gateway is disabled, the portals router avoids it and notifies the relevent NAL so it can clear any connections it may have with the gateway. When it is re-enabled, the portals router resumes using it.

Note that the portals routing tables are not affected (i.e. no topology information is changed), just which entries in the routing table will be considered for any particular routing decision.

Also note that redundant notifications are harmless.

(2) When a NAL detects a peer has died, it notifies the router. If the dead peer is a gateway, the router notifies the world via an upcall.

This upcall should be used to notify all nodes about the failed gateway using the lctl command described in (1). Any nodes, which haven't already detected the death of the gateway for themselves will now avoid it in future.

Note it uses the standard portals upcall (set via /proc/sys/portals/upcall and '/usr/lib/lustre/portals_upcall' by default) as follows....

/usr/lib/lustre/portals_upcall ROUTER_NOTIFY <nal> <nid> {up|down}

...where <nal> is a decimal number and <nid> is the NID of the gateway in hex with a leading 0x.

(3) When the dead gateway reboots and reconfigures, it should notify all nodes that it is back online using the lctl command described in (1). All notified nodes will now be able to use the gateway again.

(4) NAL peer death detection is invariably accompanied by the loss of one or more messages, so lustre will experience a transmit failure or a timeout when a gateway dies.

The notification described in (2) above will heal the network after a gateway dies, but lustre will still have to take its own recovery action when it sees an error caused by gateway failure.

Lustre's first attempt at recovery should be a simple retry, on the assumption that the network failed momentarily, and healed itself.

## 29.6. Recovery APIs

In this section we will give an overview of the API s provided in the recovery daemon. The callback functions are subsystem specific as mentioned before, so those are not discussed here.

### 29.6.1. ptlrpc_fail_import.

29.6.1.1. *Prototype.*

```
void ptlrpc_ fail_import(struct obd_import *imp, int generation)
```

29.6.1.2. *Parameters.*

**input: imp:** The import associated with the connection to the target.
**input: generation:** generation number in the lock request that caused a timeout and triggered recovery.

29.6.1.3. *Return Values.*

29.6.1.4. *Description.* This function is called when a lock request made by the client times out. Here, we first verify if the client is already in recovery by checking the value of imp_level in the obd_import structure for the connection that is used for the request that triggered recovery. If the import is already at an imp_level of LUSTRE_CONN_RECOVD, it implies that we are already in recovery. The next step is to compare the generation number in the import with the generation number passed as an input to this function, this is the generation number for the request that triggered recovery. If the generation number in the import is greater than that in the request, it means that the service is newer and the recovery is already complete, otherwise set the imp_level to LUSTRE_CONN_RECOVD to indicate that we are heading into recovery. This would cause the usermode recovery upcall to be invoked to pass the information required for recovery to proceed.

### 29.6.2. ptlrpc_recover_import.

29.6.2.1. *Prototype.*

```
int ptlrpc_ recover_import(struct obd_import *imp, char *new_uuid)
```

29.6.2.2. *Parameters.*

**input: imp:** The import associated with the connection involved in recovery.
**input: new_uuid:** UUID of the new server in case of a failover. In case of a recovery and reconnect to the same target, this value is not needed

29.6.2.3. *Return Values.* Upon successful completion ptlrpc_recover_import will return 0, otherwise one of the following error codes will be returned: EINVAL: If a ptlrpc_connection structure does not exist for the connection to the target node

29.6.2.4. *Description.* This function is called when the OBD_IOC_CLIENT_RECOVER ioctl is invoked to pass on the information of the new target ( only required in case of a failover) and to initialize the connection to this target. It uses the ptlrpc_get_connection to verify if a ptlrpc_connection structure exists for this connection in the active or unused connection list. If it doesn t, one is allocated and initialized. The connection in obd_omport and DLMexports is switched to refer to the new connection. The next step is to re-connect to the target node, this is followed by the required steps needed to complete recovery before entering normal request processing. For example, if the client has some requests that were sent, saw a reply but were not committed to persistent store, the client will have to replay it. Siimilarly, the client might have to resend requests in certain cases. Once the recovery is complete, the connection level is marked to be LUSTRE_CONN_FULL and normal request processing is started.

### 29.6.3. ptlrpc_reconnect_import.

29.6.3.1. *Prototype.*

```
int ptlrpc_ reconnect_import(struct obd_import *imp, ptlrpc_ req **reqptr)
```

29.6.3.2. *Parameters.*

**input: imp:** The import associated with the connection involved in recovery.
**output: reqptr:** Pointer to a ptlrpc_request structure that is sent to the target to initiate a new connection and used to return the reply to the calling function

29.6.3.3. *Return Values.* Upon successful completion ptlrpc_reconnect_import will return 0, otherwise one of the following error codes will be returned: ENOMEM: Not enough memory to allocate for a new ptlrpc_request

29.6.3.4. *Description.* This function increments the generation number in the obd_import to indicate that a new connection is being initiated, it then sends out a request to the server to establish a new connection. The reply to this request is returned to the calling function. It passes the imp_export handle, which is the client s DLM callback service handle, to the server so that the server knows where to send the ASTs.

### 29.6.4. ptlrpc_resend.

29.6.4.1. *Prototype.*

```
int ptlrpc_resend(struct obd_import *imp)
```

29.6.4.2. *Parameters.*

**input: imp:** The import associated with the connection involved in recovery.

29.6.4.3. *Return Values.* Upon successful completion ptlrpc_resend will return 0, otherwise it will return an error code depending on the ptlrpc_replay_req function.

29.6.4.4. *Description.* This function walks through the list of requests in the imp_sending_list and resend all the requests.

### 29.6.5. ptlrpc_run_recovery_upcall.

29.6.5.1. *Prototype.*

```
void ptlrpc_ run_recovery_upcall(struct obd_import *imp)
```

29.6.5.2. *Parameters.*

**input: imp:** The import associated with the connection involved in recovery.

29.6.5.3. *Return Values.*

29.6.5.4. *Description.* This function is used to invoke the usermode helper/upcall, the path and name of the upcall script should be setup as a part of the initial cluster configuration. The required setup for the new target is done by the upcall, for example in the failover scenario, the old OSC has to be cleaned up and a new one setup for the new target server.

### 29.6.6. ptlrpc_replay.

29.6.6.1. *Prototype.* int ptlrpc_replay(struct obd_import *imp)

29.6.6.2. *Parameters.*

**input: imp:** The import associated with the connection involved in recovery.

29.6.6.3. *Return Values.* Upon successful completion ptlrpc_replay will return 0, otherwise it will return the error codes from ptlrpc_replay_req.

29.6.6.4. *Description.* This function is invoked after the recovery has been initiated, it is called to cleanup the committed requests and replay the requests that were not committed before the server failed. The committed requests, where the request transaction number is less than or equal to the committed transaction number (maintained in the obd_import), are cleaned up by calling ptlrpc_free_committed. All the requests that are sent but not commited are left hanging off the replay_list in obd_import structure for the connection. The ptlrpc_replay_req is then invoked for every request that needs to be replayed.

### 29.6.7. ptlrpc_replay_req.

29.6.7.1. *Prototype.*

```
int ptlrpc_replay_req(struct ptlrpc_ request *req)
```

29.6.7.2. *Parameters.*

**input: imp:** The import associated with the connection involved in recovery.

29.6.7.3. *Return Values.* Upon successful completion ptlrpc_replay_req will return 0, otherwise one of the following error codes will be returned: EINTR: received an unknown (unneccessary) interrupt -EPROTO: Unpack reply failed

29.6.7.4. *Description.* This function essentially resends the requests that need to be replayed.

### 29.6.8. target_handle_connect.

29.6.8.1. *Prototype.*

```
int target_handle_connect(struct ptlrpc_ request *req, svc_handler_t handler)
```

29.6.8.2. *Parameters.*

**input: req:** The request for initiating a new connection to the OST/MDS. input: handler service handler for the server(ost_handle for OST, mds_handle for MDS)

29.6.8.3. *Return Values.* The function returns 0 when successful, otherwise it returns one of the following errors: EINVAL: a bad target or client UUID have been passed in the request ENODEV: there is no device with the specified target UUID 29.6.8.4. *Description.* The function first checks if there is an existing export for the client that issued the connect request. If it does, then call target_handle_reconnect to reconnect. If the device is marked as starting recovery and if this client is the first one, start the recovery timer. If an export is not available for the client, and the device has completed the recovery cycle, a new export will be set up for the client. On the other hand, the device will not accept connect requests from new clients during recovery.

### 29.6.9. target_handle_reconnect.

29.6.9.1. *Prototype.* int target_handle_reconnect(struct lustre_ handle *conn, struct obd_export *exp, struct obd_uuid *cluuid)

29.6.9.2. *Parameters.* input: conn The connection handle passed by the client input: exp Export for the client input: cluuid UUID of the client that asked for reconnect

29.6.9.3. *Return Values.* The function returns 0 when successful, otherwise it returns one of the following errors: EALREADY: the handle passed by the client matches the remote_handle the export knows of -EALREADY: the handle passed by the client does not match the export s remote_ handle

29.6.9.4. *Description.* The function checks if the connection handle specified by the client is already known to the export, if it is then probably the client is trying to re-connect after a partition. If the two do not match, then disconnect the client by clearing the connection. If the export does not have any information, it passes the cookie from export handle to the client.

### 29.6.10. target_start_recovery_timer.

29.6.10.1. *Prototype.*

```
void target_ start_recovery_timer(struct obd_device *obd, svc_handler_t handler)
```

29.6.10.2. *Parameters.*

**input: obd_device:** the device that has been marked for recovery input: handler service handler for the device

29.6.10.3. *Return Values.*

29.6.10.4. *Description.* This function is called if a device is ready to start recovery, its called only for the first client that tries to connect/reconnect. A recovery timer is set, any clients that do not reconnect within this period will be evicted from the cluster when the timer expires.

### 29.6.11. target_handle_disconnect.

29.6.11.1. *Prototype.*

```
int target_handle_disconnect(struct ptlrpc_ request *req)
```

29.6.11.2. *Parameters.*

**input: :** req the request from client for disconnect

29.6.11.3. *Return Values.* Returns 0 if successful, otherwise it will return: -ENOMEM: no memory available to allocated a lustre message structure

29.6.11.4. *Description.* This function is called to disconnect a target.

### 29.6.12. target_abort_recovery.

29.6.12.1. *Prototype.*

```
void target_abort_recovery(void *data)
```

29.6.12.2. *Parameters.*

**input: data:** device that is aborting recovery

29.6.12.3. *Return Values.*

29.6.12.4. *Description.* As soon as the recovery timer expires, this function is called to stop the recovery process, drop all the delayed replies and return an ENOTCONN error to the clients for whom the reply was intended for. All the clients that did not reconnect during the recovery window, they are evicted from the cluster.

### 29.6.13. target_queue_recovery_request.

29.6.13.1. *Prototype.*

```
int target_queue_recovery_request(struct ptlrpc_ request *req, struct obd_device *obd)
```

29.6.13.2. *Parameters.*

**input: req:** request
**input: obd:** device where the requests should be queued up

29.6.13.3. *Return Values.*

29.6.13.4. *Description.* Hang the request off of the obd_recovery_queue for the device.

### 29.6.14. target_queue_final_reply.

29.6.14.1. *Prototype.*

```
int target_queue_final_reply(struct ptlrpc_ request *req, int rc)
```

29.6.14.2. *Parameters.*

**input: req:** request input: rc return code

29.6.14.3. *Return Values.*

29.6.14.4. *Description.*

### 29.6.15. target_send_reply.

29.6.15.1. *Prototype.*

```
void target_send_reply(struct ptlrpc_ request *req, int rc, int fail_id)
```

29.6.15.2. *Parameters.*

**input: req:** request input: rc return code input: fail_id

29.6.15.3. *Return Values.*

29.6.15.4. *Description.*

### 29.6.16. target_handle_ping.

### 29.6.17. Prototype.

```
int target_handle_ping(struct ptlrpc_ request *req)
```

29.6.17.1. *Parameters.*

**input: req:** ping request sent by the client

29.6.17.2. *Return Values.* -ENOMEM: no memory available to allocate for a lustre_msg for reply

29.6.17.3. *Description.*

## 29.7. Changelog

**Version 2.5 (06/19/2003)**

(1) Radhika Vullikanti - Added details about orphan recovery.

**Version 2.0 (05/05/2003)**

(1) Radhika Vullikanti - Recovery has changed tremendously over the last few months, so added description for the new recovery procedure and new APIs.

**Version 1.5 (11/13/2002)**

(1) Radhika Vullikanti - Added some more explainations and pictures to section 19.1 and 19.2. Section 19.5 has descriptions for all the recovery APIs.

CHAPTER 30

# Lustre File System Repair

## 30.1. Overview

Lustre has journal recovery, including distributed recovery of orphaned objects to cover all aspects of recovery during normal operation.

In the presence of hardware failures or software bugs file systems may incur unusual damage that needs to be controlled. Lustre follows a two phronged approach to this:

(1) Metadata server and OST storage is on file systems for which advanced and reliable file system checkers are available. The best available tool of this kind is e2fsck for ext3 file systems. This will render the OST and MDS storage into a consistent state. During this part of the file system repair it is offline.
(2) Storage objects may be orphaned from their inodes: a scanning tools will find orphaned objects and remove them, or collect them in a lost and found directory.
(3) Inodes may lose storage data objects. The system will automatically replace such objects with new objects but will report errors and a scan of the namespace will reveal all such objects.

## 30.2. The basic offline lfsck implementation

We want to use this tool in case there are possibly multiple disk failures or other damage from which log recovery doesn't recover. The tool can also be used as an occasional sanity check. This tool is maximally simple, to make sure it is solid when disaster strikes.

During e2fsck, which is part of all these checks/recovery events, we build a few databases. The –mdsdb=<filename) option on e2fsck could, when run on the MDS walk the inodes (it does this anyway). If we hit a regular file we look at the EA. We build a few entries:

MDS_table: (fid, object id, ost index, flag) (flag is initially set to 0)

E2fsck also walks the directories and can build a table:

Directory membership table: (fid, name, dir_fid)

On the OST's of index "index" we can list the O/R/* directories and build a table:

OST_table(index) <object id, flags> for all objects present. (flag is initially set to 0)

When e2fsck is done you can mount the file system. This produces databases which for MCR I estimate would be a few GB (20 million files, 10 stripes on average). If we now use these databases in conjunction we can easily construct:

- the objects for which there is no entry in the MDS table (orphans)
- MDS inodes for which the objects no longer exist (and their pathnames)

If we now use these databases in conjunction we can easily construct:

- the objects for which there is no entry in the MDS table (orphans)
- MDS inodes for which the objects no longer exist (and their pathnames)

It is then relatively simple to remove the orphaned objects (that can be done later when the file system is mounted again) and to create "lost+found" inodes on the MDS where we brutally set the objects to those found in the orphan collection.

**30.2.1. Dealing with pre-created objects.** The MDS uses pre-created objects and these require some further synchronization.

(1) If the last id found on the MDS is > > last id found on the OST, then we have lost some objects that could be-recreated at write time and reported.
(2) If the last id found on the MDS is < < last id found on the OST then
  (a) move non-empty objects beyond MDS last ID to lost+found
  (b) remove emptry objects (the corresponding inodes were lost)

**30.2.2. The llog's for file deletions.** After it completes the check, lfsck should remove all llogs related to file deletions.

**30.2.3. Additional consistency checks.**

**Duplicate objects:** During the MDS scan it is possible to perform an additional integrity check, namely to check for duplicate objects.
**EA integrity:** The checker could try to assess if the EA's are still in tact.

**30.2.4. The case of RAID1 OST's.** This case is not significantly different. It is possible that one or both of the objects are lost. In case only one is lost, the other can be-recreated and data copied, using the mechanisms described above. In case both objects exist but the data is different, the repair tool can make a choice to synchronize data either:

(1) always from the *primary* to a secondary OBD. This is probably good in the case of one-way replication
(2) check mtime and synchronize from device with the latest mtime on the object to the device with the older mtime object.

The RAID1 OBD is already able to avoid further object creation once it has gone into degraded mode. This means that the collection of damaged objects will not change, so the syncing could be done while the system is mounted, however, it is probably better to sync offline.

A group lock can help with synchronization and ordinary I/O but it does not solve the problem entirely because I/O will continue to go to one OST until the failed OST is brought back online.

### 30.3. Changes for an Online File System Checker

The approach described in this section leads to an on-line checker. However this checker requires metadata changes in the EA area of the inodes.

The key issue is that we want to repair the file system as much as possible while it is online. There is no reason to scan objects that were introduced since the last re-mount, since a running system will not create new orphans. The new objects introduced since the current mount are subject to races when the checks for orphaned objects are applied to them. So I have argued we should use a snapshot.

I did this to avoid races and locks altogether. Sure, one can lock, but the locks would have to be subject to recovery for crashing systems and crashing lfsck applications, would have to obey ordering constraints and it would slow down the scan very very much, unnecesarily.

I found a refinement of the solution that avoids LVM level snapshots, but still does what I proposed.

The OST's have at present an ordering among the object id's. This is key implementation feature that aids this solution. However, our planned future file system based OST / MDS snapshots would enable an equally efficient and simple solution.

The object id's consist of a 64bit number that strictly increases and is transactionally consistent (let's call this the object number, analogous to the inode number), together with a generation number.

Objects, at the moment, are internally created as entries in a directory and their names are a stringified version of the object number. There is some distribution of these over several directories but that does not change the argument below and will be removed from LLP anyway, because we have refined the locks.

This was done intentionally and has two important consequences:

- the object id's are unique. In particular even after using an object number in one boot cycle, losing it and re-using it in another the (esstially random) generation will assure us that the object id remains unique. This avoids having wrong refereneces in MDS inodes to re-created objects. This is not so important for orphaned OST objects, but it is obviously very important in the orphan recovery scheme.
- relevant for lfsck is that almost trivially the OST can limit an iteration over all objects to the objects that were present as of the last reboot. Such objects may be removed if the node containing a reference to them is destroyed, but new objects which are the subject of all the races we discussed would have higher object numbers.

This last object number that is used is internally also written to a file that we can expose under the root only accessiblye OST file system view on the OST system.

So the iteration for the OST pass of lfsck can, without an LVM snapshot easily scan all the objects, except new ones. This means that effectively we do have a sort of a snapshot (it is not quite because objects can be removed), without LVM.

It does hinge on the current implementation of Lustre's Linux OST's (and would not apply to BlueArc's). We may gradually change this and use a more optimized protocol for objects, but we can postpone that until we also have snapshotting capabilities inside the file systems.

I also want to point out that we will expose the MDS fid namespace through a user level file system. All fid entries can be found in a directory __iopen__. Similarly I plan to expose the object id's on the OST in a directory so readdir can be used normally. Secondly, all extended attributes, for the MDS fid on the OST and for the object id's on the MDS can be access from user space with the normal EA api's for LLP. Our new staff member, Alex Tomas, has now implemented it and it has also been accepted in 2.5.

I believe that lfsck becomes very beautiful and simple now. All that is needed is that we get the last object number from the previous boot cycle, which again can be read from an ordinary file that we will expose in the OST namespace. Then we iterate over the OST object namespace, the directory with object numbers, and we stat the __iopen__ directory on the MDS for the fid found. If it is not found the existing obd_echo client interface in user space can remove the object.

### 30.3.1. Steps for implementation.

- remove the subdirectory code for object storage. The directory locks have been refined and all objects can again live in one directory.
- add a directory to the OST and MDS configuration ioctl's (the setup one) and make the user visible, read-only mount point there.
- make minor change to retain the last used object number in the previous mount.
- write a wrapper around the raw EA api to obtain extended attributes for objects and inodes
- make changes to llite and api's to propagate the entire fid (ino group + ino + generation) and store it in the object.
- make change to format the OST file system such the the MDS fid fits into enlarged inode.
- re-write the obd iterator api method to have a readdir cookie parameter as Don suggests
- implement the iterator method in terms of readdir using open/close
- we are required to report full pathnames of inodes with affected objects. For this a (in the case of hard links not necessarily unique) MDS parent directory fid should be made part of the inode EA.

CHAPTER 31

# Lustre Configuration

## 31.1. Architecture

Figure 1.1.1 shows the architecture of Lustre system configuration. A eXtended Markup Language (XML) file is used to describe Lustre system configuration. The entries that can be made in the XML file are described in a Data Type Description (DTD) file. Three configuration utilities are provided to make the configuration: lmc, lconf and lctl. LMC produces a XML file. LCONF helps in doing the low level configuration based upon the XML configuration file generated using lmc. LCTL is a low level config tool that can be used for configuration as well as troubleshooting/debugging.

The whole configuration process is as below:

1. The user of Lustre uses lmc to enter the configuration information.

2. LMC commands then create an XML configuration file with these information. All the entities of XML configuration file, produced by lmc, must fit the format described in DTD file.

3. LCONF reads the XML file and generates the appropriate lctl commands. It also loads all the modules needed for the Lustre system.

4. LCTL does the low level device specific configuration and mount a Lustre filesystem.

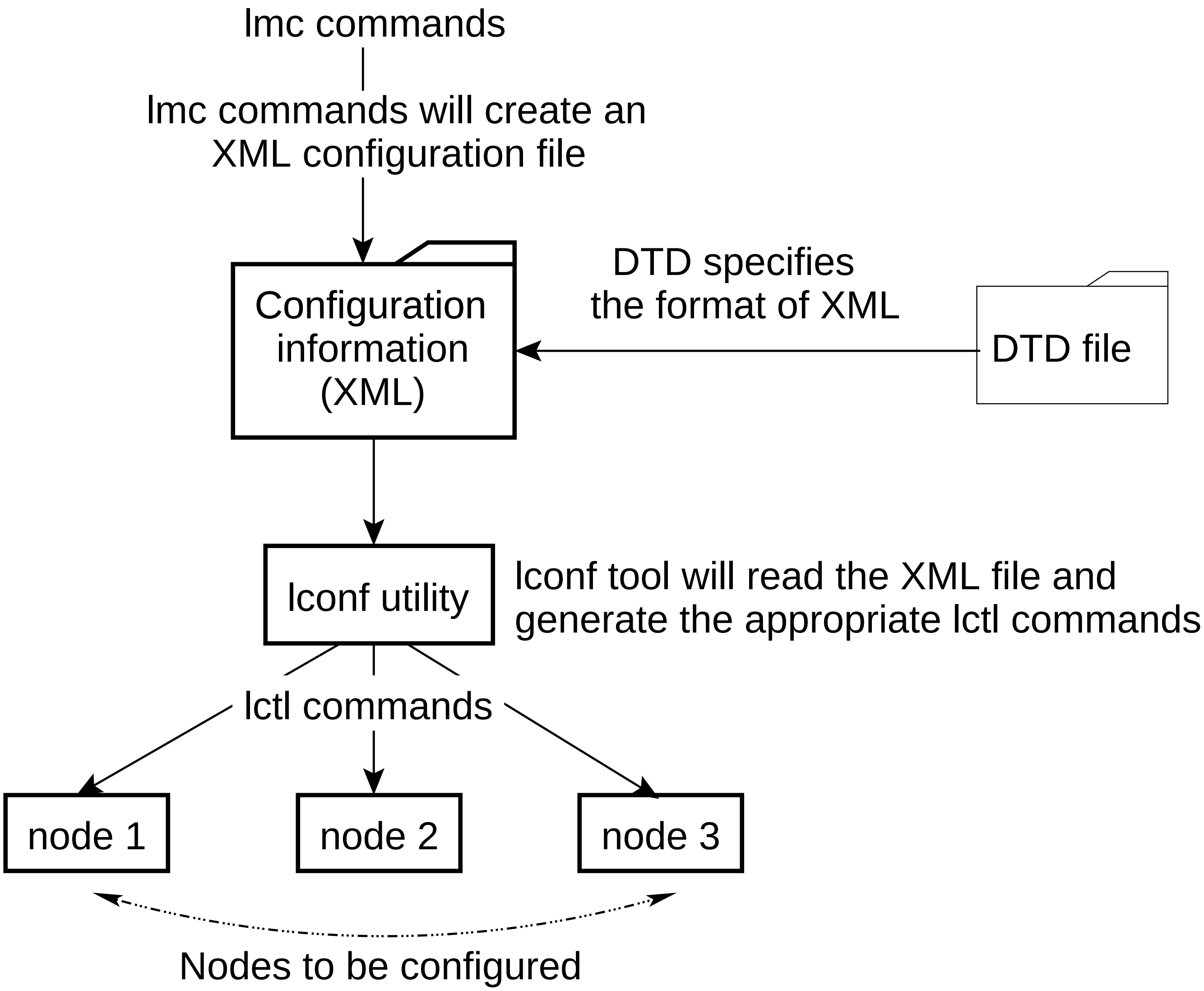

FIGURE 1.1.1 Lustre Configuration

### 31.2. High-level design

This section introduces the high-level design of the Lustre configuration. It includes the structure of XML and DTD files, the main loop of LMC, the class inheritance and main loop of LCONF, and the main loop of lctl.

**31.2.1. XML configuration files and configuration DTD.** As mentioned in “Architecture” section, we uses XML to describe the lustre configurations. Such XML configuration files can be easily generated using *lmc* and implemented using *lconf* . The use of XML configuration files can eliminate misconfiguration errors by checking the syntax against the Lustre configuration Data Type Definition (DTD). Figure 1.2.1 shows the whole structure of the DTD file.

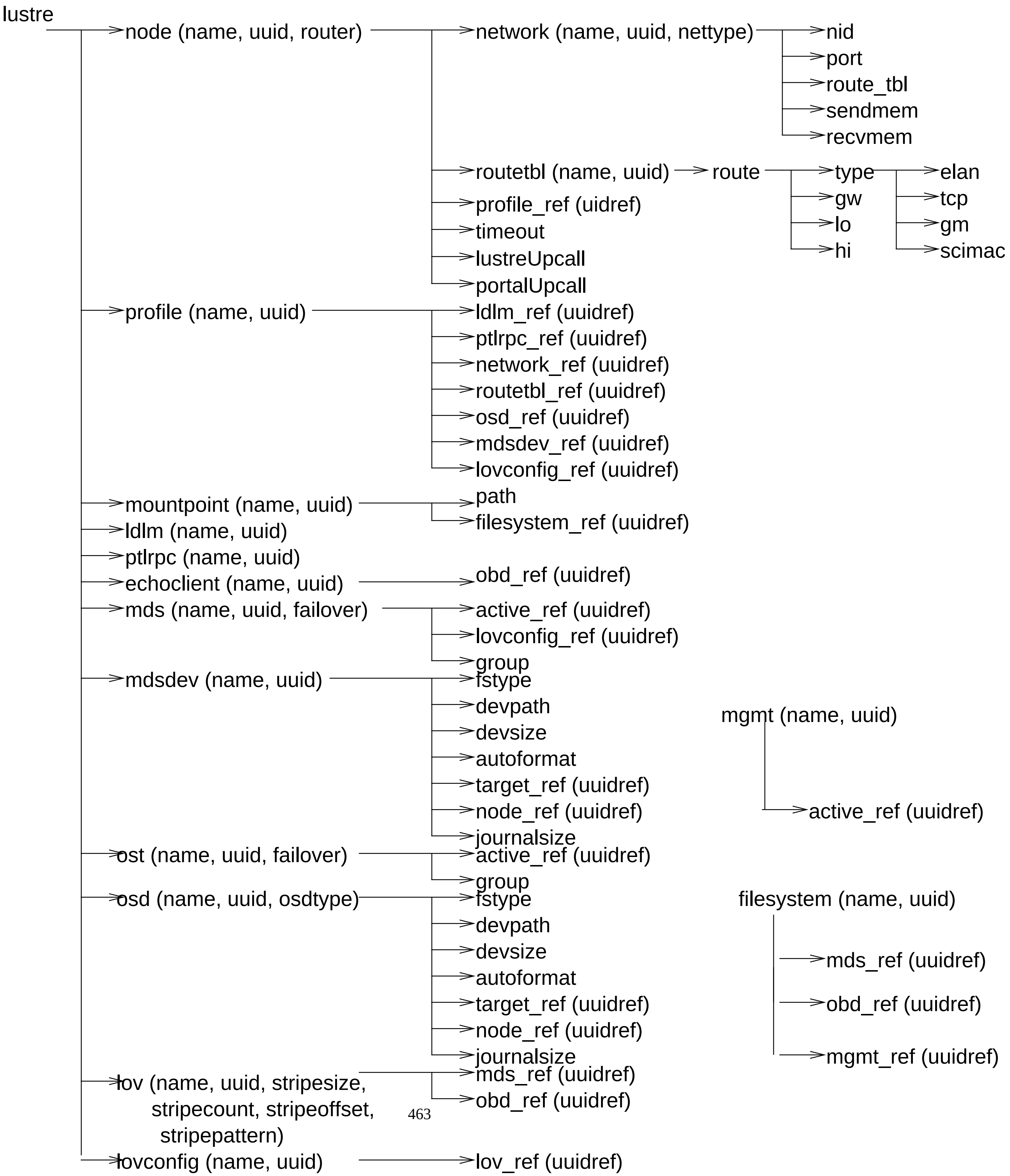

FIGURE 1.2.1 Structure of DTD

From the figure, we can see that the entities in XML file follow the following rules:

1. The structure of configuration information is organized in levels, as showed in figure 1.2.1.

2. Some entities(marked "*" in dtd file) may appear many times in the XML file, such as ost and osd, which have appeared three times in lov.xml for three OSTs.

3. Since the information structure is organized in levels, most of the top level entities are compound structures.

The entries defined in DTD file, the items in XML file produced by lmc and the commands generated by lconf are consistent. For example, one of the entries is for network device. It defines the name, uuid and the network type required to describe it:

```
<!ELEMENT network (nid | port | route_tbl | sendmem | recvmem)*>
<!ATTLIST network %object.attr;
                  nettype (tcp | elan | gm) 'tcp'>
```

So, the relative section in XML file is:

```
<network name='NET_localhost_tcp' nettype='tcp'  uid='NET_localhost_tcp_UUID'>
    <nid>localhost</nid>
    <clusterid>0</clusterid>
    <port>988</port>
</network>
```

When the XML parser (*lconf* for Lustre) has seen the OBD elements in the XML configuration file, it invokes *obdcontrol* commands equivalent to:

```
lctl „ EOF
    network %net
    mynid %nid
    quit
    EOF
```

**31.2.2. lmc.** *LMC* is a utility written in Python; it produces or merges XML descriptors. The main call flow of lmc is showed in figure 1.2.2. XXX is the element which will be added to XML file, and it may be node, network, route, mds, mgmt, ost, cobd, echo_client, lov, filesystem and mtpt, etc..

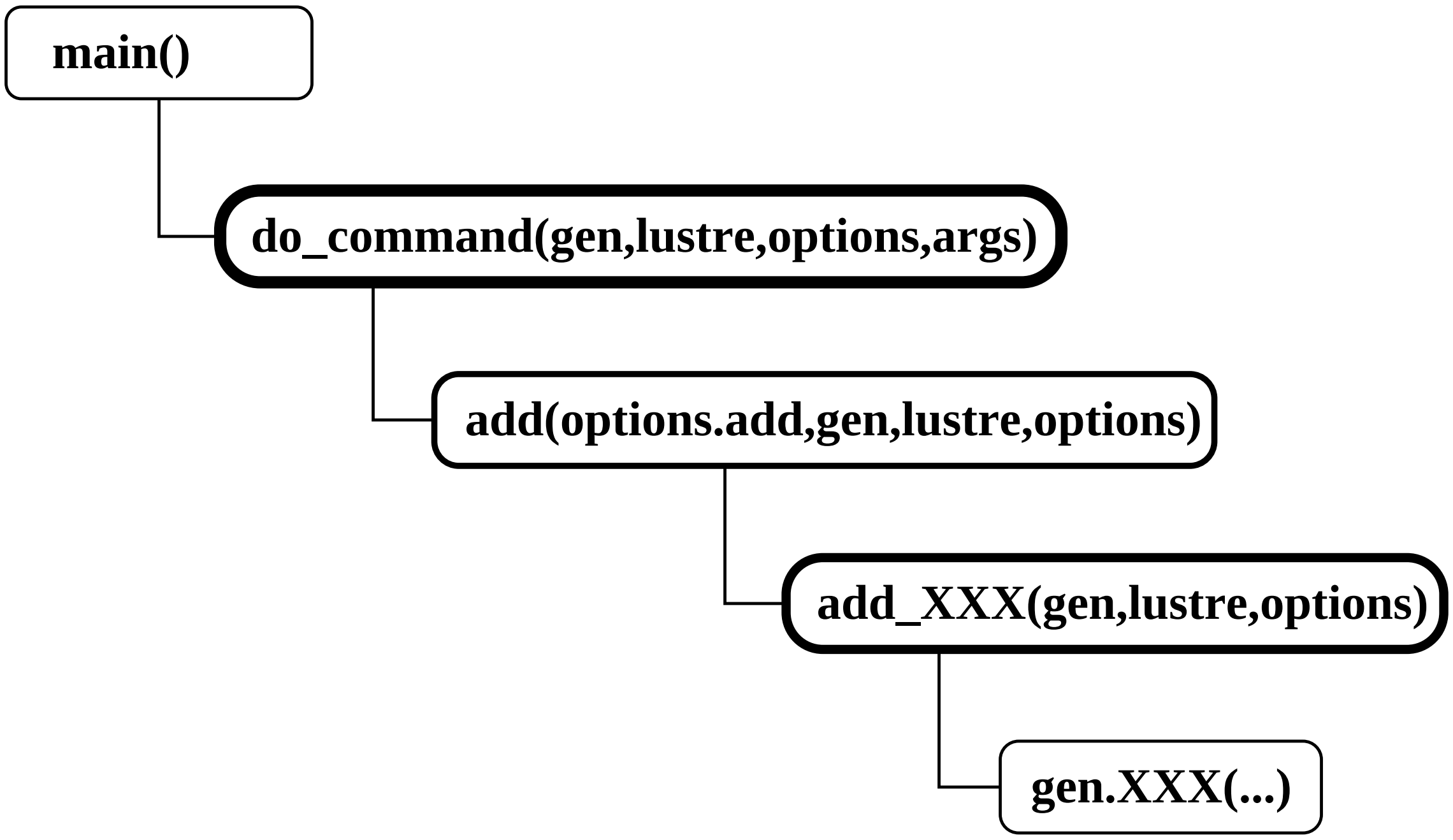

FIGURE 1.2.2 Call Flow of LMC

The following is an example to show the detail flow, it adds an element for mds:

The command is:

```
lmc -m $config --add mds --node localhost --mds mds1 --fstype $FSTYPE --dev $MDSDEV --
```

1. Since this command has -m $config option, the program uses the $config file as the output XML file.

2. The main() will call do_command() with the parameters from the command line.

3. do_command() will call add(“mds”,gen,lustre,options).

4. add() will call add_mds().

5. add_mds() will call gen.mds() and gen.mdsdev().

6. gen.mds will write mds relative information to $config file, and gen.mdsdev() will write mdsdev relative information to $config file.

The call flow for other elements is the same as that for mds.

**31.2.3. lconf.** *LConf* (Lustre Configuration) utility is another tool that helps in doing the low level configuration based upon the XML configuration file generated using *lmc*. The configuration information can be fed to *lconf* as an XML file or in the form of a URL that indicates the location of all the configuration information. If a system consists of multiple nodes, the nodename can be specified in *lconf*. Based on the nodename specified or using the default nodename, *lconf* determines the configuration information for the node being configured. Then it invokes *lctl* to do the low level device specific configuration.This utility automatically loads all the modules needed for the Lustre system.

lconf is written in Python. It defines several classes to deal with different devices, the classes and their inheritance are showed in figure 1.2.3.

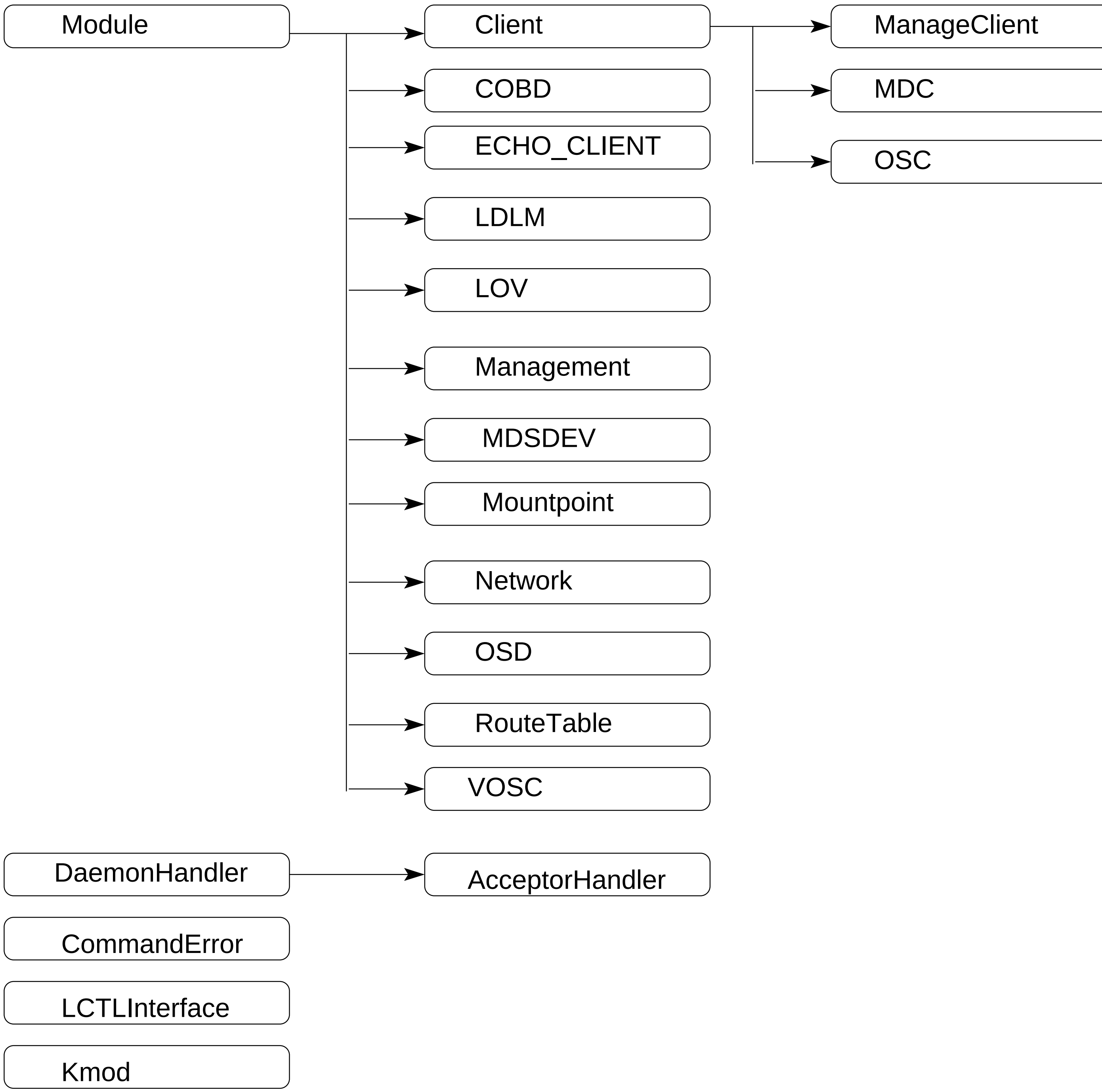

FIGURE 1.2.3 Classes and their inheritance in lconf

The main call flow of lconf is as figure 1.2.4.

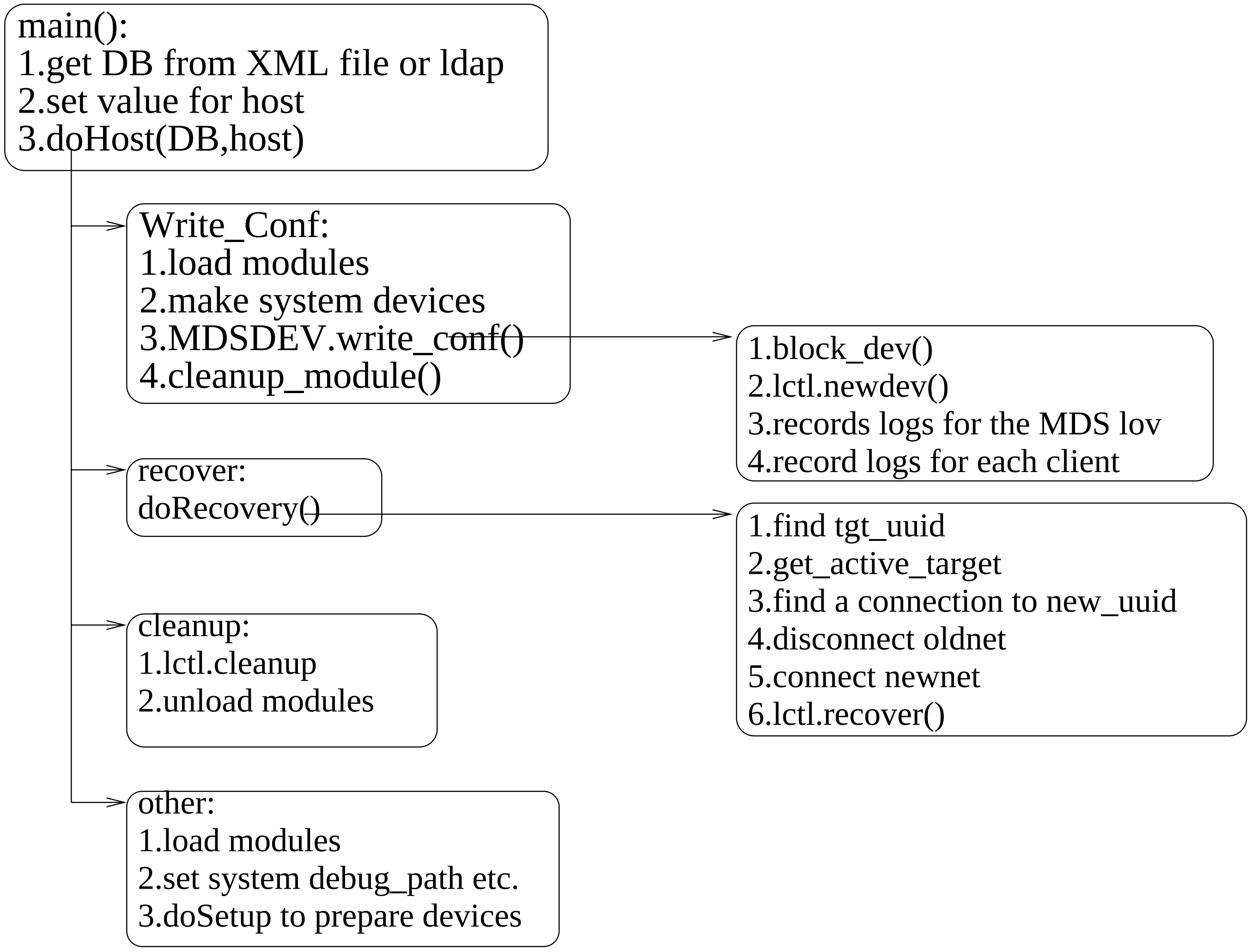

FIGURE 1.2.4 Main Loop of Lconf

### 31.2.4. lctl.

## 31.3. low-level design

In this section, we introduce the low-level design of Lustre configuration. As the previous section, it is divided into several subsection:the detail structure of xml and dtd, the main content of lmc, lconf and lctl.

**31.3.1. DTD.** A Data Type Description (DTD) file is used to describe the format of the XML configuration files used for Lustre. In this section we present the *lustre.dtd* file used as a basis for the lustre XML files.

First, the basic entities are described. All the main elements(except lustre) have object.attr attribute, that is, they have name and uuid attribute. All the object reference tag elements have objref.attr attribute, that is, they have uuidref attribute.

```
<!-- Lustre Management DTD -->
<!ENTITY % object.content "(#PCDATA)">
<!ENTITY % object.attr "
             name CDATA #REQUIRED
             uuid CDATA #REQUIRED">
<!ENTITY % objref.attr    "uuidref CDATA #REQUIRED">
```

The main elements are usually compound structures, which follow one by one. The first element is lustre, which can contain no or any number of node, profile, mountpoint, ldlm, ptlrpc,echoclient, mds, mdsdev,ost, osd, lov and lovconfig. This element has "version" attribute.

```
<!-- main elements -->
<!ELEMENT lustre (node | profile | mountpoint | ldlm |
                 ptlrpc |echoclient | mds | mdsdev|
                 ost | osd | lov | lovconfig)*>
<!ATTLIST lustre version CDATA #REQUIRED>
<!ATTLIST node %object.attr;
               router CDATA #IMPLIED>
```

The next element is network, which can contain no or any number of nid, port, route_tbl, sendmem and recvmem. This element has "name", "uuid" and "nettype" attributes, and the network type is tcp, elan, or gm, by default, it's tcp.

```
<!ELEMENT network (nid | port | route_tbl | sendmem | recvmem)*> <!ATTLIST network %ob
                 nettype (tcp | elan | gm) 'tcp'>
<!ATTLIST routetbl %object.attr;>
<!ELEMENT route %object.content;>
<!ATTLIST route type (elan | tcp | gm) #REQUIRED                                gw CDA
               lo CDATA #REQUIRED
               hi CDATA #IMPLIED >
<!ELEMENT profile (ldlm_ref | ptlrpc_ref | network_ref | routetbl_ref |
                    filesystem_ref #REQUIRED >
<!ATTLIST mountpoint %object.attr;>
<!ATTLIST echoclient %object.attr;>
<!ATTLIST ldlm %object.attr;>
<!ATTLIST mgmt %object.attr;>
```

```
<!ATTLIST ptlrpc %object.attr;>
<!ATTLIST osd %object.attr;
              osdtype (obdfilter | obdecho) 'obdfilter'>
<!ATTLIST ost %object.attr;
              failover ( 1 | 0 ) #IMPLIED>
                     obd_ref #REQUIRED
                     (mgmt_ref)* >
<!ATTLIST filesystem %object.attr;>
<!ATTLIST mds %object.attr;
              failover ( 1 | 0 ) #IMPLIED>
<!ATTLIST mdsdev %object.attr;>
<!ATTLIST lov %object.attr;
              stripesize    CDATA #REQUIRED
              stripecount   CDATA #REQUIRED
              stripeoffset  CDATA #IMPLIED
              stripepattern CDATA #REQUIRED>
<!ATTLIST lovconfig %object.attr;>
```

The following elemens are the basic elements.

```
<!-- basic elements -->
<!ELEMENT recoveryUpcall %object.content;>
<!ELEMENT timeout       %object.content;>
<!ELEMENT journalsize   %object.content;>
<!ELEMENT fstype        %object.content;>
<!ELEMENT nid           %object.content;>
<!ELEMENT port          %object.content;>
<!ELEMENT sendmem       %object.content;>
<!ELEMENT recvmem       %object.content;>
<!ELEMENT autoformat    %object.content;>
<!ELEMENT activetarget  %object.content;>
<!ELEMENT devpath       %object.content;>
<!ELEMENT devsize       %object.content;>
<!ELEMENT path          %object.content;>
<!ELEMENT fileset       %object.content;>
```

The following elemens are the object reference tag elements, and everyone has a “uuidref” attribute.

```
<!-- object reference tag elements -->
<!ELEMENT network_ref     %objref.content;>
<!ATTLIST network_ref     %objref.attr;>
<!ELEMENT routetbl_ref    %objref.content;>
<!ATTLIST routetbl_ref    %objref.attr;>
```

```
<!ELEMENT node_ref        %objref.content;>
<!ATTLIST node_ref        %objref.attr;>
<!ELEMENT profile_ref     %objref.content;>
<!ATTLIST profile_ref     %objref.attr;>
<!ELEMENT osd_ref         %objref.content;>
<!ATTLIST osd_ref         %objref.attr;>
<!ELEMENT mds_ref         %objref.content;>
<!ATTLIST mds_ref         %objref.attr;>
<!ELEMENT mdsdev_ref      %objref.content;>
<!ATTLIST mdsdev_ref      %objref.attr;>
<!ELEMENT obd_ref         %objref.content;>
<!ATTLIST obd_ref         %objref.attr;>
<!ELEMENT ost_ref         %objref.content;>
<!ATTLIST ost_ref         %objref.attr;>
<!ELEMENT active_ref      %objref.content;>
<!ATTLIST active_ref      %objref.attr;>
<!ELEMENT target_ref      %objref.content;>
<!ATTLIST target_ref      %objref.attr;>
<!ELEMENT lov_ref         %objref.content;>
<!ATTLIST lov_ref         %objref.attr;>
<!ELEMENT lovconfig_ref   %objref.content;>
<!ATTLIST lovconfig_ref   %objref.attr;>
<!ELEMENT mgmt_ref        %objref.content;>
<!ATTLIST mgmt_ref        %objref.attr;>
<!ELEMENT mountpoint_ref  %objref.content;>
<!ATTLIST mountpoint_ref  %objref.attr;>
<!ELEMENT filesystem_ref  %objref.content;>
<!ATTLIST filesystem_ref  %objref.attr;>
<!ELEMENT echoclient_ref  %objref.content;>
<!ATTLIST echoclient_ref  %objref.attr;>
<!ELEMENT failover_ref    %objref.content;>
<!ATTLIST failover_ref    %objref.attr;>
<!ELEMENT ldlm_ref        %objref.content;>
<!ATTLIST ldlm_ref        %objref.attr;>
<!ELEMENT ptlrpc_ref      %objref.content;>
<!ATTLIST ptlrpc_ref      %objref.attr;>
```

**31.3.2. XML.** The XML file is created by lmc utility and used by lconf utility according to the defininition of DTD file. Its structure is organized in levels(as described in figure 1.2.1). The following is an example:

```
<?xml version='1.0' encoding='UTF-8'?>
<!DOCTYPE lustre>
```

```
<lustre version='2003070801'>
  <ldlm name='ldlm' uuid='ldlm_UUID'/>
  <node name='localhost' uuid='localhost_UUID'>
    <profile_ref uuidref='PROFILE_localhost_UUID'/>
    <network name='NET_localhost_tcp' nettype='tcp'
            uuid='NET_localhost_tcp_UUID'>
       <nid>localhost.localdomain</nid>
       <clusterid>0</clusterid>
       <port>988</port>
    </network>
  </node>
  <profile name='PROFILE_localhost' uuid='PROFILE_localhost_UUID'>          <ldlm_ref
    <network_ref uuidref='NET_localhost_tcp_UUID'/>
    <mdsdev_ref uuidref='MDD_mds1_localhost_UUID'/>
    <osd_ref uuidref='OSD_OST_localhost_localhost_UUID'/>                  <mountpoin
  </profile>
  <mds name='mds1' uuid='mds1_UUID'>
    <active_ref uuidref='MDD_mds1_localhost_UUID'/>
    <lovconfig_ref uuidref='LVCFG_lov1_UUID'/>
    <filesystem_ref uuidref='FS_fsname_UUID'/>
  </mds>
  <mdsdev name='MDD_mds1_localhost' uuid='MDD_mds1_localhost_UUID'>         <fstype>ex
    <devpath>/tmp/mds1-localhost.localdomain</devpath>                     <autoforma
    <devsize>100000</devsize>
    <journalsize>0</journalsize>
    <nspath>/mnt/mds_ns</nspath>
    <mkfsoptions>-I 128</mkfsoptions>
    <node_ref uuidref='localhost_UUID'/>
    <target_ref uuidref='mds1_UUID'/>
  </mdsdev>
  <lov stripesize='65536' stripepattern='0' stripecount='0'                   uuid='l
    <mds_ref uuidref='mds1_UUID'/>
    <obd_ref uuidref='OST_localhost_UUID'/>
  </lov>
  <lovconfig uuid='LVCFG_lov1_UUID' name='LVCFG_lov1'>
    <lov_ref uuidref='lov1_UUID'/>
  </lovconfig>
  <ost name='OST_localhost' uuid='OST_localhost_UUID'>
    <active_ref uuidref='OSD_OST_localhost_localhost_UUID'/>
  </ost>
  <osd osdtype='obdfilter' name='OSD_OST_localhost_localhost'                 uuid='O
    <target_ref uuidref='OST_localhost_UUID'/>
    <node_ref uuidref='localhost_UUID'/>
```

```
      <fstype>ext3</fstype>                                                   <devpath>/
      <autoformat>no</autoformat>
      <devsize>200000</devsize>
      <journalsize>0</journalsize>
      <nspath>/mnt/ost_ns</nspath>
    </osd>
    <filesystem uuid='FS_fsname_UUID' name='FS_fsname'>
      <mds_ref uuidref='mds1_UUID'/>
      <obd_ref uuidref='lov1_UUID'/>
    </filesystem>
    <mountpoint uuid='MNT_localhost_UUID' name='MNT_localhost'>             <filesyste
      <path>/mnt/lustre</path>
    </mountpoint>
  </lustre>
```

**31.3.3. lmc.** lmc provides a general config class, GenConfig, to add configuration information to XML file. It also presents several top level functions(add_XXX()), used to call the member functions(gen.XXX()) of GenConfig. These top level functions are used by the main loop of lmc, as showed in figure 1.2.2. In addition, several name and uuid relative functions and utilities are provided for using in the top level functions and GenConfig class.

3.3.1 GenConfig

It's the most important class used to add configuration information to XML file. It's prototype is as blow:

```
class GenConfig:
    doc = None
    dom = None
    def __init__(self, doc): ...
    def ref(): ...
    def newService(): ...
    def addText(): ...
    def addElement(): ...
    def network(): ...
    def routetbl(): ...
    ... ...
    def echo_client(): ...
```

3.3.2 top-level functions

The following functions are provided:

```
set_node_options()
do_add_node()
add_node()
```

```
add_network()
add_route()
add_mds()
add_mgmt()
add_ost()
add_cobd()
add_echo_client()
add_lov()
new_filesystem()
get_fs_uuid()
add_mtpt()
```

3.3.3 some useful utilities

```
new_name()
new_uuid()
new_lustre()
init_names()
get_format_flag()
getName()
getUUID()
findByName()
lookup()
name2uuid()
lookup_filesystem()
get_net_uuid()
lov_add_obd()
ref_exists()
node_add_profile()
getattr()
```

**31.3.4. lconf.** As showed in figure 1.2.3, lconf provides a lot of classes used to make the configuration easy. lconf also provides many system-level functions, miscellaneous query functions, routing functions and other functions. In this section, we will list them in details.

3.4.1 the basic class module

The class module is the parent class of 12 child classes; it provides several common functions to its child classes. The prototype of module is as below:

```
class module:
    def __init__(): ...
    def info(): ...
    def cleanup(): ...
    def add_portals_module(): ...
```

```
def add_lustre_module(): ...
def load_module(): ...
def cleanup_module(): ...
def safe_to_clean(): ...
def saft_to_clean_modules(): ...
```

3.4.2 other classes

3.4.3 system-level functions

```
runcmd()
run()
run_daemon()
```

#### 31.3.5. lctl.

## 31.4. Lustre configuration scenarios

**31.4.1. Hostid/clusterid.** To describe the usage of clusterid, hostaddr, port, gw_clusterid and target_clusterid, we can use a routed cluster example.

The below is the structure of this example:

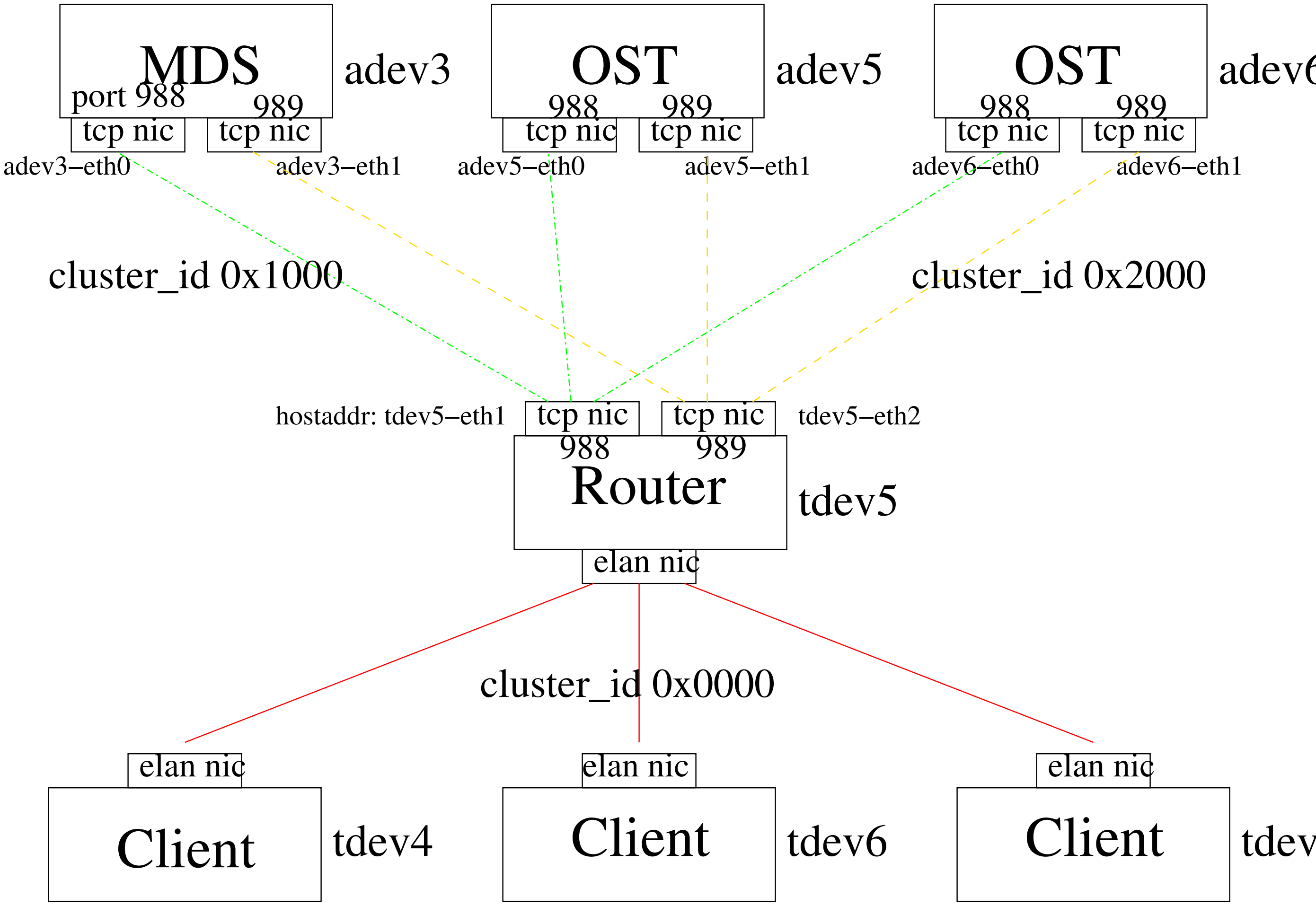

utils/lmc will create a XML file with this lmc batch script:

```
--node adev3 --add net --nid adev3 --cluster_id 0x1000 --nettype tcp --hostaddr adev3-eth0 --port 988 --tcpbuf 1048576
--node adev5 --add net --nid adev5 --cluster_id 0x1000 --nettype tcp --hostaddr adev5-eth0 --port 988 --tcpbuf 1048576
--node adev6 --add net --nid adev6 --cluster_id 0x1000 --nettype tcp --hostaddr adev6-eth0 --port 988 --tcpbuf 1048576
--node adev3 --add net --nid adev3 --cluster_id 0x2000 --nettype tcp --hostaddr adev3-eth1 --port 989 --tcpbuf 1048576
--node adev5 --add net --nid adev5 --cluster_id 0x2000 --nettype tcp --hostaddr adev5-eth1 --port 989 --tcpbuf 1048576
--node adev6 --add net --nid adev6 --cluster_id 0x2000 --nettype tcp --hostaddr adev6-eth1 --port 989 --tcpbuf 1048576
--node tdev5 --add net --nid 5 --cluster_id 0x1000 --nettype tcp --hostaddr tdev5-eth1 --port 988 --tcpbuf 1048576 --router --irq_affin
--node tdev5 --add net --nid 5 --cluster_id 0x2000 --nettype tcp --hostaddr tdev5-eth2 --port 989 --tcpbuf 1048576 --router --irq_affin
--node tdev5 --add net --nid 5 --cluster_id 0x0000 --nettype elan --router --irq_affinity 1
--node client --add net --nid '*' --cluster_id 0x0000 --nettype elan
--node adev3 --add mds --mds mds_mnt_lustre --dev /tmp/mds --size 2000000 --fstype ext3
--add lov --lov lov1 --mds mds_mnt_lustre --stripe_sz 65536 --stripe_cnt 2 --stripe_pattern 0
--node adev5 --add ost --lov lov1 --dev /dev/sdc --fstype ext3
--node adev6 --add ost --lov lov1 --dev /dev/sdc --fstype ext3
--node adev3 --add route --nettype tcp --gw 5 --gateway_cluster_id 0x1000 --target_cluster_id 0x1000 --lo 4 --hi 7
--node adev5 --add route --nettype tcp --gw 5 --gateway_cluster_id 0x1000 --target_cluster_id 0x1000 --lo 4 --hi 7
--node adev6 --add route --nettype tcp --gw 5 --gateway_cluster_id 0x1000 --target_cluster_id 0x1000 --lo 4 --hi 7
--node adev3 --add route --nettype tcp --gw 5 --gateway_cluster_id 0x2000 --target_cluster_id 0x2000 --lo 4 --hi 7
--node adev5 --add route --nettype tcp --gw 5 --gateway_cluster_id 0x2000 --target_cluster_id 0x2000 --lo 4 --hi 7
--node adev6 --add route --nettype tcp --gw 5 --gateway_cluster_id 0x2000 --target_cluster_id 0x2000 --lo 4 --hi 7
```

```
--node tdev5 --add route --nettype tcp --gw 5 --gateway_cluster_id 0x1000 --target_cluster_id 0x1000 --lo adev3
--node tdev5 --add route --nettype tcp --gw 5 --gateway_cluster_id 0x1000 --target_cluster_id 0x1000 --lo adev5
--node tdev5 --add route --nettype tcp --gw 5 --gateway_cluster_id 0x1000 --target_cluster_id 0x1000 --lo adev6
--node tdev5 --add route --nettype tcp --gw 5 --gateway_cluster_id 0x2000 --target_cluster_id 0x2000 --lo adev3
--node tdev5 --add route --nettype tcp --gw 5 --gateway_cluster_id 0x2000 --target_cluster_id 0x2000 --lo adev5
--node tdev5 --add route --nettype tcp --gw 5 --gateway_cluster_id 0x2000 --target_cluster_id 0x2000 --lo adev6
--node tdev5 --add route --nettype elan --gw 5 --gateway_cluster_id 0x0000 --target_cluster_id 0x0000 --lo 4 --hi 7
--node client --add mtpt --path /mnt/lustre --mds mds_mnt_lustre --lov lov1
```

In the lmc command, user will assign cluster_id, hostaddr, port, gateway_cluster_id and target_cluster_id, which will be written to proper section of XML file by lmc.

The generated XML file is as below, and the related section has been highlighted:

```
<?xml version='1.0' encoding='UTF-8'?>
<!DOCTYPE lustre> <lustre version='2003070801'>
  <ldlm name='ldlm' uuid='ldlm_UUID'/>
  <node name='adev3' uuid='adev3_UUID'>
    <profile_ref uuidref='PROFILE_adev3_UUID'/>
    <network name='NET_adev3_tcp' nettype='tcp' uuid='NET_adev3_tcp_UUID'>
      <nid>adev3</nid>
      <clusterid>0x1000</clusterid>
      <hostaddr>adev3-eth0</hostaddr>
      <port>988</port>
      <sendmem>1048576</sendmem>
      <recvmem>1048576</recvmem>
    </network>
    <network name='NET_adev3_tcp_2' nettype='tcp' uuid='NET_adev3_tcp_2_UUID'>
      <nid>adev3</nid>
      <clusterid>0x2000</clusterid>
      <hostaddr>adev3-eth1</hostaddr>
      <port>989</port>
      <sendmem>1048576</sendmem>
      <recvmem>1048576</recvmem>
    </network>
    <routetbl name='RTBL_adev3' uuid='RTBL_adev3_UUID'>
      <route gw='5' hi='7' gwclusterid='0x1000' lo='4' tgtclusterid='0x1000' type='tcp'/>
      <route gw='5' hi='7' gwclusterid='0x2000' lo='4' tgtclusterid='0x2000' type='tcp'/>
    </routetbl>
  </node>
  <profile name='PROFILE_adev3' uuid='PROFILE_adev3_UUID'>
    <ldlm_ref uuidref='ldlm_UUID'/>
    <network_ref uuidref='NET_adev3_tcp_UUID'/>
    <network_ref uuidref='NET_adev3_tcp_2_UUID'/>
    <mdsdev_ref uuidref='MDD_mds_mnt_lustre_adev3_UUID'/>
    <routetbl_ref uuidref='RTBL_adev3_UUID'/>
  </profile>
  <node uuid='adev5_UUID' name='adev5'>
    <profile_ref uuidref='PROFILE_adev5_UUID'/>
    <network name='NET_adev5_tcp' nettype='tcp' uuid='NET_adev5_tcp_UUID'>
      <nid>adev5</nid>
      <clusterid>0x1000</clusterid>
      <hostaddr>adev5-eth0</hostaddr>
      <port>988</port>
      <sendmem>1048576</sendmem>
      <recvmem>1048576</recvmem>
    </network>
```

```
  <network name='NET_adev5_tcp_2' nettype='tcp' uuid='NET_adev5_tcp_2_UUID'>
    <nid>adev5</nid>
    <clusterid>0x2000</clusterid>
    <hostaddr>adev5-eth1</hostaddr>
    <port>989</port>
    <sendmem>1048576</sendmem>
    <recvmem>1048576</recvmem>
  </network>
  <routetbl uuid='RTBL_adev5_UUID' name='RTBL_adev5'>
    <route gw='5' hi='7' gwclusterid='0x1000' lo='4' tgtclusterid='0x1000' type='tcp'/>
    <route gw='5' hi='7' gwclusterid='0x2000' lo='4' tgtclusterid='0x2000' type='tcp'/>
  </routetbl>
</node>
<profile uuid='PROFILE_adev5_UUID' name='PROFILE_adev5'>
  <ldlm_ref uuidref='ldlm_UUID'/>
  <network_ref uuidref='NET_adev5_tcp_UUID'/>
  <network_ref uuidref='NET_adev5_tcp_2_UUID'/>
  <osd_ref uuidref='OSD_OST_adev5_adev5_UUID'/>
  <routetbl_ref uuidref='RTBL_adev5_UUID'/>
</profile>
<node name='adev6' uuid='adev6_UUID'>
  <profile_ref uuidref='PROFILE_adev6_UUID'/>
  <network name='NET_adev6_tcp' nettype='tcp' uuid='NET_adev6_tcp_UUID'>
    <nid>adev6</nid>
    <clusterid>0x1000</clusterid>
    <hostaddr>adev6-eth0</hostaddr>
    <port>988</port>
    <sendmem>1048576</sendmem>
    <recvmem>1048576</recvmem>
  </network>
  <network name='NET_adev6_tcp_2' nettype='tcp' uuid='NET_adev6_tcp_2_UUID'>
    <nid>adev6</nid>
    <clusterid>0x2000</clusterid>
    <hostaddr>adev6-eth1</hostaddr>
    <port>989</port>
    <sendmem>1048576</sendmem>
    <recvmem>1048576</recvmem>
  </network>
  <routetbl name='RTBL_adev6' uuid='RTBL_adev6_UUID'>
    <route gw='5' hi='7' gwclusterid='0x1000' lo='4' tgtclusterid='0x1000' type='tcp'/>
    <route gw='5' hi='7' gwclusterid='0x2000' lo='4' tgtclusterid='0x2000' type='tcp'/>
  </routetbl>
</node>
<profile name='PROFILE_adev6' uuid='PROFILE_adev6_UUID'>
  <ldlm_ref uuidref='ldlm_UUID'/>
  <network_ref uuidref='NET_adev6_tcp_UUID'/>
  <network_ref uuidref='NET_adev6_tcp_2_UUID'/>
  <osd_ref uuidref='OSD_OST_adev6_adev6_UUID'/>
  <routetbl_ref uuidref='RTBL_adev6_UUID'/>
</profile>
<node router='1' name='tdev5' uuid='tdev5_UUID'>
  <profile_ref uuidref='PROFILE_tdev5_UUID'/>
  <network name='NET_tdev5_tcp' nettype='tcp' uuid='NET_tdev5_tcp_UUID'>
    <nid>5</nid>
    <clusterid>0x1000</clusterid>
```

```
    <hostaddr>tdev5-eth1</hostaddr>
    <port>988</port>
    <sendmem>1048576</sendmem>
    <recvmem>1048576</recvmem>
    <irqaffinity>1</irqaffinity>
  </network>
  <network name='NET_tdev5_tcp_2' nettype='tcp' uuid='NET_tdev5_tcp_2_UUID'>
    <nid>5</nid>
    <clusterid>0x2000</clusterid>
    <hostaddr>tdev5-eth2</hostaddr>
    <port>989</port>
    <sendmem>1048576</sendmem>
    <recvmem>1048576</recvmem>
    <irqaffinity>1</irqaffinity>
  </network>
  <network name='NET_tdev5_elan' nettype='elan' uuid='NET_tdev5_elan_UUID'>
    <nid>5</nid>
    <clusterid>0x0000</clusterid>
  </network>
  <routetbl name='RTBL_tdev5' uuid='RTBL_tdev5_UUID'>
    <route gw='5' hi='adev3' gwclusterid='0x1000' lo='adev3' tgtclusterid='0x1000' type='tcp'/>
    <route gw='5' hi='adev5' gwclusterid='0x1000' lo='adev5' tgtclusterid='0x1000' type='tcp'/>
    <route gw='5' hi='adev6' gwclusterid='0x1000' lo='adev6' tgtclusteridd='0x1000' type='tcp'/>
    <route gw='5' hi='adev3' gwclusterid='0x2000' lo='adev3' tgtclusterid='0x2000' type='tcp'/>
    <route gw='5' hi='adev5' gwclusterid='0x2000' lo='adev5' tgtclusterid='0x2000' type='tcp'/>
    <route gw='5' hi='adev6' gwclusterid='0x2000' lo='adev6' tgtclusterid='0x2000' type='tcp'/>
    <route gw='5' hi='7' gwclusterid='0x0000' lo='4' tgtclusterid='0x0000' type='elan'/>     </routetbl>
</node>
<profile name='PROFILE_tdev5' uuid='PROFILE_tdev5_UUID'>
  <ldlm_ref uuidref='ldlm_UUID'/>
  <network_ref uuidref='NET_tdev5_tcp_UUID'/>
  <network_ref uuidref='NET_tdev5_tcp_2_UUID'/>
  <network_ref uuidref='NET_tdev5_elan_UUID'/>
  <routetbl_ref uuidref='RTBL_tdev5_UUID'/>
</profile>
<node uuid='client_UUID' name='client'>
  <profile_ref uuidref='PROFILE_client_UUID'/>
  <network name='NET_client_elan' nettype='elan' uuid='NET_client_elan_UUID'>
    <nid>*</nid>
    <clusterid>0x0000</clusterid>
  </network>
</node>
<profile uuid='PROFILE_client_UUID' name='PROFILE_client'>
  <ldlm_ref uuidref='ldlm_UUID'/>
  <network_ref uuidref='NET_client_elan_UUID'/>
  <mountpoint_ref uuidref='MNT_client_UUID'/>
</profile>
<mds name='mds_mnt_lustre' uuid='mds_mnt_lustre_UUID'>
  <active_ref uuidref='MDD_mds_mnt_lustre_adev3_UUID'/>
  <lovconfig_ref uuidref='LVCFG_lov1_UUID'/>
  <filesystem_ref uuidref='FS_fsname_UUID'/>
</mds>
<mdsdev name='MDD_mds_mnt_lustre_adev3' uuid='MDD_mds_mnt_lustre_adev3_UUID'>
  <fstype>ext3</fstype>
  <backfstype>ext3</backfstype>
```

```
        <devpath>/tmp/mds</devpath>
        <autoformat>no</autoformat>
        <devsize>2000000</devsize>
        <journalsize>0</journalsize>
        <inodesize>0</inodesize>
        <node_ref uuidref='adev3_UUID'/>
        <target_ref uuidref='mds_mnt_lustre_UUID'/>
      </mdsdev>
      <lov stripesize='65536' stripepattern='0' stripecount='2' uuid='lov1_UUID' name='lov1'>
        <mds_ref uuidref='mds_mnt_lustre_UUID'/>
        <obd_ref uuidref='OST_adev5_UUID'/>
        <obd_ref uuidref='OST_adev6_UUID'/>
      </lov>
      <lovconfig uuid='LVCFG_lov1_UUID' name='LVCFG_lov1'>
        <lov_ref uuidref='lov1_UUID'/>
      </lovconfig>
      <ost name='OST_adev5' uuid='OST_adev5_UUID'>
        <active_ref uuidref='OSD_OST_adev5_adev5_UUID'/>
      </ost>
      <osd osdtype='obdfilter' name='OSD_OST_adev5_adev5' uuid='OSD_OST_adev5_adev5_UUID'>
       <target_ref uuidref='OST_adev6_UUID'/>
        <node_ref uuidref='adev6_UUID'/>
        <fstype>ext3</fstype>
        <backfstype>ext3</backfstype>
        <devpath>/dev/sdc</devpath>
        <autoformat>no</autoformat>
        <devsize>0</devsize>
        <journalsize>0</journalsize>
        <inodesize>0</inodesize>
      </osd>
      <filesystem uuid='FS_fsname_UUID' name='FS_fsname'>
        <mds_ref uuidref='mds_mnt_lustre_UUID'/>
        <obd_ref uuidref='lov1_UUID'/>
      </filesystem>
      <mountpoint uuid='MNT_client_UUID' name='MNT_client'>
        <filesystem_ref uuidref='FS_fsname_UUID'/>
        <path>/mnt/lustre</path>
      </mountpoint>
    </lustre>
```

>From the XML file, we can found:

1. clustreid, hostaddr, and port are only used in <network></network> section to identify which network the current node belongs to.

2. Every server node has a <routebl></routebl>, which specified the packets for “which” node in “which”cluster should pass through “which” gateway(router).

3. clusterid is only used in lconf to identify current node belonging to which cluster, it’s not used in the below command/function, such as lctl.add_route, lctl.add_autoconn.

- The usage of hostaddr:

hostaddr in XML will be transfer to the real host address(such as 192.168.1.20 for tcp) in lconf.Then only lctl.connect() and lctl.disconnect will use this parameter to add/del_route.

```
lctl.connect-->add_autoconn(nid,hostaddr,port, flags)-->jt_ptl_add_autoconnect()-->pcf
```

flags will be irq_affinity/share/eager.

- The usage of cluster_id/gw_cluster_id/tgt_cluster_id

The three parameters are only used in lconf to determine which route entry from xml should be add to current node by lctl.connect(that is ksocknal_add/del_route).

- The usage of lo/hi

The gateway will use lo/hi to add route by RouteTable()–>add_route–>jt_ptl_add_route(gateway_nalid, gateway_nid, lo_nid, hi_nid). lo-hi is used to identify a range.

# Part 3

# Manuals

CHAPTER 32

# Manpages

## 32.1. lstripe

### 32.1.1. NAME.

**lstripe:** Lustre utility to create a file with specific striping pattern

### 32.1.2. SYNOPSIS. lstripe <file_name> <stripe_size> <start_ost> <stripe_cnt>

### 32.1.3. DESCRIPTION.

This utility can be used to create a new file with a specific striping pattern. The arguments are explained below:

**file_name** The name and complete path of the new file to be created.
**stripe_size** Size of each stripe (in Bytes) for this file.
**start_ost** The OST number to start the striping on.
**stripe_cnt** The number of stripes for this file.

The number of stripes would determine the number of OSTs the file would be striped on. The OST numbering begins at 0, so 0 refers to OST1 and 1 refers to OST2.

### 32.1.4. EXAMPLES.

**Creating a file on a single OST:**

```
$lstripe /mnt/lustre/file1 131072 0 1
```

This creates the *file1* with stripes only on *OST1*.

```
$lstripe /mnt/lustre/file2 131072 1 1
```

This will create the *file2* only on *OST2*.

**Creating a file striped on two OSTs:**

```
$lstripe /mnt/lustre/file12 131072 0 2
```

This will create *file12* on two targets *OST1* and *OST2*. So, if *file12* has size 256k, it will have 2 stripes, on *OST1* and *OST2*, the first stripe will be on *OST1*.

```
$lstripe /mnt/lustre/file21 131072 1 2
```

This will create *file21* with two stripes with the first stripe on *OST2*.

### 32.1.5. BUGS.

None are known.

## 32.2. llanalyze

### 32.2.1. NAME.

**llanalyze:** Lustre utility useful for analysis of debug logs

### 32.2.2. SYNOPSIS. llanalyze [-pid <pid_num>] [-trace] [-silent] [–rpctrace] [-nodlm] [-nonet] < [logfile] | less -r

**32.2.3. DESCRIPTION.** This utility expects the log file input from stdin. The ***less -r*** should be used to get the output with indentations and coloring.

**-pid pid_num:** This option allows to extract debugs logs based on the *pid*
**-nodlm:** This option can be used to exclude the locking related logs from the output
**-trace:** This option can be used to extract only the Entry/Exit markers from the debug logs
**-nonet:** This option allows to exclude any logs related to network communication from the output
**-rpctrace:** Can be used to obtain a trace of the RPCs sent/received

**32.2.4. EXAMPLES.** Assume that the debug logs are saved in ***/tmp/lustre/full_log***, this utility can be used as shown below:

```
llanalyze -pid 3901 -trace </tmp/lustre/full_log >/tmp/lustre/trace_log
```

**32.2.5. BUGS.** None are known.

CHAPTER 33

# Lustre Specific API calls

This chapter gives some details about ioctls and other special interfaces to the Lustre file system.

## 33.1. File Locking

The Lustre file system allows a few ioctl(2) calls on open file descriptors. These can be used to change locking and striping behavior on files.

Remember the prototype of the ioctl call:

```
int ioctl(int fd, int request, char *argp);
```

We desribe the ioctl's here using the variable names above.

### 33.1.1. LL_IOC_GETFLAGS.

33.1.1.1. *Description.* This returns the flags set for the file. Flags that may be set are:

**LL_FILE_IGNORE_LOCK:** do not lock file extents for Posix semantics.

33.1.1.2. *Parameters.*

**fd:** File descriptor of the file to obtain flags for
**request:** should be set to LL_IOC_GETFLAGS
**argp:** should point to an *int* in the process address space. The flags will be place into that integer.

### 33.1.2. LL_IOC_SETFLAGS.

33.1.2.1. *Description.* This *or's* the flags set for the file. Flags that may be set are:

**LL_FILE_IGNORE_LOCK:** do not lock file extents for Posix semantics.

33.1.2.2. *Parameters.*

**fd:** File descriptor of the file to obtain flags for
**request:** should be set to LL_IOC_GETFLAGS
**argp:** should point to an *int* in the process address space. The flags to be or'd with should be stored in this integer.

### 33.1.3. LL_IOC_CLEARFLAGS.

33.1.3.1. *Description.* This *and's* the flags set for the file. Flags that may be set are:

**LL_FILE_IGNORE_LOCK:** do not lock file extents for Posix semantics.

33.1.3.2. *Parameters.*

**fd:** File descriptor of the file to obtain flags for
**request:** should be set to LL_IOC_GETFLAGS
**argp:** should point to an *int* in the process address space. The flags to be and'd with should be stored in this integer.

### 33.1.4. LL_IOC_GROUPLOCK.

33.1.4.1. *Description.* This obtains a group lock on the file, to be used by a group of cooperating processes.

33.1.4.2. *Parameters.*

**fd:** File descriptor of the file to obtain flags for
**request:** should be set to LL_IOC_GROUPLOCK
**argp:** should point to an *int* in the process address space. This integer is the group id. The group id must be established by the application. Group locks can only be granted for one group id at any time. Group locks for further group id's will be queued.

## 33.2. Striping Information

Lustre Lite will support a limited form of extended attributes, accessible through the standard Linux EA API's, *attr_get, attr_getf, attr_set, attr_setf*.

The extended attributes that are relevant for Lustre Lite are

**system.lustre_mds_objid:** This attribute contains in binary form a *struct lov_mds_md*:

```
struct lov_mds_md {
 __u32 lmd_magic;
 __u32 lmd_easize;   /* packed size of extended */
 __u64 lmd_object_id;     /* lov object id */
 __u64 lmd_stripe_offset; /* offset of the stripe */
 __u64 lmd_stripe_size;     /* size of the stripe */
 __u32 lmd_stripe_count;  /* how many objects are being striped */
 __u32 lmd_stripe_pattern;  /* per-lov object stripe pattern */
 struct lov_object_id lmd_objects[0];
};
```

This can be used to determine in which OST's an inode has stripes.

Setting this attribute can make the object believe it does not have a certain stripe anymore.

## 33.3. Lustre-conf

### 33.3.1. NAME.

**Lustre-conf:** Lustre configuration examples.

**33.3.2. SYNOPSIS.** This manual page contains examples how to generate Lustre configuration files for a variety of clusters.

### 33.3.3. EXAMPLES.

**Single system:** The following shell scripts create XML output suitable to configure a node as a client, MDS, and OST. Notice that the UUID of the object *id* passed to the client is the hostname prefixed with *OSC_*.

```
  #!/bin/bash
config=${1:-local.xml}
LMC=${LMC-../utils/lmc}
# create nodes
${LMC} -o $config --add node --node localhost || exit 10
${LMC} -m $config --add net --node localhost --nid localhost
--nettype tcp || exit 11
# configure mds server
${LMC} -m $config --add mds --node localhost --mds mds1 --dev /tmp/mds1
--size 50000 || exit 20
# configure ost
${LMC} -m $config --add ost --node localhost --ost ost1 --dev /tmp/ost1
--size 100000 || exit 30
# create client config
${LMC} -m $config --add mtpt --node localhost --path /mnt/lustre --mds mds1
--ost ost1 || exit 40
```

**LOV configuration:** The following configuration creates a logical volume striped over two OST's and an MDS and client on a single host. This configuration would cause files created to be striped over both the OSTs unless specified differently using *lstripe*.

```
  #!/bin/bash
config=${1:-lov.xml}
LMC=../utils/lmc
# create nodes
${LMC} -o $config --add node --node localhost ||exit 10
${LMC} -m $config --add net --node localhost --nid localhost --nettype tcp
||exit 11
# configure mds server
${LMC} -m $config --format --add mds --node localhost --mds mds1
--dev /tmp/mds1 --size 50000 || exit 20
```

```
# configure ost
${LMC} -m $config --add lov --lov lov1 --mds mds1 --strip_sz 4096
--stripe_cnt 0 --stripe_pattern 0 || exit 30
${LMC} -m $config --add ost --node localhost --lov lov1 --ost ost1
--dev /tmp/ost1 --size 100000 || exit 31
${LMC} -m $config --add ost --node localhost --lov lov1 --ost ost2
--dev /tmp/ost2 --size 100000 || exit 32
# create client config
${LMC} -m $config --add mtpt --node localhost --path /mnt/lustre
--mds mds1 --lov lov1 || exit 40
```

**Multiple client configuration:** The following configuration can be used to setup clusters with multiple clients using the generic *client* node name.

```
  #!/bin/bash
config=${1:-multi_client.xml}
LMC=../utils/lmc
# create nodes
${LMC} -o $config --add net --node client --nettype tcp --nid '*'||exit 10
${LMC} -m $config --add net --node mds_hostname --nid mds_hostname --nettype tcp
||exit 11
${LMC} -m $config --add net --node ost_hostname --nid ost_hostname --nettype tcp
# configure mds server
${LMC} -m $config --add mds --node mds_hostname --mds mds1
--dev /tmp/mds1 --size 50000 || exit 20
# configure ost
${LMC} -m $config --add lov --lov lov1 --mds mds1 --strip_sz 4096
--stripe_cnt 0 --stripe_pattern 0 || exit 30
${LMC} -m $config --add ost --node ost_hostname --lov lov1 --obd obd1
--dev /tmp/ost1 --size 100000 || exit 31
${LMC} -m $config --add ost --node ost_hostname --lov lov1 --obd obd2
--dev /tmp/ost2 --size 100000 || exit 32
# create client config
${LMC} -m $config --add mtpt --node client --path /mnt/lustre
--mds mds1 --lov lov1 || exit 40
```

**MCR with_routers:** Builds an XML for the whole MCR cluster with 10 GW nodes, each pointing to two BlueArc OST's, a single MDS, and client nodes 40-96. This script uses batchmode.

```
  #!/bin/bash
config=${1:-mcr.xml}
UUIDLIST=/home/bluearc/UUID.0916
LMC="save_cmd"
LMC_REAL="../../lustre/utils/lmc -m $config"
```

```
# TCP/IP servers
SERVER_START=41
SERVER_CNT=20
GW_START=11
GW_CNT=10
MDS=mcr23

# this is needed for to create route for elan network
CLIENT_LO=mcr50
CLIENT_HI=mcr96
PORT=2432
TCPBUF=1048576
STRIPE_SIZE=65536

# the following functions compute net addresses (nid's)
h2elan () { echo $1 | sed 's/[^0-9]*//g' }
h2ip () { echo "${1}" }

# map gwNN to mcrNN
# assumes /etc/hosts looks like this:
# 192.168.40.11 emcr10 mcr10-eth0 emcr-r2-s1 gw10
gw2mcr()
{
awk '$5 = /'$1'$/ {print substr($2,2)}' /etc/hosts
}
BATCH=/tmp/lmc-batch.$$

save_cmd()
{
echo "$@" >> $BATCH
}
[ -f $config ] && rm $config

# MDS node
${LMC} --add net --node $MDS --nid `h2elan $MDS` --nettype elan
|| exit 1
${LMC} --add mds --node $MDS --mds mds1 --dev /tmp/mds1 --size 100000
|| exit 1
${LMC} --add lov --lov lov1 --mds mds1 --stripe_sz $STRIPE_SIZE
--stripe_cnt 0 --stripe_pattern 0

# Client node
```

```
${LMC} --add net --node client --nid '*' --nettype elan || exit 1
${LMC} --add mtpt --node client --path /mnt/lustre --mds mds1
--lov lov1
# this is crude, but effective
let server_per_gw=($SERVER_CNT / $GW_CNT )
let tot_server=$server_per_gw*$GW_CNT
echo "Allocating $server_per_gw per gateway."
echo "For a total of $tot_server Blue Arcs"

let gw=$GW_START
let server=$SERVER_START
while (( $gw < $GW_CNT + GW_START ));
do
   echo "gw$gw"
   gwnode=`gw2mcr gw$gw`
${LMC} --add net --router --node $gwnode --tcpbuf $TCPBUF
--nid `h2ip $gwnode` --nettype tcp --port $PORT || exit 1
${LMC} --add net --node $gwnode --nid `h2elan $gwnode`
--nettype elan|| exit 1
${LMC} --add route --node $gwnode --nettype elan --gw `h2elan $gwnode`
--lo `h2elan
$CLIENT_LO` --hi `h2elan $CLIENT_HI` || exit 2
let i=0
while (( $i < $server_per_gw ));
do
   echo "server: $server"
   ba=ba$server
OBD_UUID=`awk "/$OST / { print \\$2 }" $UUIDLIST`
[ "$OBDUUID" ] && OBDUUID="--obduuid=$OBDUUID" || echo "$OST: no UUID"
# server node: networking info.
${LMC} --add net --node $ba --tcpbuf $TCPBUF --nid $ba --nettype tcp
--port $PORT || exit 1
# the device on the server
# note how it is asked to "join" the LOV
${LMC} --add lov --lov lov1 --node $ba $OBDUUID --ost bluearc
|| exit 3
# route to server
${LMC} --add route --node $gwnode --nettype tcp --gw `h2ip $gwnode`
--tgt $ba || exit 2
let server=$server+1
let i=$i+1
done
let gw=$gw+1
```

```
done
# now execute all the commands we have selected
$LMC_REAL --batch $BATCH
rm -f $BATCH
```

**Client testing multiple OST's:** The following configures a client node to talk with each BlueArc OST as a different device over IP. This was used with lctl addressing each BA OST to determine the throughput and led to the graphic overview of the performance.

```
  #!/bin/bash
config=${1:-echo-no-gw.xml}
LMC="save_cmd"
LMC_REAL="../../lustre/utils/lmc -m $config"

# TCP/IP servers
SERVER_START=0
SERVER_CNT=62
PORT=2432
TCPBUF=1048576

UUIDLIST=/home/bluearc/UUID.0916

h2ip ()
{
echo "${1}"
}
BATCH=/tmp/lmc-batch.$$
save_cmd()
{
echo "$@" >> $BATCH
}
[ -f $config ] && rm $confi


# Client node
${LMC} --add net --node client --tcpbuf $TCPBUF --nid '*' --nettype tcp
--port $PORT || exit 1
# this is crude, but effective
let server_per_gw=($SERVER_CNT / $GW_CNT )
let tot_server=$server_per_gw*$GW_CNT
let server=$SERVER_START
while (( $server < $SERVER_CNT + SERVER_START ));
do
   echo "server: $server"
```

```
  ba=ba$server
OBD_UUID=`awk "/$OST / { print \\$2 }" $UUIDLIST`
[ "$OBDUUID" ] && OBDUUID="--obduuid=$OBDUUID" || echo "$OST: no UUID"

# server node
${LMC} --add net --node $ba --tcpbuf $TCPBUF --nid $ba --nettype tcp
--port $PORT || exit 1
# the device on the server:
# the --add ost flags tells the client this is an OST!!
# ... and if $ba isn't a BA it sets up echo
${LMC} --node $ba --osdtype=obdecho --add ost || exit 3
# osc on client
${LMC} --node client --ost OSC_$ba
let server=$server+1
done
$LMC_REAL --batch $BATCH
rm -f $BATCH
```

**MDS with failover:** Notice how the active MDS is found with lustre-query and given to the client mount configuration. This script should NOT be run on the standby until failover takes place, in which case Kimberlite is responsible for setting up the new MDS.

```
  #!/bin/sh
LMC=/usr/local/cfs/lustre/utils/lmc

# LMC="echo lmc"
CONFIG=mcr-mds-failover.xml
LUSTRE_QUERY=/usr/local/cfs/lustre-failover/lustre-query
GW_NODE=mcr21
CLIENT_ELAN=`hostname | sed s/[^0-9]*//;`
OST_BA=ba50
OST_UUID=10400010-5dec-11c2-0b5f-00301700041a
MDS_DEVICE=/dev/sda3
MDS_SIZE=500000
TCPBUF=1048576
TCPPORT=988

MDSNODES=`$LUSTRE_QUERY -h emcri -s id=mds -f`
ACTIVEMDS=`$LUSTRE_QUERY -h emcri -s id=mds -a`

echo "MDS nodes: $MDSNODES, active: $ACTIVEMDS"
```

```
h2elan ()
{
echo $1 | sed 's/[^0-9]*//g'
}

h2ip ()
{
echo "${1}"
}

# create client node
$LMC -o $CONFIG --add net --node client --nid '*' --nettype elan
$LMC -m $CONFIG --add net --router --node mcr21 --tcpbuf $TCPBUF
--nid `h2ip $GW_NODE` --nettype tcp
$LMC -m $CONFIG --add net --router --node mcr21 --nid `h2elan $GW_NODE`
--nettype elan
$LMC -m $CONFIG --add route --node $GW_NODE --nettype elan
--gw `h2elan $GW_NODE` --tgt $CLIENT_ELAN
# create MDS node entries
for mds in $MDSNODES;
do
   elanaddr=`$LUSTRE_QUERY -h emcri -s id=$mds -e`
$LMC -m $CONFIG --add net --node $mds --nid $elanaddr --nettype elan
$LMC -m $CONFIG --add mds --node $mds --mds mds_$mds\ --dev $MDS_DEVICE
--size $MDS_SIZE
done

# create OST node entry
$LMC -m $CONFIG --add net --node $OST_BA --tcpbuf $TCPBUF --nid $OST_BA
--nettype tcp --port $TCPPORT
$LMC -m $CONFIG --add ost --node $OST_BA --obduuid $OST_UUID --obd bluearc
$LMC -m $CONFIG --add route --node $GW_NODE --nettype tcp --gw `h2ip $GW_NODE`
--tgt $OST_BA
# mount
$LMC -m $CONFIG --add mtpt --node client --path /mnt/lustre
--mds mds_$ACTIVEMDS --obd OSC_$OST_BA
```

## 33.4. Changelog

(1) Radhika Vullikanti (04/03/2003) - Updated the lmc manpage to reflect the new journal size option, added some other options that were missing (timeout, recovery_upcall, nid_exchange etc). Made corrections to some of the sample scripts available in this chapter.
(2) Radhika Vullikanti (01/03/2003) - Added a new manpage for llanalyze, lstripe
(3) Radhika Vullikanti (12/20/2002) - Corrected the examples in lustre-conf to reflect the new lmc
(4) Radhika Vullikanti (10/24/2002) - Updated the lmc manpage to reflect the changes made to the tool
(5) Radhika Vullikanti (10/24/2003) - Removed manpage for lstripe and added manpage for lfs.

# Part 4

# Appendices

CHAPTER 34

# Lustre Management Design Specifications

## 34.1. Introduction

In some of the earlier chapters of the book, we have described the various configuration management tools that are provided by Lustre - lmc (lustre make config), lconf (lustre configuration) and lctl (lustre control). The Lustre system configurations are described using eXtended Markup Language (XML) files. The entries that can be made in the XML file are described in a Data Type Description (DTD) file. The lconf can obtain configuration information for a node either from a file or from an LDAP server, the LDAP repository is specially convenient for large clusters. The configuration information stored on LDAP follows a schema similar to the XML configuration file. Lustre also provides Simple Network Management Protocol (SNMP) support.

In this chapter we will describe the Lustre DTD, LDAP schema and the SNMP schema. It can be used as a reference for Lustre configuration management.

## 34.2. Lustre DTD

A Data Type Description (DTD) file is used to describe the format of the XML configuration files used for Lustre. In this section we present the *lustre.dtd* file used as a basis for the lustre XML files.

First the top level entities are described:

```
<!-- Lustre Management DTD -->
<!-- basic entities -->
<!ENTITY % tag.content "(#PCDATA)">
<!ENTITY % tag.ref "
num CDATA #IMPLIED
name CDATA #IMPLIED
uuidref CDATA #REQUIRED">
<!ENTITY % tag.attr "
name CDATA #REQUIRED
uuid CDATA #REQUIRED">
<!-- main elements -->
<!ELEMENT lustre (node | mountpoint | ldlm |
mds | mdc | obd | ost | osc | lov | lovconfig)*>
```

The top level entities are usually compound structures, which follow one by one.

```
   <!ELEMENT node (network | profile)*>
<!ATTLIST node router CDATA #IMPLIED
               %tag.attr;>
<!ELEMENT network (server | port | route_tbl |
send_mem | recv_mem)*>
<!ATTLIST network type (tcp | elan | gm) 'tcp'
                %tag.attr;>
   <!ELEMENT route_tbl (route)*>
<!ELEMENT route %tag.content;>
<!ATTLIST route type (elan | tcp | gm) #REQUIRED
                gw CDATA #REQUIRED
                lo CDATA #REQUIRED
                hi CDATA #IMPLIED >
<!ELEMENT profile (ldlm_ref | network_ref | obd_ref | ost_ref |
osc_ref | mds_ref | mdc_ref | lov_ref | lovconfig_ref| mountpoint_ref)*>
<!ATTLIST profile >
<!ELEMENT mountpoint (path | fileset | mds_ref | osc_ref)*>
<!ATTLIST mountpoint %tag.attr;>
<!ELEMENT ldlm EMPTY>
<!ATTLIST ldlm %tag.attr;>
```

The next one is an *obd.* It defines the name and UUID of the object device and the file system type required to mount it.

```
   <!ELEMENT obd (fstype | device | autoformat)*>
<!ATTLIST obd %tag.attr; type (obdfilter | obdecho) 'obdfilter'>
<!ELEMENT ost (network_ref | obd_ref | failover_ref)*>
<!ATTLIST ost %tag.attr;>
<!ELEMENT mds (network_ref | fstype | device | autoformat |
          server_ref | failover_ref | node_ref )*>
<!ATTLIST mds %tag.attr;>
```

Our next element is an OSC, it requires the name and UUID to be defined:

```
   <!ELEMENT osc (ost_ref | obd_ref)*>
<!ATTLIST osc %tag.attr;>
<!ELEMENT mdc (network_ref | mds_ref)*>
<!ATTLIST mdc %tag.attr;>
<!ELEMENT lov (devices | mds_ref)*>
<!ATTLIST lov %tag.attr;>
<!ELEMENT lovconfig (lov_ref)>
<!ATTLIST lovconfig %tag.attr;>
<!ELEMENT devices (osc_ref)+>
```

```
<!ATTLIST devices stripesize CDATA #REQUIRED
                  stripecount CDATA #REQUIRED
                  stripeoffset CDATA #IMPLIED
                  pattern CDATA #REQUIRED>
   <!-- basic elements -->
<!ELEMENT fstype %tag.content;>
<!ELEMENT device %tag.content;>
<!ATTLIST device size CDATA #IMPLIED>
<!ELEMENT server %tag.content;>
<!ELEMENT port %tag.content;>
<!ELEMENT send_mem %tag.content;>
<!ELEMENT recv_mem %tag.content;>
<!ELEMENT autoformat %tag.content;>
<!ELEMENT path %tag.content;>
<!ELEMENT fileset %tag.content;>
<!-- id tag elements -->
<!ELEMENT network_ref %tag.content;>
<!ATTLIST network_ref %tag.ref;>
<!ELEMENT node_ref %tag.content;>
<!ATTLIST node_ref %tag.ref;>
<!ELEMENT profile_ref %tag.content;>
<!ATTLIST profile_ref %tag.ref;>
<!ELEMENT obd_ref %tag.content;>
<!ATTLIST obd_ref %tag.ref;>
<!ELEMENT mds_ref %tag.content;>
<!ATTLIST mds_ref %tag.ref;>
<!ELEMENT osc_ref %tag.content;>
<!ATTLIST osc_ref %tag.ref;>
<!ELEMENT ost_ref %tag.content;>
<!ATTLIST ost_ref %tag.ref;>
<!ELEMENT lov_ref %tag.content;>
<!ATTLIST lov_ref %tag.ref;>
<!ELEMENT lovconfig_ref %tag.content;>
<!ATTLIST lovconfig_ref %tag.ref;>
<!ELEMENT mdc_ref %tag.content;>
<!ATTLIST mdc_ref %tag.ref;>
<!ELEMENT mountpoint_ref %tag.content;>
<!ATTLIST mountpoint_ref %tag.ref;>
<!ELEMENT server_ref %tag.content;>
<!ATTLIST server_ref %tag.ref;>
<!ELEMENT failover_ref %tag.content;>
<!ATTLIST failover_ref %tag.ref;>
<!ELEMENT ldlm_ref %tag.content;>
```

```
<!ATTLIST ldlm_ref %tag.ref;>
```

### 34.3. LDAP Schema

As described earlier, the Lustre file system may use the LDAP server as a central repository for configuration information. The configuration information stored in LDAP follows the layout given in luste.schema file. The LDAP server allows adding, deleting, modifying and querying the database. The LDAP schema is described using objects and a set of attributes associated with them. In this section we give an example of the Lustre LDAP schema for a node with three services - client, MDS, OST. An LDAP schema consists of some fundamental object types that can be used to store the required configuration information for Lustre clients, we will now describe these object types with examples.

The example of a node entry for sys4.goober.org machine :

```
   dn: id=sys4.goober.org,
type=node,
fs=lustre
objectClass: lustreNode
nodeUUID: sys4UUID
id: sys4.goober.org
netUUIDs: sys4NetUUID
ldlmUUID: sys4LdlmUUID
profileUUID: sys4ProfUUID
routerUUID: sys4RouteUUID
fs: lustre
```

The sys4.goober.org node has a Profile entry. The profile shows what are services are available from this node. Which are client, ost and mds services:

```
   dn: profileUUID=sys4ProfUUID,
type=profile,
fs=lustre
objectClass: lustreNodeProfile
profileUUID: sys4ProfUUID
ostUUIDs: sys4OstUUID
mdsUUIDs: sys4MdsUUID
clientUUID: sys4ClntUUID
fs: lustre
```

ldlm configuration information

```
   dn: ldlmUUID=lustre1
LdlmUUID,
type=LDLM,
fs=lustre
```

```
objectClass: lustreLDLM
ldlmUUID: lustre1LdlmUUID
devName: lustre1Ldlm
fs: lustre
Network information: fnetUUID is the failover net UUID.
dn: netUUID=sys4NetUUID,
type=net,
fs=lustre
objectClass: lustreNetwork
netUUID: sys4NetUUID
id: sys4.goober.org
fnetUUID: sys4NetUUID
netType: tcp
netAddress: 192.168.30.253
port: 1234
recvMem: 2048
sendMem: 2048 fs: lustre
```

The sys4.goober.org have client service, so the LDAP server should have client specific information

```
   dn: clientUUID=sys4ClntUUID,
type=client,
fs=lustre
objectClass: lustreClient
clientUUID: sys4ClntUUID
mountUUIDs: sys4MntUUID
netUUID: sys4netUUID
fs: lustre
```

An example of MDS configuration information:

```
   dn: mdsUUID=sys4MdsUUID,
type=MDS,
fs=lustre
objectClass: lustreMds
mdsUUID: sys4MdsUUID
devName: sys4Mds
devUUID: sys4MdsDevUUID
lovUUID: sys4LovUUID
fs: lustre
```

A sample OBD configuration information:

```
   dn: obdUUID=sys4ObdUUID,
type=OBD,
```

```
fs=lustre
objectClass: lustreOBD
obdUUID: sys4ObdUUID d
evName: sys4ObdDev
devUUID: sys4OstDevUUID
fs: lustre
```

OSC configuration information

```
   dn: oscUUID=sys4OscUUID,
type=OSC,
fs=lustre
objectClass: lustreOSC
oscUUID: sys4OscUUID
devName: sys4OscDev
obdUUID: sys4ObdUUID
fs: lustre
```

OST configuration information

```
   dn: ostUUID=sys4OstUUID,
type=OST,
fs=lustre
objectClass: lustreOst
ostUUID: sys4OstUUID
devName: sys4Ost o
bdUUID: sys4ObdUUID
fs: lustre
```

Mount configuration information

```
   dn: mountUUID=sys4MntUUID,
type=mountPoint,
fs=lustre
objectClass: lustreMount
mountUUID: sys4MntUUID
mdcUUID: sys4MdcUUID
lovUUID: sys4LovUUID
mountPath: /mnt/lustre default: yes
fs: lustre
```

MDS configuration information

```
   dn: mdsUUID=sys4MdsUUID,
type=MDS,
fs=lustre
```

```
objectClass: lustreMds
mdsUUID: sys4MdsUUID
devName: sys4Mds
devUUID: sys4MdsDevUUID
lovUUID: sys4MdsLovUUID
fs: lustre
```

MDC configuration information

```
   dn: mdcUUID=sys4MdcUUID,
type=MDC,
fs=lustre
objectClass: lustreMDC
mdcUUID: sys4MdcUUID
devName: sys4MdcDev
mdsUUID: sys4MdsUUID
fs: lustre
```

Device configuration information: Here two device config info available.

OST device information

```
   dn: devUUID=sys4OstDevUUID,
type=device,
fs=lustre
objectClass: lustreDevice
devUUID: sys4OstDevUUID
id: sys4.goober.org
fid: sys4.goober.org
netUUID: sys4NetUUID
fnetUUID: sys4NetUUID
device: /dev/ost
auto: yes
fsType: extN
size: 50000
fs: lustre
```

MDS device information

```
   dn: devUUID=sys4MdsDevUUID,
type=device,
fs=lustre
objectClass: lustreDevice
devUUID: sys4MdsDevUUID
id: sys4.goober.org
fid: sys4.goober.org
```

```
netUUID: sys4NetUUID
fnetUUID: sys4NetUUID
device: /dev/mds
auto: yes
fsType: extN
size: 50000
fs: lustre
```

Lov configuration information

```
   dn: lovUUID=sys4LovUUID,
type=LOV,
fs=lustre
objectClass: lustreLOV
lovUUID: sys4LovUUID
oscUUIDs: sys4OscUUID
stripeOffset: 0
stripeSize: 4096
StripeCount: 1
pattern: 0
fs: lustre
```

**34.3.1. LDAP Access API's.** The MDS and OST server failover mechanism execute queries and updates to LDAP. When the failover system, such as RedHat's cluman has reactivated the service on the failover node, it updates the *active server* field in the configuration.

**lustre-modify:** -d dn -a attr=val [-h host | -p port |\n" " -u user | -w passwd | -v?]

- -h host LDAP hostname (default %s)
- -p port LDAP port (default %d)
- -u user LDAP user (default none)
- -w passwd LDAP port (default none)
- -d dn LDAP dn\n" " -a attr=val Modify attribs, attr1=val1+attr2=val2,val3+...
- -v Verbose output
- -? Show this help

The client system will initiate recovery after a timeout event takes place. It will query ldap to find the currently active MDS or OST node. This is done with the lustre-query command.

**lustre-query:** [-h host | -p port | -u user | -w passwd | -s search | -vimdneaf?]

- -h host LDAP hostname (default %s)
- -p port LDAP port (default %d)
- -u user LDAP user (default none)
- -w passwd LDAP port (default none)
- -s search LDAP filter (default %s)
- -v Verbose output

- -i Output the lustreCommonID attribute
- -m Output the lustreUUID attribute
- -d Output the lustreDesc attribute
- -n Output the lustreIPAddress attribute
- -e Output the lustreElanAddress attribute
- -a Output the lustreActiveUUID attribute
- -f Output the lustreFailoverUUID attribute
- -? Show this help

## 34.4. SNMP Schema

The SNMP support for Lustre consists of two major components - SNMP MIB, SNMP Agent. The SNMP MIB (Management Information Base)is a text file written in SMIv2 language, it describes the schema for the Lustre specific information that would be made available to the SNMP agent. The MIB would describe the various components of Lustre - OST, Client. MDS.

It is important that the Lustre Lite SNMP MIB presents information that is generally useful to system administrators and end users of Lustre Lite. This information should represent the logical IP network-level view of Lustre Lite, and needs to be implementation independent.

The SNMP Agent support for Lustre Lite will be implemented as an extension of the UCD-SNMP Agent. The SNMP Agent support will be implemented as a script (most likely Python) that is separate from UCD-SNMP. The script will be called by the SNMP Agent with an OID(object identifier) as a parameter. The script will retrieve the appropriate information that matches the given OID from the /proc/lustre tree, reformat that information, and feed it back to the SNMP Agent.

**34.4.1. Lustre filesystem MIB.** In this section, we present a sample of the MIB used in Lustre SNMP support.

```
=====================================================================
 Object Storage Targets
=====================================================================
ostNumber OBJECT-TYPE
SYNTAX Unsigned32
MAX-ACCESS read-only
STATUS current
DESCRIPTION         "The number of Object Storage Targets on this system."
     ::= { objectStorageTargets 1 }
ostTable OBJECT-TYPE
SYNTAX SEQUENCE OF OstEntry
MAX-ACCESS not-accessible
STATUS current
DESCRIPTION         "A table listing the Object Storage Targets available on
```

```
  this system. The number of entries in this table is available in ostNumber."
    ::= { objectStorageTargets 2 }
ostEntry OBJECT-TYPE
SYNTAX OstEntry
MAX-ACCESS not-accessible
STATUS current
DESCRIPTION         "Table entry with information an Object Storage Target
 on this system."
INDEX { ostIndex }
::= { ostTable 1 }
OstEntry ::=
SEQUENCE { ostIndex    Unsigned32,
           ostUUID     DisplayString,
           ostCommonName    DisplayString,
           ostCapacity      Counter64,
           ostFreeCapacity  Counter64,
           ostRowStatus     RowStatus     }
ostIndex OBJECT-TYPE
SYNTAX Unsigned32 (1..2147483647)
MAX-ACCESS not-accessible
STATUS current
DESCRIPTION         "Index into the table of Object Storage Targets on this
 system."     ::= { ostEntry 1 }
ostUUID OBJECT-TYPE
SYNTAX DisplayString
MAX-ACCESS read-only
STATUS current
DESCRIPTION         "The Lustre Universally Unique Identifier (UUID) for the
Object Storage Target."
::= { ostEntry 2 }
ostCommonName OBJECT-TYPE
SYNTAX DisplayString
MAX-ACCESS read-only
STATUS current
DESCRIPTION         "The common name of the Object Storage Target."

::= { ostEntry 3 }
ostCapacity OBJECT-TYPE
SYNTAX Counter64
MAX-ACCESS read-only
STATUS current
DESCRIPTION         "The capacity of the Object Storage Target in bytes."
:= { ostEntry 4 }
```

```
ostFreeCapacity OBJECT-TYPE
SYNTAX Counter64
MAX-ACCESS read-only
STATUS current
DESCRIPTION         "The remaining free capacity of the Object Storage
 Target in bytes."
::= { ostEntry 5 }
ostRowStatus OBJECT-TYPE
SYNTAX RowStatus
MAX-ACCESS read-write
STATUS current
DESCRIPTION         "The status of the row."
::= { ostEntry 6 }
====================================================================

   Metadata Servers
====================================================================
mdsNumber OBJECT-TYPE
SYNTAX Unsigned32
MAX-ACCESS read-only
STATUS current
DESCRIPTION         "The number of Metadata Servers on this system."
     ::= { metaDataServers 1 }
mdsTable OBJECT-TYPE
SYNTAX SEQUENCE OF MdsEntry
MAX-ACCESS not-accessible
STATUS current
DESCRIPTION    "A table listing the Object Storage Targets available on this
 system. The number of entries in this table is available in
mdsNumber."     ::= { metaDataServers 2 }
mdsEntry OBJECT-TYPE
SYNTAX MdsEntry
MAX-ACCESS not-accessible
STATUS current
DESCRIPTION     "Table entry with information an Object Storage Target on
 this system."
INDEX { mdsIndex }
::= { mdsTable 1 }
MdsEntry ::=
SEQUENCE { mdsIndex   Unsigned32,
           mdsUUID    DisplayString,
           mdsCommonName    DisplayString,
           mdsCapacity      Counter64,
```

```
          mdsFreeCapacity  Counter64,
          mdsRowStatus     RowStatus     }
mdsIndex OBJECT-TYPE
SYNTAX Unsigned32 (1..2147483647)
MAX-ACCESS not-accessible
STATUS current
DESCRIPTION   "Index into the table of Object Storage Targets on this
 system."
::= { mdsEntry 1 }
mdsUUID OBJECT-TYPE
SYNTAX DisplayString
MAX-ACCESS read-only
STATUS current
DESCRIPTION     "The Lustre Universal Unique Identifier (UUID) for the
Object Storage Target."
::= { mdsEntry 2 }
mdsCommonName OBJECT-TYPE
SYNTAX DisplayString
MAX-ACCESS read-only
STATUS current
DESCRIPTION    "The common name of the Metadata Server."
::= { mdsEntry 3 }
mdsCapacity OBJECT-TYPE
SYNTAX Counter64
MAX-ACCESS read-only
STATUS current
DESCRIPTION    "The capacity of the Metadata Server in bytes."
::= { mdsEntry 4 }
mdsFreeCapacity OBJECT-TYPE
SYNTAX Counter64
MAX-ACCESS read-only
STATUS current
DESCRIPTION    "The remaining free capacity of the Metadata Server in
 bytes."
::= { mdsEntry 5 }
mdsRowStatus OBJECT-TYPE
SYNTAX RowStatus
MAX-ACCESS read-write
STATUS current
DESCRIPTION    "The status of the row."
::= { mdsEntry 6 }
==================================================================
   Lustre Clients
```

```
==================================================================
cliNumber OBJECT-TYPE
SYNTAX Unsigned32
MAX-ACCESS read-only
STATUS current
DESCRIPTION    "The number of lustre clients (regardless of their current
 state) which are currently on this system."
::= { lustreClients 1 }
cliTable OBJECT-TYPE
SYNTAX SEQUENCE OF CliEntry
MAX-ACCESS not-accessible
STATUS current
DESCRIPTION    "A table listing the lustre clients and their configurations.
  The current number of entries is specified by cliNumber."
  ::= { lustreClients 2 }
cliEntry OBJECT-TYPE
SYNTAX CliEntry
MAX-ACCESS not-accessible
STATUS current
DESCRIPTION    "Information about a single client."
INDEX { cliIndex }
::= { cliTable 1 }
CliEntry ::=
SEQUENCE {cliIndex     Unsigned32,
          cliUUID      DisplayString,
          cliCommonName      DisplayString,
          cliMDSUUID         DisplayString,
          cliMDSCommonName   DisplayString,
          cliUsesLOV         TruthValue,
          cliLOVStripeDepth  Unsigned32,
          cliLOVStripeFactor Unsigned32,
          cliMountPoint      DisplayString,
          cliRowStatus       RowStatus     }
cliIndex OBJECT-TYPE
SYNTAX Unsigned32 (1..2147483647)
MAX-ACCESS not-accessible
STATUS current
DESCRIPTION     "Index into the table of lustre clients on this system."
     ::= { cliEntry 1 }
cliUUID OBJECT-TYPE
SYNTAX DisplayString
MAX-ACCESS read-only
STATUS current
```

```
DESCRIPTION     "The Lustre Universal Unique Identifier (UUID) for the
 Client File System."
::= { cliEntry 2 }
cliCommonName OBJECT-TYPE
SYNTAX DisplayString
MAX-ACCESS read-only
STATUS current
DESCRIPTION      "The Lustre Common Name for the Client File System."
    ::= { cliEntry 3 }
cliMDSUUID OBJECT-TYPE
SYNTAX DisplayString
MAX-ACCESS read-only
STATUS current
DESCRIPTION      "The UUID of the Metadata Server to which this client is
 connected."
::= { cliEntry 4 }
cliMDSCommonName OBJECT-TYPE
SYNTAX DisplayString
MAX-ACCESS read-only
STATUS current
DESCRIPTION      "The Common Name of the Metadata Server to which this
 client is connected."
::= { cliEntry 5 }
cliUsesLOV OBJECT-TYPE
SYNTAX TruthValue
MAX-ACCESS read-only
STATUS current
DESCRIPTION      "This variable is true(1) if this client is using a Logical
 Object volume (LOV), and false(2) otherwise."
::= { cliEntry 6 }
cliLOVStripeDepth OBJECT-TYPE
SYNTAX Unsigned32
MAX-ACCESS read-only
STATUS current
DESCRIPTION      "The size of the blocks which are distributed over the OSTs
in the Logical Object Volume. The value should be zero if cliUsesLOV
 is    false."
::= { cliEntry 7 }
cliLOVStripeFactor OBJECT-TYPE
SYNTAX Unsigned32
MAX-ACCESS read-only
STATUS current
DESCRIPTION      "The number of OSTs over which the stripe of the Logical
```

```
 Object Volume are distributed. The value should be zero if cliUsesLOV is
 false."
::= { cliEntry 8 }
cliMountPoint OBJECT-TYPE
SYNTAX DisplayString
MAX-ACCESS read-only
STATUS current
DESCRIPTION      "The local mount point on this system of the filesystem
 associated with this client."
::= { cliEntry 9 }
cliRowStatus OBJECT-TYPE
SYNTAX RowStatus
MAX-ACCESS read-write
STATUS current
DESCRIPTION      "The status of the row."
::= { cliEntry 10 }
cliOSTTable OBJECT-TYPE
SYNTAX SEQUENCE OF CliOSTEntry
MAX-ACCESS not-accessible
STATUS current
DESCRIPTION     "A table all of the Object Storage Targets accessed by all
 of the clients on this system."
::= { lustreClients 3 }
cliOSTEntry OBJECT-TYPE
SYNTAX CliOSTEntry
MAX-ACCESS not-accessible
STATUS current
DESCRIPTION     "Information about the Object Storage Target(s) used by a
 single client on this system."
INDEX { cliIndex, cliOSTIndex }
::= { cliOSTTable 1 }
CliOSTEntry ::=
SEQUENCE {cliOSTIndex     Unsigned32,
          cliOSTUUID      DisplayString,
          cliOSTCommonName      DisplayString,
          cliOSTRowStatus       RowStatus     }
cliOSTIndex OBJECT-TYPE
SYNTAX Unsigned32 (1..2147483647)
MAX-ACCESS not-accessible
STATUS current
DESCRIPTION    "Index into the table of OSTs used by the lustre clients on
 this system."
::= { cliOSTEntry 1 }
```

```
cliOSTUUID OBJECT-TYPE
SYNTAX DisplayString
MAX-ACCESS read-only
STATUS current
DESCRIPTION    "The Lustre Universally Unique Identifier (UUID) of an Object
 Storage Target accessed by this client."
::= { cliOSTEntry 2 }
cliOSTCommonName OBJECT-TYPE
SYNTAX DisplayString
MAX-ACCESS read-only
STATUS current
DESCRIPTION     "The Lustre Common Name of an Object Storage Target accessed
 by this client."
::= { cliOSTEntry 3 }
cliOSTRowStatus OBJECT-TYPE
SYNTAX RowStatus
MAX-ACCESS read-only
STATUS current
DESCRIPTION     "The status of the row."
::= { cliOSTEntry 4 }
```

ACKNOWLEDGEMENT 34.4.1. This document was written with the help of information from Christopher Morrone(LLNL) - SNMP, and S. Ravi(HP-India) - LDAP .

CHAPTER 35

# Lustre procfs interface

## 35.1. Introduction

The Lustre file system consists of many modules that can be stacked in multiple ways to create devices that serve a specific functionality. There are, broadly, three classes of devices that compose a Lustre filesystem, the **CLIENT**, the **MDS** and the **OST**. And all of these, or a subset of these can be co-located on the same physical machine. The stack for each device consists of many Lustre modules and multiple combinations exist for creating each class of Lustre device. For e.g. a simple Lustre filesystem may consist of, apart from a metadata server, a single **CLIENT** talking to a single **OST**, or a client striping data over multiple **OSTs**. During setting up of the client for the former case, there would be no **Logical Object Volume (LOV)**, while in the latter case, the **CLIENT** stack requires a **LOV**. Similarly, the **OST** stack requires that it be built over a journalling file system. There are multiple candidates for the backend filestore over which an **OST** stack is built. This backend filestore currently is **extN**, but it could also be any other journalling file system (reiserFS, for example).

In short, the Lustre stack is highly configurable and modular, and it would be useful to expose to the user space, for every Lustre mount point, how the stack underneath that mount point is built. The **/proc/fs/lustre** interface (hereafter referred to as LprocFS) exposes this information to the user space. In addition, it also exposes various file system statistics, like disk free space, total disk space, number of files free, number of files total, basic block size for every Lustre mount point, which changes as files are read/written on the client.

## 35.2. LprocFS

**35.2.1. LProcFS variables.** LProcFS provides an infrastructure for materializing variables under **/proc/fs/lustre**. These variables can be classified as:

(1) **Static variables :** These variables are created when an obd_device or a module registers itself with LProcFS. They are known as “static” because the names of these variables are known at compile time, and these get created in the LProcFS hierarchy during the insertion of the module or when the obd device instance is attatched.

(2) **Dynamic variables:** These variables are added to the LProcFS hierarchy "after" a module has been inserted, or after a device has been attached. These are added by invoking the call to **lprocfs_add_vars()** described later.

In addition, the variables can also be classified as module level variables, obd instance variables or "other" variables.

(1) **Module variables**: These are variables that are specified when a module is inserted using insmod, or are added later under the LProcFS root for that module. These remain in existence until the module has been removed using rmmod. These are created under **/proc/fs/lustre/<module-name>/**

(2) **Obd instance variables:** These are variables that are present for every OBD instance. These are created under /proc/fs/lustre/<obd-module>/<obd-instance>. These variables can be either static, in which case, they are created when the obd device attaches itself (i.e. along with the above directory creation) in o_attach, or may be added later using **lprocfs_add_vars** at any depth below the instance's root directory. Note that there may be multiple obd instances for an obd module, and they will materialize under different names in the LProcFS hierarchy.

(3) **Other variables:** There are variables that do not strictly fall under either of the above categories (e.g. mount point variables etc). These can also be created, either statically (by invoking the generic API **lprocfs_register()**) or dynamically by the call to **lprocfs_add_vars.** For example, the mount point variables for every Lustre lite mount point are created in the above fashion.

### 35.2.2. LProcFS flow

**.** LProcFS is available only if *lproc* support has been compiled into the kernel (i.e. *CONFIG_PROC_FS* is defined and true) and the macro **LPROCFS** (in include/linux/lprocfs_status.h) is defined. This is true, by default. The rest of the discussion assumes this to be the case.

35.2.2.1. *Initialization.* The LprocFS module gets initialized when the **obdclass** module is inserted using insmod. This module **must** be inserted before any services offered by LProcFS can be used. Inserting this module creates **/proc/fs/lustre** and the global variable **proc_lustre_root** (which is the root for the hierarchy) is instantiated.

35.2.2.2. *Creating hierarchies.* After the LProcFS module initialization is complete, directories/variables can be added below the root of the tree (i.e. below **/proc/fs/lustre**). These hierarchies can be created by invoking any of the LprocFS registration APIs ( **lprocfs_obd_attach** for OBD devices and **lprocfs_register** for everything else ). LProcFS hierarchy for an OBD device instance is created by invoking **lprocfs_obd_attach** in the overloaded o_attach function for the OBD device. Similarly, for all obd modules, the registration with LprocFS is performed during module insertion via insmod. Once an LProcFS hierarchy is created, variables can be added below them by the call to **lprocfs_add_vars()**. LProcFS variables and registration APIs are discussed in the section on data structures.

35.2.2.3. *Cleanup.* Removing hierarchies follows exactly the same design paradigm as registration. There are two APIs provided, **lprocfs_obd_detach()** for removing OBD device LProcFS hierarchies (created using lprocfs_obd_attach) and **lprocfs_remove()** for removing other hierarchies (created using lprocfs_register).

## 35.3. Data Structures

This section discusses the APIs provided by LProcFS to enable registration/deregistration of hierarchies, and any important interface data structures that LprocFS uses.

**35.3.1. LProcFS data structures.** Every variable that needs to materialize in the lprocfs hierarchy needs to define an instance of the following structure:

```
struct lprocfs_vars {
char *name; /* Name of the variable.*/
read_proc_t *read_fptr; /*Function pointer to get value of name */
write_proc_t *write_fptr; /*Function pointer to set value of name */
void *data; /*Context sensitive data for this variable.  */
};
```

- name: The "name" variable defines the name by which the variable must appear below the root of the hierarchy. The token "/" in the name demarcates a directory. Hence a name of "a/b/c/d" will create (if not present already) the proc directory entries a, b, c below the specified root, and a proc variable entry with the name"d" below the directory "c".
- read_fptr, write_fptr: The "read_fptr" and the "write_fptr" are function pointers that need to be provided for reading/writing into this LProcFS variable. Currently all LProcFS variables are read only.
- data: "data" allows the caller to add context to the newly created proc entry. During reading/writing of this variable, the value assigned to "data" is passed into the read function call invoked by the user space process trying to read the value of this variable.

A null-terminated list of such variables may be specified during regsitration of a new hierarchy (for OBD devices, modules, mount points). These create the root of the hierarchy, and add all such variables below the newly created root. The API discussion provides details.

**35.3.2. LProcFS APIs.** The following APIs are provided by LProcFS to allow users to setup/remove arbitrarily deep proc hierarchies below **/proc/fs/lustre**

- `struct proc_dir_entry *lprocfs_register(const char *name, struct proc_dir_entry *parent, struct lprocfs_vars *list, void *data)`

  **Parameters:**

  **name:** Name for root of the tree that will be newly created. Can be arbitrary. For modules, it is usually the module name.

  **parent:** The proc entry for the parent. For OBD modules, this could be (and is indeed the case currently) **proc_lustre_root.**

  **list:** A null terminated "list" of **lprocfs_vars** variables. These will appear as a hierarchy below the directory entry for **"name"**created during the invocation of this call.

  **data:** If "data" is non-NULL and no "data" variable is defined for an individual lprocfs_vars variable, then this is the "data' variable in the proc structure created for that specific lprocfs_vars variable in "list"

  **Returns:**

  The root of the newly created hierarchy on success, NULL otherwise.

  **Description:**

  The above API is invoked during module registration and during mount point registration. It is expected that the pointer returned by an invocation to this call is stored by the caller and returned back during deregistration of this hierarchy. OBD modules store the pointer in the field **typ_procroot** in the per-obd module structure **struct obd_type.** Also, OBD modules provide their common name for the "name" parameter. Hence all OBD modules appear under **/proc/fs/lustre/<obd_module_name>/** after successfully invoking this call.

- `int lprocfs_obd_attach(struct obd_device *dev, struct lprocfs_vars *list)`

  **Parameters:**

  **dev:** Pointer to the obd_device for which the new hierarchy is being created

  **list:** Null terminated list of **lprocfs_vars** variables.

  **Returns:**

  zero on success, non-zero otherwise

  **Description::**

This API is invoked when an OBD device wants to create variables under the LProcFS hierarchy. By default, these variables are created under **/proc/lustre/root/<obd-module-name>/<obd-device-name>/.** This indirectly invokes the more generic **lprocfs_register** call and stores its return proc pointer in the field **obd_proc_entry** in struct **obd_device**.
The call to this API is made, usually from the overloaded o_attach method for the OBD device.

- `int lprocfs_add_vars(struct proc_dir_entry *root, struct lprocfs_vars *list, void *data)`

  **Parameters:**

  **root:** The root value returned during registration, or (for OBD devices) stored in **obd_proc_entry** in **obd_device instance.** (populated during the previously invoked lprocfs_obd_attach)

  **list:** Null terminated list of lprocfs_vars variables.

  **data:** If "data" is non-NULL and no "data" variable is defined for an individual lprocfs_vars variable, then this is the "data' variable in the proc structure created for that specific lprocfs_vars variable. in "list".

  **Returns:**

  zero on success, non zero otherwise.

  **Description::**

  This API is used by users of LProcFS to add, post-initialization, new LProcFS variables to an existing hierarchy. These are used for adding the so called "dynamic" variables defined previously. The "root" value passed in must be the one returned from the generic lprocfs_register, or for OBD devices which have already been registered (i.e for which lprocfs_obd_attach has been called) must be the **obd_proc_entry** in the struct **obd_device** instance for that OBD device.

- `int lprocfs_obd_detach(struct obd_device *dev)`

  **Parameters:**

  **dev:** The instance of the OBD device for which LProcFS deregistration needs to be performed.

  **Returns:**

  zero on success, non-zero otherwise.

  **Description:**

  It is expected that "dev" had registered successfully with LProcFS before this API is invoked. For OBD devices, LProcFS deregistration is currently being performed in the o_detach method for the OBD device. In addition to removing all LprocFS variables below root, it also removes root and sets the **obd_proc_entry** for dev toNULL.

- `void lprocfs_remove(struct proc_dir_entry* root)`

**Parameters:**

**root:** The root of the LProcFS subtree that needs to be entirely removed.

**Returns:**

Zero on success, non-zero otherwise.

**Description:**

This function is the generic counterpart of lprocfs_obd_detach. It removes all proc entries under **root, including root.** It is expected that the caller will set the root pointer (points to garbage after the function successfully returns) to NULL after invocation of this call.

### 35.4. Changelog

**1.:** Radhika Vullikanti (02/28/2003) - Added this document made available by Intel describing the lprocfs infrastructure.

CHAPTER 36

# Lustre Kernel Modifications

Except for some trivial kernel modifications to export symbols, Lustre introduces three somewhat significant pathces for the Linux kernel.

## 36.1. Meta Data Locking

The Linux kernel uses *struct dentry* data structures to traverse the file system. The Linux VFS is very careful in pinning and releasing these entries from a cache. In a distributed file system the validity of these cached items is something that needs to be guaranteed with locks.

In order to do so we made two modifications to the Linux VFS:

(1) Involve the file system not only upon first access to a *dentry* but also let it know when the *dentry* is no longer used by the VFS.
(2) Indicate to the file system what the *intent* of the use of the dentry is, when the dentry is the for the component of a pathname.

Issue 1 assists Lustre to keep dentries valid with locks, while item 2 allows it to select the correct locks.

The Linux 2.4 implementation of these changes was done in such a way to not require any changes to other file systems, to only minimally affect the stability of the system. In 2.5 we expect to do a more elegant version of the same mechanism.

This patch has been discussed with Linus, Stephen Tweedie and Al Viro and we expect that it will go into the 2.5 kernel with some improvements and discussion but effectively without protest.

The Linux 2.5 version of this patch is much similar but simpler than the 2.4 version, since the intents can now be embedded in the nameidata.

## 36.2. Secure Pointers

It is very attractive to send a pointer to a remote system, which it can use in a future RPC. As an example, when a file is opened on the MDS, the MDS may want the client file system to know the file pointer so that it can supply that when the file needs to be closed.

There are two problems with this approach:

(1) When the file pointer is received by the MDS, how can the system be certain that the pointer is in fact a valid pointer?
(2) How can the system be sure that the pointer is the one originally sent to the client?

In order to address these issues we have taken two steps:

(1) Given a slab cache and a pointer, we have a validation pointer that checks that the pointer is pointing to a valid, allocated object in the slab cache.
(2) When the pointer is first given to a remote system a random number is also given to the remote station. When the pointer is found to be valid, access to the data is granted only if the random numbers sent in by the remote and held on the owner match. [This is not a kernel patch.]

I don't expect significant difficulties in getting this patch into the kernel, although it may well see some modifications before it is accepted.

There are no fundamental changes in this patch for Linux 2.5.

### 36.3. Socket Data Sent Callbacks

When the Linux TCP stack sends a packet out, it does not have a callback when that packet has left the system and can be freed again. This forces networking to use synchronous writes to socket file descriptors.

We have added a method to receive such notification from the stack.

This patch is not likely to be accepted in the 2.5 kernel as is, but a more general asynchronous socket I/O implementation will eliminate the need for the callback.

CHAPTER 37

# System Evolution

Version: 0.1, 6/8/02

## 37.1. Evolution of Lustre modules

The following table gives an overview of the expected evolution of Lustre modules.

The following markers indicate how serious the changes are that we anticipate:

**no marker:** not a very serious change
***:** a serious change to a code module, but still clearly an addition of features, not a re-write
****:** the module will undergo very substantial changes. Existing features will be preserved, but new functionality will likely displace most existing features.

| | LLite | LL Perf | Performance & Clustered MD | Security & T10 | GNS & Mgmt |
|---|---|---|---|---|---|
| **Portals** | Scalable allocation of objects | | Performance tweaks | | |
| **NALs** | DMA support, GM Nal | ** 0-copy TCP, DMA support, async TCP | Performance tweaks | | |
| **OBDclass** | Global naming of devices, LDAP related support | | | Configuration security | Dynamic reconfiguration of module stacks |
| **ptlrpc** | More support for recovery, statistics | * Event driven request processing | Adaptation to run over interrupt free networks | ** significant changes for security | |
| **OBDfilter** | | * File extents, preallocation of objects | ? Optimization for parallel I/O | * NASD or GSS-API compatible security | |
| **ldlm** | | Performance tweaks | * Distributed instead of server managed resources | Robustness of distributed recovery | |
| **OSC/OST** | Recovery from failed clients, OST/MDS recovery | | * Parallel I/O changes, interaction with MDS cluster | | * Data migration support, snapshot support, dynamic reconfiguration |
| **MDC/MDS** | Recovery and metadata locking support | | ** Load balancing over block allocation groups, cluster recovery | Auditing filter file system and auditing of authorization | Fileset & snapshot support |
| **LLite** | Recovery and metadata locking support, pagecache | ** Metadata writeback caching, page cache improvements | * Exploit clustered metadata, parallel I/O and page cache improvements | * ACL and file data crypting support | Snapshot support |
| **Resource DB** | Full configuration information, support for redundant failover OST | More options for file striping | ** Cluster resource database, more support for recovery, group membership in MDS cluster | * Kerberos infrastructure, group membership data bases | |
| **Other** | Basic SNMP support | | ** Several new daemons to support cluster events, membership and recovery | ** T10 OSC/OST modules, Kernel level PAG support | * Filtering GNS file module |

TABLE 1. Evolution of Lustre modules

CHAPTER 38

# Project Uncertainties

Version: 0.1, 6/8/02

Version: 0.2, 6/27/02 - added Terry's questions about quorum.

## 38.1. Design/performance uncertainties and alternatives

The Lustre design has matured to a roadmap that can be executed with the expectation that a stable system will be achieved. In several areas we are entering new terrain and this appendix gives an overview of the areas where we feel the greatest uncertainty. The items have been ordered first by chronological order then by quality.

**Lustre Lite:**

- **Recovery:** Will the recovery of the cluster scale to 1,000 node clusters without introducing complex timeout problems? I anticipate that the number of interactions between systems is fairly large and this might force rather long timeouts. A more scalable solution can be implemented by using a hierarchical structure in the cluster that distributes the load in a tree like fashion, but this adds considerable complexity to the system.
- **Metadata locks:** Metadata locking has always been plagued by numerous unanticipated deadlock problems etc. Hopefully we haven't made too many mistakes. The remedy is unfortunately merely sweat.
- **Elan networking:** While clearly offering excellent performance, the overall stability of the system is insufficient. This will require serious collaboration from the vendor to cure the problems.

**Lustre Lite Performance:**

- **Performance of Linux kernel:** I fear that Linux 2.4 does not nearly have the page cache sophistication needed to get top notch throughput in the I/O and networking paths. Almost certainly Lustre Lite performance will have to run on an early 2.5 kernel to show things like 95% of the raw performance. A remedy exists by making serious changes to the Linux kernel.
- **Metadata writeback cache:** InterMezzo pioneered the write back caches for metadata. One area of concern is a flow control, back-pressure mechanism to prevent situations where too many outstanding changes can start to seriously interfere with lock revocations.

**Collaborative cache:** A collaborative cache can lead to un-precedented scalability improvements for reads. However, failure recovery and collaboration management are delicate problems that have seen relatively little study.

**Lustre Clustered Metadata & Performance:**

**Clustered metadata:** We believe that we have good solutions to distributing metadata updates over a cluster. However a number of serious questions remain in the area of recovery, which is very complex even without clustering.

**Generic clustering infrastructure:** Our first thought about the metadata cluster was to give it a standard cluster infrastructure, involving quorum, membership etc. Questions have been asked about the wisdom of this approach, given experiences with such subsystems in larger clusters. An alternative approach is to have a failover nodes for every service unit thtat is being offered.

**Lustre T10/Security:**

**Object Security:** The precise implementation roadmap of the object security protocol remains somewhat open ended. The fundamental choice falls between a Kerberized service and a NASD style security model.

**Security Performance:** Can we limit the consequences for performance of strict security in our system. Industry experience here hasn't been good.

**Lustre Mgmt/Global Namespaces:**

**Management:** Can SNMP based management scale effectively to many 1000's of nodes. If not what are the options to remedy the problems.

CHAPTER 39

# Design Alternatives

During the design of Lustre many alternative solutions were considered. In some cases the choices were major architectural alternatives, such as symmetric versus non-symmetric cluster file systems. In other cases API's were chosen, such as for message passing. Often, existing software was rejected or used. This chapter describes some of the choices we have made.

## 39.1. Architectural Choices

**39.1.1. Networking.** Many architectural alternatives for the networking used by Lustre have been proposed. Early on we realized that we needed:

(1) An API that could support multiple networks.
(2) It should include support for remote DMA.
(3) It should allow advanced request processing.
(4) It should be perfect or somewhat mallable, e.g. not completely standardized

The following candidates were seriously considered:

**Roll your own:** Many parties suggested we roll our own networking library. We found the Portals API well-thought-through and found that it addressed many issues we had faced in other file system projects.

**VIA:** VIA, or a subset, has been proposed for Lustre networking. This was rejected on the grounds that (i) the event delivery API was insufficiently rich to handle request processing smoothly, and (ii) the NAL and forwarding abstractions found in Portals which allow us to easily run over many networks, were not present.

**DAFS RPC protocol:** DAFS was a serious candidate. Lustre network format is quite aligned with DAFS packaging. However DAFS was not defined in sufficient generality (just for the DAFS file system). Also, DAFS lacked the NAL and forwarding abstractions offered by Portals.

**Quadrics Kernelcomms API:** This is a good API which we could have used. It is not that dissimilar from Portals.

**Infiniband:** This was rejected because it was too complex and large.

**39.1.2. Clustering.** Many questions still remain in the area of what the best clustering arrangement is for a file system like Lustre. Alternatives that have been rejected are:

**One Symmetric cluster:** The clustering algorithms for membership and quorum have shown poor scalability and are very complex. In large clusters, node failures might occur relatively frequently and it would be burdensome to initiate a cluster recovery all the time. Finally, the trust models proposed make it attractive to limit the true cluster to a small set of metadata servers and to regard clients as a relatively powerless outside system.

Still under consideration are the following choices:

**Failover only:** Given the many complications in fully fledged clustering software, is it possible to build a metadata cluster that merely uses failover pairs for redundancy? This could considerably reduce the complexity of the Lustre software.

**Clustering Toolkits:** Ensemble and Spread are two well known toolkits. These toolkits are in use in cluster file systems, for example in the Polyserve file system. The code quality of both toolkits is good and the complexity of the algorithms is considerable. The team at present feels that if we could end up with something much simpler for the MDS cluster, that would be very desirable.

**39.1.3. Metadata.** Many alternatives have entered our mind for metadata updates:

**Symmetric metadata:** One might argue that limiting the metadata handling to an MDS cluster limits the scalability. If metadata is distributed over many systems two problems surface: (i) the cluster infrastructure of that cluster becomes more complicated; membership, quorum, and recovery will be harder to arrange when a very large collection of metadata servers are present, and (ii) distributed metadata updates are a challenge. Most existing systems have gone for two phase commit style updates which is very slow and prone to indefinite hangs. It is possible that our intent lists and orphan handling is sufficient to overcome these issues, but we have not yet explored this. It is interesting that IBM GPFS has migrated towards a metadata server model (see FAST proceedings).

**Directory structures:** Many file systems have used various kinds of btree's as directory storage. We considered this, but found that the hashing to blocks as is offered by the HTree directories in Ext3 is more amenable to loadbalancing resources over a cluster. The subdivision of directory data over multiple compute nodes seems difficult unless a mapping to blocks can be computed.

**File System Layout:** One might argue that the metadata layout of ext3 is primitive in comparison with XFS. The reason we have rejected the XFS choice is that the primitive nature of the ext3 data layout allows us to subdivide the file system in block groups and spread the block groups over multiple systems in the metadata cluster. Historically it is known that matching disk layout to cluster requirements is an important aspect of cluster file system design and we hope that the track we are on is correct.

**39.1.4. File I/O.**

**39.1.5. Protocols.** We have contemplated keeping the client server protocol close to DAFS or NFSv4. Unfortunately there were a number of issues with that approach. The NFSv4 provides little or no support to recover in the presence of metadata updates that are write behind or flushed asynchronously to disk. Secondly, the storage networking did not integrate well with Portals. In addition, the object storage components of Lustre would have to be separated from the NFS protocol and the adaptations to the core protocol required for this looked cumbersome.

Instead we have implemented a simple new protocol documented in earlier chapters.

## 39.2. Component choices

**39.2.1. Lock Manager.** We have decided not to use the open source IBM DLM. It appeared unnecessarily complex and not sufficiently stable.

**39.2.2. File systems.** XFS and ReiserFS have often been mentioned as good candidates underlying Lustre. We have chosen to primarily use ext3 for two reasons: (i) performance differences are minimal if present at all and ext3 performance is improving far more than that of other file systems, and (ii) the code size of ext3 is 10% that of Reiser and 3% that of XFS. Keeping things simple seems worth it.

CHAPTER 40

# Portals Diffs

On May 18, 2003 Lustre started to maintain its own Portals version. This page documents important differences for NAL writers and lustre developers. These differences were made to allow portals to operate efficiently as a kernel service and to provide guarantees that all network operations can be made to complete in finite time.

At this time (19th Feb 2004) the differences documented hare are implemented on CVS branch b_cray_portals_merge.

The first section documents differences at the API. The next section documents differences in the lib/NAL interface. The next section describes the router/NAL interface and the final section contains some notes about compatibility with Cray portals.

## 40.1. Differences at the API

Differences in API prototypes and overall functionality between lustre portals and the portals specification are summarised below. The following sections provide a greater level of detail on the non-trivial topics.

- **Message Ordering**
  Applications should not rely on lustre portals to preserve message order.
- **Handles**
  Lustre portals handles are valid only for the lifetime of the relevant object.
- **Event Completion Status**
  Completion status is determined by the value of ptl_event_t::ni_fail_type as described in the 3.3 spec.
- **Start Events**
  Lustre portals does not support START events and END events cannot be disabled. However if an MD is created with PTL_MD_NONE, no events will be created for the MD.
- **Event Callbacks**
  Lustre portals supports event callbacks as described in the 3.3 spec with the exception that it does *not* support the calling of data movement functions from these callbacks.
- **PtlMDUnlink(), ptl_event_t::unlinked**
  Explicit MD unlinking in lustre portals is asynchronous. Unlink completion is flagged by an event with its `unlinked` flag set.

- **ptl_size_t**
  The 3.3 spec says ptl_size_t should be a 64 bit unsigned integer. Lustre portals implements it as a 32 bit unsigned integer.
- **ptl_pid_t**
  Lustre NALs initialise themselves with a pid of 0. Apps should not expect or require any other pid.
- **ptl_uid_t, PTL_UID_ANY, PtlGetUid(), ptl_event_t::uid**
  User ID is not implemented. The corresponding event field is absent.
- **ptl_jid_t, PTL_JID_ANY, PtlGetJid(), ptl_event_t::jid**
  Job ID is not implemented. The corresponding event field is absent.
- **PtlACEntry()**
  Access control lists are not implemented in lustre portals.
- **PtlMEAttachAny()**
  New in 3.3 and not implemented in lustre portals.
- **PltPutRegion(), PtlGetRegion(), PtlGetPut()**
  New in 3.3 and not implemented in lustre portals.
- **Error Codes**
  The portals spec s
- **PTL_MTU, PTL_MD_MAX_IOV**
  Lustre portals requires its NALs to support a maximum message size of at least PTL_MTU, and a maximum number of fragments of at least PTL_MD_MAX_IOV.
- **ptl_ni_limits_t**
  The ptl_ni_limits_t type is not implemented. Equivalent limits are described in the table below. The userspace limits are defined in <portals/lib-p30.h>.

| 3.3 Spec | Userspace lustre portals | Kernel lustre portals |
|---|---|---|
| max_mes | `#define MAX_MES 2048` | dynamic |
| max_mds | `#define MAX_MDS 2048` | dynamic |
| max_eqs | `#define MAX_EQS 512` | dynamic |
| max_ac_index | N/A | |
| max_pt_index | determined by NAL | |
| max_me_list | dynamic | |
| max_getput_md | N/A | |

**40.1.1. Message Order.** Message ordering in lustre portals depends on the NALs doing the actual communication. Some NALs do *not* preserve message order, and gateway routers used in multi-cluster configurations make no guarantees about message order.

As a general rule applications are safer and more portable if they do not rely on the network to preserve message order. Lustre itself does not require any such guarantees.

**40.1.2. Completion in Finite Time.** Lustre portals guarantees that network operations can be completed in finite time and it provides a final completion event in all cases. The success or failure of the network operation is signalled in Once any kind of network operation is underway, it can be caused to complete in finite time by unlinking the MD associated with it. For example, if the target of a GET (i.e. the data source) crashes, the GET will not complete. A robust application, coded to recover from network failures, will time out and unlink the GET's MD. Lustre portals guarantees that this will cause a completion event to be generated in finite time. When the application receives this event it may now re-use the data sink buffers.

It works like this...

(1) Lustre portals handles are unique for the life of the network interface they are for. This ensures that once an MD has been unlinked, its handle becomes invalid and it will never become valid again.[1]

    Consider the case of using PtlMDUnlink() to terminate a network operation that itself will unlink the MD on completion. The call to PtlMDUnlink() and completion of the network operation are now racing. Whoever wins this race will unlink the MD and invalidate the loser's handle. Any further attempts to unlink the same handle will fail. Furthermore, since handles are never re-used, there is no possibility that this handle could now refer to a new valid MD.

(2) Lustre portals PtlMDUnlink() is asynchronous. It "launches" the unlink, but does not guarantee that the unlink has completed before it returns, since the network (i.e. the relevant NAL) might be actively moving data in or out of the memory described by the MD. In this case, the MD is marked for automatic unlinking once the network has finished.

(3) Lustre portals NALs must guarantee that operations that access application memory will complete in finite time. This ensures that an MD cannot remain busy indefinitely.

(4) The application is notified when an MD has been unlinked when it receives an event with its 'unlinked' flag set. If the unlink was performed by PtlMDUnlink() on an idle MD, an UNLINK event is generated. Otherwise, the actual unlink is delayed until all active network operations have completed and the 'unlinked' flag is set in the event generated at this time.

    Another way to think about it is to imagine that lustre portals always creates an UNLINK event when an MD is actually unlinked. However if a data-movement event is created at the same time, lustre portals optimises event delivery by merging the final UNLINK event with the data-movement event. If the MD is idle at the time of unlink (only possible in PtltheirMDUnlink()), the UNLINK event can't be optimised out and it is received by the application. However if the MD is busy, a data-movement event will be generated when the MD becomes idle, and the UNLINK event gets merged into it.

### 40.1.3. Events.

[1] Lustre portals wire handles are unique for all time. This ensures that a new instance of lustre portals, say on a node that reboots, cannot be confused by replies requested by its previous incarnation.

START, END and FAIL events. Lustre portals does not support enabling START events or disabling END events. The only way to disable event delivery in lustre portals is to specify PTL_EQ_NONE when the MD is created. Otherwise, an attempt to create an MD without PTL_MD_EVENT_START_D or with PTL_MD_EVENT_END_DISABLE will result in an assertion failure. This is to leave the application in no doubt as to what is supported. Note that applications that do not handle completion events cannot expect to function correctly in a "real" environment where peer NIDs may crash or shutdown and restart.

Completion Status. The success or failure of the network operation associated with an event may be determined by looking at the event's `ni_fail_type` field, as described in the portals 3.3 spec. FAIL event, which appeared briefly in the portals 3.2 spec are not supported. Lustre portals actually uses the same error codes for reporting completion status in events, as it returns from API functions; i.e. ptl_err_t and ptl_ni_fail_t are the same type, and PTL_NI_OK andPTL_OK are the same value.

"unlinked" flag. As described in section 40.1.2 above, lustre portals events have an "unlinked" flag that allows the final UNLINK event generated by an automatic unlink to be optimised out, and allows PltMDUnlink() to be asynchronous.

"arrival time". Lustre portals events include a "struct timeval arrival_time" field. This records the time at which the event was queued. Applications may use this to gather time-to-service stats.

Callbacks. Lustre portals events support callbacks as described in the 3.3 spec, with the exception that data movement functions (PtlPut(), PltGet()) may *not* be called from these callbacks. Lustre portals serialises callbacks on a particular network interface, with other callbacks and most API functions, therefore calling back into a data-movement function will hang the callback.

Note that in the kernel, lustre portals may call the callback at IRQ priority. An applications should therefore use an IRQ spinlock to synchronise with its threads.

**40.1.4. Scatter/Gather.** Lustre portals supports fragmented data buffers and the PTL_MD_IOVEC option for describing them with struct iovec as described in the 3.3 spec.

In the kernel, it also supports the PTL_MD_KIOV option. This uses the ptl_kiov_t type to describe each buffer fragment with a page, and the offset and size of a contiguous fragment within that page.

```
typedef struct {
        struct page  *kiov_page;
        unsigned int  kiov_len;
        unsigned int  kiov_offset;
} ptl_kiov_t;
```

This option has the great advantage that it allows lustre portals to do I/O on any address space, without having to map buffers into kernel virtual addresses.

Lustre portals supports a maximum message size of at least PTL_MTU, fragmented over at least PTL_MD_MAX_IOV fragments. In the current implementation, this provides for a maximum single message of 0.5MBytes spread over 128 fragments.

#### 40.1.5. Initialisation.

The portals specification expects the application to initialise each portals network interface it uses by calling PtlNIInit(). The interface handle is returned by this call, and effectively the app owns the interface.

Lustre portals implements this somewhat differently, since in the kernel, portals is a shared service and NALs are implemented by kernel modules. The NAL itself calls PtlNIInit() when it loads, and PtlNIFini() when it is unloaded.

```
ptl_handle_ni_t *kportal_get_ni (int nal);
void             kportal_put_ni (int nal);
```

An app wishing to use a particular network interface in lustre portals, calls kportal_get_ni(), with the interface's enum (e.g. QSWNAL, SOCKNAL). This returns the interface's handle and takes a reference on the NAL's module to ensure it remains loaded. When it has finished with the interface, it calls kportal_put_ni().

The NAL itself calls PtlNIInit() when it is loaded and PtlNIFini() when it is unloaded.

```
int  PtlNIInit(ptl_interface_t  interface,
               ptl_pt_index_t   ptl_size,
               ptl_ac_index_t   acl_size,
               ptl_pid_t        requested_pid,
               ptl_handle_ni_t *handle);
void PtlNIFini(ptl_handle_ni_t  handle);
```

The `interface` parameter is actually a callback provided by the NAL. `ptl_size`, `acl_size` and `requested_pid` are simply passed to this callback which typically, will call lib_init() if it is going to use the lib/NAL interface. `handle` should be a pointer to the global ptl_handle_ni_t that the NAL exports. The NAL writer must add a new NAL enum (e.g. QSWNAL, SOCKNAL etc) and add a new case in kportal_get_ni() to "publish" the interface.

## 40.2. Differences at The Lib/NAL interface

This section describes differences between lustre portals and sandia portals at the lib/NAL interface.

#### 40.2.1. Paged I/O.

Lustre portals kernel NALs implement I/O to non-contiguous buffers described by arrays of type ptl_kiov_t (see section 40.1.4 above). The NAL's cb_send_pages() and cb_recv_pages() procedures work completely analagously to cb_send() and cb_receive(). In fact an implementation that simply mapped the given pages and constructed a struct iovec to pass to the "old" routines would be correct, if not the most efficient.[2]

[2] Note that this implementation would also have to guarantee it did not deadlock the "kmem window" somehow.

**40.2.2. Completion with failure.** Lustre portals requires it NAL's implementations of cb_send(), cb_send_pages(), cb_recv() and cb_recv_pages() to return an error code of type ptl_err_t. lib_finalize() takes an additional 'status' parameter, also of type ptl_err_t and is void (i.e. returns no value).

When lustre portals calls into a NAL to send or receive data, and the NAL encounters some problem that means it can't even begin to start receiving or transmitting, the NAL should return immedatiately with a failure code (i.e. anything other than PTL_OK).

Otherwise, if the NAL returns PTL_OK, the lib layer takes that to mean that sending or receiving the message has been initiated successfully. It expects the NAL to call lib_finalize() **in finite time**, when the message has completed and the NAL has stopped accessing the memory associated with the message.

When the NAL calls lib_finalize(), it passes an extra 'status' parameter. This should be PTL_OK if the communication completed successfully, and any other value, if something screwed up (e.g. network error, timeout etc).

Until then, the lib layer assumes the NAL may be accessing the memory associated with the message. Lustre portals **requires** the NAL to commit to calling lib_finalize() **in finite time** to ensure that it can abort communications with crashing peers over crashing networks.

**40.2.3. Message Memory.** Lustre portals passes and additional 'offset' parameter when int calls into a NAL with cb_send(), cb_send_pages(), cb_recv() or cb_recv_pages(). Also, the scatter/gather descriptor passed to them must be treated as READ ONLY.

Portals messages can be sent from or received into a sub-section of an MD. The initial implementation of scatter/gather messages in sandia portals attempted to simplify the NAL interface by always calculating a temporary scatter/gather descriptor for this subset, and passing that down to the NAL. Unfortunately, this compounds memory pressure problems when these temporary descriptors are allocated dynamically.

Lustre portals therefore passes the NAL the actual scatter/gather descriptor belonging to the lib layer's MD and the 'offset' parameter tells the NAL where to start.

Lustre portals exports lib_extract_iov() and lib_extract_kiov() for the NAL's convenience if it still needs a scratch scatter/gather descriptor starting at the given offset. Also, lib_copy_iov2buf(), lib_copy_buf2iov(), lib_copy_kiov2buf() and lib_copy_buf2kiov() have got an 'offset' parameter for compatibility with this usage.

**40.2.4. Optimised GETs.** The "normal" implementation of PtlGet() is for the data source (i.e. the remote node) to return the requested data in a REPLY message. When the REPLY message header is parsed in lib_parse(), the NAL is called back to copy the data from the network into the sink buffers via cb_recv() or cb_recv_pages() as appropriate.

NALs for networks that implement RDMA may be able to optimize the implementation of PtlGet(), by passing appropriate network-specific descriptors for the sink buffers and completion notification in the outgoing GET message. The receiving NAL can then initiate an RDMA that completely bypasses the "normal" portals REPLY message to provide better bandwidth and reduced latency.

```
lib_msg_t *lib_create_reply_msg(nal_cb_t  *nal,
                                ptl_nid_t  peer_nid,
                                lib_msg_t *get_msg);
```

When the NAL sending the GET message decides to optimize the GET, it calls lib_create_reply_msg() with the GET message it was passed, to simulate an incoming REPLY message. This marks the sink buffer as accessible to the network.

When the NAL receives notification from that the RDMA has completed, or gives up on it (and can guarantee no futher network access to the sink buffer), it may call lib_finalize() in the normal way to signal completion.

**40.2.5. Other changes.** Lustre portals' lib_parse() is void (i.e. it returns no value).

The return code for cb_map() and cb_map_pages() is of type of ptl_err_t. PTL_OK means success, anything else means failure.

cb_callback() is optional. If it is NULL, the lustre portals defaults to calling the event queue's callback if it has one.

## 40.3. Multiple Networks

The lustre portals router is a kernel module that provides gateway lookup and message forwarding services to kernel NALs. This allows portals to span different clusters with different cluster fabrics by using nodes with more than one network interface as inter-cluster gateways. It supports redundant gateways to provide higher performance and resilience.

The router presents two different interfaces. The router's NAL interface allows a NAL to connect to the router to participate in message forwarding. The router's control interface is used to manipulate the router's route table and notify it of gateway state. This interface is exported to userspace using ioctls, and is typically driven by the lustre *lctl* utility.

**40.3.1. Route Tables.** The portals router maintains three tables. The route table, the gateway table and the NAL table.

The route table describes the multi-net's topology. It is simply a set of route table entries, where each entry specifies a range of target NIDs, and a gateway NAL and NID. The NIDs in the target range is a set of nodes that can be reached via the route table entry. They must all be direct peers of the gateway. The gateway NAL is the local interface on which to send messages to NIDs in the target range. If the NIDs in the target range are direct peers of the node, the gateway NID is the NID of the gateway NAL. Otherwise, the gateway NID specifies a node that can forward messages to NIDs in the target range.

Arbitrary sets of target nodes, reachable via any particular gateway, can be built up by creating multiple route table entries, each with the same gateway NID and NAL. Redundant gateways can be described using route table entries with overlapping target NID ranges and different gateway NIDs. These can be used for load sharing and failover.

The gateway table is derived from the route table. It contains one entry for each unique gateway. It is where state about each gateway is recorded (e.g. is it running). The NAL table lists NALs that have registered with the router.

When the router is asked to find a gateway for a given destination NID, it simply searches for route table entries that have the following properties...

(1) The destination NID falls in the target NID range.
(2) The gateway NAL is registered and running and
  (a) *is* the requesting NAL, if the node is the source of the message and that NAL will send it, or
  (b) *is not* the requesting NAL, if that NAL has received the message and now wants it forwarded.
(3) The gateway node is believed to be up and running.

Candidate gateways meet all three conditions. If there is only a single candidate, it is the obvious choice. Otherwise the router must make a load-balancing decision.

Route table entries are added, removed and listed using the lustre lctl utility. Please consult the lctl documentation for the specific commands and options. The route table is expected to remain relatively static in the lifetime of any node. It describes all the nodes that can possible be reached and how to reach them. This is moderated by the gateway and NAL tables since they determine which route table entries can actually be used.

**40.3.2. NAL registration.** At startup, NALs register with the portals router with **kpr_register()**. This creates an entry in the router's NAL table with two callbacks. One is used to notify the NAL of failing gateways and the other is used to pass messages to the NAL for forwarding.

Shutdown is a two-stage process, since both the NAL and the router may be actively calling into each other or have such callbacks pending. First, the NAL calls **kpr_shutdown()** to start the process of disnegagement. On return, further calls into the router by other NAL threads, or calls into the NAL by the router will fail immediately. This ensures the router and the NAL cannot generate any more work for each other after shutdown has commenced.

Finally, the NAL calls **kpr_deregister()**. This blocks until all outstanding callbacks have completed before freeing NAL resources in the router.

**40.3.3. Message Forwarding.** A NAL can only deliver messages to peer NIDs on its own network. When a NAL is asked to send a message to a NID that is *not* a peer, it calls **kpr_lookup()** to ask the router if it knows of a suitable gateway NID. The lookup must satisfy condition 2a above, since the application, by its choice of interface, has already decided the sending NAL. [3]

When the message arrives at a gateway, with a (final) destination NID that is *not* the NID of the gateway, the NAL will ask its router to forward it. However, before it can be forwarded, it first has to ensure the complete message has been buffered. Before forwarding, it prepares a router forwarding descriptor with **kpr_fwd_init()** to describe the message buffers as an array of **ptl_kiov_t** page fragments, and a callback to run when forwarding has completed. When the message is ready for forwarding, the NAL calls **kpr_fwd_start()** to ask the router to handle it.

[3]NB the application must choose the correct interface. Consider a source node with NALs on networks A and B. A gateway on network A can reach the destination on network C, but no gateways to network C exists on network B. Attempts to send to a node in network C will succeed if sent via NAL A, but fail if sent via NAL B.

The router now searches for a NAL and gateway to forward the message to. This lookup must satisfy condition 2b above, since the message should have gone direct to the gateway if that gateway was on the same network as the receiving NAL. If the router finds a suitable gateway NID and NAL, it calls the gateway NAL's forwarding callback to actually get it sent. When sending has completed (possibly with failure), the gateway NAL calls **kpr_fwd_done()** to run the callback set up in the forwarding descriptor by the receiving NAL.

Note that the receiving NAL's completion callback will be called, no matter how forwarding was completed or who completed it. For example, if the router can find no gateway, or if the gateway's NAL has not registered with the router, the router signals completion immediately.

**40.3.4. Handling Gateway Failure.** When the lustre portals router first encounters a new gateway (i.e. a new route table entry was created specifying an as yet unknown gateway), it assumes the gateway is up and running. If and when the gateway's NAL has registered, the gateway will become a candidate when a NID is looked up that matches its target NID range.

If the gateway crashes, or becomes unresponsive, the NAL trying to send to it will detect an error or eventually time out. When it does, it notifies the router of peer failure with **kpr_notify()**. The router marks the gateway down, makes an upcall to notify userland, and stops using the gateway until further notice.

The upcall command is set via /proc/sys/portals/upcall, and is "/usr/lib/lustre/portals_upcall" by default. The upcall parameters are...

```
<cmd> ROUTER_NOTIFY <NAL> <NID> <when>
```

...where <cmd> is the upcall command, <NAL> is the NAL type as a decimal integer, <NID> is the gateway NID as a 64 bit hexadecimal integer and <when> is unix "seconds-since-the-epoch" as a decimal integer.

Lustre's **lctl** utility can also notify the router explicitly, not only of gateway failure, but also when a gateway comes back online...

```
lctl [--net <type>] set_route <NID> up|down [<when>]
```

The formats of the <type>, <NID> and optional <when> fields are quite flexible, but compatible with the ROUTER_NOTIFY upcall format. See the lctl documentation for details.

The router is insensitive to redundant notification; it keeps track of the most recent "news" and disregards anything out of date. Provided system clocks are maintained in synch to within a minute or so[4], *anything* noticing gateway failure, including system watchdogs beyond the scope of lustre/portals, can use site-scripts to broadcast the appropriate lctl notification to all portals routers.

[4]or at least less than the time it takes to reboot a gateway.

The ROUTER_NOTIFY upcall can be just one mechanism that triggers this. Such notification ensures early avoidance of the failing gateway and minimise message loss. When the gateway reboots, its own boot scripts should trigger a similar notification, this time of gateway health.

## 40.4. Lustre Compatibility with Cray Portals

There are some importand differences between Cray portals and lustre portals. This section describes how lustre (both liblustre in userspace lustre in the kernel) can inter-operate with Cray portals.

**40.4.1. Scatter/Gather.** Cray portals does not support the PTL_MD_KIOV option for describing fragmented kernel buffers. Instead, lustre passes Cray portals a struct iovec with the physical addresses of each buffer fragment[5], along with PTL_MD_KIOVEC and PTL_MD_PHYS options.

Cray portals does not support fragmented buffers at all on compute nodes. In this case, when lustre build a bulk I/O descriptor, it merges all the page fragments, and passes a single contiguous MD to Cray portals. Note that this requires the readv() and writev() implementations to process each fragment separately in liblustre.

**40.4.2. ptl_event_t::arrival_time.** Cray portals does not support this field. Lustre "fakes up" the arrival time by sampling the time when an event is dequeued, rather than when it is enqueued.

**40.4.3. Event Callbacks.** On compute nodes, lustre's event queue callback is used to call into lustre to make progress on incoming lustre messages when the application is not itself blocking in a filesystem call. When the application does a network op and the Cray runtime notices an outstanding event on the lustre event queue, it simple calls the callback. This "bends the rules" on callbacks not doing PtlEQGet(), since lustre *does* dispose of outstanding events at this time.

**40.4.4. Completion Semantics.** Cray portals mainly supports the 3.2 portals specification. However it provides the special MD option PTL_MD_LUSTRE_COMPLETION_SEMANTICS, if the creator of an MD requires the completion behaviour described in section 40.1.2 above. Lustre uses this option in all MDs it creates. This option is a no-op in lustre portals.

[5]NB this only works on architectures where physical addresses can be represented with a `void *`.